%% file: all.tex
\documentclass[11pt]{cernrep}
\usepackage{mcite}
\usepackage{psfig}
\usepackage{epsfig}
\usepackage{axodraw}
\usepackage{graphicx}
\usepackage{here}
\usepackage{a4}
\usepackage{graphics}
\usepackage{dcolumn}
\usepackage{color}
\usepackage{psfrag}
\usepackage{cite}
\usepackage{rotating}
\usepackage{amssymb}
\usepackage{amsbsy}
\usepackage{colortbl}
\usepackage{multirow}
\usepackage{amsmath}
\usepackage{amsfonts}
\usepackage{epsf,colordvi,helvet,subfigure}
\usepackage{url}
\usepackage{floatflt}
\usepackage{a4wide}

\newcommand{\ssx}{\hspace*{0.6cm}}
\renewcommand{\textfraction}{-0.5}

\begin{document}

\pagestyle{myheadings}
\thispagestyle{empty}

\begin{center}
%\vspace*{.5cm}

{\Large\sc \bf THE HIGGS WORKING GROUP: }

\vspace*{0.3cm}

{\Large\sc \bf Summary Report}

\vspace*{.7cm}

Conveners: \\[0.2cm]
{\sc K.A.\,Assamagan$^1$, M.\,Narain$^2$, A.\,Nikitenko$^3$,
M.\,Spira$^4$ and D.\,Zeppenfeld$^5$}

\vspace*{0.5cm}

 Working Group: \\[0.2cm]
{\sc
J.\,Alwall$^{6}$,
C.\,Bal\'azs$^{7}$,
T.\,Barklow$^{8}$,
U.\,Baur$^{9}$,
C.\,Biscarat$^{10}$,
M.\,Bisset$^{11}$,
E.\,Boos$^{12}$,
G.\,Bozzi$^{13}$,
O.\,Brein$^{14}$,
J.\,Campbell$^{7}$,
S.\,Catani$^{13}$,
M.\,Ciccolini$^{15}$,
K.\,Cranmer$^{16}$,
A.\,Dahlhoff$^{17}$,
S.\,Dawson$^{1}$,
D.\,de Florian$^{18}$,
A.\,De Roeck$^{19}$,
V.\,Del Duca$^{20}$,
S.\,Dittmaier$^{21}$,
A.\,Djouadi$^{22}$,
V.\,Drollinger$^{23}$,
L.\,Dudko$^{12}$,
M.\,D\"uhrssen$^{17}$,
U.\,Ellwanger$^{24}$,
M.\,Escalier$^{25}$,
Y.Q.\,Fang$^{16}$,
S.\,Ferrag$^{25}$,
J.R.\,Forshaw$^{26}$,
M.\,Grazzini$^{19}$,
J.\,Guasch$^{4}$,
M.\,Guchait$^{27}$,
J.F.\,Gunion$^{28}$,
T.\,Hahn$^{21}$,
R.\,Harlander$^{5}$,
H.-J.\,He$^{29}$,
S.\,Heinemeyer$^{19}$,
J.\,Heyninck$^{30}$,
W.\,Hollik$^{21}$,
C.\,Hugonie$^{31}$,
C.\,Jackson$^{32}$,
N.\,Kauer$^{14}$,
N.\,Kersting$^{11}$,
V.\,Khoze$^{33}$,
N.\,Kidonakis$^{34}$,
R.\,Kinnunen$^{35}$,
M.\,Kr\"amer$^{15}$,
Y.-P.\,Kuang$^{11}$,
B.\,Laforge$^{25}$,
S.\,Lehti$^{35}$,
M.\,Lethuillier$^{36}$,
J.\,Li$^{11}$,
H.\,Logan$^{16}$,
S.\,Lowette$^{30}$,
F.\,Maltoni$^{37}$,
R.\,Mazini$^{38}$,
B.\,Mellado$^{16}$,
F.\,Moortgat$^{39}$,
S.\,Moretti$^{40}$,
Z.\,Nagy$^{41}$,
P.\,Nason$^{42}$,
C.\,Oleari$^{43}$,
S.\,Paganis$^{16}$,
S.\,Pe{\~n}aranda$^{21}$,
T.\,Plehn$^{19}$,
W.\,Quayle$^{16}$,
D.\,Rainwater$^{44}$,
J.\,Rathsman$^{6}$,
O.\,Ravat$^{37}$,
L.\,Reina$^{32}$,
A.\,Sabio Vera$^{34}$,
A.\,Sopczak$^{10}$,
Z.\,Tr\'ocs\'anyi$^{45}$,
P.\,Vanlaer$^{46}$,
D.\,Wackeroth$^{9}$,
G.\,Weiglein$^{33}$,
S.\,Willenbrock$^{47}$,
Sau~Lan~Wu$^{16}$,
C.-P.\,Yuan$^{48}$
and
B.\,Zhang$^{11}$. }
\vspace*{.7cm}

{\small
$^1$ Department of Physics, BNL, Upton, NY 11973, USA. \\
$^2$ FNAL, Batavia, IL 60510, USA. \\
$^3$ Imperial College, London, UK. \\
$^4$ Paul Scherrer Institut, CH-5232 Villigen PSI, Switzerland. \\
$^5$ Institut f\"ur Theoretische Teilchenphysik, Universit\"at
Karlsruhe, Germany \\
$^6$ Uppsala University, Sweden \\
$^7$ High Energy Physics Division, Argonne National Laboratory, Argonne,
Il~~60439, USA \\
$^8$ Stanford Linear Accelerator Center, Stanford University, Stanford,
California, USA \\
$^9$ Physics Department, State University of New York at Buffalo, Buffalo,
NY 14260, USA \\
$^{10}$ Lancaster University, UK \\
$^{11}$ Tsinghua University, PR China \\
$^{12}$ Skobeltsyn Institute of Nuclear Physics, MSU, 119992 Moscow, Russia \\
$^{13}$ Florence University and INFN, Florence, Italy \\
$^{14}$ Institut f\"ur Theoretische Physik, RWTH Aachen, Germany \\
$^{15}$ School of Physics, The University~of~Edinburgh, Edinburgh~EH9
3JZ, Scotland \\
$^{16}$ Department of Physics, University of Wisconsin, Madison, WI 53706, USA. \\
$^{17}$ Physics Department, Universit\"at Freiburg, Germany \\
$^{18}$ Universidad de Buenos Aires, Argentina\\
$^{19}$ CERN, 1211 Geneva 23, Switzerland \\
$^{20}$ Istituto Nazionale di Fisica Nucleare, Sezione di Torino, via P.
Giuria 1, 10125 Torino, Italy \\
$^{21}$ Max-Planck-Institut f\"ur Physik (Werner-Heisenberg-Institut),
M\"unchen, Germany \\
$^{22}$ Laboratoire de Physique Math\'ematique et Th\'eorique,
Universit\'e de Montpellier II, France \\
$^{23}$ Dipartimento di Fisica "Galileo Galilei", Universit\`a di
Padova, Italy \\
$^{24}$ LPTHE, Universit\'e de Paris XI, B\^atiment 210, F091405 Orsay
Cedex, France \\
$^{25}$ LPNHE-Paris, IN2P3-CNRS, Paris France \\
$^{26}$ Department of Physics {\&} Astronomy, Manchester University,
Oxford Rd., Manchester M13 9PL, UK \\
$^{27}$ TIFR, Mumbai, India \\
$^{28}$ Department of Physics, U.C. Davis, Davis, CA 95616, USA \\
$^{29}$ University of Texas at Austin, USA\\
$^{30}$ Vrije Universiteit Brussel Inter-University Institute for High
Energies, Belgium \\
$^{31}$ AHEP Group,
I. de F\'isica Corpuscular -- CSIC/Universitat de
Val\`encia, Edificio Institutos de Investigaci\'on,
Apartado de Correos 22085, E-46071, Valencia, Spain \\
$^{32}$ Physics Department, Florida State University, Tallahassee, FL
32306, USA \\
$^{33}$ Department of Physics and Institute for
Particle Physics Phenomenology, University of Durham, DH1 3LE, UK \\
$^{34}$ Cavendish Laboratory, University of Cambridge,
Madingley Road, Cambridge CB3 0HE, UK \\
$^{35}$ Helsinki Institute of Physics, Helsinki, Finland \\
$^{36}$ IPN Lyon, Villeurbanne, France \\
$^{37}$ Centro Studio e ricerche ``Enrico Fermi'', vis
Panisperna,89/A--00184 Rome, Italy \\
$^{38}$ University of Toronto, Canada \\
$^{39}$ Department of Physics, University of Antwerpen, Antwerpen,
Belgium \\
$^{40}$ Southampton University, UK \\
$^{41}$ Institute of Theoretical Science,
5203 University of Oregon, Eugene, OR 97403-5203, USA \\
$^{42}$ INFN, Sezione di Milano, Italy \\
$^{43}$ Dipartimento di Fisica "G. Occhialini", Universit\`a di
Milano-Bicocca, Milano, Italy\\
$^{44}$ DESY, Hamburg, Germany \\
$^{45}$ University of Debrecen and Institute of Nuclear Research of the
Hungarian Academy of Science, Debrecen, Hungary \\
$^{46}$ Universit\'e Libre de Bruxelles
Inter-University Institute for High Energies, Belgium \\
$^{47}$ Department of Physics, University of Illinois at
Urbana-Champaign, Urbana, IL~61801, USA \\
$^{48}$ Michigan State University, USA
}

\vspace*{.5cm}

{\it Report of the HIGGS working group for the Workshop \\[0.1cm]
``Physics at TeV Colliders", Les Houches, France, 26 May -- 6 June 2003.}
\end{center}

%\newpage
%%%%%%%%%%%%%%%%%%%%%%%%%%%%%%%%%%%%%%%%%%%%%%%%%%%%%%%%%%%%%%%%%%%%%%
\vspace*{.5cm}
\begin{center}
{\bf \large CONTENTS}
\end{center}

\vspace*{0.1cm}

\noindent {\bf \ssx Preface} \hfill 4 \\

\noindent {\bf A. Theoretical Developments}
\hfill 5 \\[0.2cm] \hspace*{0.5cm}
C.\,Bal\'azs, U.\,Baur, G.\,Bozzi, O.\,Brein, J.\,Campbell, S.\,Catani,
M.\,Ciccolini, A.\,Dahlhoff, \\ \ssx
S.\,Dawson, D.\,de Florian, A.\,De Roeck, V.\,Del Duca, S.\,Dittmaier,
A.\,Djouadi, V.\,Drollinger, \\ \ssx
M.\,Escalier, S.\,Ferrag, J.R.\,Forshaw, M.\,Grazzini, T.\,Hahn,
R.\,Harlander, H.-J.\,He, S.\,Heinemeyer, \\ \ssx
W.\,Hollik, C.\,Jackson, N.\,Kauer, V.\,Khoze, N.\,Kidonakis, M.\,Kr\"amer,
Y.-P.\,Kuang, B.\,Laforge, \\ \ssx
F.\,Maltoni, R.\,Mazini, Z.\,Nagy, P.\,Nason, C.\,Oleari, T.\,Plehn, D.\,Rainwater,
L.\,Reina, A.\,Sabio Vera,  \\ \ssx
M.\,Spira, Z.\,Tr\'ocs\'anyi, D.\,Wackeroth, G.\,Weiglein, S.\,Willenbrock, 
C.-P.\,Yuan, D.\,Zeppenfeld \\ \ssx
and B.\,Zhang \\

\noindent {\bf B. Higgs Studies at the Tevatron}
\hfill 69 \\[0.2cm] \hspace*{0.5cm}
E.\,Boos, L.\,Dudko, J.\,Alwall, C.\,Biscarat, S.\,Moretti, J.\,Rathsman and
A.\,Sopczak \\

\noindent {\bf C. Extracting Higgs boson couplings from LHC data}
\hfill 75 \\[0.2cm] \hspace*{0.5cm}
M.\,D\"uhrssen, S.\,Heinemeyer, H.\,Logan, D.\,Rainwater, G.\,Weiglein
and D.\,Zeppenfeld \\

\noindent {\bf D. Estimating the Precision of a tan$\beta$ Determination with
H/A~$\to\tau\tau$ and H$^{\pm} \to \tau\nu$ in CMS}
\hfill 84 \\[0.2cm] \hspace*{0.5cm}
R.\,Kinnunen, S.\,Lehti, F.\,Moortgat, A.\,Nikitenko and M.\,Spira \\

\noindent {\bf E. Prospects for Higgs Searches via VBF at the LHC with
the ATLAS Detector}
\hfill 96 \\[0.2cm] \hspace*{0.5cm}
K.\,Cranmer, Y.Q.\,Fang, B.\,Mellado, S.\,Paganis, W.\,Quayle and
Sau~Lan~Wu \\

\noindent {\bf F.  Four-Lepton Signatures at the LHC of heavy neutral MSSM 
Higgs Bosons via \\ Decays into Neutralino/Chargino Pairs}
\hfill 108 \\[0.2cm] \hspace*{0.5cm}
M.\,Bisset, N.\,Kersting, J.\,Li, S.\,Moretti and F.\,Moortgat \\

\noindent {\bf G. The $H\to\gamma\gamma$ in associated production
channel}
\hfill 114 \\[0.2cm] \hspace*{0.5cm}
O.\,Ravat and M.\,Lethuillier \\

\noindent {\bf H. MSSM Higgs Bosons in the Intense-Coupling Regime at the
LHC}
\hfill 117 \\[0.2cm] \hspace*{0.5cm}
E.\,Boos, A.\,Djouadi and A.\,Nikitenko \\

\noindent {\bf I. Charged Higgs Studies}
\hfill 121 \\[0.2cm] \hspace*{0.5cm}
K.A.\,Assamagan, J.\,Guasch, M.\,Guchait, J.\,Heyninck, S.\,Lowette,
S.\,Moretti, \\ \ssx
S.\,Pe{\~n}aranda and P.\,Vanlaer \\

\noindent {\bf J. NMSSM Higgs Discovery at the LHC}
\hfill 138 \\[0.2cm] \hspace*{0.5cm}
U.\,Ellwanger, J.F.\,Gunion, C.\,Hugonie and S.\,Moretti \\

\noindent {\bf K. Higgs Coupling Measurements at a 1 TeV Linear Collider}
\hfill 145 \\[0.2cm] \hspace*{0.5cm}
T.\,Barklow \\

\newpage

%%%%%%%%%%%%%%%%%%%%%%%%%%%%%%%%%%%%%%%%%%%%%%%%%%%%%%%%%%%%%%%%%%%%

\input preface.tex

\newpage

\noindent
{\Large \bf A. Theoretical Developments} \\[0.5cm]
{\it C.\,Bal\'azs, U.\,Baur, G.\,Bozzi, O.\,Brein, J.\,Campbell,
S.\,Catani, M.\,Ciccolini, A.\,Dahlhoff, S.\,Dawson, \\ 
D.\,de Florian,
A.\,De Roeck, V.\,Del Duca, S.\,Dittmaier, A.\,Djouadi, V.\,Drollinger,
M.\,Escalier, S.\,Ferrag, J.R.\,Forshaw, M.\,Grazzini, T.\,Hahn,
R.\,Harlander, H.-J.\,He, S.\,Heinemeyer, W.\,Hollik, C.\,Jackson, \\
N.\,Kauer, V.\,Khoze, N.\,Kidonakis, M.\,Kr\"amer, Y.-P.\,Kuang,
B.\,Laforge, F.\,Maltoni, R.\,Mazini, 
Z.\,Nagy, P.\,Nason, C.\,Oleari, T.\,Plehn, D.\,Rainwater,
L.\,Reina, A.\,Sabio Vera, M.\,Spira, Z.\,Tr\'ocs\'anyi, D.\,Wackeroth,
G.\,Weiglein, S.\,Willenbrock, C.-P.\,Yuan, D.\,Zeppenfeld and B.\,Zhang}

\begin{abstract}
Theoretical progress in Higgs boson production and background processes
is discussed with particular emphasis on QCD corrections at and beyond
next-to-leading order as well as next-to-leading order electroweak
corrections. The residual theoretical uncertainties of the investigated
processes are estimated in detail. Moreover, recent investigations of
the MSSM Higgs sector and other extensions of the SM Higgs sector are
presented. The potential of the LHC and a high-energy linear $e^+e^-$
collider for the measurement of Higgs couplings is analyzed.
\end{abstract}

\input willenbrock.tex
\input grazzini1.tex
\input grazzini2.tex
\input kraemer.tex
\input oleari.tex
\input djouadi.tex
\input baur.tex
\input kidonakis.tex
\input heinemeyer.tex
\input sabio.tex
\input yuan.tex
\input escalier.tex
\input rainwater.tex
\input duca.tex
\input kauer.tex
\input drollinger.tex
\input roeck.tex

\setcounter{section}{0}
\newpage

\noindent
{\Large \bf B. Higgs Studies at the Tevatron} \\[0.0cm]
{\it E.\,Boos, L.\,Dudko, J.\,Alwall, C.\,Biscarat, S.\,Moretti, J.\,Rathsman and
A.\,Sopczak}

\begin{abstract}
An optimal choice of proper kinematical variables is one of the main
steps in using neural networks (NN) in high energy physics.  An
application of an improved method to the Higgs boson search at the
Tevatron leads to an improvement in the NN efficiency by a factor of
1.5-2 in comparison to previous NN studies. \\
The $\rm p\bar p \rightarrow tbH^\pm$ production process with Monte
Carlo simulations in HERWIG and PYTHIA is studied at the Tevatron,
comparing expected cross sections and basic selection variables.
\end{abstract}

\input dudko.tex
\input sopczak.tex

\setcounter{section}{0}
\newpage

\input zeppenfeld.tex

\setcounter{section}{0}
\newpage

\input lehti.tex

\setcounter{section}{0}
\newpage

\input mellado.tex

\setcounter{section}{0}
\newpage

\input bisset.tex

\setcounter{section}{0}
\newpage

\input ravat.tex

\setcounter{section}{0}
\newpage

\input boos.tex

\setcounter{section}{0}
\newpage

\noindent
{\Large \bf I. Charged Higgs Studies} \\[0.0cm]
{\it K.A.\,Assamagan, J.\,Guasch, M.\,Guchait, J.\,Heyninck, 
S.\,Lowette, S.\,Moretti, S.\,Pe{\~n}aranda and P.\,Vanlaer}

\begin{abstract}
The existence of charged Higgs bosons is a central prediction of many
extensions of the Higgs sector. Recent results for the discrimination
between different models are presented. If the charged Higgs mass is
comparable to the top quark mass, previous analyses have to be refined.
The results of this special case are discussed. Finally, the discovery
reach of heavy charged MSSM Higgs bosons is investigated in the $H^+\to
t\bar b$ channel, tagging three $b$-quarks.
\end{abstract}

\input guasch.tex
\input guchait.tex
\input lowette.tex

\setcounter{section}{0}
\newpage

\input gunion.tex

\setcounter{section}{0}
\newpage

\input barklow.tex

\vspace*{1.0cm}

\noindent
{\large \bf Acknowledgements.} \\

M.~Kr\"amer would like to thank the DESY Theory Group for their
hospitality and financial support. This work has been supported in part
by the European Union under contract HPRN-CT-2000-00149. M.~L.~Ciccolini
is partially supported by ORS Award ORS/2001014035. R.~Harlander is
supported by DFG, contract HA 2990/2-1.  This research was supported in
part by the National Science Foundation under grant No.~PHY-0139953.
and by the Hungarian Scientific Research Fund grants OTKA T-038240, by
the US Department of Energy, contract DE-FG0396ER40969, by the European
Community's Human Potential Programme under contract HPRN-CT-2002-00326,
[V.D.], by the INTAS 00-0679, CERN-INTAS 99-377, YSF 02/239 and
Universities of Russia UR.02.03.002 grants and by U.S. Department of
Energy contract DE-AC03-76SF00515.  During this work, C.\,Balazs was
supported by the U.S.  Department of Energy HEP Division under contracts
DE-FG02-97ER41022 and W-31-109-ENG-38, and by LPNHE-Paris.
N.\,Kidonakis' research has been supported by a Marie Curie Fellowship
of the European Community programme ``Improving Human Research
Potential'' under contract number HPMF-CT-2001-01221.  A.\,Sabio Vera
acknowledges the support of PPARC (Postdoctoral Fellowship:
PPA/P/S/1999/00446).  E.~Boos thanks the Humboldt Foundation for the
Bessel Research Award.  J.F.\,Gunion is supported by the U.S. Department
of Energy and the Davis Institute for High Energy Physics. S.\,Moretti
thanks the UK-PPARC and the Royal Society (London, U.K.) for financial
support.  C.\,Hugonie is supported by the European Commission RTN grant
HPRN-CT-2000-00148.  J.F.\,Gunion, C.\,Hugonie and U.\,Ellwanger thank
the France-Berkeley fund for partial support of this research.
A.\,Sopczak thanks the Particle Physics and Astronomy Research Council
for financial support.

\end{document}

%% file: preface.tex
\noindent
{\Large\bf PREFACE} \\

\noindent
This working group has investigated Higgs boson searches at the Tevatron
and the LHC. Once Higgs bosons are found their properties have to be
determined. The prospects of Higgs coupling measurements at the LHC and
a high-energy linear $e^+e^-$ collider are discussed in detail within
the Standard Model and its minimal supersymmetric extension (MSSM).
Recent improvements in the theoretical knowledge of the signal and
background processes are presented and taken into account. The residual
uncertainties are analyzed in detail.

Theoretical progress is discussed in particular for the gluon-fusion
processes $gg\to H (+j)$, Higgs-bremsstrahlung off bottom quarks and the
weak vector-boson-fusion (VBF) processes. Following the list of open
questions of the last Les Houches workshop in 2001 several background
processes have been calculated at next-to-leading order, resulting in a
significant reduction of the theoretical uncertainties. Further
improvements have been achieved for the Higgs sectors of the MSSM and
NMSSM.

This report summarizes our work performed before and after the workshop
in Les Houches. Part A describes the theoretical developments for signal
and background processes. Part B presents recent progress in Higgs boson
searches at the Tevatron collider. Part C addresses the determination of
Higgs boson couplings, part D the measurement of $\tan\beta$ and part E
Higgs boson searches in the VBF processes at the LHC. Part F summarizes
Higgs searches in supersymmetric Higgs decays, part G photonic Higgs
decays in Higgs-strahlung processes at the LHC, while part H
concentrates on MSSM Higgs bosons in the intense-coupling regime at the
LHC. Part I presents progress in charged Higgs studies and part J the
Higgs discovery potential in the NMSSM at the LHC. The last part K
describes Higgs coupling measurements at a 1 TeV linear $e^+e^-$
collider. \\

\noindent
{\bf Acknowledgements.} \\
We thank the organizers of this workshop for the friendly and
stimulating atmosphere during the meeting. We also thank our colleagues
of the QCD/SM and SUSY working group for the very constructive
interactions we had. We are grateful to the ``personnel'' of the Les
Houches school for enabling us to work on physics during day and night
and their warm hospitality during our stay.

%% file: willenbrock.tex
{

\catcode`@=11
\def\citer{\@ifnextchar
[{\@tempswatrue\@citexr}{\@tempswafalse\@citexr[]}}

% \citer as abbreviation for 'citerange' replaces the ',' by a '--'
%

\def\@citexr[#1]#2{\if@filesw\immediate\write\@auxout{\string\citation{#2}}\fi
  \def\@citea{}\@cite{\@for\@citeb:=#2\do
    {\@citea\def\@citea{--\penalty\@m}\@ifundefined
       {b@\@citeb}{{\bf ?}\@warning
       {Citation `\@citeb' on page \thepage \space undefined}}%
\hbox{\csname b@\@citeb\endcsname}}}{#1}}
\catcode`@=12

\newcommand{\lsim}{\raisebox{-0.13cm}{~\shortstack{$<$ \\[-0.07cm] $\sim$}}~}
\newcommand{\gsim}{\raisebox{-0.13cm}{~\shortstack{$>$ \\[-0.07cm] $\sim$}}~}

\section[]{Higgs Boson Production in Association with Bottom
Quarks\footnote{J. Campbell, S. Dawson, S. Dittmaier, C. Jackson, M.
Kr\"amer, F. Maltoni, L.  Reina, M. Spira, D. Wackeroth and S.
Willenbrock}}

\subsection{Introduction}

In the Standard Model, the production of a Higgs boson in association
with $b$ quarks is suppressed by the small size of the Yukawa
coupling, $g_{bbh}=m_b/v\sim 0.02$. However, in a supersymmetric
theory with a large value of $\tan \beta$, the $b$-quark Yukawa
coupling can be strongly enhanced, and Higgs production in association
with $b$ quarks becomes the dominant production
mechanism.

In a four-flavor-number scheme with no $b$ quarks in the initial
state, the lowest order processes are the tree level contributions
$gg\rightarrow b {\overline b} h$ and $q {\overline q}\rightarrow
b {\overline b} h$, illustrated in Fig.~\ref{fg:ggbbh_feyn}.  The
inclusive cross section for $gg\rightarrow b {\overline b} h$
develops potentially large logarithms proportional to $L_b\equiv
\log(Q^2/m_b^2)$ which arise from the splitting of gluons into
$b\bar b$ pairs.\footnote{It should be noted that the $b$ mass in the
argument of the logarithm arises from collinear $b\bar b$
configurations, while the large scale $Q$ stems from $b$ transverse
momenta of this order, up to which factorization is valid.  The scale
$Q$ is the end of the collinear region, which is expected to be of the
order of $M_h/4$~\cite{rsz,msw,bp}.} Since $Q\gg m_b$, the splitting
is intrinsically of ${\cal O}(\alpha_s L_b)$, and because the
logarithm is potentially large, the convergence of the perturbative
expansion may be poor.  The convergence can be improved by summing the
collinear logarithms to all orders in perturbation theory through the
use of $b$ quark parton distributions (the five-flavor-number
scheme)~\cite{dw} at the factorization scale $\mu_F=Q$.  This approach
is based on the approximation that the outgoing $b$ quarks are at
small transverse momentum.  Thus the incoming $b$ partons are given
zero transverse momentum at leading order, and acquire transverse
momentum at higher order.  In the five-flavor-number scheme, the
counting of perturbation theory involves both $\alpha_s$ and $1/L_b$.
In this scheme, the lowest order inclusive process is $b {\overline
b}\rightarrow h$, see Fig.~\ref{fg:bbh_feyn}. The first order
corrections contain the ${\cal O}(\alpha_s)$ corrections to $b
{\overline b}\rightarrow h$ and the tree level process $g b\rightarrow
b h$, see Fig.~\ref{fg:bghb_feyn}, which is suppressed by ${\cal
O}(1/L_b)$ relative to $b {\overline b}\rightarrow h$~\cite{dszw}.  It
is the latter process which imparts transverse momentum to the $b$
quarks.  The relevant production mechanism depends on the final state
being observed.  For inclusive Higgs production it is $b {\overline
b}\rightarrow h$, while if one demands that at least one $b$ quark be
observed at high-$p_T$, the leading partonic process is $gb\rightarrow
b h$.  Finally, if two high-$p_T$ $b$ quarks are required, the leading
subprocess is $g g\rightarrow b {\overline b} h$.

The leading order (LO) predictions for these processes have large
uncertainties due to the strong dependence on the
renormalization/factorization scales and also due to the scheme
dependence of the $b$-quark mass in the Higgs $b$-quark Yukawa
coupling. The scale and scheme dependences are significantly reduced
when higher-order QCD corrections are included.

Section 2 describes the setup for our analysis, and in Section 3 we
compare the LO and NLO QCD results for the production of a Higgs boson
with two high-$p_T$ $b$ jets.  Section 4 contains a discussion of the
production of a Higgs boson plus one high-$p_T$ $b$ jet at NLO,
including a comparison of results within the four-flavor-number and
the five-flavor-number schemes. We consider the corresponding inclusive
Higgs cross sections in Section 5.  Although motivated by the MSSM and
the possibility for enhanced $b$ quark Higgs boson couplings, all
results presented here are for the Standard Model. To a very good
approximation the corresponding MSSM results can be obtained by
rescaling the bottom Yukawa coupling~\cite{dks,djrw}.

\begin{figure}[hbt]
\begin{center}
%\vspace*{0.8cm}
\hspace*{-1.5cm} \epsfxsize=7cm \epsfbox{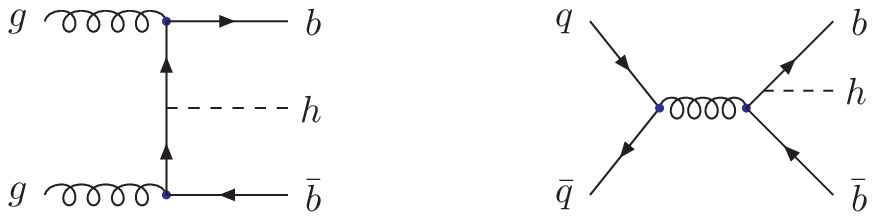}
\caption[ ]{Sample Feynman diagrams for $gg\rightarrow b {\overline b}
h$ and $q {\overline q}\rightarrow b {\overline b} h$ production.}
\label{fg:ggbbh_feyn}
\vspace*{-0.65cm}
\end{center}
\end{figure}

\begin{figure}[hbt]
\begin{center}
%\vspace*{0.8cm}
\hspace*{-1.5cm} \epsfxsize=2.8cm \epsfbox{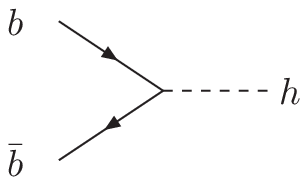}
\caption[ ]{Feynman diagram for $b {\overline b}\rightarrow h$ production. }
\label{fg:bbh_feyn}
\vspace*{-0.65cm}
\end{center}
\end{figure}

\begin{figure}[hbt]
\begin{center}
%\vspace*{0.8cm}
\hspace*{-1.5cm} \epsfxsize=7cm \epsfbox{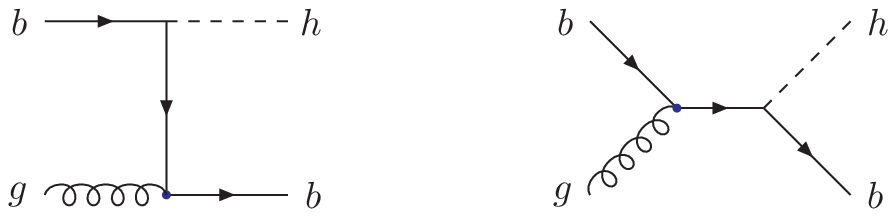}
\caption[ ]{Feynman diagrams for $g b\rightarrow b  h$ production. }
\label{fg:bghb_feyn}
\vspace*{-0.65cm}
\end{center}
\end{figure}

\subsection{Setup}

All results are obtained using the CTEQ6L1 parton distribution
functions (PDFs)~\cite{cteq} for lowest order cross sections and
CTEQ6M PDFs for NLO results.  The top quark is decoupled from the
running of $m_b(\mu)$ and $\alpha_s(\mu)$ and the NLO (LO) cross
sections are evaluated using the $2$~($1$)-loop evolution of
$\alpha_s(\mu)$ with $\alpha_s^{NLO}(M_Z)=0.118$. We use the
${\overline{\rm MS}}$ running $b$ quark mass, $m_b(\mu)$, evaluated at
$2$~($1$)-loop for $\sigma_{NLO}$ ($\sigma_{LO}$), with the $b$ pole
mass taken as $m_b=4.62$~GeV. 
The dependence of the rates on the
renormalization ($\mu_R$) and factorization $(\mu_F)$ scales is
investigated~\cite{dszw,dks,djrw,cemw,hk} in order to
estimate the uncertainty of the predictions for the inclusive Higgs
production channel and for the Higgs plus $1~b$-jet channel.
The dependence of the Higgs plus $2~b$- jet rates on the
renormalization ($\mu_R$) and factorization $(\mu_F)$ scales has been
investigated elsewhere~\cite{dks,djrw} and here we fix
$\mu=\mu_R=\mu_F=(2m_b+M_h)/4$, motivated by the studies in
Refs.~\cite{rsz,msw,bp,dszw,dks,djrw,cemw,hk}.

In order to reproduce the experimental cuts as closely as possible for
the case of Higgs plus 1 or 2 high-$p_T$ $b$ quarks, we require the
final state $b$ and ${\overline b}$ to have a pseudorapidity $\mid
\eta \mid < 2$ for the Tevatron and $\mid \eta \mid < 2.5$ for the LHC.
To better simulate the detector response, the gluon and the
$b/{\overline b}$ quarks are treated as distinct particles only if the
separation in the azimuthal angle-pseudorapidity plane is $\Delta
R>0.4$. For smaller values of $\Delta R$, the four-momentum vectors of
the two particles are combined into an effective $b$/${\overline b}$
quark momentum four-vector.  All results presented in the
four-flavor-number scheme have been obtained independently by two
groups with good agreement~\cite{dks,djrw,dks2,djrw2}.

\subsection{Higgs + 2 \boldmath{$b$} Jet Production}

Requiring two high-$p_T$ bottom quarks in the final state reduces
the signal cross section with respect to that of the zero and one
$b$-tag cases, but it also greatly reduces the background.  It
also ensures that the detected Higgs boson has been radiated off a
$b$ or ${\overline b}$ quark and the corresponding cross section
is therefore unambiguously proportional to the square of the
$b$-quark Yukawa coupling at leading order, while at
next-to-leading order this property is mildly violated by closed
top-quark loops~\cite{dks,djrw}. The parton level processes
relevant at lowest order are $gg\rightarrow b {\overline b} h$ and
$q {\overline q}\rightarrow b {\overline b} h$, as illustrated in
Fig.~\ref{fg:ggbbh_feyn}.  Searches for the neutral MSSM Higgs
bosons $h,H,A$ produced in association with $b$ quarks have been
performed at the Tevatron~\cite{cdf}.

The rate for Higgs plus 2 high-$p_T$ $b$ jets has been computed at NLO
QCD in Refs.~\cite{dks,djrw} and is shown in Fig.~\ref{fg:2b_sigtot}
for both the Tevatron and the LHC. The NLO QCD corrections modify the
LO predictions by $\lsim 30\%$ at the Tevatron and $\lsim 50\%$ at the
LHC. The total cross section plots include a cut on $p_T^{b/{\overline
b}} > 20$ GeV, which has a significant effect on the cross sections.
We show the dependence of the cross section on this cut in
Fig.~\ref{fg:2b_ptcut}. The NLO corrections are negative at large
values of the cut on $p^{b/\bar b}_T$ and tend to be positive at small
values of $p^{b/\bar b}_T$.

\begin{figure}[htb]
\begin{center}
\vspace*{5mm}
\includegraphics[bb=50 250 580 600,scale=0.4]{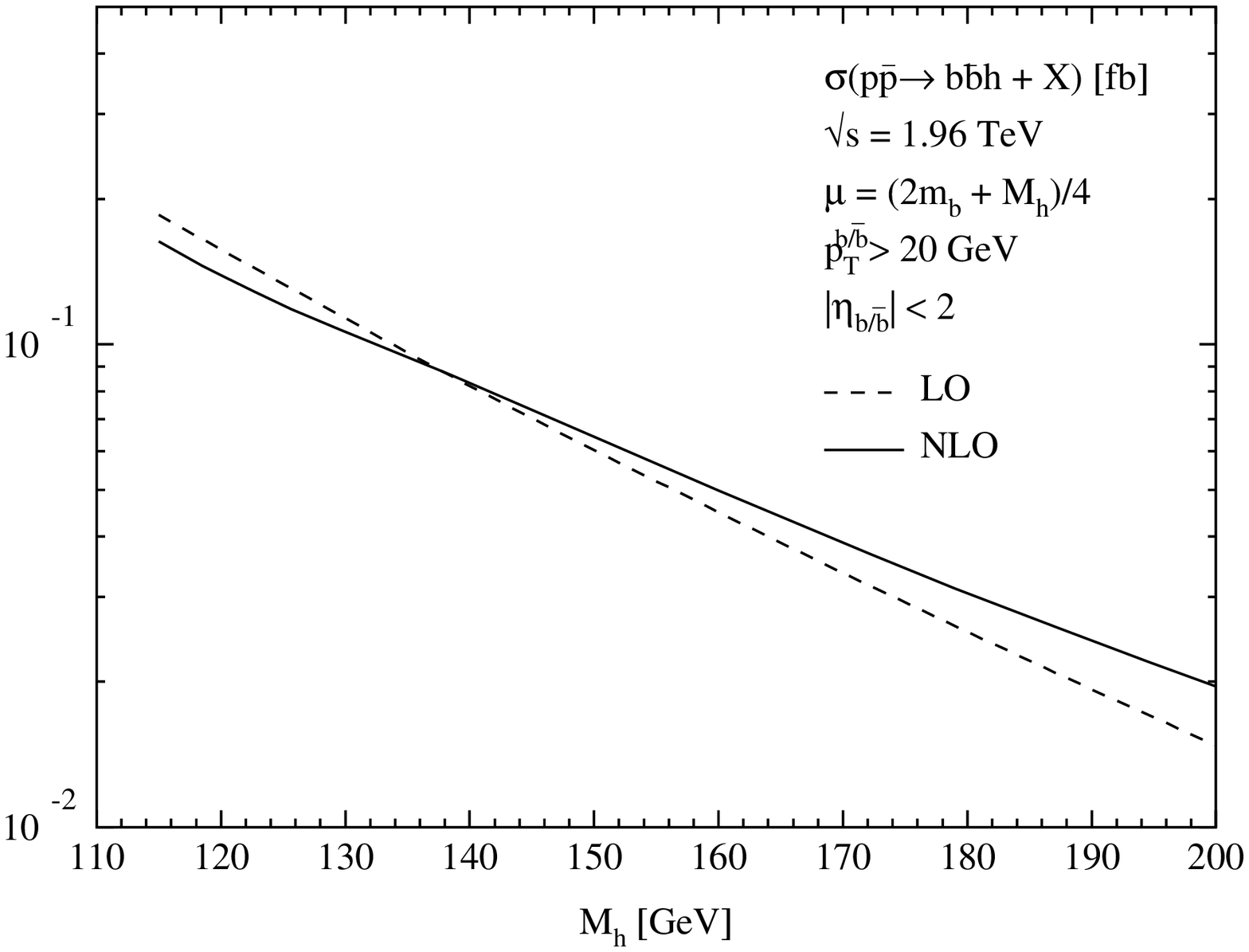}
\includegraphics[bb=50 250 580 600,scale=0.4]{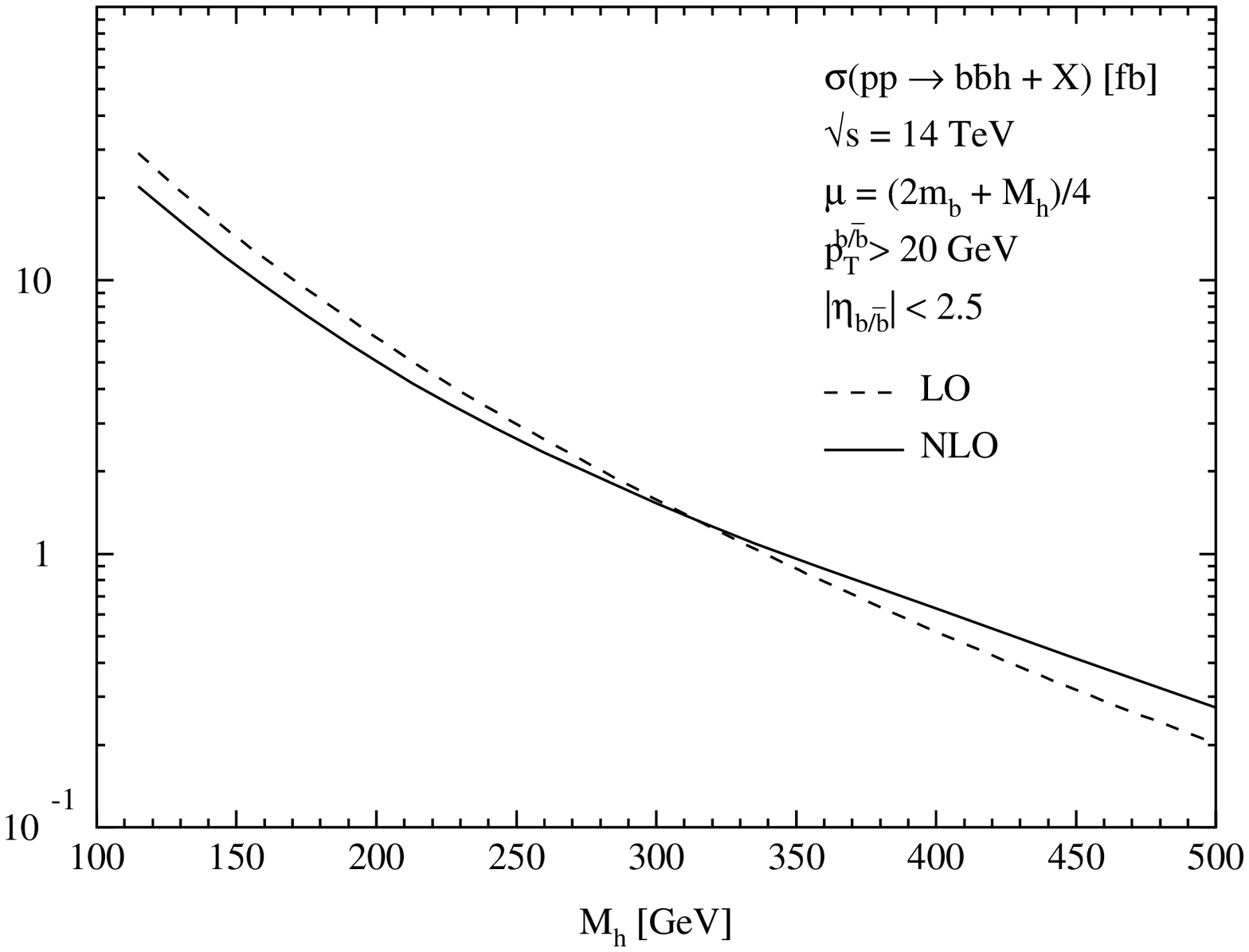}
\caption[]{Total cross sections for $p{\overline p} (pp) \rightarrow b
{\overline b} h+X$ at the Tevatron and the LHC as a function of the
Higgs mass $M_h$ with two high-$p_T$ $b$ jets identified in the final
state.  The $b/\bar b$ quarks are required to satisfy $p_T^{b/\bar
b}>20~GeV$.  We fix $\mu=\mu_R=\mu_F=(2m_b+M_h)/4$.}
\label{fg:2b_sigtot}
\end{center}
\end{figure}
\begin{figure}[htb]
\begin{center}
\includegraphics[bb=50 250 580 600,scale=0.4]{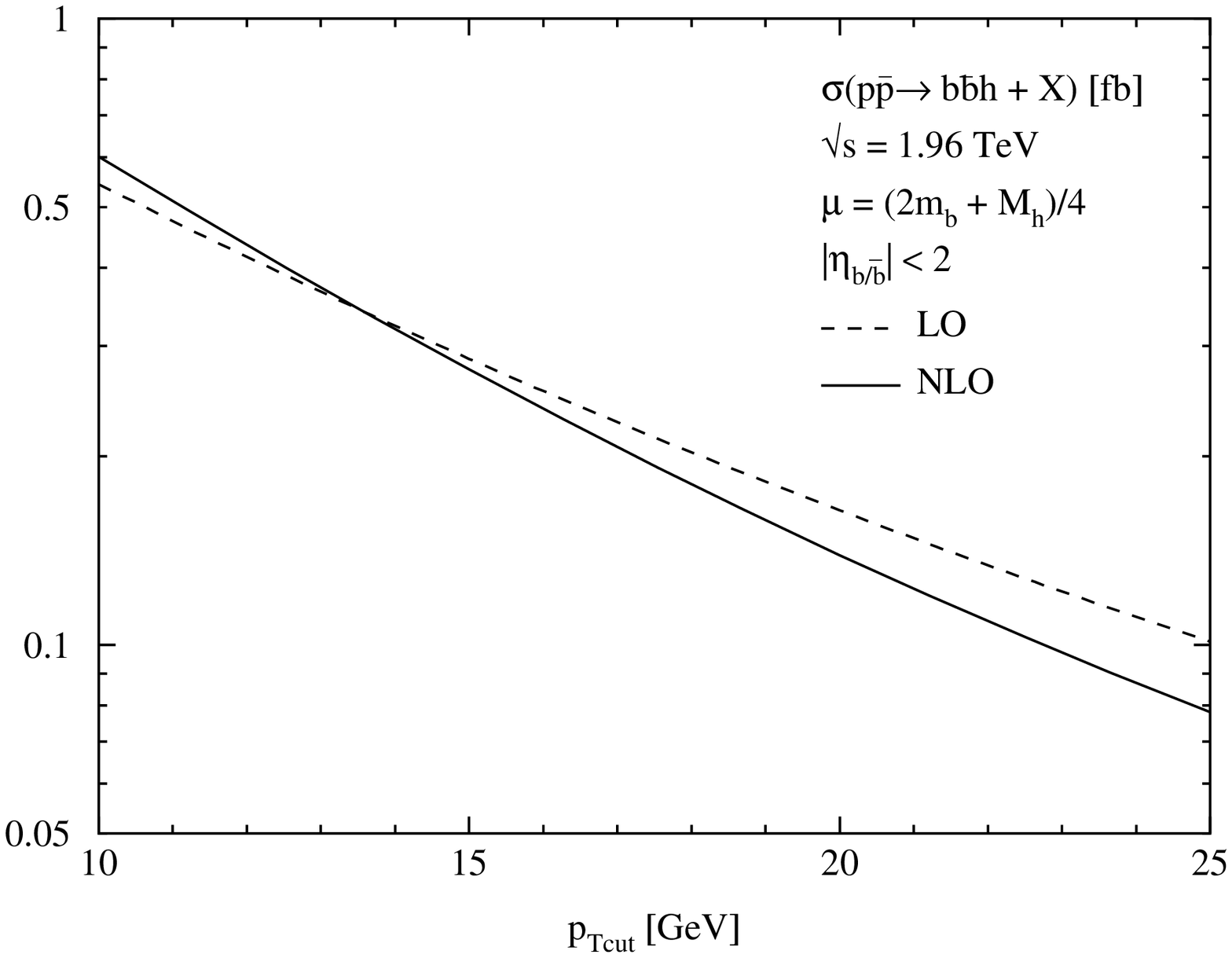}
\includegraphics[bb=50 250 580 600,scale=0.4]{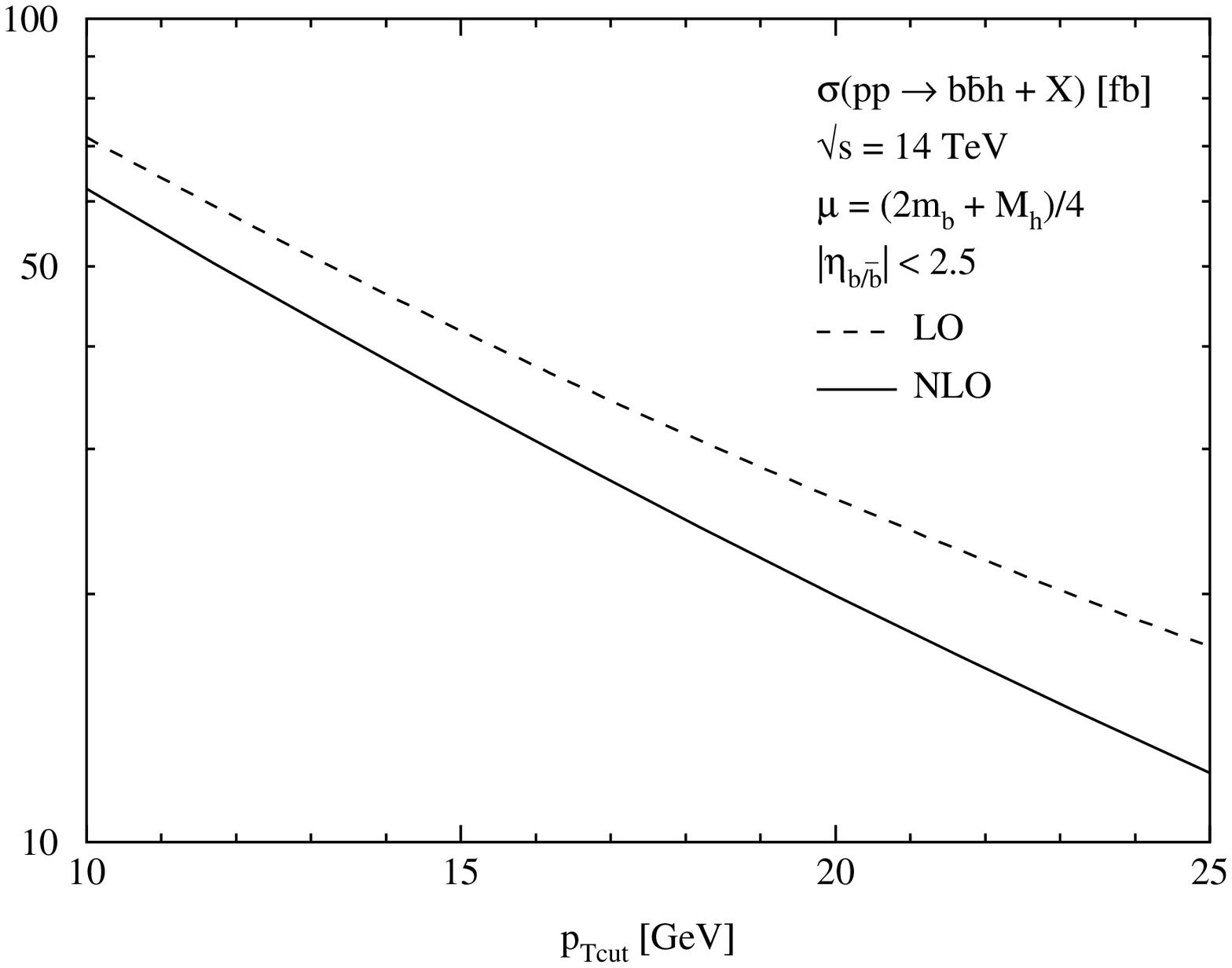}
\caption[]{Total cross sections for $p{\overline p} (pp) \rightarrow b
{\overline b} h+X$ at the Tevatron and the LHC as a function of the
cut $p_{T {\rm cut}}$ in $p_T^{b/\bar b}$ for a Higgs mass $M_h=120$
GeV with two high-$p_T$ $b$ jets identified in the final state.  We
fix $\mu=\mu_R=\mu_F=(2m_b+M_h)/4$.}
\label{fg:2b_ptcut}
\end{center}
\end{figure}

\subsection{Higgs + 1 \boldmath{$b$} Jet Production}

The associated production of a Higgs boson plus a single $b$ quark (or
$\bar b$ quark) is a promising channel for Higgs production in models
with enhanced $b {\overline b} h$ couplings.  The cross section is an
order of magnitude larger than that for Higgs plus 2 high-$p_T$ $b$
jet production for the cuts imposed in our analysis.

In the four-flavor-number scheme, this process has been computed to
NLO, with the momentum of one of the $b$ quarks integrated
over~\cite{dks,dks2,djrw2}.  This integration yields a potentially
large factor $L_b$. Both the total cross sections and the dependence
on the $p_T^{b,{\overline b}}$ cut at the Tevatron and the LHC are
shown in Figs.~\ref{fg:1b_sigma} and \ref{fg:1b_ptcut}. The NLO
corrections increase the cross section by $\lsim 50\%$ at the Tevatron
and $\lsim 80\%$ at the LHC.
The renormalization/factorization scales are
varied around the central 
value $\mu=\mu_R=\mu_F\equiv (2m_b+M_h)/4$.  
At the Tevatron, the upper bands of
the curves for the four-flavor-number scheme
 in Figs.~\ref{fg:1b_sigma} and
\ref{fg:1b_ptcut} correspond to $\mu_R=\mu_F=2\mu$, while
the lower bands correspond to $\mu_R=\mu_F=\mu/2$.
  The scale dependence is more interesting at the
LHC, where the upper bands are obtained with
$\mu_R=\mu/2$ and $\mu_F=2\mu$, while
the lower bands correspond to $\mu_R=2 \mu$ and $\mu_F=\mu/2$.
At both the Tevatron and the LHC, the width of the error band below
the central value ($\mu=\mu_R=\mu_F$) is larger than above.

In the five-flavor-number scheme, the NLO result consists of the
lowest order process, $b g\rightarrow b h$, along with the ${\cal
O}(\alpha_s)$ and ${\cal O}(1/L_b)$ corrections, which are of moderate
size for our scale choices~\cite{cemw}. The potentially large
logarithms $L_b$ arising in the four-flavor-number scheme have been
summed to all orders in perturbation theory by the use of $b$ quark
PDFs.  
In the five-flavor-number scheme, the upper bands of
the curves for the Tevatron  in Figs.~\ref{fg:1b_sigma} and
\ref{fg:1b_ptcut} correspond to $\mu_R=\mu$ and $\mu_F=2\mu$, while
the lower bands correspond to $\mu_R=\mu/2$ and $\mu_F=\mu$.
At the
LHC, the upper bands are obtained with
$\mu_R=\mu$ and $\mu_F=2\mu$, while
the lower bands correspond to $\mu_R=2 \mu$ and $\mu_F=\mu/2$.
 The two approaches agree within their scale uncertainties, but
the five-flavor-number scheme tends to yield larger cross sections as
can be inferred from Figs.~\ref{fg:1b_sigma} and
\ref{fg:1b_ptcut}.

Contributions involving closed top-quark loops
have not been included in the five-flavor-number scheme
calculation of Ref.~\cite{cemw}.  This contribution is negligible
in the MSSM
for large $\tan\beta$.  In the four-flavor scheme, the closed top-quark
loops have been included and in the Standard Model
 reduce the total cross section for the production of a
Higgs boson  plus a single $b$ jet by 
$\sim -7\%$ at the Tevatron and $\sim -13~\%$ at 
the LHC for $M_h=120$~GeV \cite{dks2,djrw2}.  

\begin{figure}[hbt]
\begin{center}
\vspace*{5mm}
\includegraphics[bb=50 250 580 600,scale=0.4]{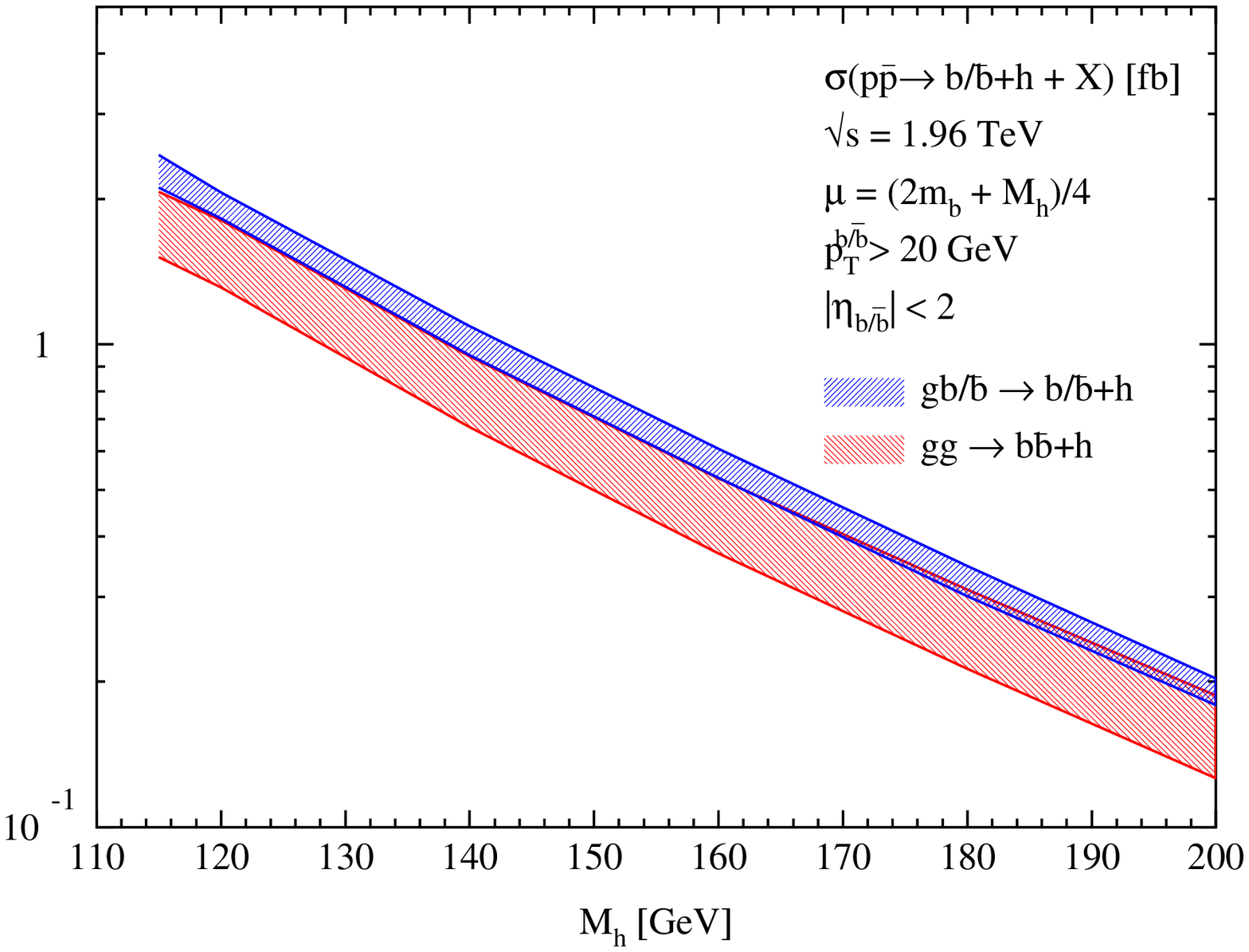}
\includegraphics[bb=50 250 580 600,scale=0.4]{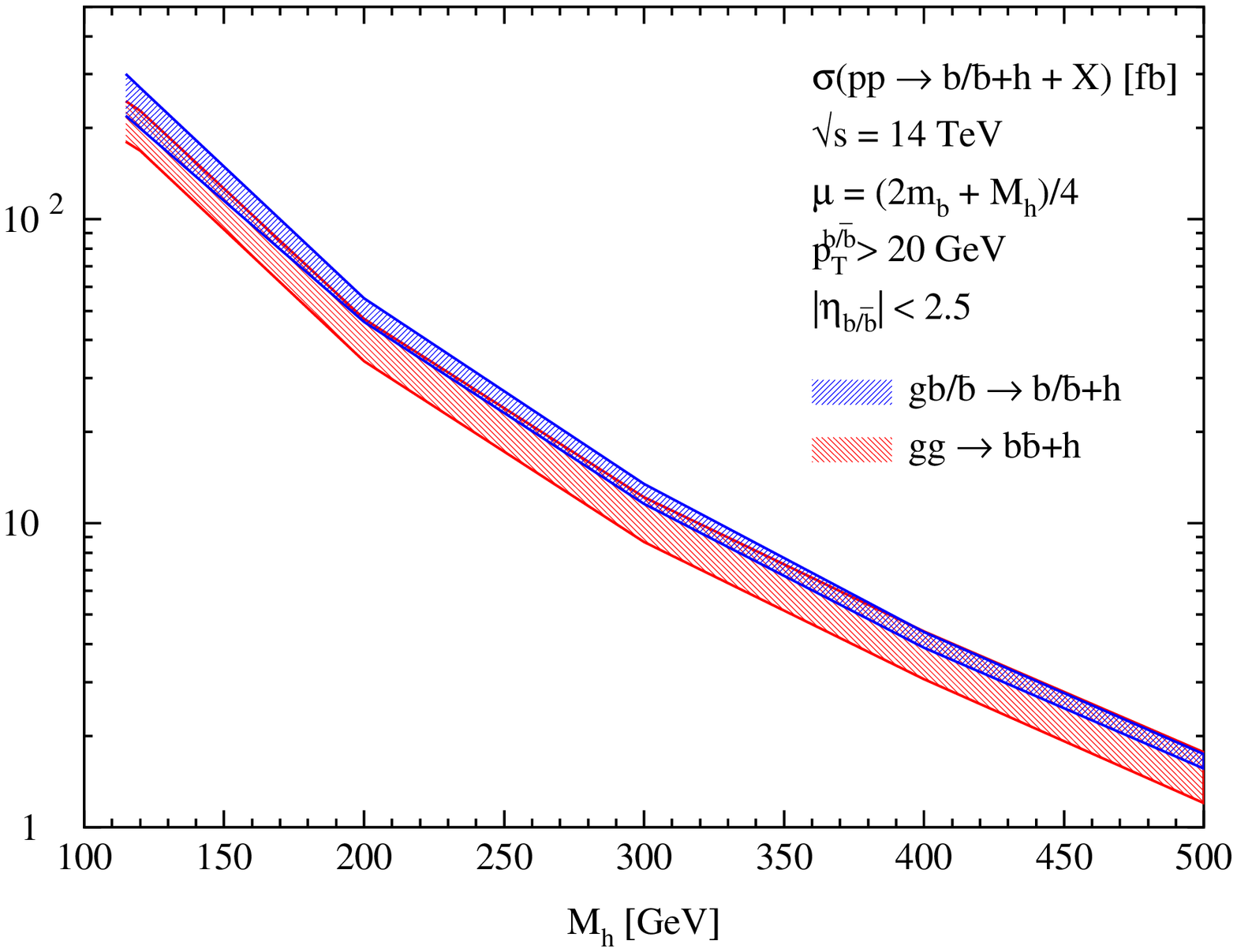}
\caption[]{Total cross sections for $p{\overline p} (pp) \rightarrow b
{\overline b} h+X$ at the Tevatron and the LHC as a function of the
Higgs mass $M_h$ with one high-$p_T$ $b$ jet identified in the final
state.  The $b(\bar b)$ quark is required to satisfy $p_T^{b/\bar
b}>20~\mbox{GeV}$.  We vary the renormalization/factorization
scales around the central 
value $\mu=\mu_R=\mu_F=(2m_b+M_h)/4$ as described in the text.}
\label{fg:1b_sigma}
\end{center}
\end{figure}
\begin{figure}[hbt]
\begin{center}
\includegraphics[bb=50 250 580 600,scale=0.4]{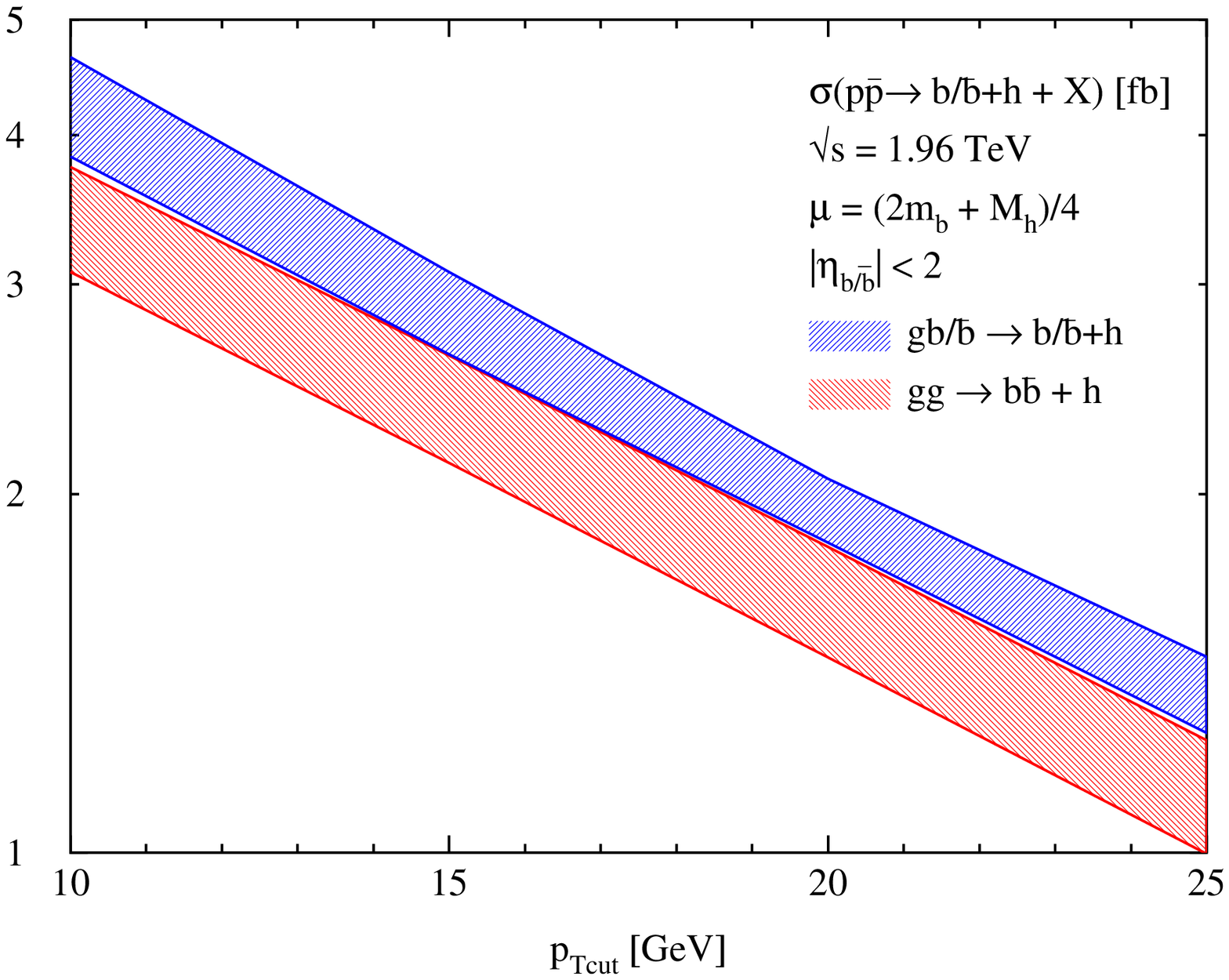}
\includegraphics[bb=50 250 580 600,scale=0.4]{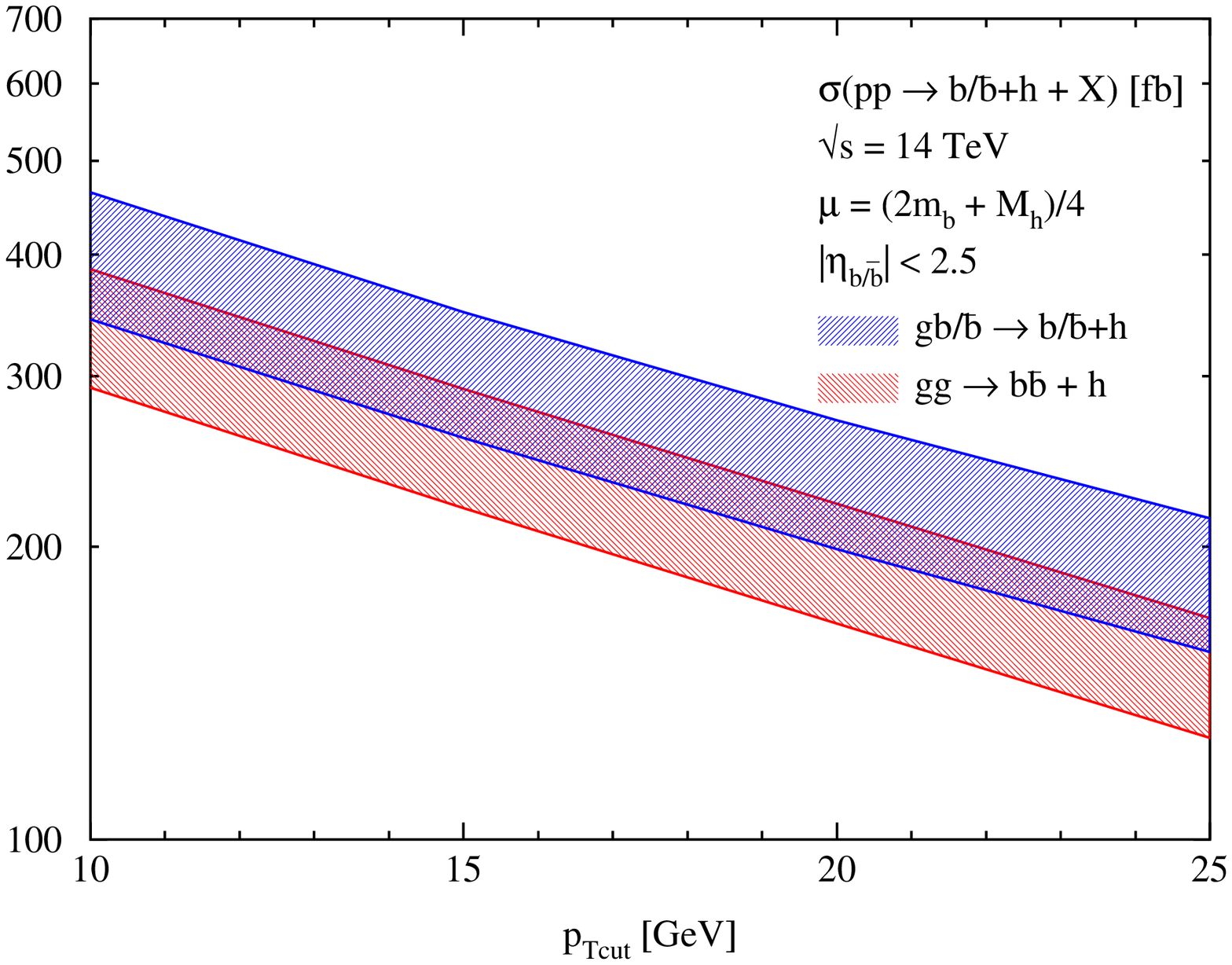}
\caption[]{Total cross sections for $p{\overline p} (pp) \rightarrow b
{\overline b} h+X$ at the Tevatron and the LHC as a function of the cut
$p_{T {\rm cut}}$ in $p_T^{b/\bar b}$ for a Higgs mass $M_h=120$ GeV with one
high-$p_T$ $b$ jet identified in the final state.  
We vary the renormalization/factorization scales around the central 
value $\mu=\mu_R=\mu_F=(2m_b+M_h)/4$ as described in the text.}
\label{fg:1b_ptcut}
\end{center}
\end{figure}

\subsection{Inclusive Higgs Boson Production}

If the outgoing $b$ quarks are not observed, then the dominant process
for Higgs production in the five-flavor-number scheme at large values
of $\tan\beta$ is $b {\overline b} \rightarrow h$.  This final state
contains two spectator $b$ quarks (from the gluon splittings) which
tend to be at low transverse momentum.  At the LHC this state can be
identified through the decays into $\mu^+\mu^-$ and $\tau^+\tau^-$ for
the heavy Higgs bosons $H,A$ at large values of $\tan\beta$ in the
MSSM~\cite{atlascms}.  The $b {\overline b}\rightarrow h$ process has
been computed to NLO~\cite{dszw} and NNLO~\cite{hk} in perturbative
QCD.  The rate depends on the choice of renormalization/factorization
scale $\mu_{R/F}$, and at NLO a significant scale dependence remains.
The scale dependence becomes insignificant at NNLO.  It has been
argued that the appropriate factorization scale choice is
$\mu_F=(M_h+2m_b)/4$~\cite{msw,bp} and it is interesting to note that at
this scale, the NLO and NNLO results nearly coincide~\cite{hk}.

An alternative calculation is based on the processes $gg\to b\bar b h$
and $q\bar q\to b\bar b h$ (four-flavor-number scheme), which has been
calculated at NLO~\cite{dks,dks2,djrw2}. Despite the presence of the
logarithms $L_b$ in the calculation based on $gg\to b\bar bh$, which
are not resummed, it yields a reliable inclusive cross section, as
evidenced by Fig.~\ref{fg:0b_sigma}. A sizeable uncertainty due to the
renormalization and factorization scale dependence remains which might
reflect that the logarithms $L_b$ are not resummed in this approach,
so that the perturbative convergence is worse than in the
corresponding case of $t\bar th$ production~\cite{tth}.
In the Standard Model,
the closed top-quark loops have been included
in the four-flavor-number calculation and
reduce the inclusive NLO total
cross section for $p p (p{\overline p})
\rightarrow b {\overline b} h$
 by $\sim -4\%$ at the Tevatron  and  $\sim -9\%$ at the LHC for
$M_h=120$~GeV \cite{dks2,djrw2}.
In the MSSM, the closed top quark loops are
negligible for large
$\tan\beta$~\cite{dks,djrw}.

The NLO four-flavor-number scheme calculation is compared with the
NNLO calculation of $b\bar b\to h$ (five-flavor-number scheme) in
Fig.~\ref{fg:0b_sigma}.  The two calculations agree within their
respective scale uncertainties for small Higgs masses, while for large
Higgs masses the five-flavor-number scheme tends to yield larger cross
sections. Note that closed top-quark loops have not been included in
the NNLO calculation of $b\bar b\to h$~\cite{hk}.  

To all orders in perturbation theory the four- and five-flavor number
schemes are identical, but the way of ordering the perturbative
expansion is different and the results do not match exactly at finite
order. The quality of the approximations in the two calculational
schemes is difficult to quantify, and the residual uncertainty of the
predictions may not be fully reflected by the scale variation
displayed in Fig.~\ref{fg:0b_sigma}.

\begin{figure}[htb]
\begin{center}
\vspace*{5mm}
\includegraphics[bb=50 250 580 600,scale=0.4]{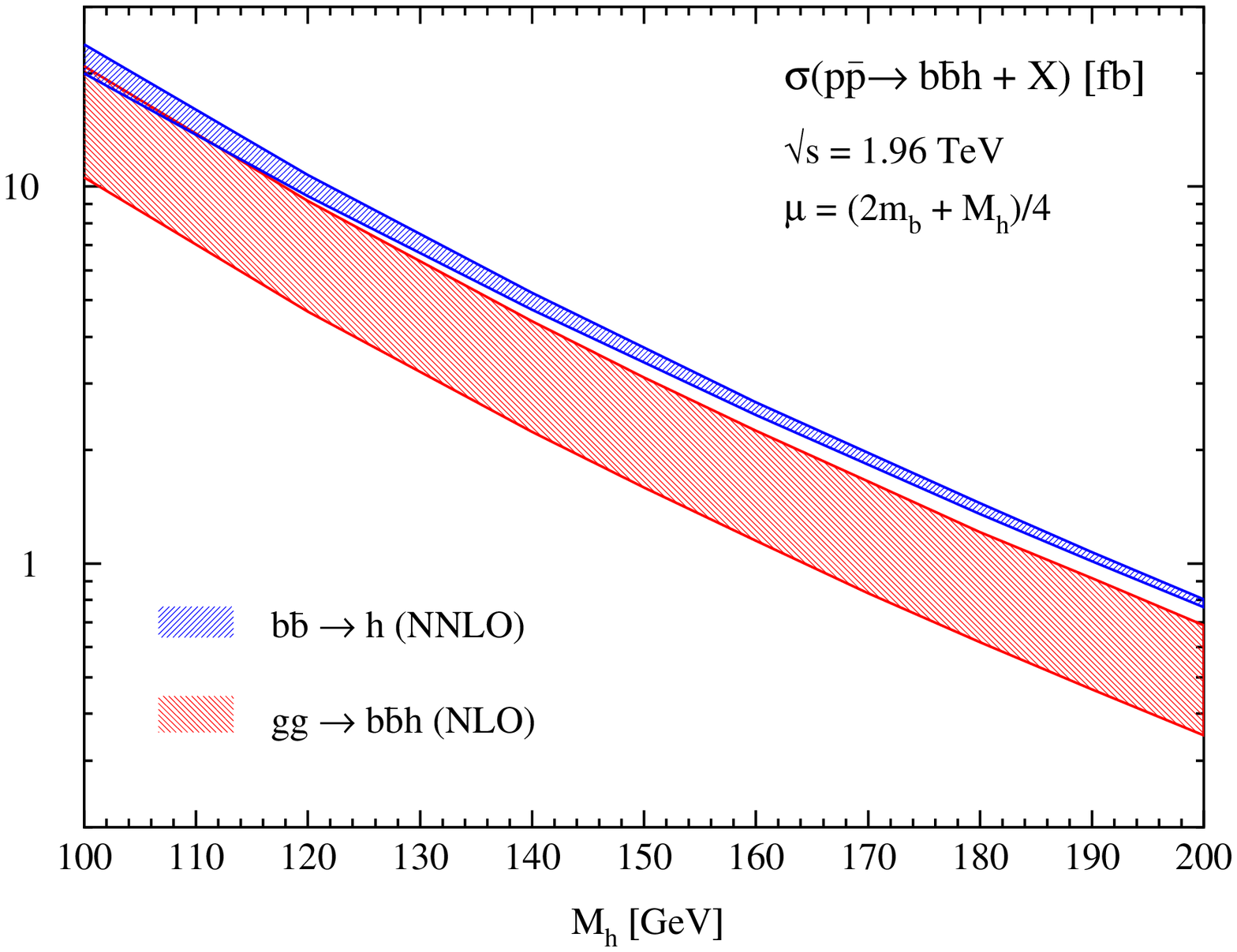}
\includegraphics[bb=50 250 580 600,scale=0.4]{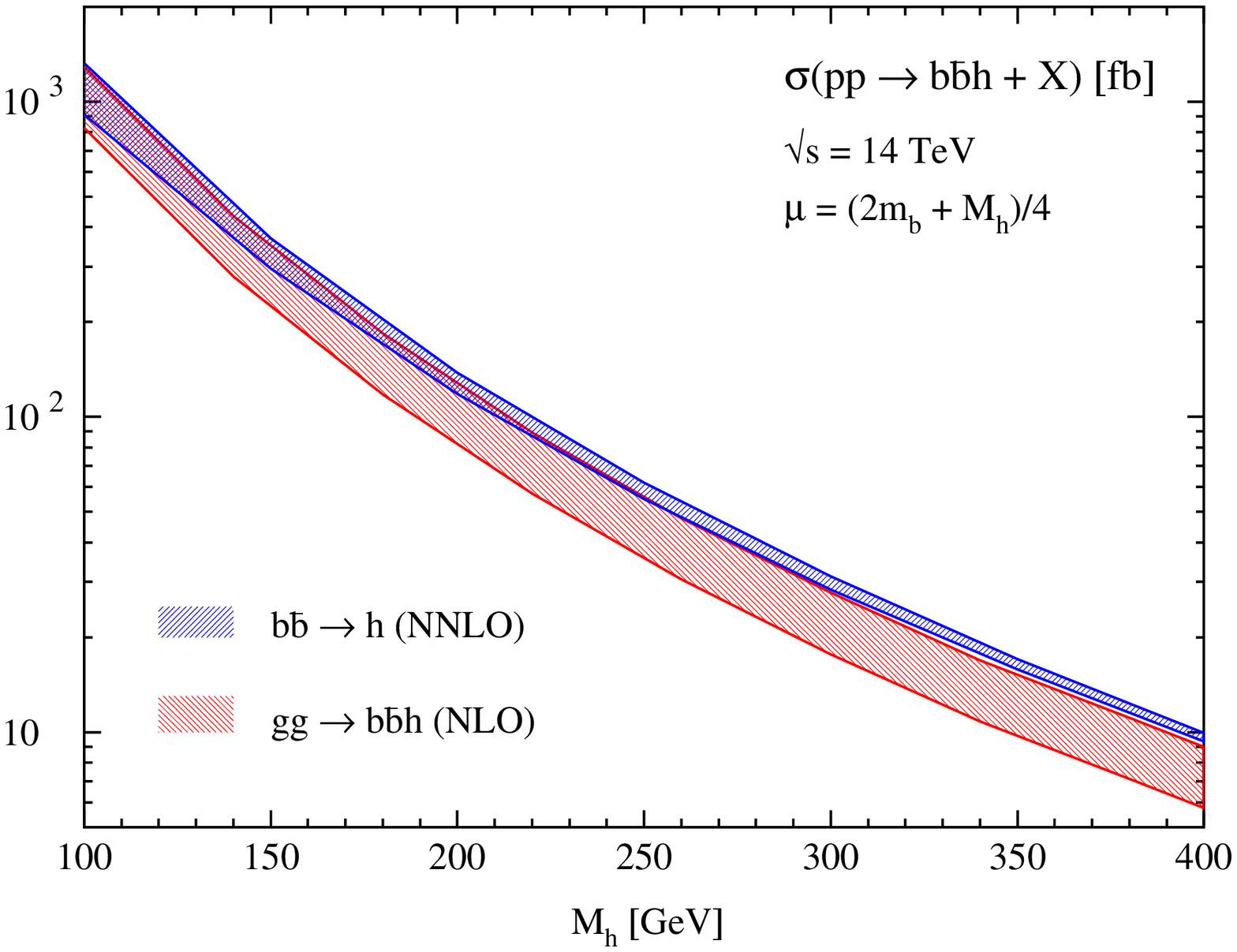}
\caption[]{Total cross sections for $p{\overline p} (pp) \rightarrow b
{\overline b} h+X$ at the Tevatron and the LHC as a function of the
Higgs mass $M_h$ with no $b$ jet identified in the final state.  The
error bands correspond to varying the scale from $\mu_R=\mu_F=(2m_b+M_h)/8$ to
$\mu_R=\mu_F=(2m_b+M_h)/2$.  The NNLO curves are from Ref.~\cite{hk}.}
\label{fg:0b_sigma}
\end{center}
\end{figure}

\subsection{Conclusions}

We investigated $b\bar bh$ production at the Tevatron and the LHC,
which is an important discovery channel for Higgs bosons at large
values of $\tan\beta$ in the MSSM, where the bottom Yukawa coupling is
strongly enhanced~\cite{cdf,atlascms}.  Results for the cross sections
with two tagged $b$ jets have been presented at NLO including
transverse-momentum and pseudorapidity cuts on the $b$ jets which are
close to the experimental requirements. The NLO corrections modify the
predictions by up to $50\%$ and reduce the theoretical uncertainties
significantly.  For the cases of one and no tagged $b$ jet in the
final state we compared the results in the four- and
five-flavor-number schemes.  Due to the smallness of the $b$ quark
mass, large logarithms $L_b$ might arise from phase space integration
in the four-flavor-number scheme, which are resummed in the
five-flavor-number scheme by the introduction of evolved $b$ parton
densities. The five-flavor-number scheme is based on the approximation
that the outgoing $b$ quarks are at small transverse momentum. Thus the
incoming $b$ partons are given zero transverse momentum at leading
order, and acquire transverse momentum at higher order.  The two
calculational schemes represent different perturbative expansions of
the same physical process, and therefore should agree at sufficiently
high order.  It is satisfying that the NLO (and NNLO) calculations
presented here agree within their uncertainties.  This is a major
advance over several years ago, when comparisons of $b\bar b\to h$ at
NLO and $gg\to b\bar bh$ at LO were hardly encouraging~\cite{rsz,tev}.

}

%% file: grazzini1.tex
{
%\setlength{\topmargin}{-1.5 cm} 
%\setlength{\evensidemargin}{.0 cm} 
%\setlength{\oddsidemargin}{-.5 cm} 
%\setlength{\textheight}{24cm} 
%\setlength{\textwidth}{17.5cm} 
%\parskip = 2ex 

%definizione slash 
\def\slash#1{\ooalign{$\hfil/\hfil$\crcr$#1$}} %\esempio: \slash\partial 

\def\ltap{\raisebox{-.4ex}{\rlap{$\,\sim\,$}} \raisebox{.4ex}{$\,<\,$}} 
\def\gtap{\raisebox{-.4ex}{\rlap{$\,\sim\,$}} \raisebox{.4ex}{$\,>\,$}} 
\def\lra{\leftrightarrow} 
\def\naive{na\"{\i}ve} 
\newcommand\as{\alpha_{\mathrm{S}}} 
\newcommand\f[2]{\frac{#1}{#2}} 
\def\dO{{\cal D}_{0}} 
\def\dl{{\cal D}_{1}} 
\def\dll{{\cal D}_{2}} 
\def\dlll{{\cal D}_{3}} 
\def\ee{$e^+e^-$} 
\def\la{\lambda} 
\def\LN{\ln N} 
\def\beq{\begin{equation}} 
\def\eeq{\end{equation}} 
\def\bom#1{{\mbox{\boldmath $#1$}}} 
\def\to{\rightarrow} 
\def\nn{\nonumber} 
\def\arrowlimit#1{\mathrel{\mathop{\longrightarrow}\limits_{#1}}} 
\def\qt{q_{\perp}} 
\def\res{{\rm res.}} 
\def\ms{${\overline {\rm MS}}$} 
\def\msbar{{\overline {\rm MS}}} 
\def\asp{{\alpha_s}\over{\pi}} 
\def\b0{b_0}
\def\bone{b_1}
\def\btwo{b_2}
\def\GE{\gamma_E}

\newcommand{\ccaption}[2]{
    \begin{center}
    \parbox{0.85\textwidth}{
      \caption[#1]{\small{{#2}}}
      }
    \end{center}
    }

\newcommand\mlbl[1]{{\mbox{\footnotesize #1}}} 

\newcommand\add[1]{{\color{blue} $[$#1$]$}}
\newcommand\delete[1]{{\color{red} $<$#1$>$}}
\newcommand\comment[1]{{\color{green} $\{$#1$\}$}}

\section[ ]{The total Cross Section $gg\to H$ at Hadron
Colliders\footnote{S.\,Catani, D.\,de Florian, M.\,Grazzini and P.\,Nason}}

The most important mechanism for Standard Model (SM)
Higgs boson production at hadron colliders
is gluon--gluon fusion through a heavy (top) quark loop \cite{Georgi:1978gs}.
Next-to-leading order (NLO) QCD corrections to this process were found  
%are known 
to be large
\cite{Dawson:1991zj,Djouadi:1991tk,Spira:1995rr}: their effect increases
the leading order (LO) cross section by about 80--100$\%$, 
thus leading to very uncertain
predictions and, possibly, casting
doubts on the reliability of the perturbative QCD expansion. 

Recent years have witnessed a substantial improvement of this situation. 
The NLO corrections are well
approximated \cite{Kramer:1996iq}
by the large-$M_t$ ($M_t$ being the mass of the top quark) limit.
Using this approximation considerably simplifies the evaluation
of higher-order terms, and 
%First,
the calculation of the next-to-next-to-leading order (NNLO) corrections 
has been completed \cite{Harlander:2000mg,Catani:2001ic,Harlander:2001is,
Harlander:2002wh,Anastasiou:2002yz,Ravindran:2003um}.
Moreover,
the logarithmically-enhanced contributions from multiple soft-gluon emission
have been consistently included, up to next-to-next-to-leading
logarithmic (NNLL) accuracy, in the calculation \cite{Catani:2003zt}.
An important point is that the origin of the dominant perturbative 
contributions has been identified and understood:
the bulk of the radiative corrections is due to virtual and soft-gluon
terms. Having those terms under control allows to reliably predict the value 
of the cross section and, more importantly, to reduce the associated 
perturbative 
(i.e. excluding the uncertainty form parton densities)
uncertainty below about $\pm 10\%$~\cite{Catani:2003zt}, 
as discussed below.

In this contribution we present QCD predictions for the total cross section,
including soft-gluon resummation, and we discuss the present theoretical
uncertainties.
Denoting the Higgs boson mass by $M_H$ and the collider centre--of--mass
energy by $\sqrt s$,
the resummed cross section can be written as~\cite{Catani:2003zt}
\vspace*{-3mm}
\begin{equation}
\label{eq}
\sigma^{\rm (res)}(s,M_H^2) = \sigma^{\rm (SV)}(s,M_H^2)
+ \sigma^{\rm (match.)}(s,M_H^2) \;\;,
\end{equation}
where $\sigma^{\rm (SV)}$ contains the virtual and soft-gluon terms, 
and $\sigma^{\rm (match.)}$ includes the remaing hard-radiation terms.
$\sigma^{\rm (SV)}$, which gives the bulk of the QCD radiative corrections
at the Tevatron and the LHC, is obtained through
the resummation of the large logarithmic soft-gluon contributions.  
$\sigma^{\rm (match.)}$ is given by the fixed-order cross section minus
the corresponding fixed-order truncation of $\sigma^{\rm (SV)}$.
%the soft-gluon resummed terms.
The order of magnitude of the relative contribution 
from $\sigma^{\rm (match.)}$ is of ${\cal O}(10\%)$
and of ${\cal O}(1\%)$ at NLO and NNLO, respectively. Therefore, 
%the fixed-order cross section 
$\sigma^{\rm (match.)}$ quantitatively behaves
as naively expected from a power series expansion
whose expansion parameter is $\as \sim 0.1$.
We expect that the presently unknown (beyond NNLO) corrections to  
$\sigma^{\rm (match.)}$ have no practical quantitative impact on
the QCD predictions for Higgs boson production at the Tevatron and the LHC.

The NNLO and NNLL cross sections at the LHC (Tevatron)
are plotted in Fig.~\ref{fig:xslhc} (Fig.~\ref{fig:xstev})
in the mass range $M_H=100$--300~GeV ($M_H=100$--200~GeV).
The central curves are obtained by fixing the factorization ($\mu_F$)
and renormalization ($\mu_R$) scales at the default value
$\mu_F=\mu_R=M_H$. 
The bands are obtained by varying $\mu_F$ and $\mu_R$
simultaneously and independently
in the range $0.5M_H\leq \mu_F,\mu_R\leq 2M_H$ with the constraint
$0.5 \leq \mu_F/\mu_R \leq 2$.
The results in Figs.~\ref{fig:xslhc} and \ref{fig:xstev} are obtained
by using the NNLO densities of the 
MRST2002 \cite{mrst2002} set of parton distributions.
Another NNLO set (set A02 from here on) of parton densities 
has been released in Ref.~\cite{Alekhin:2002fv}.
Tables with detailed numerical values of Higgs boson cross sections
%(obtained by 
(using both MRST2002 and A02 parton densities)
can be found in Ref.~\cite{Catani:2003zt}.
The NNLL cros sections are larger than the NNLO ones;
the increase is of about $6\%$ at the LHC and varies from about 
$12\%$ (when $M_H=100$~GeV) to about $15\%$ (when $M_H=200$~GeV)
at the Tevatron.

%%====================================
\begin{figure}[htb]
\begin{center}
\begin{tabular}{c}
\epsfxsize=12truecm
\epsffile{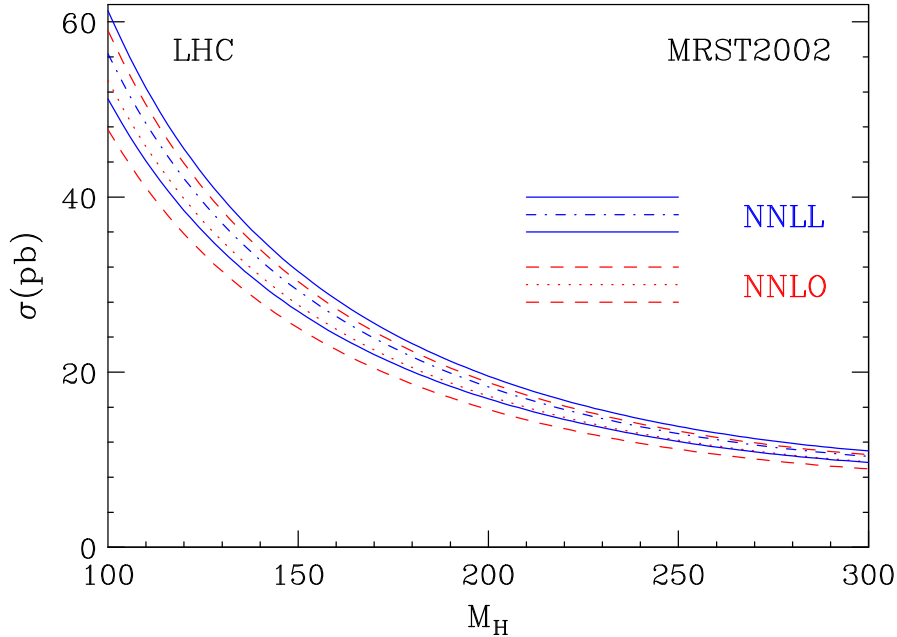}\\
\end{tabular}
\end{center}
\vspace*{-5mm}
\caption{\label{fig:xslhc}{\em NNLL and NNLO cross sections at the LHC,
using MRST2002 parton densities.}}
\end{figure}
%%====================================

%%====================================
\begin{figure}[htb]
\begin{center}
\begin{tabular}{c}
\epsfxsize=12truecm
\epsffile{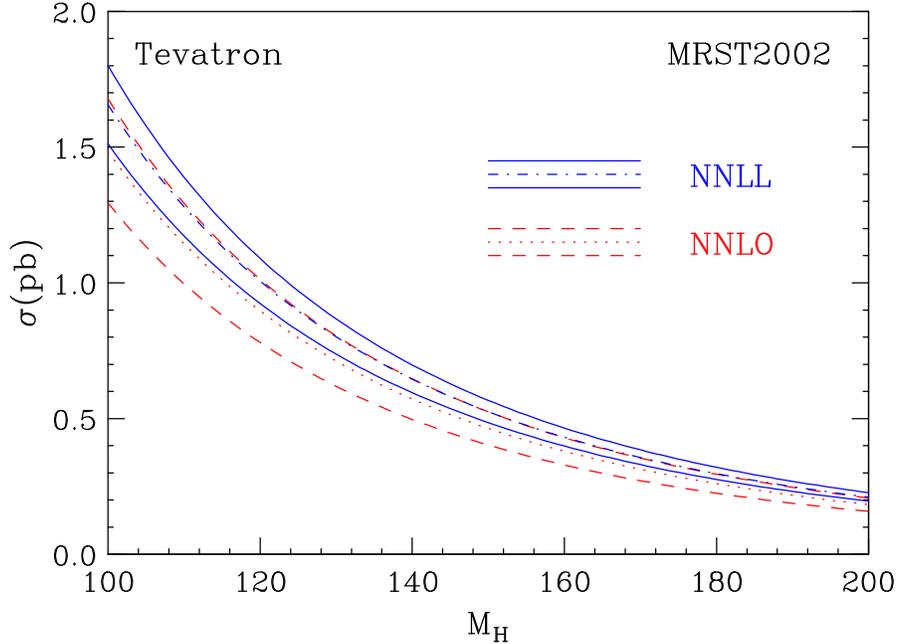}\\
\end{tabular}
\end{center}
\vspace*{-5mm}
\caption{\label{fig:xstev}{\em NNLL and NNLO cross sections at the Tevatron
($\sqrt{s}=1.96$~TeV), using MRST2002 parton densities.}}
\end{figure}
%%====================================

We now would like to summarize the various sources
of uncertainty that
still affect the theoretical prediction of the Higgs production cross section,
focusing on the low-$M_H$ region ($M_H \ltap 200$~GeV).
The uncertainty has basically two origins:
the one originating from still uncalculated 
%unknown
(perturbative) 
radiative corrections,
% QCD contributions,
and the one due to our limited knowledge of the parton distributions.

Uncalculated higher-order QCD contributions
%radiative corrections 
are 
the most important source of uncertainty on the 
radiative corrections.
%coefficients.
A method, which is customarily used in perturbative QCD calculations,
to estimate their size is to vary
the renormalization and factorization scales
around the hard scale $M_H$.
In general, this procedure can only give
a lower limit on the `true'
uncertainty. In fact, the LO and NLO bands do not overlap 
\cite{Catani:2001ic,Catani:2003zt}. 
However, the NLO and NNLO bands and, also,
the NNLO and NNLL bands do overlap.
Furthermore, the central value of the
NNLL bands lies inside the corresponding NNLO bands 
(see Figs.~\ref{fig:xslhc} and \ref{fig:xstev}).
This gives us confidence in using scale
variations to estimate the uncertainty at NNLO and at NNLL order.

Performing scale variations
we find the following results.
At the LHC, the NNLO scale dependence ranges from about $\pm 10 \%$ 
when $M_H=120$~GeV, to about $\pm 9 \%$ when $M_H=200$~GeV.
At NNLL order, it is about $\pm 8\%$ when $M_H\ltap 200$~GeV.
At the Tevatron, when $M_H\ltap 200$~GeV,
the NNLO scale dependence is about $\pm 13 \%$, whereas
the NNLL scale dependence is about $\pm 8\%$.

Another method to estimate the size of higher-order corrections is
to compare the results at the highest order that is available
with those at the previous order. Considering the differences between the
NNLO and NNLL cross sections, we obtain results that are consistent with
the uncertainty estimated from scale variations.

To estimate higher-order contributions, we also investigated the impact 
of collinear terms, which are subdominant with respect to the soft-gluon 
contributions. Performing the resummation of the leading collinear terms,
we found negligible numerical effects \cite{Catani:2003zt}. The uncertainty
coming from these terms can thus be safely neglected.

%We may also consider other possible sources of higher-order uncertainty.
%The uncertainty from the not complete knowledge of the resummation coefficient $A^{(3)}$ can safely be neglected.
%Similar conclusions can be drawn about the inclusion of the dominant collinear
%contributions in the resummed formula \cite{Catani:2003zt}.

A different and relevant
%The other (important) 
source of perturbative QCD uncertainty
comes from the use of the
large-$M_t$ approximation. 
%in the computation of the coefficient function beyond the LO.
The comparison \cite{Kramer:1996iq}
between the exact NLO cross section with the one obtained in
the large-$M_t$ approximation  
(but rescaled with the full Born result, including its
exact dependence on the masses of the top and bottom quarks)
%$M_t$ and $M_b$) 
shows that the approximation works well also for $M_H\gtap M_t$.
This is not accidental. In fact,
the higher-order contributions to the cross section are dominated
by the radiation of soft partons, 
%(i.e. the first term in Eq.~(\ref{eq}),
which is weakly sensitive to mass of the heavy quark in the 
loop at the Born level.
%In other words, as for the size of the QCD radiative
%corrections, what matters is that the heavy-quark mass, $M_t$ is actually
%bigger than the soft-gluon scale, $M_H/N$, rather than the Higgs boson scale $M_H$.
%This feature, i.e. 
The dominance of soft-gluon effects 
persists at NNLO \cite{Catani:2001ic} and it is thus natural to assume that,
having normalized our cross sections with the exact Born result,
the uncertainty ensuing from the large-$M_t$ approximation should be
of order of few per cent for $M_H\ltap 200$~GeV, as it is at NLO.

Besides QCD radiative corrections, electroweak corrections 
have also to be considered.
The ${\cal O}(G_F M_t^2)$  
dominant corrections in the large-$M_t$ limit
have been computed 
and found to give
a very small effect \cite{ewc}.

The other independent and relevant
source of theoretical uncertainty in the cross section
is the one coming from parton distributions. 

The most updated sets of parton distributions are MRST2002 \cite{mrst2002},
A02 \cite{Alekhin:2002fv}
and CTEQ6 \cite{Pumplin:2002vw}.
However, the CTEQ collaboration does not provide a NNLO set,
so that a consistent comparison with MRST2002 and A02 can be performed
only at NLO.
At the LHC, we find that the CTEQ6M results are slightly larger than 
the MRST2002 ones,
the differences decreasing from about $2\%$ at $M_H=100$~GeV to below $1\%$
at $M_H=200$~GeV. The A02 results are instead slightly smaller than the 
MSRT2002 ones,
the difference being below $3\%$ for $M_H\ltap 200$~GeV.
At the Tevatron, CTEQ6 (A02) cross sections are smaller than the MRST2002 ones,
the differences increasing from $6\%$ ($7\%$) to $10\%$ ($9\%$) when $M_H$ 
increases from $100$~GeV to $200$~GeV.
These discrepancies arise because the gluon density (in particular, its
behaviour as a function of the momentum fraction $x$)
is different in the three sets of parton distributions.
The larger discrepancies at the Tevatron
are not unexpected, since here the gluon distribution is
probed at larger values of $x$, where differences between the three sets
are maximal.

All three NLO sets include a study of the effect of the experimental 
uncertainties in the extraction of the parton densities from fits
of hard-scattering data. The ensuing uncertainty on the Higgs cross section
at NLO is studied in Ref.~\cite{Djouadi:2003jg} (note that 
Ref.~\cite{Djouadi:2003jg} uses the MRST2001 set \cite{Martin:2002aw}, 
while we use the MRST2002 set).
The cross section differences that we find at NLO
are compatible with this experimental uncertainty,
which is
about $\pm 3$--$5\%$ (depending on the set) at the LHC
and about $\pm 5$--$15\%$ (in the range $M_H=100$--200~GeV) at the Tevatron.

In summary, the NLO Higgs boson cross section has an uncertainty
from parton densities that is smaller than the perturbative uncertainty,
which (though difficult to quantify with precision) is 
of the order of many tens of per cent.

We now consider the NNLL (and NNLO) cross sections. The available NNLO parton
densities are from the MRST2002 and A02 sets, but only the A02 set
includes an estimate of the corresponding experimental errors.
Computing the effect of these errors on the cross section,
we find \cite{Catani:2003zt} that 
the A02 results have an uncertainty of about $\pm 1.5\%$ at the LHC
and from about $\pm 3\%$ to about $\pm 7\%$ (varying $M_H$ from 100 to 200~GeV)
at the Tevatron.

Comparing the cross sections obtained by using the A02 and MSRT2002 sets,
we find relatively large 
differences \cite{Catani:2003zt} that cannot be
accounted for by the errors provided in the A02 set.
At the LHC, the A02 results are larger than the MRST2002 results,
and the differences go from about $8\%$
at low masses to about $2\%$ at $M_H=200$~GeV.
At the Tevatron, the A02 results are smaller than the MRST2002 results,
with a difference going from about $7\%$ at low $M_H$ to about $14\%$
at $M_H=200$~GeV.

The differences in the cross sections are basically
due to differences in the gluon--gluon luminosity,
which are typically larger than
the estimated uncertainty of experimental origin.
In particular, the differences between the $gg$ luminosities appear to
increase with the perturbative order, i.e. going from LO to NLO and to NNLO
(see also Figs.~13 and 14 in Ref.~\cite{Catani:2003zt}).
We are not able to trace the origin of these differences.
References~\cite{Alekhin:2002fv} and \cite{mrst2002} use the same
(though approximated) NNLO evolution kernels, but
the A02 set is obtained through a fit to deep-inelastic scattering (DIS) 
data only, whereas the MRST2002 set is based on a fit of DIS, Drell--Yan 
and Tevatron jet data (note that not all these observables 
are known to NNLO accuracy). 

The extraction of the parton distributions is also affected by uncertainties 
of theoretical origin, besides those of experimental origin.
These `theoretical' errors are more difficult to quantify. 
Some sources of theoretical errors have recently been investigated by 
the MRST collaboration \cite{Martin:2003sk}, showing that they
can have non neglible effects on the parton densities and, correspondingly,
on the Higgs cross section. At the Tevatron
these effect can be as large as $5\%$,
but they are only about $2\%$ at the LHC.

As mentioned above, the MRST2002 and A02 sets use 
approximated NNLO evolution kernels \cite{vnvogt}, 
which should be sufficiently accurate.
This can be checked as soon the exact NNLO kernels are available
\cite{vermaseren}.

%The recent study of the MRST collaboration \cite{Martin:2003sk}
%shows that different cuts in $x$ and $Q^2$ in the global fit
%to the parton distributions may result in non neglible
%differences in the partons and, correspondingly,
%on the Higgs cross section. At the Tevatron
%this effect can be of ${\cal O}(5\%)$,
%but its only about ${\cal O}(1\%)$ at the LHC.
%It is conceivable that the use of parton densities
%obtained from global fits to
%very different data sets may result in more significant
%discrepancies.

%From the discussion presented so far, 
We conclude
 that the theoretical uncertainties of perturbative origin in the calculation
of the Higgs production cross section, after inclusion of both NNLO corrections
and soft-gluon resummation
at the NNLL level, are below 10\% 
in the low-mass range ($M_H \ltap 200$~GeV). 
%We note that 
This amounts to an improvement in the accuracy of 
almost
%about 
one order of magnitude with respect to the predictions that were available
just few years ago.
%It is nevertheless clear that 
Nonetheless,
there are uncertainties in the
(available) parton densities alone that can reach values larger
than 10\%, and that are not fully understood
at present.

}

%% file: grazzini2.tex
{
%\input{epsf} 
%\setlength{\topmargin}{-1.5 cm} 
%\setlength{\evensidemargin}{.0 cm} 
%\setlength{\oddsidemargin}{-.5 cm} 
%\setlength{\textheight}{24cm} 
%\setlength{\textwidth}{17.5cm} 
%\parskip = 2ex 

%definizione slash 
\def\slash#1{\ooalign{$\hfil/\hfil$\crcr$#1$}} %\esempio: \slash\partial 

\def\ltap{\raisebox{-.4ex}{\rlap{$\,\sim\,$}} \raisebox{.4ex}{$\,<\,$}} 
\def\gtap{\raisebox{-.4ex}{\rlap{$\,\sim\,$}} \raisebox{.4ex}{$\,>\,$}} 
\def\lra{\leftrightarrow} 
\def\naive{na\"{\i}ve} 
\newcommand\as{\alpha_{\mathrm{S}}} 
\newcommand\f[2]{\frac{#1}{#2}} 
\def\dO{{\cal D}_{0}} 
\def\dl{{\cal D}_{1}} 
\def\dll{{\cal D}_{2}} 
\def\dlll{{\cal D}_{3}} 
\def\ee{$e^+e^-$} 
\def\la{\lambda} 
\def\LN{\ln N} 
\def\beq{\begin{equation}} 
\def\eeq{\end{equation}} 
\def\bom#1{{\mbox{\boldmath $#1$}}} 
\def\to{\rightarrow} 
\def\nn{\nonumber} 
\def\arrowlimit#1{\mathrel{\mathop{\longrightarrow}\limits_{#1}}} 
\def\qt{q_{\perp}} 
\def\res{{\rm res.}} 
\def\ms{${\overline {\rm MS}}$} 
\def\msbar{{\overline {\rm MS}}} 
\def\asp{{\alpha_s}\over{\pi}} 
\def\b0{b_0}
\def\bone{b_1}
\def\btwo{b_2}
\def\GE{\gamma_E}

\newcommand{\ccaption}[2]{
    \begin{center}
    \parbox{0.85\textwidth}{
      \caption[#1]{\small{{#2}}}
      }
    \end{center}
    }

\newcommand\mlbl[1]{{\mbox{\footnotesize #1}}} 

\newcommand\add[1]{{\color{blue} $[$#1$]$}}
\newcommand\delete[1]{{\color{red} $<$#1$>$}}
\newcommand\comment[1]{{\color{green} $\{$#1$\}$}}

\section[ ]{The $q_T$ Spectrum of the Higgs Boson at the
LHC\footnote{G.\,Bozzi, S.\,Catani, D.\,de Florian, M.\,Grazzini}}

An accurate theoretical prediction of the transverse-momentum ($q_T$)
distribution of the Higgs boson at the LHC can be important to enhance the 
statistical significance of the signal over the background.
In fact, a comparison of signal and backround $q_T$ spectra may suggest
cuts to improve background rejection \cite{unknown:1999fr,Bayatian:1994pu}.
In what follows we focus on the most relevant production mechanism: 
the gluon--gluon fusion process via a top-quark loop. 

It is convenient to treat separately
the large-$q_T$ and small-$q_T$ regions of the spectrum.
Roughly speaking, the large-$q_T$ region is identified by the condition $q_T \gtap M_H$.
In this region, the perturbative series is controlled by a small expansion
parameter, $\as(M_H^2)$, and a calculation based on the truncation
of the series at a fixed-order in $\as$ is theoretically justified.
The LO calculation ${\cal O}(\as^3)$ was reported in Ref.~\cite{Ellis:1987xu};
it shows that the large-$M_t$ 
approximation ($M_t$ being the mass of the top quark) works well as long
as both $M_H$ and $q_T$ are smaller than $M_t$.
In the framework of this approximation, the NLO QCD corrections
were computed first numerically \cite{deFlorian:1999zd}
and later analytically \cite{Ravindran:2002dc}, \cite{Glosser:2002gm}.

The small-$q_T$ region ($q_T\ll M_H$) is the most important, because
it is here that the bulk of events is expected. In this region
the convergence of the fixed-order expansion is spoiled, since
the coefficients of the perturbative series in $\as(M_H^2)$ are enhanced
by powers of large logarithmic terms, $\ln^m (M_H^2/q_T^2)$. To obtain
reliable perturbative predictions, these terms have 
to be systematically resummed to all orders in $\as$ \cite{Dokshitzer:hw},
(see also the list of references in Sect.~5 of
Ref.~\cite{Catani:2000jh}).
To correctly enforce transverse-momentum conservation,
the resummation has to be carried out in $b$ space, where the impact parameter
$b$ is the variable conjugate to $q_T$ through a Fourier transformation.
In the case of the Higgs boson,
the resummation has been explicitly worked out at
leading logarithmic (LL), next-to-leading logarithmic (NLL) 
\cite{Catani:vd}, \cite{Kauffman:cx}
and next-to-next-to-leading logarithmic (NNLL) \cite{deFlorian:2000pr} level.
The fixed-order and resummed approaches have then
to be consistently matched at intermediate values of $q_T$, to obtain
a prediction which is everywhere as good as the fixed order result, but much
better in the small-$q_T$ region.

In the following
we compute the Higgs boson $q_T$ distribution at the LHC
with the formalism described in Ref.~\cite{Catani:2000vq}.
In particular, we include the
most advanced perturbative information that is available at present:
NNLL resummation at small $q_T$ and NLO calculation at large $q_T$.
An important feature of our formalism is that a unitarity constraint
on the total
cross section is automatically enforced, such that
the integral of the spectrum reproduces the
known results at NLO \cite{Dawson:1991zj,Djouadi:1991tk,Spira:1995rr}
and NNLO \cite{NNLOtotal}.
More details can be found in Ref.~\cite{Bozzi:2003jy}.

Other recent phenomenological predictions can be found in \cite{recent}.

We are going to present quantitative results at NLL+LO and NNLL+NLO
accuracy. 
At NLL+LO (NNLL+NLO) accuracy the NLL (NNLL) resummed result is matched
to the LO (NLO) perturbative calculation valid at large $q_T$. 
As for the evaluation of the fixed order results, the Monte Carlo program 
of Ref.~\cite{deFlorian:1999zd} has been used.
The numerical results are obtained by choosing $M_H=125$~GeV and using 
the MRST2002 set of parton distributions \cite{Martin:2003es}.
They slightly differ from those presented in \cite{Bozzi:2003jy}, 
where we used the MRST2001 set \cite{Martin:2001es}.
At NLL+LO, LO parton densities and 
1-loop $\as$ have been used, whereas at NNLL+NLO
we use NLO parton densities 
and 2-loop $\as$.

%%====================================
\begin{figure}[htb]
\begin{center}
\begin{tabular}{c}
\epsfxsize=12truecm
\epsffile{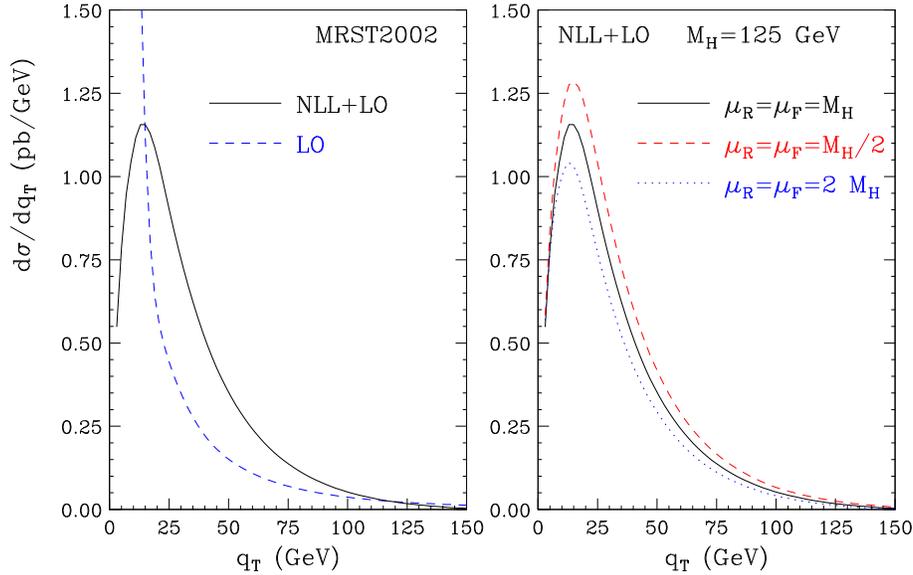}\\
\end{tabular}
\end{center}
\caption{\label{fig1}
{\em 
LHC results at NLL+LO accuracy.}}
\end{figure}
%%====================================

The NLL+LO results at the LHC are shown in Fig.~\ref{fig1}.
In the left panel, the full NLL+LO result (solid line)
is compared with the LO one (dashed line)
at the default scales $\mu_F=\mu_R=M_H$.
We see that the LO calculation diverges to $+\infty$ as $q_T\to 0$. 
The effect of the resummation is relevant below $q_T\sim 100$~GeV.
In the right panel we show the NLL+LO band that is obtained
by varying $\mu_F=\mu_R$ between $1/2 M_H$ and $2M_H$.
The scale dependence increases from about $\pm 10\%$ at the peak
to about $\pm 20\%$ at $q_T=100$~GeV.

%%====================================
\begin{figure}[htb]
\begin{center}
\begin{tabular}{c}
\epsfxsize=12truecm
\epsffile{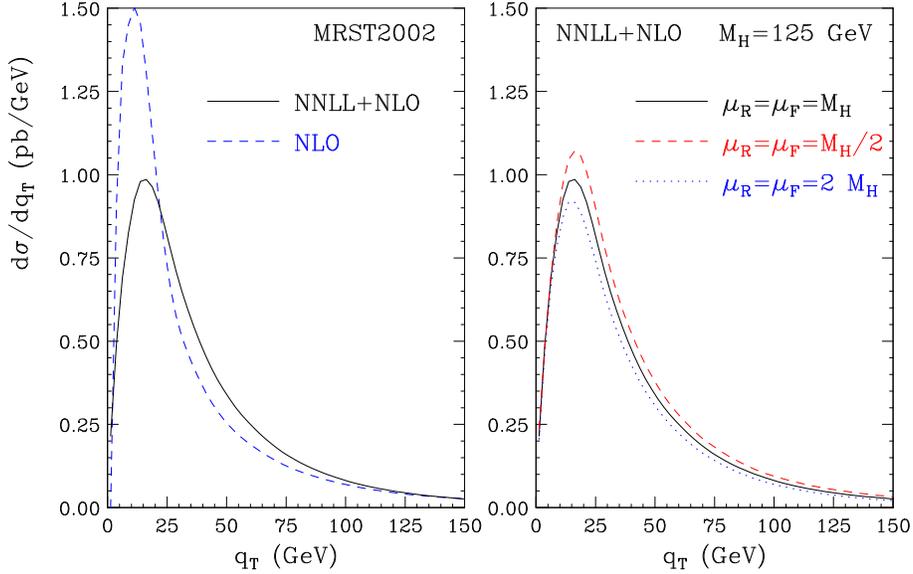}\\
\end{tabular}
\end{center}
\caption{\label{fig2}
{\em 
LHC results at NNLL+NLO accuracy. }}
\end{figure}
%%====================================

The NNLL+NLO results at the LHC are shown in Fig.~\ref{fig2}.
In the left panel, the full result (solid line)
is compared with the NLO one (dashed line) at the
default scales $\mu_F=\mu_R=M_H$.
The NLO result diverges to $-\infty$ as $q_T\to 0$ and, at small values of 
$q_T$, it has an unphysical peak (the top of the peak is close to the vertical
scale of the plot) which is produced by the numerical compensation of negative
leading logarithmic and positive subleading logarithmic contributions.
It is interesting to compare the LO and NLL+LO curves in Fig.~\ref{fig1}
and the NLO curve in Fig.~\ref{fig2}. At $q_T \sim 50$~GeV, the 
$q_T$ distribution sizeably increases when going from LO to NLO and from NLO
to NLL+LO. This implies that in the intermediate-$q_T$ region there are
important contributions that have to be resummed to all orders rather than
simply evaluated at the next perturbative order.
The $q_T$ distribution is (moderately) harder at NNLL+NLO
than at NLL+LO accuracy.
The height of the NNLL peak is a bit lower than the NLL one.
This is mainly due to 
the fact that the total NNLO cross section
(computed with NLO parton densities and 2-loop $\as$),
which fixes the value of the $q_T$ integral of our resummed result,
is slightly smaller than the NLO one, whereas the high-$q_T$ tail is higher at NNLL order,
thus leading to a reduction of the cross section at small $q_T$.
The resummation effect starts to be visible below $q_T\sim 100$~GeV, and 
it increases the NLO result by about $40\%$ at $q_T=50$~GeV.
The right panel of Fig.~\ref{fig2} shows the scale dependence computed as
in Fig.~\ref{fig1}. The scale dependence is now about $\pm 8\%$ at the peak
and increases to $\pm 20\%$ at $q_T=100$~GeV.
Comparing Figs.~1 and 2, we see that the NNLL+NLO band is smaller 
than the NLL+LO one and overlaps with the latter at $q_T \ltap 100$~GeV.
This suggests a good convergence of the resummed perturbative expansion.

The predictions presented so far are obtained in a
purely perturbative framework.
It is known that the transverse momentum distribution
is affected by non-perturbative (NP)
effects, which become important as $q_{T}$ becomes small.
These effects are associated to the large-$b$ region in impact parameter.
In our study the integral over the impact parameter turns out to have
support for $b\ltap 0.1-0.2$ GeV$^{-1}$. We thus do not anticipate particularly
large NP effects in the case of Higgs boson production at the LHC.

The standard way of modelling NP effects
in the case of Drell-Yan (DY) lepton-pair production
is to modify the form factor for
$b\gtap b_{max}$.
There exist several parametrizations 
in literature: in the following
we consider the DSW \cite{DSW}, LY \cite{LY}, and BLNY \cite{BLNY}.
The corresponding coefficients are obtained
from global fits to DY data.
To estimate the size of the NP effects
in the case of Higgs boson production
we define the relative deviation from the purely perturbative resummed result
\begin{equation}
\Delta=\f{d\sigma_{NLL}^{PT}-d\sigma_{NLL}^{PT+NP}}{d\sigma_{NLL}^{PT}}.
\end{equation}
In Fig.~\ref{fig3} we plot $\Delta$ for the DSW, LY and BLNY
parametrizations,
assuming either the same coefficients fitted for DY
(as updated in Ref.~\cite{BLNY}) or rescaling them
with the factor $C_A/C_F$. 
We also test a purely gaussian NP factor of the form
$\exp{(-g b^2)}$, with the coefficient $g$
in the range suggested by the study (KS) of Ref.~\cite{KS}.
We see that the impact of NP effects is below $10\%$ for $q_T\gtap 10$ GeV.

%%====================================
\begin{figure}[htb]
\begin{center}
\begin{tabular}{c}
\epsfxsize=12truecm
\epsffile{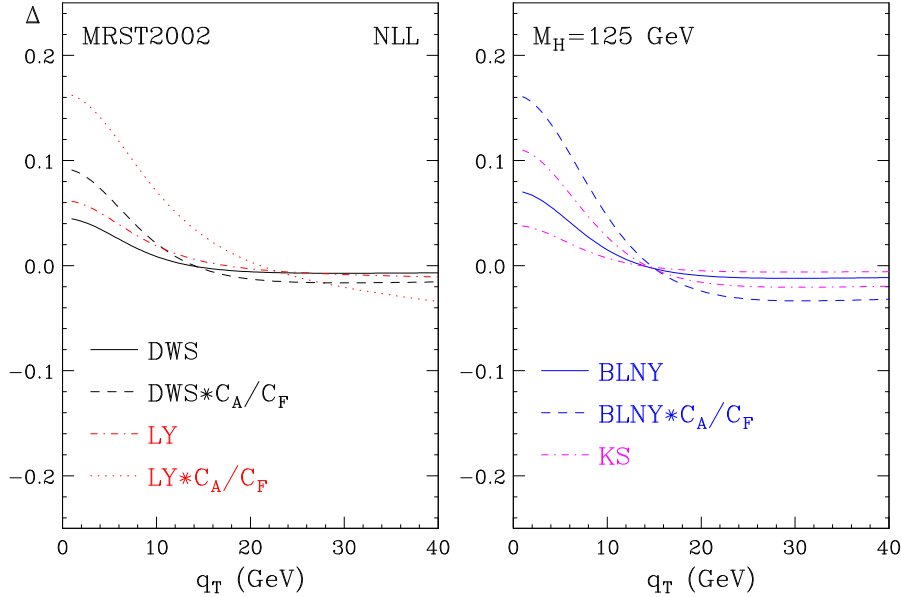}\\
\end{tabular}
\end{center}
\caption{\label{fig3}
{\em Relative size of NP effects at NLL accuracy.}}
\end{figure}
%%====================================

}

%% file: kraemer.tex
{
% some abbreviations
\def\refeq#1{\mbox{(\ref{#1})}}
\def\reffi#1{\mbox{Fig.~\ref{#1}}}
\def\reffis#1{\mbox{Figs.~\ref{#1}}}
\def\refta#1{\mbox{Table~\ref{#1}}}
\def\reftas#1{\mbox{Tables~\ref{#1}}}
\def\refse#1{\mbox{Sect.~\ref{#1}}}
\def\refses#1{\mbox{Sects.~\ref{#1}}}
\def\refapp#1{\mbox{App.~\ref{#1}}}
\def\refapps#1{\mbox{Apps.~\ref{#1}}}
\def\citere#1{\mbox{Ref.~\cite{#1}}}
\def\citeres#1{\mbox{Refs.~\cite{#1}}}
\newcommand{\GF}{\mathswitch {G_\mu}}
\newcommand{\sw}{\mathswitch {s_{\scriptstyle  W}}}
\newcommand{\cw}{\mathswitch {c_{\scriptstyle  W}}}

% particles
\def\mathswitch#1{\relax\ifmmode#1\else$#1$\fi}
\def\mathswitchr#1{\relax\ifmmode{\mathrm{#1}}\else$\mathrm{#1}$\fi}
\newcommand{\PB}{\mathswitchr B}
\newcommand{\PW}{\mathswitchr W}
\newcommand{\PZ}{\mathswitchr Z}
\newcommand{\Pg}{\mathswitchr g}
\newcommand{\PH}{\mathswitchr H}
\newcommand{\Pe}{\mathswitchr e}
\newcommand{\Pne}{\mathswitch \nu_{\mathrm{e}}}
\newcommand{\Pane}{\mathswitch \bar\nu_{\mathrm{e}}}
\newcommand{\Pnmu}{\mathswitch \nu_\mu}
\newcommand{\Pd}{\mathswitchr d}
\newcommand{\Pf}{f}
\newcommand{\Ph}{\mathswitchr h}
\newcommand{\Pl}{l}
\newcommand{\Pu}{\mathswitchr u}
\newcommand{\Ps}{\mathswitchr s}
\newcommand{\Pb}{\mathswitchr b}
\newcommand{\Pc}{\mathswitchr c}
\newcommand{\Pt}{\mathswitchr t}
\newcommand{\Pp}{\mathswitchr p}
\newcommand{\Pq}{\mathswitchr q}
\newcommand{\Pep}{\mathswitchr {e^+}}
\newcommand{\Pem}{\mathswitchr {e^-}}
\newcommand{\Pmum}{\mathswitchr {\mu^-}}
\newcommand{\PWp}{\mathswitchr {W^+}}
\newcommand{\PWm}{\mathswitchr {W^-}}
\newcommand{\PWpm}{\mathswitchr {W^\pm}}

\section[ ]{Precision Calculations for associated 
$WH$ and $ZH$ Production at Hadron Colliders\footnote{O.\,Brein,
M.\,Ciccolini, S.\,Dittmaier, A.\,Djouadi, R.\,Harlander and M.\,Kr\"amer}}

\subsection{Introduction}

At the Tevatron, 
Higgs-boson production in association with $W$ or
$Z$~bosons, $p\bar p \to WH/ZH+X$,
is the most promising discovery channel for a SM Higgs
particle with a mass below about 135 GeV, where decays into
$b\bar{b}$ final states are dominant~\cite{Carena:2000yx,Stange:ya}.
At the $pp$ collider LHC other Higgs-production mechanisms play the leading 
role~\cite{unknown:1999fr,Bayatian:1994pu,Djouadi:2000gu,Cavalli:2002vs}, 
but nevertheless these Higgs-strahlung processes should be observable.

At leading order (LO), the production of a Higgs boson in association
with a vector boson, $p\bar p \to VH+X, (V=W,Z)$ proceeds through
$q\bar{q}$ annihilation~\cite{Glashow:ab}, $q\bar{q}' \to V^* \to VH$.
The next-to-leading order (NLO) QCD corrections coincide with those to
the Drell-Yan process and increase the cross section by about
30\%~\cite{Han:1991ia}.  Beyond NLO, the QCD corrections to $VH$
production differ from those to the Drell-Yan process by contributions
where the Higgs boson couples to a heavy fermion loop.  The impact of
these additional terms is, however, expected to be small in
general~\cite{Dicus:1985wx}.  Moreover, for $ZH$ production the
one-loop-induced process $gg\to ZH$ contributes at
next-to-next-to-leading order (NNLO).  The NNLO corrections
corresponding to the Drell-Yan mechanism as well as the $gg\to ZH$
contribution have been calculated in \citere{Brein:2003wg}.  These
NNLO corrections further increase the cross section by the order of
5--10\%.  Most important, a successive reduction of the
renormalization and factorization scale dependence is observed when
going from LO to NLO to NNLO. The respective scale uncertainties are
about 20\% (10\%), 7\% (5\%), and 3\% (2\%) at the Tevatron (LHC).  At
this level of accuracy, electroweak corrections become significant and
need to be included to further improve the theoretical prediction.  In
\citere{Ciccolini:2003jy} the electroweak ${\cal O}(\alpha)$
corrections have been calculated; they turn out to be negative and
about --5\% or --10\% depending on whether the weak couplings are
derived from $G_\mu$ or $\alpha(M_Z^2)$, respectively.  In this paper
we summarize and combine the results of the NNLO corrections of
\citere{Brein:2003wg} and of the electroweak ${\cal O}(\alpha)$
corrections of \citere{Ciccolini:2003jy}.

The article is organized as follows.
In \refses{se:QCD} and \ref{se:EW} we describe the salient features
of the QCD and electroweak corrections, respectively.
Section~\ref{se:numres} contains explicit numerical results on the
corrected $WH$ and $ZH$ production cross sections, including a
brief discussion of the theoretical uncertainties originating from the
parton distribution functions (PDFs).
Our conclusions are given in \refse{se:concl}

\subsection{QCD Corrections}
\label{se:QCD}

The NNLO corrections, i.e.\ the contributions at ${\cal O}(\alpha_{\rm
  s}^2)$, to the Drell-Yan process $p\bar p/pp \to V^*+X$ consist of
the following set of radiative corrections:
\begin{itemize}
\item 
two-loop corrections to $q\bar{q} \to V^*$, which have to be
multiplied by the Born term,
\item 
one-loop corrections to the processes $qg \to qV^*$ and $q\bar{q}
\to gV^*$, which have to be multiplied by the tree-level $g q$ and $q\bar{q}$
terms,
\item 
tree-level contributions from $q\bar{q}, qq,qg, gg \to V^*+$ 2
partons in all possible ways; the sums of these diagrams for a given initial
and final  state have to be squared and added.
\end{itemize}
These corrections have been calculated a decade ago in 
\citere{Hamberg:1990np} and have recently been updated \cite{Harlander:2002wh}.
They represent a basic building block in the NNLO corrections to
$VH$ production. There are, however, two other sources of 
${\cal O}(\alpha_s^2)$ corrections:
\begin{itemize}
\item 
irreducible two-loop boxes for $q\bar{q}'\to VH$
where the Higgs boson couples via heavy-quark loops to two gluons that
are attached to the $q$ line,
\item 
the gluon--gluon-initiated mechanism $gg \to ZH$ \cite{Barger:1986jt}
at one loop;
it is mediated by closed quark loops which induce $ggZ$ and $ggZH$
couplings and contributes only to $ZH$ but not to $WH$ production.
\end{itemize}
In \citere{Brein:2003wg} the NNLO corrections to $VH$ production
have been calculated from the results \cite{Harlander:2002wh} on
Drell-Yan production and completed by the (recalculated) contribution
of $gg \to ZH$. The two-loop contributions with quark-loop-induced $ggZ$ 
or $ggH$ couplings are expected to be very small and have been neglected.

The impact of higher-order (HO) QCD corrections is usually quantified by
calculating the $K$-factor, which is defined as the ratio between the cross sections
for the process at HO (NLO or NNLO), with the value of $\alpha_{\rm s}$ and the PDFs
evaluated also at HO, and the cross section at LO,  with $\alpha_{\rm s}$
and the PDFs consistently  evaluated also at LO:
$K_{\rm HO}=\sigma_{\rm HO}(p\bar p/pp \to VH+X) / \sigma_{\rm LO}(p\bar p/pp\to VH+X)$.
A $K$-factor for the LO cross section, $K_{\rm LO}$, may also be
defined by  evaluating the latter at given factorization and renormalization
scales and normalizing to the LO cross sections evaluated at the central scale,
which, in our case, is given by $\mu_F=\mu_R=M_{VH}$, where
$M_{VH}$ is the invariant mass of the $VH$ system.

The $K$-factors at NLO and NNLO are shown in \reffi{fig:Kfact} 
(solid black lines)
for the LHC and the Tevatron  as a function of the Higgs  mass
$M_H$ for the process $p\bar p/pp \to WH+X$; they are practically the same 
for the process $p\bar p/pp \to ZH+X$ when the
contribution of the $gg \to ZH$ component is not included. 
Inclusion of this contribution adds substantially to the uncertainty 
of the NNLO prediction for $ZH$ production. This is because
$gg \to ZH$ appears at $\cal O(\alpha_{\rm s}^{\rm 2})$ in LO.

\begin{figure}
\begin{center}
{ \unitlength 1cm
\begin{picture}(15.5,6.0)
\put(-2.2,-5.7){\includegraphics{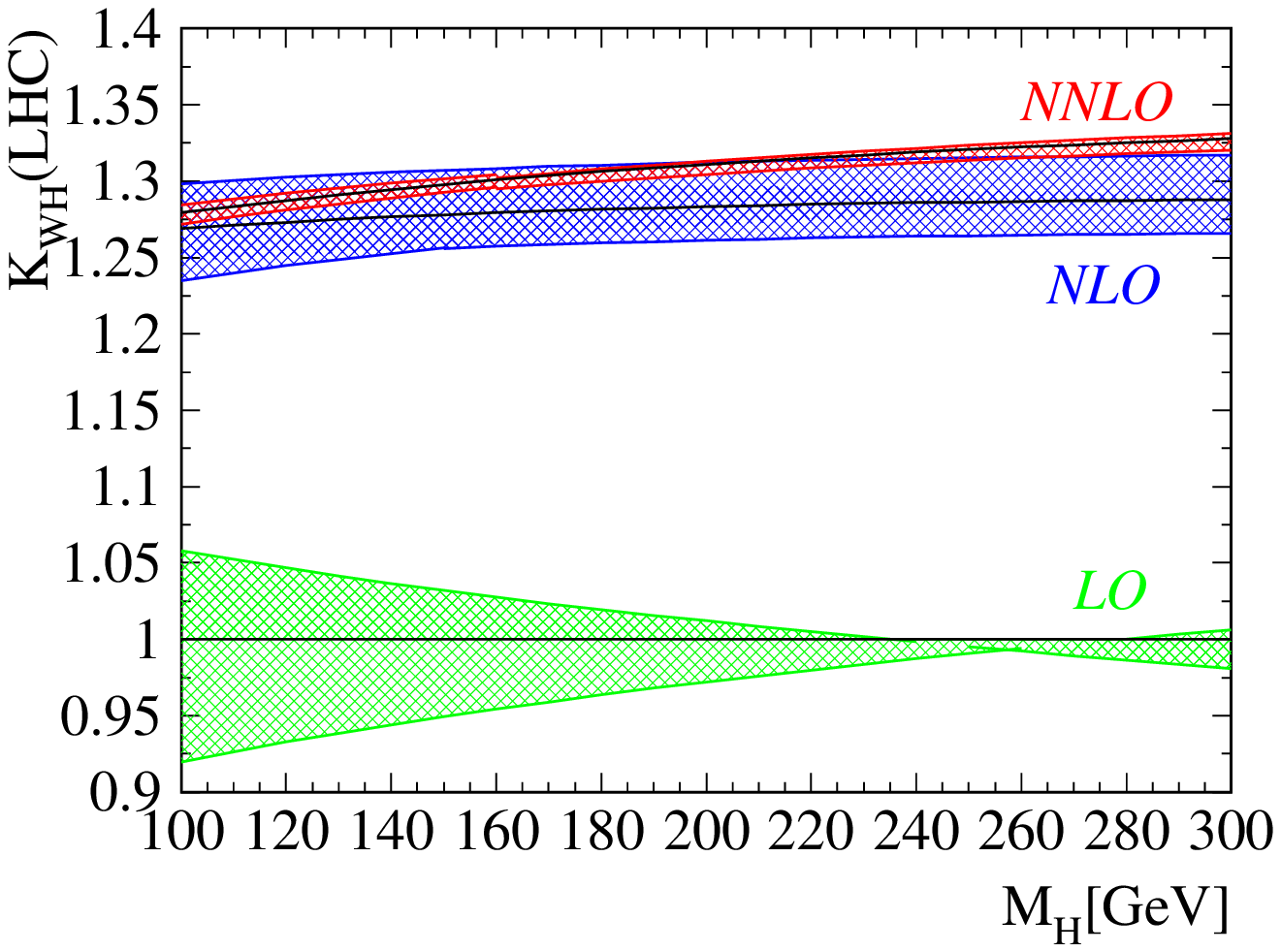}}
\put( 6.0,-5.7){\includegraphics{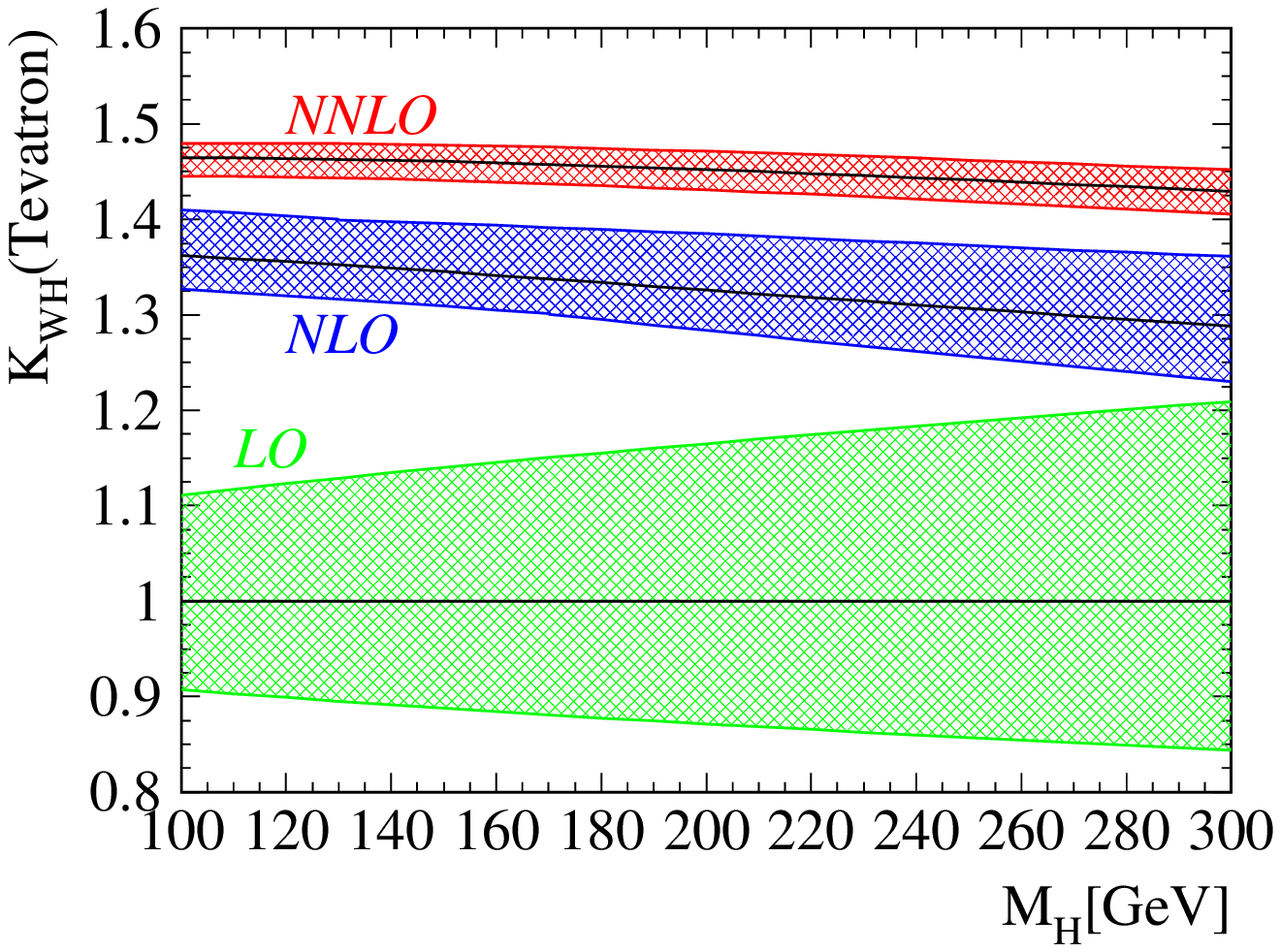}}
\end{picture} }
\caption{QCD $K$-factors for $WH$ production 
(i.e.\ from the sum of $W^+H$ and $W^-H$ cross sections) at the LHC
(l.h.s.)  and the Tevatron (r.h.s.). The bands represent the spread of
the cross section when the renormalization and factorization scales
are varied in the range $\frac{1}{3}M_{VH} \leq \mu_R\, (\mu_F) \leq
3M_{VH}$, the other scale being fixed at $\mu_F (\mu_R)= M_{VH}$.
(Taken from \citere{Brein:2003wg}.)}
\label{fig:Kfact}
\end{center}
\end{figure}
The  scales have
been fixed to $\mu_F=\mu_R=M_{VH}$, and the MRST sets of PDFs for each
perturbative order (including the NNLO PDFs of \citere{Martin:2002dr})
are used in a consistent manner.

The NLO $K$-factor is practically constant at the LHC, increasing only from
$K_{\rm NLO}=1.27$ for $M_H=110$ GeV to $K_{\rm NLO}=1.29$ for $M_H=300$ GeV.
The NNLO contributions increase the $K$-factor by a mere 1\% for the low $M_H$
value and by 3.5\% for the high value. At the Tevatron, the NLO $K$-factor is
somewhat higher than at the LHC, enhancing the cross section between  $K_{\rm
NLO}=1.35$ for $M_H=110$ GeV and $K_{\rm NLO}=1.3$ for $M_H=300$ GeV with a
monotonic decrease.  The NNLO corrections increase the $K$-factor uniformly by
about 10\%. Thus, these NNLO corrections are more important at the Tevatron
than at the LHC. 

\begin{figure}[thb]
\begin{center}
\mbox{
\epsfig{file=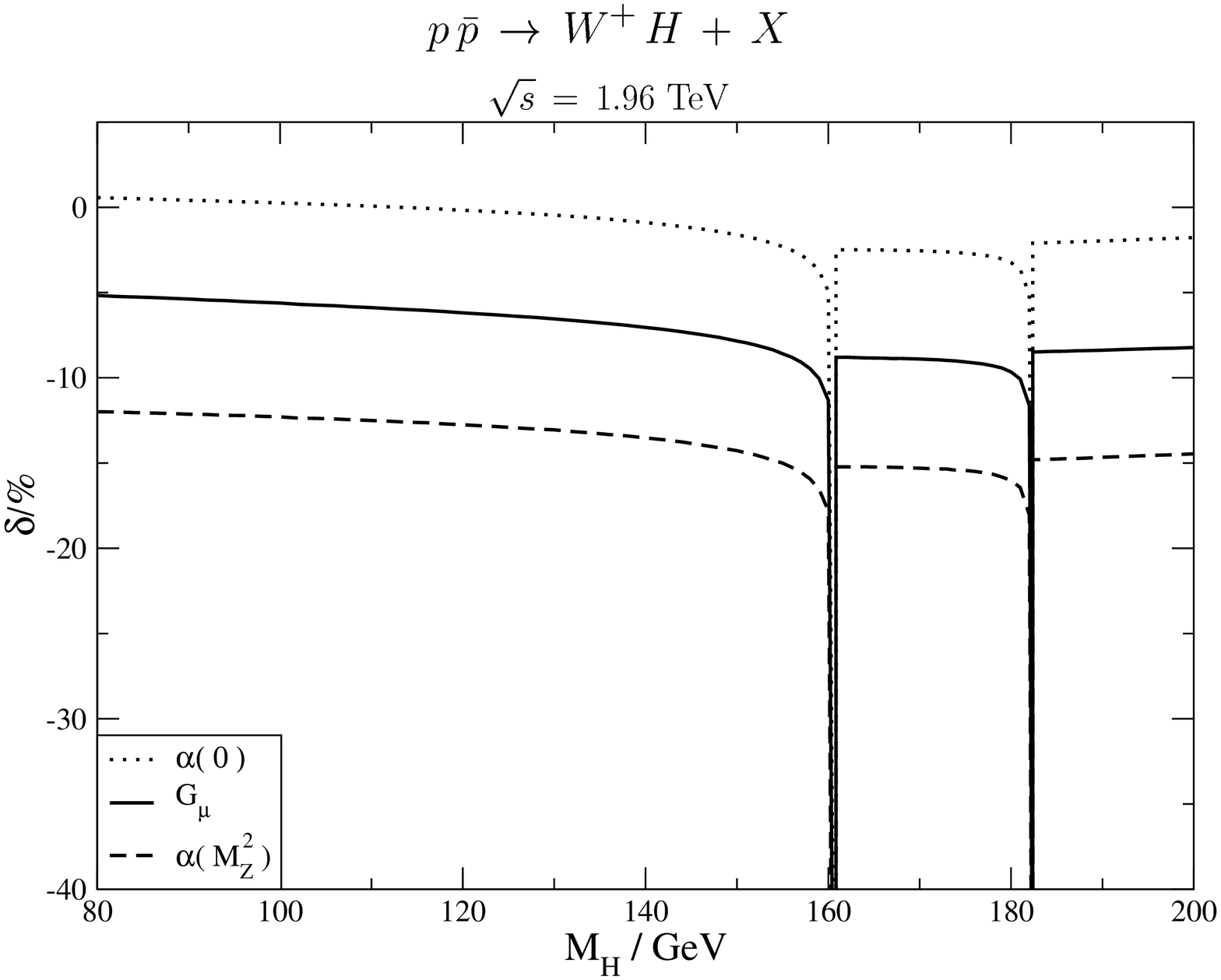,%
        bbllx=35pt,bblly=50pt,bburx=719pt,bbury=582pt,scale=0.33}
\hspace{.5em}
\epsfig{file=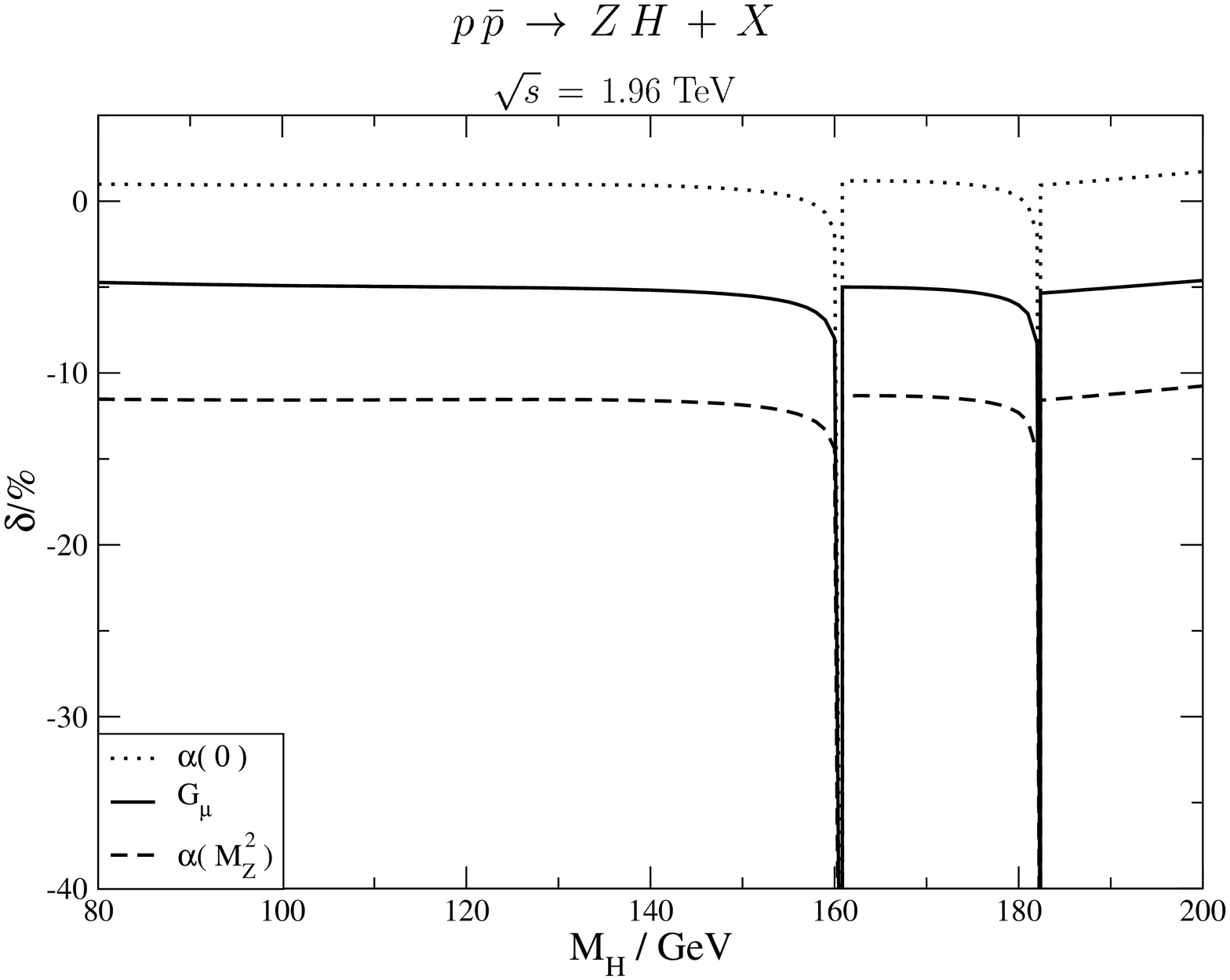,%
        bbllx=35pt,bblly=50pt,bburx=719pt,bbury=582pt,scale=0.33}
}
\vspace{.0em}
\caption{Relative electroweak correction $\delta$ as
 a function of $M_H$ for the total cross section of
 $p\bar{p}\to W^+H+X$ (l.h.s.) and $p\bar{p}\to ZH+X$ (r.h.s.)
at the Tevatron in various input-parameter schemes. 
(Taken from \citere{Ciccolini:2003jy}.)}
\label{fig:tevvh}
\end{center}
\vspace*{-0.2in}
\end{figure}

The bands around the $K$-factors represent the cross section uncertainty due
to the variation of either the renormalization or factorization scale from
$\frac{1}{3} M_{VH} \leq \mu_F \, (\mu_R) \leq 3M_{VH}$, with the other scale
fixed at $\mu_R \, (\mu_F) = M_{VH}$; the
normalization is provided by the production cross section evaluated at scales
$\mu_F=\mu_R=M_{VH}$. As can be seen, except from the accidental cancellation
of the scale dependence of the LO cross section at the LHC, the decrease of the
scale variation is strong when going from LO to NLO and then to NNLO. For
$M_H=120$ GeV, the uncertainty from the scale choice at the LHC drops from 10\%
at LO, to 5\% at NLO, and to 2\% at NNLO. At the Tevatron and for the same
Higgs boson mass, the scale uncertainty drops from 20\% at LO, to 7\% at NLO,
and to 3\% at NNLO. If this variation of the cross section with the two scales
is taken as an indication of the uncertainties due to the not yet calculated
higher-order corrections, one concludes that once the NNLO QCD contributions are
included in the prediction, the QCD corrections to the cross section 
for the $p\bar p/pp \to VH+X$ process are known at the rather accurate level 
of 2 to 3\% relative to the LO.

\subsection{Electroweak Corrections}
\label{se:EW}

The calculation of the electroweak ${\cal O}(\alpha)$ corrections,
which employs established standard techniques, is described in
detail in \citere{Ciccolini:2003jy}. The virtual one-loop corrections
involve a few hundred diagrams, including self-energy, vertex, and
box corrections. In order to obtain IR-finite corrections, real-photonic 
bremsstrahlung has to be taken into account.
In spite of being IR finite, the ${\cal O}(\alpha)$ corrections
involve logarithms of the initial-state quark masses which are due to
collinear photon emission. These mass singularities are absorbed
into the PDFs in exactly the same way as in QCD, viz.\
by $\overline{\mbox{MS}}$ factorization.
As a matter of fact, this requires also the
inclusion of the corresponding ${\cal O}(\alpha)$ corrections into the
DGLAP evolution of these distributions and into their fit to
experimental data. At present, this full incorporation of ${\cal
O}(\alpha)$ effects in the determination of the quark distributions
has not been performed yet. However, an approximate inclusion of the
${\cal O}(\alpha)$ corrections to the DGLAP evolution shows
\cite{Kripfganz:1988bd} that the impact of these corrections on the
quark distributions in the $\overline{\mbox{MS}}$ factorization scheme
is well below 1\%, at least in the $x$ range that is relevant for
associated $VH$ production at the Tevatron and the LHC.  
This is also supported by a recent analysis of the MRST collaboration
\cite{Stirling} who took into account the ${\cal O}(\alpha)$ effects
to the DGLAP equations.

The size of the ${\cal O}(\alpha)$ corrections depends on the
employed input-parameter scheme for the coupling $\alpha$. 
This coupling can, for instance, be derived from
the fine-structure constant $\alpha(0)$, from the effective
running QED coupling $\alpha(M_Z^2)$ at the Z~resonance, or from
the Fermi constant $\GF$ via $\alpha_{\GF}=\sqrt{2}\GF M_W^2\sw^2/\pi$.
The corresponding schemes are known as $\alpha(0)$-, $\alpha(M_Z^2)$-,
and $\GF$-scheme, respectively.
In contrast to the $\alpha(0)$-scheme, where the ${\cal O}(\alpha)$ 
corrections are sensitive to the non-perturbative regime of the
hadronic vacuum polarization, in the $\alpha(M_Z^2)$- and $\GF$-schemes
these effects are absorbed into the coupling constant $\alpha$.
In the $\GF$-scheme large renormalization effects
induced by the $\rho$-parameter are absorbed in addition
via $\alpha_{\GF}$.
Thus, the $\GF$-scheme is preferable over the two other schemes
(at least over the $\alpha(0)$-scheme).

Figure~\ref{fig:tevvh} shows the relative size of the ${\cal
  O}(\alpha)$ corrections as a function of the Higgs-boson mass for
$p\bar p \to W^+ H + X$ and $p\bar p \to ZH + X$ at the Tevatron. The
numerical results have been obtained using the
CTEQ6L1~\cite{Pumplin:2002vw} parton distribution function, but the
dependence of the relative electroweak correction $\delta$ displayed
in Fig.~\ref{fig:tevvh} on the PDF is insignificant.
Results are presented for the three
different input-parameter schemes. The corrections in the $\GF$- and
$\alpha(M_Z^2)$-schemes are significant and reduce the cross section
by 5--9\% and by 10--15\%, respectively. The corrections in the
$\alpha(0)$-scheme differ from those in the $\GF$-scheme by $2\Delta
r\approx 6\%$ and from those in the $\alpha(M_Z^2)$-scheme by
$2\Delta\alpha(M_Z^2)\approx 12\%$. 
The quantities $\Delta r$ and $\Delta\alpha(M_Z^2)$ denote, respectively, the
radiative corrections to muon decay and the correction describing
the running of $\alpha(Q^2)$ from $Q=0$ to $M_Z$
(see \citere{Ciccolini:2003jy} for details).
The fact that the relative
corrections in the $\alpha(0)$-scheme are rather small results from
accidental cancellations between the running of the electromagnetic
coupling, which leads to a contribution of about
$2\Delta\alpha(M_Z^2)\approx +12\%$, and other (negative) corrections
of non-universal origin.  Thus, corrections beyond ${\cal O}(\alpha)$
in the $\alpha(0)$-scheme cannot be expected to be suppressed as well.
In all schemes, the size of the corrections does not depend strongly
on the Higgs-boson mass.

For the LHC the corrections are similar in size
to those at the Tevatron and reduce the cross section by 5--10\% in
the $\GF$-scheme and by 12--17\% in the $\alpha(M_Z^2)$-scheme
(see Figs.~13 and 14 in \citere{Ciccolini:2003jy}).
%The electroweak corrections to $p p \to W^- H + X$ at the
%LHC differ from those to $p p \to W^+ H + X$ by less than about $2\%$.

\begin{figure}[bht]
\vspace*{1.3cm}
\begin{center}
{ \unitlength 1cm
\begin{picture}(15.5,8.5)
\put(-2.2,-2){\includegraphics{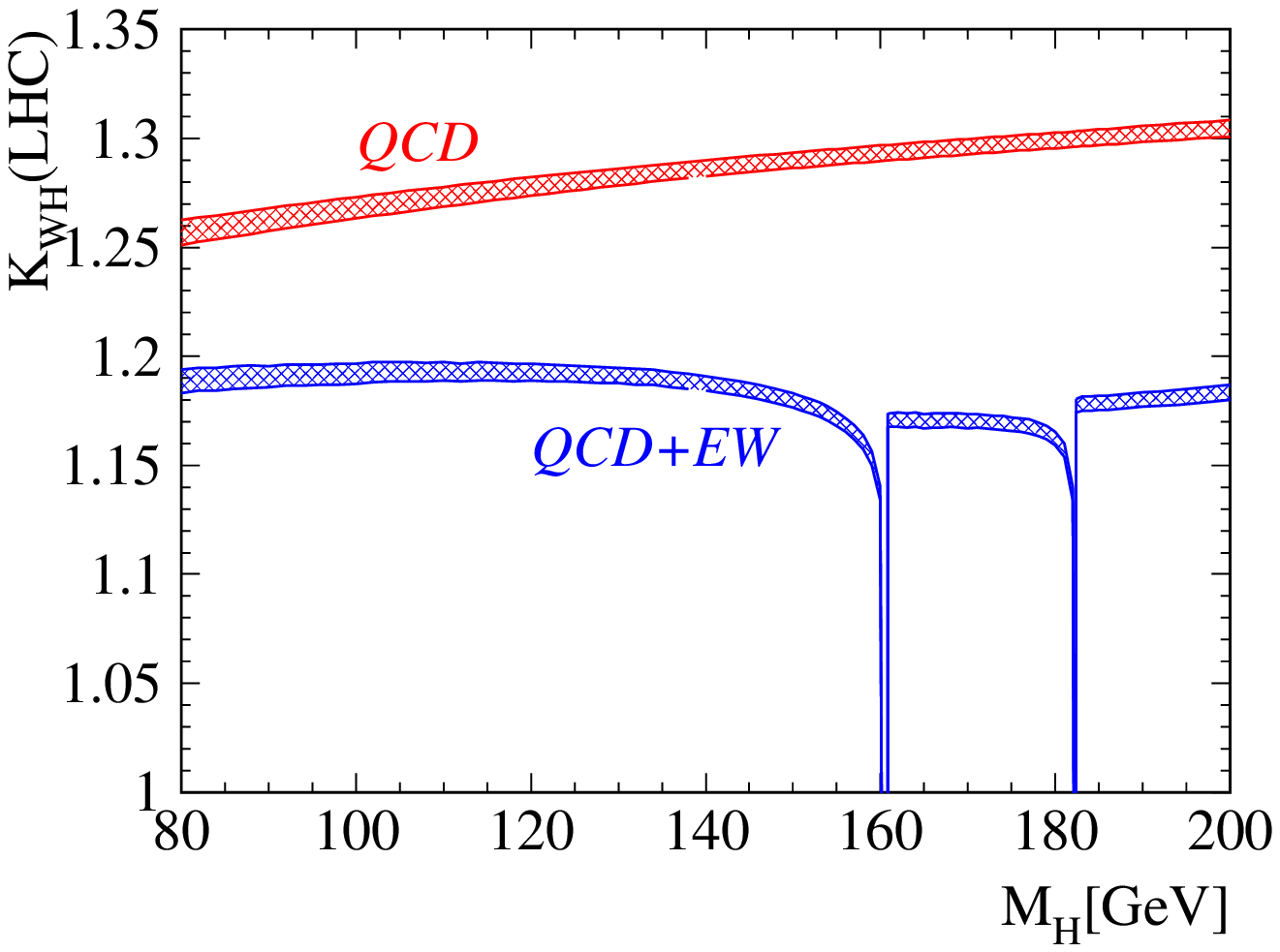}}
\put( 6.0,-2){\includegraphics{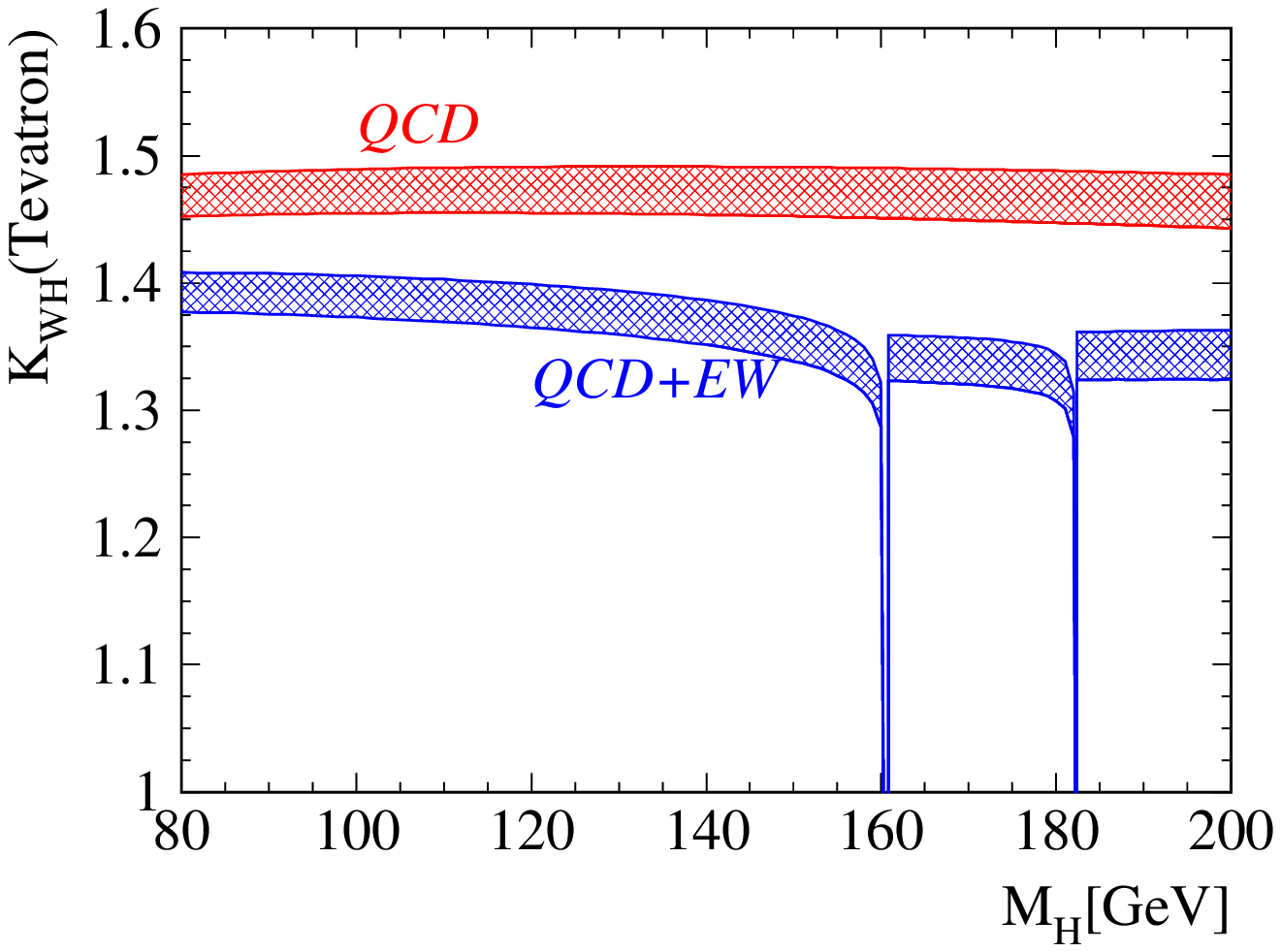}}
\put(-2.2,-8.5){\includegraphics{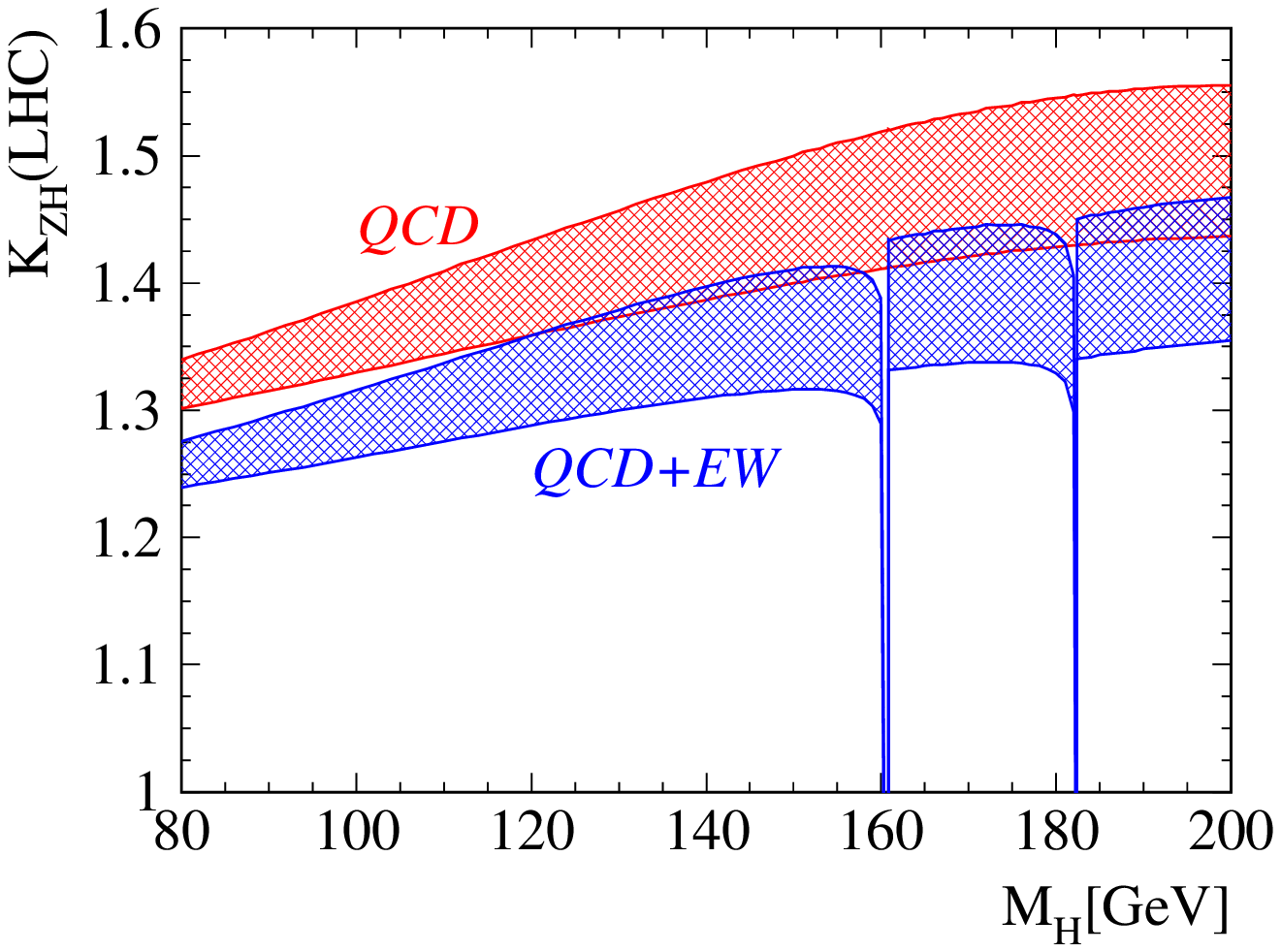}}
\put( 6.0,-8.5){\includegraphics{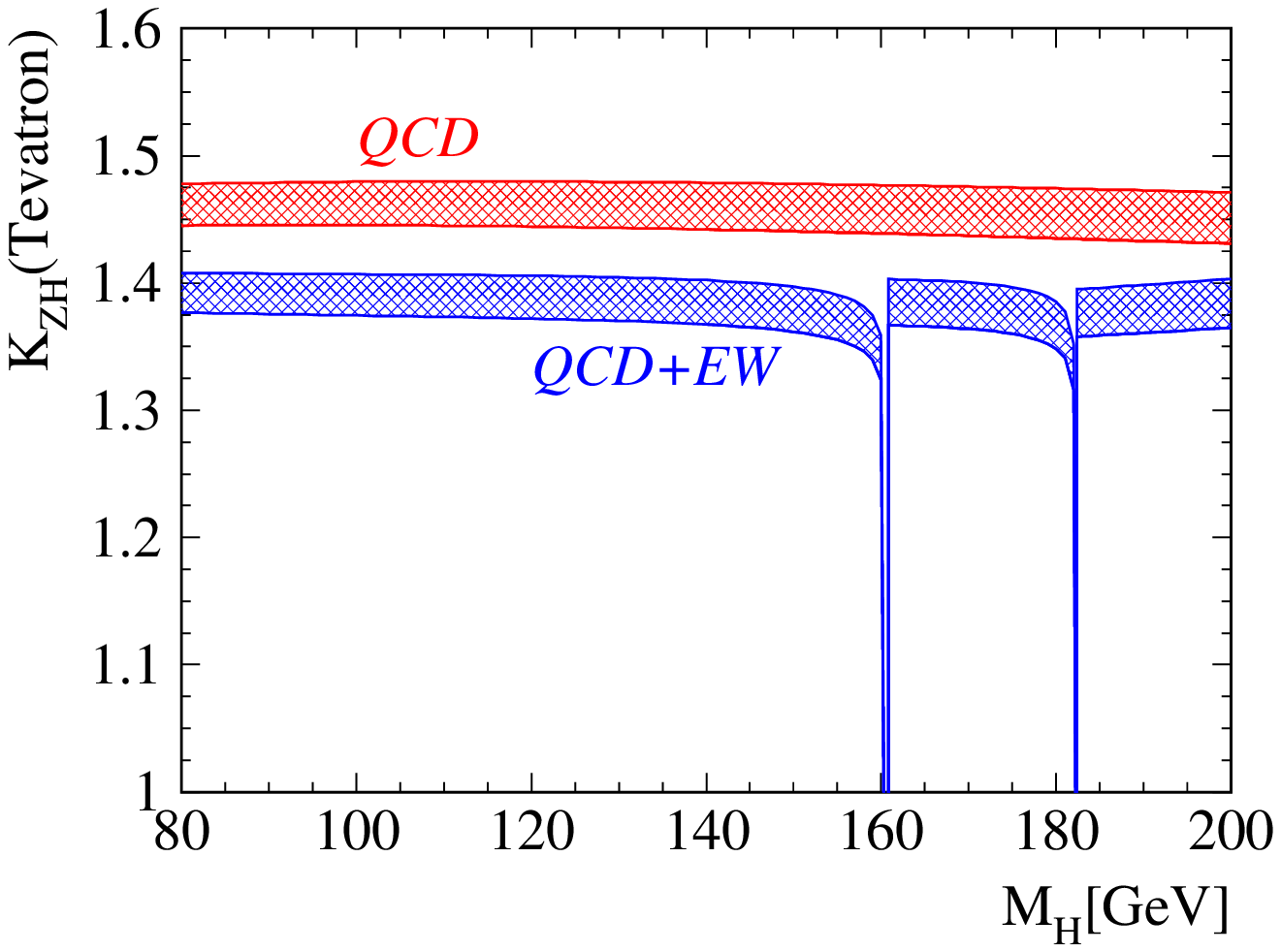}}
\end{picture} }
\vspace*{30mm}
\caption{
\label{fig:Kfactor}
$K$-factors for $WH$ production and $ZH$ production at the LHC
(l.h.s.) and the Tevatron (r.h.s.) after inclusion of the NNLO QCD and
electroweak ${\cal O}(\alpha)$ corrections. Theoretical errors as
described in Figure~\ref{fig:Kfact}.}
\end{center}
\end{figure}

In \citere{Ciccolini:2003jy} the origin of the electroweak corrections
was further explored by separating gauge-invariant building blocks.
It turns out that fermionic contributions (comprising all diagrams
with closed fermion loops) and remaining bosonic corrections partly
compensate each other, but the bosonic corrections are dominant.  The
major part of the corrections is of non-universal origin, i.e.\ the
bulk of the corrections is not due to coupling modifications, photon
radiation, or other universal effects.

Figure~\ref{fig:Kfactor} shows the $K$-factor after inclusion of both
the NNLO QCD and the ${\cal O}(\alpha)$ electroweak corrections for
$p\bar p/pp \to W H + X$ and $p\bar p/pp \to ZH + X$ at the Tevatron
and the LHC.  The larger uncertainty band for the $ZH$ production
process at the LHC is due to the contribution of $gg \to HZ$.

\subsection{Cross-Section Predictions}
\label{se:numres}

Figure~\ref{fig:xsection} shows the predictions for the cross sections
of $WH$ and $ZH$ production at the LHC and the Tevatron, 
including the NNLO QCD and electroweak ${\cal O}(\alpha)$ corrections
as discussed in the previous sections.
\begin{figure}[hbt]
\begin{center}
{ \unitlength 1cm
\begin{picture}(15.5,7)
\put(-4.4,-8.5){\includegraphics{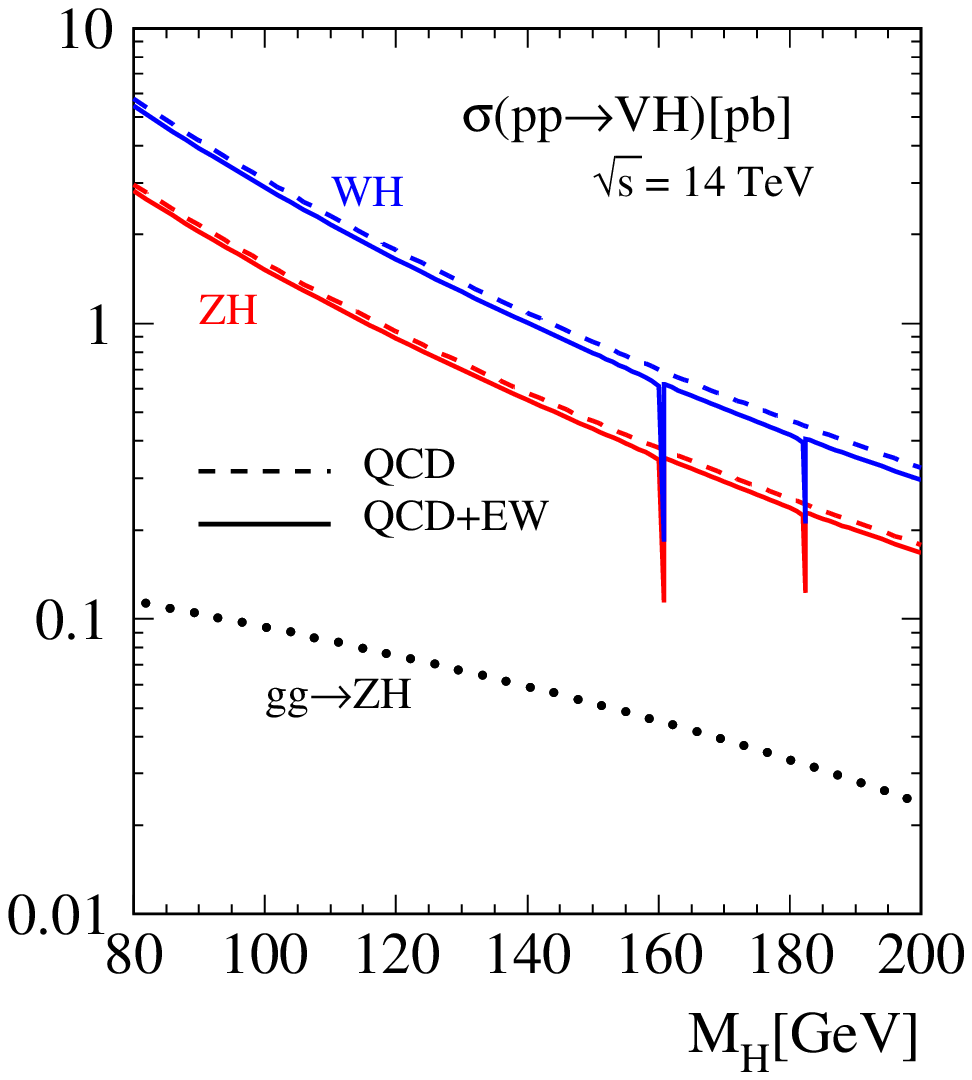}}
\put( 3.8,-8.5){\includegraphics{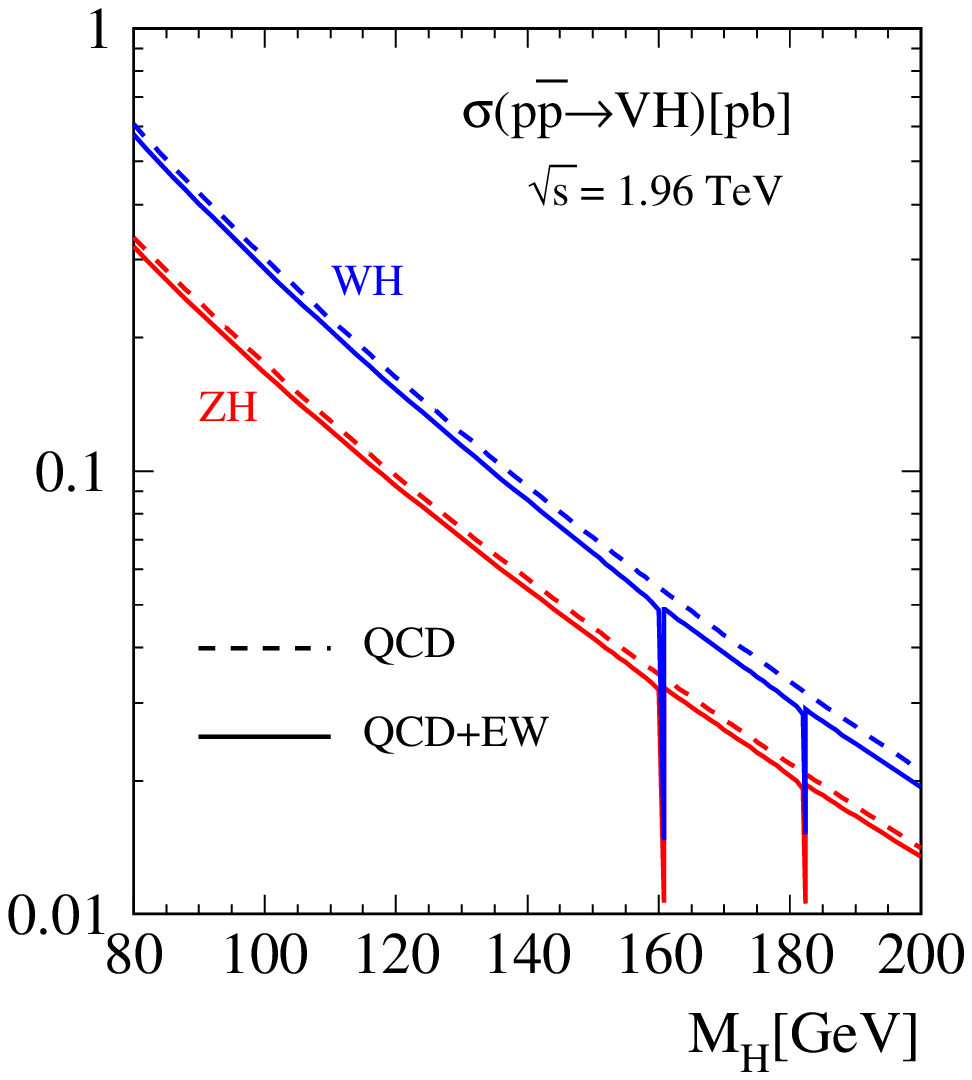}}
%\put(-2.0,-15.7){\special{psfile=xsection.ps hscale=90 vscale=90
%                  angle=0 hoffset=0 voffset=0}}
\end{picture} }
\vspace*{15mm}
\caption{
\label{fig:xsection}
Cross-section predictions (in the $\GF$-scheme)
for $WH$ and $ZH$ production 
at the LHC (l.h.s.) and the Tevatron (r.h.s.), including NNLO QCD
and electroweak ${\cal O}(\alpha)$ corrections.}
\end{center}
\end{figure}
At the LHC the process $gg\to ZH$ adds about 10\% to the $ZH$
production cross section, which is due to the large gluon flux;
at the Tevatron this contribution is negligible.

Finally, we briefly summarize the discussion \cite{Ciccolini:2003jy}
of the uncertainty in the cross-section predictions due to the error
in the parametrization of the parton densities (see also
\cite{Djouadi:2003jg}).  To this end the NLO cross section evaluated
using the default CTEQ6~\cite{Pumplin:2002vw} parametrization with the
cross section evaluated using the MRST2001~\cite{Martin:2002aw}
parametrization are compared. The results are collected in
\reftas{tab:pdfvhtev} and \ref{tab:pdfvhlhc}. Both the CTEQ and MRST
\begin{table}
\caption{\label{tab:pdfvhtev} Total cross sections (in fb) at
 the Tevatron ($\sqrt{s}=1.96\;\mathrm{TeV}$) including NLO QCD and
 electroweak corrections in the $\GF$-scheme for different sets of
 PDFs. The results include an estimate of the uncertainty due to the
 parametrization of the PDFs as obtained with the
 CTEQ6~\cite{Pumplin:2002vw} and MRST2001~\cite{Martin:2002aw}
 eigenvector sets. The renormalization and factorization scales have
 been set to the invariant mass of the Higgs--vector-boson pair, $\mu
 = \mu_0 = M_{VH}$. (Taken from \citere{Ciccolini:2003jy}.)}
\vspace{.5em}
\centerline{
\begin{tabular}{|c||c|c||c|c|}
\hline 
& \multicolumn{2}{c||}{$p\bar{p} \to WH+X$}
& \multicolumn{2}{c|}{$p\bar{p} \to ZH+X$}
\\
\hline 
$M_H/\mathrm{GeV}$ &
CTEQ6M~\cite{Pumplin:2002vw} & 
MRST2001~\cite{Martin:2002aw} &
CTEQ6M~\cite{Pumplin:2002vw} & 
MRST2001~\cite{Martin:2002aw} \\ \hline \hline
100.00  & 268.5(1)  $\pm$ 11    & 269.8(1) $\pm$ 5.2  &
          158.9(1)  $\pm$ 6.4   & 159.6(1) $\pm$ 2.0  \\ \hline
120.00  & 143.6(1)  $\pm$ 6.0   & 143.7(1) $\pm$ 3.0  &
           88.20(1) $\pm$ 3.6   & 88.40(1) $\pm$ 1.1  \\ \hline
140.00  &  80.92(1) $\pm$ 3.5   & 80.65(1) $\pm$ 1.8  &
           51.48(1) $\pm$ 2.1   & 51.51(1) $\pm$ 0.66 \\ \hline
170.00  &  36.79(1) $\pm$ 1.7   & 36.44(1) $\pm$ 0.91 &
           24.72(1) $\pm$ 1.0   & 24.69(1) $\pm$ 0.33 \\ \hline
190.00  &  22.94(1) $\pm$ 1.1   & 22.62(1) $\pm$ 0.60 &
           15.73(1) $\pm$ 0.68  & 15.68(1) $\pm$ 0.21 \\ \hline
\end{tabular}}
\vspace{2em}
\caption{\label{tab:pdfvhlhc} Same as in \refta{tab:pdfvhtev},
but for the LHC ($\sqrt{s}=14\;\mathrm{TeV}$)
(Taken from \citere{Ciccolini:2003jy}.)}
\vspace{.5em}
\centerline{
\begin{tabular}{|c||c|c||c|c|}
\hline 
& \multicolumn{2}{c||}{$pp \to WH+X$}
& \multicolumn{2}{c|}{$pp \to ZH+X$}
\\
\hline 
$M_H/\mathrm{GeV}$ 
& CTEQ6M~\cite{Pumplin:2002vw} & MRST2001~\cite{Martin:2002aw} 
& CTEQ6M~\cite{Pumplin:2002vw} & MRST2001~\cite{Martin:2002aw} 
\\ \hline \hline
100.00  & 2859(1)  $\pm$  96   & 2910(1)  $\pm$ 35  &
          1539(1)  $\pm$  51   & 1583(1)  $\pm$ 19  \\ \hline
120.00  & 1633(1)  $\pm$  55   & 1664(1)  $\pm$ 21  &
          895(3)   $\pm$  30   & 9217(3)  $\pm$ 11  \\ \hline
140.00  & 989(3)   $\pm$  34   & 1010(1)  $\pm$ 12  &
          551(2)   $\pm$  19   & 568.1(2) $\pm$ 6.7 \\ \hline
170.00  & 508(1)   $\pm$  18   & 519.3(1) $\pm$ 6.3 &
          290(1)   $\pm$  10   & 299.4(1) $\pm$ 3.6 \\ \hline
190.00  & 347(1)   $\pm$  12   & 354.7(2) $\pm$ 4.3 &
          197.8(1) $\pm$   6.9 & 204.5(1) $\pm$ 2.5 \\ \hline
\end{tabular}}
\end{table}
parametrizations include parton-distribution-error packages which
provide a quantitative estimate of the corresponding uncertainties in
the cross sections.%
\footnote{In addition, the MRST~\cite{Martin:2001es} parametrization allows to
study the uncertainty of the NLO cross section due to the variation of
$\alpha_{\rm s}$. For associated $WH$ and $ZH$ hadroproduction, the
sensitivity of the theoretical prediction to the variation of
$\alpha_{\rm s}$ ($\alpha_{\rm s}(M_Z^2) = 0.119\pm 0.02$) turns out
to be below $2\%$.}  Using the parton-distribution-error packages and
comparing the CTEQ and MRST2001 parametrizations, we find that the
uncertainty in predicting the $WH$ and $ZH$ production processes 
at the Tevatron and the LHC due to the
parametrization of the parton densities is less than approximately
$5\%$.

\subsection{Conclusions}
\label{se:concl}

After the inclusion of QCD corrections up to NNLO and of the
electroweak ${\cal O}(\alpha)$ corrections, the cross-section
predictions for $WH$ and $ZH$ production are by now the most
precise for Higgs production at hadron colliders. 
The remaining uncertainties
should be dominated by renormalization and factorization scale
dependences and uncertainties in the parton distribution
functions, which are of the order of 3\% and 5\%, respectively.
These uncertainties may be reduced by forming the ratios
of the associated Higgs-production cross section with the
corresponding Drell-Yan-like W- and Z-boson production channels,
i.e.\ by inspecting $\sigma_{p\bar{p}/pp\to VH+X}/\sigma_{p\bar{p}/pp\to
V+X}$, rendering their measurements particularly interesting 
at the Tevatron and/or the LHC.

}

%% file: oleari.tex
{
\newcommand\WBFNLO{VBFNLO}

\section[ ]{NLO CORRECTIONS FOR VECTOR BOSON FUSION
PROCESSES\footnote{R.\,Mazini, C.\,Oleari, D.\,Zeppenfeld}}
%\institute{Istituto Nazionale di Fisica Nucleare,
%Sezione di Genova, Italy}

%\subsection{INTRODUCTION}
The vector-boson fusion (VBF) process, $qQ\to qQH$, is expected to provide a
copious source of Higgs bosons in $pp$-collisions at the LHC.
Together with gluon fusion, it represents the most promising
production process for Higgs boson
%discovery~\cite{CMS,ATLAS}.
discovery~\cite{Bayatian:1994pu,Kinnunen:1999ak,Drollinger:2001ym,unknown:1999fr,
Richter-Was:1999sa,Kersevan:2002vu}.  
Beyond discovery and determination of its mass, the measurement of Higgs
boson couplings to gauge bosons and fermions will be a primary goal of the 
LHC. Here, VBF will be crucial for separating the contributions of different
decay 
modes of the Higgs boson, as was first pointed out during the 1999 Les Houches
workshop~\cite{Zeppenfeld:2000td} and as discussed in 
the Higgs boson coupling section of this report. 

VBF rates (given by the cross section times the branching ratios,
$\sigma\times B$) can be measured at the LHC with  
statistical accuracies reaching 5
to 10\%~~\cite{Zeppenfeld:2000td,Zeppenfeld:2002ng,Belyaev:2002ua}.
In order to extract the Higgs boson couplings
with this full statistical power, a theoretical prediction of the
SM production cross section with error well below 10\% is
required, and this clearly entails knowledge of the NLO QCD corrections.

For the total Higgs boson production cross section via VBF, these NLO
corrections have been available for a decade~\cite{Han:1992hr}
and they are
relatively small, with $K$-factors around 1.05 to 1.1. These modest
$K$-factors are another reason for the importance of Higgs boson production
via VBF: theoretical uncertainties will not limit the precision of the
coupling measurements. This is in contrast to the dominant gluon fusion
channel where the $K$-factor is larger than 2 and residual uncertainties of
10-20\% remain, even after the 2-loop corrections have been
%evaluated~\cite{HggNLO,H2loop}.
evaluated~\cite{Djouadi:1991tk,Spira:1995rr,Dawson:1991zj,Giele:2002hx,
Catani:2001ic,Harlander:2001is,Harlander:2002wh,Anastasiou:2002yz,
Ravindran:2003um}.
To distinguish the VBF Higgs boson signal from backgrounds,
stringent cuts are required on the Higgs boson decay products as well as on
the two forward quark jets which are characteristic for VBF. Typical cuts
have an acceptance of less than 25\% of the starting value for $\sigma\times
B$. With such large reduction factors, NLO cross sections within these
acceptance cuts are needed for a precise extraction of coupling information.

Analogous to Higgs boson production via VBF, the production of $Wjj$ and
$Zjj$ events via vector-boson fusion will proceed with sizable cross sections
at the LHC.  These processes have been considered previously at
leading order for the study of rapidity gaps at hadron
colliders~\cite{Chehime:1993ub,Rainwater:1996ud, Khoze:2002fa}, 
as a background to Higgs boson searches in
VBF~\cite{Rainwater:1998kj,Plehn:1999xi,Rainwater:1999sd,Kauer:2000hi,
Eboli:2000ze,Cavalli:2002vs}, 
or as a probe of anomalous triple-gauge-boson couplings~\cite{Baur:1993fv}, 
to name but a few examples. 
In addition, one would like to exploit $W$ and $Z$ production via VBF as
calibration processes for Higgs boson production, namely as a tool to
understand the tagging of forward jets or the distribution and veto of
additional central jets in VBF.  The precision needed for Higgs boson studies
then requires the knowledge of NLO QCD corrections also for $Wjj$ and $Zjj$
production.

In order to address the theoretical uncertainties 
not only of total cross sections but also of cross sections within cuts and of
distributions, we have written a fully flexible NLO parton-level Monte Carlo 
program (called \WBFNLO\ below) that
computes NLO QCD corrections to $Hjj$, $Zjj$ and $Wjj$ production channels,
in the kinematic configurations where typical VBF cuts are applied (see
Refs.~\cite{Figy:2003nv,Oleari:2003tc} for a detailed description of the
calculation and further results). Here we give only a brief overview of
results. 
For $Hjj$ production via VBF, an independent Monte Carlo program for the NLO
cross  
section is available within the MCFM package~\cite{MCFM}. Results from these
two NLO programs for Higgs production are compared below.

%\subsection{VBF CUTS AND RESULTS FOR THE LHC}
%\label{sec:pheno_MOZ}

\begin{figure}[thb]
\begin{center}
\includegraphics[width=14.5cm]{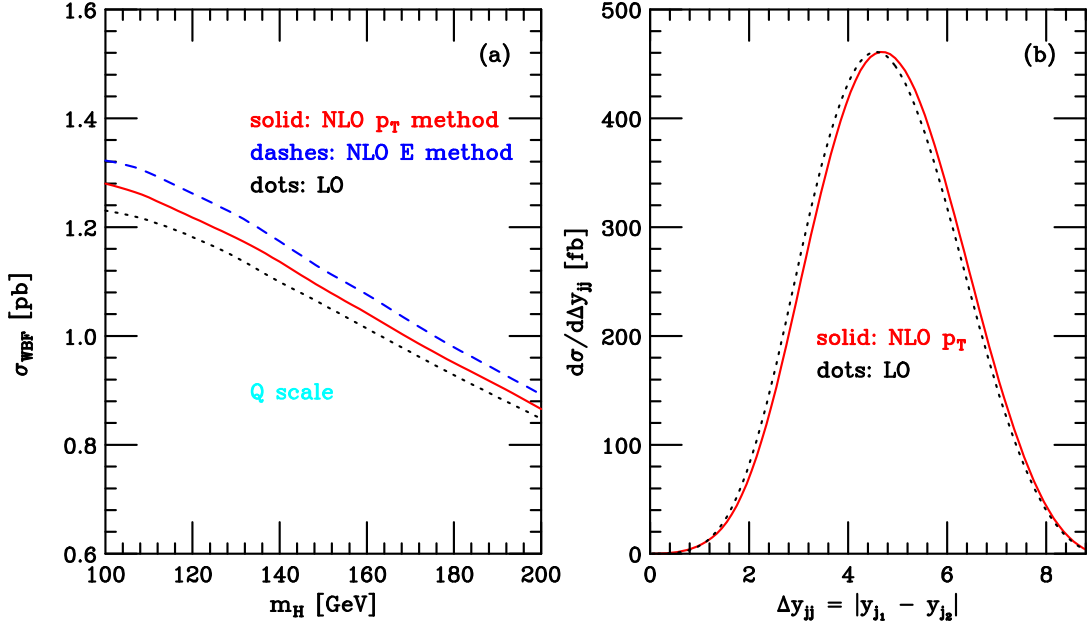} 
\caption{Effect of QCD radiative corrections on the Higgs boson
production cross section via VBF. Results are given at LO (black dotted) 
and at NLO for the $p_T$ method (solid red) and the $E$ method (dashed blue) 
of defining tagging jets. Panel (a) gives the cross section within the cuts of
Eqs.~(\protect\ref{eq:cuts1_MOZ})--(\protect\ref{eq:cuts4_MOZ}) as a function 
of the Higgs boson mass, {\protect $m_H$}. Panel (b) shows the rapidity separation 
of the two tagging jets for $m_H=120$~GeV.
}
\label{fig:sigtot_MOZ}
\end{center}
\end{figure}

In order to reconstruct jets from final-state partons, the $k_T$
algorithm~\cite{Catani:1992zp,Catani:1993hr,Ellis:1993tq}, 
as described in Ref.~\cite{Blazey:2000qt}, 
is used, with resolution parameter $D=0.8$. We calculate the partonic cross
sections for events with at least two hard jets, which are required to have
\begin{equation}
\label{eq:cuts1_MOZ}
p_{Tj} \geq 20~{\rm GeV} \, , \qquad\qquad |y_j| \leq 4.5 \, .
\end{equation}
Here $y_j$ denotes the rapidity of the (massive) jet momentum which is
reconstructed as the four-vector sum of massless partons of pseudorapidity
$|\eta|<5$. The two reconstructed jets of highest transverse momentum are 
called ``tagging jets'' and are
identified with the final-state quarks which are characteristic for
vector-boson fusion processes. We call this method of choosing the tagging
jets the ``$p_T$ method'', as opposed to the ``$E$ method'' which identifies
the two jets with the highest lab energy as tagging jets.

The Higgs boson decay products (generically 
called ``leptons'' in the following) are required to fall between the
two tagging jets in rapidity and they should be well observable. While 
cuts for the Higgs boson decay products will depend 
on the channel considered, we here substitute such specific requirements
by generating isotropic Higgs boson decay into two massless ``leptons'' 
(which represent $\tau^+\tau^-$ or $\gamma\gamma$ or $b\bar b$ final states)
and require
\begin{equation}
\label{eq:cuts2_MOZ}
p_{T\ell} \geq 20~{\rm GeV} \,,\qquad |\eta_{\ell}| \leq 2.5  \,,\qquad 
\triangle R_{j\ell} \geq 0.6 \, ,
\end{equation}
where $\triangle R_{j\ell}$ denotes the jet-lepton separation in the
rapidity-azimuthal 
angle plane. In addition the two ``leptons'' are required to fall between the
two tagging jets in rapidity
\begin{equation}
\label{eq:cuts3_MOZ}
y_{j,min}  < \eta_{\ell_{1,2}} < y_{j,max} \, .
\end{equation}
When considering the decays $Z\to \ell^+\ell^-$ and $W\to \ell\nu$, we apply
the same cuts of Eqs.~(\ref{eq:cuts2_MOZ}) and~(\ref{eq:cuts3_MOZ}) to the
charged leptons (we have used here a slightly smaller jet-lepton separation
$\triangle R_{j\ell} \geq 0.4$).

Backgrounds to vector-boson fusion are significantly suppressed by requiring
a large rapidity separation of the two tagging jets (rapidity-gap cut)
\begin{equation}
\label{eq:cuts4_MOZ}
\Delta y_{jj}=|y_{j_1}-y_{j_2}|>4\; .
\end{equation}

\begin{figure}[thb]
\begin{center}
\includegraphics[width=14.5cm]{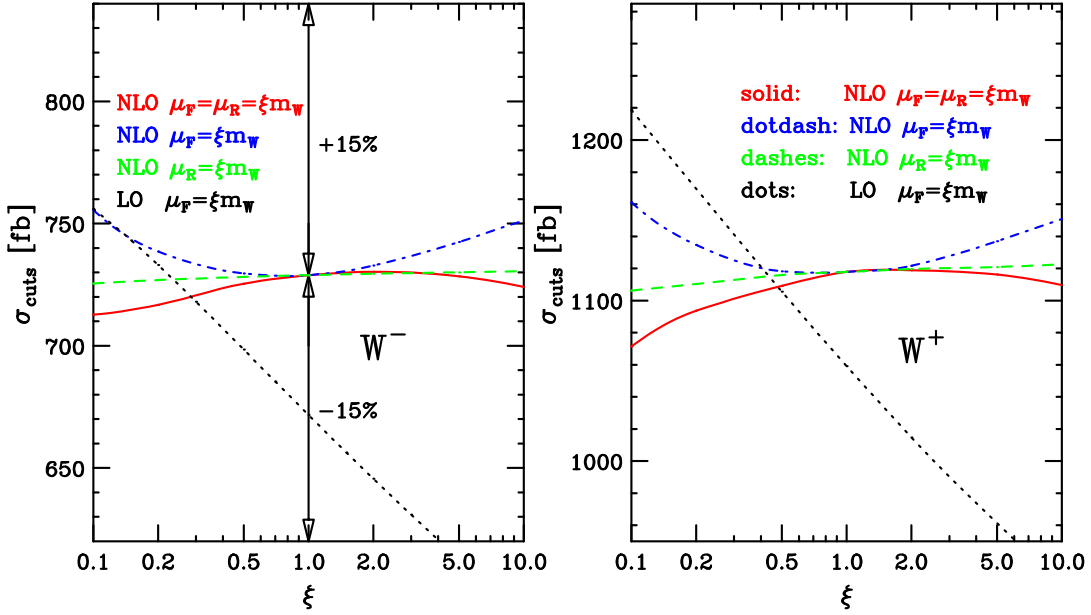} 
\caption{Scale dependence of the total cross section at LO and NLO within the
cuts of Eqs.~(\ref{eq:cuts1_MOZ})--(\ref{eq:cuts4_MOZ}) for $W^-$ and $W^+$
production at the LHC.  The factorization scale $\mu_F$ and/or the
renormalization scale $\mu_R$ have been taken as multiples of the
vector-boson mass, $\xi\, m_W$, and $\xi$ is varied in the range $0.1 < \xi <
10$. The NLO curves are for $\mu_F=\mu_R=\xi m_W$ (solid red line),
$\mu_F=m_W$ and $\mu_R=\xi\, m_W$ (dashed green line) and $\mu_R=m_W$ and
$\mu_F$ variable (dot-dashed blue line).  The dotted black curve shows the
dependence of the LO cross section on the factorization scale. At LO,
$\alpha_s(\mu_R)$ does not enter.}
\label{fig:scale_depW_MOZ} 
%\vspace*{-5mm}
\end{center}
\end{figure}

Cross sections, within the cuts of
Eqs.~(\ref{eq:cuts1_MOZ})--(\ref{eq:cuts4_MOZ}), are shown in
Fig.~\ref{fig:sigtot_MOZ}(a) as a function of the Higgs boson mass,
$m_H$. As for the total VBF cross section, the NLO effects are modest for the
cross section within cuts, amounting to a 3-5\% increase for the $p_T$ method
of selecting tagging jets  and a 6-9\% increase when the
$E$ method is used. The differential cross section as function of the rapidity 
separation between the two tagging jets is plotted in Fig.~\ref{fig:sigtot_MOZ}(b).
The wide separation of the tagging jets, which is important for rejection of QCD 
backgrounds, slightly increases at NLO. This example also shows that the 
$K$-factor,
the ratio of NLO to LO differential cross sections, is strongly phase space 
dependent, i.e.\ an overall constant factor will not be adequate to simulate 
the data.

A comparison of our \WBFNLO\ program with the MCFM Monte Carlo shows good
agreement for predicted Higgs boson cross sections and also for those jet
distributions which we have investigated. As an example, 
Table~\ref{tab:sigmas_MOZ} shows cross sections within the cuts of
Eqs.~(\ref{eq:cuts1_MOZ}) and~(\ref{eq:cuts4_MOZ}).  No cuts on Higgs decay
products are imposed because MCFM does not yet include Higgs boson
decays. Cross sections agree at the 2\% level or better, which is more than
adequate for LHC applications. The results in the table were obtained with 
fixed scales $\mu_R=\mu_F=m_H$ and have Monte Carlo statistical errors of 
less than 0.5\%.

%\begin{table}[ht]
%\caption{Higgs production cross sections (in pb) from MCFM and \WBFNLO\ after
%jet cuts.}
%\begin{center}
%%\vskip0.2cm
%\begin{tabular}{|c|c|c|c|c|c|c|c|}
%\hline
%      &  m$_H$ (GeV)        & 100   & 120   &  140  & 160   & 180    & 200  \\
%\hline
%MCFM  & $\sigma_{LO}$ & 1.91 & 1.68 & 1.48 & 1.32& 1.17 & 1.05\\
%\cline{2-8}
%  & $\sigma_{NLO}$  & 2.00 & 1.78 & 1.58 & 1.42 & 1.27 & 1.14 \\
%\hline
%\WBFNLO & $\sigma_{LO}$ &  1.88 & 1.66 & 1.47 & 1.31 & 1.16 & 1.04 \\
%\cline{2-8}
%  & $\sigma_{NLO}$  & 1.95 & 1.74 & 1.55 & 1.40 & 1.25 & 1.13 \\
%\hline
%\end{tabular}
%\label{tab:sigmas_MOZ}
%\end{center}
%\end{table}    
%

\begin{table}[b]
\caption{Higgs production cross sections (in pb) from MCFM and \WBFNLO\ after
jet cuts.}
\begin{center}
%\vspace*{-2mm}
%\vskip0.2cm
\begin{tabular}{|c|c|c|c|c|c|c|c|}
\hline
      &  m$_H$ (GeV)        & 100   & 120   &  140  & 160   & 180    & 200  \\
\hline\hline
$\sigma_{LO}$  & MCFM & 1.91 & 1.68 & 1.48 & 1.32& 1.17 & 1.05\\
\cline{2-8}
 & \WBFNLO &  1.88 & 1.66 & 1.47 & 1.31 & 1.16 & 1.04 \\
\hline\hline
$\sigma_{NLO}$  &  MCFM  & 2.00 & 1.78 & 1.58 & 1.42 & 1.27 & 1.14 \\
\cline{2-8}
  & \WBFNLO  & 1.95 & 1.74 & 1.55 & 1.40 & 1.25 & 1.13 \\
\hline
\end{tabular}
\label{tab:sigmas_MOZ}
\end{center}
\end{table}    

Cross sections for $W^-jj$ and $W^+jj$ production, within the cuts listed
above,  
are shown in Fig.~\ref{fig:scale_depW_MOZ}.  In both panels, the
scale dependence of cross sections is shown for fixed
renormalization and factorization scales, $\mu_R = \xi_R\,m_W$ and 
$\mu_F= \xi_F\,m_W$.
The LO cross sections only depend on $\mu_F$. At NLO we show
three cases: (a) $\xi_F=\xi_R=\xi$ (red solid line); (b) $\xi_F=\xi$,
$\xi_R=1$ (blue dot-dashed line); and (c) $\xi_R=\xi$, $\xi_F=1$ 
(green dashed line). While the factorization-scale dependence of the LO result
is sizable, the NLO cross sections are quite insensitive to scale variations:
allowing a factor 2 variation in either directions, i.e., considering the range
$0.5<\xi <2$, the NLO cross sections change by less than 1\% in all cases.
Similar results were found for the VBF Higgs production 
cross section~\cite{Figy:2003nv}. Alternative scale choices, like the virtuality 
of the exchanged electroweak bosons, also lead to cross sections changes of order
1-2\% at NLO. Also for distributions, scale variations rarely exceed this 
range~\cite{Figy:2003nv,Oleari:2003tc}. These results indicate very
stable NLO predictions for VBF cross sections with generic acceptance cuts.

In addition to these quite small scale uncertainties, we have estimated the
error of the $Hjj$ and $W^\pm jj$ cross sections due to uncertainties in the
determination of the PDFs, and  we have found a total PDF uncertainty of $\pm
4\%$ with the CTEQ PDFs, and of roughly $\pm 2\%$ with the MRST set.

Summarizing, QCD corrections to distributions in VBF processes are
in general of modest size, of order 10\%, but occasionally they reach larger
values.  These corrections are strongly phase-space dependent for jet
observables and an overall $K$-factor multiplying the LO distributions is not
an adequate approximation. Within the phase-space region relevant for Higgs
boson searches, we find differential $K$-factors as small as 0.9 or as large
as 1.3. The residual combined QCD and PDF uncertainties of the NLO VBF cross 
sections are about 4\%.

%\section*{ACKNOWLEDGMENTS}

%This research was supported in part by the University
%of Wisconsin Research Committee with funds granted by the Wisconsin Alumni
%Research Foundation and in part by the U.~S.~Department of Energy under
%Contract No.~DE-FG02-95ER40896.
%C.O. thanks the UK Particle Physics and Astronomy Research Council for
%supporting his research.

}

%% file: djouadi.tex
{

\newcommand{\lsim}{\raisebox{-0.13cm}{~\shortstack{$<$ \\[-0.07cm] $\sim$}}~}
\newcommand{\gsim}{\raisebox{-0.13cm}{~\shortstack{$>$ \\[-0.07cm] $\sim$}}~}
\newcommand{\dx}{\mbox{\rm d}}
\newcommand{\ra}{\rightarrow}
\newcommand{\s}{\smallskip}
\newcommand{\nn}{\noindent}
\newcommand{\non}{\nonumber}
\newcommand{\beq}{\begin{eqnarray}}
\newcommand{\eeq}{\end{eqnarray}}

\section[ ]{PDF uncertainties in Higgs production at the
LHC\footnote{A.\,Djouadi and S.\,Ferrag}}

Parton distribution functions (PDFs), which describe the momentum distribution
of a parton in the proton,  play a central role at hadron colliders.  A precise
knowledge of the PDFs over a wide range of the proton momentum fraction $x$
carried by the parton and the squared centre-of-mass  energy $Q^2$ at which the
process takes place, is mandatory to precisely predict the production cross
sections of the various signals and background hard processes. However, they
are plagued by uncertainties, which arise either from the starting
distributions obtained from a global fit  to the available data from
deep-inelastic scattering, Drell--Yan and hadronic data, or from the DGLAP
evolution  to the higher $Q^2$ relevant to the LHC
scattering processes.   Together with the effects of unknown perturbative
higher order corrections, these uncertainties dominate the theoretical error on
the predictions of the production cross sections. 

PDFs with intrinsic uncertainties became available in 2002. Before that date,
to quantitatively estimate the uncertainties due to the structure functions, it
was common practice to calculate the production cross sections using the
``nominal fits" or reference set of the PDFs provided by different
parametrizations and to consider the dispersion between the various predictions
as being the ``uncertainty" due to the PDFs. However, the comparison between
different parametrizations cannot be regarded as an unambiguous way to estimate
the uncertainties since the theoretical and experimental errors spread into
quantitatively different intrinsic uncertainties following  their treatment in
the given parametrization. The CTEQ and MRST collaborations and Alekhin
recently introduced new schemes, which provide the possibility of estimating
the  intrinsic uncertainties and the spread  uncertainties on the prediction 
of physical observables at hadron colliders\footnote{Other sets of PDFs with
errors are available in the literature, but they will not be
discussed here.}.

In this short note, the spread uncertainties on the Higgs boson production cross
sections at the LHC, using the CTEQ6 \cite{CTEQ6}, 
MRST2001 \cite{MRST2001E} and ALEKHIN2002 \cite{ALEKHIN} sets of PDFs, 
are investigated and compared. For more details, we refer to \cite{Samir}. 

The scheme introduced by both the CTEQ and MRST collaborations is based on the
Hessian matrix method.  The latter enables a characterization of a parton
parametrization in the neighbourhood of the global $\chi^2$ minimum fit and 
gives an access to the uncertainty estimation through a set of PDFs that
describes this neighbourhood. Fixed target Drell--Yan data as well as $W$
asymmetry and jet data from the  Tevatron are used in the fit procedure. 

The corresponding PDFs are constructed as follows: 
(i)  a global fit of the data is performed using the free parameters
$N_{\rm PDF}=20$ for CTEQ and $N_{\rm PDF}=15$ for MRST; this provides the
nominal PDF (reference set) denoted by $S_0$ and corresponding to CTEQ6M and
MRST2001E, respectively; (
(ii) the global $\chi^2$ of the fit  is increased by $\Delta \chi^2=100$
for CTEQ and $\Delta \chi^2=50$ for MRST, to obtain the error matrix [note that
the choice of an allowed tolerance is only intuitive for a global analysis
involving a number of different experiments and processes];  
(iii) the error matrix is diagonalized to obtain $N_{\rm PDF}$ eigenvectors
corresponding to $N_{\rm PDF}$ independent directions in the parameter space; 
(iv) for each eigenvector, up and down excursions are performed in the 
tolerance gap, leading to $2N_{\rm PDF}$ sets of new  parameters, corresponding
to 40 new sets of PDFs for CTEQ and 30 sets for MRST. They are denoted by $S_i$,
with $i=1, 2N_{\rm PDF}$.

To built the Alekhin PDFs \cite{ALEKHIN}, only light-target  deep-inelastic
scattering data [i.e. not the Tevatron data] are used. This PDF set involves 14
parameters, which are fitted simultaneously with $\alpha_s$ and the structure
functions.  To take into account the experimental errors and their
correlations,  the fit is performed by minimizing a $\chi^2$ functional based
on a covariance matrix.  Including the uncertainties on the $\alpha_s$ fit, one
then obtains $2N_{\rm PDF}=30$  sets of PDFs for the uncertainty estimation. 

These three sets of PDFs are used to calculate the uncertainty on
a cross section $\sigma$ in the following way \cite{Samir}: one first evaluates
the cross section with the nominal PDF $S_0$ to obtain  the central value
$\sigma_0$. One then calculates the cross section with  the $S_i$ PDFs, giving 
$2N_{\rm PDF}$ values $\sigma_i$, and defines, for each $\sigma_i$ value, the 
deviations  $\Delta \sigma_i^\pm =\mid \sigma_i -\sigma_0\mid$ when $\sigma_i \
^{>}_{<}  \sigma_0$. The uncertainties are summed quadratically to calculate
{\bf $\Delta  \sigma^\pm  = \sqrt{ \sum_i \sigma_i^{\pm 2} }$}.  The cross
section, including the error, is then given by $\sigma_0|^{+\Delta \sigma^+}_{-
\Delta \sigma^-}$.   \s

This procedure is applied to estimate the cross sections for the production of
the Standard Model Higgs boson in the following four main mechanisms:
\beq
{\rm associate~production~with}~W/Z: & & q\bar{q} \ra VH \\
{\rm massive~vector~boson~fusion}: & & qq \ra   Hqq \\
{\rm the~gluon~gluon~fusion~mechanism}: & & gg  \ra H \hspace*{2cm} \\
{\rm associate~production~with~top~quarks}: & & gg,q\bar{q}\ra t \bar{t} H
\eeq
We will use the Fortran codes {\tt V2HV, VV2H, HIGLU} and {\tt  HQQ} of 
Ref.~\cite{Michael} for the evaluation of the production cross sections of 
processes (1) to (4), respectively, at the LHC. A few remarks are to be 
made in this context:
(i) the NLO QCD corrections to the Higgs-strahlung processes
\cite{NLOHV,M+A}  are practically the same for $WH$ and $ZH$ final states; we
thus simply concentrate on the dominant $q\bar{q} \to WH$ process;
the corrections to $qq \to Hqq$ have been obtained in 
Ref.~\cite{Han:1992hr,Figy:2003nv}
(ii) for the gluon fusion process, $gg \to H$, we include the full  dependence
on the top and bottom quark masses of the NLO cross section \cite{NLOggfull} 
and not only the result in the infinite top  quark mass limit \cite{NLOgg};
(iii) for the $pp \to Ht\bar{t}$  production process, the NLO corrections have
been calculated only recently \cite{NLOHtt} and the programs 
for calculating the cross sections
are not yet publicly available.  However, we choose a scale for which the LO
and NLO cross sections are approximately equal and use the program {\tt HQQ}
for the LO cross section that we fold with the NLO PDFs; 
(iv) finally, we note that the NNLO corrections are also known in the case of 
$q\bar{q} \to HV$  \cite{Brein:2003wg} and $gg \to H$ [in the
infinite top quark mass limit] \cite{NNLOgg} processes. We do not
consider these higher order corrections since the CTEQ and MRST PDFs with
errors are not available at this order. 

\begin{figure}[thbtp]
\begin{center}
\vspace*{-4cm}
\hspace*{-2cm}
\psfig{figure=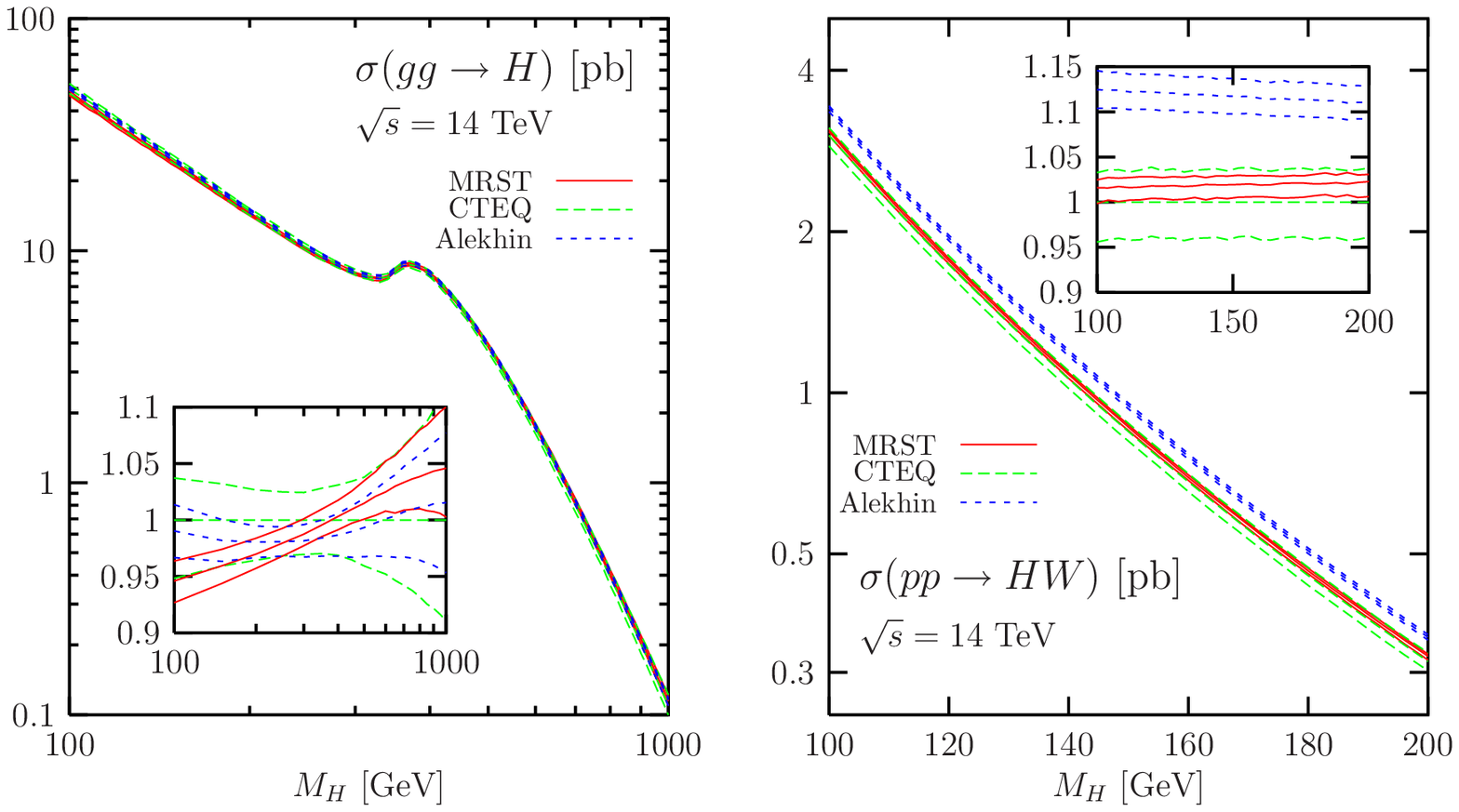,width=20cm}\\[-18cm]
\hspace*{-2cm}
\psfig{figure=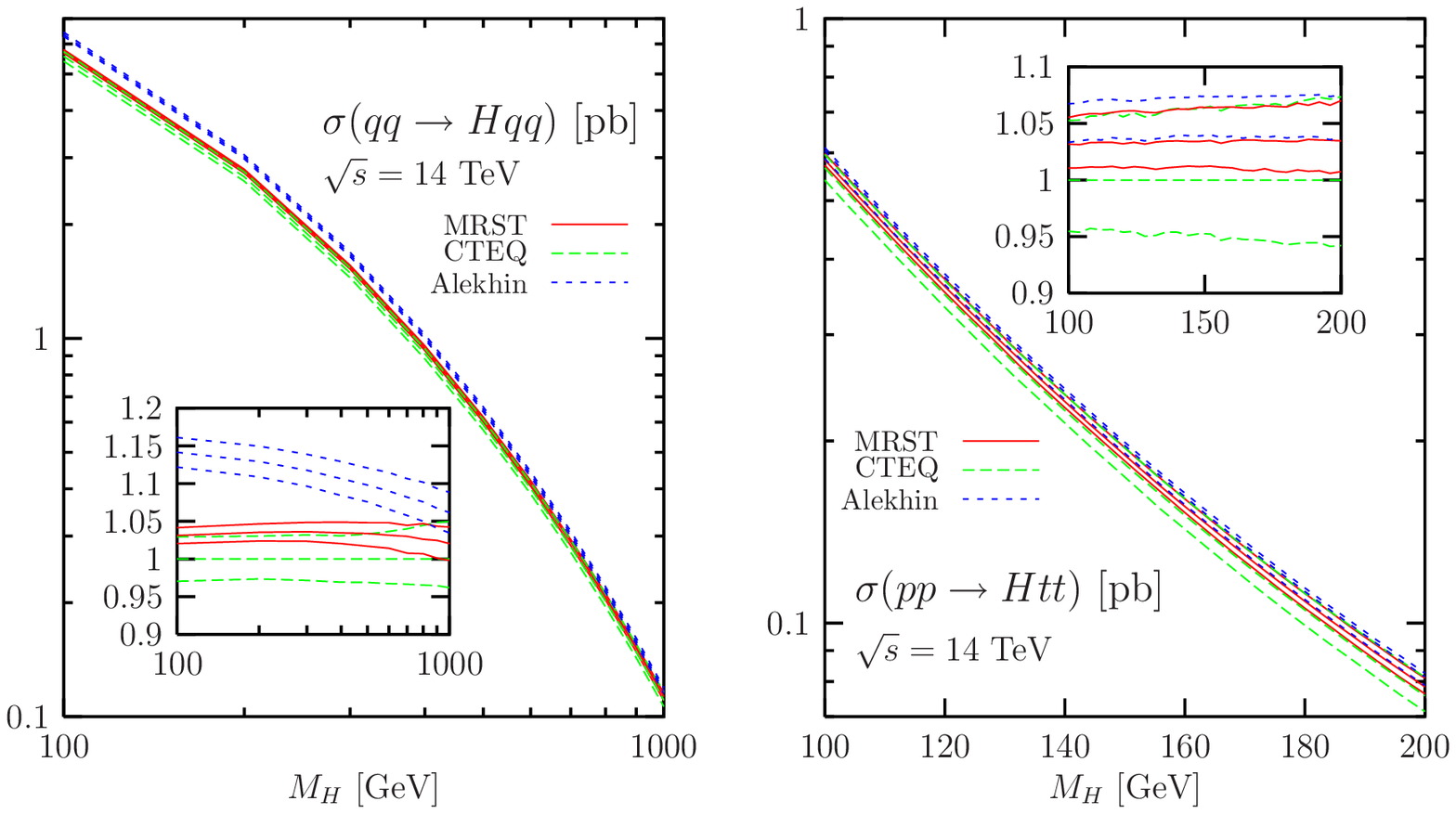,width=20cm}
\vspace*{-17.2cm}
\caption[]{\it 
The CTEQ, MRST and Alekhin PDF uncertainty bands for the NLO
cross sections for the production of the Higgs boson at the LHC 
in the four production processes.  The insets 
show the spread in the predictions, when the NLO cross sections are
normalized to the prediction of the reference CTEQ6M set.}
\label{fig:djouadi1}
\end{center}
%\vspace*{-2mm}
\end{figure}

The behaviour of the Higgs production cross sections and their uncertainties
depends on the considered partons and their $x$ regime discussed above.  In
Fig.~\ref{fig:djouadi1}, 
we present the cross sections for the four production processes 
at the LHC. The  central values and the 
uncertainty band limits of the NLO cross sections are shown for the CTEQ, MRST
and Alekhin parameterizations. In the insets to these figures, we show
the spread uncertainties in the predictions for the NLO cross sections, when
they are normalized to the prediction of the reference CTEQ6M set. Note that
the three sets of PDFs do not use the same value for  $\alpha_s$: at NLO, the
reference sets CTEQ6M, MRST2001C and A02 use, respectively, the values
$\alpha_s^{\rm NLO}(M_Z)= 0.118$, $0.119$  and 0.117. 

By observing Fig.~\ref{fig:djouadi1},
we see that the uncertainties for the Higgs  cross
sections obtained using the CTEQ6 set are two times larger than those using the
MRST2001 sets. This is mainly due to two reasons: first, as noted previously,
the CTEQ collaboration increased the global $\chi^2$ by $\Delta\chi^2=100$ to
obtain the error matrix, while the  MRST collaboration used only
$\Delta\chi^2=50$; second, 2$\times$20 parameter uncertainties are summed
quadratically in CTEQ6, while only 2$\times$15 are used in the MRST case. The
uncertainties from  the Alekhin PDFs are larger than the MRST ones and smaller
than the CTEQ ones. In the subsequent discussion,  the magnitude of the
uncertainty band is expressed  in terms of the CTEQ6 set.  \s

$\bullet$ $q\bar{q} \ra VH$:  the uncertainty band is almost
constant  and is of the order of 4\% [for CTEQ] over a Higgs masse range
between 100 and 200 GeV.  To produce a vector plus a Higgs boson in this 
mass range,  the incoming quarks originate from the intermediate-$x$ regime.
The  different magnitude of the cross sections, $\sim 12$\% larger
in the Alekhin case than for CTEQ, is due to the larger
quark and antiquark densities. 

$\bullet$ $gg  \ra H$: the uncertainty band for the CTEQ set of
PDFs  decreases from the level of about 5\% at $M_{H} \sim 100$ GeV, down to
the 3\% level at $M _H \sim$ 300 GeV.  This is because Higgs bosons with
relatively small masses  are mainly  produced by  asymmetric  low-$x$--high-$x$
gluons  with a low effective c.m. energy; to produce heavier Higgs bosons, a
symmetric process in which the participation of intermediate-$x$ gluons with
high density, is needed, resulting in a smaller  uncertainty band. At higher
masses, $M_H \gsim 300$ GeV, the participation of  high-$x$ gluons becomes more
important, and the uncertainty band increases, to reach the 10\% level at 
Higgs masses of about 1 TeV. 

$\bullet$ $gg/q\bar{q}\ra t\bar{t}H$: at the LHC, the associated production of
the Higgs boson  with a top quark pair is dominantly generated by the 
gluon--gluon fusion mechanism. Compared with the  process $gg  \ra H$ discussed
previously  and for a fixed Higgs boson mass, a larger $Q^2$ is needed for this
final state;  the initial gluons should therefore have higher $x$ values. In
addition, the quarks that are involved in the subprocess $q\bar{q}\ra
t\bar{t}H$, which is also contributing,  are still in  the intermediate regime
because of the higher value $[x \sim 0.7$] at which the quark high-$x$ regime
starts.  This explains why the uncertainty band increases smoothly from 5\% to
7\% when the $M_H$ value increases from 100 to 200 GeV.\s

$\bullet$ $qq \ra   Hqq$: in the entire Higgs boson mass range from 100 GeV  to
1 TeV, the incoming quarks involved in this process originate from the
intermediate-$x$ regime and the uncertainty band is almost constant, ranging
between 3\% and 4\%. 
When using the Alekhin  set of PDFs, the behaviour is different, 
because the quark PDF behaviour is different, as discussed in the case of the
$q\bar{q}  \to HV$  production channel. The decrease in the central value with
higher Higgs boson mass [which is absent in the $q\bar{q} \to HV$ case, since
we stop the $M_H$ variation at 200 GeV] is due to the fact that we reach here
the high-$x$ regime, where the Alekhin $\bar{u}$ PDF drops steeply. \\

In summary,  we have considered three sets of PDFs with uncertainties provided
by the CTEQ and MRST collaborations and by Alekhin. We evaluated their impact
on the total cross sections at next-to-leading-order for the production of the
Standard Model Higgs boson at the LHC. Within a given set
of PDFs, the deviations of the cross sections from the values obtained with the
reference PDF sets are rather small, ${\cal O}(5$\%), in  the case of the
Higgs-strahlung, vector boson fusion and associated  $t\bar{t}H$ production
processes, but they can reach the level of 10\% at the LHC in
the case of the gluon--gluon fusion process for large enough Higgs boson
masses, $M_H \sim 1$ TeV. However, the relative differences
between the cross sections evaluated with different sets of PDFs can be much
larger. Normalizing to the values  obtained with the CTEQ6M set, for instance, 
the cross sections can be different by up to 15\% for the four production
mechanisms.

}

%% file: baur.tex
{
\section[ ]{Measuring the Higgs Self-Coupling\footnote{U.\,Baur, A.\,Dahlhoff,
T.\,Plehn and D.\,Rainwater}}

\subsection{Introduction}

The LHC is widely regarded as capable of directly observing the agent
responsible for electroweak symmetry breaking and fermion mass
generation. This is generally believed to be
a light Higgs boson with mass $m_H<219$~GeV~\cite{lepewwg}. The LHC will
easily find a light Standard Model 
(SM) Higgs boson with very moderate
luminosity~\cite{Dittmar:1997ss,Rainwater:1999sd,Kauer:2000hi,wbf_ww}.
Moreover, the LHC will have significant capability to determine many
of its properties~\cite{unknown:1999fr,Bayatian:1994pu}, such as its fermionic and
bosonic decay modes and couplings~\cite{ATL-PHYS-2003-030,Rainwater:2000fm,Drollinger:2002uj,Yt,Maltoni:2002jr,Belyaev:2002ua}. 
An $e^+e^-$ linear
  collider with a center of mass energy of 350~GeV or more can
  significantly improve these preliminary measurements, in some cases
  by an order of magnitude in precision, if an integrated luminosity
  of 500~fb$^{-1}$ can be achieved~\cite{LC}.

Perhaps the most important measurement after a Higgs
boson discovery is of the Higgs potential itself, which requires
measurement of the trilinear and quartic Higgs boson self-couplings.
Only multiple Higgs
boson production can probe these directly~\cite{Dicus:1988ic,Glover:1988nx,Boudjema:1996cb,Ilyin:1996iy,Djouadi:1999gv,Miller:1999ct,Battaglia:2001nn}.
Phenomenologically one should write an effective Lagrangian that does
not already assume SM couplings, as the object at hand could be a
radion or other Higgs boson-like field that has different tree-level
self-couplings. Only after the potential is measured can it be decided
what the candidate actually is. We take the Lagrangian as the
effective potential
\begin{equation}
\label{eq:Hpot}
V(\eta_H) \, = \, 
{1\over 2}\,m_H^2\,\eta_H^2\,+\,\lambda\, v\,\eta_H^3\,+\,{1\over 4}\,
\tilde\lambda\,\eta_H^4 ,
\end{equation}
where $\eta_H$ is the physical Higgs field, $v=(\sqrt{2}G_F)^{-1/2}$ is the
vacuum expectation value, and $G_F$ is the Fermi constant. In the SM, 
\begin{equation}
\label{eq:lamsm}
\tilde\lambda=\lambda=\lambda_{SM}={m_H^2\over 2v^2}\,.
\end{equation}
Since future collider
experiments likely cannot probe $\tilde\lambda$, we concentrate on the
trilinear coupling $\lambda$ in the following.  The quartic Higgs
coupling does not affect the Higgs pair production processes we
consider.

There are numerous quantitative sensitivity limit analyses of Higgs
boson pair production in $e^+e^-$ collisions ranging from 500~GeV to
3~TeV center of mass energies~\cite{Boudjema:1996cb,Ilyin:1996iy,Djouadi:1999gv,Miller:1999ct,Battaglia:2001nn,LC_HH4}. In the
past two years, several studies exploring the potential of
the LHC, a luminosity-upgraded LHC (SLHC) with roughly
ten times the amount of data expected in the first run, and a
Very Large Hadron Collider (VLHC), were carried 
out~\cite{SLHC,Baur:2002rb,Baur:2002qd,blondel,Baur:2003gp,Baur:2003gp1}. In the following
we briefly summarize our~\cite{Baur:2002rb,Baur:2002qd,Baur:2003gp,Baur:2003gp1} studies of
Higgs pair production at 
hadron colliders, concentrating on the LHC and SLHC, and compare the
capabilities of future hadron and 
lepton colliders to measure $\lambda$. We also present a new estimate of
the $t\bar{t}j$ background,
and the results of a first study of how QCD corrections affect the
signal in Higgs pair production for $m_H>140$~GeV.

\subsection{Higgs Pair Production at Hadron Colliders}

At LHC energies, inclusive Higgs boson pair production is dominated by
gluon fusion~\cite{lhc_hh,Djouadi:1999rc}.  Other processes, such as weak boson
fusion, $qq\to qqHH$~\cite{Dobrovolskaya:1991kx,Dicus:1988ez,Abbasabadi:1988bk,Keung:1987nw}, associated production with heavy
gauge bosons, $q\bar{q}\to WHH, ZHH$~\cite{assoc}, or
associated production with top quark pairs, $gg,\,q\bar{q}\to
t\bar{t}HH$~\cite{SLHC}, yield cross sections which are factors of
10--30 smaller than that for $gg\to HH$.  Since $HH$
production at the LHC is generally rate limited, we consider only the
gluon fusion process.

For $m_H<140$~GeV, the dominant decay mode of the SM Higgs boson is
$H\to b\bar{b}$. In this region, the $b\bar{b}\gamma\gamma$ final state
offers the best prospects to measure the Higgs
self-coupling~\cite{Baur:2003gp1}; other final states such as $4b$,
$b\bar{b}\tau^+\tau^-$ and $b\bar{b}\mu^+\mu^-$ are overwhelmed by
backgrounds~\cite{Baur:2003gp,Baur:2003gp1}. 

For $m_H>140$~GeV, $H\to W^+W^-$
dominates, and the $W^+W^-W^+W^-$ final state has the largest
individual branching ratio. Here, the
$(jj\ell^\pm\nu)(jj{\ell'}^\pm\nu)$ final state offers the best chance to
extract information on $\lambda$~\cite{SLHC,Baur:2002rb,Baur:2002qd,blondel}.

\subsubsection{$m_H<140$~GeV: The $b\bar{b}\gamma\gamma$ decay channel}
\label{sec:light}

The Feynman diagrams contributing to $gg\to HH$ in the SM consist of
fermion triangle and box diagrams~\cite{Dicus:1988ic,Glover:1988nx}. 
%(see Fig.~\ref{fig:fig1})
%
%\begin{figure}[t] 
%\begin{center}
%\includegraphics[width=13cm]{fig1.eps}
%\vspace*{2mm}
%\caption[]{\label{fig:fig1} 
%Representative Feynman diagrams for the process $gg\to HH$. }
%\end{center}
%\end{figure}
%
Non-standard Higgs boson self-couplings only affect the triangle
diagrams with a Higgs boson exchanged in the $s$-channel. We calculate
the $gg\to HH\to b\bar{b}\gamma\gamma$ cross section using exact loop
matrix elements~\cite{Dicus:1988ic,Glover:1988nx}. Signal results are computed
consistently to leading order QCD with the 
top quark mass set to $m_t=175$~GeV and the renormalization and
factorization scales are taken to be $m_H$. The effects of
next-to-leading order (NLO) QCD corrections are included via a multiplicative
factor $K=1.65$~\cite{hh_nlo}.

The kinematic acceptance cuts for events are:
\begin{eqnarray}
\label{eq:cuts1}
\nonumber &
p_T(b) > 45~{\rm GeV} \; , \qquad
|\eta(b)| < 2.5 \; , \qquad
\Delta R(b,b) > 0.4 \; , \\
\nonumber &
m_H-20~{\rm GeV} \, < \, m_{b\bar{b}} \, < \, m_H+20~{\rm GeV} \; , \\
\nonumber &
p_T(\gamma) > 20~{\rm GeV} \; , \qquad
|\eta(\gamma)| < 2.5 \; , \qquad
2.0> \Delta R(\gamma,\gamma) > 0.4 \; , \qquad \Delta R(\gamma,b) > 1.0 \; ,\\
\nonumber &
m_H-2.3~{\rm GeV} \, < \, m_{\gamma\gamma} \, < \, m_H+2.3~{\rm GeV} \; , 
\end{eqnarray}
motivated by the requirement that the events can pass the
ATLAS and CMS triggers with high efficiency~\cite{unknown:1999fr,Bayatian:1994pu},
and that the $b$-quark and photon pairs reconstruct to windows around
the known Higgs boson mass. We take the identification
efficiency for each photon to be $80\%$ and
assume that $b$-quarks are tagged with an efficiency of
$\epsilon_b=50\%$. The $\Delta R(\gamma,\gamma)$ and $\Delta
R(\gamma,b)$ cuts help to reduce the background such 
that $S/B\sim 1/2$ (1/1) is possible at the LHC
(SLHC)~\cite{Baur:2003gp1}. 

The only irreducible background processes are
QCD $b\bar{b}\gamma\gamma$, $H(\to\gamma\gamma)b\bar{b}$ and $H(\to
b\bar{b})\gamma\gamma$ production.
However, there are multiple QCD reducible backgrounds
resulting from jets faking either $b$-jets or photons, such as 4~jet
production (one or two fake $b$-jets, two fake photons) or
$b\bar{b}j\gamma$ production (one fake photon)~\cite{Baur:2003gp1}. We
simulate these backgrounds assuming a misidentification probability of
light jets as $b$-quarks of 
$P_{j\to b}=1/140$ (1/23) at the LHC (SLHC), and a jet -- photon
misidentification probability in the range
$1/2500<P_{j\to\gamma}<1/1600$. With these parameters, most of the
background originates from reducible sources.

Almost all reducible backgrounds depend on whether one requires one or both
$b$-quarks to be tagged.  Requiring only one tagged $b$-quark results
in a signal cross section which is a factor $(2/\epsilon_b-1)=3$
larger than the one with both $b$-quarks tagged.  This larger
signal rate comes at the expense of a significantly increased
reducible background. The small $gg\to HH\to
b\bar{b}\gamma\gamma$ rate forces us to require only a single
$b$-tag at the LHC in order to have an observable signal.  At the
SLHC, on the other hand, the much higher probability to misidentify a
light jet as a $b$-jet translates into an increase of the background
which more than compensates the signal gain from using only a single
$b$-tag.  In the following we therefore require a double $b$-tag at
the SLHC. 

To discriminate between signal and background, we use the visible
invariant mass, $m_{\rm vis}$, which for this final state is the invariant
mass of the Higgs boson pair, corrected for energy loss of the
$b$-jets.  We show this in Fig.~\ref{fig:mvis-LHC} for $m_H=120$~GeV
at the LHC and SLHC.
\begin{figure}[t!]
\begin{center}
\begin{tabular}{cc}
\includegraphics[width=7.7cm]{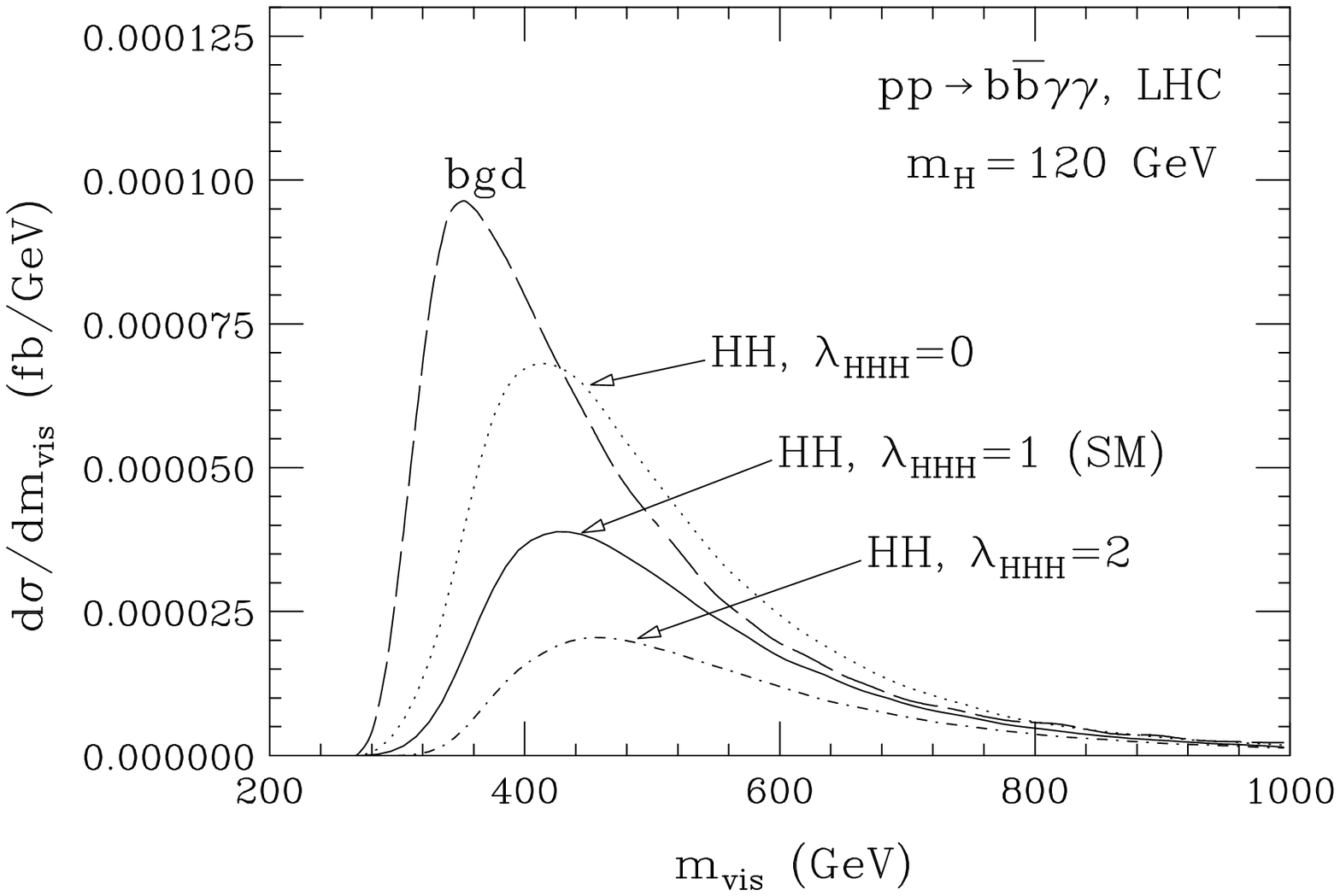} &
\includegraphics[width=7.7cm]{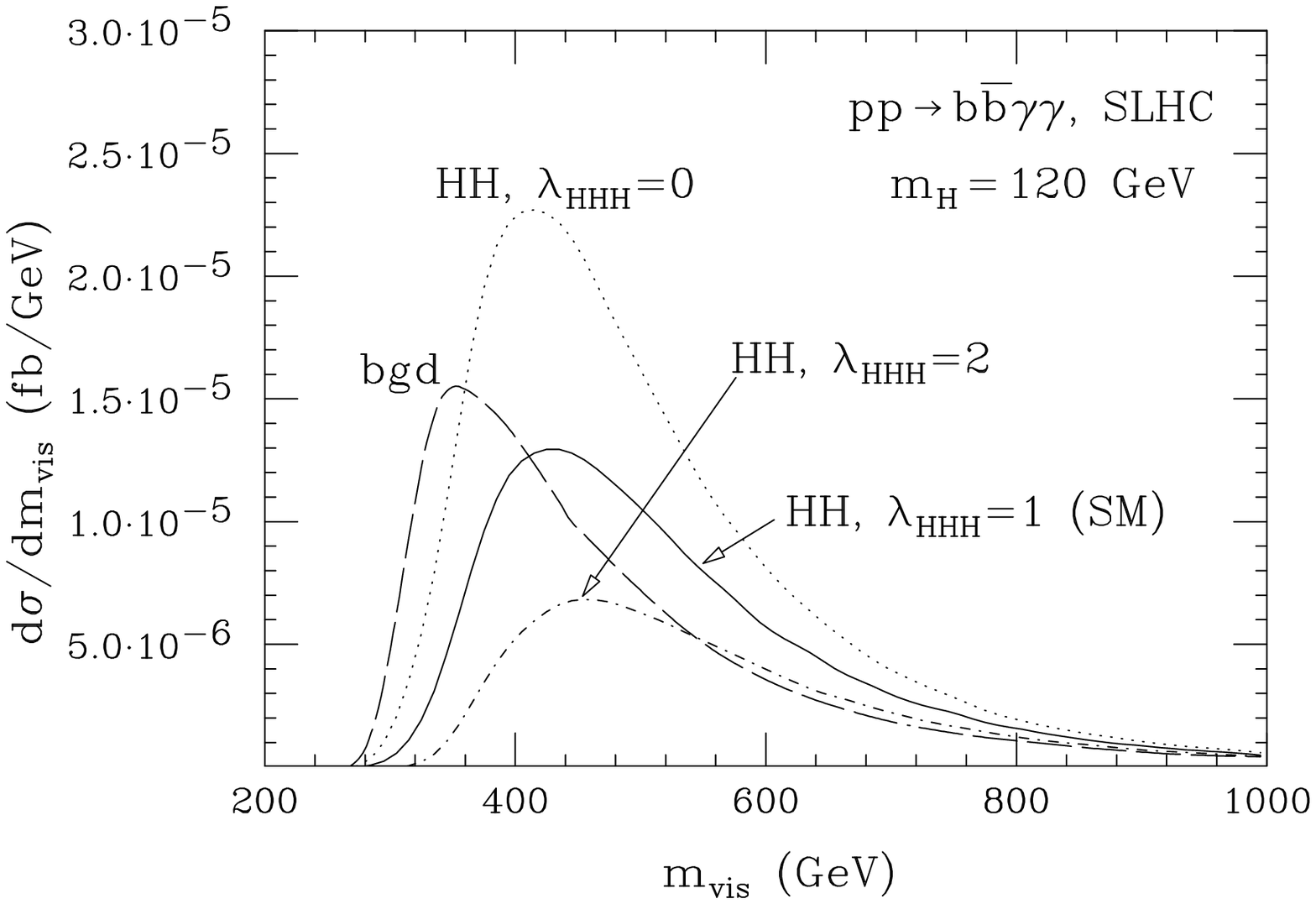} 
\end{tabular}
\caption[]{\label{fig:mvis-LHC}
  The visible invariant mass distribution, $m_{\rm vis}$, in $pp\to
  b\bar{b}\gamma\gamma$, after all kinematic cuts, for
  the QCD backgrounds (long dashed) and a SM signal of $m_H=120$~GeV
  (solid) at the LHC and SLHC. We assume $P_{j\to\gamma}=1/2500$ in the
background calculation. 
  The dotted and short dash-dotted lines show the signal cross section
  for $\lambda_{HHH}=\lambda/\lambda_{SM}=0$ and 2, respectively. }  
\vspace{-7mm}
\end{center}
\end{figure}
Performing a $\chi^2$ test of the $m_{\rm vis}$ distribution, one finds
the following $1\sigma$ sensitivity bounds for $m_H=120$~GeV:
\begin{eqnarray}
\label{eqn:one}
-1.1 <\Delta\lambda_{HHH}<1.6 & {\rm LHC~600~fb}^{-1}, \\
-0.62 <\Delta\lambda_{HHH}<0.74 & {\rm SLHC~6000~fb}^{-1}, \label{eqn:two}
\end{eqnarray}
where $\Delta\lambda_{HHH}=\lambda/\lambda_{SM}-1$. For $m_H=140$~GeV,
the SLHC could obtain bounds which are about a factor~2 weaker than
those for $m_H=120$~GeV; there are not enough signal events at
the LHC to derive sensitivity limits for $m_H=140$~GeV. If the
photon--jet and light jet--$b$ 
misidentification probabilities can be independently measured in
other processes, one can subtract large parts of the reducible
backgrounds which do not involve charm quarks. This may improve the
sensitivity limits by a factor $1.5-2$. Due to the small number of
events, the LHC and SLHC sensitivity limits 
depend significantly on the SM cross section normalization
uncertainty.  The bounds listed in Eqs.~(\ref{eqn:one})
and~(\ref{eqn:two}) have been calculated assuming a normalization
uncertainty of $10\%$ for the SM (signal plus background) cross section. 
This depends critically on knowledge of
the signal QCD corrections and the ability to determine the
background normalization.  The NLO QCD corrections to $gg\to HH$ are
currently known only in the infinite top quark mass
limit~\cite{hh_nlo}.  To ensure the $10\%$ required precision on 
differential cross sections we would need the NLO rates for finite top 
quark masses, as well as the NNLO corrections at least in the heavy top quark 
mass limit. For the background normalization one can either rely on 
calculations of the QCD corrections (which do not exist yet) or perform
a sideband analysis of the data. 

We should compare the bounds listed in Eqs.~(\ref{eqn:one}) 
and~(\ref{eqn:two}) with those 
achievable at $e^+e^-$ linear colliders.  A linear collider with
$\sqrt{s}=500$~GeV and an integrated luminosity of 1~ab$^{-1}$ can
determine $\lambda$ with a precision of about $20\%$ ($50\%$) in $e^+e^-\to
ZHH$ for $m_H=120(140)$~GeV~\cite{LC_HH4,Baur:2003gp}.  From
Eq.~(\ref{eqn:one}) it is clear 
that the LHC will be able to provide only a first rough measurement of
the Higgs self-coupling for $m_H=120$~GeV.  The SLHC
would be able to make a more precise measurement.  However, the
sensitivity bounds on $\lambda$ obtained from $b\bar{b}\gamma\gamma$
production for $m_H=120(140)$~GeV will be a factor~2 --~4
(1.2 --~3) weaker than those achievable at a linear collider. Although
a luminosity-upgraded LHC cannot compete with a linear collider for
Higgs masses $m_H<140$~GeV, a Higgs self-coupling measurement at the
SLHC would still be interesting if realized before a linear collider
begins operation.

We finally note that the $b\bar{b}\gamma\gamma$ final state is
particularly interesting in 
the MSSM framework, because it is the only channel in which we can hope 
to observe a heavy Higgs state for small values of $\tan\beta$. However,
we do not include any detailed analysis, because the MSSM search
strategy does not differ significantly from the SM case described below.

\subsubsection{$m_H>140$~GeV: The same sign dilepton channel}

A thorough analysis of $gg\to HH\to (W^+W^-)(W^+W^-)\to
(jj\ell^\pm\nu)(jj{\ell'}^\pm\nu)$ was presented in
Refs.~\cite{Baur:2002rb,Baur:2002qd}. After a brief review of the main results of
this analysis, we present a reevaluation of the $t\bar{t}j$ background
and the results of a preliminary study of how initial state gluon
radiation affects the $m_{vis}$ distribution of the signal which is used
to extract limits on $\lambda$.

In our analysis~\cite{Baur:2002rb,Baur:2002qd}, we perform the
calculation of the $gg\to HH\to (W^+W^-)(W^+W^-)\to 
(jj\ell^\pm\nu)(jj{\ell'}^\pm\nu)$ signal cross section as in
Sec.~\ref{sec:light}. The kinematic acceptance cuts are:
\begin{eqnarray}
\label{eq:cuts3}
&p_T(j) > 30,\, 30,\, 20,\, 20~{\rm GeV} , \qquad 
p_T(\ell) > 15,\,15~{\rm GeV}    ,        \\
&|\eta(j)| < 3.0     ,        \qquad \qquad \qquad \qquad 
|\eta(\ell)| < 2.5   ,              \\
&\Delta R(jj) > 1.0 ,   \qquad 
\Delta R(j \ell) > 0.4 , \qquad 
\Delta R(\ell \ell) > 0.2 .
\label{eq:cuts4}
\end{eqnarray}
In
addition we require the four jets to combine into two pseudo-$W$ pairs
with invariant masses $50~{\rm GeV} < m(jj) < 110~{\rm GeV}$,
and assume that this captures $100\%$ of the signal and backgrounds.
We do not impose a missing transverse momentum cut which would remove
a considerable fraction of the signal events; it is unnecessary for
this analysis.

The relevant SM backgrounds are those that produce two same-sign
leptons and four well-separated jets which reconstruct to two pairs, each in
a window around the $W$ boson mass. The largest contribution
originates from $W^{\pm}W^+W^-jj$ production (which includes 
$W^\pm H(\to W^+W^-)jj$), followed by
$t\bar{t}W^\pm$ where one top quark decays leptonically, the other
hadronically, and neither $b$ quark jet is tagged. Other backgrounds
are: $W^\pm W^\pm jjjj$;
$t\bar{t}t\bar{t}$, where none of the $b$ quark jets are
tagged, and additional jets or leptons are not observed; $W^\pm Z
jjjj$, $t\bar{t}Z$ and $W^+W^-Zjj$ with leptonic $Z$ decay
(including off-shell photon interference) where one lepton is not
observed; and $t\bar{t}j$ events where one $b$ quark decays
semileptonically with good hadronic isolation and the other is not
tagged. In Ref.~\cite{Baur:2002qd} we found that the $t\bar{t}j$ channel
contributes $10\%$ or less of the total background.

To discriminate signal from background we again use the $m_{vis}$ distribution,
i.e. the distribution of the $\ell^\pm{\ell'}^\pm+4j$ invariant 
mass, which we show for $m_H=180$~GeV in Fig.~\ref{fig:three}a. 
\begin{figure}[t!]
\begin{center}
\begin{tabular}{cc}
\includegraphics[width=7.7cm]{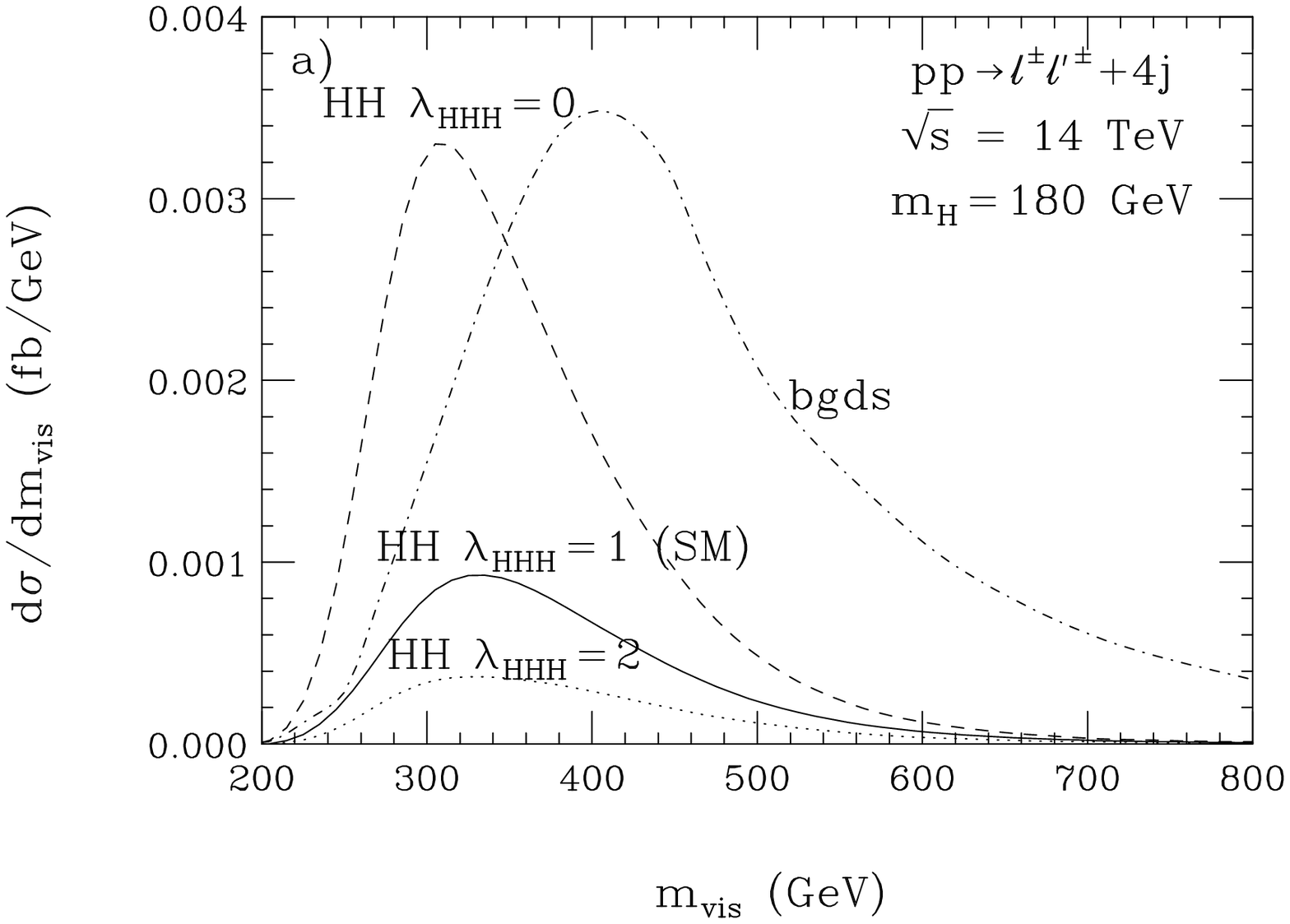} &
\includegraphics[width=7.cm]{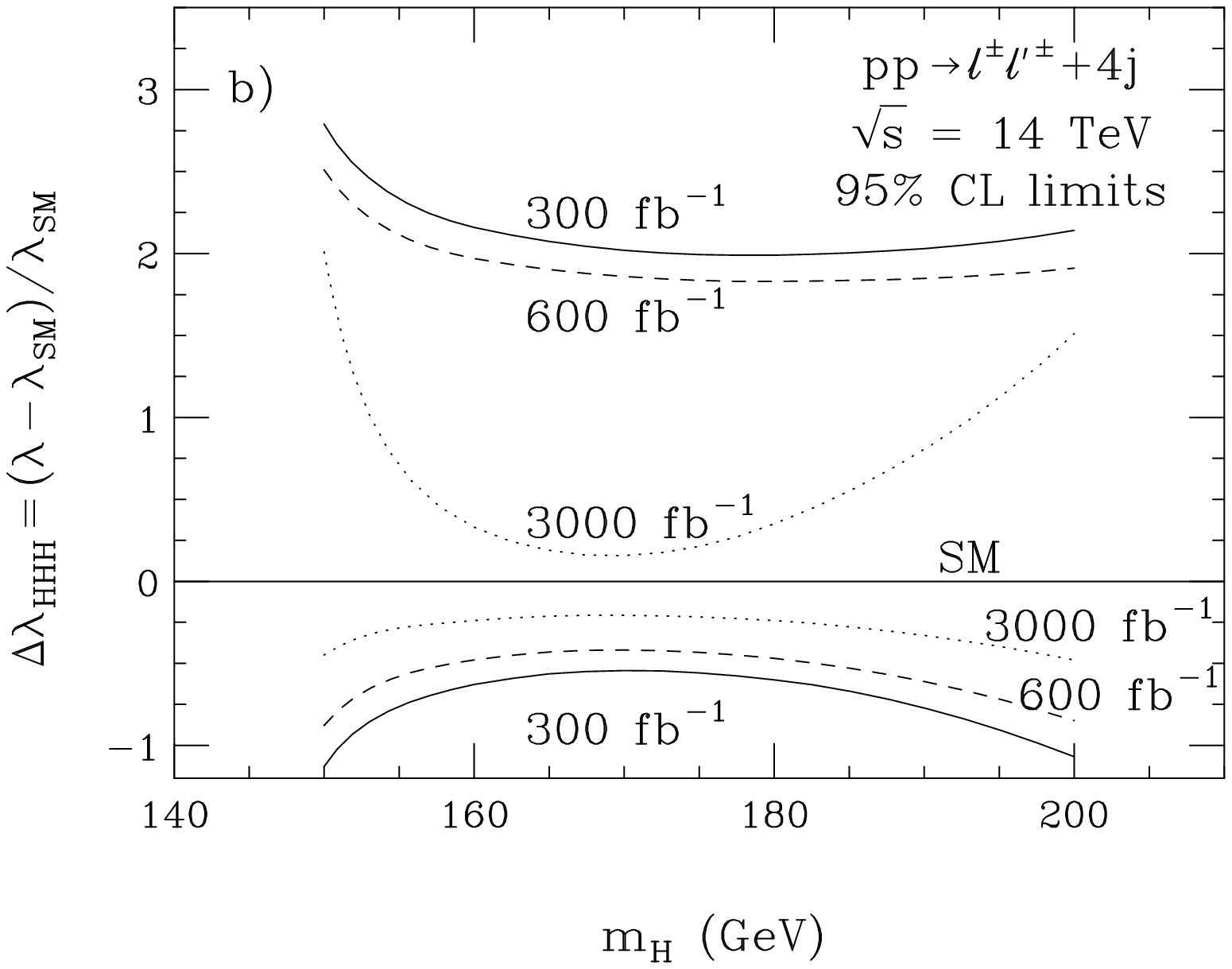} 
\end{tabular}
\caption[]{\label{fig:three}
a)  The $m_{vis}$ distribution of the signal for $pp\to
  \ell^\pm{\ell'}^\pm+4j$ and $m_H=180$~GeV at the LHC, in the SM (solid
curve), for 
  $\lambda_{HHH}=\lambda/\lambda_{SM}=0$ (dashed line) and for
  $\lambda_{HHH}=2$ (dotted line).  The dot-dashed line shows the
  combined $m_{vis}$ distribution of all background processes.
  We obtain qualitatively similar results for other values of
  $m_H$. \\
b)   Limits achievable at $95\%$ CL for
  $\Delta\lambda_{HHH}=(\lambda-\lambda_{SM})/\lambda_{SM}$ in
  $pp\to\ell^\pm{\ell'}^\pm+4j$ at the (S)LHC. Bounds are shown for
  integrated luminosities of 300~fb$^{-1}$ (solid lines),
  600~fb$^{-1}$ (dashed lines) and 3000~fb$^{-1}$ (dotted lines). The
  allowed region is between the two lines of equal texture.  The Higgs
  boson self-coupling vanishes for $\Delta\lambda_{HHH}=-1$.
  The figures are taken from Ref.~\cite{Baur:2002qd}.}  
\vspace{-7mm}
\end{center}
\end{figure}
The background distribution peaks at a significantly
higher value than the signal. Performing a $\chi^2$ test for the
$m_{vis}$ distribution, we find the $95\%$ CL limits shown in
Fig.~\ref{fig:three}b. The results of Ref.~\cite{Baur:2002rb,Baur:2002qd}
demonstrate that, with 300~fb$^{-1}$ at the LHC, 
one will be able to perform a first, albeit not very
precise, measurement of the Higgs boson self-coupling. The
non-vanishing of $\lambda$, however, can be established at $95\%$ CL
or better for $150~{\rm GeV}\lesssim m_H\lesssim 200$~GeV.  {\it This alone is an
  important, non-trivial test of spontaneous symmetry breaking}; the
exact non-zero value of $\lambda$ may vary depending on the way nature
chooses to spontaneously break the electroweak symmetry.  At the SLHC,
for 3000~fb$^{-1}$, a measurement with a precision of up to $20\%$ at
$95\%$ CL would be possible; the SLHC could determine $\lambda$ with an
accuracy of $10-30\%$ at the $1\sigma$ level for Higgs boson masses
between 150 and 200~GeV. 

Due to phase space restrictions, a center of mass energy of at least
800~GeV would be needed to search for Higgs pair production in
$e^+e^-$ collisions if $m_H\geq 150$~GeV. For $\sqrt{s}=0.8-1$~TeV,
$e^+e^-\to ZHH\to 10~{\rm jets},\,\ell\nu+8~{\rm jets}$ via Higgs boson
decays into weak boson pairs are the dominant Higgs pair production
channels. The main contributions to the background originate from
$t\bar{t}\:+$~jets and $WW+$~jets production, with cross sections
several orders of magnitude larger than the signal. As a result, it will be
difficult to determine the Higgs boson self-coupling at a linear
collider with $\sqrt{s}=0.8-1$~TeV with a precision equal to that
which can be reached at the LHC with 300~fb$^{-1}$~\cite{Baur:2003gp}. 

\subsubsection{Toward a more complete simulation of the
$(jj\ell^\pm\nu)(jj{\ell'}^\pm\nu)$ channel}

In Ref.~\cite{Baur:2002qd}, the $t\bar{t}j$ background was estimated from a 
parton level calculation which took only $b\to c\ell\nu$ 
decays into account; $b\to u\ell\nu$ decays were
ignored. Furthermore, the $\chi^2$ analysis used to extract sensitivity
bounds for $\lambda$ in Ref.~\cite{Baur:2002rb,Baur:2002qd} assumed
that higher order QCD corrections do not significantly alter the shape
of the  $m_{vis}$ distribution for signal and background. In the
following, we present a new estimate of the $t\bar{t}j$
background which includes the contribution from $b\to u\ell\nu$ decays
and address the question of how higher order QCD corrections 
change the shape of the $m_{vis}$ distribution of the signal.

\paragraph{\it Re-evaluation of the $t\bar{t}j$ background} ~\\[-0.3cm]

Since the $t\bar{t}j$ cross section is several orders of magnitude larger
than the $HH$ signal, it is crucial to suppress it sufficiently. This is
accomplished by requiring isolation of the lepton originating from
semileptonic $b$-decay. Due to phase space restrictions, 
the $t\bar{t}j$ background is extremely sensitive to
the $p_T(\ell)$ cut and the lepton isolation criteria. A typical lepton
isolation cut limits the energy fraction carried by the charm or
$u$-quark from $b$-decay in a cone around the lepton, or imposes an
upper limit on its transverse momentum. It is easy to show~\cite{dieter}
that, for $b\to c\ell\nu$ decays, the kinematics limits the transverse
momentum of the lepton to 
\begin{equation}
\label{eq:ptmax}
p_T(\ell)\leq {m_B^2\over m_D^2}~p_{Tmax}(c),
\end{equation}
where $p_{Tmax}(c)$ is the maximal transverse momentum of the charm
quark allowed in the isolation cone, and $m_B$ and $m_D$ are the $B$ and
$D$ meson masses which are used to approximately obtain the correct
kinematics. In Ref.~\cite{Baur:2002qd}, $p_{Tmax}(c)=3$~GeV
and a cone size of 
$\Delta R=0.4$ were used, implying that $p_T(\ell)<24$~GeV. 
The large suppression of the $t\bar{t}j$ background noted in our previous
analysis, and 
its extreme dependence on the  $p_T(\ell)$ cut, thus is entirely due to
phase space suppression. In fact, for $p_{Tmax}(c)=1$~GeV which was used in
Ref.~\cite{blondel}, the $t\bar{t}j$ cross section would vanish for the cuts
listed in Eq.~(\ref{eq:cuts3}).

From Eq.~(\ref{eq:ptmax}) it is obvious that the phase space 
is much less suppressed for $b\to u\ell\nu$ decays. There, $m_D$ in
Eq.~(\ref{eq:ptmax}) has to be replaced by either the $\pi$ or $\rho$
mass, allowing much larger lepton transverse momenta. Although $b\to
u\ell\nu$ decays are suppressed by the ratio $(V_{ub}/V_{cb})^2\approx
8\times 10^{-3}$ with respect to $b\to c\ell\nu$, there are regions of
phase space where contributions from $b\to u\ell\nu$ decays dominate
over those from $b\to c\ell\nu$ in the $t\bar{t}j$ background. 

We estimate the
$t\bar{t}j$ background with $b\to u\ell\nu$ decays using the approach
described in Ref.~\cite{Baur:2002qd}, the measured
$B\to\pi\ell\nu$ branching fractions~\cite{PDG}, and assuming that all
remaining $b\to u\ell\nu$ decays result in final state hadrons with an
invariant mass $\geq m_\rho$. Taking into account the uncertainties in
$V_{ub}$ ($V_{ub}=0.0036\pm 0.0007$~\cite{PDG}) we find, imposing the
cuts listed 
in Eqs.~(\ref{eq:cuts3}) and~(\ref{eq:cuts4}),
\begin{equation}
\label{eq:ttj}
\sigma(t\bar{t}j, b\to u\ell\nu)= 0.76\pm 0.28~{\rm fb}
\end{equation}
if one requires $p_T(u)<3$~GeV in a cone of $\Delta R=0.4$
around the charged lepton. 
This should be compared with $\sigma(t\bar
tj, b\to c\ell\nu)=0.08$~fb for $p_T(c)<3$~GeV obtained in
Ref.~\cite{Baur:2002qd}. Taking into account $b\to u\ell\nu$ decays thus
increases the $t\bar{t}j$ background cross section by a factor $6-12$ and
the total background by a factor $1.4-2.2$. This is expected to weaken
the limits on $\lambda$ by a factor $1.2-1.4$. If the $p_T$ threshold is
lowered to 
$p_T(u)<1$~GeV, one finds $\sigma(t\bar{t}j, b\to u\ell\nu)=0.33\pm
0.12$~fb. In this case the sensitivity limits worsen by a
factor~1.2 at most. 

We also note that the $t\bar{t}j$, $b\to u\ell\nu$ background cross
section significantly depends on the size of the cone used in the
isolation of the lepton. Reducing the cone size from $\Delta R=0.4$ to 
$\Delta R=0.2$, for example, increases the $t\bar{t}j$ cross section by
approximately a factor~6 for $p_T(u)<1$~GeV. On the other hand, if the
isolation cone is 
increased to $\Delta R=0.5$, the $t\bar tj$ cross section is reduced by
a factor~2.

The cross sections for the $t\bar{t}j$ background listed in
Eq.~(\ref{eq:ttj}) assume vertex tagging of the hadronically decaying
$b$-quark, rejecting events with a factor~2, which approximates the
fraction where the $b$-quark would be tagged. 
We made no such assumption for the semileptonically decaying
$b$. Requiring a small impact parameter for the lepton originating from
$b$-decay may result in an additional suppression of the $t\bar{t}j$
background.  

We emphasize that our matrix element-based estimate of the $t\bar{t}j$
background should be viewed with some caution. Our treatment of the
phase space in $b\to u\ell\nu$ clearly is oversimplified. Furthermore, 
effects from
hadronization, event pileup and extra jets from initial or final state
radiation, as well as detector resolution effects may significantly
affect the rejection. For a reliable estimate of the background, a
full detector simulation is required, which is best carried out by
interfacing a matrix element based calculation of $t\bar{t}j$ production
with an event generator such as {\sc pythia}. This is now underway.

\paragraph{\it Initial state gluon radiation in $HH$ production} ~\\[-0.3cm]

To investigate how extra jets from initial state gluon
radiation affect the shape of the $m_{vis}$ distribution of the signal,
we used an interface of the {\sc hpair} program with 
{\sc pythia}~\cite{lafaye}. The 
results for $m_H=200$~GeV are shown in Fig.~\ref{fig:four}.
\begin{figure}[t] 
\begin{center}
\includegraphics[width=8.4cm]{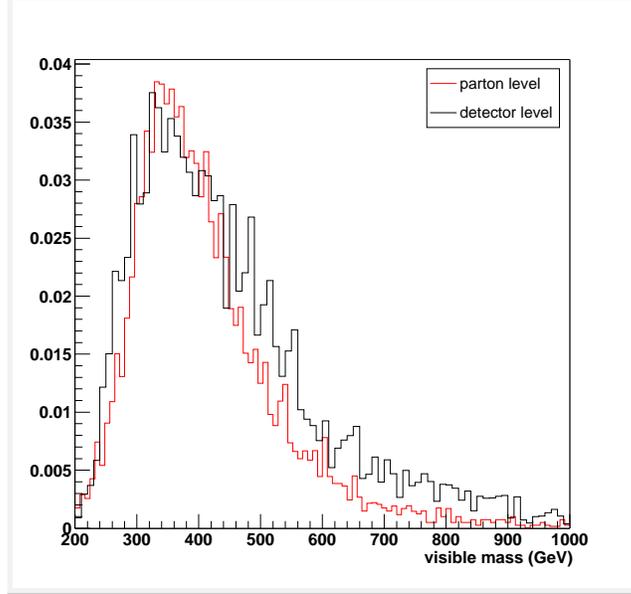}
\caption[]{\label{fig:four} 
The $m_{vis}$ distribution for $gg\to HH+X\to
(jj\ell^\pm\nu)(jj{\ell'}^\pm\nu)+X$ with $m_H=200$~GeV at the LHC
obtained from interfacing the {\sc hpair} program with {\sc pythia}. The
red histogram (``parton level'') shows the result obtained using the
momenta of the four 
jets from $W$ decays in calculating the visible invariant mass. In the
black histogram, the momenta of the 
four jets with the highest transverse momenta in the event are used in
computing $m_{vis}$. The units on the vertical axis are arbitrary.}
\end{center}
\end{figure}
The red histogram, labeled ``parton level'', displays the result obtained
when the momenta of the four 
jets which originate from $W$ decays are used to calculate $m_{vis}$, 
and corresponds to the result of a lowest order calculation. For the
black histogram, the momenta of the 
four jets with the highest transverse momenta in the event are used to
compute 
$m_{vis}$. Frequently, one of these jets originates from initial state
gluon radiation. Fig.~\ref{fig:four} demonstrates that, while QCD corrections
broaden the $m_{vis}$ distribution somewhat, the location of the peak
and the shape of the distribution remain essentially unchanged. 

Similar calculations have to be carried out for the main background
channels, $WWWjj$ and $t\bar{t}W$ production, before firm conclusions 
how QCD corrections affect the sensitivity limits for $\lambda$ can be
drawn. 

\subsection{Conclusions}

After discovery of an elementary Higgs boson and
tests of its fermionic and gauge boson couplings, experimental
evidence that the shape of the Higgs potential has the form required
for electroweak symmetry breaking will complete the proof that fermion
and weak boson masses are generated by spontaneous symmetry breaking.
One must determine the Higgs self-coupling to probe the shape of the
Higgs potential.

Only Higgs boson pair production at colliders can accomplish this.
Numerous studies~\cite{Boudjema:1996cb,Ilyin:1996iy,Djouadi:1999gv,Miller:1999ct,Battaglia:2001nn,LC_HH4} have established
that future $e^+e^-$ machines can measure $\lambda$ at the $20-50\%$
level for $m_H<140$~GeV. A recent study has shown that a measurement of
the Higgs self-coupling at a luminosity upgraded LHC in this mass range
using the $b\bar{b}\gamma\gamma$ final state is also possible. Although
the SLHC cannot compete with a linear collider in this mass range, a 
Higgs self-coupling measurement at the
SLHC will still be interesting if realized before a linear collider
begins operation.

While a measurement of the Higgs self-coupling at a linear collider
for $m_H>140$~GeV  requires a center of mass energy larger than
1~TeV~\cite{albert}, the LHC may rule out the case of vanishing $\lambda$ for
$150~{\rm GeV}<m_H<200$~GeV at $95\%$ CL. At the SLHC,
for 3000~fb$^{-1}$, a measurement with a precision of up to $20\%$ at
$95\%$ CL is possible. These sensitivity limits were derived from
a parton level analysis of the visible invariant mass distribution in
$pp\to\ell^\pm{\ell'}^\pm+4j$. Major uncertainties in this analysis are
the size of the $t\bar{t}j$ background and how initial state gluon
radiation affects the $m_{vis}$ distribution. First, but still preliminary,
results indicate that the shape of the signal $m_{vis}$ distribution
is broadened slightly by QCD corrections. Since $b\to u\ell\nu$ decays
were originally neglected, the $t\bar{t}j$ background was underestimated
by a factor $3-6$ in Ref.~\cite{Baur:2002qd}. This is expected to weaken the
sensitivity bounds on $\lambda$ by up to $20\%$. A more complete
calculation of the $t\bar{t}j$ background, including detector effects, is
needed before realistic sensitivity limits for the Higgs self-coupling
at the LHC for $m_H>140$~GeV can be obtained.

}

%% file: kidonakis.tex
{
 \section[ ]{Charged Higgs Production via
$bg \longrightarrow t H^-$\footnote{N.\,Kidonakis}}

\subsection{Introduction}

The Minimal Supersymmetric Standard Model (MSSM) 
introduces charged Higgs bosons 
in addition to the Standard Model neutral Higgs boson.
The discovery of the Higgs is one of the main aims of the current
Run II at the Tevatron and the future program at the LHC.
The charged Higgs would be an important signal of new physics
beyond the Standard Model. 

An important partonic channel for charged Higgs discovery
at hadron colliders is associated production with
a top quark via bottom-gluon fusion, $bg \longrightarrow t H^-$.
The complete next-to-leading order (NLO) corrections to this process
have been studied in Refs. \cite{Zhu:2001nt,Plehn:2002vy}.
Here, I discuss soft-gluon corrections to charged Higgs
production, which are expected to be important near threshold, 
the kinematical region where the charged Higgs may be discovered 
at the LHC. Threshold corrections have been shown to be important 
for many processes in hadron colliders 
\cite{Kidonakis:1999hq,Kidonakis:2003bh,Kidonakis:2000gi,
Kidonakis:2000ui,Kidonakis:2003qe,Kidonakis:2003sc,Kidonakis:2003xm}.

In the next section I first discuss the NLO
soft-gluon corrections for $bg \longrightarrow t H^-$,
and then I present the NNLO corrections at next-to-leading
logarithmic (NLL) accuracy.

\subsection{Threshold NNLO-NLL Corrections}

For the process $b(p_b)+g(p_g) \rightarrow t(p_t)+H^-(p_H)$,
we define the kinematical invariants $s=(p_b+p_g)^2$,
$t=(p_b-p_t)^2$, $u=(p_g-p_t)^2$, and $s_4=s+t+u-m_t^2-m_H^2$,
where $m_H$ is the charged Higgs mass, $m_t$ is the top quark mass,
and we ignore the bottom quark mass $m_b$. 
Note that near threshold, i.e. when we have just enough
partonic energy to produce the $tH^-$ final state,  $s_4 \rightarrow 0$.
Threshold corrections appear as $[\ln(s_4/m_H^2)/s_4]_+$.

The differential Born cross section is 
$d^2{\hat\sigma}^{bg \rightarrow t H^-}_B/(dt \; du)
=F^{bg \rightarrow t H^-}_B \delta(s_4)$
where 
\begin{eqnarray}
F^{bg \rightarrow t H^-}_B&=&
\frac{\pi \alpha \alpha_s (m_b^2 \tan^2\beta
+m_t^2 \cot^2\beta)}{12 s^2 m_W^2 \sin^2\theta_W}
\left\{\frac{s+t-m_H^2}{2s} \right. 
\nonumber \\ && \hspace{-5mm} \left.
{}-\frac{m_t^2(u-m_H^2)+m_H^2(t-m_t^2)+s(u-m_t^2)}{s(u-m_t^2)}
-\frac{m_t^2(u-m_H^2-s/2)+su/2}{(u-m_t^2)^2}\right\} \, ,
\end{eqnarray}
where $\alpha_s$ is the strong coupling, 
$\alpha=e^2/(4\pi)$, $\tan \beta=v_2/v_1$ is the ratio
of the vacuum expectation values of the two Higgs doublets 
in the MSSM, and we have kept $m_b$ non-zero
only in the coupling.

The NLO soft-gluon corrections for $bg \rightarrow tH^-$ are
\begin{equation}
\frac{d^2{\hat\sigma}^{(1)}_{bg\rightarrow t H^-}}{dt \, du}
=F^{bg \rightarrow t H^-}_B
\frac{\alpha_s(\mu_R^2)}{\pi} \left\{
c^{bg \rightarrow tH^-}_{3} \left[\frac{\ln(s_4/m_H^2)}{s_4}\right]_+
+c^{bg \rightarrow tH^-}_{2} \left[\frac{1}{s_4}\right]_+
+c^{bg \rightarrow tH^-}_{1}  \delta(s_4)\right\} \, .
\label{NLObgtH}
\end{equation}

Here $c^{bg \rightarrow tH^-}_{3}=2(C_F+C_A)$, where $C_F=(N_c^2-1)/(2N_c)$
and $C_A=N_c$ with $N_c=3$ the number of colors,  and
\begin{eqnarray}
c^{bg \rightarrow tH^-}_{2}&=&2 {\rm Re} {\Gamma'}_S^{(1)}
-C_F-C_A-2C_F\ln\left(\frac{-u+m_H^2}{m_H^2}\right)
-2C_A\ln\left(\frac{-t+m_H^2}{m_H^2}\right)
\nonumber \\ 
&& -(C_F+C_A)\ln\left(\frac{\mu_F^2}{s}\right)
\equiv T^{bg \rightarrow tH^-}_{2}-(C_F+C_A)
\ln\left(\frac{\mu_F^2}{m_H^2}\right) \, ,
\end{eqnarray}
where $\mu_F$ is the factorization scale, and we have defined 
$T^{bg \rightarrow tH^-}_{2}$ as the scale-independent part
of $c^{bg \rightarrow tH^-}_{2}$.
The term ${\rm Re} {\Gamma'}_S^{(1)}=C_F \ln[(-t+m_t^2)/(m_t\sqrt{s})]
+(C_A/2) \ln[(-u+m_t^2)/(-t+m_t^2)]+C_A/2$ denotes the real part
of the one-loop soft anomalous dimension, which describes
noncollinear soft-gluon emission 
\cite{Kidonakis:1996aq,Kidonakis:1997gm,Kidonakis:1999ze}. 
Also
$c^{bg \rightarrow tH^-}_{1}=[
C_F \ln((-u+m_H^2)/m_H^2)+C_A \ln((-t+m_H^2)/m_H^2)
-3C_F/4-\beta_0/4]\ln(\mu_F^2/m_H^2)
+(\beta_0/4) \ln(\mu_R^2/m_H^2)$, 
where $\mu_R$ is the renormalization scale and $\beta_0=(11C_A-2n_f)/3$ 
is the lowest-order $\beta$ function, 
with $n_f=5$ the number of light quark flavors.
Note that $c^{bg \rightarrow tH^-}_{1}$ represents the scale-dependent
part of the $\delta(s_4)$ corrections. We do not calculate the full virtual
corrections here. Our calculation of the soft-gluon corrections
includes the leading and next-to-leading logarithms (NLL) of $s_4$
and is thus a NLO-NLL calculation. 

\begin{figure}
\begin{center}
\includegraphics[width=7.95cm]{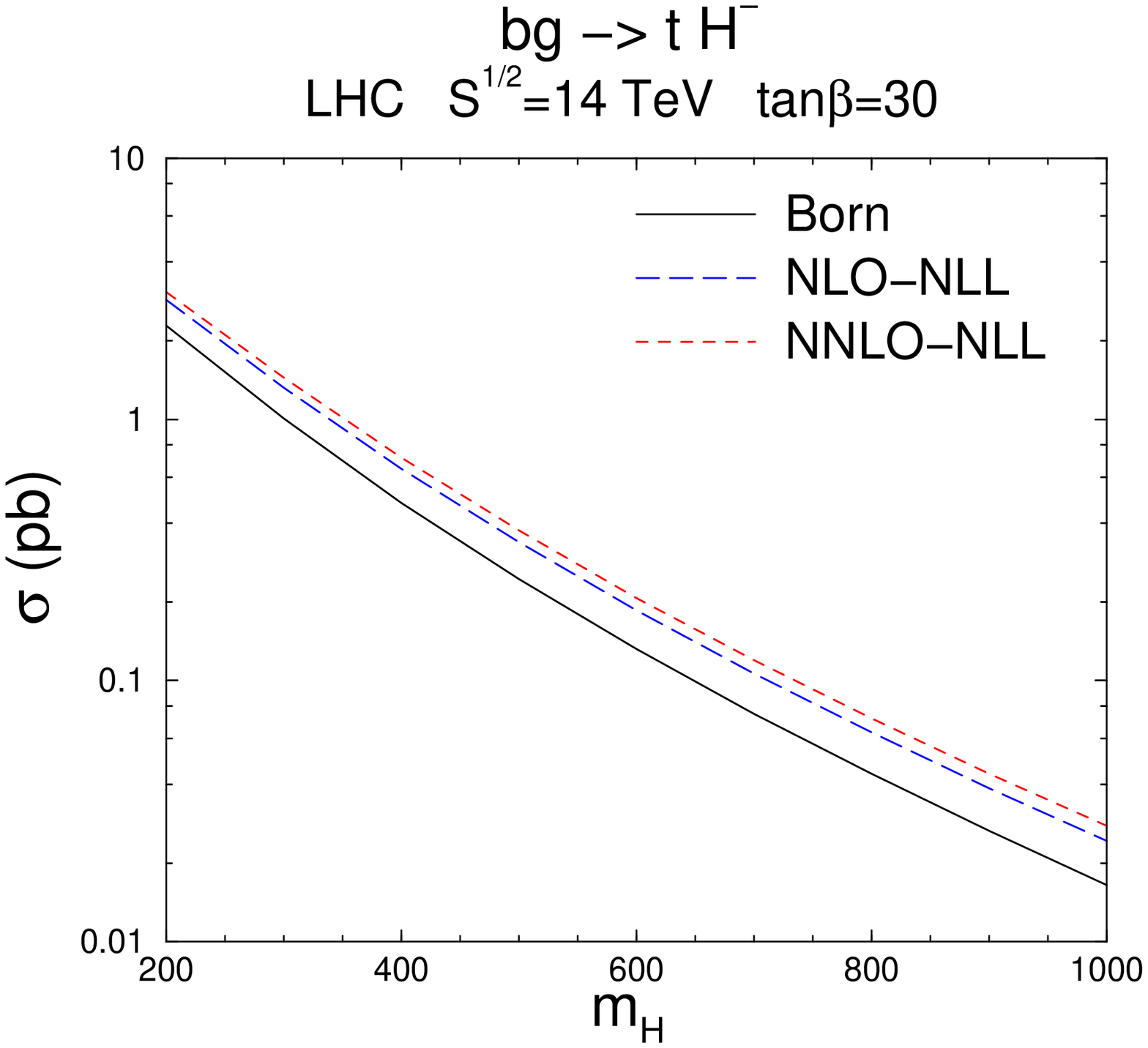} 
\includegraphics[width=7.95cm]{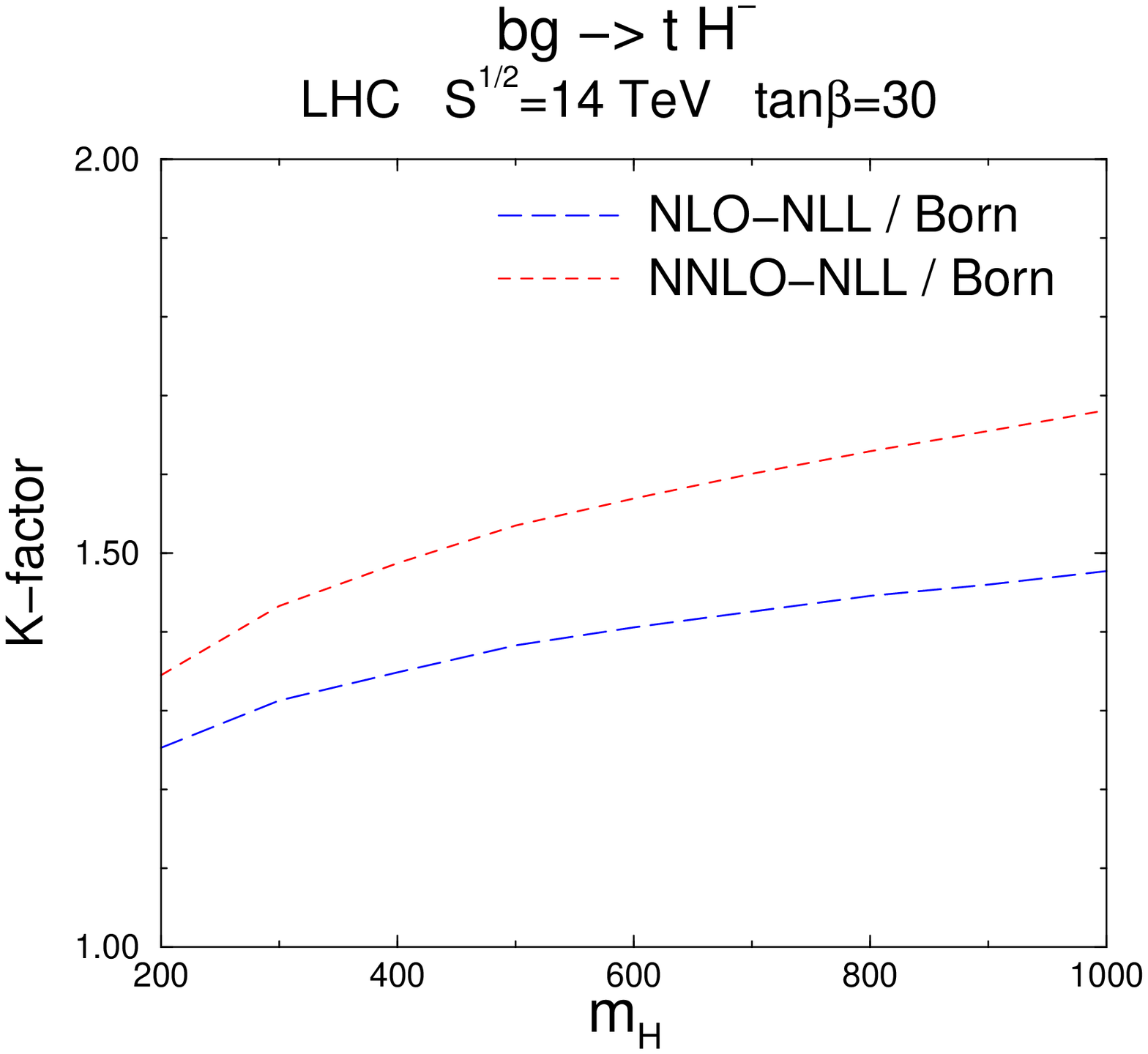} 
\caption{Charged Higgs production at the LHC.
Left: The total cross section.
Right: The $K$-factors.}
\end{center}
\end{figure}

We next calculate the NNLO soft-gluon corrections for $bg \rightarrow tH^-$
using the methods and master formulas of Ref. \cite{Kidonakis:2003tx}:
\begin{eqnarray}
&& \hspace{-5mm}\frac{d^2{\hat\sigma}^{(2)}_{bg \rightarrow tH^-}}
{dt \, du}
=F^{bg \rightarrow tH^-}_B \frac{\alpha_s^2(\mu_R^2)}{\pi^2} 
\left\{\frac{1}{2} \left(c^{bg \rightarrow tH^-}_{3}\right)^2 
\left[\frac{\ln^3(s_4/m_H^2)}{s_4}\right]_+ \right.
\nonumber \\ && \hspace{-5mm}
{}+\left[\frac{3}{2} c^{bg \rightarrow tH^-}_{3} \, c^{bg 
\rightarrow tH^-}_{2}
-\frac{\beta_0}{4} c^{bg \rightarrow tH^-}_{3} \right] 
\left[\frac{\ln^2(s_4/m_H^2)}{s_4}\right]_+ 
\nonumber \\ && \hspace{-5mm}
{}+\left[c^{bg \rightarrow tH^-}_{3} \, c^{bg \rightarrow tH^-}_{1}
+(C_F+C_A)^2\ln^2\left(\frac{\mu_F^2}{m_H^2}\right)
-2(C_F+C_A) T_2^{bg \rightarrow tH^-}\ln\left(\frac{\mu_F^2}{m_H^2}\right)
\right.
\nonumber \\ && \quad \left.
{}+\frac{\beta_0}{4} c^{bg \rightarrow tH^-}_{3} 
\ln\left(\frac{\mu_R^2}{m_H^2}\right)
-\zeta_2 \, \left(c^{bg \rightarrow tH^-}_{3}\right)^2 \right]
\left[\frac{\ln(s_4/m_H^2)}{s_4}\right]_+
\nonumber \\ && \hspace{-5mm} 
{}+\left[-(C_F+C_A) \ln\left(\frac{\mu_F^2}{m_H^2}\right)
c^{bg \rightarrow tH^-}_{1}
-\frac{\beta_0}{4} (C_F+C_A) \ln\left(\frac{\mu_F^2}{m_H^2}\right) 
\ln\left(\frac{\mu_R^2}{m_H^2}\right) \right.
\nonumber \\ && \quad \left. \left.
{}+(C_F+C_A)\frac{\beta_0}{8} \ln^2\left(\frac{\mu_F^2}{m_H^2}\right)
-\zeta_2 \, c^{bg \rightarrow tH^-}_{2} \, c^{bg \rightarrow tH^-}_{3}
+\zeta_3 \, \left(c^{bg \rightarrow tH^-}_{3}\right)^2\right]
\left[\frac{1}{s_4}\right]_+ \right\} \, ,
\label{NNLObgtH}
\end{eqnarray}
where $\zeta_2=\pi^2/6$ and $\zeta_3=1.2020569...$.
We note that only the coefficients of the leading ($\ln^3$) and 
next-to-leading logarithms ($\ln^2$) of $s_4$ are complete.
Hence this is a NNLO-NLL calculation.
Consistent with a NLL calculation we have also kept all logarithms of the 
factorization and renormalization scales in the
$[\ln(s_4/m_H^2)/s_4]_+$ terms, and squares of scale logarithms
in the $[1/s_4]_+$ terms, as well as $\zeta_2$ and $\zeta_3$ terms
that arise in the calculation of the soft corrections.

We now convolute the partonic cross sections with parton distribution
functions (we use {MRST2002 NNLO ~\cite{Martin:2002aw})
to obtain the hadronic cross section in $pp$ collisions at the LHC.
We use $\mu_F=\mu_R=m_H$ for our numerical results.
In the left frame of Fig. 24 we plot the cross section for charged 
Higgs production 
at the LHC with $\sqrt{S}=14$ TeV versus the charged Higgs mass.
We use $m_t=175$ GeV, $m_b=4.5$ GeV, and $\tan \beta=30$.
The Born, NLO-NLL, and NNLO-NNLL results are shown.
Both the NLO and the NNLO threshold corrections are important
as can be more clearly seen in the right frame of Fig. 24 where
we plot the $K$-factors, i.e. the ratios of the NLO-NLL over Born
and the NNLO-NLL over Born cross sections. As expected, the corrections
increase for larger charged Higgs mass since then 
we get closer to threshold. Finally, we note that the cross section
for ${\bar b} g \longrightarrow {\bar t} H^+$ is the same 
as for $bg \longrightarrow t H^-$.

\subsection{Conclusion}

The soft-gluon threshold corrections for the process
$bg \longrightarrow t H^-$ have been calculated
through next-to-next-to-leading order and next-to-leading logarithmic
accuracy. We have seen that numerically both the NLO and
NNLO threshold corrections to charged Higgs production at the LHC
are important.

}

%% file: heinemeyer.tex
{
%%%%%%%%%%%%%%%%%%%%%%%%%%%%%%%%%%%%%%%%%%%%%%%%%%%%%%%%%%%%%%
%%%%%%%%%%%%%%%%%%%%%%%%%%%%%%%%%%%%%%%%%%%%%%%%%%%%%%%%%%%%%%

\newcommand{\cp}{{\cal CP}}
\newcommand{\mgl}{m_{\tilde{g}}}
\newcommand{\MHp}{M_{H^\pm}}
\newcommand{\tb}{\tan\beta}
\newcommand{\drbar}{$\overline{\rm{DR}}$}
\newcommand{\msbar}{$\overline{\rm{MS}}$}
\newcommand{\br}{{\rm BR}}
\def\citere#1{\mbox{Ref.~\cite{#1}}}
\def\citeres#1{\mbox{Refs.~\cite{#1}}}
\def\order#1{${\cal O}(#1)$}
\newcommand{\gev}{\,\, \mathrm{GeV}}
\newcommand{\FH}{\emph{FeynHiggs 2.1}}

%%%%%%%%%%%%%%%%%%%%%%%%%%%%%%%%%%%%%%%%%%%%%%%%%%%%%%%%%%%%%%
%%%%%%%%%%%%%%%%%%%%%%%%%%%%%%%%%%%%%%%%%%%%%%%%%%%%%%%%%%%%%%

\section[ ]{FeynHiggs 2.1: High Precision Calculations in the MSSM Higgs
Sector\footnote{T.\,Hahn, S.\,Heinemeyer, W.\,Hollik and G.\,Weiglein}}

%%%%%%%%%%%%%%%%%%%%%%%%%%%%%%%%%%%%%%%%%%%%%%%%%%%%%%%%%%%%%%
%%%%%%%%%%%%%%%%%%%%%%%%%%%%%%%%%%%%%%%%%%%%%%%%%%%%%%%%%%%%%%

\subsection{Introduction}

The search for the lightest Higgs boson is a crucial test of
Supersymmetry (SUSY) which can be performed with the present and the
next generation of accelerators.  Especially for the Minimal
Supersymmetric Standard Model (MSSM) a precise prediction for the masses
of the Higgs bosons and their decay widths to other particles in terms
of the relevant SUSY parameters is necessary in order to determine the
discovery and exclusion potential of the upgraded Tevatron, and for
physics at the LHC and future linear colliders.

In the case of the MSSM with complex parameters (cMSSM) the task is even
more involved.  Several parameters can have non-vanishing phases.  In
particular, these are the Higgs mixing parameter, $\mu$, the trilinear
couplings, $A_f$, $f = t, b, \tau, \ldots$, and the gaugino masses 
$M_1$, $M_2$, and $M_3 \equiv \mgl$ (the gluino mass).  Furthermore the 
neutral Higgs bosons are no longer $\cp$-eignestates, but mix with each 
other once loop corrections are taken into account~\cite{mhiggsCPXgen}.
\begin{equation}
(h, H, A) \to h_1, h_2, h_3 ~~~{\rm with}~~~
m_{h_1} \le m_{h_2} \le m_{h_3}\,.
\end{equation}
The input parameters within the Higgs sector are then (besides the
Standard Model (SM) ones) $\tb$, the ratio of the two vacuum expectation
values, and the mass of the charge Higgs boson, $\MHp$.

%%%%%%%%%%%%%%%%%%%%%%%%%%%%%%%%%%%%%%%%%%%%%%%%%%%%%%%%%%%%%%
%%%%%%%%%%%%%%%%%%%%%%%%%%%%%%%%%%%%%%%%%%%%%%%%%%%%%%%%%%%%%%

\subsection{The Code \FH}

\FH~\cite{feynhiggs} is a Fortran code for the evaluation of masses and
mixing angles in the MSSM with real or complex parameters.  The
calculation of the higher-order corrections is based on the
Feynman-diagrammatic (FD) approach~\cite{mhiggsCPXFD}.  
At the one-loop level, it consists a complete evalutaion, including the
full momentum 
dependence.  The renormalization has been performed in a hybrid
\msbar~/on-shell scheme~\cite{feynhiggs1.2}.  At the two-loop level all
existing corrections from the real MSSM have been included
(see~\citere{mhiggsAEC} for a review). They are supplemented by the
resummation of the leading effects from the (scalar)~$b$ sector
including the full complex phase dependence.

Besides the evaluation of the Higgs-boson masses and mixing angles,
the program also includes the evaluation of all relevant Higgs-boson
decay widths. 
These are in particular:

\begin{itemize}
\item
the total width for the three neutral and the charged Higgs boson,

\item
the BR's of the Higgs bosons to SM fermions (see also \citere{hff}),
$\br(h_i \to f \bar f)$, $\br(H^+ \to f \bar f')$,

\item
the to SM gauge bosons (possibly off-shell),
$\br(h_i \to \gamma\gamma, ZZ^*, WW^*, gg)$,

\item 
the decay into gauge and Higgs bosons,
$\br(h_i \to Z h_j)$, $\br(h_i \to h_j h_k)$, $\br(H^+ \to h_i W^+)$,

\item
the decay to scalar fermions,
$\br(h_i \to \tilde{f} \bar{\tilde{f}})$,
$\br(H^+ \to \tilde{f} \bar{\tilde{f}'})$,

\item
the decay of the Higgs bosons to gauginos,
$\br(h_i \to \chi^\pm_k \chi^\mp_j)$,
$\br(h_i \to \chi^0_l \chi^0_m)$,\\
$\br(H^+ \to \chi^+_k \chi^0_l)$.
\end{itemize}
For comparisons with the SM the following quantities are also evaluated
for SM Higgs bosons with the same mass as the three neutral MSSM Higgs
bosons:
\begin{itemize}
\item
the total decay widths,

\item
the BR's of a SM Higgs boson to SM fermions,

\item
the BR's of a SM Higgs boson to SM gauge bosons (possibly off-shell).

\end{itemize}
In addition, the following couplings and cross sections are evaluated
\begin{itemize}
\item
the coupling of Higgs and gauge bosons, 
$g_{VVh_i}$, $g_{Vh_ih_j}$,

\item
the Higgs-boson self couplings,
$g_{h_i h_j h_k}$,

\item
the Higgs-boson production cross section at a $\gamma\gamma$~collider,
$\sigma(\gamma \gamma \to h_i)$.

\end{itemize}
Finally as external constraints are evaluated
\begin{itemize}
\item
the $\rho$-parameter up to the two-loop level~\cite{deltarho} that
indicates disfavored scalar top and bottom masses

\item
the anomalous magnetic moment of the muon, including a full one-loop
calculation as well as leading and subleading two-loop
corrections~\cite{gm2letter}. 

\end{itemize}

Comparing our results to existing codes like Hdecay~\cite{Djouadi:1998yw} (for the
real case) or CPsuperH~\cite{cpsh} (for the cMSSM), we find differences
in the mass evaluations for the lightest Higgs boson of \order{4 \gev}.
These are due to the inclusion of higher-order corrections in \FH\ that
shift the lightest Higgs-boson mass upwards. Concerning the BR
evaluation (and compensating for the effects from the different
Higgs-boson masses) we find quantitative and qualitative agreement.  For
more details see \citere{feynhiggs}.

\noindent
\FH\ possesses some further features that can be summarizes as,
\begin{itemize}

\item
transformation of the input parameters from the \drbar\ to the
on-shell scheme (for the scalar
top and bottom parameters), including the full \order{\alpha_s} and
\order{\alpha_{t,b}} corrections.

\item
processing of Les Houches Accord (LHA) data~\cite{lha}. \FH\ reads the
output of a spectrum generator file and evaluates the Higgs boson
masses, brachning ratios etc. The results are written in the LHA
format to a new output file.

\item
the SPS~benchmark scenarios~\cite{sps} and the Les Houches benchmarks
for Higgs boson searches at hadron colliders~\cite{LHbenchmark} are
given as a possibly predefined input

\item
detailed information about all the features of \FH\ (see also the next
section) are provided in man pages.

\end{itemize}

%%%%%%%%%%%%%%%%%%%%%%%%%%%%%%%%%%%%%%%%%%%%%%%%%%%%%%%%%%%%%%
%%%%%%%%%%%%%%%%%%%%%%%%%%%%%%%%%%%%%%%%%%%%%%%%%%%%%%%%%%%%%%

\subsection{How to install and use \FH}

To take advantage of all features of \FH, the LoopTools library 
\cite{lt} needs to be installed, which can be obtained from 
www.feynarts.de/looptools.  Without this library, \FH\ will still 
compile, but not all branching ratios will be available.
\begin{itemize}
\item Download the package from www.feynhiggs.de.
\item Say {\tt ./configure} and {\tt make}.  This creates libFH.a and the
      command-line frontend.
\item To build also the Mathematica frontend, say {\tt make all}.
\end{itemize}
There are three different ways to use \FH.

\subsubsection{The Fortran library}

The libFH.a library can be linked directly to other Fortran programs. 
To avoid naming conflicts, all externally visible symbols have been
prefixed with ``fh.''  No include files are needed since the user calls
only subroutines (no functions).  Detailed descriptions of the 
invocations of the subroutines are given in the respective man pages.

\subsubsection{The command-line frontend}

The FeynHiggs executable is a command-line frontend to the libFH.a 
library.  It reads the parameters from an ASCII input file and writes 
the output in a human-readable form to the screen.  Alternatively, this 
output can be piped through a filter to yield a machine-readable 
version appropriate for plotting etc.  The parameter file is fairly 
flexible and allows to define also loops over parameters.  Also the 
Les-Houches-Accord file format can be read and written.

\subsubsection{The Mathematica frontend}

The MFeynHiggs executable provides access to the libFH.a functions from
Mathematica via the MathLink protocol.  This is particularly convenient
both because \FH\ can be used interactively this way and because
Mathematica's sophisticated numerical and graphical tools, e.g.\ 
FindMinimum, are available.

%%%%%%%%%%%%%%%%%%%%%%%%%%%%%%%%%%%%%%%%%%%%%%%%%%%%%%%%%%%%%%
%%%%%%%%%%%%%%%%%%%%%%%%%%%%%%%%%%%%%%%%%%%%%%%%%%%%%%%%%%%%%%

%\section*{ACKNOWLEDGEMENTS}

%%%%%%%%%%%%%%%%%%%%%%%%%%%%%%%%%%%%%%%%%%%%%%%%%%%%%%%%%%%%%%
%%%%%%%%%%%%%%%%%%%%%%%%%%%%%%%%%%%%%%%%%%%%%%%%%%%%%%%%%%%%%%
%%%%%%%%%%%%%%%%%%%%%%%%%%%%%%%%%%%%%%%%%%%%%%%%%%%%%%%%%%%%%%
%%%%%%%%%%%%%%%%%%%%%%%%%%%%%%%%%%%%%%%%%%%%%%%%%%%%%%%%%%%%%%
}

%% file: sabio.tex
{
\section[ ]{Mass Bounds for a SU(2) Triplet Higgs\footnote{J.R.\,Forshaw,
A.\,Sabio Vera}}

\subsection{Introduction}

The precision high--energy measurements of electoweak observables by 
LEP, SLC and Tevatron have confirmed the Glashow--Salam--Weinberg model to 
a great certainty. The remaining challenge is to pin down the nature of the 
breaking of the electroweak symmetry. If it occurs via the Standard Model 
Higgs mechanism, with a complex isospin doublet, the mass of the lightest 
Higgs is $81^{+52}_{-33}$ GeV~\cite{1}, with 
similar bounds for the simplest supersymmetric models. In this contribution 
we present the study of a simple extension of the Standard Model where the 
new feature is the addition of a real Higgs triplet. This model is compatible 
with precision measurements as will be shown below, and allows for the 
lightest Higgs boson mass to go up to about 500 GeV. This review is based on 
the work in Ref.~\cite{2,3}. 

\subsection{The Triplet Higgs Model}

The lagrangian of the model in terms of the usual Standard
Model Higgs, $\Phi_1$, and the new triplet, $\Phi_2$, reads
\begin{eqnarray}
{\cal L} = (D_\mu \Phi_1)^{\dagger} D^\mu \Phi_1 
+ \frac{1}{2} (D_\mu \Phi_2)^{\dagger} D^\mu \Phi_2 
- V_0(\Phi_1 , \Phi_2)\,, \nonumber
\end{eqnarray}
with a scalar potential 
$V_0(\Phi_1 , \Phi_2) = \mu_1^2 |\Phi_1|^2 + \frac{\mu_2^2}{2} 
|\Phi_2|^2 + \lambda_1 |\Phi_1|^4+ \frac{\lambda_2}{4} |\Phi_2|^4 
+ \frac{\lambda_3}{2} |\Phi_1|^2 |\Phi_2|^2 
+ \lambda_4 \, {\Phi_1}^\dagger \sigma^\alpha \Phi_1
{\Phi_2}_\alpha$.  
$\sigma^\alpha$ are the Pauli matrices.  The expansion of the field
components is
\begin{eqnarray}
\Phi_1 = \begin{pmatrix} \phi^+ \cr
\frac{1}{\sqrt{2}}\left(h_c^0 + h^0 + i \phi^0\right) \end{pmatrix}_{Y=1}\,,
\qquad
\Phi_2 = \begin{pmatrix} \eta_1 \cr \eta_2 \cr \eta_c^0 + \eta^0 \end{pmatrix}_{Y=0} \nonumber
\end{eqnarray}
where $\eta^\pm = ( \eta_1 \mp i \eta_2) / \sqrt{2}$ and $\phi^0$ is
the Goldstone boson which is eaten by the $Z^0$. In the neutral Higgs sector 
we have two CP-even states which mix 
with angle $\gamma$. The mass eigenstates $\{H^0, N^0\}$ are  
\begin{eqnarray}
\begin{pmatrix} H^0\cr N^0 \end{pmatrix} = 
\begin{pmatrix} \cos{\gamma}&-\sin{\gamma}\cr \sin{\gamma}&\cos{\gamma} \end{pmatrix}
\begin{pmatrix} h^0\cr \eta^0 \end{pmatrix}. \nonumber
\end{eqnarray}
There is also mixing in the charged Higgs sector. We define the mass
eigenstates $\{g^\pm, h^\pm\}$ by 
\begin{eqnarray}
\begin{pmatrix} g^\pm\cr h^\pm \end{pmatrix} = 
\begin{pmatrix} \cos{\beta}&-\sin{\beta}\cr \sin{\beta}&\cos{\beta} \end{pmatrix}
\begin{pmatrix} \phi^\pm\cr \eta^\pm \end{pmatrix}. \nonumber
\end{eqnarray}
The $g^\pm$ are the Goldstone bosons corresponding to $W^\pm$ and,
at tree level, the mixing angle~is 
$\tan{\beta} = 2 \frac{\eta^0_c}{h^0_c}$. 
The model violates custodial symmetry at tree level giving a
prediction for the $\rho$-parameter of
$\rho = 1 + 4 \left(\frac{\eta^0_c}{h^0_c}\right)^2$. 
We will show below how it is precisely this violation of custodial symmetry 
what allows for the mass of the lightest Higgs to be large in this model.

\subsection{Comparison with Data from oblique Corrections}

Predictions for the oblique corrections to EW observables can be written in 
terms of the \( S \), \( T \) and \( U \) parameters which can quantify the 
effect
of varying the Higgs mass. The TM contributions, to leading order 
in \( \beta  \), are~\cite{2}
\begin{eqnarray}
S_{TM}&=&  0, \nonumber \\
T_{TM} &=& \frac{1}{8\pi} \, \frac{1}{s^{2}_{W} c^{2}_{W}} 
\left[ \frac{m^{2}_{N^0} + m^{2}_{h^\pm}}{m^{2}_{Z}} \;
- \; \frac{2 m^{2}_{h^\pm} m^{2}_{N^0}}{m^{2}_{Z}(m^{2}_{N^0} 
- m^{2}_{h^\pm})} \log\left(\frac{m^{2}_{N^0}}{m^{2}_{h^\pm}}\right)
\right] 
\simeq   \frac{1}{6\pi} \, \frac{1}{s^{2}_{W} c^{2}_{W}} \; 
\frac{(\Delta m)^{2}}{m^{2}_{Z}}. \nonumber \\ 
U_{TM} &=& -\frac{1}{3 \pi} \left( m_{N^0}^4 
\log \left( \frac{m_{N^0}^2}{m_{h^\pm}^2} \right) 
\frac{ (3 m_{h^\pm}^2-m_{N^0}^2)}{(m_{N^0}^2-m_{h^\pm}^2)^3} 
+ \frac{5(m_{N^0}^4+m_{h^\pm}^4)-22 
m_{N^0}^2 m_{h^\pm}^2}{6(m_{N^0}^2 - m_{h^\pm}^2)^2} \right) 
+ O\left(\frac{m_Z}{m_{h^\pm}}\right) \nonumber \\ &\simeq & 
\frac{\Delta m}{3 \pi m_{h^\pm}}. \nonumber
\end{eqnarray}
The TM contribution to \( S \) is zero to this order since $Y=0$. Apart from 
the loop correction, there is also a tree level contribution
which arises in all observables as a result of the tree level deviation of
the rho parameter from unity. This contribution leads effectively to a
positive contribution to T. The TM contribution
to \( T \) is positive and, in the approximation of 
\( \Delta m=m_{N^0}-m_{h^\pm}\ll m_{h^\pm} \),
has the rough power dependence shown above. \( U \) also vanishes when 
\( \Delta m\to 0 \),
and falls to zero at large triplet masses. In particular, it has a negligible
effect on all the results we shall subsequently show provided 
\( m_{N^0},\, m_{h^\pm}>1 \) TeV.

\begin{figure}[!thb]
\vspace*{5.0cm}
\begin{center}
\includegraphics{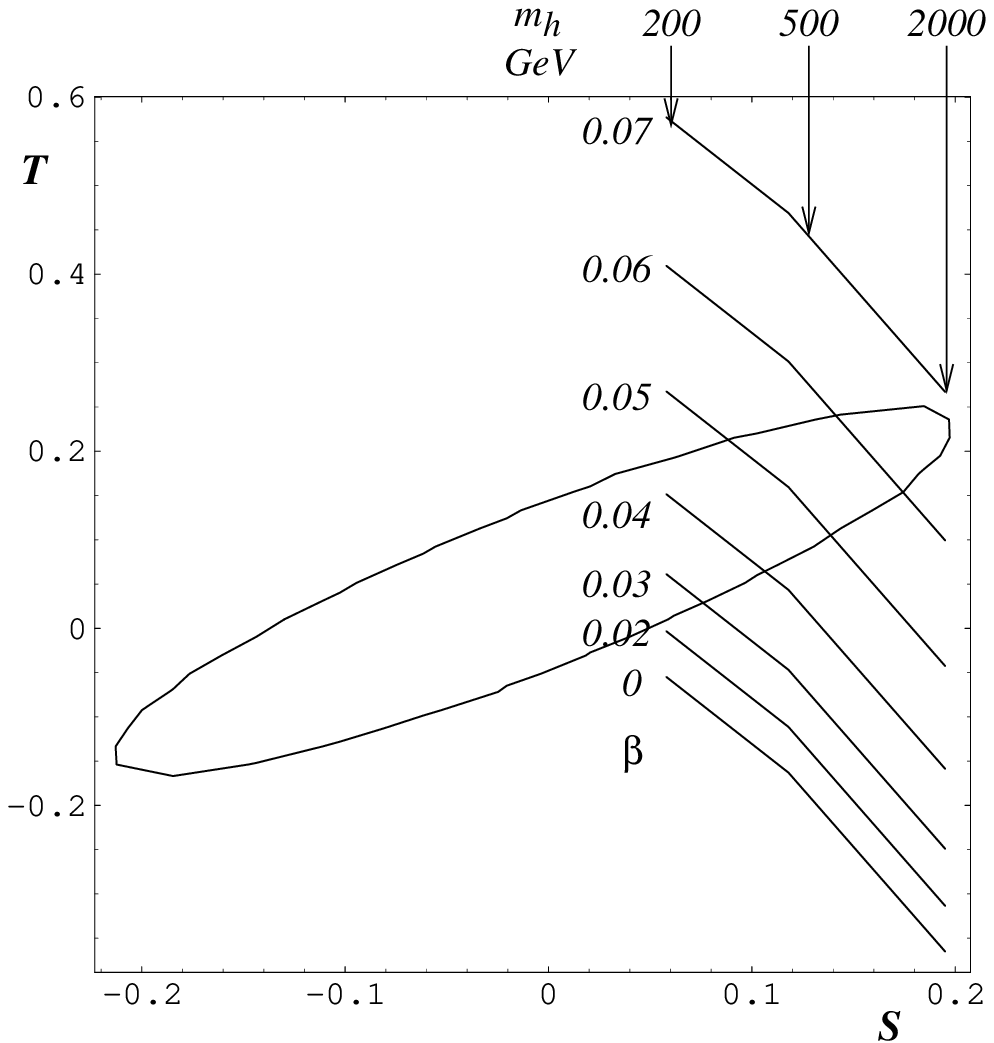}
\includegraphics{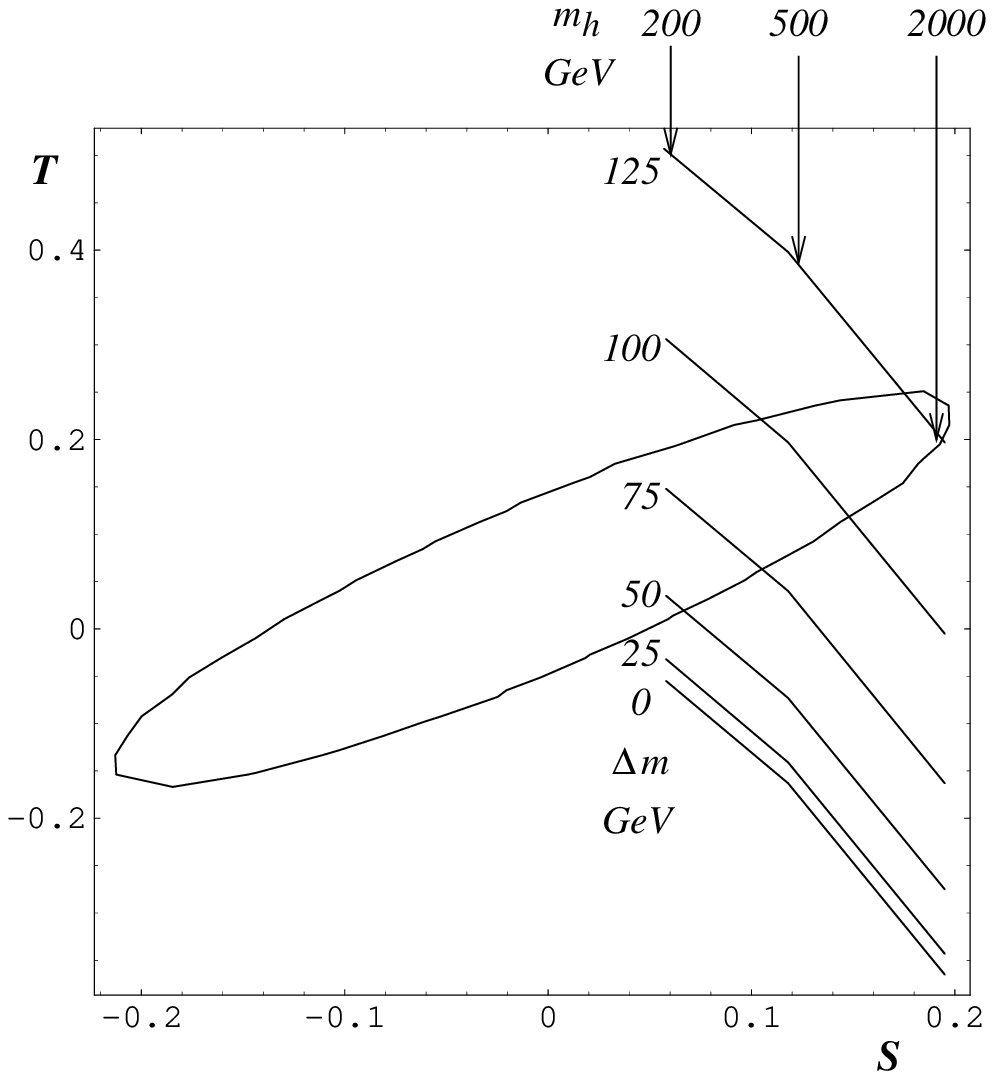}
\caption[*]{Ellipse encloses the region allowed by data. Left: Curves show 
results in the TM for various values of \protect\( \beta \protect \) and 
various doublet Higgs masses with \protect\( \Delta m=0\protect \) and 
\protect\( U=0\protect \). Right: Curves show results in the TM
for various mass splittings and various doublet Higgs masses with 
\protect\( \beta \protect \) and \protect\( U\protect \) assumed to be 
negligible.}
\label{TripletHiggsFig1}
\end{center}
\end{figure}

\label{sec_zfitter}Using the program \texttt{ZFITTER}~\cite{4} we
compute a total of 13 standard observables~\cite{1}  with 
$m_{h}^{ref}=100~{\rm GeV}$, 
\( m_{t}=174.3 \) GeV, \( G_{\mu }=1.16639\times 10^{-5} \)
GeV\( ^{-2} \), \( m_{Z}=91.1875 \) GeV, \( \alpha _{s}= \)0.119 and 
\( \Delta \alpha _{had}^{(5)}(m_{Z}) \)=
0.02804. These results then determine the allowed region in $S$--$T$ parameter
space. This is represented by the interior of the ellipses shown in 
Fig.~\ref{TripletHiggsFig1} (the ellipse corresponds to a 95$\%$ confidence limit). 

In Fig.~\ref{TripletHiggsFig1} (left) each line shows the TM at a particular value of
\( \beta  \) for \( \Delta m=0 \) (which turns off the quantum corrections)
and \( m_{h} \) varying from 200 GeV to 2 TeV. We see that even in the absence
of quantum corrections the TM is able to accommodate any \( m_{h} \) up to 
2 TeV and the mixing angle \( \beta  \) must be less than 0.07. In 
Fig.~\ref{TripletHiggsFig1} (right) each line shows the TM result as \( m_{h} \) is 
varied,
as before, at fixed \( \Delta m \). \( \beta  \) is assumed to be negligibly
small in this plot (which turns off the tree-level correction)
and as a result the \( \Delta m=0 \) line is identical to that which would
arise in the SM. Clearly the quantum corrections contribute to \( T \) so as
to allow any \( m_{h} \) up to 2 TeV and the mass splitting \( \Delta m \)
must be less than 125 GeV.

\subsection{Renormalisation Group Evolution}

We would like to examine the RG flow of the
couplings and hence establish bounds on the scalar masses under the
assumption that the triplet model remain perturbative and have a stable vacuum 
up 
to some scale
$\Lambda= 1$\,TeV. 

In Ref.~\cite{3} the beta functions for the couplings were 
calculated using the one--loop effective potential 
\cite{5,6,7,8,9,10} 
with $\overline{\rm MS}$ renormalization in 't~Hooft-Landau gauge and 
the anomalous dimensions for $h^0$ and $\eta^0$, the results read
\begin{eqnarray}
\beta_{\mu_1} &=&
\frac{1}{16 \pi^2}\,\left(6 \, \lambda_4^2 + 12 \, \lambda_1 \mu_1^2 
+ 3 \, \lambda_3 \, \mu_2^2\right)
+ \frac{1}{8 \pi^2}\,\left(3\,h_t^2 
-\frac{9}{4}\,g^2 - \frac{3}{4}\,{g'}^2\right)\,\mu_1^2,\nonumber\\
\beta_{\mu_2} &=& 
\frac{1}{16 \pi^2}\,\left(4\,\lambda_4^2 + 4\, \lambda_3 \, \mu_1^2 + 10 \,
\lambda_2 \, \mu_2^2 \right)
-\frac{3}{4 \pi^2}\,g^2 \, \mu_2^2 ,\nonumber\\
\beta_{\lambda_1} &=& \frac{1}{8 \pi^2}\left(\frac{9}{16}\,g^4 - 3\,{{h_t}}^4 
+12 \, \lambda_1^2 + \frac{3}{4}\,\lambda_3^2 + \frac{3}{8}\,g^2\,{g'}^2
+ \frac{3}{16}\,{g'}^4\right) \nonumber\\
&+&\frac{1}{4 \pi^2}\,\left(3\,h_t^2 
-\frac{9}{4}\,g^2 - \frac{3}{4}\,{g'}^2\right)\,\lambda_1,\nonumber\\
\beta_{\lambda_2} &=& \frac{1}{8\pi^2}
\left(6\,g^4 + 11\,\lambda_2^2 + \lambda_3^2 \right)
-\frac{3}{2 \pi^2}\,g^2 \, \lambda_2,\nonumber\\
\beta_{\lambda_3} &=& \frac{1}{8\pi^2}
\left(3 \, g^4 + 6\,\lambda_1 \, \lambda_3 + 5 \, \lambda_2 \, \lambda_3 
+ 2 \, \lambda_3^2\right)
+\frac{3 \lambda_3}{8 \pi^2}\,\left(h_t^2 
-\frac{11}{4}\,g^2 - \frac{1}{4}\,{g'}^2\right),\nonumber\\
\beta_{\lambda_4}&=& \frac{1}{4\pi^2}\,\lambda_4 \, 
\left(\lambda_1 + \lambda_3 \right)+\frac{3}{32 \pi^2}\,\left(4\,h_t^2 
- 7\,g^2 - \,{g'}^2\right)\,\lambda_4.\nonumber
\end{eqnarray}
In the gauge and top quark 
sector the beta functions for the $U(1)$, $SU(3)$ and Yukawa 
couplings are the same as in the Standard Model and only the $SU(2)$ 
coupling is modified due to the extra Higgs triplet in the adjoint 
representation, i.e. $\beta_g = - \frac{5}{32 \pi^2}\,g^3$. 

Working with the tree-level effective potential with couplings
evolved using the one-loop $\beta$ and $\gamma$ functions we are able to 
resum the leading logarithms to all orders in the effective potential. 
To carry out the RG analysis we first introduce the 
parameter $t$, related to $\mu$ through  $\mu (t) = m_Z \exp{(t)}$. 
The RG equations are coupled differential equations in the 
set $\left\{g_s, ~g, ~g',~h_t,~\mu_1,~\mu_2,~\lambda_1,~\lambda_2,~\lambda_3
,~\lambda_4\right\}$. 
We choose rather to use the set of variables  
$\left\{\alpha_s,~m_Z,~\sin^2{\theta_W},~m_t,~m_{h^\pm},~m_{H^0},~m_{N^0},~v,~
\tan{\beta},~\tan{\gamma} \right\}$.

Within the accuracy to which we are working, the values of the 
couplings at $t=0$ can be obtained from the input set
using the appropriate tree--level expressions. 
Inverting the tree level relations for the masses we can fix the $t=0$ boundary
conditions for the subsequent evolution. To ensure that the system remains in 
a local minimum we impose the condition 
that the squared masses should remain positive.
We also impose that the couplings remain
perturbative, insisting that 
$|\lambda_i (t)| < 4 \pi$ for $i={1,2,3}$ and $|\lambda_4| < 4 \pi v$.
We run the evolution from $t = 0$ to $t_{\rm max}= \log{(\Lambda/m_Z)}$, with 
$\Lambda = 1$ TeV.

In the non-decoupling regime the triplet cannot be arbitrarily heavy. 
In Fig.~\ref{TripletHiggsFig2} (left) we show the range 
of Higgs masses allowed when there is no mixing in the neutral 
Higgs sector, $\gamma = 0$, for $\beta = 0.04$. Such a value is 
interesting because it allows a rather heavy lightest Higgs, e.g. for 
$\beta = 0.04,\; m_{H^0}>150$~GeV and for 
$\beta = 0.05, \; m_{H^0}>300$~GeV (see Fig.~\ref{TripletHiggsFig1} (left) where $\Delta m \sim 0$ 
and the perturbativity of $\lambda_2$ implies negligible quantum 
corrections).
The strong correlation between the $h^\pm$ and $N^0$ masses arises in order
that $\lambda_2$ remain perturbative. The upper bound on the triplet Higgs 
masses 
($\approx 550$~GeV) comes
about from the perturbativity of $\lambda_3$ whilst that on $H^0$ 
($\approx 520$~GeV) comes
from the perturbativity of $\lambda_1$.  
The hole at low masses is due to vacuum stability. 
The non--zero $\gamma$ case has been considered in Ref. \cite{2}.
\begin{figure}
\vspace*{2.5cm}
\begin{center}
\includegraphics{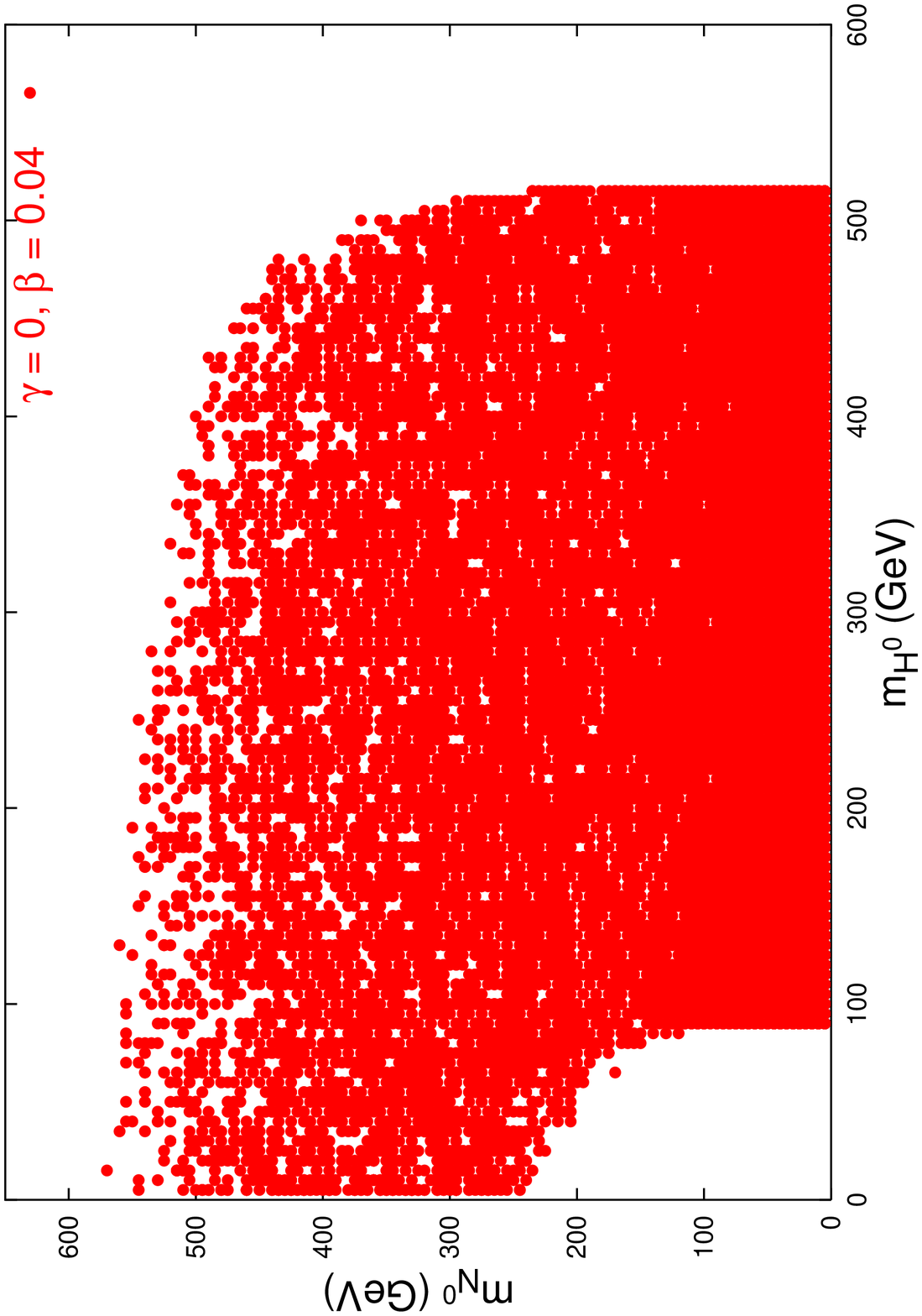}
\includegraphics{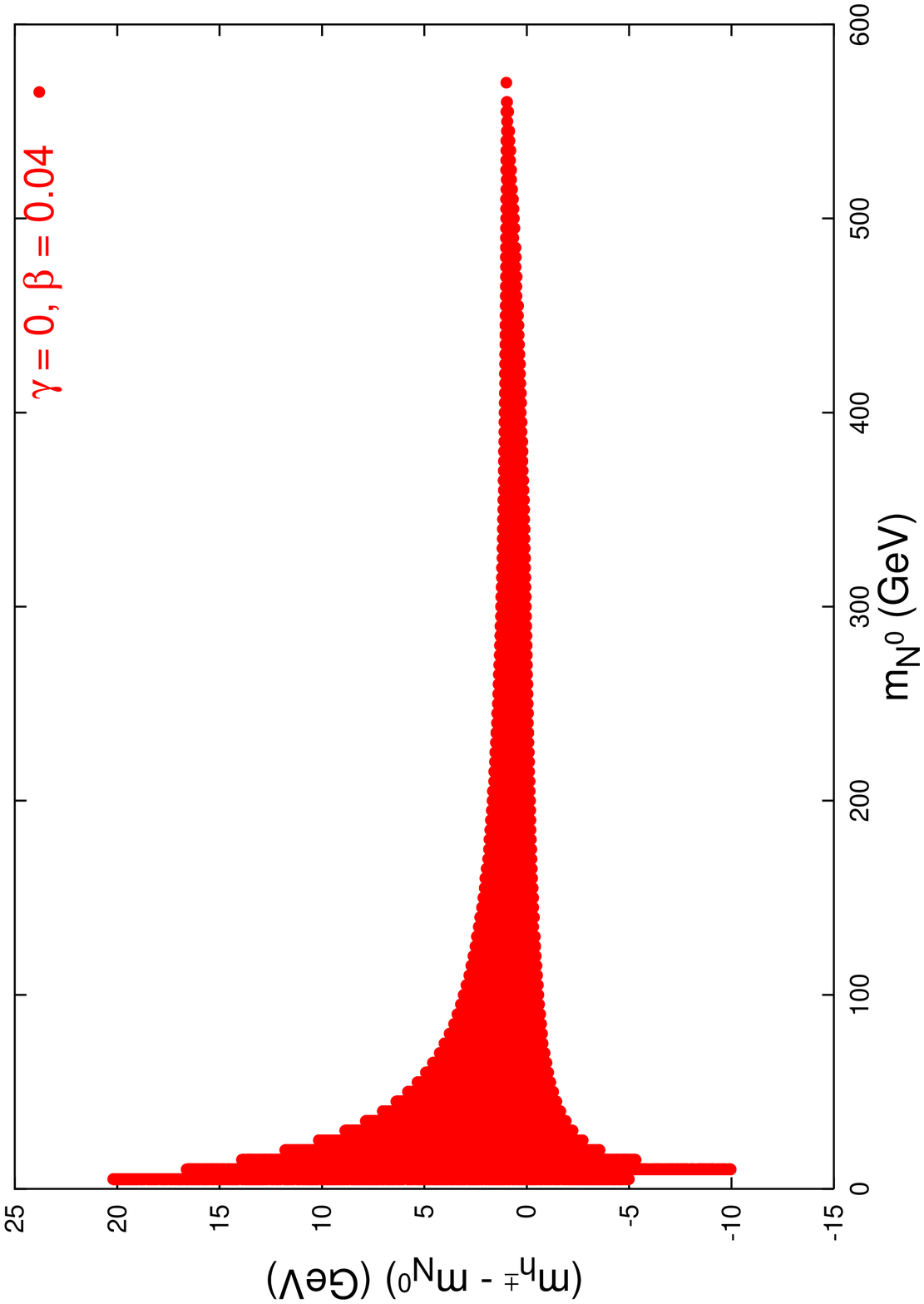}
\vspace*{2.2cm}
\caption[*]{Allowed values of scalar masses for $\gamma=0$}
\label{TripletHiggsFig2}
\end{center}
\vspace*{-1.2cm}
\end{figure}
For $\beta = \gamma = 0$ there is no doublet-triplet mixing and no bound on 
the triplet
mass. This is a special case of the more general decoupling scenario,
which occurs when $|\beta + \gamma| \ll \beta$. For small mixing angles, the 
triplet Higgs has mass squared
$\sim \lambda_4 v / \beta$ and it is possible to have 
$\lambda_4 \sim v$ by keeping $\mu_2^2$ large. In this case 
$\beta+\gamma \approx 0$. This is the decoupling
limit in which the triplet mass lies far above the mass of the
doublet and the low energy model looks identical to the Standard Model. 

\subsection{Conclusions}
We have shown that it is quite natural in the triplet model for the lightest 
Higgs boson to have a mass as large as 500 GeV. Through an analysis of the 
oblique corrections it is possible to see that the model is compatible with 
precision electroweak data. Through a computation of the one--loop beta 
functions for the scalar couplings, and considerations of perturbativity of 
the couplings and vacuum stability we have identified the
allowed masses of the Higgs bosons in the non-decoupling regime. In the
decoupling regime, the model resembles the SM.
 The near degeneracy of the triplet Higgs
masses ensures that, at least for small $\gamma$, the quantum corrections to 
the $T$ parameter are 
negligible (the $S$ parameter vanishing since the triplet has zero 
hypercharge). This means that the lightest Higgs
boson can have a mass as large as 500 GeV as a result of the compensation 
arising
from the explicit tree-level violation of custodial symmetry. Since the 
hypercharge of the triplet is zero, there are no associated problems with 
unwanted phenomenology and thus it is 
possible to be in a regime where all the scalars are $\sim 500$ GeV without
any other deviation from the Standard Model.

}

%% file: yuan.tex
{
%\include{rlFig}
%\floatsep 0cm \textfloatsep 0.2cm
%\renewcommand{\textfraction}{0.3}
%\renewcommand{\topfraction}{0.7}

\def    \nn             {\nonumber}
\def    \=              {\;=\;}
\def    \frac           #1#2{{#1 \over #2}}
\def    \ret            {\\[\eqskip]}
\def    \ie             {{\em i.e.\/} }
\def    \eg             {{\em e.g.\/} }
\def    \lsim           {\raisebox{-3pt}{$\>\stackrel{<}{\scriptstyle\sim}\>$}}
\def    \gsim           {\raisebox{-3pt}{$\>\stackrel{>}{\scriptstyle\sim}\>$}}
\def    \gtrsim         {\raisebox{-3pt}{$\>\stackrel{>}{\scriptstyle\sim}\>$}}
\def    \esim           {\raisebox{-3pt}{$\>\stackrel{-}{\scriptstyle\sim}\>$}}
\newcommand     \be     {\begin{equation}}
\newcommand     \ee     {\end{equation}}
\newcommand     \ba     {\begin{eqnarray}}
\newcommand     \ea     {\end{eqnarray}}
\newcommand     \sst            {\scriptstyle}
\newcommand     \sss            {\scriptscriptstyle}
\newcommand     \avg[1]         {\left\langle #1 \right\rangle}
\newcommand     \Ca             {{C_{\rm A}}}
\newcommand     \Cf             {{C_{\rm f}}}
\newcommand     \lambdamsb     {\ifmmode
          \Lambda_4^{\rm \scriptscriptstyle \overline{MS}} \else
         $\Lambda_4^{\rm \scriptscriptstyle \overline{MS}}$ \fi}
\newcommand     \MSB            {\ifmmode {\overline{\rm MS}} \else
                                 $\overline{\rm MS}$  \fi}
\newcommand     \nf             {n_{\rm f}}
\newcommand     \nlf            {n_{\rm lf}}
\newcommand     \ptmin     {\ifmmode p_{\scriptscriptstyle T}^{\sss min} \else
                           $p_{\scriptscriptstyle T}^{\sss min}$ \fi}
\def     \muf           {\mbox{$\mu_{\sss F}$}}
\def     \mur            {\mbox{$\mu_{\sss R}$}}
\def    \muo            {\mbox{$\mu_0$}}
\newcommand\as{\alpha_{\sss S}}
\newcommand\astwo{\alpha_{\sss S}^2}
\newcommand\asthree{\alpha_{\sss S}^3}
\newcommand\asfour{\alpha_{\sss S}^4}
\newcommand\epb{\overline{\epsilon}}
\newcommand\aem{\alpha_{\rm em}}
\newcommand\QQb{{Q\overline{Q}}}
\newcommand\qqb{{q\overline{q}}}
\newcommand\cb{\overline{c}}
\newcommand\bb{\overline{b}}
\newcommand\tb{\overline{t}}
\newcommand\Qb{\overline{Q}}
\newcommand\qq{{\scriptscriptstyle Q\overline{Q}}}
\def \asopi{\mbox{$\frac{\as}{\pi}$}}
\def \oacube {\mbox{${\cal O}(\asthree)$}}
\def \oatwo {\mbox{${\cal O}(\astwo)$}}
\def \oas   {\mbox{${\cal O}(\as)$}}
\def \ppbar {\mbox{$p \bar p$}}
\def \ttbar {\mbox{$t \bar t$}}
\def \bbbar {\mbox{$b \bar b$}}
\def \ccbar {\mbox{$c \bar c$}}
\def \mtt   {\mbox{$M_{\scriptscriptstyle t\bar t}$}}
\def \pt   {\mbox{$p_{\scriptscriptstyle T}$}}
\def \ptpair   {\mbox{$p_{\scriptscriptstyle T}^{\scriptscriptstyle
                t\bar t}$}}
\def \et   {\mbox{$E_{\scriptscriptstyle T}$}}
\def \etsq {\mbox{$E_{\scriptscriptstyle T}^2$}}
\def \rap   {\mbox{$\eta$}}
\def \deltar {\mbox{$\Delta R$}}
\def \dphi {\mbox{$\Delta \phi$}}
\def \to   {\mbox{$\rightarrow$}}
\def    \mb             {\mbox{$m_b$}}
\def    \mc             {\mbox{$m_c$}}
\def    \mt             {\mbox{$m_t$}}
\newcommand \jpsi{\ifmmode{J/\psi
    }\else{$J/\psi$}\fi}
\def\calF{{\cal F}}
\def\calP{{\cal P}}
\def\calM{{\cal M}}
\def\calO{{\cal O}}
\newcommand{\Tr}{{\mbox{\rm Tr}}}
\newcommand{\mn}{{\mu\nu}}
\newcommand{\half}{{1\over 2}}
\newcommand{\bea}{\begin{eqnarray}}
\newcommand{\eea}{\end{eqnarray}}

\section[ ]{W$^+$W$^+$ Scattering as a Sensitive Test of the 
Anomalous Gauge Couplings of the Higgs Boson at the LHC\footnote{H.-J.\,He,
Y.-P.\,Kuang, C.-P.\,Yuan and B.\,Zhang}}

The electroweak symmetry breaking  mechanism (EWSBM) is one of the most
profound puzzles in particle physics. Since the Higgs
sector of the standard model (SM) suffers the well-known problems of triviality
and unnaturalness
, there has to be new
physics beyond the SM above certain high energy scale $\Lambda$.
If a light Higgs boson candidate ($H$)
is found in future collider experiments,
the next important task is to experimentally measure the gauge
interactions of this Higgs scalar and explore the nature of the
EWSBM. Let $~V=W^\pm,~Z^0~$ be the electroweak (EW) gauge bosons.
The detection of the anomalous $HVV$ couplings (AHVVC) will point to new physics beyond the SM
underlying the EWSBM.
\begin{figure}[h]
\vskip -0.8cm
\begin{center}
\includegraphics[width=0.90\textwidth,clip]{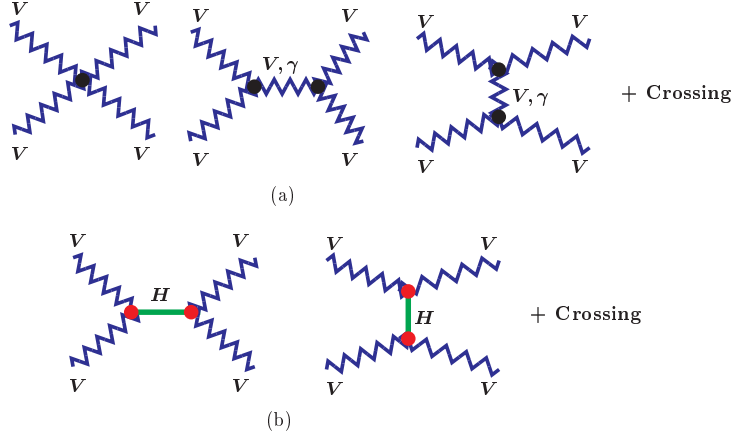}
\end{center}
\vskip -1.2cm
\caption{Illustration of Feynman diagrams for $VV$
scatterings in the SM: (a) diagrams contributing to $T(V,\gamma)$,
(b) diagrams contributing to $T(H)$.} \label{Fig.1}
\end{figure}

Before knowing the correct new physics, the effect of new physics
at energy below $\Lambda$ can be parametrized as
effective operators in an effective theory.
This is a model-independent
description.
Testing the AHVVC relative to that of the SM can
discriminate the EWSBM in the new physics model
from that of the SM. In Ref. \cite{HKYZ03}, we propose
a sensitive way of testing the
AHVVC via $VV$ scatterings, especially the $W^+W^+$
scatterings, at the LHC \cite{HKYZ03}. This includes the test of
either the dim-3 AHVVC in a nonlinearly realized
Higgs model (NRHM) \cite{CK} or the dim-6 AHVVC
in the linearly realized effective interactions (LREI) \cite{linear}.
The reason for the sensitiveness is the following.
 The scattering amplitude
contains two parts: (i) the amplitude $T(V,\gamma)$ related only
to $V$ and $\gamma$ (Fig.\,1(a)), and (ii)
the amplitude $T(H)$ related to the Higgs boson (Fig.\,1(b)).
At high energies, both $T(V,\gamma)$ and $T(H)$
%contain a piece
increase with the center-of-mass energy ($E$) as
$E^2$ in the NRHM and as $E^4$ in the LREI.
In the SM, though individual diagrams in Fig.\,1(a) may behave as
$E^4$, the sum of all diagrams in Fig.\,1(a) can have
at most $E^2$-dependent contribution.
The $HVV$ coupling constant
in the SM is just the non-Abelian gauge coupling constant. This causes the two
$E^2$-dependent pieces to precisely cancel with each other in
$T(V,\gamma)+T(H)$, resulting in the expected $E^0$-behavior for
the total amplitude, as required by the unitarity of the
$S$ matrix. If there is AHVVC due to new physics effect, $T(V,\gamma)+T(H)$
can grow as $E^2$ or $E^4$ in the high energy regime.
Such deviations from the $E^0$ behavior of the SM
amplitude can provide a rather sensitive test of the AHVVC in high energy
$VV$ scattering experiments.  This type
of tests do not require
the measurement of the $H$ decay branching
ratios, and is thus of special interest, especially if the AHVVC
are very large or very small \cite{HKYZ03}.

We take such enhanced $VV$ scatterings as the signals for
testing the AHVVC. To avoid the large hadronic
backgrounds at the LHC \cite{Chanowitz}, We choose the gold-plated pure leptonic
decay modes of the final state $V$s as the tagging modes.
Even so, there are still several kinds of backgrounds to be
eliminated \cite{WW94,WW95}.
We take all the kinematic cuts given in Ref. \cite{WW95} to suppress the
backgrounds, and calculate the complete tree level
contributions to the process
\begin{eqnarray}                        %(1)
pp\to VVjj\to llll(\nu\nu)jj,
\label{VVjj}
\end{eqnarray}
where $j$ is the forward jet that is tagged
to suppress the large background rates.
Our calculation shows that, for
not too small AHVVC, all the
backgrounds can be reasonably suppressed by such kinematic cuts.
In the case of the SM, there are still considerably large remaining backgrounds
contributed by the transverse component $V_T$. We
shall call these the {\it remaining SM backgrounds}
(RSMB) after taking the above treatment. Our calculation shows
that the signals can be considerably larger than the RSMB even
with not very large AHVVC.

We first consider the NRHM.
The effective Lagrangian below $\Lambda$, up to dim-4 operators,
respecting the EW gauge symmetry,
charge conjugation, 
%{\tt C} and parity {\tt P},
parity,
and the custodial $SU(2)_c$ symmetry, is \cite{CK}:.
\begin{eqnarray}                            %(2)
{\cal L}&=&-\frac{1}{4}{\overrightarrow
W}_{\mu\nu}\cdot{\overrightarrow
W}^{\mu\nu}-\frac{1}{4}B_{\mu\nu}B^{\mu\nu}%\nonumber\\
%&&
+\frac{1}{4}(v^2+2\kappa vH+\kappa^\prime H^2){\rm
Tr}(D_\mu\Sigma^\dagger D^\mu\Sigma)\nonumber\\
&&+\frac{1}{2}\partial_\mu H\partial^\mu H
-\frac{m_H^2}{2}H^2-\frac{\lambda_3 v}{3!}H^3+\frac{\lambda_4}{4!}H^4,
\label{Lagrangian}
\end{eqnarray}
where $\overrightarrow W_{\mu\nu}$ and $B_{\mu\nu}$ are field
strengths of the EW gauge fields, $v\simeq 246$\,GeV is the
vacuum expectation value (VEV) breaking the EW gauge symmetry,
$(\kappa,\,\lambda_3)$ and
$(~\kappa^\prime,\,\lambda_4)$ are, respectively,
dimensionless coupling constants from the dim-3 and
dim-4 operators,
$\Sigma=\exp\{i{\overrightarrow\tau}\cdot{\overrightarrow\omega}/
{v}\}$,
and $~D_\mu\Sigma=
\partial_\mu\Sigma+ig\frac{\overrightarrow\tau}{2}\cdot{\overrightarrow
W}_\mu\Sigma -ig'B_\mu\Sigma\frac{\tau_3}{2}$.
The SM corresponds to
$\kappa=\kappa^\prime=1$ and
$\displaystyle\lambda_3=\lambda_4 =\lambda={3 m_H^2}/{v^2}$.

At the tree level, only the dim-3 operator
$\frac{1}{2}\kappa vHD_\mu\Sigma^\dagger D^\mu\Sigma$
contributes to the $VV$ scatterings in Fig.\,1. Therefore,
$VV$ scatterings can test
$\kappa$, and $\Delta\kappa\equiv\kappa -1$ measures the deviation from the
SM value $\kappa =1$.

In Ref. \cite{HKYZ03}, the full tree level cross sections for all the processes
in (\ref{VVjj}) are calcuated for $115~{\rm GeV}\le m_H\le 300$ GeV.
The results show that the most sensitive channel
is ~$pp\to W^+W^+jj\to l^+\nu l^+\nu jj$ \cite{HKYZ03}. With an integrated
luminosity of 300 fb$^{-1}$, there are more than 20 events for
$\Delta\kappa\ge 0.2$ or $\Delta\kappa\le -0.3$, while there are only
about 15 RSMB events (see Ref. \cite{HKYZ03} for details).
Considering only the statistical errors,
the LHC can limit $\Delta\kappa$ to the range
\begin{eqnarray}                    %(3)
-0.3<\Delta\kappa<0.2
\label{constraint}
\end{eqnarray}
at roughly the $(1-3)\sigma$ level if data is consistent with the
SM prediction \cite{HKYZ03}.

Other constraints on $\Delta\kappa$ from the precision EW data,
the requirement of the unitarity of the $S$-matrix, etc. were
studied in Ref.~\cite{HKYZ03}, which are either weaker than Eq.
(\ref{constraint}) or of the similar level \cite{HKYZ03}.

Next, we consider the LREI. In this theory, the leading AHVVC are from the effective
operators of dim-6 \cite{linear,G-G}.
As is shown in Refs.~\cite{linear,G-G}, the {\tt C} and {\tt P}
conserving effective Lagrangian up to dimension-6 operators
containing a Higgs doublet $\Phi$ and the weak bosons $V^a$ is
given by
\begin{equation}                    %(4)
{\cal L}_{\mbox{eff}} ~\,=~\, \sum_n \frac{f_n}{\Lambda^2} {\cal O}_n \,,
\label{l:eff}
\end{equation}
where ${\cal O}_n$'s are dim-6 operators composed of $\Phi$ and the EW gauge
fields (cf. Ref. \cite{G-G}), $f_n/\Lambda^2$'s are the AHVVC.

The precision EW data and the requirement of the unitarity of the $S$-matrix
give certain constraints on the $f_n$'s. The constraints on
$f_{WWW}/\Lambda^2$, $f_{WW}/\Lambda^2$, $f_{BB}/\Lambda^2$,
$f_W/\Lambda^2$, and $f_B/\Lambda^2$ from the presently available
experimental data are rather weak \cite{HKYZ03}.
A better test of them is to study the $VV$ scatterings.
In ${\cal L}_{\mbox{eff}}$, the operator ${\cal
O}_{WWW}$ contributes to the triple and quartic $V$ boson
self-interactions which may not be directly related to the EWSBM,
we assume $f_{WWW}/\Lambda^2$ is small in the analysis.
and concentrate on the test of $f_{WW}/\Lambda^2$,
$f_{BB}/\Lambda^2$, $f_W/\Lambda^2$,
and $f_B/\Lambda^2$ . They are related to the following AHVVC
in terms of $H$, $W^\pm$, $Z$, and $\gamma$ \cite{G-G}:
\begin{eqnarray}                         %(5)
{\cal L}^H_{\rm eff}&=&g_{H\gamma\gamma}HA_{\mu\nu}A^{\mu\nu}
+g^{(1)}_{HZ\gamma}A_{\mu\nu}Z^\mu\partial^\nu H%\nonumber\\
%&&\hspace{0.4cm}
+g^{(2)}_{HZ\gamma}HA_{\mu\nu}Z^{\mu\nu}
+g^{(1)}_{HZZ}Z_{\mu\nu}Z^\mu
\partial^\nu H\nonumber\\
&&%\hspace{0.4cm}
+g^{(2)}_{HZZ}HZ_{\mu\nu}Z^{\mu\nu}
+g^{(1)}_{HWW}(W^+_{\mu\nu} W^{-\mu}\partial^\nu H+{\rm h.c.})
+g^{(2)}_{HWW}HW^+_{\mu\nu}W^{-\mu\nu},
\label{LHeff}
\end{eqnarray}
where
\begin{eqnarray}                           %(6)
\displaystyle
&&g^{}_{H\gamma\gamma}=-\bigg(\frac{gm_W}{\Lambda^2}\bigg)\frac{s^2(f_{BB}
+f_{WW})}{2},%\nonumber\\
%&&
~~g^{(1)}_{HZ\gamma}=\bigg(\frac{gm_W}{\Lambda^2}\bigg)\frac{s(f_W-f_B)}{2c},
\nonumber\\
&&g^{(2)}_{HZ\gamma}=\bigg(\frac{gm_W}{\Lambda^2}\bigg)\frac{s[s^2f_{BB}
-c^2f_{WW}]}{c},%\nonumber\\
%&&
~~g^{(1)}_{HZZ}=\bigg(\frac{gm_W}{\Lambda^2}\bigg)\frac{c^2f_W+s^2f_B}{2c^2},
\nonumber\\
&&g^{(2)}_{HZZ}=-\bigg(\frac{gm_W}{\Lambda^2}\bigg)\frac{s^4f_{BB}
+c^4f_{WW}}{2c^2},%\nonumber\\
%&&
~g^{(1)}_{HWW}=\bigg(\frac{gm_W}{\Lambda^2}\bigg)\frac{f_W}{2},%\nonumber\\
%&&
~g^{(2)}_{HWW}=-\bigg(\frac{gm_W}{\Lambda^2}\bigg)f_{WW},\,\,\,\,\,
\label{g}
\end{eqnarray}
with $s\equiv \sin\theta_W$ and $c\equiv \cos\theta_W$.

The test of these AHVVC via $VV$ scatterings is quite different from that of
$\Delta\kappa$. The relevant operators ${\cal O}_n$'s contain two derivatives.
so, at high energies, the interaction vertices themselves behave
as $E^2$, and thus
the longitudinal $VV$ scattering amplitudes, $V_LV_L \to V_LV_L$,
grows as $E^4$, and those containing $V_T$
grow as $E^2$. Hence the scattering processes containing
$V_T$ actually behave as {\it signals} rather than backgrounds.

It is shown in Ref. \cite{HKYZ03} that the most sensitive channel is still
$pp\to W^+W^+jj\to l^+\nu l^+\nu jj$. Detailed calculations show that the
contributions of $f_B$ and $f_{BB}$
are small even if they are of the same order of magnitude
as $f_W$ and $f_{WW}$ \cite{HKYZ03}.
Hence, we take account of only $f_W/\Lambda^2$ and
$f_{WW}/\Lambda^2$ in the analysis.
If they are of the same order of magnitude,
the interference between them may be significant, depending on
their relative phase which undoubtedly complicates the
analysis. Hence, we perform a single parameter analysis,
i.e., assuming only one of them is dominant at a time.
In the case that $f_W$ dominates, the obtained numbers of events
in $pp\to W^+W^+jj$ $\to l^+\nu l^+\nu jj$
with an integrated luminosity of $300$ fb$^{-1}$ are more than 20
for $f_W/\Lambda^2\ge 0.85$ TeV$^{-2}$ or $f_W/\Lambda^2\le -1.0$ TeV$^{-2}$
, and the number of the RSMB events are still around 15 (see
Ref. \cite{HKYZ03} for details). If no AHVVC
effect is found at the LHC, we can set the following bounds on $f_W/\Lambda^2$
(in units of TeV$^{-2}$) when taking into account of only the statistical error:
\begin{eqnarray}                       %(7)
1\sigma:~~-1.0< f_W/\Lambda^2< 0.85,~~~~~~~~~~~
2\sigma:~~-1.4< f_W/\Lambda^2\leq 1.2. \,\,\,\,\,\label{fW}
\end{eqnarray}
In the case that $f_{WW}$ dominates,
the corresponding bounds are (in
units of TeV$^{-2}$):
\begin{eqnarray}                       %(8)
1\sigma:~~-1.6\leq f_{WW}/\Lambda^2<1.6,
~~~~~~~~~~2\sigma:~~-2.2\leq f_{WW}/\Lambda^2< 2.2. \label{fWW}
\end{eqnarray}
These are somewhat weaker than those in Eq.~(\ref{fW}).
From  Eqs. (\ref{fW}) and (\ref{fWW})
we obtain the corresponding bounds on $g^{(i)}_{HVV},~i=1,2$ (in units of TeV$^{-1}$):
\begin{eqnarray}                           %(9)
&&1\sigma:\nonumber\\
&&\hspace{0.2cm}-0.026< g^{(1)}_{HWW}< 0.022,
~~~-0.026< g^{(1)}_{HZZ}< 0.022,
~~~-0.014< g^{(1)}_{HZ\gamma}< 0.012,\nonumber\\
&&\hspace{0.2cm}
-0.083\leq g^{(2)}_{HWW}< 0.083,
~~~~~~~~0.032\leq g^{(2)}_{HZZ}< 0.032,
~~~-0.018\leq g^{(2)}_{HZ\gamma}< 0.018,
\label{1sigmabounds}
\end{eqnarray}
\begin{eqnarray}                     %(10)
&&2\sigma:\nonumber\\
&&\hspace{0.4cm}-0.036< g^{(1)}_{HWW}\leq 0.031,
~~~~~~~0.036< g^{(1)}_{HZZ}\leq 0.031,
~~~~~~~0.020< g^{(1)}_{HZ\gamma}\leq 0.017,\nonumber\\
&&\hspace{0.4cm}-0.11~\leq g^{(2)}_{HWW}< ~0.11,
~~~~-0.044\leq g^{(2)}_{HZZ}< 0.044,
~~~-0.024\leq g^{(2)}_{HZ\gamma}< 0.024.
\label{2sigmabounds}
\end{eqnarray}
These bounds are to be compared with the $1\sigma$ bound on
$g^{(2)}_{HWW}$ obtained from studying the on-shell Higgs boson production
via weak boson fusion at the LHC given in
Ref.~\cite{PRZ}, where $g^{(2)}_{HWW}$ is
parametrized as $g^{(2)}_{HWW}$$=1/\Lambda_{5}=g^2
v/\Lambda^2_{6}$. The obtained $1\sigma$ bound on $\Lambda_6$
for an integrated luminosity of 100 fb$^{-1}$ is about $\Lambda_6
\ge$ 1 TeV \cite{PRZ}, which corresponds to $g^{(2)}_{HWW}=1/\Lambda_5\le
0.1~ {\rm TeV}^{-1}$. We see that the $1\sigma$ bounds in
Eq. (\ref{1sigmabounds}) are all stronger than the above bound
given in Ref.~\cite{PRZ}. For an
integrated luminosity of 300 fb$^{-1}$, the bound
in Ref.~\cite{PRZ} corresponds roughly to a 1.7$\sigma$ level accuracy.
Comparing it with the results in Eq. (\ref{2sigmabounds}), we conclude that our 2$\sigma$ bound on
$g^{(2)}_{HWW}$ is at about the same level of accuracy,
while our 2$\sigma$ bounds on
the other five $g^{(i)}_{HVV}~(i=1,2)$ are all stronger than
those given in Ref.~\cite{PRZ}.

It has been shown in Ref.~\cite{HZZ} that the anomalous $HZZ$
coupling constants $g^{(1)}_{HZZ}$ and $g^{(2)}_{HZZ}$ can be
tested rather sensitively at the Linear Collider (LC) via the
Higgs-strahlung process $e^+e^-\to Z^*\to Z+H$ with $Z\to
f\bar{f}$. The obtained limits are $g^{(1)}_{HZZ}\sim
g^{(2)}_{HZZ}\sim O(10^{-3}\--10^{-2})$ TeV$^{-1}$ \cite{HZZ}.
Although the bounds shown in Eqs. (\ref{1sigmabounds}) and
(\ref{2sigmabounds}) are weaker than these LC bounds, $W^+W^+$
scattering at the LHC can provide the bounds on
$g^{(i)}_{HWW},~i=1,2$ which are not easily accessible at the LC.
So that the two experiments are complementary to each other.

Further discrimination of the effect of the AHVVC
from that of a strongly interacting EW symmetry breaking sector
with no light resonance will eventually demand a multichannel
analysis at the LHC by searching for the light Higgs resonance
through all possible on-shell production channels including
gluon-gluon fusion. Once the light Higgs resonance is confirmed,
$VV$ scatterings, especially the $W^+W^+$ channel, can provide
rather sensitive tests of various AHVVC for
probing the EWSB mechanism. So $VV$ scatterings are not only
important for probing the strongly interacting EWSBM when there is
no light Higgs boson, but can also provide sensitive test of the
AHVVC (especially the anomalous $HWW$
couplings) at the LHC for discriminating new physics from the SM
when there is a light Higgs boson.

}

%% file: escalier.tex
{
\section[ ]{Higgs boson and diphoton production at the
LHC\footnote{C.\,Bal\'azs, M.\,Escalier and B.\,Laforge}}

\subsection{Introduction}

In the Standard Model (SM) the Higgs field plays a central role in giving mass 
to the electroweak (EW) gauge bosons and the fermions. Despite the thorough 
experimental search, the Higgs boson has not been discovered up to date.
The CERN Large Electron Positron collider (LEP) imposes a lover limit on the 
Higgs mass: $m_H>114.3$ GeV at $95\%$ confidence level \cite{Hagiwara:2002fs}. 
The global fit to the EW measurements, using all LEP, SLC and Tevatron 
data, lead to a Higgs mass of $m_H = 91^{+58}_{-37}$ GeV \cite{Monig:2003qu}. 
This might suggest new physics or indicate a light SM Higgs boson.
In the region $m_H<150$ GeV, the $H\to\gamma\gamma$ decay mode is the most 
promising channels to discover the Higgs boson. Despite a very small branching 
ratio of a few $10^{-3}$, the clean final state can be identified with relative 
ease. This channel has been analysed in the past \cite{unknown:1990fr, 
unknown:1992fr, Eynard:1998jn, unknown:1999fr}, but since then new theoretical 
calculations have been published \cite{Binoth:1999qq, Balazs:1998hv, 
Balazs:1998bm, Balazs:1999yf, Balazs:1999gh, Balazs:2000wv, Balazs:2000sz, 
Balazs:2000rg}.

Diphoton production is an irreducible background for the Higgs search, while 
misidentified final state particles seen as photons contribute to a reducible 
background. (The latter is mainly due to one or two jets misidentification.)
In this work the 
\vspace{-0.1cm}
\begin{floatingtable}[r]{
\begin{tabular}{|r|c|l|}
\hline
\rowcolor[gray]{.9}
ISUB &process($pp\rightarrow H^0$) &$\sigma$(mb)\\
\hline
102& $gg\rightarrow H^0$ &$1.82\times10^{-8}$\\
\hline
124 &$qq\rightarrow qqH^0(W^+W^- fusion)$  &$3.04\times10^{-9}$\\
\hline
26 &$q\bar{q}\rightarrow W^+H^0$ &$1.26\times10^{-9}$ \\
\hline
123 &$qq\rightarrow qqH^0 (Z^0Z^0 fusion)$ &$1.22\times10^{-9}$\\
\hline
3 &$q\bar{q}\rightarrow H^0$ &$1.03\times10^{-9}$ \\
\hline
24 &$q\bar{q}\rightarrow Z^0H^0$ &$7.37\times10^{-10}$ \\
\hline
121& $gg\rightarrow Q\bar{Q} H^0$ &$ 4.24\times10^{-10}$\\
\hline
122& $q\bar{q}\rightarrow Q\bar{Q} H^0$ &$1.75\times10^{-10}$\\
\hline
\end{tabular}}
\caption{Leading order production cross sections at the LHC for a Standard Model 
Higgs boson of mass 120 GeV by PYTHIA 6.210 with the CTEQ6L1 PDF.
\label{section_eff_PYTHIA}}
\end{floatingtable}
\noindent
irreducible diphoton background is studied in detail. First, predictions for the 
signal are briefly compared using the Monte Carlo (MC) codes PYTHIA 
\cite{Sjostrand:2000wi} and ResBos \cite{Balazs:1997xd}, the next-to-leading-%
order (NLO) program Hi\-Glu \cite{Spira:1995mt}, and one of the recent NLO 
calculations \cite{Ravindran:2003ia}.
Then the background is scrutinized using the MC codes Diphox 
\cite{Binoth:1999qq}, PY\-THIA, ResBos and a recent NLO QCD calculation 
\cite{Bern:2002jx}.
The reducible background will be computed using PYTHIA and presented in a 
later work.

\subsection{The Higgs signal}

Higgs boson production at the LHC mainly proceeds via gluon fusion, which is 
illustrated by Table \ref{section_eff_PYTHIA}. The second highest channel is 
vector boson fusion (VBF). 
In order to assess the signal, we computed the LO and NLO gluon fusion cross 
sections for the LHC and a light Higgs boson ($m_H = 120$ GeV) using HiGlu with 
CTEQ6M parton density functions (PDFs) \cite{Kretzer:2003it}. The NLO/LO K-factor
turns out to be $K_{NLO/LO}=1.8$, which is consistent with the one quoted 
in Refs. \cite{Harlander:2002wh,Anastasiou:2002yz,Ravindran:2003um}. Recent NNLO 
calculations \cite{Ravindran:2003um} report a NNLO\-/NLO K-factor $K_{NNLO/NLO} 
= 1.16$, which shows the good convergence of the perturbative expansion of the 
cross section.
Similarly, we used HiGlu to evaluate the K-factor in VBF. The results give 
$K_{NLO/LO} = 1.04$ which is consistent with Ref.\cite{Figy:2003nv}.
We also computed the uncertainties due to the PDFs in the HiGlu NLO gluon 
fusion, using 40 parameterizations of CTEQ6 \cite{Stump:2002yv}. Our results show 
$\delta\sigma_{NLO}=_{-4.64}^{+3.44}\% $, in good agreement with 
Ref.\cite{Djouadi:2003jg}.

%\newpage

%\paragraph{\it Transverse momentum distributions} ~\\[-0.3cm]
\paragraph{\it Transverse momentum distributions} ~\\

\begin{floatingfigure}[r]{9cm}
\vspace{-0.5cm}
\begin{center}
\setlength\unitlength{1cm}
\includegraphics[width=8cm]{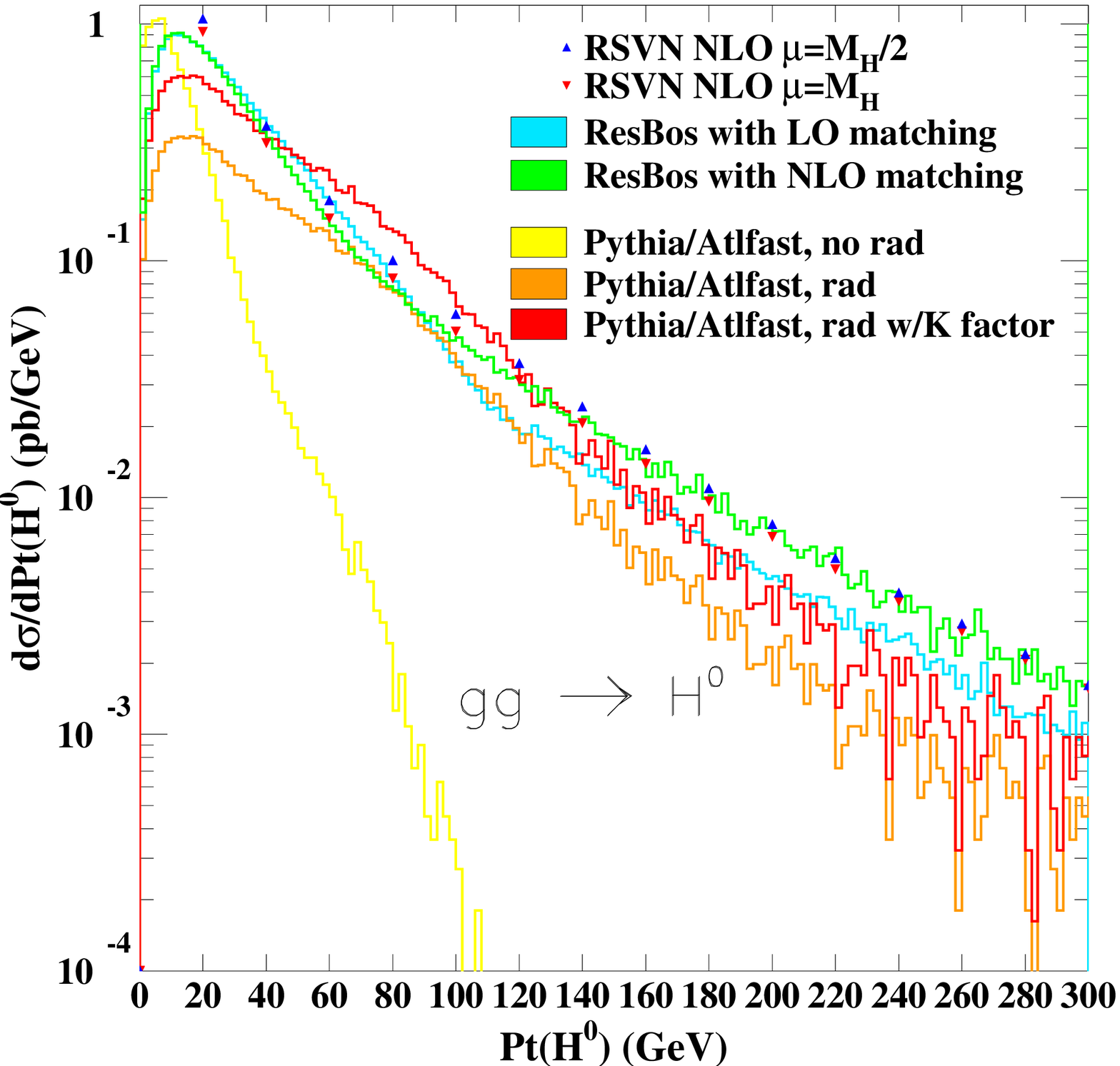}
\put(-4.5,0.8){\includegraphics[width=2.5cm]{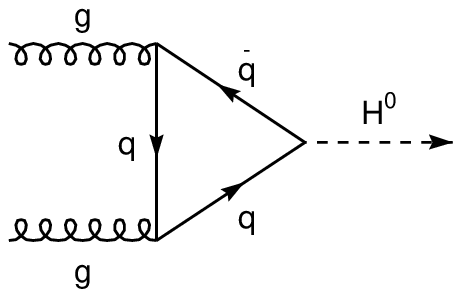}}
\end{center}
\vspace{-0.5cm}
\caption{Transverse momentum distributions of 120 GeV Higgs bosons predicted 
by PYTHIA 6.210, ResBos, and the NLO calculation of Ravindran, Smith and 
Van-Neerven.}
\label{gg_H_resbos}
\end{floatingfigure}

\noindent
To realistically estimate the acceptance and other detector effects, we need 
predictions for NLO differential cross sections, especially for the transverse 
momentum ($P_t$) distribution of Higgs bosons.  We computed these with ResBos, 
which calculates the total cross section and invariant mass distribution at NLO 
and the $P_t$ distribution shape at the resummed level with NLO matching 
and normalization \cite{Balazs:2000wv,Balazs:2000rg}. 
ResBos decays the Higgs boson to two photons, in a very good agreement with 
HDecay \cite{Djouadi:1998yw}. These photons are reconstructed after smearing the 
final states photon momenta according to the ATLAS detector performance 
\cite{Wielers_conf,ATLFast}. This distribution is compared to other predictions 
in Fig.~\ref{gg_H_resbos}.

The ResBos distribution, with NLO matching and renormalization and factorization 
scales $\mu_R=\mu_F=M_H$, shows a very good agreement with the recent NLO 
calculation \cite{RSVN_private} at the intermediate to high $P_t$. The deviation 
between the NLO matched ResBos (light green) histogram and the NLO prediction
(red triangles) is in the order of the MC statistics, which is less than about 
10\% for $P_t \sim M_H$.
In the low $P_t$, without initial and final state radiation, PYTHIA 
predicts a much softer spectrum than ResBos. But if we allowed for radiation in 
PYTHIA, we were not able to reproduce the shape of the $P_t$ distribution 
predicted by the other calculations. The ResBos total rate is found to be 
$\sigma=36.7$ pb, in excellent agreement with the HiGlu prediction of 
$\sigma=36.4$ pb.

\subsection{The diphoton background}

In this section we compare results of various calculations for the irreducible 
diphoton background of Higgs production.\footnote{A study of issues related to 
the reducible background, which are beyond the scope of this work, can be found 
in Ref.\cite{Wielers_rejection}.} At the lowest order the most important 
contributors are the $q{\bar q} \to \gamma \gamma$ (Born) and the $gg \to \gamma 
\gamma$ (box) subprocesses. At the LHC, with CTEQ6L1, PYTHIA predicts 35.0 and 
37.8 pb total cross sections, respectively. (Here $p_t>25$ GeV required for 
each photon.) Since PYTHIA yields a very soft $P_t$ distribution for the photon 
pair, we use two other MC codes to compute the diphoton $P_t$: Diphox and 
ResBos.

% \subsection{Diphox}

The Diphox code implements the NLO QCD calculation of the $q{\bar q}+{\bar 
q}^{\hspace{-0.24cm}(~\,)}g\to\gamma\gamma X$, and the LO calculation of the $gg 
\to \gamma \gamma$ subprocess. (These are referred to as direct processes). It 
also includes the single and double fragmentation processes as schematically 
depicted in Table~\ref{tab:TotXsec}(a). The NLO singularities, related to parton 
emission collinear to one of the photons, are regulated by a cone of radius $R$ 
around each photons. Since Diphox slices the phase space in the $P_t$ variable, 
its prediction for the $P_t$ distribution depends on non- physical parameters 
which handicaps its use in the experimental analysis.

% \subsubsection{ResBos}

ResBos implements a generalized factorization formalism applied to diphoton 
production \cite{Balazs:1998hv,Balazs:1998bm,Balazs:1999yf}. As a result, it 
includes the direct processes up to NLO for the $q{\bar q}+{\bar q}^{\hspace{-
0.24cm}(~\,)} g\to\gamma\gamma X$ and the $gg\to\gamma\gamma X$ subprocesses 
(except the two loop virtual corrections to the gluon fusion subprocess), 
and also the resummation of $\log(P_t/M_{\gamma\gamma})$ (where 
$M_{\gamma\gamma}$ is the mass of the photon pair). ResBos matches the resummed 
low $P_t$ prediction to the high $P_t$ NLO distribution \cite{Balazs:1998hv}. 
This feature enables ResBos to give a correct prediction of the full $P_t$ 
distribution which is important at the Higgs search when a likelihood ratio 
method is used to reject background. Finally, ResBos implements fragmentation 
at LO.

The full NLO QCD corrections to the gluon fusion subprocess, including the two 
loop $gg\to\gamma\gamma$ virtual corrections were recently calculated 
\cite{Bern:2002jx}. The NLO K-factor $K_{NLO/LO}$ was found to be about 1.6 for 
an invariant mass of $M_{\gamma\gamma} = 120$ GeV. The authors also demonstrated 
the reduction of the scale dependence in the NLO calculation. 

\begin{table}[h]
\begin{tabular}{cc|c|r|r|}
\multirow{6}{8cm}{\includegraphics[width=7cm]{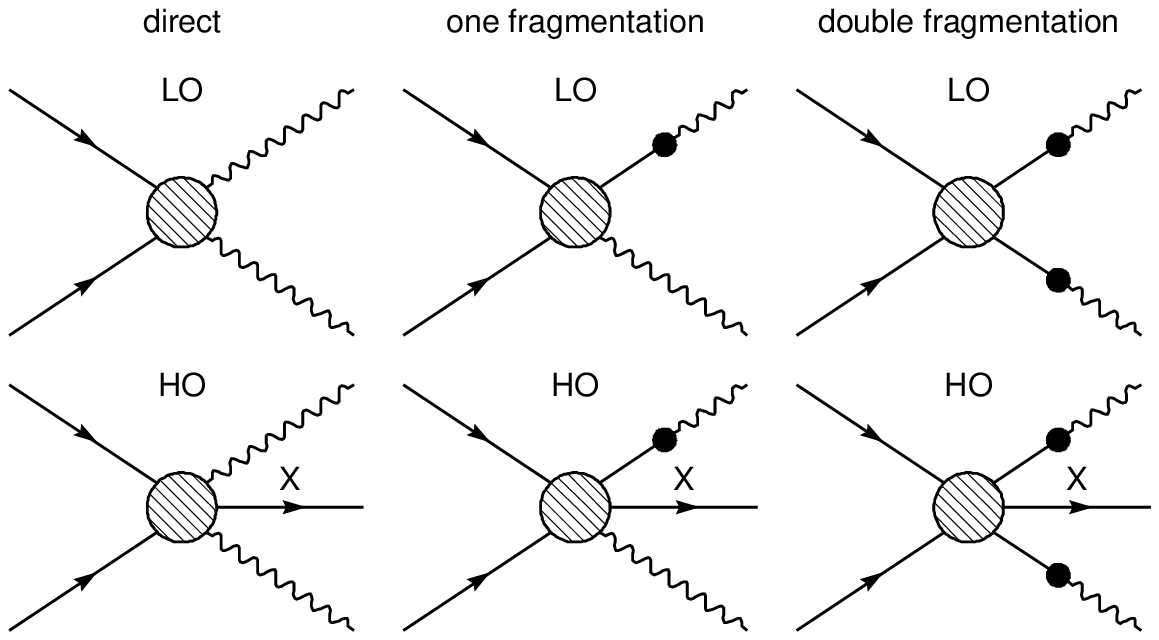}}
&\multicolumn{4}{c}{$80 < M_{\gamma\gamma} < 140$ GeV, ~~~~ $p_{t_\gamma}\ge 25$ GeV,} \\ 
&\multicolumn{4}{c}{$|\eta|_{\gamma}<2.4$, ~~~~ $R=0.4$, ~~~~ ${E_T}_{cone}\le 15$ GeV.} \\ 
&\multicolumn{4}{l}{\small \hspace{-0.2cm} (b) Cuts on the diphoton final state.} \\ 
\\
\cline{2-5} % \rowcolor[gray]{.9}
&\multicolumn{1}{|c|}{Process}            &Contribution            &Diphox &ResBos\\
\cline{2-5}
&\multicolumn{1}{|c|}{$q\bar{q}+{\bar q}^{\hspace{-0.24cm}(~\,)}g$}&direct &$ 9.23$ &$13.2$\\
\cline{3-5}                                
&\multicolumn{1}{|c|}{$\to\gamma\gamma$}  &total                   &$19.1$&$18.5$\\
\cline{2-5}                                     
&\multicolumn{1}{|c|}{$gg\to\gamma\gamma$}&direct                  &$ 5.47$ &$6.85$\\
\cline{2-5}
\end{tabular}
\caption{
(a) Schematic nomenclature used by Diphox for  \hspace{1.10cm} (c) Diphoton~ production~ cross~ sections,~ in~ picobarns,~ at
direct, one, and two fragmentation processes.  \hspace{2.60cm} the LHC with CTEQ6M and cuts given above. 
}
\label{tab:TotXsec}
\end{table}

\newpage

\paragraph{\it Numerical results} ~\\[-0.3cm]

In this section we present numerical results for the total cross section, the 
invariant mass ($M_{\gamma\gamma}$) and the transverse momentum ($P_t$) 
distributions of the diphoton system calculated by various codes. All these 
results are computed for the LHC ($pp$ collisions at $\sqrt{S}=14$ TeV) with 
CTEQ6M PDF (unless stated otherwise) and with equal renormalization and 
factorization scales $\mu_F=\mu_R=M_{\gamma\gamma}$. 
Since our interest is the background to a light Higgs boson, we limit all the 
events in the mass region $80 < M_{\gamma\gamma} < 140$ GeV. Besides, we request 
the standard ATLAS cuts given in Table \ref{tab:TotXsec}(b). 
For $q{\bar q}+{\bar q}^{\hspace{-0.24cm}(~\,)}g\to\gamma\gamma X$ we use Diphox 
and ResBos while for $gg\to\gamma\gamma X$ we also show results of the NLO 
calculation by Bern, Dixon and Schmidt \cite{Bern_Dixon_Schmidt_private}. 

The total cross sections, obtained by Diphox and ResBos, are summarized in Table 
\ref{tab:TotXsec}(c). ResBos tend to predict higher direct cross sections since 
in addition to the NLO contribution the infinite tower of $\alpha_S^n 
\log^m(P_t/M_{\gamma\gamma})$ is also included. On the other hand, Diphox 
implements NLO fragmentation while ResBos does this at LO. The higher ResBos 
rate is also due in part to the fact that the resummation calculation integrates 
over the soft radiation and thus provides no information about the soft partons 
around the photons. As a result, at low $P_t$, the ${E_T}_{cone}\le 15$ GeV cut 
is not imposed on ResBos.
In spite of the above listed differences, the total cross sections computed 
by the two codes agree within 4\% after summed over all the final states.
As Table \ref{tab:TotXsec}(c) shows, at the LHC, the dominant contribution 
is coming from the $q{\bar q}+{\bar q}^{\hspace{-0.24cm}(~\,)}g$ initiated 
subprocesses. These subprocesses yield about 75-80\% of the total rate.

\begin{figure}[t]
\setlength\unitlength{1cm}
\begin{picture}(-2,7)
\includegraphics[width=7.5cm]{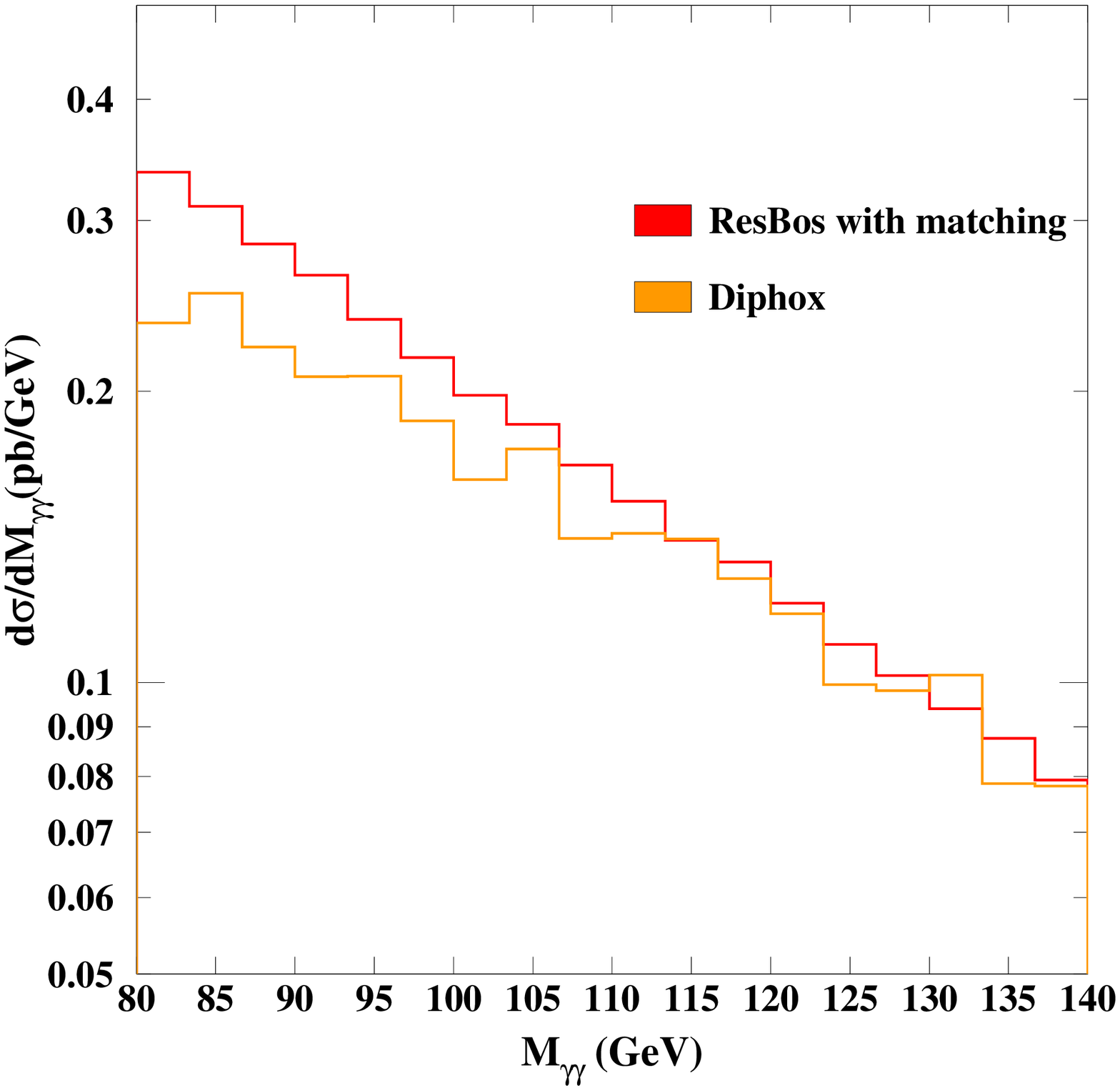}
\put(-6.2,2.6){\includegraphics[width=2.3cm]{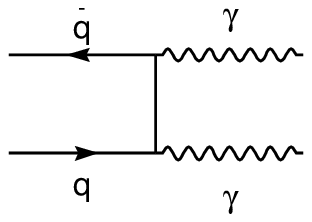}}
\put(-6.2,0.8){\includegraphics[width=2.3cm]{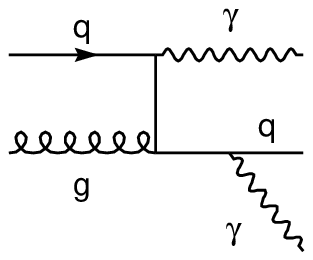}}
\hspace{0.5cm}
\includegraphics[width=7.5cm]{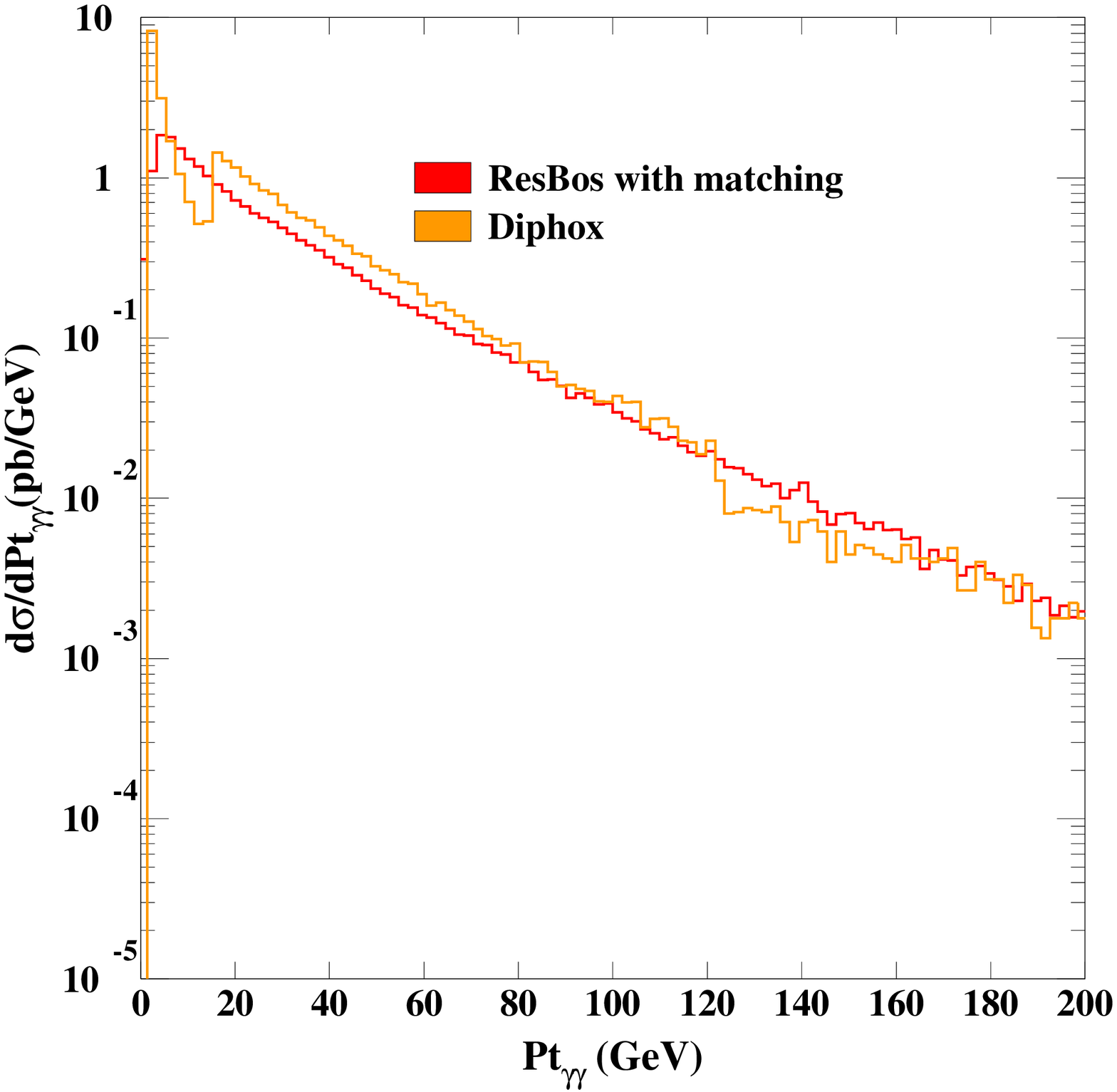}
\put(-6.2,2.6){\includegraphics[width=2.3cm]{born.eps}}
\put(-6.2,0.8){\includegraphics[width=2.3cm]{bremstrah.eps}}
\end{picture}
\caption{
Invariant mass and transverse momentum distributions for the 
$q\bar{q}+{\bar q}^{\hspace{-0.24cm}(~\,)}g\to\gamma\gamma$ subprocesses 
% predicted 
by Diphox and ResBos.}
\label{diphox_resbos_born_brem_distribution}
\end{figure}

Fig.~\ref{diphox_resbos_born_brem_distribution} shows the invariant mass and 
transverse momentum distributions for processes initiated by $q\bar{q}$ or 
${\bar q}^{\hspace{-0.24cm}(~\,)}g$. Overall there is a good agreement between 
Diphox and ResBos in the relevant kinematic regions. The two codes somewhat 
disagree for low values of $M_{\gamma\gamma}$ but, fortunately, they agree for 
$M_{\gamma\gamma} > 110$ GeV within the statistical uncertainties. 
The low and mid $P_t$ prediction of Diphox suffers from the problem of phase 
space separation and agreement with ResBos is not expected. 
On the other hand, according to expectations, in the high $P_t$ region the 
two predictions agree within the statistical errors.

In Fig.~\ref{diphox_box_distribution} the $M_{\gamma\gamma}$ and $P_t$ 
distributions are shown for the $gg\to\gamma\gamma X$ subprocess. 
Here we used CTEQ6L1 for the LO and CTEQ6M for the NLO distributions.
For $M_{\gamma\gamma}$ the Diphox and the LO analytic calculation 
\cite{Bern_Dixon_Schmidt_private} are in agreement as expected. The ResBos rate 
is somewhat higher due to the additional logarithmic contributions which are 
enhanced for the $gg$ initial state. ResBos also overestimates the NLO 
results, which is probably due to the fact that the $gg\to\gamma\gamma$ two loop 
virtual corrections are missing from it. For low invariant masses PYTHIA tend to 
agree with ResBos while it favors the NLO result at larger $M_{\gamma\gamma}$.
On the one hand, comparison of the LO and NLO QCD calculation emphasizes the 
importance of the implementation of the full NLO corrections in the MC codes. On 
the other, comparison of the LO QCD calculation and PYTHIA shows the importance 
of the initial state radiation. The almost 30\% scatter of the various 
predictions shows that each of them is missing an important part of the known 
QCD corrections.

For the $P_t$ distribution, the Diphox prediction  peaks at zero since this 
process is LO in Diphox. ResBos and PYTHIA provide soft gluon radiation so their 
spectra are broader. ResBos also contains the $gg\to\gamma\gamma g$ real 
emission contribution. On the other hand, as it was shown in Ref. 
\cite{Balazs:1999yf}, this contribution is almost negligible, so we do not 
include it here.

\begin{figure}[t]
%\begin{center}
\setlength\unitlength{1cm}
\begin{picture}(-2,7)
\includegraphics[width=7.5cm]{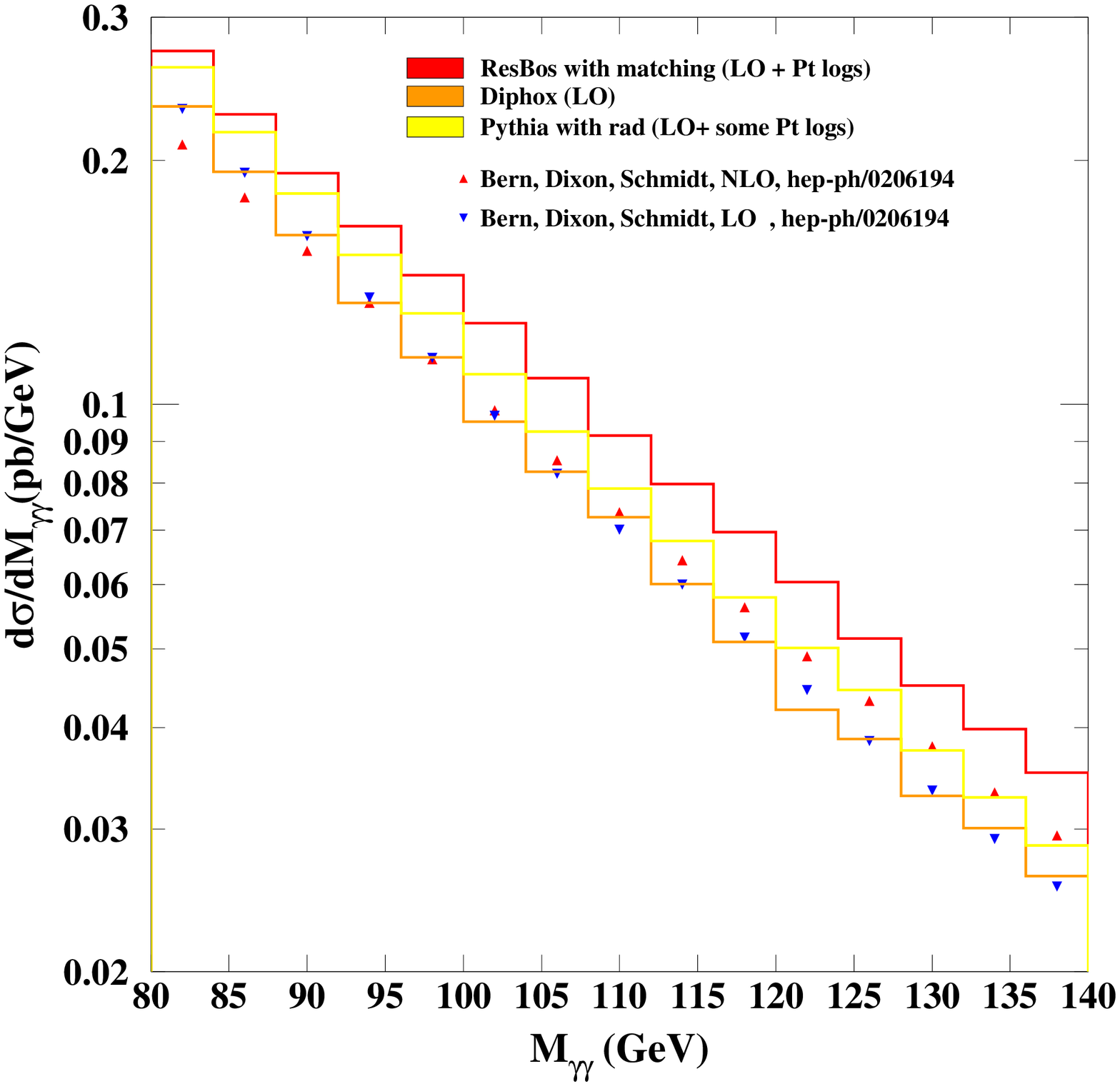}
\put(-6.2,0.8){\includegraphics[width=3.5cm]{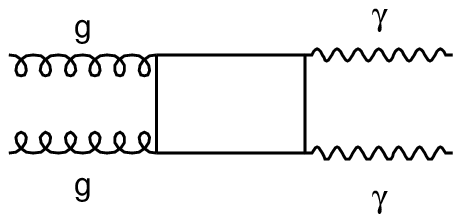}}
\hspace{0.5cm}
\includegraphics[width=7.5cm]{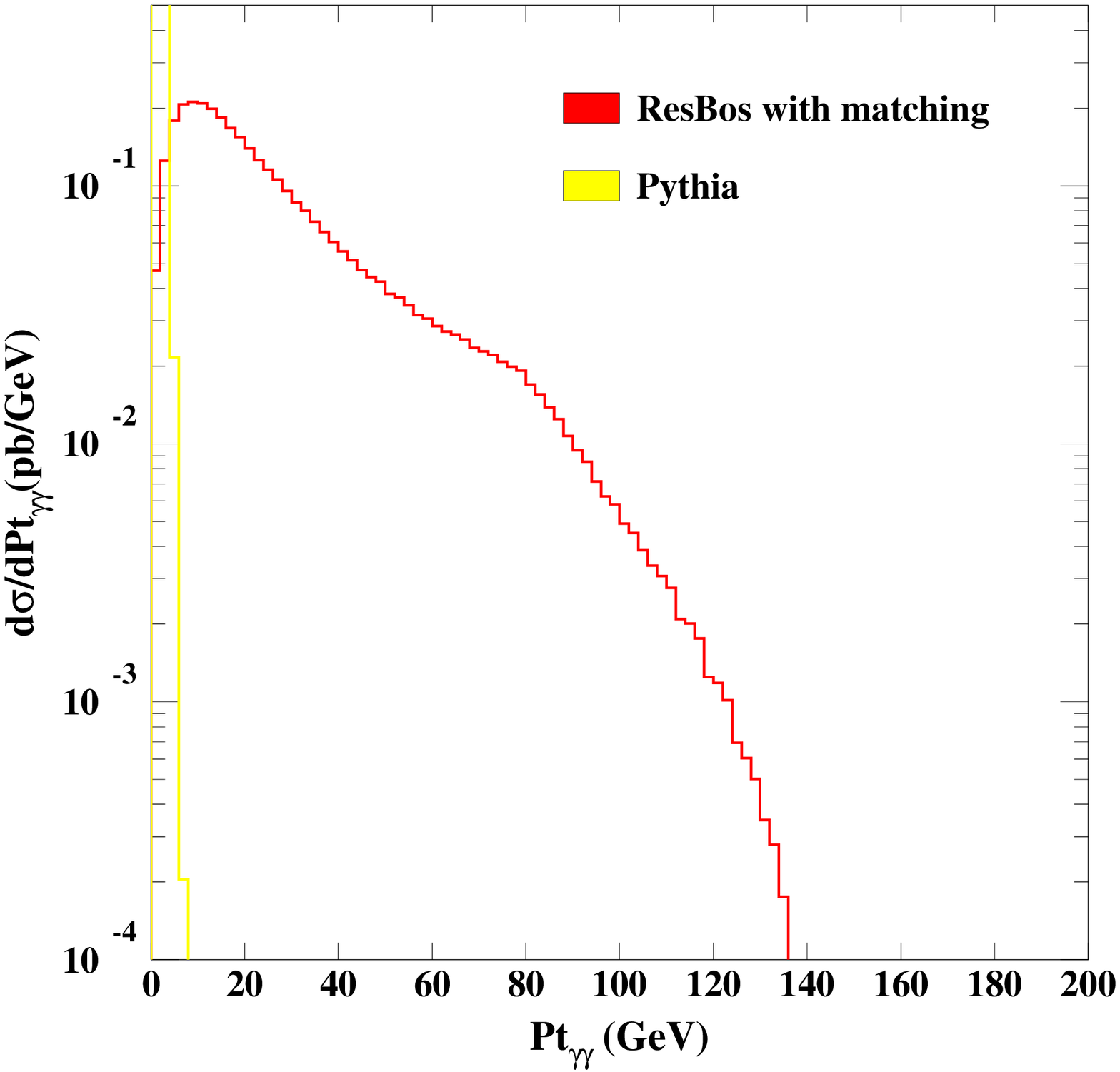}
\put(-6.2,0.8){\includegraphics[width=3.5cm]{box.eps}}
\end{picture}
\caption{
Distributions for the $gg\to\gamma\gamma X$ subprocesses predicted by Diphox, 
PYTHIA, ResBos and a NLO analytic calculation. The $P_t$ spread in the PYTHIA 
spectrum reflects the resolution of the ATLAS detector.
}
\label{diphox_box_distribution}
%\end{center}
\end{figure}

\subsection{Summary}

This work describes the present status of the diphoton cross section 
calculations at LHC. The total cross section, invariant mass and transverse 
momentum distributions have been compared using various codes. 
First, we considered the $pp(gg)\to H(\to\gamma\gamma)X$ process and compared 
predictions of PYTHIA and ResBos to a recent NLO calculation. We found an 
excellent agreement between ResBos and the NLO code in the mid to high $P_t$.
Then, for the irreducible background, we computed both the $q{\bar q}+{\bar 
q}^{\hspace{-0.24cm}(~\,)}g\to\gamma\gamma X$ and the $gg\to\gamma\gamma X$ 
subprocesses. 
For $q{\bar q}+{\bar q}^{\hspace{-0.24cm}(~\,)}g\to\gamma\gamma X$ comparison of 
Diphox and ResBos showed encouraging agreement within the generated statistics.
In case of the less explored gluon fusion subprocess ($gg\to\gamma\gamma X$), we 
found significant (${\mathcal O}$(30\%)) differences between the MC and recent 
analytic QCD calculations. We believe that the implementation of the NLO gluon 
fusion process in the MC codes would benefit the Higgs search at the LHC.
The impact of these results on the Higgs research will be analyzed in a
separate work.

}

%% file: rainwater.tex
{
\newcommand{\sla}[1]{/\!\!\!#1}

\section[ ]{NLO corrections to $V+$jets\footnote{D.\,Rainwater}}

%%%%%%%%%%%%%%%%%%%%%%%%%%%%%%%%%%%%%%%%%%%%%%%%%%%%%%%%%%%%%%%%%%%%%%%%%%%%%%%%

\subsection{Introduction}

The Higgs search is one of the highest priorities for the LHC physics
program. While a plethora of phenomenological and detector studies
have shown that the LHC has significant capability to discover a Higgs
of any allowed mass in multiple channels, for many of those channels
the background rates are known only at LO. This implies large
uncertainties in how much luminosity it will take to detect a Higgs
candidate in some channel, but more importantly it affects the more
important task of measuring all the quantum properties of that
candidate. We must improve our knowledge of the SM rates containing
the same final states as the Higgs channels to as high a degree as
possible, either by calculating quantum corrections to those
processes, or establishing a reliable technique of measuring the SM
rates in sidebands where we are confident there is no Higgs.

Our results here are improvements on the theoretical predictions. We
have taken the previous results of NLO QCD corrections to $Wjj$ and
$Zjj$ production~\cite{Campbell:2002tg,Campbell:2003hd}, where the
jets may or may not be heavy flavor, and applied them to two cases of
interest in the Higgs program: $Zjj$ production as a background to
weak boson fusion (WBF) Higgs production, where the jets are very far
forward/backward in the detector; and $Wb\bar{b}$ production, one of
the principal backgrounds to $WH$ production in the $H\to b\bar{b}$
decay mode.

\subsection{$Zjj$}

The first channel, WBF Higgs production with subsequent decay to tau
leptons~\cite{Rainwater:1998kj,Plehn:1999xi}, $pp\to Hjj\to
\tau^+\tau^-jj$, is also one of the most important. It is the only
fermionic Higgs decay channel with both large rate and large signal to
background ratio. Thus it is a crucial input necessary for extracting
%Higgs couplings~\ref{sec:ewsb;Hcoup}. It is also a powerful channel in
Higgs couplings. It is also a powerful channel in
MSSM scenarios, providing a No-Lose Theorem for observing at least one
of the CP-even states~\cite{Plehn:1999nw}. Finally, any of the WBF
channels provide information on the gauge coupling strength and tensor
structure of the Higgs to weak bosons, via the azimuthal angular
distribution of the two ``tagging'' jets~\cite{Plehn:2001nj}.

Because the Higgs is emitted with significant $p_T$ in this production
mode, the taus will generally not lie back-to-back in the transverse
plane, and their very nearly collinear decays allow for complete
reconstruction, such that the Higgs resonance is visible with a width
on the order of 10-15~GeV~\cite{Azuelos:2001yw,CMS-NOTE-2001/016}. The
nearby $Z$ resonance in taus has a significant tail in the tau
invariant mass distribution which overlaps the Higgs signal. We thus
need a precise prediction for this contribution. QCD $Zjj$ production
is several orders of magnitude larger than the signal, but it is
greatly reduced in the phase space region populated by $W$-scattered
forward quarks in WBF. Because of this, we require a separate NLO
calculation in just this phase space region of high invariant mass,
forward-scattered jets.

Our previous work~\cite{Campbell:2003hd} established the basic size of
the NLO corrections in the WBF Higgs search region. For a fixed
factorization and renormalization scale of the $Z$ mass,
$\mu\equiv\mu_f=\mu_r=M_Z$, the corrections are about $15\%$ positive
and have little residual combined scale dependence. At LO, a lower
scale increases the cross section. All previous studies of this
channel used lower scales, but implemented them dynamically, typically
as the minimum or average $p_T$ of the tagging jets. As a dynamical
scale is theoretical inconsistent at NLO, we cannot easily make
comparison to the existing studies; they should be reperformed.
However, we can surmise from the scale dependence that they
overestimated the QCD $Zjj$ contribution by a modest amount.

Also important is the azimuthal angular distribution of the tagging
jets. This distribution changes significantly at NLO, as shown in
Fig.~\ref{fig:Zjj.dphijj}. Specifically, it flattens out considerably
relative to the LO expectation. However, this does not take into
account the minijet veto used in all WBF analyses: additional central
jets of $p_T>20$~GeV cause the event to be vetoed, as these tend to
appear only in QCD backgrounds and not the Higgs signal.
Unfortunately, a veto for such a low $p_T$ range at NLO produces
nonsense: the distrubution actually becomes negative for some angles.
This is because of the delicate cancellations between the 2-jet
virtual component and the soft singularities in the 3-jet real
emission contribution, which cancel. Without a resummation calculation
to correct the $p_T$ distrbution of the soft central jet, we cannot
predict the tagging jet angular distribution at NLO with a minijet
veto.

We can, however, subtract from the NLO distribution the LO $Zjjj$
distribution where the third jet satisfies the minijet veto criteria.
The total rate $Zjjj$ is normalized to the NLO total rate via the
truncated shower approximation~\cite{Rainwater:1996ud}, which
approximates a NLO-resummed result. The result is that the rate is
greatly reduced, as expected with the minijet veto, but the tagging
jets' azimuthal angular distribution remains mostly flat, closer to
the NLO result than the LO result. This is a significant result, and
currently state of the art, but clearly a full resummation calculation
is highly desirable.

\begin{figure}
\begin{center}
\includegraphics[width=13cm,height=8cm]{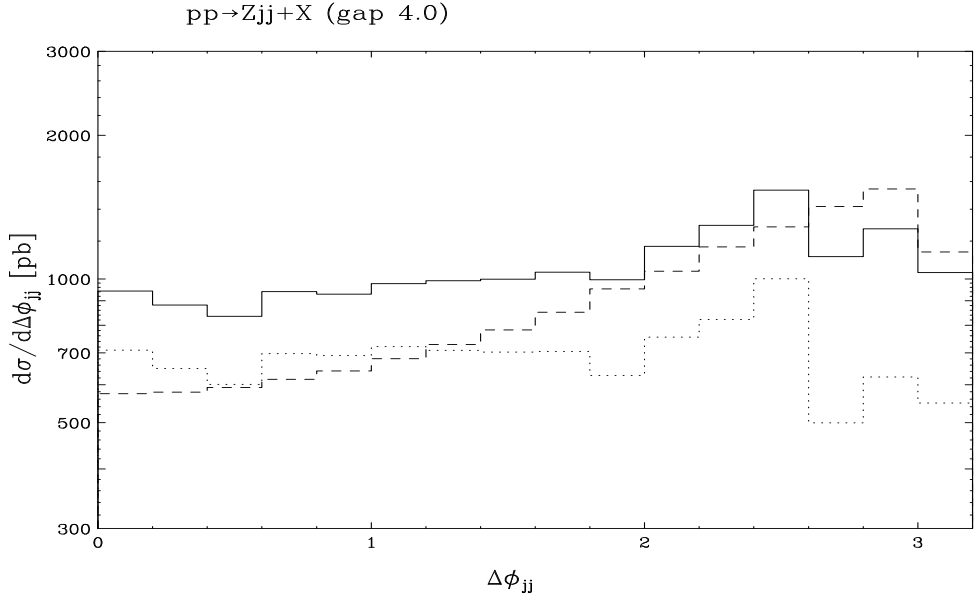} 
\caption{Azimuthal angular distribution of the tagging jets in QCD
  $Zjj$ production at LO (dashed) and NLO (solid) in the phase space
  region applicable to WBF Higgs searches. The gap between the two
  tagging jets is greater than 4 units of rapidity. The dotted curve
  is the NLO ``vetoed'' rate as explained in the text.}
\label{fig:Zjj.dphijj}
\vspace{-8mm}
\end{center}
\end{figure}

\subsection{$Wb\bar{b}$}

The second channel, $W$-associated production with subsequent decay to
a pair of bottom quarks, $pp\to WH\to\ell\nu b\bar{b}$, is a weak
channel statistically~\cite{ATL-PHYS-2000-023,CMS-NOTE-2002/006}, but
highly desirable as it could give access to the $b$ Yukawa coupling.

The NLO QCD corrections to the background, QCD $Wb\bar{b}$ production,
are surprisingly large, about a factor 2.4~\cite{Campbell:2003hd} in
the relevant phase space region of the Higgs resonance. We show the
$b$-pair invariant mass distribution in Fig.~\ref{fig:Wbb.mbb}. The
large enhancement at NLO comes from additional Feynman diagrams that
enter at NLO in the real emission, which are gluon-initiated; only
quark-initiated processes exist at LO. The extra jet is typically hard
and can be vetoed, but as in the $Zjj$ case this is unreliable at NLO.
Therefore the NLO+veto curve in Fig.~\ref{fig:Wbb.mbb} has again a
large uncertainty, which can be reduced only be performing a NLO
calculation of the $Wb\bar{b}j$ rate. As a result, the impact of these
corrections on the $WH;H\to b\bar{b}$ channel are uncertain, but not
optimistic.

\begin{figure}[h]
\begin{center}
\includegraphics[width=13cm,height=8cm]{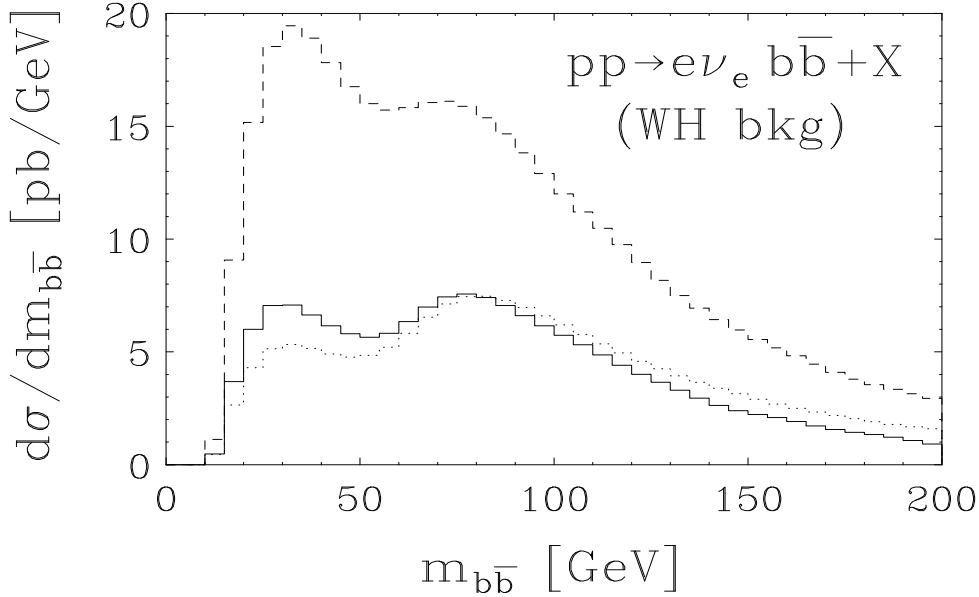} 
\caption{Bottom quark pair invariant mass distribution at LO (dashed) 
  and NLO (solid), and with a veto on the additional light jet
  (dotted), for kinematic cuts corresponding to a Higgs search in $WH$
  production.}
\label{fig:Wbb.mbb}
\vspace{-8mm}
\end{center}
\end{figure}

%\subsection*{ACKNOWLEDGEMENTS}

}

%% file: duca.tex
{
%%%%%%%%%%%%%%%%%%%%%%%%%%%%%%%%%%%%%%%%%%%%%%%%%%%%%%%%%%%%%%%%%%%
\newcommand{\beq}{\begin{equation}}
\newcommand{\eeq}{\end{equation}}
\def\eqn#1{Eq.~(\ref{#1})}
\def\ord{{\cal O} }
\def\eps{\epsilon}
\def\bom#1{{\mbox{\boldmath $#1$}}}
\renewcommand{\perp}{{\rm T}}
\newcommand\et{E_\perp}
\newcommand\pt{p_\perp}
\newcommand\ptg{p_{\gamma\perp}}
\newcommand\ptgg{p_{\gamma\gamma\perp}}
\newcommand\jet{+ {\rm jet}}
\newcommand\etj{E_{{\rm jet}\perp}}
\newcommand\etag{\eta_{\gamma}}
\newcommand\etaj{\eta_{\rm jet}}
\newcommand\drgg{\Delta R_{\gamma\gamma}}
\newcommand\drgj{\Delta R_{\gamma{\rm jet}}}
\newcommand\mgg{M_{\gamma\gamma}}
\newcommand\dmgg{\Delta M_{\gamma\gamma}}
\newcommand\sss{\scriptscriptstyle}
\newcommand\as{\alpha_{\sss S}}
\newcommand\mh{m_{\sss H}}
\newcommand\mur{\mu_{\sss R}}
\newcommand\muf{\mu_{\sss F}}
\def\lsim{\mathrel{\raisebox{-.6ex}{$\stackrel{\textstyle<}{\sim}$}}} 
\def\gsim{\mathrel{\raisebox{-.6ex}{$\stackrel{\textstyle>}{\sim}$}}}
%%%%%%%%%%%%%%%%%%%%%%%%%%%%%%%%%%%%%%%%%%%%%%%%%%%%%%%%%%%%%%%%%%%

\section[ ]{Radiative Corrections to Di--Photon $+$ 1 Jet
Production\footnote{V.\,Del Duca, F.\,Maltoni, Z.\,Nagy and Z.\,Tr\'ocs\'anyi}}

\subsection{Introduction}

Higgs production in association with a jet of high transverse energy $\et$ 
with a subsequent decay into two photons, 
$pp\to H\ \jet \to \gamma\gamma\jet$, 
is considered a very promising discovery channel for
a Higgs boson of intermediate mass (100\,GeV $\lsim \mh \lsim$ 
140\,GeV)~\cite{Dubinin:1997rq,Abdullin:1998er}. In fact if a
high-$\pt$ jet is present in the final state, the photons are
more energetic than in the inclusive channel and the
reconstruction of the jet allows for a more precise determination of
the interaction vertex.  Moreover, the presence of the jet offers the
advantage of being more flexible with respect to choosing suitable
acceptance cuts to curb the background.  These advantages compensate
for the loss in the production rate. The analysis presented in
Refs.~\cite{Dubinin:1997rq,Abdullin:1998er} was done at the parton level,
for LHC operating at low luminosity (30 fb$^{-1}$ of accumulated data).
It included only leading-order perturbative predictions for both the
signal~\cite{Ellis:1988xu} and for the background~\cite{Boos:1994xb},
although anticipating large radiative corrections which were taken into
account by using a constant $K^{\rm NLO} = 1.6$ factor for both the
signal and the background processes. 

In the analysis of Refs.~\cite{Dubinin:1997rq,Abdullin:1998er},
two photons with $\ptg \ge 40 \mbox{~GeV}$ and $|\eta_\gamma| \le 2.5$,
and a jet with $\etj \ge 30 \mbox{~GeV}$ and $|\etaj| \le 4.5$
were selected, with a photon-photon distance $\drgg \ge 0.3$
and a jet-photon distance $\drgj \ge 0.3$. The binning of
the photon-photon invariant mass $\mgg$ was taken to be
$\dmgg = 3.25~(2.0)~\mbox{GeV}$ for ATLAS~(CMS), with a photon
identification efficiency of 73\%. A cut over the parton centre-of-mass
energy, $\sqrt{\hat s} \ge 300$~GeV, was used in order to improve the
signal-over-background ratio, $S/B$. It was found that for a Higgs mass of
$\mh = 120$~GeV, the significance $S/\sqrt{B}$ was well above the discovery 
limit, both for the ATLAS and CMS detectors.

Using CompHEP~\cite{Boos:1994xb} and 
PYTHIA~\cite{Sjostrand:2000wi,Sjostrand:2001yu}, the analysis of 
Refs.~\cite{Dubinin:1997rq,Abdullin:1998er} was repeated at the hadron 
level~\cite{Zmushko:2002} for the ATLAS detector. The same cuts as in
the analysis of Refs.~\cite{Dubinin:1997rq,Abdullin:1998er} were used,
except for $\drgj \ge 0.4$ and a different size in the binning of
the photon-photon invariant mass $\dmgg = 3.64 \mbox{GeV}$, with a photon
identification efficiency of 80\%. At the parton level, the analysis of 
Ref.~\cite{Zmushko:2002} was consistent with 
Refs.~\cite{Dubinin:1997rq,Abdullin:1998er}, and at the hadron level it 
found that the significance $S/\sqrt{B}$ was about at the discovery 
limit\footnote{As a caveat, it should be noted that in the analysis of 
Ref.~\cite{Zmushko:2002}, the background, and thus the significance,
depends significantly on the evolution scale which is chosen for the 
parton showering
in PYTHIA.}. In the analysis of Ref.~\cite{Zmushko:2002}, no $K$~factor
was used either in the signal or in the background.

Since the first analyses of Refs.~\cite{Dubinin:1997rq,Abdullin:1998er}
were made, the next-to-leading order (NLO) radiative corrections to the
signal~\cite{deFlorian:1999zd,Ravindran:2002dc,Glosser:2002gm,Campbell:2000bg} 
and to the background~\cite{DelDuca:2003uz}
have been computed. The main contribution to the signal comes from Higgs
production via gluon fusion, where the interaction between the gluons
and the Higgs is mediated by a loop of heavy quarks (in the Standard Model,
with an accuracy of the order of 0.1\%, one can limit oneself to consider 
only the top quark circulating in the loop). 
$H\ \jet$ production via gluon fusion at leading order was evaluated
in Ref.~\cite{Ellis:1988xu}. As regards the NLO contribution,
only the bremsstrahlung corrections
are known~\cite{DelDuca:2001fn}. However, the full NLO 
corrections~\cite{deFlorian:1999zd,Ravindran:2002dc,Glosser:2002gm,
Campbell:2000bg} have been evaluated in the large $m_{\rm top}$ 
limit~\cite{Shifman:1979eb,Ellis:1976ap}, which for
$H\ \jet$ production is valid as long as $\mh \lsim 2m_{\rm top}$ and
the transverse energy is smaller than the top-quark mass 
$\et\lsim m_{\rm top}$~\cite{Baur:1990cm}.
Furthermore, the large $m_{\rm top}$ limit is insensitive
to the jet-Higgs invariant mass becoming larger than 
$m_{\rm top}$~\cite{DelDuca:2003ba}. For a Higgs mass of
$\mh = 120$~GeV, the NLO corrections were found to increase the
leading-order prediction by about 
60\%~\cite{deFlorian:1999zd,Ravindran:2002dc,Glosser:2002gm,Campbell:2000bg}.

The NLO corrections to the background, $pp\to \gamma\gamma\jet$, have
been computed in Ref.~\cite{DelDuca:2003uz}, using the known NLO matrix 
elements~\cite{DelDuca:1999pa,Bern:1995fz,Signer:1995nk}.
The quark-loop mediated $gg\to g \gamma\gamma$ sub-process, which is
$\ord(\as^3)$ and thus formally belongs to the 
NNLO corrections, and might have been significant due to the large gluon 
luminosity, had been computed previously and
found to yield a modest contribution~\cite{deFlorian:1999tp,Balazs:1999yf}.  
In Ref.~\cite{DelDuca:2003uz} the same cuts for photons and jets, 
$\pt > 40$~GeV and $|\eta| < 2.5$, as in Ref.~\cite{deFlorian:1999tp}
were used. The midpoint cone algorithm~\cite{Blazey:2000qt} 
with a cone size of $R = \sqrt{\Delta \eta^2 + \Delta \phi^2} = 1$
was used in order to find the jet. Furthermore, both photons were isolated
from the partons in a cone of size $R_\gamma = 0.4$.

At NLO the isolated photon cross section is not infrared safe. To
define an infrared safe cross section, one has to allow for some
hadronic activity inside the photon isolation cone. In a parton level
calculation it means that soft partons up to a predefined maximum
energy are allowed inside the isolation cone.
The standard way of defining an isolated prompt photon cross section,
that matches the usual experimental definition, is to allow for
transverse hadronic energy inside the isolation cone up to
$E_{\perp,{\rm max}}$, where $E_{\perp,{\rm max}}$ is either a fixed energy
value or it is equal to $\varepsilon p_{\gamma\perp}$, with
typical values of $\varepsilon$ between 0.1 and 0.5, and where
$p_{\gamma\perp}$ is taken either to be the photon transverse momentum
on an event-by-event basis or to correspond to the minimum value in the
$p_{\gamma\perp}$ range. In perturbation
theory this isolation requires the splitting of the cross section into
a direct and a fragmentation contribution.

\begin{figure}
\includegraphics*[width=.48\linewidth]{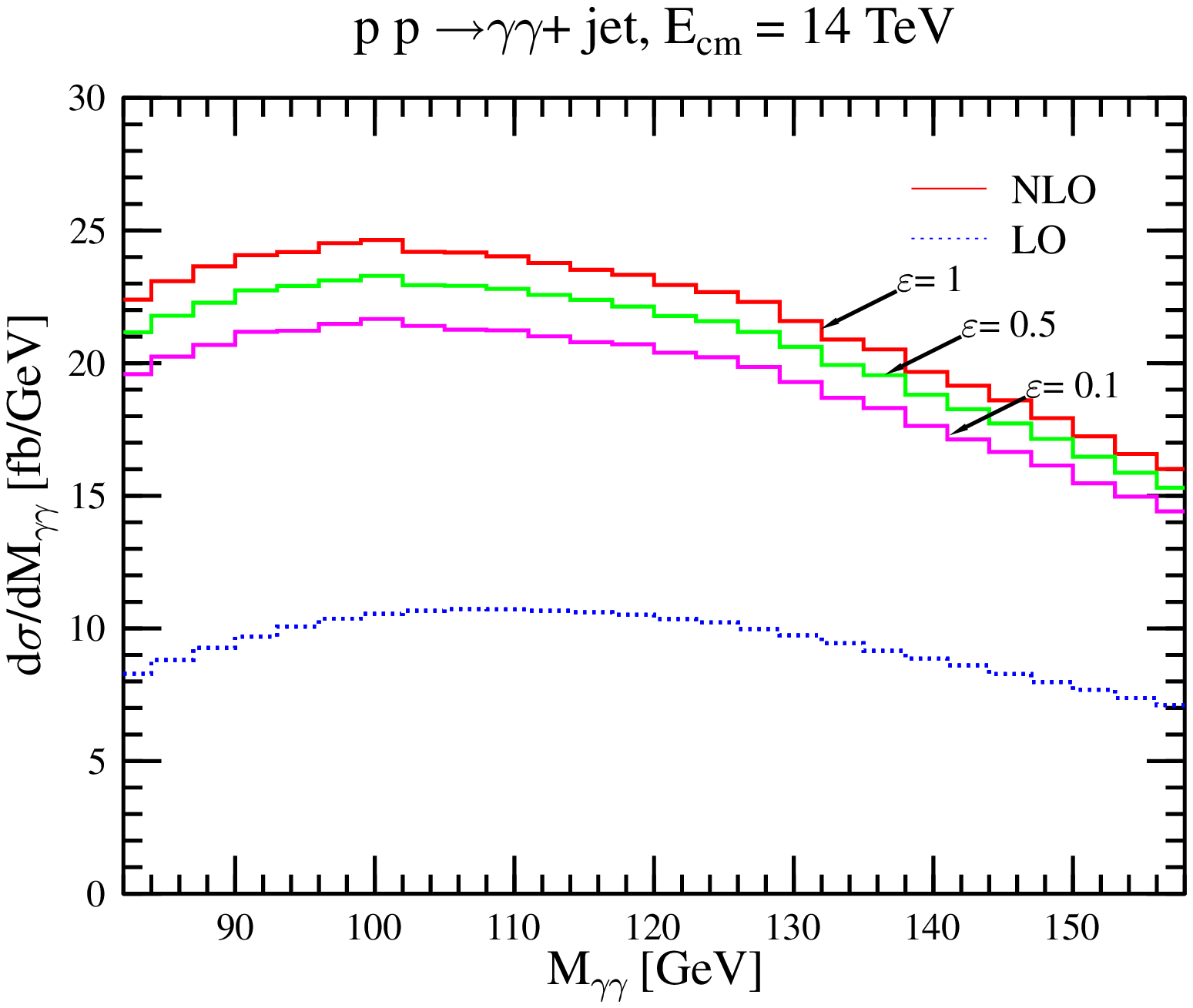} \quad
\includegraphics*[width=.48\linewidth]{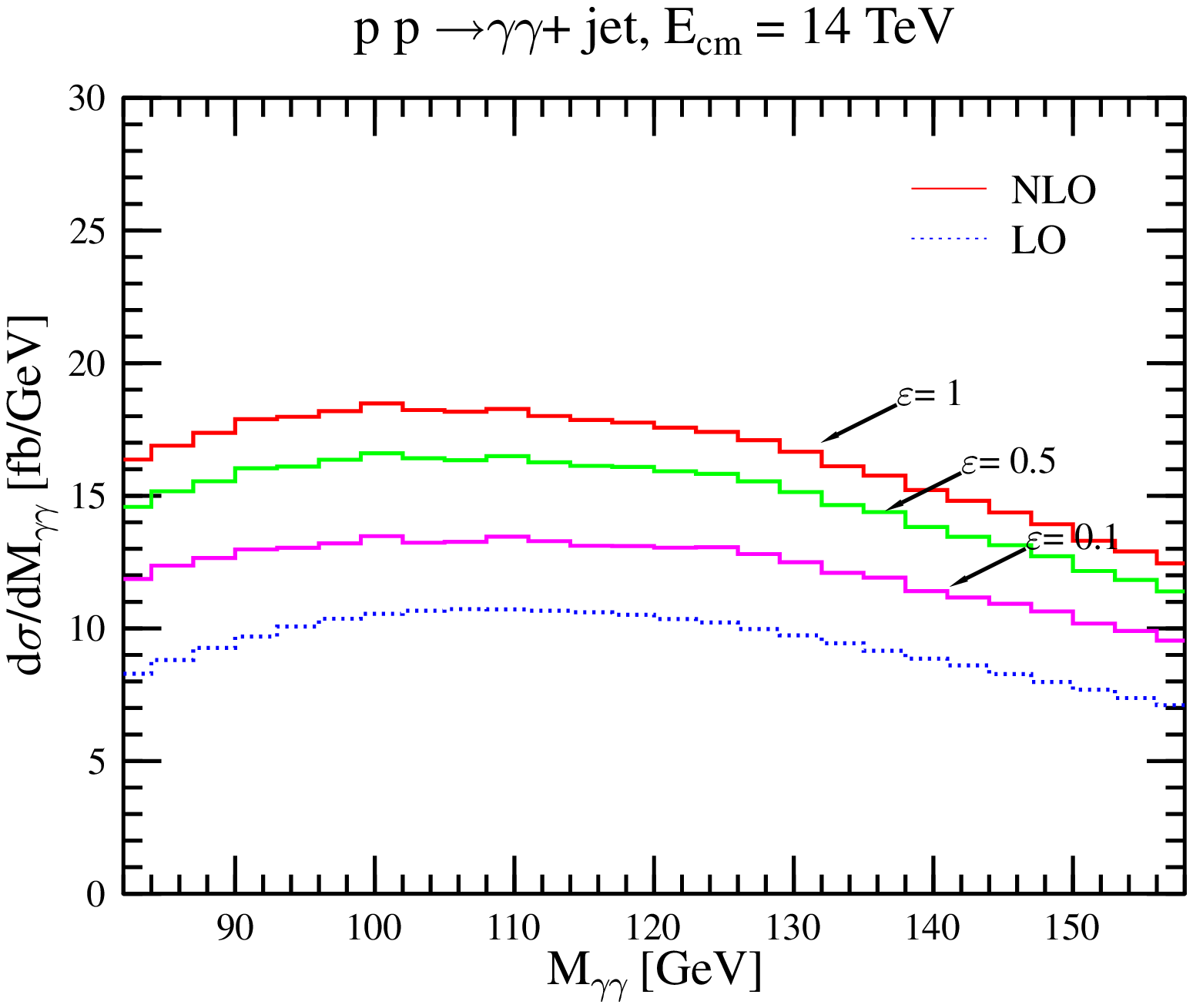}
\caption{Dependence of the invariant mass distribution of the photon pair 
on the photon isolation parameter $\eps$ for $R_\gamma = 0.4~(1.0)$ on
the left(right)-hand-side panel. The photons and the jet are required to 
have transverse momentum
$|p_\perp| \ge 40$~GeV and lie in the central rapidity region of
$|\eta| \le 2.5$. In addition, it is required that 
$\drgj \ge 1.5$ and $p_{\gamma\gamma,\perp} \ge 40$\,GeV.
The dashed curve is the leading order prediction. The solid curves are
the NLO corrections.}
\label{fig:depmgg}
\end{figure}

In Ref.~\cite{DelDuca:2003uz}, only the direct contribution to the 
production of two photons was included. That was possible thanks to
a ``smooth'' photon-isolation prescription which does not require a
fragmentation contribution~\cite{Frixione:1998jh}. This isolation prescribes
that the energy of the soft parton inside the isolation cone has to
converge to zero smoothly if the distance in the $\eta-\phi$ plane
between the photon and the parton vanishes. Explicitly, the amount of hadronic
transverse energy  $E_\perp$ (which in a NLO partonic computation is
equal to the transverse momentum of the possible single parton in the
isolation cone) in all cones of radius $r < R_\gamma$ must be less than
\beq
E_{\perp,{\rm max}} = \varepsilon p_{\gamma\perp}
\left(\frac{1 - \cos r}{1 - \cos R_\gamma}\right)^n\:,
\label{eqn:frixione}
\eeq
In Ref.~\cite{DelDuca:2003uz}, $n = 1$ and $\varepsilon = 0.5$ were used
as default values, and $p_{\gamma\perp}$ was taken to be the photon 
transverse momentum on an event-by-event basis.

In Fig.~\ref{fig:depmgg}, the distribution of the invariant mass $\mgg$
of the photon pair is analysed as a function of the photon isolation 
parameter $\eps$, for $R_\gamma = 0.4$ and 1.
Firstly, we note that the smaller
$R_\gamma$ the larger the NLO correction. 
In fact, for $R_\gamma = 0.4$ the $K$-factor is typically above 2.
In addition, the larger
$\eps$ the larger the NLO correction, with the effect being larger if
$R_\gamma$ is larger.  
%This behaviour is in agreement with \eqn{eqn:frixione},
%according to which smaller $R_\gamma$ and larger $\epsilon$ imply
%larger amounts of soft-parton energy that is allowed inside the cone.
Another remarkable feature of Fig.~\ref{fig:depmgg} is
that with the applied cuts, the two-photon plus jet background for the
search of a Higgs boson with mass in the 120--140\,GeV range is rather
flat, therefore, well measurable from the sidebands around the
hypothetical Higgs signal. This feature is very different from the shape
of the background to the inclusive $pp\to H \to \gamma\gamma$ channel,
which is steeply falling.

\begin{figure}
\begin{center}
\includegraphics*[width=.7\linewidth]{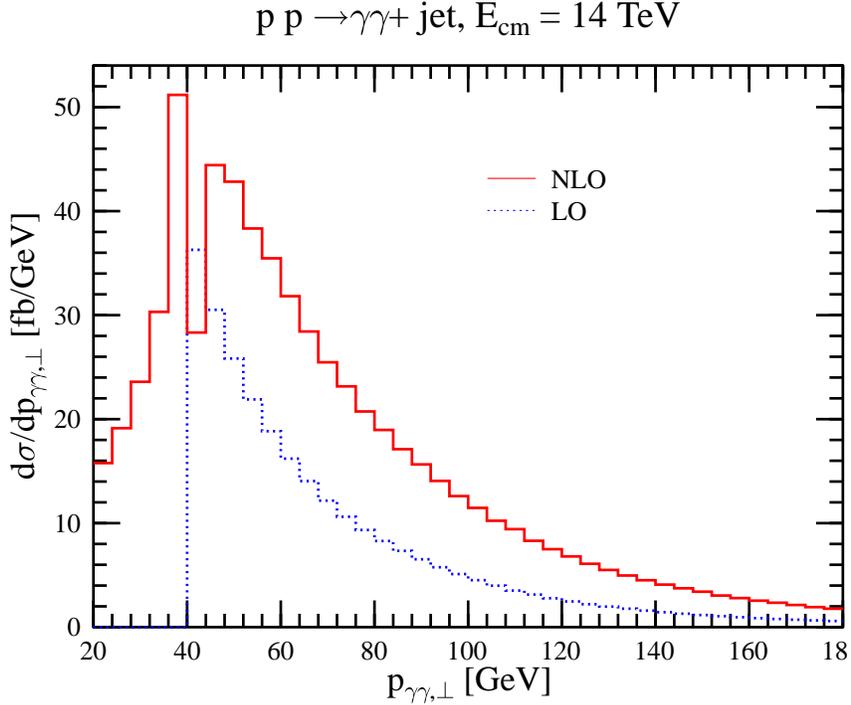}
\caption{Transverse momentum distribution of the photon pair.
The same selection cuts are used as in Fig.~1, with $R_\gamma = 0.4$ 
and $\epsilon = 0.5$.}
\label{fig:ptgaga}
\end{center}
\end{figure}

In Fig.~\ref{fig:ptgaga}, we plot the differential distribution of the
transverse momentum of the photon pair,
$\ptgg = |\bom{p}_{\gamma_1,\perp} + \bom{p}_{\gamma_2,\perp}|$, with a
cut at $p_{{\rm jet} \perp}^{\rm min} = 40$~GeV. In accordance with the
current experimental analyses, the photon isolation radius was taken to be
$R_\gamma = 0.4$. At leading order the
jet recoils against the photon pair and the respective jet and photon
pair $p_\perp$ distributions are identical.  At NLO the extra parton
radiation opens the
part of the phase space with $\ptgg < p_{{\rm jet} \perp}^{\rm min}$.
The double peak around 40\,GeV is an artifact of the fixed-order
computation, similar to the NLO prediction at $C = 0.75$ for the
$C$-parameter distribution in electron-positron annihilation.  The
fixed-order calculation is known to be unreliable in the vicinity of
the threshold, where an all-order resummation is 
necessary~\cite{Catani:1997xc}. That would result in a structure, 
called Sudakov shoulder, which is
continuous and smooth at $\ptgg = p_{{\rm jet} \perp}^{\rm min}$. 
Without the resummation, we must introduce a cut, $p_{\gamma\gamma,\perp}
\ge 40$\,GeV to avoid regions in the phase space where the fixed-order
prediction is not reliable. Accordingly, in Fig.~1 we have required
that $p_{\gamma\gamma,\perp} \ge 40$\,GeV.

One of the goals of computing the radiative corrections to a production
rate is to examine the behaviour of the cross section under variations
of the renormalization $\mur$ and factorization $\muf$ scales. The analysis of
Ref.~\cite{DelDuca:2003uz} showed that the dependence of the cross section 
under variations of $\mur$ and/or $\muf$ remains about the same (in relative
size) in going from the
leading order to the NLO prediction, if the default value of the radius 
$R_\gamma = 0.4$ (with $\epsilon = 0.5$) is used, while it slightly decreases
if the photon isolation parameters are taken to be $R_\gamma = 1$
and $\eps = 0.1$.

In conclusion, the analysis of
Ref.~\cite{DelDuca:2003uz} found large radiative corrections,
however these are strongly dependent on the selection cuts and the photon
isolation parameters. Choosing a small isolation cone radius $R_\gamma = 0.4$ 
(which is nowadays the experimental preferred choice), with relatively large
hadronic activity allowed in the cone, results in more than 100\,\%
correction with a residual scale dependence which is larger at NLO than at 
leading order. 
Increasing the cone radius to $R_\gamma = 1$ and decreasing the
hadronic activity in the cone reduces both the magnitude of the
radiative corrections as well as the dependence on the renormalization
and factorization scales. This result shows that a constant $K^{\rm NLO}
= 1.6$ factor, as used in Refs.~\cite{Dubinin:1997rq,Abdullin:1998er}, 
is certainly not
appropriate for taking into account the radiative corrections to the
irreducible background of the $pp \to H\jet \to \gamma\gamma\jet$
discovery channel at the LHC.  

}

%% file: kauer.tex
{
\def\sla#1{\ifmmode%
\setbox0=\hbox{$#1$}%
\setbox1=\hbox to\wd0{\hss$/$\hss}\else%
\setbox0=\hbox{#1}%
\setbox1=\hbox to\wd0{\hss/\hss}\fi%
#1\hskip-\wd0\box1 }

\section[ ]{Top background extrapolation for $H\to WW$
  searches\protect\footnote{N.\,Kauer}}

\subsection{Introduction}

Recent studies indicate that the LHC will be able to discover
a Standard Model (SM) Higgs boson with mass between 100 and 200 GeV
with an integrated luminosity of only 10 to 30 fb$^{-1}$ if 
weak boson fusion (WBF) followed by $H\to\tau\tau$ and $H\to WW$
channels are taken into account (see Ref.~\cite{SN-ATLAS-2003-024} and
refs. therein).  This intermediate mass range is currently
favored in light of a lower bound of 114.1 GeV from direct 
searches at LEP2 and an upper bound of 196 GeV from a SM
analysis of electroweak precision data (at 95\% CL)
\cite{Barate:2003sz,Grunewald:2003ij}.  
As discussed in detail in Ref.~\cite{Cavalli:2002vs},
Sec.~A.1, the precise knowledge of the
significance of any observed Higgs signal will require an
accurate determination of the SM backgrounds.
The WBF and gluon fusion $H\to WW\to ll\sla{p}_T$ channels
are particularly challenging, because missing momentum prevents the
observation of a narrow mass peak that would allow an interpolation
of the backgrounds from side bands.

In this section we demonstrate how the extrapolation approach proposed in
Ref.~\cite{Cavalli:2002vs} can in fact be used to reduce the uncertainty of the 
dominant $t\bar{t}$ + 1 jet background
to the $H\to W^+W^-\to l_1^\pm l_2^\mp\sla{p}_T$ search in WBF
and the large $t\bar{t}$ background to the same Higgs decay mode
in gluon fusion.  To be specific, we consider the $t\bar{t}j$ background in the 
WBF selection cuts of Ref.~\cite{Kauer:2000hi} and the $t\bar{t}$ background
in the selection cuts suggested for the inclusive $H\to WW$ search
in Ref.~\cite{unknown:1999fr},
Sec.~19.2.6, with a transverse mass window cut based on a Higgs mass of 170 GeV.
All cross sections are calculated using the parton-level Monte Carlo programs of 
Refs.~\cite{Kauer:2001sp} and \cite{Kauer:2002sn},
which include finite width effects and
the complete leading order (LO) matrix elements for 
$l_1^\pm l_2^\mp\nu\bar{\nu}b\bar{b}$ (+ jets)
final states.  The calculations take into account finite resolution effects
and a suboptimal $b$ tagging efficiency
based on expectations for the ATLAS detector.

To investigate the scale uncertainty of these backgrounds and how it
can be reduced we apply the following definitions for the
renormalization and factorization scales $\mu_R$ and $\mu_F$.
A factor $\xi = \mu/\mu_0$ is then used to vary the scales around the central values.
The suggestive scale choice for top production is the top mass $m_t = 175$ GeV:
\begin{equation}
\label{topscheme}
\mu_R = \mu_F = \xi m_t\,.
\end{equation}
Results for this scale choice are shown as solid curves in Figs.~\ref{wbf-scalevariation} and \ref{inclusive-scalevariation}.
For WBF, due to forward tagging selection cuts, the dominant
background arises from $t\bar{t}$ production with one additional hard jet.
To avoid double counting in this case, we alternatively calculate with
scales based on the minimal transverse mass:
\begin{equation}
\label{transscheme}
\mu_F = \xi \min(m_{T,t}, m_{T,\bar{t}}, p_{T,j}) \quad\text{and}\quad
\alpha_s^3 = \alpha_s(\xi m_{T,t})\alpha_s(\xi m_{T,\bar{t}})\alpha_s(\xi p_{T,j})\,.
\end{equation}
Results for this second definition are shown as dashed curves in Fig.~\ref{wbf-scalevariation}.
In principle, the renormalization and factorization scales are independent.  
We find, however, that the strongest scale variation occurs if both scales
are varied in the same direction and thus only introduce a single parameter $\xi$.
Scale-dependent quantities
are customarily condensed into the form $\hat{x} \pm\Delta \hat{x}$
based on a particular low and high scale choice.  We use the convention
\begin{equation}
\label{theoryerrorestimate}
\hat{x} = (x(\xi = \frac{1}{2})+x(\xi = 2))/2 \quad\text{and}\quad
\Delta \hat{x} = |x(\xi = \frac{1}{2})-x(\xi = 2)|/2\;,
\end{equation}
where $x$ is a cross section or cross section ratio.

Figs.~\ref{wbf-scalevariation}(a) and \ref{inclusive-scalevariation}(a)
show the large scale variation that is expected for the LO 
background cross sections in both search channels.  For the WBF search channel,
the scale scheme of Eq.~(\ref{topscheme}) yields a background cross section of
$0.27 \pm 0.11$ fb, whereas the
scheme of Eq.~(\ref{transscheme}) yields
$0.41 \pm 0.17$ fb.  The theoretical uncertainty is about 40\% in both cases.
Since the second cross section is not consistent with the first within 1$\sigma$,
it seems more appropriate to apply the prescription Eq.~(\ref{theoryerrorestimate})
to the envelope of both curves.  All subsequent WBF results will be given using 
this procedure.  Here, one obtains $0.38 \pm 0.21$ fb, with an even larger 
uncertainty of 55\%.
For the top background in the inclusive $H\to WW$ search
a somewhat smaller theoretical uncertainty is obtained, i.e.~3.7 fb with
an uncertainty of 25\%.  In both cases it is obvious that the accuracy of
theoretical background calculations at LO is insufficient to determine the
total background to an accuracy of 10\%, as assumed in
Ref.~\cite{SN-ATLAS-2003-024}.

\subsection{Extrapolation}

The extrapolation approach allows a more accurate determination
of a background cross section $\sigma_{bkg}$ if a
reference selection with a corresponding well-defined, measurable event 
rate $\sigma_{ref}\cdot{\cal L}$ can be found, so that the theoretical
uncertainty of the ratio $\sigma_{bkg}/\sigma_{ref}$ is small and
$\sigma_{ref}$ can be measured with low experimental uncertainty.
The background cross section can then be approximated through
\begin{equation}
\label{extrapolationapproximation}
\sigma_{bkg} \quad \approx \quad \underbrace{\left(
     \frac{\sigma_{bkg,\text{ LO}}}{\sigma_{ref,\text{ LO}}}\right)}_{\stackrel{\mbox{\footnotesize low theoret.}}{\mbox{\footnotesize uncertainty}}}
     \quad\cdot\quad
\underbrace{\sigma_{ref}}_{\stackrel{\mbox{\footnotesize low experim.}}{\mbox{\footnotesize uncertainty}}}\;.
\end{equation}
To derive suitable reference
selections from the corresponding background selections,
we propose the following strategy:
The WBF and inclusive $H\to WW$ search channel top
backgrounds are effectively suppressed through a central jet
veto.  Discarding this veto leads to a sizable increase of the
cross sections.  Secondly, to identify the top backgrounds in both cases,
we require that only events be considered that contain one
or more identified $b$ jets.   For our calculations we assume
that $b$ jets with $|\eta|< 2.5$ and $p_T > 15$ GeV will be
tagged with a probability of 60\%.  Finally, if the resulting
reference rate is still too small, the lepton pair cuts are also
discarded.  This is only necessary in the WBF channel,
where the reference cross section with lepton pair cuts is
10.8 fb, which, with 30 fb$^{-1}$, would result in a statistical
uncertainty for the measured rate of about 6\%.

\begin{figure}[htbp]
\begin{center}
\begin{minipage}[c]{.49\linewidth}
\flushright \includegraphics[width=6.cm, angle=90]{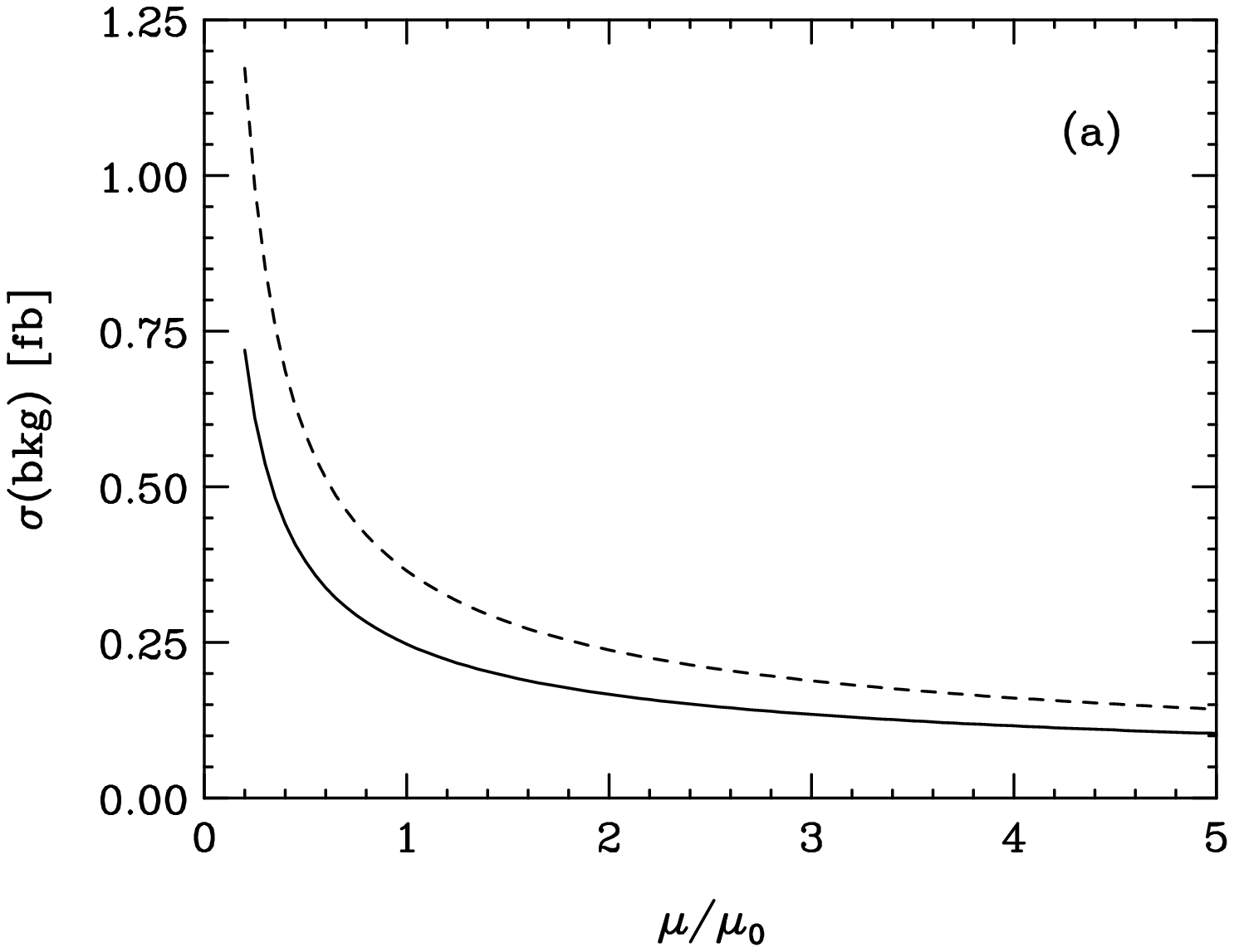}
\end{minipage} \hfill
\begin{minipage}[c]{.49\linewidth}
\flushleft \includegraphics[width=6.cm, angle=90]{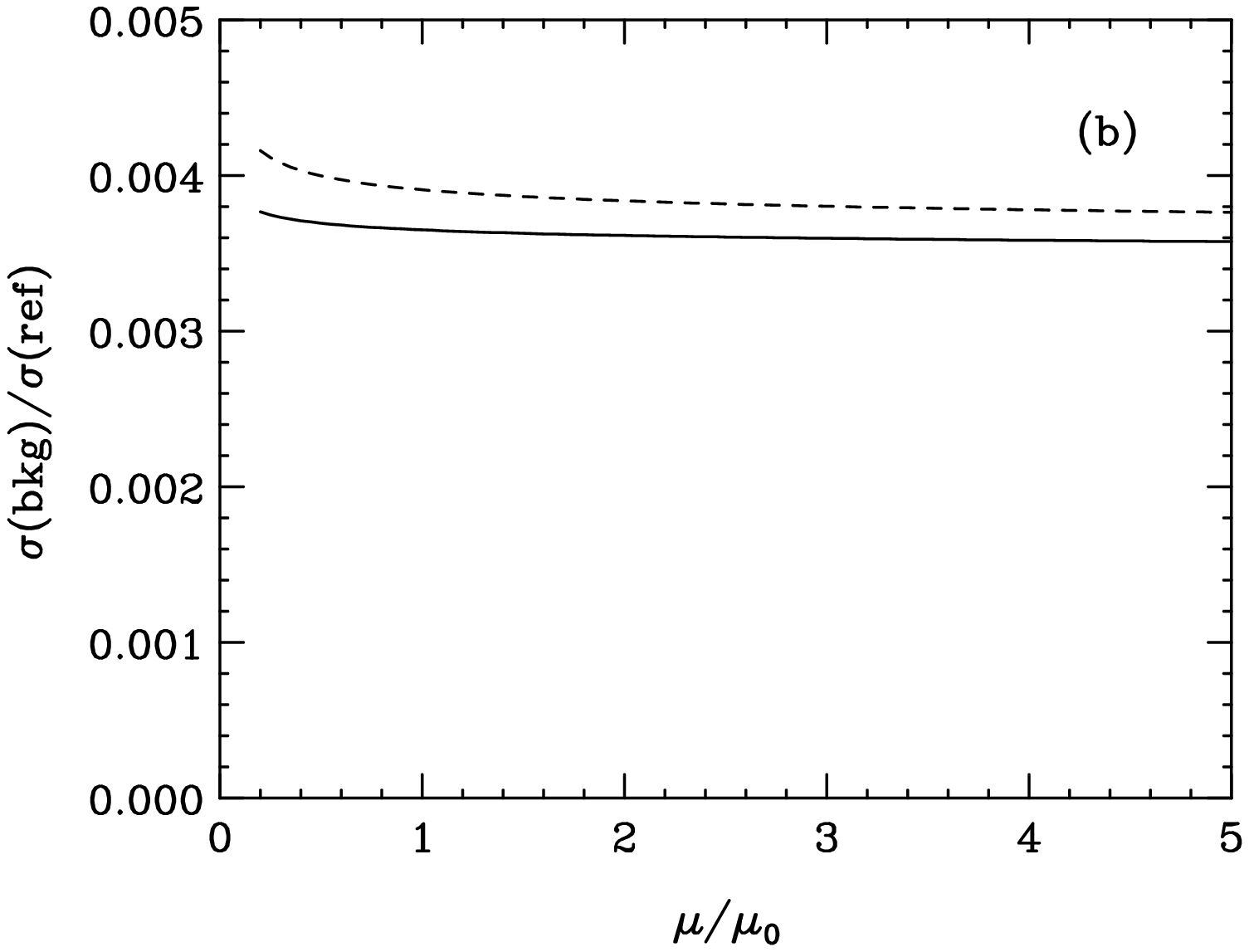} 
\end{minipage}
\caption{
  Renormalization and factorization scale variation of $t\bar{t}j$ background cross
  section (a) and ratio with reference cross section (b) to 
  $H\to W^+W^-\to l_1^\pm l_2^\mp\sla{p}_T$ search in weak boson fusion at the LHC
  for different scale definitions (see main text).
}
\label{wbf-scalevariation}
\end{center}
\end{figure}

Consequently, the reference selection cuts for the WBF channel are obtained by
imposing the identified $b$ jet requirement and discarding the veto of Eq.~(4)
and the lepton pair cuts of Eqs.~(5, 6, 8, 9) in Ref.~\cite{Kauer:2000hi}. 
The resulting reference cross section of $96 \pm 50$ fb gives rise to
a statistical error of about 2\% with 30 fb$^{-1}$ of data.
Note that the scale uncertainty of this reference cross
sections is very similar to that of the background cross section.
However, the scale variation of the corresponding ratio
$\sigma_{bkg}/\sigma_{ref}$ is significantly reduced as shown in
Fig.~\ref{wbf-scalevariation}(b).  One obtains $0.0038 \pm 0.0002$, or
a relative error of 5\%.  Combining both extrapolation factors in quadrature
yields a background estimate with an accuracy of about 5\%.

\begin{figure}[htbp]
\begin{center}
\begin{minipage}[c]{.49\linewidth}
\flushright \includegraphics[width=6.cm, angle=90]{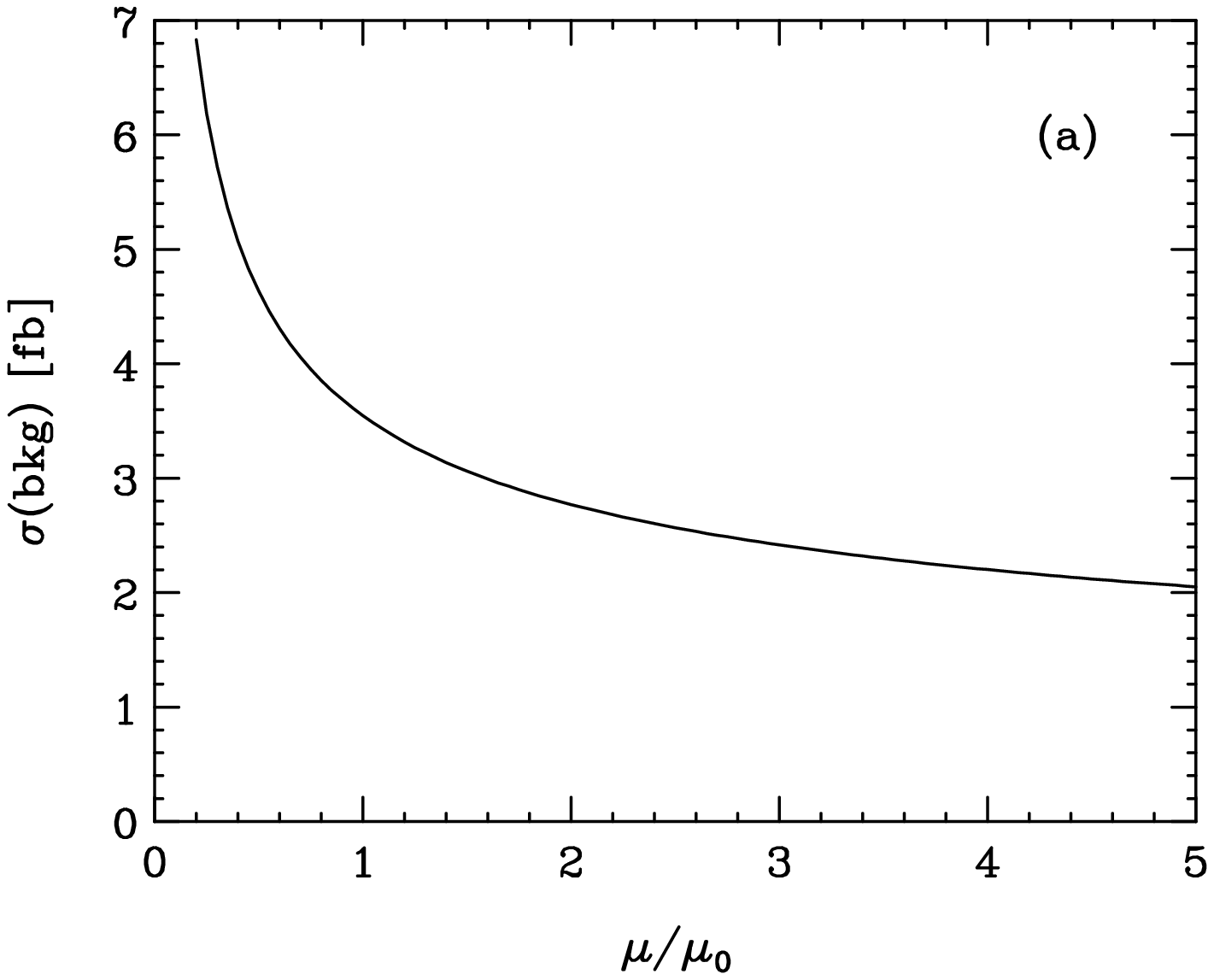}
\end{minipage} \hfill
\begin{minipage}[c]{.49\linewidth}
\flushleft \includegraphics[width=6.cm, angle=90]{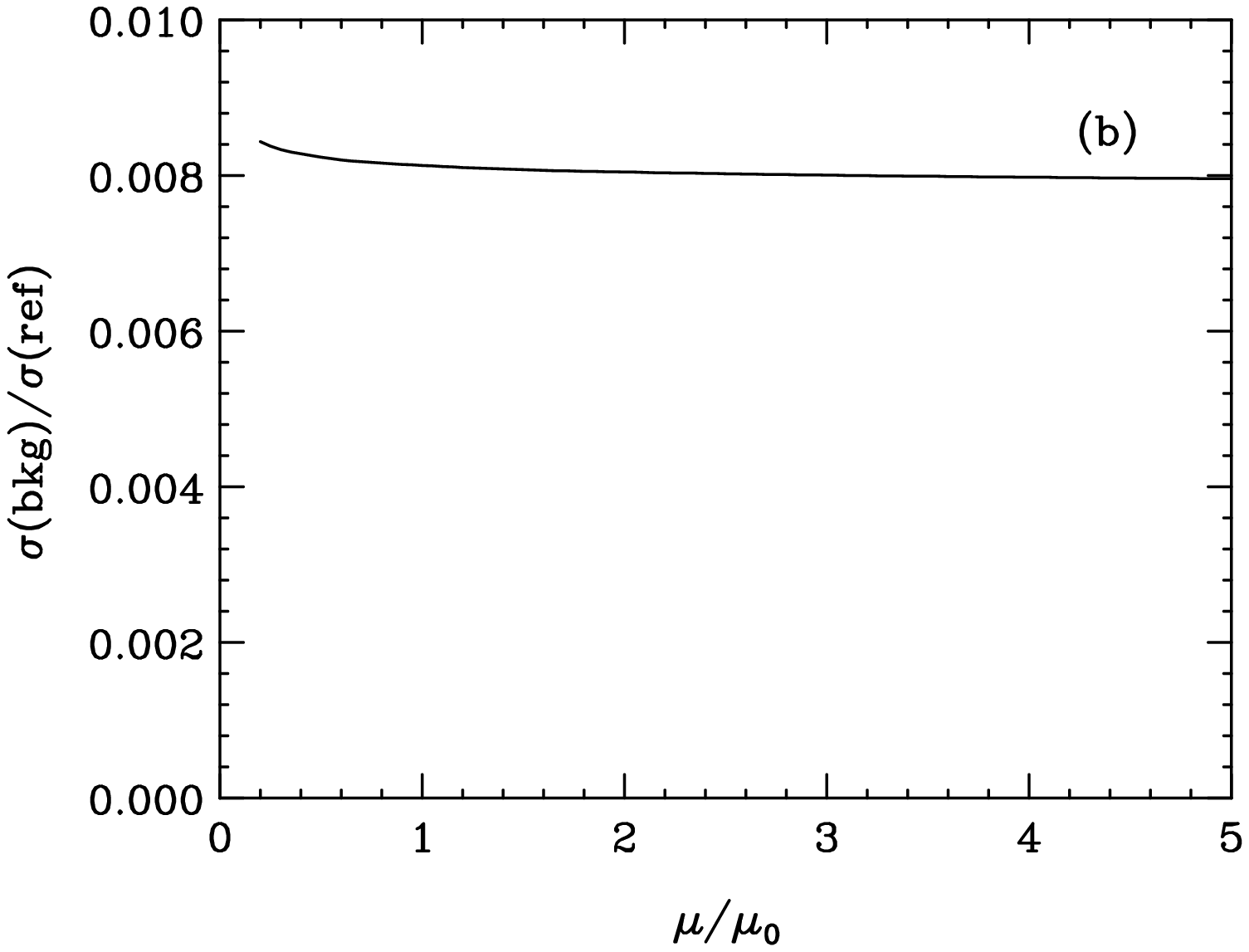} 
\end{minipage}
\caption{
  Renormalization and factorization scale variation of $t\bar{t}$ background cross
  section (a) and ratio with reference cross section (b) to 
  $H\to W^+W^-\to l_1^\pm l_2^\mp\sla{p}_T$ search in gluon fusion at the LHC.
}
\label{inclusive-scalevariation}
\end{center}
\end{figure}

For the $H\to WW$ in gluon fusion channel,
suitable reference selection cuts are obtained by requiring at least one
identified $b$ jet and applying the cuts on pp.~705-706 of Ref.~\cite{unknown:1999fr},
but without vetoing jets with $p_T < 15$ GeV and $|\eta| < 3.2$.
Here, a reference cross section of $450 \pm 88$ fb results, which 
is large enough that the lepton pair cuts can be kept.
With 30 fb$^{-1}$ of data, the reference rate could be determined
with a statistical accuracy of better than 1\%.
The scale variation of the ratio
$\sigma_{bkg}/\sigma_{ref}$ is shown in Fig.~\ref{inclusive-scalevariation}(b).
It is again significantly reduced.  One obtains $0.0081 \pm 0.0001$, or
a relative error of 1\%.  Combining both extrapolation factors yields a background
estimate with an accuracy of about 1\% in this case.

\subsection{Discussion}

The approximation Eq.~(\ref{extrapolationapproximation}) would become
an identity if the ratio $\sigma_{bkg}/\sigma_{ref}$ could be evaluated
to all orders in perturbation theory.  At fixed order in perturbation
theory, a scale dependence remains and, depending on the specific
scale choice, the result will deviate to a greater or lesser extent from
the exact result.\footnote{Note that this deviation is in addition to
any computational error made in the fixed order calculation.}
We refer to this error as residual theoretical error.  In practice, it
is commonly estimated from the scale variation using a prescription
like Eq.~(\ref{theoryerrorestimate}).  Since the scale variation typically
decreases for higher fixed order calculations, it would be instructive
to calculate $\sigma_{bkg}/\sigma_{ref}$ at NLO.  We expect the NLO ratios
and residual theoretical error estimates to be comparable to the ones obtained here,
but this should be confirmed through explicit calculation.\footnote{
At the time of writing a hadron collider program to calculate
$t\bar{t}$ + 1 jet production at NLO QCD with on-shell top quarks
is not yet available.}
We further note that a future study could take into account systematic
experimental uncertainties and parton distribution function uncertainties,
once these become available.

\subsection{Conclusions}

A LO analysis was presented that demonstrates that key top backgrounds
to $H\to W^+W^-\to l_1^\pm l_2^\mp\sla{p}_T$ decays in weak boson
fusion and gluon fusion  at the LHC can be extrapolated from experimental
data with an accuracy of ${\cal O}(5\%)$.  If LO scale variation
is accepted as proxy for the theoretical error, our parton-level results
indicate that the $t\bar{t}j$ background to the $H\to WW$ search
in WBF can be determined with a total error of about
5\%, while the $t\bar{t}$ background to the $H\to WW$ search in 
gluon fusion can be determined with a total error of about 1\%
with an integrated luminosity of 30 fb$^{-1}$.
Further details can be found in Ref.~\cite{Kauer:2004fg}.

}

%% file: drollinger.tex
{
%%%%%%%%%%%%%%%%%%%%%%%%%%%%%%%%%%%%%%%%%%%%%%%%%%%%%%%%%%%%%%%%%%%%%%%%%%%%%%%%%%%%%%

\section[ ]{Scale Dependence of $t\bar{t}b\bar{b}$
Production\footnote{V.\,Drollinger}}

The process $pp \rightarrow t\bar{t}b\bar{b}$ is the most important background to Higgs production in association with $t\bar{t}$ at the LHC \cite{ref_tth}. The cross section of $t\bar{t}b\bar{b}$ production has been calculated with CompHEP \cite{ref_chp} and ALPGEN \cite{ref_alp} at leading order. Higher order corrections for $t\bar{t}b\bar{b}$ are unknown so far. In order to estimate the theoretical uncertainties, we vary the $Q^2$ scale.

In ALPGEN, the default scale choice is $Q^2 = m_t^2$. Because we are mainly interested in $t\bar{t}b\bar{b}$ events in the phase space region which is accessible experimentally, we apply following cuts before the computation of the cross section or the event generation, respectively: $p_T(b) > $ 25 $GeV$, $|\eta(b)| < $ 2.4, and $\Delta R(b,b) > $ 0.4.

Fig.~\ref{fig_ttbb_scale} shows the cross section as a function of the prescaling factor $\xi$. In the range between 0.5 and 2.0, the cross section changes by more than a factor of two. However, the invariant mass distributions of the two $b$-quarks, not coming from a top decay, seem to be not affected. Nevertheless, one should keep in mind, that the understanding of this mass distribution is experimentally the most relevant for the search of a Higgs mass peak on top of the background in the $t\bar{t}h$ channel. Additional jets from higher order contributions complicate the jet reconstruction and can increase the combinatorial background which has a different shape.
\begin{figure}
\begin{center}
\includegraphics[width=\textwidth]{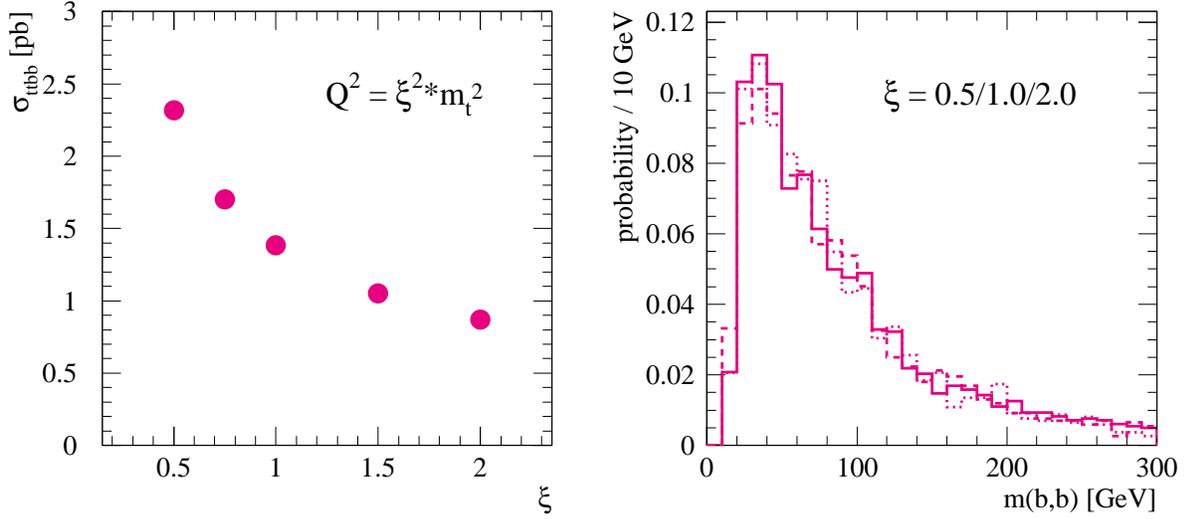} 
\caption{Left: $t\bar{t}b\bar{b}$ cross section as a function of the $Q^2$ scale prescaling factor $\xi$. Right: $b\bar{b}$ invariant mass distributions of the $b$-quarks not coming from top decays for three scale choices. All distributions are normalized to unit area. The cross sections and the mass distributions are obtained after following acceptance cuts: $p_T(b) > $ 25 $GeV$, $|\eta(b)| < $ 2.4, and $\Delta R(b,b) > $ 0.4.}
\end{center}
\label{fig_ttbb_scale}
\end{figure}

}

%% file: roeck.tex
{

\def\lapproxeq{\lower .7ex\hbox{$\;\stackrel{\textstyle <}{\sim}\;$}}
\def\gapproxeq{\lower .7ex\hbox{$\;\stackrel{\textstyle >}{\sim}\;$}}
\newcommand{\porpbar}
{\!\,^{\scriptscriptstyle(}$\mbox{$\bar{p}$}$\,^{\scriptscriptstyle)}}
\def\be{\begin{equation}}
\def\ee{\end{equation}}
\def\bea{\begin{eqnarray}}
\def\eea{\end{eqnarray}}
\def\ktbold{\mbox{\boldmath${k}$}_T}
\def\funp{{I\!\!P}}
\def\gtrsim{{ \;\raisebox{-.7ex}{$\stackrel{\textstyle >}{\sim}$}\; }}
\def\lesim{{ \;\raisebox{-.7ex}{$\stackrel{\textstyle <}{\sim}$}\; }}
\newcommand{\ksq}{k^2} \newcommand{\qsq}{q^2}
\newcommand{\epem}{e^+e^-}

\def\GeV{{\rm GeV}}
\def\MeV{{\rm MeV}}
\def\eV{{\rm eV}}
\def\ra{ \rightarrow }
\def\bb{{b\bar{b}}}
\newcommand{\eq}[1]{(\ref{eq:#1})}

\section[ ]{Studying the Higgs sector at the LHC using proton
tagging\footnote{A.\,De Roeck and V.\,Khoze}}

% ===========================================================================
\subsection{Introduction}

If the Higgs mechanism is responsible for the Electroweak Symmetry 
breaking in Nature, generally 
 at least one Higgs boson should be discovered at the LHC. In particular,
if the light Higgs predicted by the Standard Model (SM) exists
it will almost certainly be found at the LHC in the first years 
of running, but detailed studies may be challenging, see~\cite{DeRoeck:2002hk}
and references therein.
%Moreover the LHC should provide a complete coverage of the SM Higgs mass 
%range.
However, beyond the SM, various extended models predict a large diversity
of Higgs-like states with different masses, couplings and even $CP$-parities.
In these models the properties of the neutral Higgs bosons can
 differ drastically
from SM one.
The most elaborated extension is
the Minimal Supersymmetric Standard Model (MSSM), for a recent review
see~\cite{Carena:2002es}. Below we shall mainly follow this benchmark model.
The extended scenarios would complicate the study of the
Higgs sector using the conventional (semi)inclusive strategies.
%For all standard
%approaches,
%either large signals are accompanied by a huge background,
%or the processes have comparable signal and background rates for which
%the number of signal events is rather low.

After the discovery of a Higgs candidate the immediate task 
will be to establish its quantum numbers, to verify
the Higgs interpretation of the signal, 
and to make precision measurements of its properties. The separation
of different Higgs-like states will be especially challenging.
It will be an even more delicate
goal to
probe the $CP$ -parity and to
establish the nature of the  newly-discovered heavy resonance state(s).

As was shown in \cite{Kaidalov:2003fw,Kaidalov:2003ys}, the central exclusive
diffractive processes (CEDP) at the LHC can play a crucial
role in solving these vital problems.
These processes are of the form
\begin{equation}
pp\to p + \phi + p, \label{eq:cedp}
\end{equation}
where the $+$ signs denote the rapidity gaps on either
side of the Higgs-like state $\phi$. 
They have unique
advantages as compared to the traditional non-diffractive
 approaches~\cite{Khoze:2001xm,DeRoeck:2002hk}. 
In particular, if the forward protons are tagged,
then the mass of the produced central system $\phi$ can
be measured to high accuracy by the missing mass method.  Indeed,
by observing the forward protons, as well as the
$\phi\to\bb$ pairs in the central detector, 
one can match two simultaneous measurements of the $\phi$ mass:
$m_\phi =  m_{\rm missing}$ and $m_\phi = m_\bb$.
%Moreover, proton taggers allow the damaging effects of
%multiple interactions per bunch crossing (pile-up)
%to be suppressed and hence offer the possibility of studying
%CEDP processes at higher luminosities~\cite{DeRoeck:2002hk}. 
Thus, the prospects of the precise mass determination of the
Higgs-like states, and even of the 
direct measurements of their widths and $\phi\to\bb$ couplings, look feasible.
Another unique feature of the forward CEDP is that in the production vertex
the incoming gluon polarisations are
correlated, in such a way that the effective luminosity satisfies the P-even,
$J_z=0$ selection rule~\cite{Khoze:2000jm,Khoze:2001xm}. This plays a key
role in reducing the QCD background
caused by the $gg\to \bb$ subprocess. On the other hand,
this selection rule opens a
promising way of using the
forward proton taggers as a spin-($CP$)parity
analyser~\cite{Kaidalov:2003fw}.

The cross section for the production of a CEDP SM Higgs
at the LHC, with $m_h = 120$~GeV is calculated \cite{Kaidalov:2003ys} to be 
2.2~fb with an uncertainty given by the range 0.9--5.5~fb.
The inclusive diffractive process 
$pp\to p + \phi X + p$ has a larger cross 
section\cite{Boonekamp:2001vk,Boonekamp:2002vg,Khoze:2002py,Cox:2001uq} of
order 100 fb, but there is no $J_z =0$ selection rule for the background 
and the Higgs mass cannot be determined directly from the scattered 
protons. Recently also the single diffractive  channel was 
revisited and studied~\cite{Erhan:2003za} (see also \cite{Graudenz:1996jz}), 
which has a much larger cross section.

Our main goal here is to show that
forward proton tagging may significantly enlarge the potential of 
 studying the Higgs sector at the LHC.

\subsection{Potential of diffractive processes for Higgs studies} \label{sec:discoverypotential}

Over the last years such processes with rapidity gaps have attracted much
attention as a promising way to search for a Higgs boson 
in high energy proton-proton collisions, see,
 for instance,
~\cite{Bialas:1991wj,Cudell:1996ki,Levin:1999qv,Khoze:2000cy,Khoze:2001xm,Cox:2001uq,Boonekamp:2001vk,Boonekamp:2002vg,Cox:2003xp,Kaidalov:2003fw,Royon:2003ng}.

The CEDP have special advantages in the regions of the MSSM parameter space
where the partial width of the Higgs boson decay into two gluons
much exceeds the SM case. First of all ,this concerns the large 
$\tan\beta$
case,where the expected CEDP cross sections are large enough,see \cite{Kaidalov:2003ys}.
Of special interest is the so-called ``intense-coupling'' regime~\cite{Boos:2003jt},
where the masses of all three neutral Higgs bosons are close
to each other. Here the $\gamma\gamma,WW^\star,ZZ^\star$ decay modes (which
are among the main detection modes for the SM Higgs) 
are strongly suppressed. This is the regime where the variations
of all MSSM Higgs masses and couplings are very rapid. This region
is considered as one of the most troublesome
for the (conventional) Higgs searches at the LHC, see \cite{Boos:2003jt}.
On the other hand, here
the CEDP cross sections are enhanced by more than 
an order of magnitude.
Therefore the expected significance of the CEDP signal becomes quite large.
Indeed, this is evident from Fig.~1,
which shows the cross sections for the CEDP production 
of $h,H,A$ bosons as functions of their mass for
$\tan\beta=30$ and 50.
\begin{figure}[htb]
\begin{center}
\centerline{\epsfxsize=10cm\epsfbox{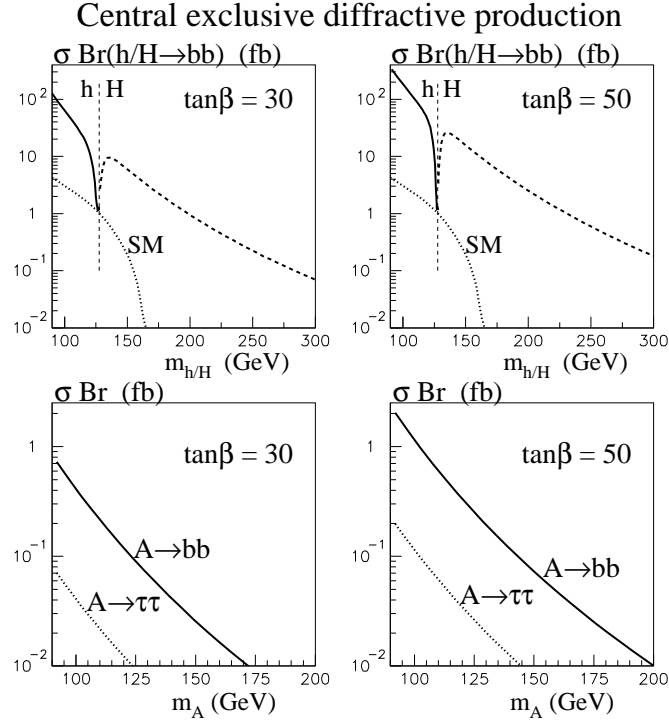}} \caption{The cross sections, 
times the appropriate $\bb$ and
$\tau^+\tau^-$ branching fractions, 
predicted (see \cite{Kaidalov:2003ys})for 
production of $h(0^+)$, $H(0^+)$
and $A(0^-)$ MSSM Higgs bosons  at the LHC. 
The dotted curve in the upper plots shows the cross section for  a SM Higgs
boson.\label{fig:a}}
\end{center}
\vspace*{-0.3in}
\end{figure}
Let us focus on the main $\phi\to \bb$ decay
mode\footnote{The studies in Ref.~\cite{DeRoeck:2002hk}
addressed mainly this mode.The use of the $\tau\tau$ decay mode 
requires an evaluation of the $pp\to
p + \tau\tau + p$ background, 
especially of the possibility of misidentifying gluon jets as $\tau$'s in the CEDP
environment}. 
%
%This mode is well suited for CEDP studies, 
%since a $P$-even, $J_z=0$ selection rule ~\cite{Khoze:2000jm}
%suppresses the QCD $\bb$ background.
%Note, that $0^-$ production is uncertain by
%almost  factor of 6, see Ref.~\cite{Kaidalov:2003ys}.

The estimates of the event rates in  Ref.~\cite{DeRoeck:2002hk} were performed
assuming 
$\sigma_{\rm CEDP}$=3~fb, a proton  tagging
efficiency of 0.6 and a $b$ jet tagging efficiency of 0.6.
Furthermore  the 
signal has been multiplied by 0.5 to account for the
jet polar angle cut and by 0.67 for the $b\bar{b}$ branching fraction.
It is expected ~\cite{DeRoeck:2002hk} that proton taggers 
can achieve a missing mass
resolution of $\Delta m_{\rm missing}\simeq 1$~GeV, 
giving a background of 4 events for an integrated luminosity of ${\cal L} 
= 30~{\rm fb}^{-1}$.
For such a luminosity, taking into account the efficiencies,
we would expect a Higgs signal of 11 events
with a favourable signal-to-background ratio  $S/B\sim3$.
Thus, to obtain a
statistical significance  of $5\sigma$ 
it is sufficient for the cross section
for a $\bb$ signal to satisfy
\begin{equation}
{\rm Br}(\bb)\cdot\sigma > 0.7~{\rm fb}~(2.7~{\rm fb}) \label{eq:cc1}
\end{equation}
for an integrated luminosity to be ${\cal L} = 300~{\rm fb}^{-1}$
(30~fb$^{-1}$).

In the MSSM case, 
as can be seen from Fig.~1, at $\tan\beta=50$  we expect that
${\rm Br}(H\to\bb)\sigma_H$, 
is greater than 0.7fb for
masses up to $m_H\sim 250$ GeV.
The situation is worse for pseudoscalar, $A$, production, because of
the $P$-even selection rule. 
 Thus, the CEDP filters out pseudoscalar production, which allows the possibility to study
pure $H$ production, see Fig.~1. 
This may be also useful in the decoupling limit 
($m_A > 2m_Z$ and $\tan\beta >5$),
where the light scalar $h$ becomes indistinguishable from the SM Higgs,
and the other two neutral Higgs states are approximately degenerate in mass. 
Here, forward proton tagging can play an
important role in searching for 
(at least) the $H$-boson, if it is not too 
heavy ($m_H \lesim 250$~GeV). For large
values of $\tan\beta$ the decoupling regime 
essentially starts at $m_A\simeq 170$~GeV. As seen in Fig.~1,
the cross section is still sufficiently large
to ensure the observation of the $H$ boson up to $m_A\simeq250$~GeV.
The possibility to use  diffractive 
processes to explore larger masses will depend on various
experiment-related factors. In particular, on the prospects 
to achieve better mass resolution, $\Delta m$, at
higher mass, $m$.
As discussed in Ref.~\cite{Kaidalov:2003ys},
CEDP may cover also the regions of MSSM parameter
space (`window' or `hole' regions,see for example,~\cite{Carena:2002es} )
where, once the $h$
boson is discovered, it is not
possible to  identify the $H$ scalar
by traditional means at the $5\sigma$ confidence
even with 300~fb$^{-1}$ of combined ATLAS+CMS luminosity.

As mentioned above,
if a candidate signal is detected it will be a challenging task to prove its Higgs
identity. Unlike the conventional approaches, the very fact of seeing the new 
state in CEDP
automatically implies that it  has the following fundamental properties. It must have zero
electric charge and be a colour singlet. Furthermore,
assuming $P$ and $C$ conservation, the dominantly produced
state has a positive natural parity, $P=(-1)^J$ and even $CP$.
The installation of forward proton
taggers 
may provide valuable additional
leverage 
in establishing the origin of the newly discovered candidate state.
In particular,
assuming CP conservation, the CEDP allow the $0^-,1^-,1^+$ states to be filtered out,
leaving only an ambiguity
between the $0^{++}$ and $2^{++}$ states. Though without further efforts 
the $2^{++}$ state  cannot  be ruled out,
this would not be the most likely choice.

\begin{figure}[bht]
\begin{center}
\centerline{\epsfxsize=10cm\epsfbox{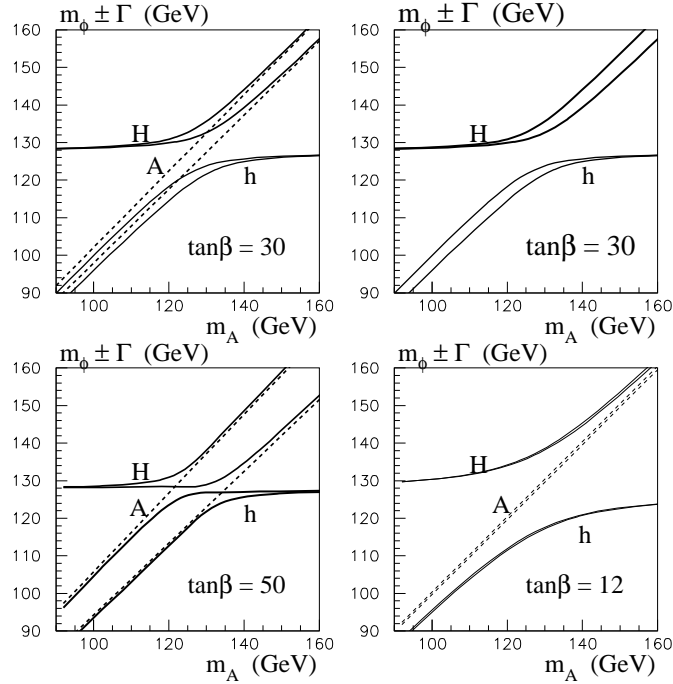}} 
\caption{The mass bands $m_\phi \pm \Gamma$
for neutral MSSM Higgs bosons as a function of $m_A$.
The upper
right hand plot shows that the $h$ and $H$ bosons are 
clearly identifiable for $\tan\beta=30$, if $A(0^-)$
production is suppressed. The lower plots show 
how the sensitivity of the widths, to variations of $\tan\beta$,
will change the profile of the peaks. \label{fig:d}}
\end{center}
\vspace*{-0.3in}
\end{figure}

As discussed in \cite{Kaidalov:2003fw,Khoze:2004rc}, 
studying of the azimuthal correlations of the outgoing protons
can allow further
spin-parity analysis.
In particular, it may be possible to isolate the $0^-$ state.
%The azimuthal distribution
%distinguishes between the production of scalar and pseudoscalar 
particles~\cite{Kaidalov:2003fw}.
Note that with the
forward protons we can determine the $CP$-properties of the Higgs boson
irrespective of the decay mode. Moreover,
CEDP allow the observation of the interference effects between the CP-even 
and CP-odd $gg\to\phi$ transitions. Their observation would signal
an explicit $CP$-violating 
mixing in the Higgs sector.

To illustrate how the CEDP can help to explore the Higgs sector
let us consider again large $\tan\beta$ case.
In the intense coupling regime it is especially difficult
to disentangle the Higgs bosons in the 
region around $m_A \sim 130$~GeV, where there is almost a mass degeneracy of
all three neutral Higgs states and their total widths can 
be quite large and  reach up to 1--2~GeV.
This can be seen from Fig.~2, where  
for numerical purposes, the same parameters as in \cite{Boos:2003jt}
were chosen. 
Since the traditional non-diffractive approaches do
not, with the exception of the $\gamma\gamma$ and $\mu\mu$ modes, 
provide a mass resolution better than
10--20~GeV, all three Higgs bosons will appear as one resonance.
Recall that in this regime the $\gamma\gamma$ decay mode is hopeless
and the dimuon Higgs decay mode is quite rare
(and, anyway, would require that the Higgs mass splitting exceeds at 3--5~GeV,
see Ref.~\cite{Boos:2003jt}).
An immediate advantage of CEDP,
for studying this  region, is that the $A$~contribution is strongly
suppressed, while the $h$ and $H$ states can be well separated 
($m_H-m_h\simeq 10$~GeV) given the anticipated
experimental mass resolution of $\Delta M \sim 1$~GeV~\cite{DeRoeck:2002hk}, 
see Fig.~2.
Note,that the forward tagging approach can provide 
a direct measurement of the width of the $h$ (for $m_h
\lesim 120$~GeV) and the $H$-boson (for $m_H \gtrsim 130$~GeV).
Outside the narrow range $m_A=130\pm 5$~GeV, the
widths of the $h$ and $H$ are quite different
(one is always much narrower than the other).
It would be instructive to observe this 
phenomenon experimentally.

For $\tan\beta =30$
the central exclusive signal should be still accessible at
the LHC up to an $H$ mass about 250~GeV. For instance, for $m_H=210$~GeV,
and LHC luminosity $30~{\rm fb}^{-1}$
($300~{\rm fb}^{-1}$), about 20 (200) $H\to\bb$ events are produced. 
If the experimental cuts and efficiencies
quoted in \cite{DeRoeck:2002hk} are imposed, then the signal is depleted
 by about a factor of 6. This leaves 3 (30) observable
events, with background of about 0.1 (1) events.

Let us make a few comments about the possibility to identify
the pseudoscalar boson, $A$, see for detailes 
Ref.~\cite{Kaidalov:2003fw,Kaidalov:2003ys}.
If the CEDP cross sections for scalar and pseudoscalar
Higgs production were comparable, it would
be possible to separate them readily by the missing mass scan,
and by the study of the azimuthal correlations
between the proton momenta.
However,the cross section for pseudoscalar Higgs exclusive
production is
strongly suppressed.For
values of $\tan\beta \sim 10$--15, the 
separation between the $H/h$ and $A$ bosons is much larger than their
widths, see Fig.~2 .  Hence it might be just 
possible to observe the pseudoscalar in CEDP.  For example, for
$\tan\beta = 15$ and $m_A=95$~GeV,
the mass separation, 3.6~GeV, between $h$ and $A$ is about 8 times
larger than the width $\Gamma_h$.  The cross section
\be {\rm Br}(A\to\bb)\cdot \sigma_A \simeq 0.15~{\rm fb}, \ee
when allowing for the large uncertainties in $\sigma_A$,
could be just sufficient to bring the process to the edge
of observability.

Probably  the best chance to identify the $A(0^-)$ boson
is to observe the double-diffractive inclusive process
\be pp  \to    X + \phi + Y,         \label{eq:A1} \ee
where both protons are destroyed. 
Process (\ref{eq:A1}) has the advantage of a much
larger cross section, see Ref.~\cite{Khoze:2001xm}.
However,here
we do not have the $J_z = 0$ selection rule to
suppress the $b\bar{b}$  background, nor do we have the possibility of
the good missing mass  resolution.
On the other hand, the $gg^{PP}$
luminosity  is more than order of magnitude larger than for the pure
exclusive case. For example, for double-diffractive inclusive
 production, with the rapidity gaps $\Delta \eta >3$, the luminosity is
20 times larger than that for the exclusive diffractive production of a Higgs boson with
mass $m_H=120$ GeV.
So, even for the $\tau\tau$ decay mode we expect a
cross section, $\sigma_{\rm incl}$, of about 20~fb for $A$
production in the MSSM with $\tan\beta=30$ and $m_A=120$~GeV.
 This looks promising, provided that the probability for gluon
misidentification as a $\tau$ is less than 1/150,
which looks feasible.
Process (\ref{eq:A1}) may be also useful in searches for
a light $CP$-violating Higgs boson, where we can study,
azimuthal correlations between the outgoing transverse energy flows
of the dissociating systems, see Ref.~\cite{Khoze:2004rc}.

\subsection{Experimental challenges}
The centrally produced Higgs particles can be measured with the 
ATLAS and CMS general purpose detectors at the LHC.
In order to tag the scattered protons these experiments will need to be 
equipped with detectors that need to be integrated with the beamline
of the LHC. 
Due to the relatively low mass of
the central (Higgs) system, the scattered protons have small $\xi$
values, in the range of $10^{-3}$--$10^{-2}$, where $\xi$ is the
momentum fraction lost by the proton in the interaction. A
classical technique to detect scattered protons at small $t$ and
with small relative momentum loss, is by using so-called roman pot
detectors. Recently a new type of detectors, called
microstations~\cite{nomokov}, has been proposed for this purpose.
Studies of the  LHC beam optics~\cite{orava} reveal that, in order
to access these small $\xi$ values, the roman pot detectors or
microstations need to be installed at about 300~m from the
interaction region.
These detectors can have an acceptance in $\xi$ down to
1--2$\times 10^{-3}$, and a parametrization of the acceptance was
included in the event estimates given before.

In order to efficiently record and measure the diffractively
scattered protons in roman pot detectors or microstations, they
have to be sufficiently separated from the beam particles. The
detectors, which are  located at 330~m and 420~m from the
interaction point, could then be used to define the proton momenta
by measuring, with respect to the beam axis, the difference in
horizontal displacement at the two locations as a function of the
average proton deflection.

We observe that a variation of $\Delta \xi = 5 \times 10^{-4}$
produces a $80\:\mu{\rm m}$ difference in the horizontal
displacement of a diffractively scattered proton. With
state-of-the-art silicon microstrip detectors this difference can
be measured with a precision of the order of 5$\mu$m. The expected
momentum spread of the beam protons is $\Delta \xi/\xi = 10^{-4}$.
For a symmetric event configuration ($\Delta  =|\xi_1-\xi_2| \le
0.04$), we then expect in the most optimistic case a mass
resolution of the order of $\Delta M_{\rm missing}/M_{\rm
missing}$ of order of 1\%~\cite{orava}. 
%This leads to the value
%$\Delta M_{\rm missing} = 1$~GeV, which is used in Table~1.

Pile-up events will also be important for the roman pot detectors.
The PYTHIA~\cite{Sjostrand:2000wi,Sjostrand:2000wi0} Monte Carlo program was used to estimate
the probability to have an additional proton accepted on one side
of the interaction region from single soft diffraction for the
different luminosities, and amounts to 8\% (medium luminosity),
and 40\% (high luminosity).
 Since
by then the mass of the Higgs  will be known to some accuracy, an
appropriate mass window can be chosen to select genuine scattered
protons that belong to the diffractive Higgs event.

The next issue is the efficiency $\varepsilon_b$ of tagging a $b$
jet. The value is correlated with the probability $P(g/b)$ to
misidentify a gluon as a $b$ jet. In
ref.~\cite{DeRoeck:2002hk} we require $P(g/b) = 0.01$ to reduce the $gg$
background to an acceptable level. For this value of $P(g/b)$, the
present estimate of the efficiency of $b$ and $\bar{b}$ tagging is
$(\varepsilon_b)^2=0.3$, but it is not inconceivable that this
could  be improved to a larger value, perhaps as large as
$(\varepsilon_b)^2=0.6$. If it turns out that this is impossible
for $P(g/b)=0.01$, then it is better to accept a worse
misidentification probability $P(g/b)$ in order to obtain a higher
value of $(\varepsilon_b)^2$. This will raise the background, but
will result only in a relatively small reduction in the
significance of the signal. For this reason we use
$(\varepsilon_b)^2 = 0.6$ in our estimates.

 The new roman pots would require also changes to the LHC machine, which 
will be a real technical challenge, if not excluded already.

The main concerns for  a project are the following
\begin{itemize}
\item How solid is the  experimental physics case: can we expect to see
a good signal over background?\\
- Are the signals sufficiently understood (cross sections)\\
- Do we have a good understanding of the background, 
in particular the inclusive one
that can feed down into exclusive peak, when smeared with the detector
resolution? Complete simulations are needed which include the experimental 
resolutions, to check if an exclusive signal remains visible.

\item The trigger: signals from 300/400 m roman pot arrive too late for the 
first level trigger  of ATLAS and CMS. E.g. the latter has a latency 
of 2.5 $\mu$sec\\
- Can we trigger with the central detector only at Level-1? The Level-1 di-jet 
trigger threshold is of order 150 GeV $E_T$ per jet, 
hence well above what can be expected 
from a low mass Higgs ($E_T$ of $\sim$ 50 GeV per jet). 
Studies which make use of the topology of the events can help to improve
see e.g. ~\cite{risto}.

\item Interference with the machine\\
- Can the detectors be integrated with the machine? Technically there is 
place at locations 330 and 420 m, but this is the cold section of the 
machine, and hence the detectors will need to be integrated with the 
cryogenic environment. This may compromise the accessibility of the 
detectors during running periods.

\item Detector choice\\
-What detectors are will be optimal for these regions?
Roman pots may be too bulky. An alternative could be the microstations 
which have the promise to be more compact.
\end{itemize}

These studies will be of interest for both ATLAS and CMS and could be 
part of a common study. The answers are needed in 2004.

\subsection{Comparison between different predictions}

There exists the plethora of predictions
from a variety of models for the cross
section for central diffractive Higgs production, which yield
answers ranging over  orders of magnitude,see,for example 
\cite{Bialas:1991wj,Cudell:1996ki,Levin:1999qv,Khoze:2000cy,Khoze:2001xm,Cox:2001uq,Boonekamp:2001vk,Boonekamp:2002vg,Cox:2003xp,Kaidalov:2003fw,Royon:2003ng,Enberg:2002id}.
One unfortunate
consequence is that this may discredit this approach as a
possible way to study a Higgs boson.A critical comparison
of these predictions and the explanation of the origin of such
wide differences was performed 
in Ref.~\cite{Khoze:2002py}.The main conclusion is that the
huge spread of predictions is {\em either} because different
diffractive processes have been considered {\em or} because
important effects have been neglected. 
Moreover,some of the models (especially those which
predict very large CEDP cross sections,are already excluded
by the existing experimental data on diffractive
dijet production at the Tevatron,see Ref.~\cite{Affolder:2000hd,Goulianos:2003gr}.

To clarify the differences we focus on the SM Higgs with
mass 120~GeV and with the dominant $H\ra\bb$ decay.
From an observational point of view, there are 
three different central diffractive
production mechanisms.

\begin{itemize}
\item[(a)] {\bf Exclusive production}:\ \ $pp\ra p + H + p$, see ~(\ref{eq:cedp}).\\ 

\item[(b)] {\bf Inclusive production}:\ \ $pp\ra X + H + Y$, see ~(\ref{eq:A1}).\\
In this case we allow both of the incoming protons to dissociate.  
The advantage is a much larger cross section. However, there is no
spin selection rule to suppress the $\bb$ background, and the
signal-to-background ratio is unfavourable.
On the other hand,this process may open a way to search for
the pseudoscalar or a light $CP$-violating Higgs boson.

\item[(c)] {\bf Central inelastic production}:\ \ $pp\ra p + (HX) + p$\\
There is additional radiation accompanying the Higgs in the
central region, which is separated from the outgoing protons by
rapidity gaps. Although this mechanism is often used for
predictions, it has, in our view, no clear advantages for Higgs
detection.
\end{itemize}

Each large rapidity gap may be associated with an
effective Pomeron exchange. It may be {\em either} a QCD Pomeron,
which at lowest order is a gluon--gluon state, {\em or} a
phenomenological Pomeron with parameters fixed by data.

Recall that, at medium and high luminosity at the LHC, the
recorded events will be plagued by overlap interactions in the
same bunch crossing. 
Hence the rapidity gaps
occurring in one interaction may be populated by particles created
in an accompanying interaction.
It is, however, possible to use
detector information to locate the vertices of the individual
interactions and, in principle, to identify hard scattering events
with rapidity gaps.
For the exclusive and central inelastic
processes the use of proton taggers makes
it much more reliable to select the rapidity gap events.Moreover,
the presence of rapidity gaps may be used as the level-1 trigger
for the central signal.

There is a price to pay for the unique advantages
of the central diffractive processes.
The cross sections are reduced by
the probabilities of the gaps not to be populated by, first, the
gluon radiation associated with a QCD Pomeron and/or the hard
$gg\ra H$ subprocess and, second, by secondaries produced in the
soft rescattering of the spectator partons. We denote these
survival probabilities by $T^2$ and $S^2$ respectively. The
probability amplitude $T$, not to radiate, can be calculated using
perturbative QCD. The expression for $T$ has the familiar
Sudakov form, see, for example, ~\cite{Khoze:2000cy,Khoze:2000jm} and references therein.
Note that the  $T$-factor plays a crucial role in providing the
infrared stability in calculations
of the Higgs cross section.
On the other hand the survival
factor, $S^2$, to soft rescattering cannot be calculated
perturbatively. The presence, and the value, of $S^2$ can be
checked experimentally by comparing the diffractive cross section
in deep inelastic reactions at HERA (where $S$ is close to 1) 
with the
cross section of diffractive dijet production at the Tevatron, for
which it turns out that $S^2 \sim 0.1$~\cite{Affolder:2000vb}. Theoretical
predictions of the survival factor, $S^2$, can be found in
Refs.~\cite{Khoze:2000wk}. Note that the factor $S^2$ is not a
universal number.Its value 
depends on the initial energy and the particular final
state. Clearly, the presence of  $S^2$ violates
factorization.The latest estimate of this factor is
$S^2$=0.026, \cite{Kaidalov:2003ys}.

A critical comparison of a  representative range of
some of the recent calculations of 
cross sections for central
diffractive production of a SM Higgs boson is given in Table~1.

\begin{table}[h]
\begin{center}
\begin{tabular}{@{} l @{\ }|@{\ } c @{\ }| c | c | c @{\ }| c | c | l @{}}
\raisebox{-1.5ex}[0ex][1ex]{Reference}  &
\raisebox{-1.5ex}[0ex][1ex]{Process}  &
\multicolumn{2}{c}{Survival factor} \vline&
\raisebox{-1.5ex}[0ex][1ex]
{Norm.}  &  \multicolumn{2}{c}{$\sigma_{\rm Higgs}$~(fb)}  \vline&  \raisebox{-1.5ex}[0ex][1ex]
{Notes} \\\cline{3-4}\cline{6-7} &  &  $T^2$  &  $S^2$  &  &  Teva.  &  LHC  &  \\
\hline%ROW 1 (TITLES) ENDS HERE---------------------
Cudell,  & excl  &  no  &  no  & $\sigma_{\rm tot}$  &  30
&  300  &  Overshoots CDF dijets \\  Hernandez~\cite{Cudell:1996ki}  &  incl  &  &  &  &  200  &  1200  &   by 1000. \\
\hline%ROW 2 ENDS HERE---------------------
\raisebox{-1.5ex}[0ex][1ex]{Levin~\cite{Levin:1999qv}} & excl & yes & yes &
$\sigma_{\rm tot}$  &
20  &    & Overshoots CDF dijets \\  &  incl  &  No DL  &&&  70  &  \raisebox{1.5ex}[0ex][1ex]{--}  &  by 300.  \\
\hline%ROW 3 ENDS HERE---------------------
\raisebox{-1.5ex}[0ex][1ex]{Khoze, Martin,} &
excl &  & &  pdf  &  0.2  &  3  &  \raisebox{-1.5ex}[0ex][1ex]{Uses unintegrated gluons.} \\
\raisebox{-1.5ex}[0ex][1ex]{Ryskin~\cite{Khoze:2001xm}}  & incl & yes & yes
& pdf  &  1  &  40  &  \raisebox{-1.5ex}[0ex][1ex]{CDF dijets OK.}
\\  &  C.inel &  & &  &  $\sim0.03$  &  50 &
\\ \hline%ROW 4 ENDS HERE---------------------
Cox, Forshaw,  &  \raisebox{-1.5ex}[0ex][1ex]{C.inel} &
\raisebox{-1.5ex}[0ex][1ex]{$T\simeq1$}  & \raisebox{-1.5ex}[0ex][1ex]{norm}  &  CDF  & \raisebox{-1.5ex}[0ex][1ex]{0.02} & \raisebox{-1.5ex}[0ex][1ex]{6} & No LO, only NLO, QCD \\
Heinemann~\cite{Cox:2001uq}  & & & &  dijet  & & &  \\
\hline
%ROW 5 ENDS HERE---------------------
Boonekamp,&&&&&&&
\\ & \raisebox{-1.5ex}[0ex][1ex]{C.inel}  &  
\raisebox{-1.5ex}[0ex][1ex]{$T\simeq1$}  &
\raisebox{-1.5ex}[0ex][1ex]{norm}  &  CDF &
\raisebox{-1.5ex}[0ex][1ex]{1.3} &
 \raisebox{-1.5ex}[0ex][1ex]{160}
& No LO, only NLO, QCD.
 \\ Peschanski,   & & & & dijet  &  & & Assume
$S^2_{\rm CDF} = S^2_{\rm LHC}\,$.  \\
Royon~\cite{Boonekamp:2003wm}  &&&&&&&  \\ \hline%ROW 6 ENDS HERE---------------------
Enberg, &&&&&&&   \\
Ingelman, & incl & \raisebox{-1.5ex}[0ex][1ex]{yes}
&  \raisebox{-1.5ex}[0ex][1ex]{yes}  &  \raisebox{-1.5ex}[0ex][1ex]{$F_2^{\rm Diff.}$}  &  \raisebox{-1.5ex}[0ex][1ex]{$<0.01$}  &  \raisebox{-1.5ex}[0ex][1ex]{0.2}  &  \raisebox{-1.5ex}[0ex][1ex]{No coherence.}\\
Kissavos,  &  C.inel  &  &  &  &  & &
\\ Timneanu~\cite{Enberg:2002id} &&&&&&&
\end{tabular}
\end{center}
\caption{\small Recent representative calculations of 
 $\sigma_{\rm Higgs}$, for exclusive, inclusive and
Central inelastic production of a Higgs boson
of mass about 120~GeV.The Norm.
column indicates the way in which the various predicted cross
sections are normalised.
``norm'' in the $S^2$ column means that $S^2$ is
determined by normalising to CDF dijet data~\cite{Affolder:2000hd}.
The cross sections for central inelastic production (C.inel)
correspond to integrating up to $M_{\rm miss}=0.1\sqrt{s}$, where
$\sqrt{s}$ is the collider energy. Note that in Ref.~\cite{Khoze:2001xm}
the C.inel cross section is 0.2~fb at the Tevatron, but this
includes the exclusive contribution. The LHC entry for Cox
et~al.~\cite{Cox:2001uq} is obtained using $S^2=0.02$.}
\end{table}

Quite recently some new results on central diffractive
Higgs production have become available, see for example~\cite{Petrov:2003yt,Erhan:2003za}

The expectation in ~\cite{Petrov:2003yt} for the CEDP SM Higgs production
are close to those in ~\cite{Khoze:2000cy,DeRoeck:2002hk}.
The background issues were not fully addressed.

Ref.~\cite{Erhan:2003za} concerns a known idea to search for the Higgs
boson in the single-diffractive events. 
In this case the event rate would be much larger.
The single-diffractive studies are certainly interesting on their own
right, as it was shown in \cite{Heyssler:1997mb}, but it 
is at present  not yet clear whether these
processes provide an additonal advantage 
in searching for the Higgs bosons over a fully inclusive one.
The main reason for this is
that hadronic activity around the Higgs boson is
practically the same as in the conventional inclusive events
at lower energy. Studies including the background are 
ongoing~\cite{schlein}.

We would like to stress that the expectations
for the exclusive cross section
can be checked experimentally. Practically all the main ingredients,
are the same for the Higgs
signal as for exclusive  central diffractive dijet production,
$pp\ra p+ {\rm dijet} + p$, where the dijet system is chosen in
the same kinematic domain as the Higgs boson, that is
$M(jj)\sim120\:{\rm GeV}$ \cite{Khoze:2001xm,Khoze:2000cy}. Therefore by observing
the larger dijet production rate, we can confirm, or correct, the
estimate of the exclusive Higgs signal.

\subsection{Conclusion}

 The central  diffractive processes promise
 a rich physics menu for studying 
the detailed properties of the Higgs sector. 
Within MSSM, that the expected
CEDP cross sections are large enough, especially
 for large $\tan\beta$, both in the intense coupling and the
decoupling regimes. 
Thus CEDP offers a way to cover those regions of MSSM parameter space which may be hard to
access with the conventional (semi) inclusive approaches. 
This considerably extends the physics potential of the
LHC and may provide studies 
which are complementary, both to the traditional non-diffractive approaches at the
LHC, and to the physics program of a future Linear $e^+e^-$ collider.
%To some extent this approach gives
%information which otherwise may have to await a Linear $e^+ e^-$ collider.

}

%% file: dudko.tex
{
%%%%%%%%%% espcrc2.tex %%%%%%%%%%
%
% $Id: espcrc2.tex 1.2 2000/07/24 09:12:51 spepping Exp spepping $
%

% change this to the following line for use with LaTeX2.09
% \documentstyle[twoside,fleqn,espcrc2]{article}

% if you want to include PostScript figures
% if you have landscape tables

% put your own definitions here:
%   \newcommand{\cZ}{\cal{Z}}
%   \newtheorem{def}{Definition}[section]
%   ...
\newcommand{\bi}{\begin{itemize}}
\newcommand{\ei}{\end{itemize}}
\newcommand{\ttbs}{\char'134}
\newcommand{\met}{\mbox{$\not\!\!E_T$}}
\newcommand{\dzero}{D\O}
\renewcommand{\AmS}{{\protect\the\textfont2
  A\kern-.1667em\lower.5ex\hbox{M}\kern-.125emS}}

% add words to TeX's hyphenation exception list
%\hyphenation{author another created financial paper re-commend-ed Post-Script}

% declarations for front matter
\section[ ]{Optimized Neural Networks to Search for Higgs Boson Production at 
the Tevatron\footnote{E.\,Boos and L.\,Dudko}}

\subsection{The basic idea}
% \vspace*{-0.3cm}
In High Energy physics a discrimination between a signal and its 
corresponding backgrounds by Neural Networks (NN) is
especially remarkable when the data statistics are limited.
In this case it is important to optimize all steps of the analysis.
One of the main questions which arises in the use of NNs 
is which, and how many variables should be chosen
for network training in order to extract 
a signal from the backgrounds in an optimal way.
The general problem is rather complicated and finding a solution
depends on having a concrete process for making the choice, because
 usually it takes a lot
of time to compare results from different sets of variables.     

One observation which helps in making the best choice of the most sensitive
variables  is to study the singularities in Feynman diagrams of the 
processes.
Let us call those kinematic variables 
in which singularities occur as "singular variables".
What is important
to stress here is that most of the rates for both the signal and for
the backgrounds come from the integration over the phase space region 
close to these singularities. 
One can compare the lists of singular variables and the positions of the 
corresponding singularities in Feynman diagrams for the 
signal process and for the  backgrounds. 
It is obvious that if some 
of the singular variables are different or the positions of 
the singularities
are different for the same variable 
for the signal and for the backgrounds the
corresponding  distributions will differ most strongly. 
Therefore, if one uses all such 
singular variables in the analysis, then the largest part of the phase space
where the signal and backgrounds differ most will be taken
into account.
One might think that it is not a simple task 
to list all the singular variables when the phase space
is very complex, for instance, for 
reactions with many particles involved.
However, in general, all singular variables can be of
only two types, either s-channel: $M_{f1,f2}^2 = (p_{f1} + p_{f2})^2$, 
where $p_{f1}$ and $p_{f2}$ are the four
momenta of the final particles  $f1$ and $f2$
or t-channel: $\hat{t}_{i,f} = (p_f-p_i)^2$, where $p_f$  
and  $p_i$ are the momenta of the final particle (or cluster)
and the initial parton. For the $\hat{t}_{i,f}$ all the needed variables
can be easily found in 
massless case: $\hat{t}_{i,f} = - \sqrt{\hat{s}} e^{Y} p_T^f e^{-|y_f|}$, 
where $\hat{s}$ 
is the total invariant mass of the
produced system, and {\it Y} is the rapidity of the total system (rapidity
of the center mass of the colliding partons), $p_T^f$ and $y_f$ 
are transverse momenta and pseudorapidity of the
final particle {\it f}.
The idea of using singular variables as the most discriminative ones is  
described in~\cite{BDO} and the corresponding method was demonstrated in 
practice in~\cite{aihenp}.

   Singular variables correspond to the structure of the denominators
 of Feynman diagrams. Another type of interesting variables corresponds to
the numerators of Feynman diagrams and reflects the spin effects and 
the corresponding
 difference in angular distributions of the final particles. 
In order to discriminate between a signal and the backgrounds, 
one should choose in addition to singular variables mentioned above 
those angular variables whose distributions are different 
for the signal and backgrounds. The set of these singular and 
angular variables will be the most efficient set for a NN analysis.

The third type of useful variables which we called "Threshold"
variables are related to the fact that various signal and background
processes may have very different thresholds. Therefore the distributions
over such kind of variables also could be very different keeping in mind
that effective parton luminosities depend strongly on 
$\hat{s}$. The variable $\hat{s}$ would be a very efficient variable of 
that kind. However, the problem is that in case of neutrinos in the final 
state one can not measure $\hat{s}$ and should use the effective
$\hat{s}$ which is reconstructed by solving t-,W-mass equations 
for the neutrino longitudinal momenta. That is why we propose to use
not only the effective variable $\hat{s}$ but the variable
$H_T^{jets}$ as well.  

To apply the method it is important to use a proper Monte-Carlo model 
of signal and background events which includes all needed spin 
correlations between production and decays. 
For the following analysis we have calculated the complete tree level
matrix elements for the background processes with all decays and 
correlations by means of the CompHEP program~\cite{comphep}.
The corresponding events are available at the FNAL Monte-Carlo events 
database~\cite{mcdb}.
% \vspace*{-0.5cm}

\subsection{Applying the method}
% \vspace{-0.3cm}
The present estimation of the expected sensitivities for the light Higgs 
boson 
search at the Tevatron by means of NNs is given 
in~\cite{prd_pushpa}. Based on the method 
described above we improve the efficiency of the NN technique. 
In the analysis we choose the Higgs boson mass to be
$M_H=115$~GeV. We model the detector smearing by the SHW 
package~\cite{SHW}.

First of all we exclude ineffective variables
from the old set~\cite{prd_pushpa}, like $P_T^e$ from the $W$-boson (shown 
at 
the left plot in Fig.~\ref{fig:pt}). After the corresponding analysis of 
Feynman diagrams and comparison of kinematical distributions we added the 
new variables for NN training. The example distribution for the new 
variable 
($cos(z_{axis},e)$) is shown in the right plot of Fig.~\ref{fig:pt}.
% \vspace*{-1.5cm}
\begin{figure}[htb]
  \hspace*{-0cm}
\begin{minipage}{.49\linewidth}
\includegraphics[width=16pc,height=14pc]{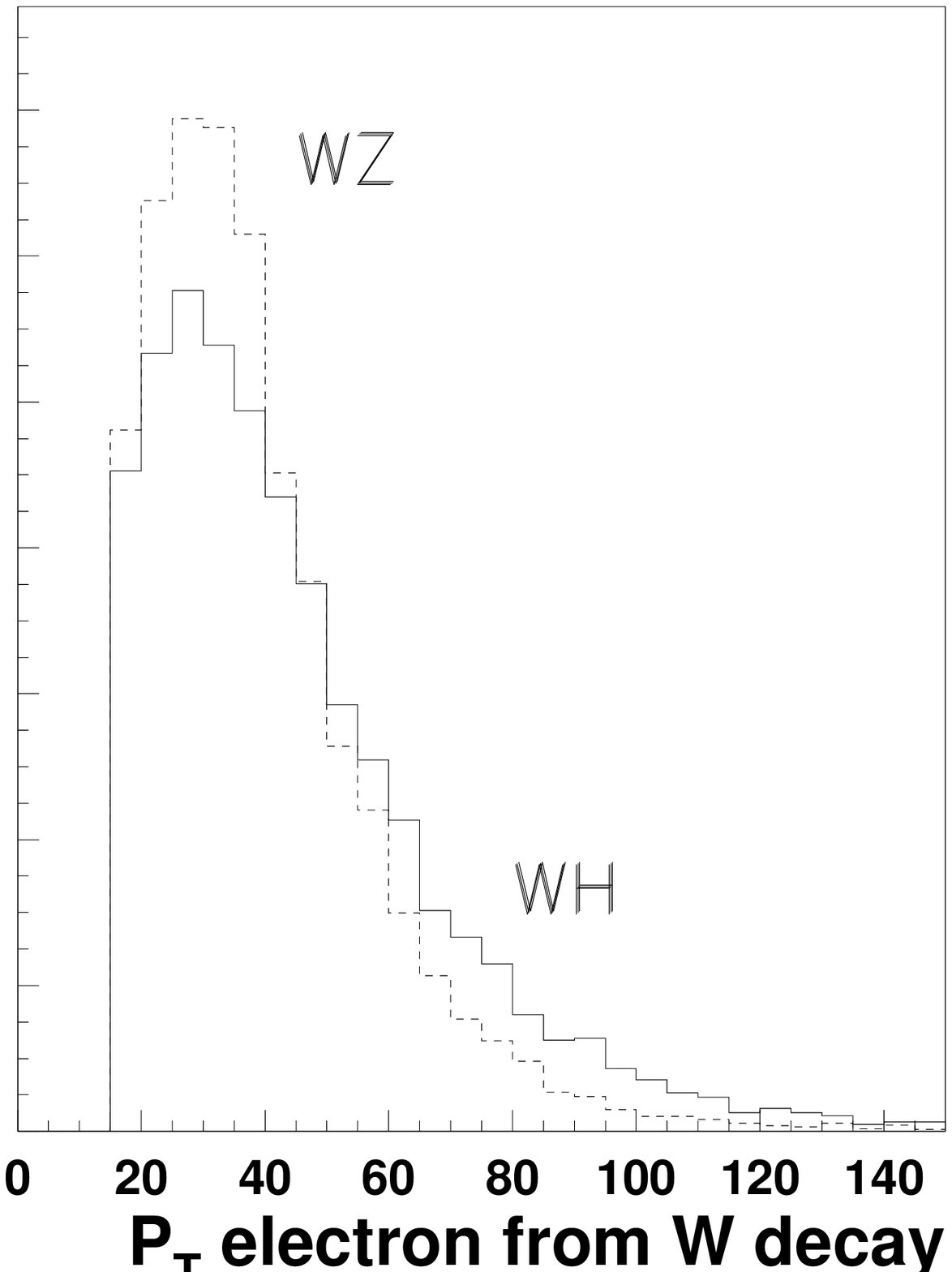}
\end{minipage}
\begin{minipage}{.49\linewidth}
\includegraphics[width=16pc,height=14pc]{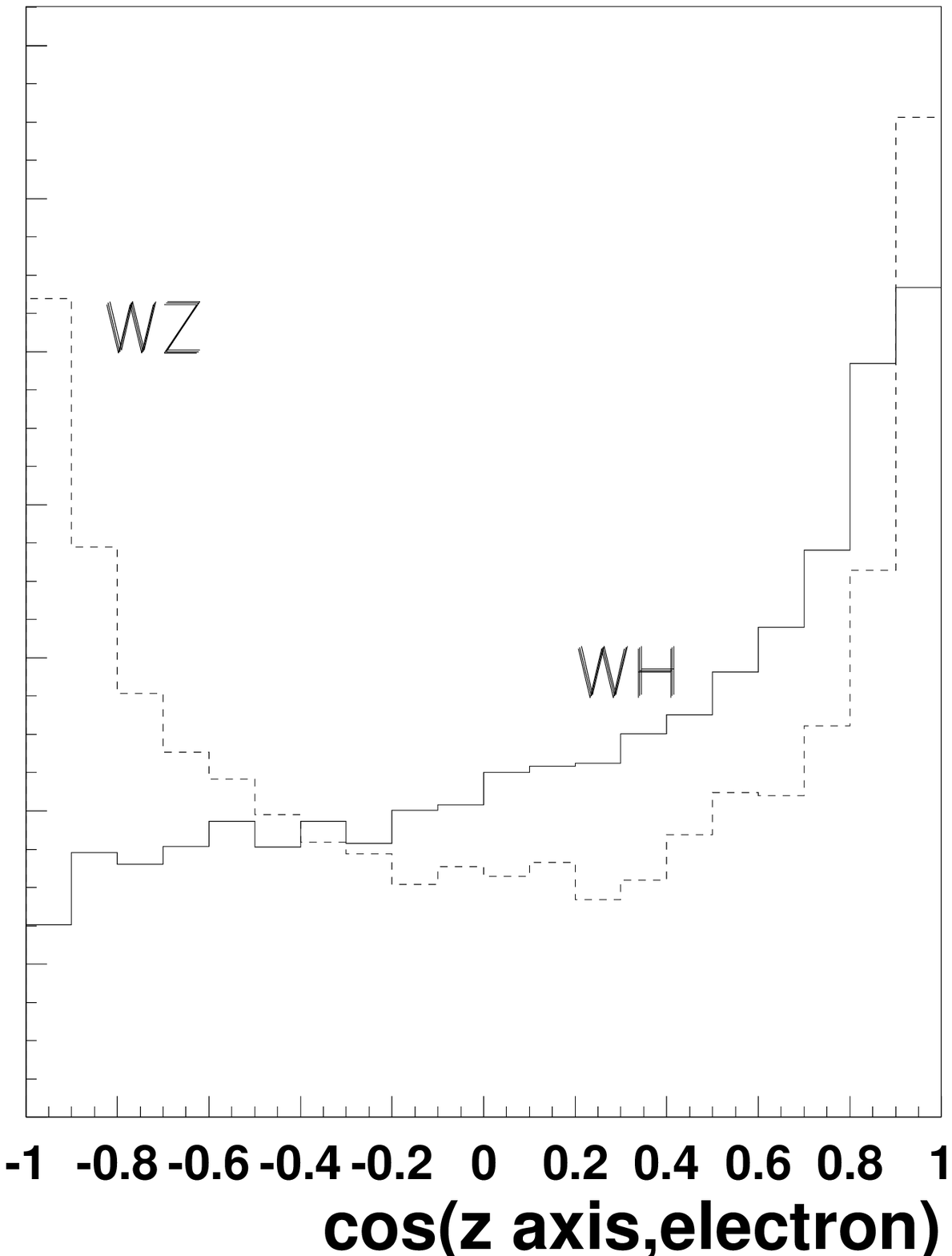}
\end{minipage}
%  \vspace*{-0.75cm}
\caption{Examples of the old kinematic variable (left plot) and the new one (right
plot)}
\label{fig:pt}
\end{figure}
%  \vspace*{-0.94cm}
  At the next step we constructed the set of NNs for pairs 
of the signal (WH) and each 
of the background from the complete set of principle backgrounds 
($Wb\bar b,\ WZ,\ t\bar t,\ tb(j)$). 

The standard steps of NN
training were used for the NNs with the old set of input variables and 
with the new one.
Efficiencies of networks with
different sets have been compared 
based on the criteria that for the better net the  ``Error function'' 
$ E = \frac{1}{N}\sum_{i=1}^{N}(d_i-o_i)^2$, where $d_i$ and $o_i$ 
are the desired and real
outputs of the net and $N$ is the number of test events, is smaller.
Two examples of distributions you can see in Fig~\ref{fig:e} 
for the $WH-t\bar t$
network (left plot) and $WH-WZ$ network (right plot). One can see  
a significant improvement for the networks with new input sets in 
comparison with old sets
 of variables, since the corresponding curves of the error 
function are significantly
 lower. 
% \vspace*{-0.3cm}
 \subsection{Results}
% \vspace*{-0.3cm}
 Based on the described method we have constructed the new NNs to search 
for a light
 Higgs boson at the Tevatron. After checking the improvement in efficiency
 of new networks we recommend the new sets of input variables for NNs, which are
 shown below:\\
%\small 
%\hspace*{-0.8cm}
\begin{itemize}
\item 
$Wb\bar b$ -- $WH$ \\
NN: 
 $M_{b\bar b},\ P_{T}^{b1},\ P_{T}^{b2},\ P_{T}^{bb},\
\hat{s},\ H_T^{jets},$\\ \hspace{0.3cm} $cos(b1,b2)|_{lab},\ cos(b1,b1b2)|_{b1b2}$

\item 
$WZ$ -- $WH$ \\
NN: 
 $M_{b\bar b},\ P_{T}^{b1},\ P_{T}^{b2},\  H_T^{jets},$\\
$cos(b1,b2)|_{lab}, \ Q\times cos(z,b1)|_{lab},\ cos(W,e)|_{W}$

\item  
$t\bar t$ -- $WH$ \\
NN: 
 $M_{b\bar b},\ M_{Wb},\ \hat{s},\ M_{Wjets-b},\
H_T^{jets},$\\ $ Q\times cos(\psi_{axis},e)|_{top},\  cos(b1,b1b2)|_{b1b2}$

\item  
$tbj,\ tb$ -- $WH$ \\
NN: 
 $M_{b\bar b},\ M_{Wb},\ P_{T}^{b2},\ \hat{s}, $\\ $M_{Wjets-b},\
P_{T}^{top},\  H_T^{jets},\ cos(z,e)|_{lab},$\\ $ Q\times cos(z,b1)|_{top},\
cos(e,j)|_{top}$
\end{itemize}
where there are three types of variables:
\bi 
\item ``Singular'' variables (denominator of Feynman diagrams):\\
   $M_{12}$ is the invariant mass of two particles and/or jets (1 and 2) and 
corresponds to s-channel singularities;\\
   $P_{T}^{f}$ (the transverse momenta of {\it f});\\ 
   $M_{Wjets-b}$ is the invariant mass of the $W$ and all jets except 
the $b$-jet for which the
   $M_t=(p_W+p_b)$ is closest to the top quark mass;
\item ``Angular'' variables (numerator of Feynman diagrams, spin effects):
   $cos(b1,b1b2)|_{b1b2}$ means the cosine of the angle between highest $P_T$ 
b-quark 
   and vector sum of the two highest $P_T$ b-quarks in the rest frame of 
these two b-quarks. 
Scalar (Higgs) and vector (gluon, Z-boson) particle decays lead to 
significantly different distributions
on this variable, this is also very much different for the case when b-quarks
come from the decay of top and anti-top quarks; \\
   $cos(b1,b2)|_{lab}$ characterizes how much two b-quarks are 
collinear;\\
   $cos(z,b1)|_{lab}$ and $cos(W,e)|_{W}$ reflect the difference 
in t-channel Z-boson  
   and s-channel Higgs-boson production topologies 
where $_{lab}$ means the laboratory rest frame
   and $z$ means the  z-axis;\\
   $cos(\psi_{axis},e)|_{top}$~\cite{tt_spin} and 
$cos(e,j)|_{top}$~\cite{single_top_spin} 
   are the top quark spin correlation 
   variables used in the analysis of the top quark 
pair and single production, the 
   lepton charge  
 $Q$ is added here to take uniformly into account the 
electron and the positron contributions 
 from the $W$-boson decays.
\item ``Threshold'' variables. 
As explained above  
 the $\hat{s}$ and $H_T^{jets}$ variables are used in our analysis. 

\ei

As one can see from the Fig.\ref{fig:e} using the new NN variables
allows to improve the NN efficiency by a factor of 1.5-2 depending on the 
background process.
It will lead to corresponding improvement in prospects to find
a light Higgs at the Tevatron. However, one needs to take into account
the ZH production channel as well as a number of detector efficiencies
in order to predict a realistic discovery limit.
%  \vspace*{-0.8cm}
\begin{figure}[htb]
%  \vspace*{-0.4cm}
  \hspace*{-0.0cm}
\begin{minipage}{.49\linewidth}
\includegraphics[width=16pc,height=14pc]{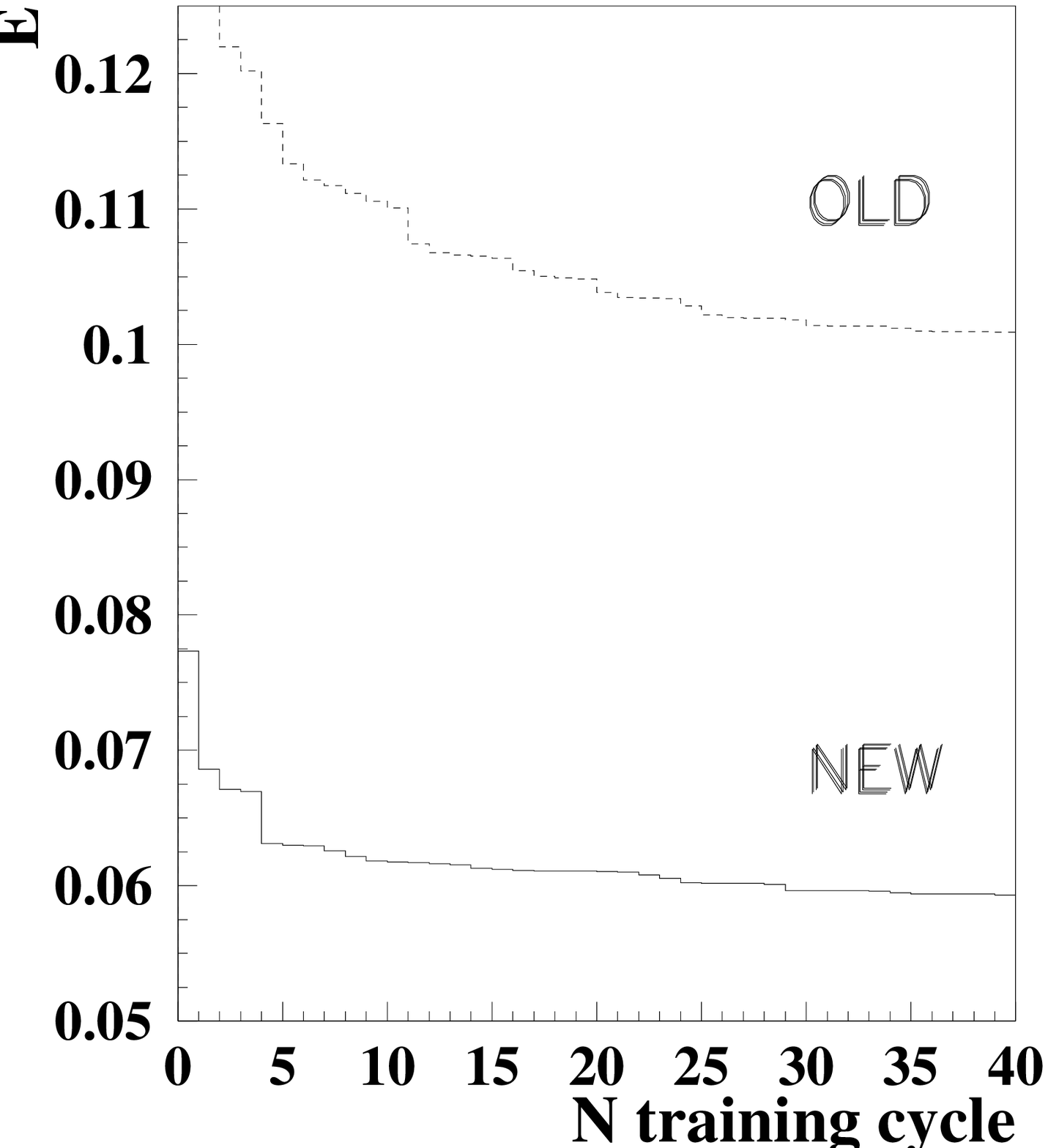}
\end{minipage}
\begin{minipage}{.49\linewidth}
  \hspace*{0.5cm}
\includegraphics[width=16pc,height=14pc]{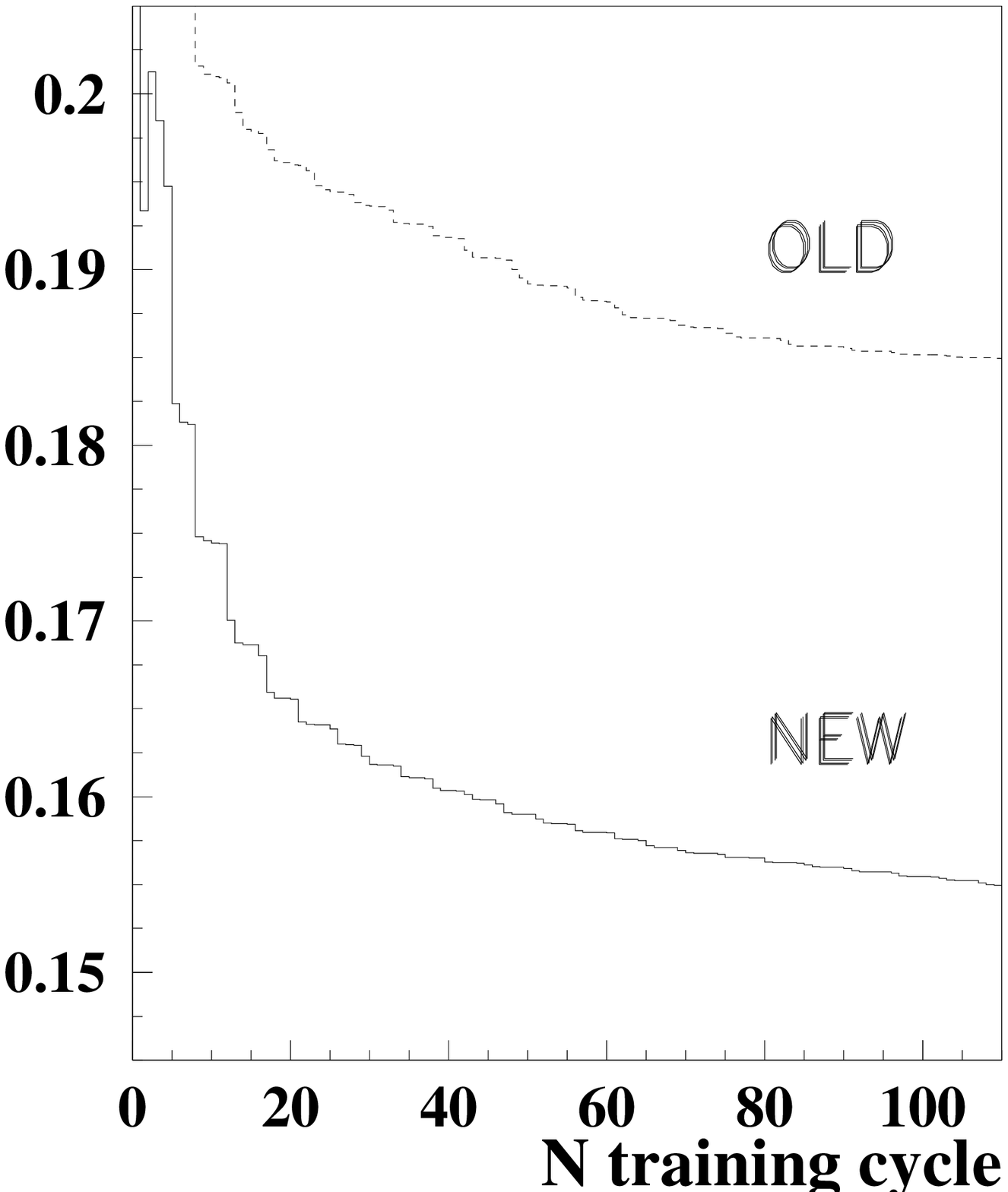}
\end{minipage}
%  \vspace*{-0.8cm}
\caption{NN Error function for the $WH-t\bar t$
(left plot) and $WH-WZ$ networks (right plot).}
\label{fig:e}
\end{figure}
%  \vspace{-1cm}

}

%% file: sopczak.tex
{
%  \let\oldthebibliography=\thebibliography
%  \let\endoldthebibliography=\endthebibliography
%  \renewenvironment{thebibliography}[1]{%
%    \begin{oldthebibliography}{#1}%
%      \setlength{\parskip}{0ex}%
%      \setlength{\itemsep}{0ex}%
%  }%
%  {%
%    \end{oldthebibliography}%
%  }

%  move the whole page up
%\topmargin=-2cm

\def\gsim{\mathrel{\raise.3ex\hbox{$>$\kern-.75em\lower1ex\hbox{$\sim$}}}}

\section[ ]{The \boldmath$\rm p\bar p \rightarrow tbH^\pm$\unboldmath\ 
        Process at the Tevatron in HERWIG and PYTHIA
Simulations\footnote{J.\,Alwall, C.\,Biscarat, S.\,Moretti, J.\,Rathsman
and A.\,Sopczak}}

\subsection{Introduction}
Charged Higgs bosons are predicted by non-standard models, for example Two-Higgs Doublet
Models such as the Minimal Supersymmetric 
Standard Model (MSSM).
Thus, their detection and the measurement of their properties 
(such as the mass which is not predicted by any model) play an important r\^ole in the 
investigation of an extended Higgs sector and in the understanding of the 
generation of particle masses.
The current limit on the charged Higgs boson mass is set by the LEP experiments
at 78.6~GeV, independent of the Higgs boson decay branching 
fractions~\cite{leplimit}.
At the Tevatron, charged Higgs bosons could be discovered for masses well 
beyond this limit.

If the charged Higgs boson mass $m_{\rm H^\pm}$ satisfies 
$m_{\rm H^\pm} < m_{\rm t} - m_{\rm b}$, where $ m_{\rm t}$ is the top quark mass and 
$ m_{\rm b}$ the bottom quark mass,
it could be produced in the decay of the top quark $\rm t \rightarrow bH^+$.
This so-called on-shell top approximation
($\rm q\bar q$, $\rm gg \rightarrow t\bar t$ with $\rm t \rightarrow bH^+$)
was previously used in the event generators. Throughout this paper this process is denoted by
$\rm p\bar p \rightarrow t\bar t \rightarrow tbH^\pm$.
Owing to the large top decay width ($\rm \Gamma_{\rm t} \simeq 1.5$~GeV) and because of 
the additional diagrams which do not proceed via direct $\rm t\bar t$ 
production~\cite{Borzumati:1999th,Miller:1999bm},
charged Higgs bosons could 
also be produced beyond the kinematic top decay threshold. 
The importance of these effects in the threshold region was emphasized in the 
previous Les Houches proceedings~\cite{Cavalli:2002vs} and the 
calculations~\cite{Borzumati:1999th,Miller:1999bm} are implemented in 
HERWIG\,\cite{Corcella:2000bw,Corcella:2002jc,Moretti:2002eu}\,and 
PYTHIA\,\cite{Sjostrand:2000wi}\footnote{HERWIG release version 6.505 and 
inclusion in a future official PYTHIA version.}. 
The full process is referred to as $\rm p\bar p \rightarrow tbH^\pm$.
Examples of the graphs contributing to the $\rm p\bar p \rightarrow \bar t bH^+$
process are~\cite{Guchait:2001pi}:

\vspace*{-1.4cm}
\begin{equation}
%Q1:
\begin{picture}(120,30)
\SetScale{0.8}
\SetWidth{0.6}
\SetOffset(2,-62.5)
\ArrowLine(30,65)(15,50)
\ArrowLine(15,80)(30,65)
%\Text(10,45)[]{$\rm \bar p$}
%\Text(10,85)[]{p}
\Gluon(30,65)(60,65){3}{3}
\ArrowLine(75,50)(60,65)
\ArrowLine(60,65)(75,80)
\DashLine(70,75)(85,75){2}
\Text(78,60)[]{\small$ \rm H^+$}
\Text(66,40)[]{\small $ \rm \bar t$}
\Text(66,70)[]{\small$ \rm b$}
\end{picture}
%G7:
\begin{picture}(120,40)
\SetScale{0.8}
\SetWidth{0.6}
\SetOffset(2,-62.5)
\Gluon(15,50)(30,65){3}{3}
\Gluon(30,65)(15,80){3}{3}
%\Text(10,45)[]{\small 2}
%\Text(10,85)[]{\small 1}
\Gluon(30,65)(60,65){3}{3}
\ArrowLine(75,50)(60,65)
\ArrowLine(60,65)(75,80)
\DashLine(70,75)(85,75){2}
\Text(78,60)[]{\small$ \rm H^+$}
\Text(66,40)[]{\small $ \rm \bar t$}
\Text(66,70)[]{\small$ \rm b$}
\end{picture}
\begin{picture}(120,30)
\SetScale{0.8}
\SetWidth{0.6}
\SetOffset(0,-60)
\Gluon(45,75)(30,90){3}{3}
\Gluon(30,40)(45,55){3}{3}
%\Text(25,95)[]{\small 2}
%\Text(25,35)[]{\small 1}
\ArrowLine(60,40)(45,55)
\ArrowLine(45,55)(45,75)
\ArrowLine(45,75)(60,90)
\DashLine(45,65)(60,65){2}
\Text(58,55)[]{\small$ \rm H^+$}
\Text(53,35)[]{\small $ \rm \bar t$}
\Text(53,75)[]{\small$ \rm b$}
\end{picture}
\vspace*{1.2cm}
\end{equation}
The t-channel 
graph is one example of a diagram which does not proceed via 
$\rm t\bar t$ production.
This graph contributes to enhanced particle production in the forward 
detector region.

A charged Higgs boson with $m_{\rm H^\pm} < m_{\rm t}$
decays predominantly into a $\tau$ lepton and
a neutrino. 
For large values of $\tan \beta$ ($\gsim$ 5), the ratio of the vacuum
expectation values of the two Higgs doublets, this branching ratio
is about 100\%.
The associated top quark decays predominantly into a W boson or a second charged Higgs boson,
and a b-quark. 
The reaction
\begin{equation}
\rm p\bar p \to  tbH^\pm~~~(t\to bW^\mp)~~~(H^\pm \to \tau ^\pm \nu_{\tau})
\label{channel}
\end{equation}
is a promising channel to search for the charged Higgs boson at the Tevatron.
Simulations are performed at the centre-of-mass energy $\sqrt s=1960$~GeV and for $\tan\beta=20$. 

%\clearpage
\subsection{Comparison of Production Cross Sections}
\vspace*{-0.1cm}

The expected production cross sections are determined using HERWIG and PYTHIA simulations, 
and are shown in Fig.~\ref{fig:xsec}. The default mass and coupling parameters of 
HERWIG version 6.5 and PYTHIA version 6.2
are used.
The cross sections depend strongly on the top decay width over the 
investigated mass range.
For the top width, the Standard Model (SM) value 
$\rm \Gamma_{\rm t} = 1.53$~GeV is used at $m_{\rm H^\pm}=210$~GeV and the width is increased 
as a function of the charged Higgs boson mass
to $\rm \Gamma_{\rm t} =1.74$~GeV at $m_{\rm H^\pm}=70$~GeV in both generators.
The production cross section in HERWIG is about a factor 2 larger compared to PYTHIA 
which can be attributed to the default choices of the standard parameters.
It is mostly driven by the different choice of the heavy quark masses entering the Higgs-quark 
Yukawa coupling.
In PYTHIA a running  b-mass is used at the tbH$^+$-vertex. For $m_{\rm H^\pm}=150$ GeV the
b-quark mass of 4.80~GeV is reduced to $m_{\rm b}=3.33$ GeV, 
while HERWIG uses $m_{\rm b}=4.95$~GeV both in the kinematics and at the vertex.
Other relevant parameters are the default Parton Distribution
Functions (PDFs) and the coupling constants $\alpha$ and $\alpha_s$,
as well as the scales used for evaluating the PDFs and couplings,
which are not the same in the default setups of the two simulation packages.

Tests comparing the total cross sections from HERWIG and PYTHIA for {\em
identical choices} of all above parameters were performed and confirmed
that the two implementations of the hard scattering matrix elements
coincide numerically. In this study, however, we maintain the default
configurations of the two simulation packages. Hence, differences
between HERWIG and PYTHIA in the various distributions shown in the next
section may be taken as an indication of systematic errors in the event
simulation.

\begin{figure}[h]
\begin{minipage}{0.4\textwidth}
\includegraphics[width=1\textwidth]{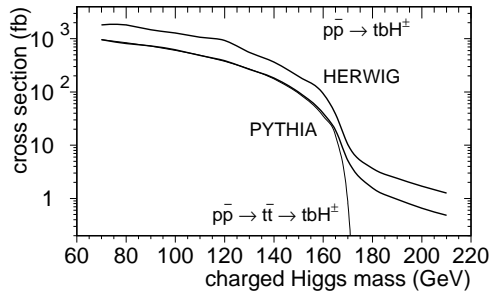}
\end{minipage} \hfill
\begin{minipage}{0.55\textwidth}
\vspace*{-1.4cm}
\caption{\label{fig:xsec}
         Charged Higgs boson production cross section at $\sqrt s=1960$~GeV
         for $\tan\beta=20$. 
         For the $\rm p\bar p\rightarrow tbH^\pm$ process (thick lines),
         the HERWIG expectation is larger by about a
         factor 2 compared to PYTHIA because of the different default setups
         as described in the text. 
The differences between the two PYTHIA curves for the 
$\rm p\bar p\rightarrow tbH^\pm$ and $\rm  p\bar p\rightarrow t\bar t \to  tbH^\pm$
processes instead result from 
top decay width effects and because of 
the additional diagrams which do not proceed via direct $\rm t\bar t$ production.}
\end{minipage}
\vspace*{-1.1cm}
\end{figure}

\subsection{Comparison of basic Selection Variables}
\vspace*{-0.1cm}

At the parton level, several distributions of variables related to the 
event topology are compared between HERWIG and PYTHIA simulations.
In addition, differences in the distributions between the $\rm p\bar p\rightarrow tbH^\pm$ process
and the $\rm  p\bar p\rightarrow t\bar t \to  tbH^\pm$ subprocess are demonstrated.
Each comparison is based on two samples of 10,000 generated events.
Effects of the different event fragmentation schemes in HERWIG and PYTHIA could influence the 
comparison and they are not considered here.
The detector simulation of the Tevatron experiments would reduce further the
sensitivity of these comparisons. 
Figure~\ref{fig:distributions} shows the transverse momentum $p_{\rm T}$ and
pseudorapidity $\eta$ of the following particles:
\smallskip
\begin{itemize}
\item[a), b)] The b-quark produced in association with the $\rm H^\pm$ 
               in the $\rm p\bar p \rightarrow tbH^\pm$ (dots) and
               $\rm p\bar p \rightarrow t\bar t \rightarrow tbH^\pm$ (solid line) processes
               in PYTHIA  
               for $m_{\rm H^\pm} = 165$~GeV.
\item[c), d)] The b-quark produced in association with the $\rm H^\pm$
               in the $\rm p\bar p \rightarrow tbH^\pm$ process
               in HERWIG (dots) and PYTHIA (solid line)
               for $m_{\rm H^\pm} = 150$~GeV.
\item[e), f)] The b-quark from the top quark decay ($\rm t\rightarrow bW^\mp$) 
               in the $\rm p\bar p \rightarrow tbH^\pm$ process
               in HERWIG (dots) and PYTHIA (solid line)
               for $m_{\rm H^\pm} = 150$~GeV.
\item[g), h)] The $\tau$ lepton from the $\rm H^\pm$ decay 
               in the $\rm p\bar p \rightarrow tbH^\pm$ process 
               in HERWIG (dots) and PYTHIA (solid line)
               for $m_{\rm H^\pm} = 150$~GeV.
\end{itemize}
\smallskip
The differences in the $p_{\rm T}$ and $\eta$ distributions
are clearly visible in Figs.~\ref{fig:distributions}~a)~and~b) 
between the processes 
$\rm p\bar p \rightarrow tbH^\pm$ and 
$\rm p\bar p \rightarrow t\bar t \rightarrow tbH^\pm$.
The HERWIG and PYTHIA simulations show good agreement in the kinematic 
distributions of Figs.~\ref{fig:distributions}~c)~to~h) 
for both b-quarks and the $\tau$ lepton.
The decay of the $\tau$ lepton is not considered here, 
but it should be noted that
spin correlations must be taken into account in the study of the final state 
particles~\cite{Guchait:2001pi,spin95}.
\vspace*{-0.3cm}

%\clearpage
\begin{figure}[thp]
\vspace*{-0.1cm}
\begin{center}
\includegraphics[width=3.9cm]{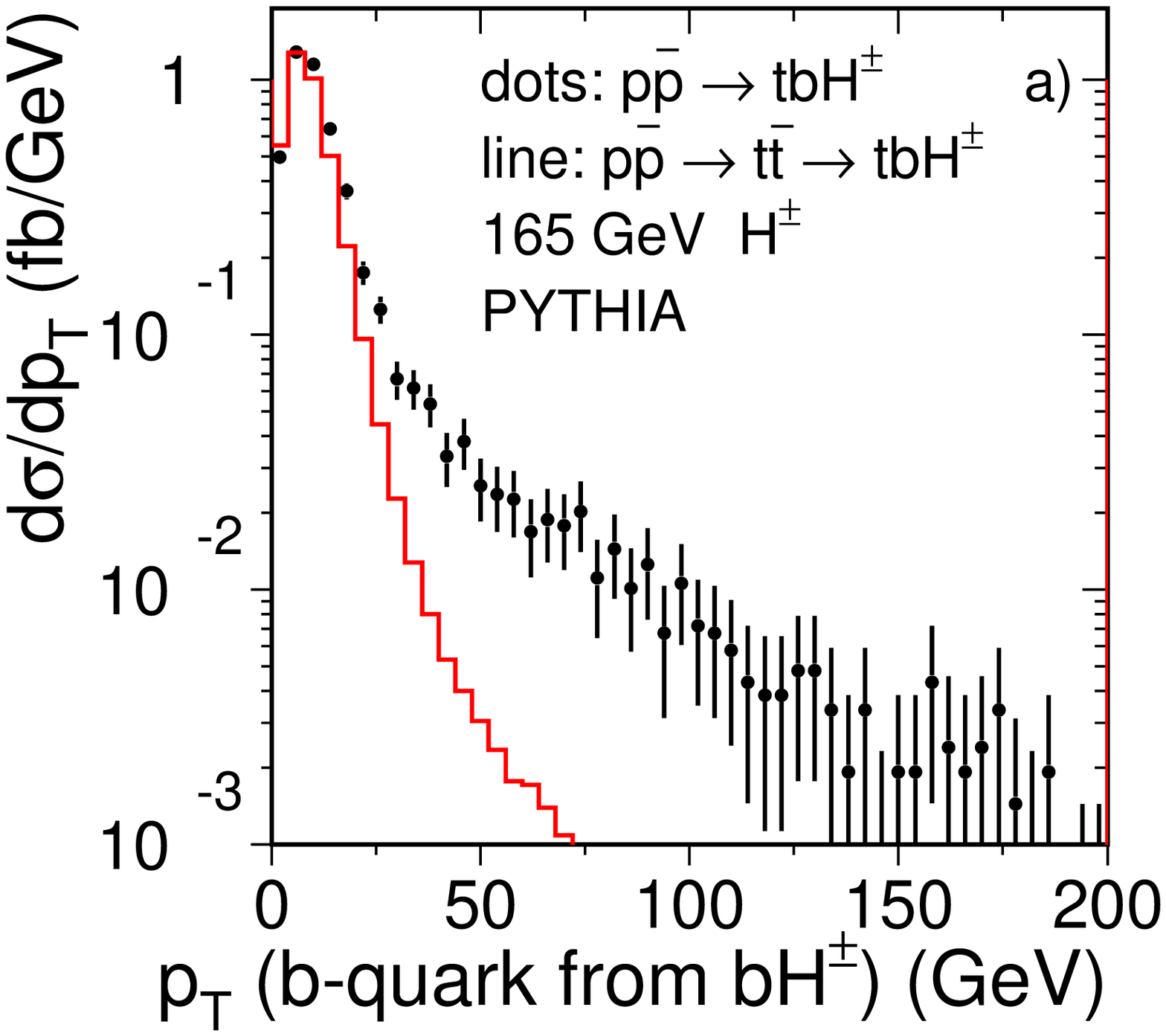}\hfill\includegraphics[width=3.9cm]{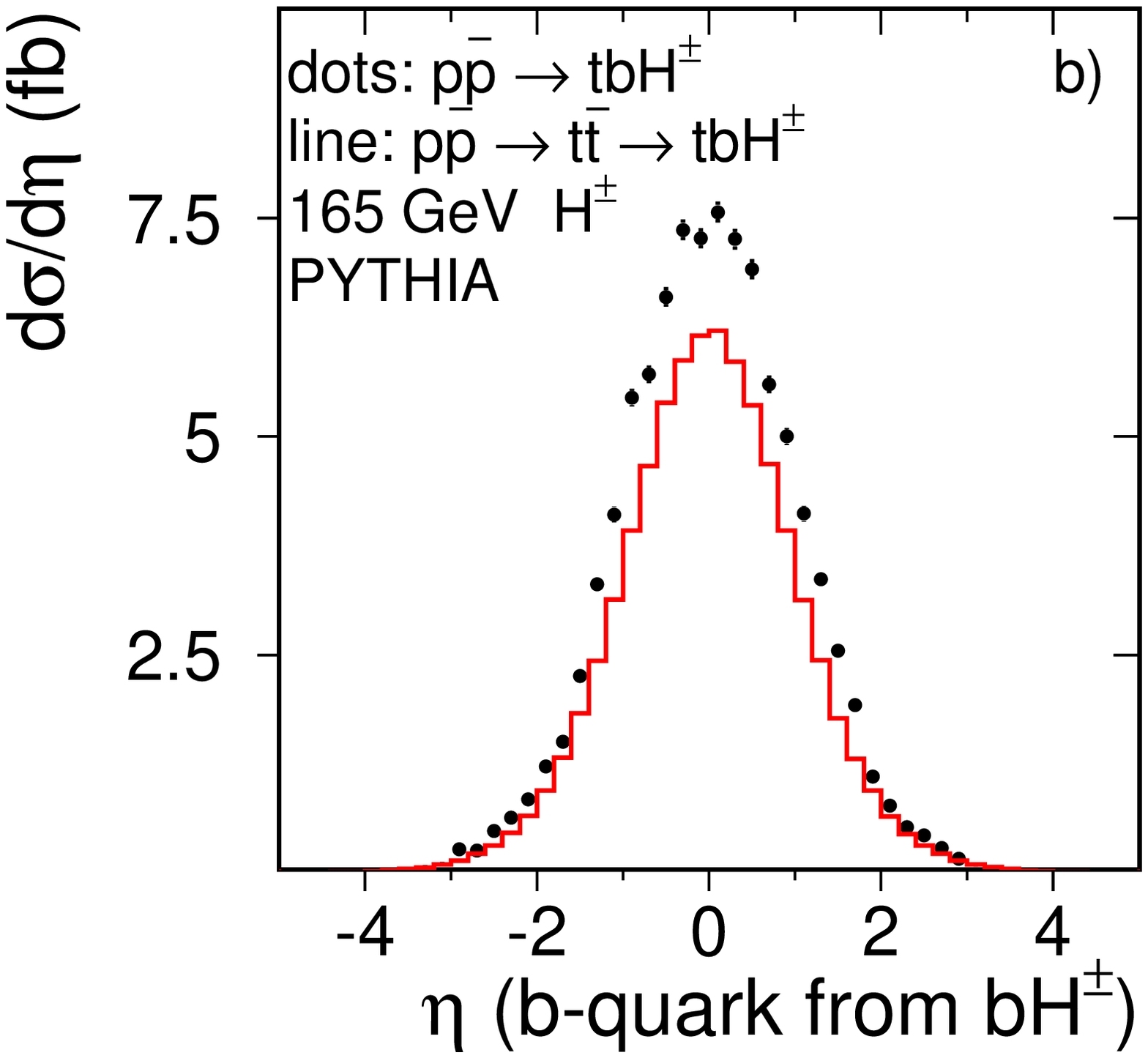}\hfill
\includegraphics[width=3.9cm]{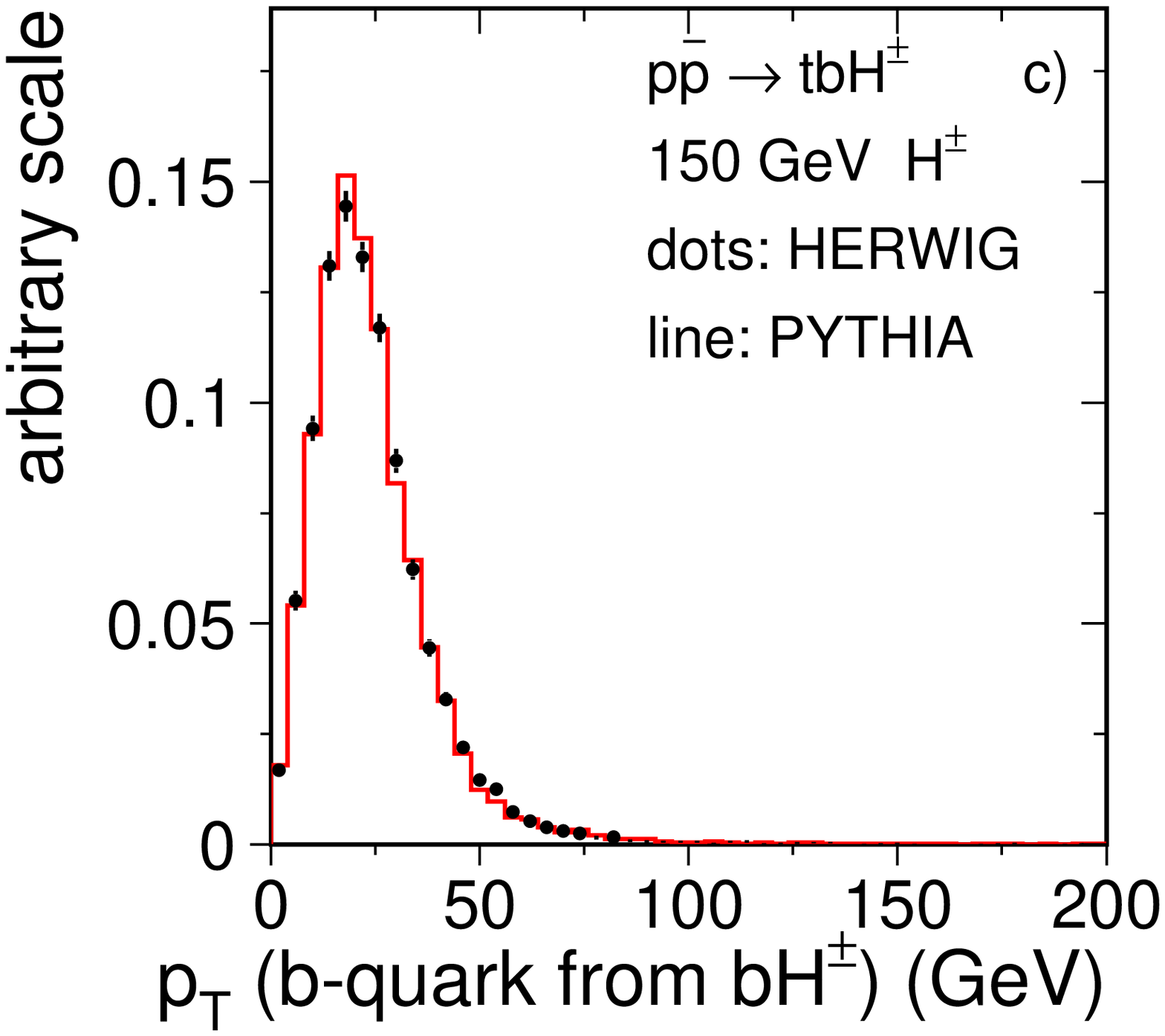}\hfill\includegraphics[width=3.9cm]{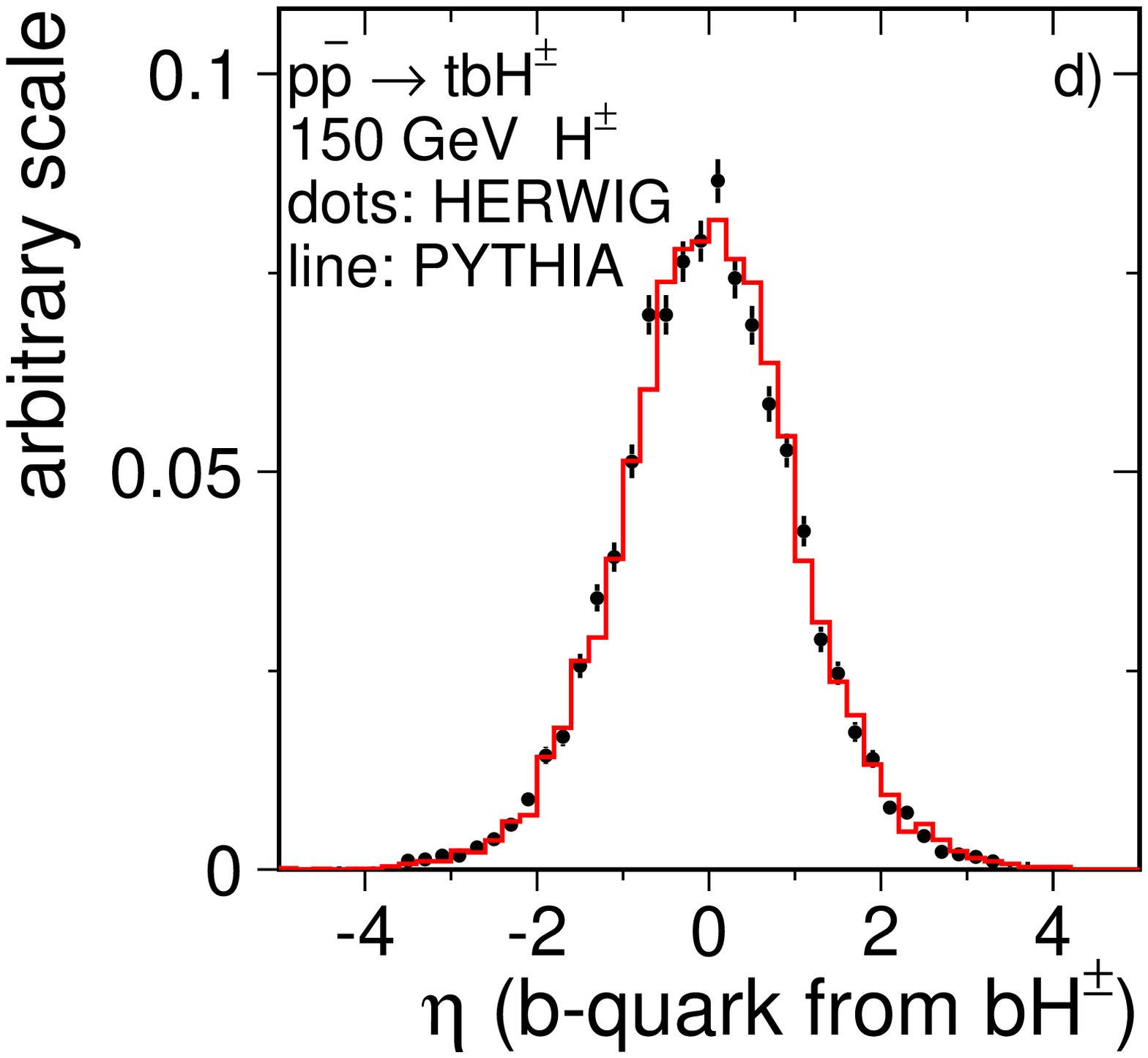}\\
\includegraphics[width=3.9cm]{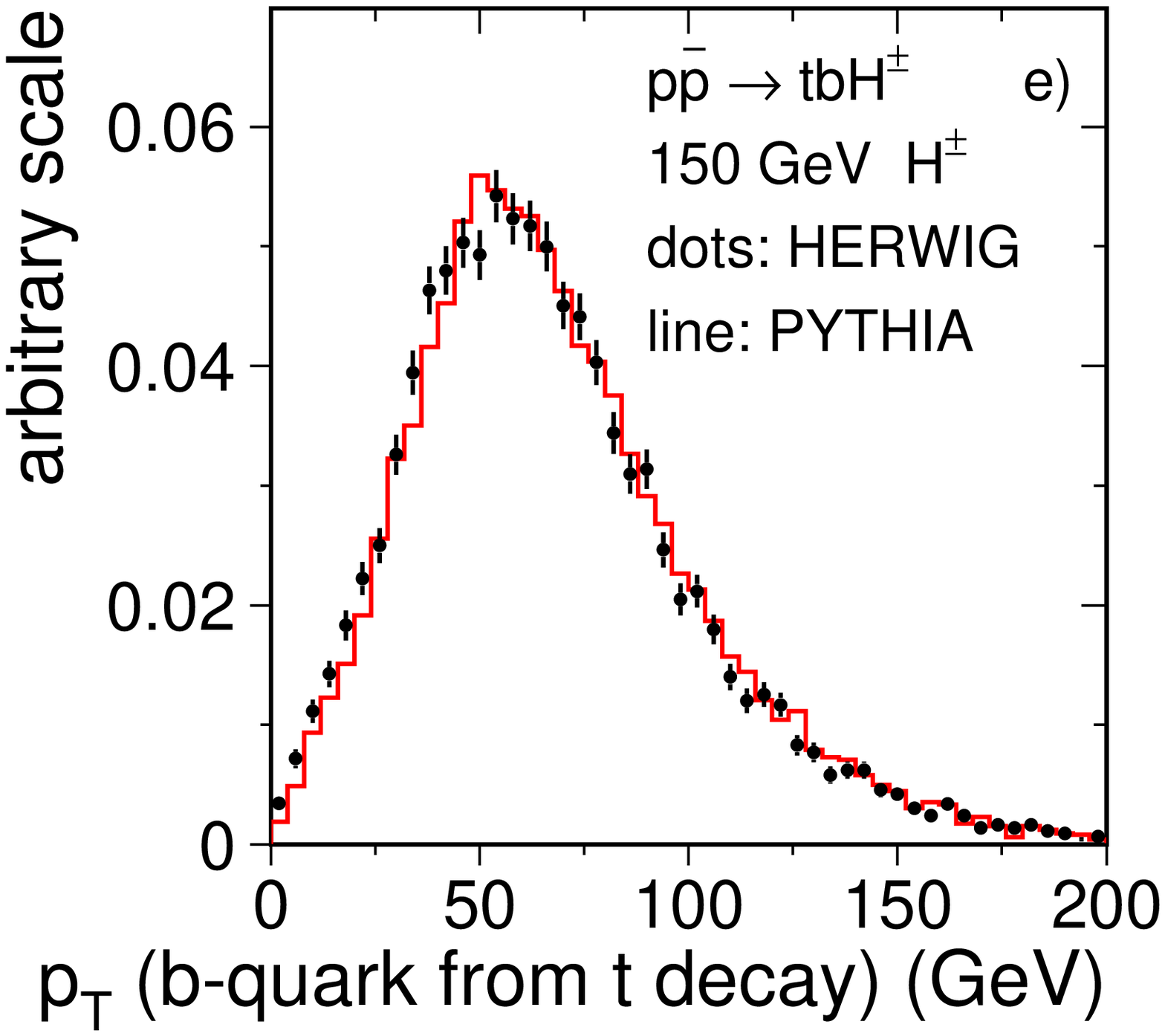}\hfill\includegraphics[width=3.9cm]{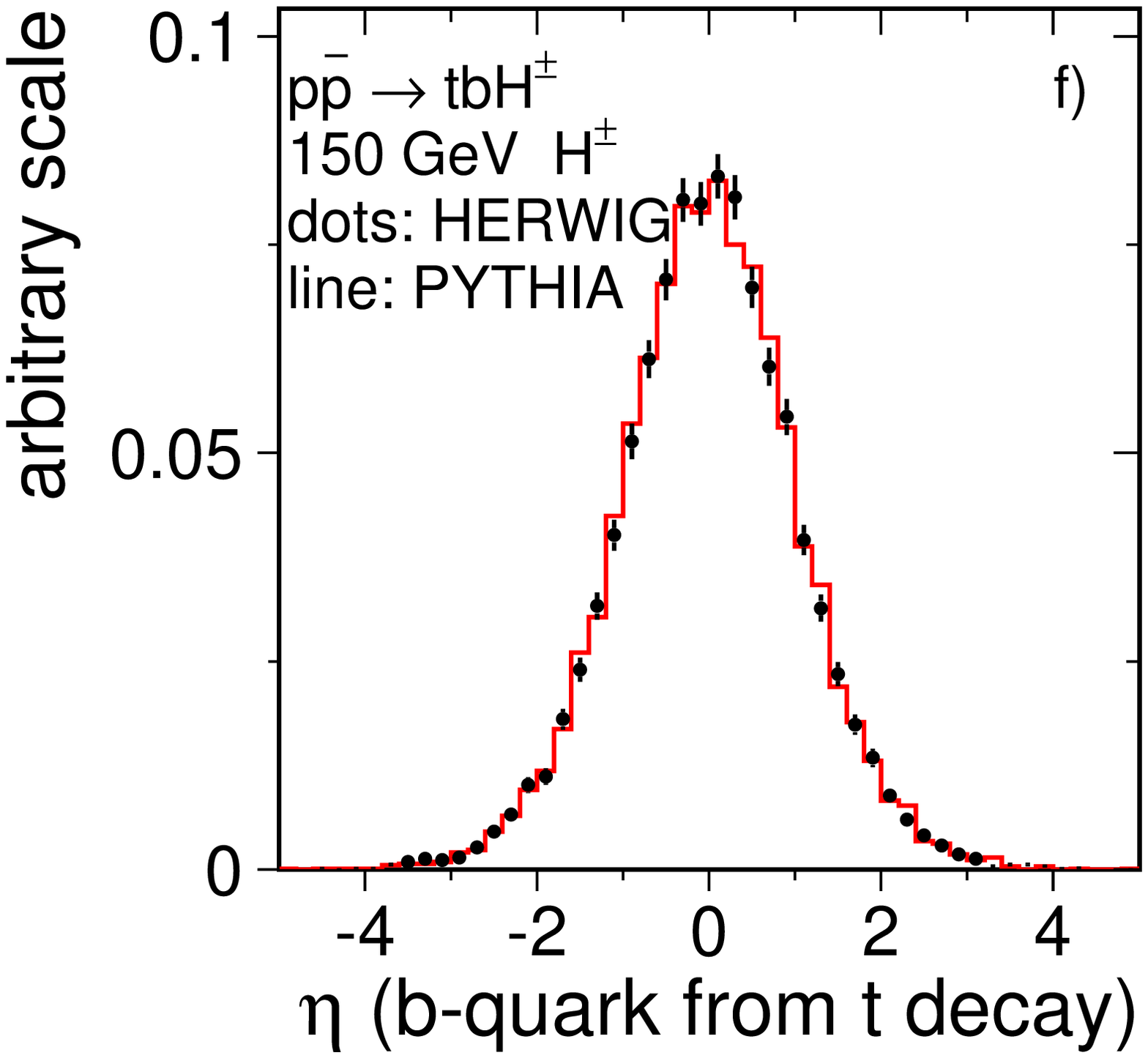}\hfill
\includegraphics[width=3.9cm]{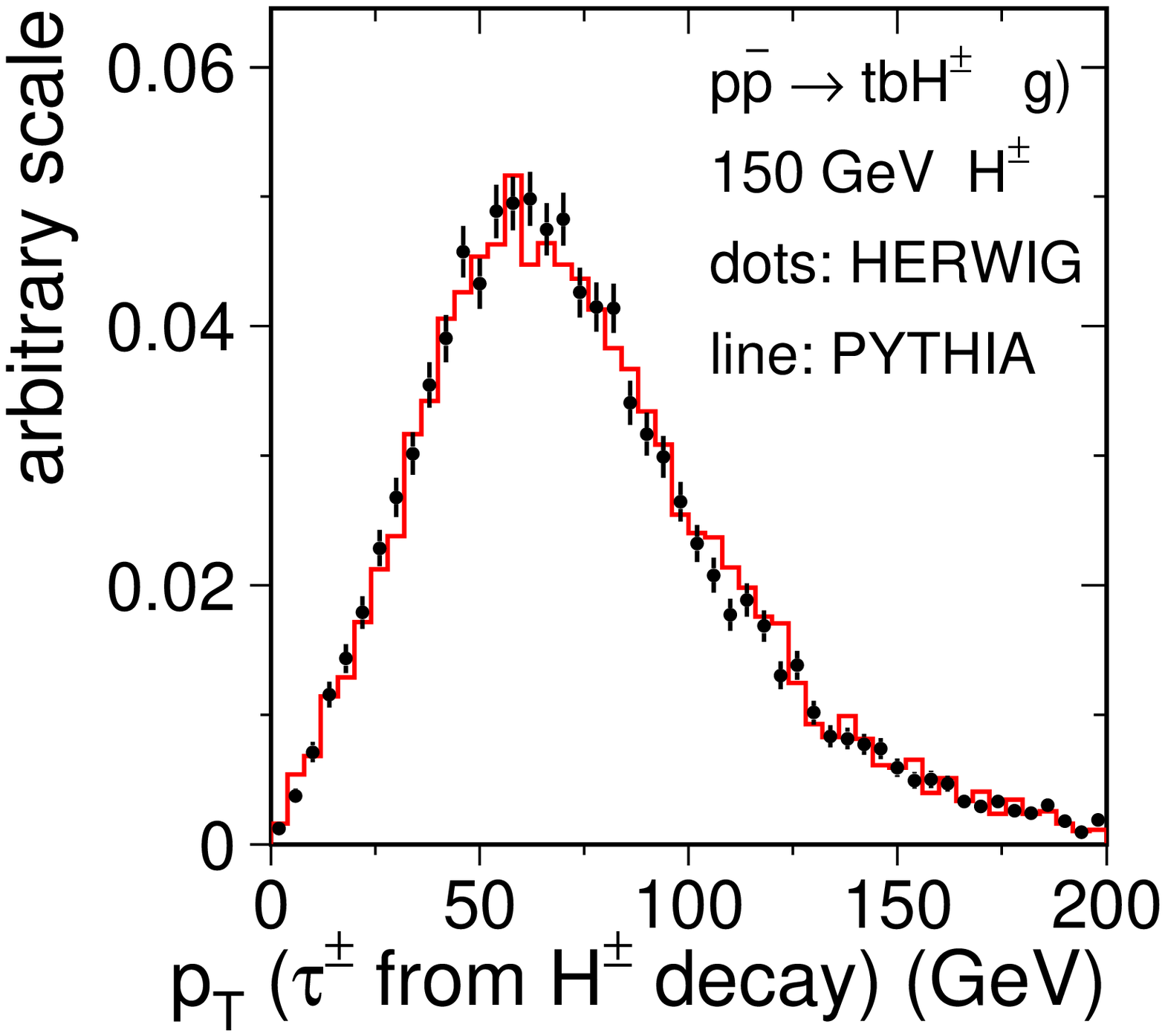}\hfill\includegraphics[width=3.9cm]{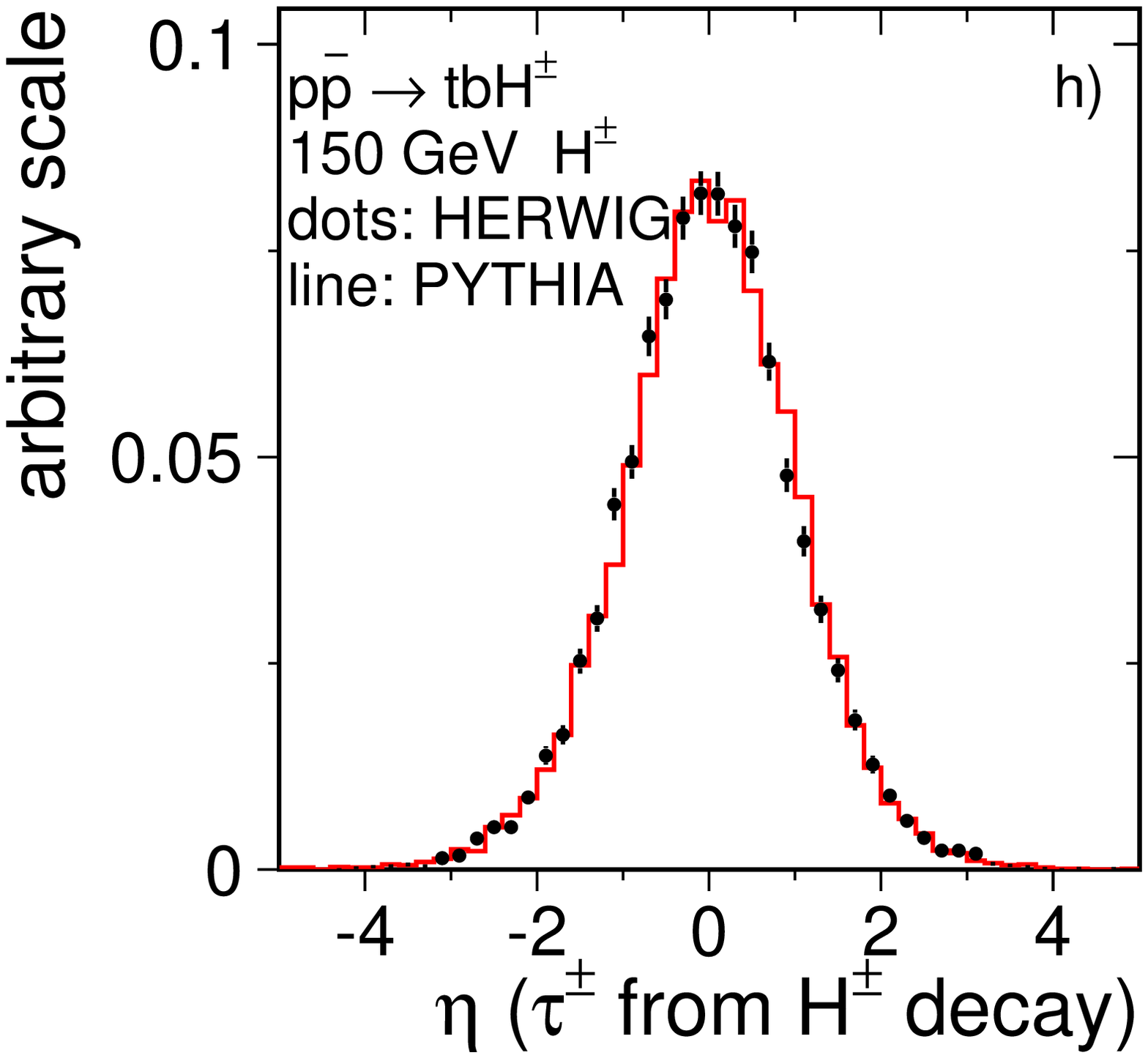}\\
\vspace*{-0.4cm}
\caption{\label{fig:distributions}
         Distributions of charged Higgs boson selection variables at the parton level
         for $\sqrt s=1960$~GeV and $\tan\beta=20$ for HERWIG and PYTHIA. 
         The variables are described in the text. 
%         In a) and b) the dominant effect is from the top off-shellness.
In a) and b) the differences are mainly from top off-shellness effects.
         In c) to h) each pair of curves is normalised to an equal area. 
         The error bars on the dots indicate the statistical\,uncertainty.
}
\end{center}
\end{figure}

%\vspace*{-0.3cm}
\subsection{Conclusions}
At Tevatron Run-II,
about 1000 $\rm p\bar p \to tbH^\pm$
events per 1~fb$^{-1}$ at $\sqrt s=1960$~GeV  
could be produced for $m_{\rm H^\pm} = 100$~GeV and $\tan\beta=20$, while about 100 events are expected for 
$m_{\rm H^\pm}=150$~GeV.
These expected event rates will strongly be reduced when selection criteria
are applied to separate signal and background events.
For the default choices of mass and coupling parameters
in HERWIG and PYTHIA we observe significant differences
in the simulated total cross sections.
We have also studied the shape of basic selection variable distributions and found 
good agreement between
the HERWIG and PYTHIA parton level predictions in the $\rm p\bar p \to tbH^\pm$ process.
In comparison with the $\rm p\bar p \rightarrow t\bar t \to tbH^\pm$ subprocess, 
which was used in previous HERWIG and PYTHIA versions, for $m_{\rm H^\pm} > 160$~GeV the 
simulation of the full process results in significantly different distributions of $\rm tbH^\pm$ 
selection variables, mainly in the $p_{\rm T}$ distribution of the b-quark produced in association 
with the $\rm H^\pm$.

}

%% file: zeppenfeld.tex
{

\newcommand{\sla}[1]{/\!\!\!#1}
\def\lsim{\mathrel{\raise.3ex\hbox{$<$\kern-.75em\lower1ex\hbox{$\sim$}}}}
\def\gsim{\mathrel{\raise.3ex\hbox{$>$\kern-.75em\lower1ex\hbox{$\sim$}}}}

\noindent
{\Large \bf C. Extracting Higgs boson couplings from LHC data} \\[0.5cm]
{\it M.\,D\"uhrssen, S.\,Heinemeyer, H.\,Logan, D.\,Rainwater,
G.\,Weiglein and D.\,Zeppenfeld}

\begin{abstract}
  We show how LHC Higgs boson production and decay data can be used to
  extract gauge and fermion couplings of Higgs bosons. Starting with 
  a general multi-Higgs doublet model, we show how successive theoretical
  assumptions overcome incomplete input data.  We also include 
  specific supersymmetric scenarios as a subset of the analysis.
\end{abstract}

%%%%%%%%%%%%%%%%%%%%%%%%%%%%%%%%%%%%%%%%%%%%%%%%%%%%%%%%%%%%%%%%%%%%%%%%%%%%%%%
%%%%%%%%%%%%%%%%%%%%%%%%%%%%%%%%%%%%%%%%%%%%%%%%%%%%%%%%%%%%%%%%%%%%%%%%%%%%%%%

\section{Introduction}

LHC experiments have the capability to observe the Higgs boson in a variety 
of channels, in particular if its mass lies in the intermediate mass region, 
114~GeV$<m_H \lsim 200$~GeV, as suggested 
by direct searches~\cite{Barate:2003sz}
and electroweak precision data~\cite{Grunewald:2003ij}. 
Once the Higgs boson is discovered, and its mass measured, one will want to
gain as much information as possible on Higgs boson couplings to both 
gauge bosons and fermions. These measurements will provide crucial tests
of the mass generation mechanism realized in nature.

The various Higgs couplings determine Higgs production cross sections and
decay branching fractions. By measuring the rates of multiple channels, 
various combinations of couplings can be determined. A principal problem 
is that there is no technique analogous to the measurement of the missing 
mass spectrum at a linear collider~\cite{Garcia-Abia:1999kv} 
which would allow for a direct 
determination of the total Higgs production cross section. In addition,
some Higgs decay modes cannot directly be observed at the LHC.
For example, $H\to gg$ or decays into light quarks
will remain hidden below overwhelming QCD dijet backgrounds. This implies that
absolute measurements of (partial) decay widths are only possible with 
additional theoretical assumptions.

One possible strategy was outlined at 
Les Houches in 1999~\cite{Djouadi:2000gu,Zeppenfeld:2000td}.
Assuming the absence of unexpected decay channels and a SM ratio of the
$H\to b\bar b$ and $H\to\tau\tau$ partial widths, absolute
measurements of $\Gamma(H\to WW/ZZ)$, $\Gamma(H\to\tau\tau)$, 
$\Gamma(H\to\gamma\gamma)$, $\Gamma(H\to gg)$ and of the top quark Yukawa 
coupling squared, $Y_t^2$, are possible, with errors in the 10--30\% range.

Here we revisit the information which can be extracted at the LHC from 
rate measurements of an intermediate mass Higgs boson.
We consider the expected accuracies at various stages of the LHC program:
after 30~fb$^{-1}$ of low luminosity running 
(at $10^{33}\,{\rm cm}^{-2}{\rm sec}^{-1}$), 300~fb$^{-1}$ of high luminosity 
running  (at $10^{34}\,{\rm cm}^{-2}{\rm sec}^{-1}$), and a mixed scenario
where the vector boson fusion channels are assumed to suffer substantially
from pile-up problems under high luminosity running conditions (making 
forward jet tagging and central jet veto fairly inefficient).

A rather model independent analysis, where only ratios of couplings  
(or partial widths) can be extracted, has been performed in 
Ref.~\cite{ATL-PHYS-2003-030}. Here we consider general 
multi-Higgs-doublet models (with or
without additional Higgs singlets), 
in which the $HWW$ and $HZZ$ couplings are bounded
from above by their SM values, i.e., we impose theoretically motivated 
constraints on these two couplings. These constraints are valid, 
in particular, for the Minimal Supersymmetric Standard Model (MSSM)
and will sharpen the 
implications of LHC data for Higgs couplings very significantly.

An alternative approach is a fit of observed rates in the Higgs sector
to specific models. Here we consider specific MSSM scenarios and use
the $m_h^{\rm max}$ scenario of Ref.~\cite{LHbenchmark} as an example.
The significance of deviations of the measured rates from SM
predictions provide a measure 
of the sensitivity of LHC measurements in the Higgs sector. We discuss 
this approach in Section~\ref{sec:mssm_specific}

%%%%%%%%%%%%%%%%%%%%%%%%%%%%%%%%%%%%%%%%%%%%%%%%%%%%%%%%%%%%%%%%%%%%%%%%%%%%%%%
%%%%%%%%%%%%%%%%%%%%%%%%%%%%%%%%%%%%%%%%%%%%%%%%%%%%%%%%%%%%%%%%%%%%%%%%%%%%%%%

\section{Summary of Higgs boson channels}
\label{sec:channels}

In order to
determine the properties of a physical state such as a Higgs boson,
one needs at least as many separate measurements as properties to be
measured, although two or more measurements can be made from the same
channel if different information is used, e.g., total rate and an
angular distribution.  Fortunately, the LHC will provide us with many
different Higgs observation channels.  In the SM there are four relevant
production modes: gluon fusion (GF; loop-mediated, dominated by the
top quark), also known as ``inclusive'' production; weak boson fusion
(WBF), which has an additional pair of hard and far
forward/backward jets in the final state; top-quark associated
production ($t\bar{t}H$); and weak boson associated production
($WH,ZH$), where the weak boson is identified by its leptonic 
decay.~\footnote{We do not consider diffractive Higgs production 
since its rate is in general small and also quite uncertain, which
limits the usefulness of this channel for Higgs coupling
determinations.}

Although a Higgs is expected to couple to all SM particles, not all
its decays to these particles would be observable.  Very rare decays
(e.g., to electrons) would have no observable rate, and other modes are
unidentifiable QCD final states (gluons or quarks lighter than
bottom).  In general, however, the LHC will be able to observe Higgs
decays to photons, weak bosons, tau leptons and $b$ quarks, in the
range of Higgs masses where the branching ratio (BR) in question is
not too small.

For a Higgs in the intermediate mass range,
the total width, $\Gamma$, 
is expected to be small enough to use the narrow-width
approximation in extracting couplings.  The rate of any
channel (with the $H$ decaying to final state particles $dd$) is, to
good approximation, given by  
\begin{equation}
\sigma(H) {\rm BR}(H\to dd) = {\sigma(H)^{\rm SM}\over \Gamma_p^{\rm SM}}
\cdot {\Gamma_p\Gamma_d \over \Gamma}\;,
\end{equation}
where $\Gamma_p$ is the Higgs partial width involving the production 
couplings and where the Higgs branching ratio for the decay is written as
${\rm BR}(H\to dd)=\Gamma_d/\Gamma$. Even with cuts, the observed rate
directly determines the product $\Gamma_p\Gamma_d/\Gamma$ 
(normalized to the calculable SM value of this product).
The LHC will have access to (or provide upper limits on) combinations of
$\Gamma_g,\Gamma_W,\Gamma_Z, \Gamma_\gamma,\Gamma_\tau,\Gamma_b$ and
the square of the top Yukawa coupling, 
$Y_t$.~\footnote{We do not write this as a
  partial width, $\Gamma_t$, because, for a light Higgs, the decay $H\to
  t\bar{t}$ is kinematically forbidden.}

Since experimental analyses are driven by the final state observed, we
classify Higgs channels by decay rather than production mode, and then
discuss the different production characteristics as variants of the
final state.  However, some initial comments on production modes are
in order.  First, experimental studies mostly do not yet include the
very large (N)NLO enhancements known for 
$gg\to H$~\cite{Harlander:2002wh,Anastasiou:2002yz,Ravindran:2003um}.  
Even if background corrections are as
large as for the signal, which they typically are not, the statistical
significance of the GF channels will be greater than estimated by the
current studies.  Second, experimental studies do not consider WBF
channels above 30~fb$^{-1}$ of integrated luminosity, because the
efficiency to tag forward jets at high-luminosity LHC running is not
yet fully understood.  This is a very conservative assumption,
which we comment on again later.

The literature on Higgs channels at LHC is extensive.  We
cite only those analyses which we use in our fits and accuracy estimates
for coupling extractions.  Mostly, these are recent experimental analyses
which contain references to the earlier phenomenological proposals. 
In the discussion below, statements about Higgs rates typically refer to
the SM-like case. Substantially suppressed branching ratios are possible
beyond the SM and may change a measurement into an upper bound.

\subsection{$H\to Z^{(*)}Z^{(*)}\to 4\ell$}

Leptons are the objects most easily identified in the final state, so
this decay is regarded as ``golden'' due to its extreme cleanliness
and very low background.  It is a rare decay due to the subdominance
of $H\to ZZ$ relative to $H\to W^+W^-$, and because of the very small BR of
$Z\to\ell^+\ell^-$.  Fortunately, due to the possible decay to off-shell
$Z$~bosons, a SM 
Higgs has non-negligible BR to $4\ell$ even for $M_H < 2M_Z$, down to
approximately 120~GeV.~\footnote{We note that for such low masses,
  doubly off-shell effects must be taken into account.}  
Due to the low event rate, current studies concentrate on inclusive 
measurements which are dominated by GF. They provide information
mainly on the product $\Gamma_g\Gamma_Z$.

The most advanced analysis for this channel~\cite{Cranmer:2003kf} was
made recently by ATLAS. (For an older CMS study,
see~\cite{Puljak.thesis}).  Its principal improvement over previous 
studies is the full use of NLO results (the only study so far to do this)
for both the dominant GF signal and its major
backgrounds. Further improvements can be expected in the inclusion of
off-shell contributions to the $gg\to ZZ^{(*)}$ background, for which 
ATLAS used an approximate K-factor.

%This is however not yet the final word, as the
%background off-shell contribution from initial gluons, $gg\to
%ZZ^{(*)}$, is not yet fully included; the ATLAS study used an
%approximate K-factor for this contribution.  A similar but older CMS
%study~\cite{Puljak.thesis} also used some NLO results.

By isolating the WBF contribution one obtains some independent 
information on the product $\Gamma_{W,Z}\Gamma_Z$, 
in particular if high-luminosity running 
can be exploited for this channel. We use the rates of 
Ref.~\cite{ATL-PHYS-2003-030} for our fits.

\subsection{$H\to\gamma\gamma$}

Photons are also readily identifiable, but are more difficult than leptons to
measure because of a large, non-trivial background from jets faking photons.
Higgs photonic decay is loop-induced and
therefore rare, even more so because of destructive interference
between the top-quark and $W$ loops.  This is in some sense
advantageous, because this decay mode is then sensitive to variations
in the weak gauge and top Yukawa couplings and additional particles in
the loop.  This decay is visible in
the SM only for the lower Higgs mass range, 110~GeV$<M_H<150$~GeV.

Despite the difficulties of identifying photons, which are not yet
fully understood for the LHC, especially for high-luminosity running,
Higgs decays to photons should be observable in both
GF~\cite{unknown:1999fr,:1997kj,Abdullin:1998er} and
WBF~\cite{ATL-PHYS-2003-036,CMS-NOTE-2001/022}, unless 
${\rm BR}(H\to\gamma\gamma)$ is substantially smaller than in the SM.  
These channels measure the
products $\Gamma_g\Gamma_\gamma$ and $\Gamma_{W,Z}\Gamma_\gamma$. The
$H\to\gamma\gamma$ signals in $t\bar{t}H,WH$ and $ZH$
production~\cite{unknown:1999fr,Dubinin:1997rn} are very weak, due to lack of
events even at high-luminosity running, but could be used as
supplemental channels, and would be especially useful if LHC observes
a non-SM Higgs.

\subsection{$H\to W^{+(*)}W^{-(*)}\to \ell^+\ell^- +\sla{p}_T$}

This decay can be observed in GF~\cite{unknown:1999fr,Dittmar:1997ss,Green:2000um}
and WBF~\cite{SN-ATLAS-2003-024,CMS-NOTE-2001/016} using
$W^+W^-\to\ell^+\ell^- +\sla{p}_T$ final states, as well as in
$t\bar{t}H$ associated production using combinations of multilepton
final states~\cite{ATL-PHYS-2002-019}.  The first two modes extract
the products $\Gamma_g\Gamma_W$ and $\Gamma^2_W$ and are extremely
powerful statistically, while the $t\bar{t}H$ mode can extract the top
Yukawa coupling with high luminosity and once $\Gamma_W$ is known.
All these channels are accessible over a wide range of Higgs masses,
approximately 120~GeV$<M_H<200$~GeV. 
%, although the GF mode has been formally studied only for $M_H>150$~GeV. 
An additional
study~\cite{ATL-PHYS-2000-013} for the $WH,H\to WW$ channel for
$M_H>150$~GeV found only a very weak signal, less than $5\sigma$ even
for 300~fb$^{-1}$ of data.

The GF mode should improve after NLO effects are included, although
the backgrounds considered did not include off-shell $gg\to WW^*$.
Also, the single-top background was conservatively overestimated.  
A reanalysis of this channel with updated simulation tools would be useful.

\subsection{$H\to\tau^+\tau^-$}

Observing Higgs decays to taus is not possible in GF because of serious 
background problems and because the invariant mass of a tau pair 
can be reconstructed only when they do not decay back-to-back, which
leaves only GF events with sizable Higgs transverse momentum.
Observation of $H\to\tau^+\tau^-$ is possible in WBF,
however~\cite{SN-ATLAS-2003-024,Azuelos:2001yw}, for Higgs masses
below about 150~GeV.  As the average Higgs $p_T$ in this production
mode is ${\cal O}(100)$~GeV, the taus are only rarely produced
back-to-back.  This is a relatively rare decay mode, since
BR($H\to\tau\tau$) is typically $5-10\%$ in this mass region and the
taus decay further. At least one tau must decay leptonically, giving
another small BR.  Fortunately, the QCD background to taus is small,
due to excellent fake jet rejection.  While not a discovery channel,
this channel is statistically quite powerful with only moderate
luminosity, and thus becomes one of the more important decay modes in
a couplings analysis.  This channel measures the product
$\Gamma_{W,Z}\Gamma_\tau$.

\subsection{$H\to b\bar{b}$}

Associated Higgs-$b$ quark production has too small a cross section in
a SM-like Higgs sector to be observable, so the decay $H\to b\bar{b}$
is the only access to the $b$ Yukawa coupling.  Because this decay
mode dominates Higgs decays at low mass ($M_H<135$~GeV within the SM), 
an accurate 
measurement of the bottom Yukawa coupling is extremely important.
Unfortunately, due to the typically large QCD backgrounds for $b$
jets, it is very difficult to observe this decay.  The production
modes
$t\bar{t}H$~\cite{ATL-PHYS-2003-024,CMS-NOTE-2001/054,Green:2001gh}
and $WH$~\cite{unknown:1999fr,CMS-NOTE-2002/006} might allow very rough
measurements for such a light Higgs, but the statistical significances
are quite low and the background uncertainties quite large; they are
definitely high-luminosity measurements.

The $t\bar{t}H$ channel measures the product $Y_t^2\Gamma_b$, and so
would require a separate, precise measurement of $Y_t$ to isolate $\Gamma_b$.
For $WH$ production, the rate is proportional to $\Gamma_W\Gamma_b$.
But here the $Wb\bar{b}$ continuum background has hitherto been 
underestimated since the NLO QCD corrections are very large and 
positive~\cite{Campbell:2003hd}.  A veto on additional jets may help 
%(see Sec.~\ref{sec:Wbb.NLO}), 
but requires
another detector-level simulation; unfortunately, it would also increase 
the background uncertainty because additional jet activity has been 
calculated at LO only.

\subsection{Other channels}

The production and decay channels discussed above refer to a single Higgs
resonance, with decay signatures which also exist in the SM. The Higgs
sector may be much richer, of course. 
The MSSM with its two Higgs doublets predicts the existence 
of three neutral and one charged Higgs boson, and the LHC may be able to 
directly observe several of these resonances. 
Within SUSY models, additional 
decays, e.g., into very light super-partners, may be kinematically allowed. 
The additional observation
of super-partners or of heavier Higgs bosons will strongly focus the 
theoretical framework and restrict the parameter space of a Higgs 
couplings analysis~\cite{schumitalkPrag}.

At the present time, even enumerating the possibilities is an open-ended task.
For our present analysis we therefore ignore the information which would be
supplied by the observation of additional new particles. Instead we ask
the better defined question of how well LHC measurements of the above
decay modes of a single Higgs resonance can determine the various Higgs boson
couplings or partial widths.

%%%%%%%%%%%%%%%%%%%%%%%%%%%%%%%%%%%%%%%%%%%%%%%%%%%%%%%%%%%%%%%%%%%%%%%%%%%%%%%
%%%%%%%%%%%%%%%%%%%%%%%%%%%%%%%%%%%%%%%%%%%%%%%%%%%%%%%%%%%%%%%%%%%%%%%%%%%%%%%

\section{Model assumptions and fits}
\label{sec:nomodel}

In spite of the many decay channels discussed above, the LHC is faced with
the challenge that not all Higgs decay modes can be detected directly
(e.g., $H\to gg$ is deemed unobservable) or that some important decay rates,
in particular $H\to b\bar b$, will suffer from large experimental 
uncertainties. In a model-independent analysis, the limited information 
which will be available then leads to strong correlations in the 
measurement of different Higgs couplings. These correlations
mask the true precision of LHC measurements when the expected errors of 
particular observables like individual partial widths or branching
ratios are considered.

The parameter correlations can be overcome
by imposing theoretical constraints.  One possible approach was suggested
in Les Houches 1999~\cite{Djouadi:2000gu,Zeppenfeld:2000td}. Fixing the ratio 
$\Gamma_b/\Gamma_\tau$ to its SM value, the $H \to \tau\tau$ measurements 
can be used to pin down the poorly measured Higgs coupling to bottom quarks.
Here we follow a different approach.
We perform general fits to the Higgs couplings with a series of 
theoretical assumptions of increasing restrictiveness, starting 
with the constraint $\Gamma_V \leq \Gamma_V^{\rm SM}$ ($V=W,Z$) 
which is justified in any model
with an arbitrary number of Higgs doublets (with or without additional 
Higgs singlets), i.e., it is true for the MSSM in particular.

Even without this constraint, the mere observation of Higgs production
puts a lower bound on the production couplings and, thereby, on the total
Higgs width. The constraint $\Gamma_V \leq \Gamma_V^{\rm SM}$, combined with
a measurement of $\Gamma_V^2/\Gamma$ from observation of $H\to VV$ in WBF,
then puts an upper bound on the Higgs total width, $\Gamma$. It is this
interplay which provides powerful constraints on the remaining Higgs couplings.

\subsection{Fitting procedure}
\label{subsec:fitproc}

Our analysis of expected LHC accuracies closely follows the work of
D\"uhrssen~\cite{ATL-PHYS-2003-030}.  
First, a parameter space (${\bf x}$) is formed 
of Higgs couplings together with additional partial
widths to allow for undetected Higgs decays and additional contributions
to the loop-induced Higgs couplings to photon pairs or gluon pairs due to
non-SM particles running in the loops.  Assuming that the measured values
correspond to the SM expectations, a log likelihood function, $L({\bf x})$,
is formed which,
for a given integrated luminosity, is based on the expected Poisson errors 
of the channels listed in Sec.~\ref{sec:channels} and on estimated 
systematic errors~\cite{ATL-PHYS-2003-030}.
% for the luminosity, detector efficiencies, background
%cross sections and QCD and pdf uncertainties~\cite{ATL-PHYS-2003-030}. 
These errors include a 5\% luminosity error,
uncertainties on the reconstruction/identification of leptons (2\%),
photons (2\%), b-quarks (3\%) and forward tagging jets and veto jets (5\%),
error propagation for background determination from side-band analyses
(from 0.1\% for $H\to\gamma\gamma$ to 5\% for
$H\to WW$ and $H\to\tau\tau$) and theoretical and parametric
uncertainties on Higgs 
boson production (20\% ggH, 15\% ttH, 7\% WH/ZH, 4\% WBF) and decays
(1\%, as a future expectation).

As an alternative, in particular for the specific MSSM scenarios discussed
in Sec.~\ref{sec:mssm_specific}, a 
Gaussian approximation to the log likelihood function is used, i.e., a 
$\chi^2$ function is constructed from the same error assumptions that enter 
the log likelihood function. We have checked that the resulting accuracy
estimates for coupling measurements are consistent for the two approaches.

Relative to SM expectations, the variation of either $2L({\bf x})$ or 
$\chi^2({\bf x})$ is then computed on this parameter space, and the 
surface of variations by one unit is traced out.  
The $1\sigma$ uncertainties in each parameter are determined
by finding the maximum deviation of that parameter from its SM value 
that lies on the $\Delta \chi^2 = 1$ ($\Delta L = 1/2$) surface.  
The procedure is repeated
for each Higgs mass value in the range 110~GeV$ \leq m_H \leq 190$ GeV in 
steps of 10 GeV.

We perform the fits under three luminosity assumptions for the LHC:
\begin{enumerate}
\item %``low luminosity'': 
      30 fb$^{-1}$ at each of two experiments, denoted 2*30 fb$^{-1}$;
\item %``medium luminosity'':
      300 fb$^{-1}$ at each of two experiments, of which only 100 fb$^{-1}$
is usable for WBF channels at each experiment, denoted 2*300 + 2*100 fb$^{-1}$;
\item %``high luminosity'':
      300 fb$^{-1}$ at each of two experiments, with the full luminosity
usable for WBF channels, denoted 2*300 fb$^{-1}$.
\end{enumerate}
The second case allows for significant degradation of the WBF channels 
in a high luminosity environment while the third case serves to motivate
additional improvements in WBF studies at high luminosity.

\subsection{General multi-Higgs-doublet model fits}

\begin{figure}[thb]
\begin{center}
\resizebox{\textwidth}{!}{
\includegraphics{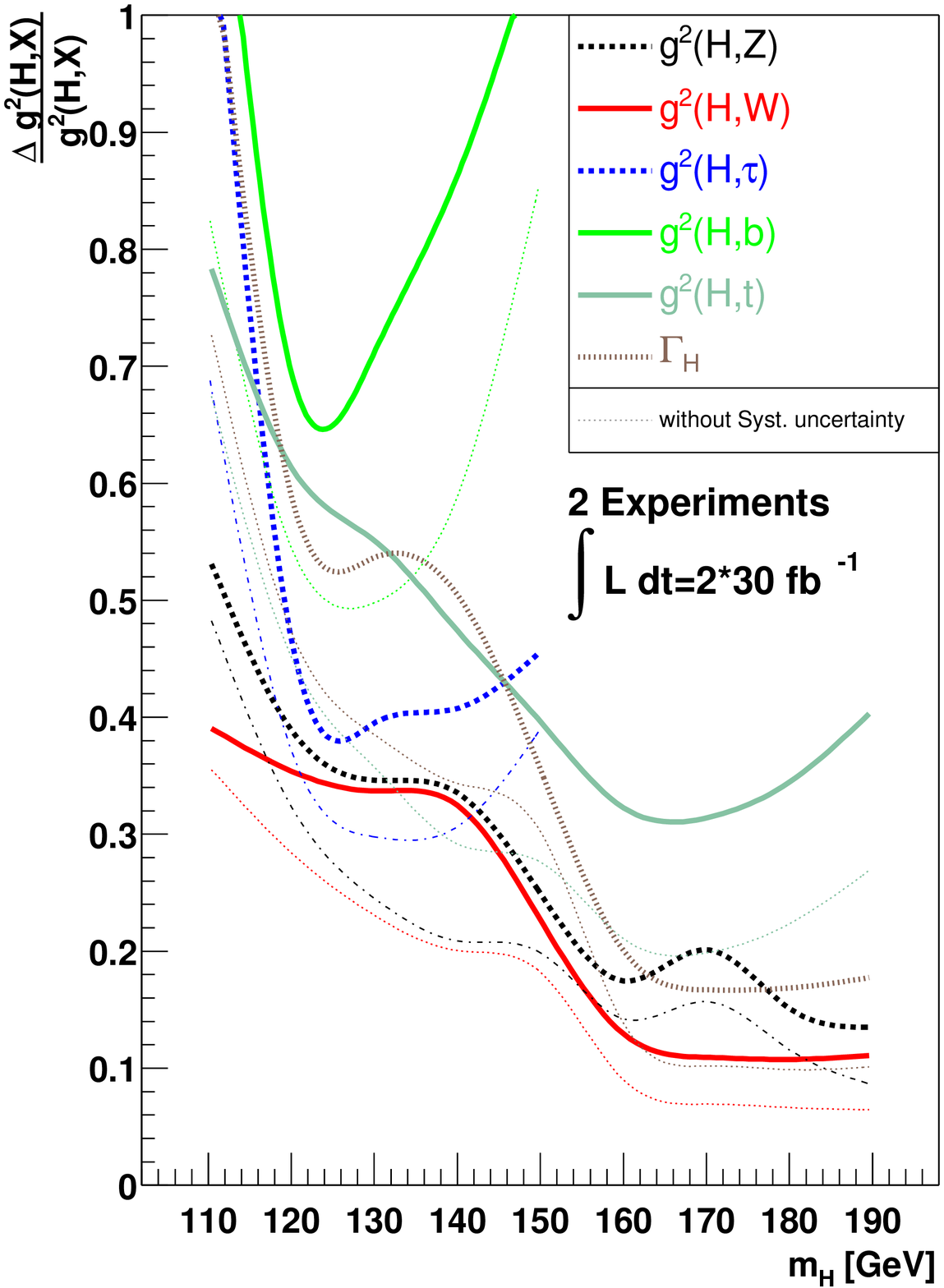}
\includegraphics{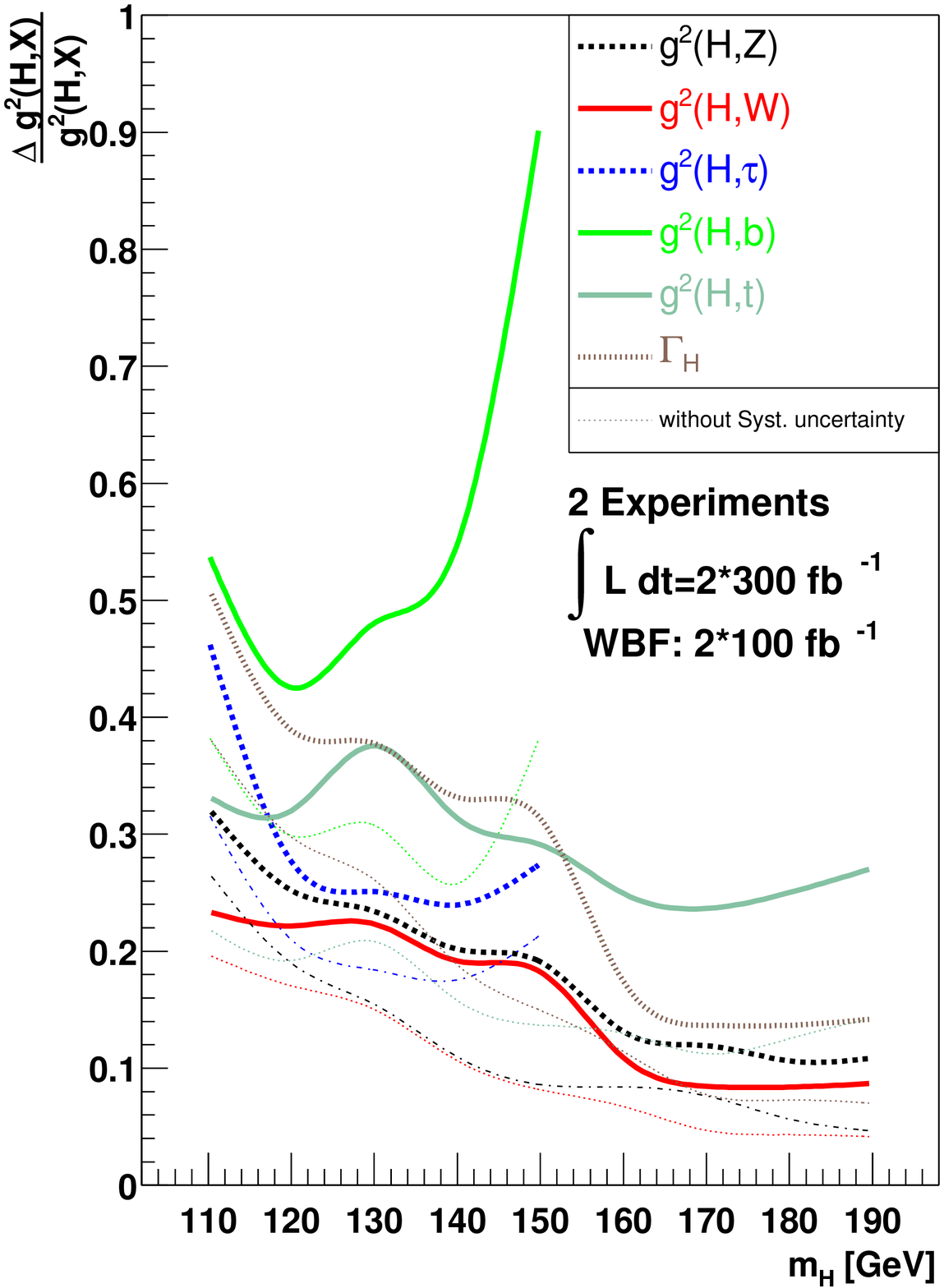}
}
\caption{Relative precisions of fitted Higgs couplings-squared as a function
of the Higgs mass assuming 30 fb$^{-1}$ at each of two experiments
(left) and 300 fb$^{-1}$ at each of two experiments for all channels 
except WBF, for which 100 fb$^{-1}$ is assumed (right).
Here we make the weak assumption that 
$g^2(H,V)<g^2(H,V,SM)+5\%$ ($V=W,Z$) but allow for new particles
in the loops for $H \to \gamma\gamma$ and $gg\to H$ and for unobservable 
decay modes. See text for details.
}
\label{fig:fit}
\end{center}
\end{figure}
	
We begin by fitting for the uncertainties in the Higgs couplings-squared 
in the most general scenario that we consider.
We assume only that $g^2(H,W)<1.05*g^2(H,W,SM)$ and 
$g^2(H,Z)<1.05*g^2(H,Z,SM)$.
Any model that contains only Higgs doublets and/or singlets will 
satisfy the relations $g^2(H,W) \leq g^2(H,W,SM)$ and 
$g^2(H,Z) \leq g^2(H,Z,SM)$.
The extra 5\% margin 
allows for theoretical uncertainties in the translation between
couplings-squared and partial widths and also for small admixtures of exotic
Higgs states, like SU(2) triplets.
We allow for the possibility of additional particles running in the
loops for $H \to \gamma\gamma$ and $gg\to H$, fitted by a positive
or negative new partial width to these contributions.
This new partial width for $H \to \gamma\gamma$ is most tightly
constrained for 120~GeV$ \lsim m_H \lsim 140$ GeV, being less than
$\pm (25-35)\%$ of $\Gamma_\gamma^{\rm SM}$ for 2*30 fb$^{-1}$ and 
$\pm (10-15)\%$ for 2*300 + 2*100 fb$^{-1}$.
The new partial width for $gg \to H$ is less well constrained, being less than
$\pm (30-90)\%$ of $\Gamma_g^{\rm SM}$ for 2*30 fb$^{-1}$ and 
$\pm (30-45)\%$ for 2*300 + 2*100 fb$^{-1}$ over the whole range of
Higgs masses.
Additional decays of the Higgs boson are fitted with a partial width
for undetected decays.
This undetected partial width can be constrained to be less than
$15-55\%$ of the total fitted Higgs width for 2*30 fb$^{-1}$ 
and $15-30\%$ for 2*300 + 2*100 fb$^{-1}$, at the $1\sigma$ level.  
This undetected partial
width is most tightly constrained for Higgs masses above 160 GeV.

The resulting parameter precisions are shown in Fig.~\ref{fig:fit} 
as a function of Higgs mass for the 2*30 fb$^{-1}$ and 
2*300 + 2*100 fb$^{-1}$ luminosity scenarios. For the latter case, typical 
accuracies range between 20 and 30\% for Higgs masses below 150~GeV. 
Above $W$-pair threshold the measurement of the then dominant $H\to WW,ZZ$
partial widths improves to the 10\% level. 
The case of 2*300 fb$^{-1}$ yields only small improvements
over the right-hand panel in Fig.~\ref{fig:fit},
%compared to the 2*300 + 2*100 fb$^{-1}$ luminosity scenario,
except in the case of $g^2(H,\tau)$ which shows a moderate improvement.
This can be understood because the $H \to \tau\tau$ decay
is measured only in WBF, and $g(H,\tau)$ does not have a large effect 
on the Higgs total width or loop-induced couplings.
%In the MSSM, $g^2(H,\tau)$ is quite slow to decouple (as is $g^2(b)$); 
%this can lead to a significant increase in the sensitivity to deviations
%from the SM between the two luminosity scenarios, as will be discussed
%in Sec.~\ref{sec:mssm_specific}.

\begin{figure}[htb]
\begin{center}
\resizebox{\textwidth}{!}{
\includegraphics{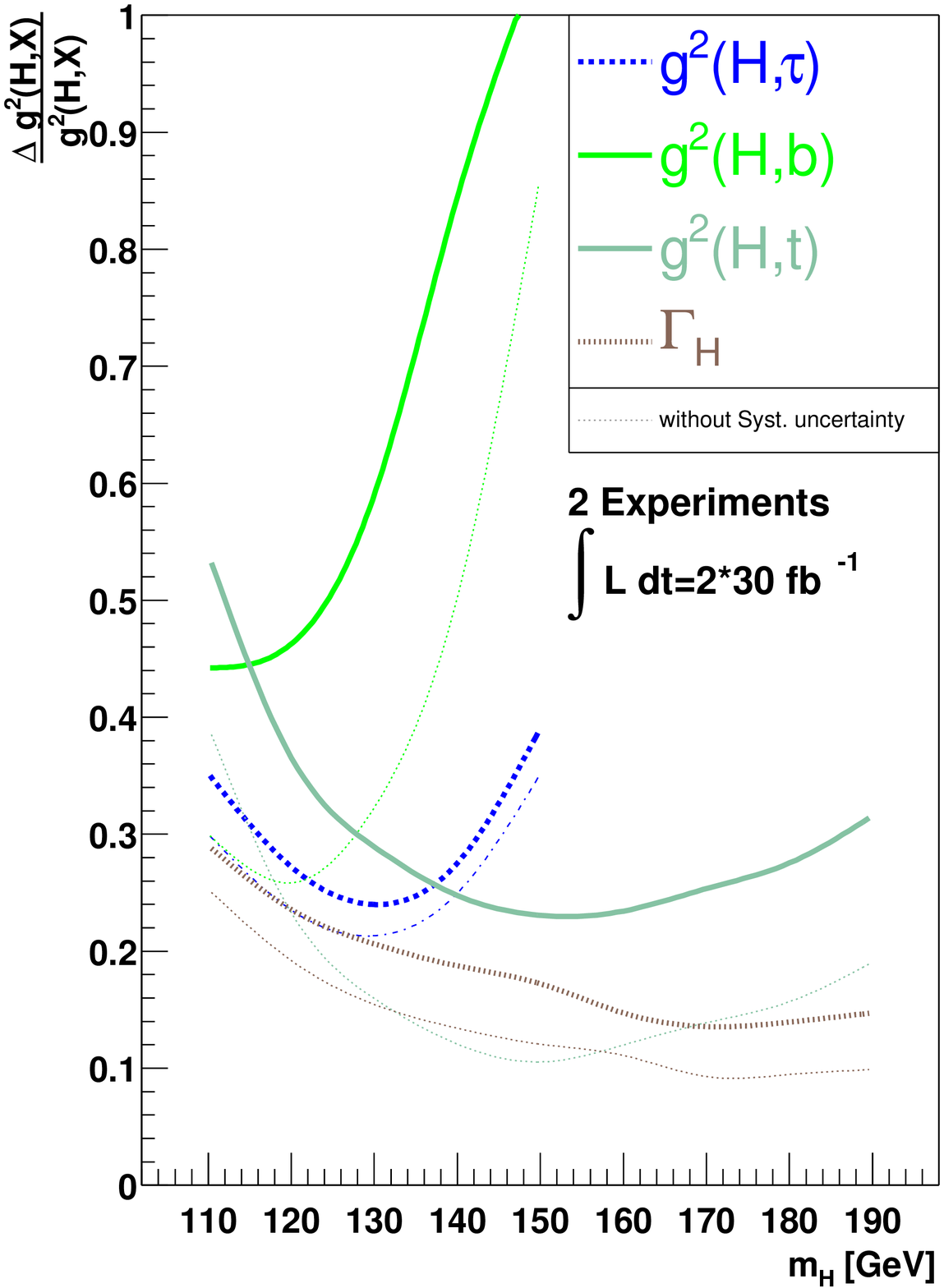}
\includegraphics{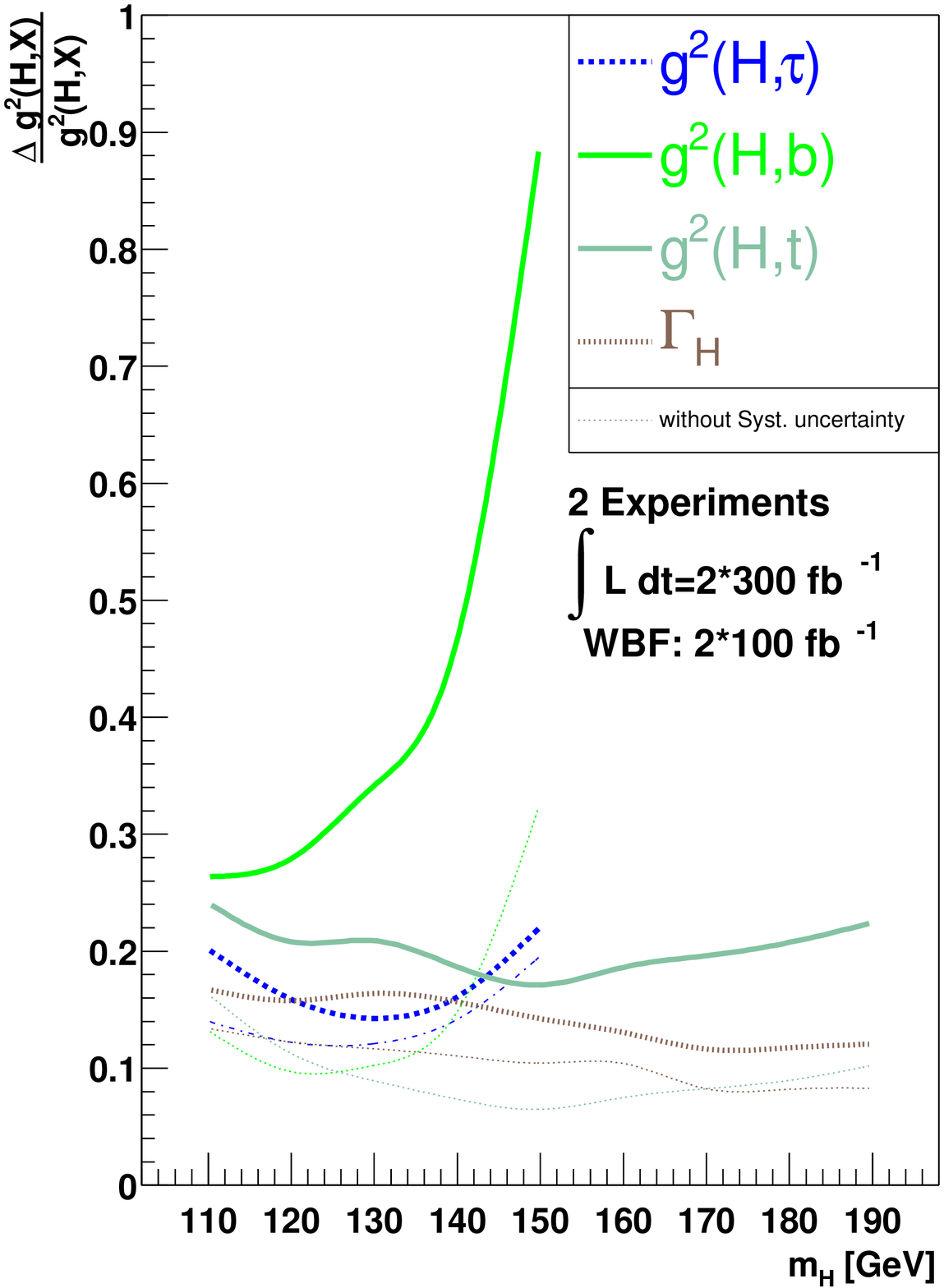}
}
\caption{As in Fig.~\ref{fig:fit}, but with more restrictive assumptions.
Here we assume that 
$g^2(H,W)=g^2(H,W,SM)\pm 5\%$ and
$g^2(H,W)/g^2(H,Z)=g^2(H,W,SM)/g^2(H,Z,SM)\pm 1\%$.
We also assume that no new particles run in the loops for
$H \to \gamma\gamma$ and $gg\to H$, so that these couplings are 
fixed in terms of the couplings of the SM particles in the loops.
As in Fig.~\ref{fig:fit}, 
additional decays of the Higgs boson are fitted with a partial width
for undetected decays (not shown).
}
\label{fig:SU2fit}
\end{center}
\end{figure}

%It is interesting to note that if the ``LEP Higgs'' at 116 GeV were
%realized~\cite{Barate:2003sz}, the accuracy in the low luminosity case 
%is rather bad. In the higher luminosity
%case, on the other hand, the region around 116 GeV is not worse than
%the rest of the Higgs mass range allowed in the MSSM.

The results shown in Fig.~\ref{fig:fit} reflect present understanding of
detector effects and systematic errors. One should note that improved 
selection and higher acceptance will decrease the statistical errors. 
At least as important is work on the reduction of systematic errors. In
Fig.~\ref{fig:fit}, the thin lines show expectations with vanishingly small
systematics: systematic errors contribute up to half the total error, 
especially at high luminosity.

\subsection{SU(2) constraints and SM loops}

The theoretical constraints used so far have been very moderate. 
If, in addition to the requirement that $g^2(H,W) < g^2(H,W,SM) + 5\%$ and 
$g^2(H,Z) < g^2(H,Z,SM) + 5\%$, we assume that no new non-SM particles run in 
the loops for $H \to \gamma\gamma$ and $gg\to H$ (which is
approximately fulfilled for the MSSM with a not too light spectrum),
the precision of the coupling measurements improves only slightly, 
with the only noticeable improvement for Higgs masses below 120 GeV.

Another small improvement is achieved by restricting the $W$ and
$Z$ couplings to their SM ratio.
Within the multi-Higgs-doublet models considered throughout, 
SU(2) symmetry relates these two couplings. It thus is natural to
forgo an independent measurement of their ratio and to rather assume that
%fix their ratio to the SM expectation, by assuming that 
\begin{equation}
\label{eq:su2}
g^2(H,W)/g^2(H,Z) = g^2(H,W,SM)/g^2(H,Z,SM) \pm 1\%\;.
\end{equation}
Within the MSSM, this coupling ratio is indeed very close to its SM value.
Over most of the MSSM parameter space even the individual $hVV$ 
couplings will be close to their SM values since decoupling sets in rapidly 
once the mass of the $CP$-odd Higgs boson becomes large, 
$m_A \gsim 200$~GeV. This motivates a fit where in addition 
to Eq.~\ref{eq:su2} we assume 
\begin{equation}
g^2(H,W) = g^2(H,W,SM) \pm 5\%\;.
\end{equation}
We again assume that no new non-SM particles run in 
the loops for $H \to \gamma\gamma$ and $gg\to H$.
However, additional decays of the Higgs boson are fitted with a partial width
for undetected decays.
The constraints on this undetected partial width are essentially
the same as in our least constrained fit.
The resulting parameter precisions are shown in Fig.~\ref{fig:SU2fit}
and reach 10--20\% over the entire intermediate Higgs mass range for the
2*300 + 2*100 fb$^{-1}$ luminosity scenario.  

Loosening assumptions slightly, by allowing non-SM particles to contribute
to the $H \to \gamma\gamma$ partial width, has a noticeable effect on the 
coupling determination only for $m_H \lsim 120$ GeV.  For example, for the 
2*300 + 2*100 fb$^{-1}$ luminosity scenario, the precision on 
$g^2(H,\tau)$, $g^2(H,b)$ and the Higgs total width at $m_H = 110$ GeV jump
to about 40\%.

%%%%%%%%%%%%%%%%%%%%%%%%%%%%%%%%%%%%%%%%%%%%%%%%%%%%%%%%%%%%%%%%%%%%%%%%%%%%%%%
%%%%%%%%%%%%%%%%%%%%%%%%%%%%%%%%%%%%%%%%%%%%%%%%%%%%%%%%%%%%%%%%%%%%%%%%%%%%%%%

\section{Higgs couplings within the MSSM}
\label{sec:mssm_specific}

A plausible scenario is that one or several Higgs bosons will be discovered at 
the LHC together with evidence for supersymmetry (SUSY) at the TeV scale.
Once SUSY has been confirmed, 
we are led to analyzing the Higgs sector in terms of a two Higgs doublet
model with MSSM constraints. 

For the sake of brevity let us assume that the pseudoscalar Higgs and the
charged Higgs are fairly heavy ($m_A\gsim 150$~GeV, and they may, but need not,
have been observed directly) and that the observed
sparticles' masses ensure that the light Higgs boson can decay into only
SM particles. Then the light Higgs that we consider here will have 
couplings to the $W$ and $Z$ which are suppressed by the same
factor $\sin(\alpha-\beta)$ compared to SM strength, and Higgs couplings
to fermions in addition depend on $\tan\beta=v_2/v_1$ and
$\Delta_b$~\cite{Hall:1994gn}, 
which incorporates non-universal loop corrections to the $h\bar bb$ coupling.
A fit of the Higgs couplings can then be performed in terms of this reduced 
parameter set. Obviously this analysis falls within the $g_V\le
g_V^{\rm SM}$ analysis described in the previous section. Upper bounds 
on the expected measurement 
errors for MSSM partial widths can hence be derived from Fig.~\ref{fig:fit}, 
while Fig.~\ref{fig:SU2fit} gives an estimate of 
errors which can be expected for $m_A\gsim 200$~GeV.

\begin{figure}[th]
\begin{center}
\resizebox{\textwidth}{!}{
\rotatebox{270}{\includegraphics[50,50][555,590]{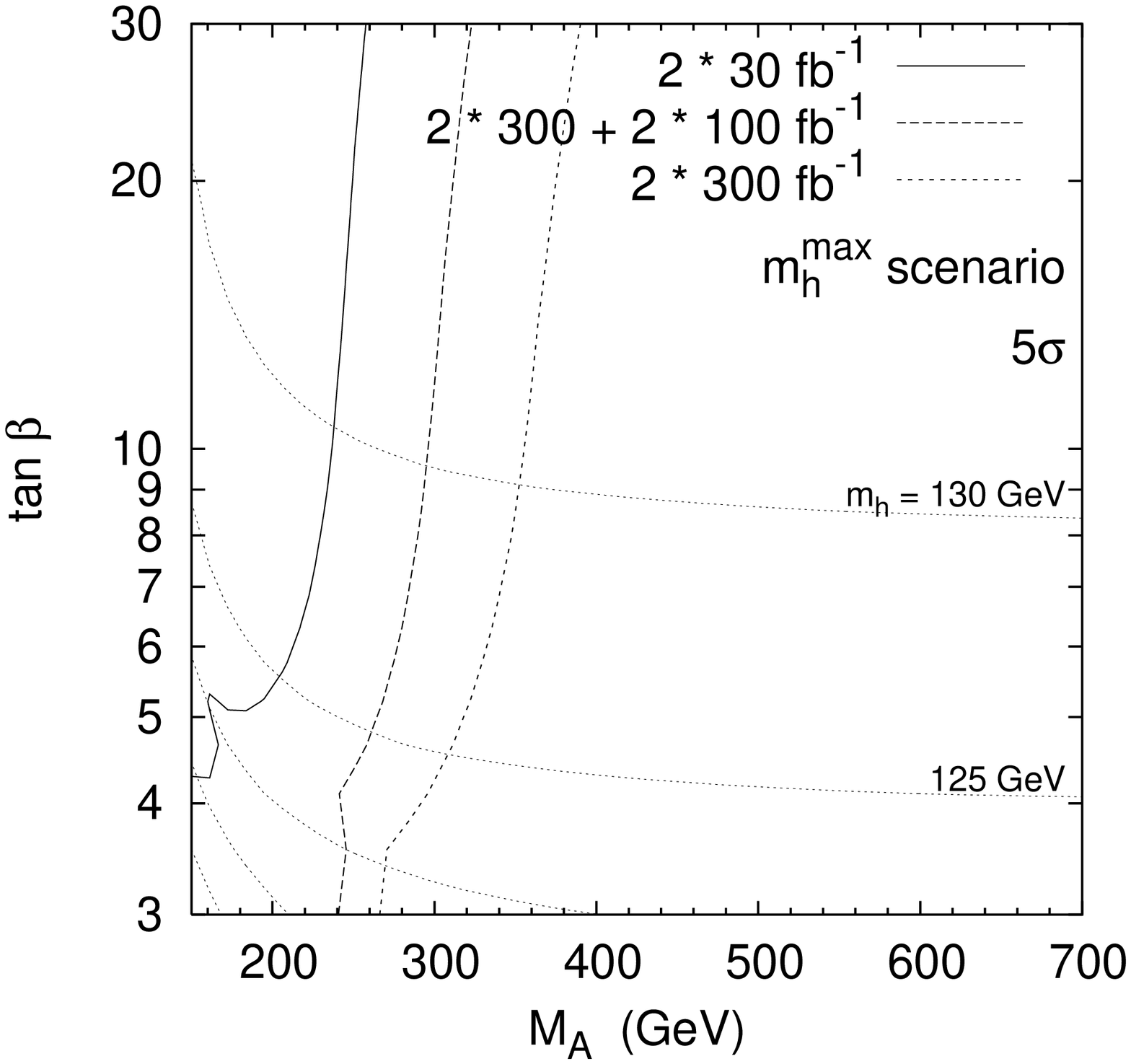}}
\rotatebox{270}{\includegraphics[50,50][555,590]{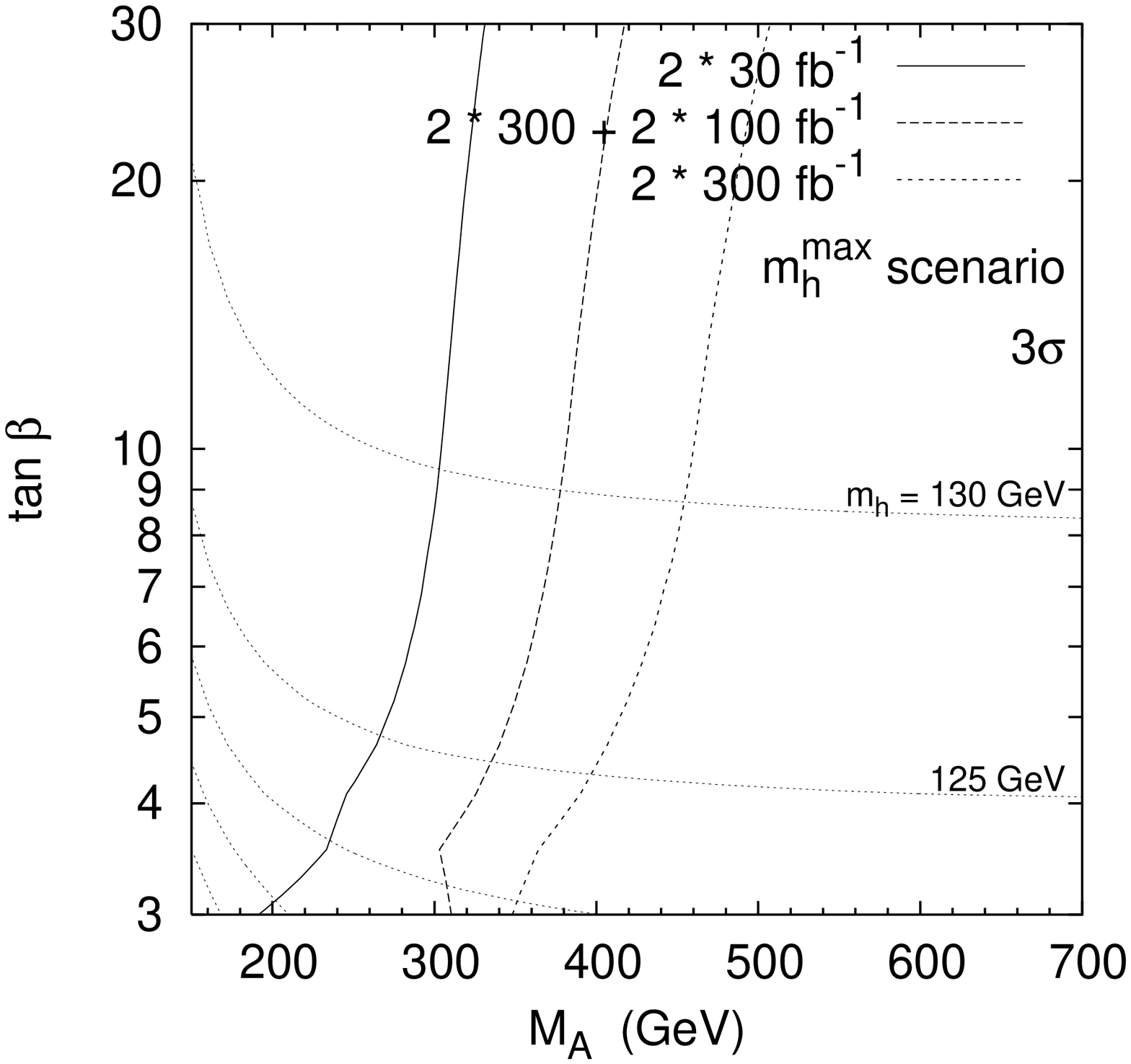}}
}
\caption{Fit within the MSSM $m_h^{\rm max}$ scenario in the 
$M_A$--$\tan\beta$ plane for three luminosity scenarios.
The two panels show the region (to the left of the curves) in which 
a $\geq 5 \sigma$ ($\Delta \chi^2 \geq 25$) or 
$\geq 3 \sigma$ ($\Delta \chi^2 \geq 9$) discrepancy
from the SM can be observed. 
The mostly-horizontal dotted lines are contours of $m_h$ in steps
of 5 GeV.
}
\label{fig:mssm_contour}
\end{center}
\end{figure}

A quantitative, global measure of how well the LHC can distinguish the SM
from a specific MSSM scenario is provided by a $\chi^2$-analysis of the
deviations expected for a specific SUSY model. As a specific example we
consider the $m_h^{\rm max}$ scenario of
Ref.~\cite{LHbenchmark}.  We calculate the mass and branching fractions
of the MSSM Higgs boson using HDECAY3.0 \cite{Djouadi:1998yw}, 
using the FeynHiggsFast1.2.2
\cite{Heinemeyer:1999be,Heinemeyer:2000nz} option to compute  
the MSSM Higgs masses and couplings. Assuming that, for a given $m_A$ and
$\tan\beta$, the corresponding SUSY model is realized in nature, we
may ask at what significance the SM can be ruled out from
$h$~measurements alone. The resulting contours 
are shown in Fig.~\ref{fig:mssm_contour} for the three luminosity
assumptions defined in Sect.~\ref{subsec:fitproc}.
%for each of the two LHC detectors: (i) 30~fb$^{-1}$ of low luminosity
%running, (ii) 300~fb$^{-1}$ have been collected, but, effectively, 
%only 100~fb$^{-1}$ can be used for the weak boson fusion channels due to 
%problems with forward jet tagging and central jet veto, and 
%(iii) 300~fb$^{-1}$ can effectively be used for all channels. 
In the areas to the left of the 
contours the SM can be rejected with more than 5$\sigma$ or 3$\sigma$
significance, respectively. 

The $\chi^2$ definition in  Fig.~\ref{fig:mssm_contour} assumes the same
systematic errors as our analysis in Sec.~\ref{sec:nomodel} 
Event rates and resulting
statistical errors, however, are those expected for the 
MSSM. It should be 
noted that the position of the contours shifts only very little if one assumes 
SM rates to be observed. 
%The resulting contours would then represent the 
%significance with which a specific MSSM scenario can be excluded by 
%LHC observations in the Higgs sector alone. 
% SH: this has been said above already
The contours do shift significantly, however,
when using different SUSY scenarios: other SUSY parameters can have 
a large effect on the relation of $m_h$, $m_A$, $\tan\beta$ and Higgs 
couplings and, hence, an indirect determination of $m_A$ from observed
deviations in light Higgs couplings is problematic without significant
input from other direct SUSY observations at the LHC.
%Furthermore the parameters within the $m_h^{\rm max}$ scenario have
%been fixed, neglecting their prospective experimental errors. 
%Taking
%them into account would weaken considerably the significance (see
%e.g. \cite{deschi-project}).

%%%%%%%%%%%%%%%%%%%%%%%%%%%%%%%%%%%%%%%%%%%%%%%%%%%%%%%%%%%%%%%%%%%%%%%%%%%%%%%
%%%%%%%%%%%%%%%%%%%%%%%%%%%%%%%%%%%%%%%%%%%%%%%%%%%%%%%%%%%%%%%%%%%%%%%%%%%%%%%

\section{Summary}

Measurements in the Higgs sector are expected to provide many complementary
signatures after several years of LHC running. Combining these measurements
allows one to extract information on Higgs partial widths and Higgs couplings 
to fermions and gauge bosons. Because significant contributions from 
unobservable channels cannot easily be ruled out at the LHC, 
model-independent analyses produce large correlations between 
extracted partial 
widths. A reduction of correlations and hence smaller errors on particular 
couplings can be achieved with a variety of theory assumptions. 
In this contribution we have analyzed the constraints expected in generic
multi Higgs doublet models, namely that $HVV$ couplings cannot be larger than 
within the SM. Within such models, the LHC can measure Higgs couplings
to top, tau, $W$ and $Z$ with accuracies in the 10--30\% range, once 
300~fb$^{-1}$ of data have been collected.

Within the MSSM, significant deviations in the Higgs sector should be 
observable at the LHC, provided that the charged and the pseudoscalar 
Higgs masses are not too heavy, i.e., that decoupling is not yet completely
reached. Within the $m_h^{\rm max}$ scenario and with 300~fb$^{-1}$ of data, 
the LHC can distinguish the MSSM and the SM at the 3$\sigma$ level up to 
$m_A\simeq 450$~GeV and with 5$\sigma$ significance up to $m_A\simeq 350$~GeV.
The LHC will thus provide a surprisingly sensitive first look at the Higgs
sector, even though it cannot match precision and the model-independence 
of analyses which are expected for a linear $e^+e^-$ 
collider~\cite{Aguilar-Saavedra:2001rg}.
 
}

%% file: lehti.tex
{
%\includeonly{tanbeta_text_v10}
\newcommand{\nc}{\newcommand}
\nc{\lsim}{\mbox{\raisebox{-.6ex}{~$\stackrel{<}{\sim}$~}}}
\nc{\gsim}{\mbox{\raisebox{-.6ex}{~$\stackrel{>}{\sim}$~}}}
\nc{\esim}{\mbox{\raisebox{-.6ex}{~$\stackrel{-}{\sim}$~}}}
\nc{\beq}{\begin{equation}}
\nc{\eeq}{\end{equation}}

\def\FIG_PATH{./}

\catcode`@=11
\def\citer{\@ifnextchar
[{\@tempswatrue\@citexr}{\@tempswafalse\@citexr[]}}

% \citer as abbreviation for 'citerange' replaces the ',' by a '--'
%

\def\@citexr[#1]#2{\if@filesw\immediate\write\@auxout{\string\citation{#2}}\fi
  \def\@citea{}\@cite{\@for\@citeb:=#2\do
    {\@citea\def\@citea{--\penalty\@m}\@ifundefined
       {b@\@citeb}{{\bf ?}\@warning
       {Citation `\@citeb' on page \thepage \space undefined}}%
\hbox{\csname b@\@citeb\endcsname}}}{#1}}
\catcode`@=12

\newcommand{\ra}{\rightarrow}

%==============================================================================
% title page for few authors

%\begin{titlepage}

% select one of the following and type in the proper number:
%  \cmsnote{2003/XXX}
%  \internalnote{2003/014}
%  \conferencereport{2005/000}
%   \date{1 December 2003}

%  \title{ESTIMATE OF TAN$\beta$ MEASUREMENT PRECISION IN H/A~$\ra\tau\tau$ AND  
%  H$^{\pm} \ra \tau\nu$ IN CMS}
\noindent
{\Large \bf D. Estimating the Precision of a tan$\beta$ Determination with
H/A~$\ra\tau\tau$ and H$^{\pm} \ra \tau\nu$ in CMS} \\[0.5cm]
{\it R.\,Kinnunen, S.\,Lehti, F.\,Moortgat, A.\,Nikitenko and M.\,Spira}

%  \end{Authlist}
%  \Instfoot{hip}{Helsinki Institute of Physics, Helsinki, Finland}
%  \Instfoot{belg}{Department of Physics, University of Antwerpen, Antwerpen} 
%  \Instfoot{impe}{Imperial College, University of London, London, UK}
%  \Anotfoot{a}{On leave from ITEP, Moscow, Russia}
%  \Instfoot{PSI}{Paul Scherrer Institut, CH-5232 Villigen PSI, Switzerland}

%\end{titlepage}

\begin{abstract}
estimated for the H/A~$\ra\tau\tau$ and H$^{\pm} \ra \tau\nu$ decay
channels in the associated production processes gg~$\ra \rm b \bar{\rm
b}\rm H/A$ and gb~$\ra$tH$^{\pm}$ at large tan$\beta$ in CMS. The value
of tan$\beta$ can be determined with better than 35\% accuracy when
statistical, theoretical, luminosity and mass measurement uncertainties
are taken into account.
\end{abstract}

%==============================================================================

\section{Introduction}

The Higgs mechanism is a cornerstone of the Standard Model (SM) and its
supersymmetric extensions. Therefore, the search for Higgs bosons is one of
the top priorities at future high-energy experiments.  Since
the minimal supersymmetric extension of the Standard Model (MSSM)
requires the introduction of two Higgs doublets in order to preserve
supersymmetry, there are five elementary Higgs particles, two CP-even
(h,H), one CP-odd (A) and two charged ones (H$^\pm$). At lowest
order all couplings and masses of the MSSM Higgs sector are determined by two
independent input parameters, which are generally chosen as
tan$\beta=v_2/v_1$, the ratio of the two vacuum expectation values
$v_{1,2}$, and the pseudoscalar Higgs-boson mass m$_{\rm A}$. At LO the light
scalar Higgs mass m$_{\rm h}$ has to be smaller than the Z-boson mass m$_{\rm Z}$.
However, this upper bound is significantly enhanced by radiative
corrections, the leading part of which grows with the fourth power of
the top mass and logarithmically with the stop masses. Including the
one-loop and dominant two-loop corrections the upper bound is increased
to m$_{\rm h}\lsim 135$ GeV$/c^2$ \cite{Carena:2002es}. The negative direct searches
for the Higgsstrahlung processes $e^+e^-\to\rm Zh/ZH$ and the associated
production $e^+e^-\to\rm Ah/AH$ yield lower bounds of m$_{\rm h,H} > 91.0$
GeV$/c^2$ and m$_{\rm A} > 91.9$ GeV$/c^2$. The range $0.5 < $tan$\beta < 2.4$
in the MSSM is excluded for m$_{\rm t}=174.3$ GeV$/c^2$ by the Higgs searches at
the LEP2 experiments \cite{lep2}.

Thus, one of the most important parameters to be determined in the
Minimal Supersymmetric Standard Model (MSSM) as well in a general
type-II Two-Higgs Doublet Model (2HDM) is tan$\beta$.  In the MSSM
tan$\beta$ plays a crucial role, since it characterizes the relative
fraction of the two Higgs boson doublets contributing to the electroweak
symmetry breaking.  Consequently, it enters in all sectors of the
theory.  For small tan$\beta$, it may be possible to determine the value
of tan$\beta$ within the sfermion or neutralino sector \cite{baer,
lali}. For large tan$\beta$ this method has not been found to be
effective. However, in this regime, there are good prospects to measure the
value of tan$\beta$ by exploiting the Higgs sector \cite{gunion}.

At large tan$\beta$ neutral and charged Higgs boson production is
dominated by the bremsstrahlung processes $\rm gg\rightarrow b\bar{\rm
b}H/A$, gb~$\ra$ tH$^{\pm}$ and gg~$\ra$ tbH$^{\pm}$. The dominant parts
of the production cross sections are proportional to
tan$^2\beta$\footnote{For the heavy scalar MSSM Higgs boson H this
behaviour is valid within 1\% for tan$\beta\gsim 10$, if the
pseudoscalar mass m$_{\rm A}$ is larger than about 200 GeV$/c^2$, while for
m$_{\rm A} > 300$ GeV$/c^2$ it is satisfied for tan$\beta\gsim 5$ already.}. Due
to this feature the uncertainty of the tan$\beta$ measurement is only
half of the uncertainty of the rate measurement. In the MSSM the
supersymmetric loop corrections introduce an additional tan$\beta$
dependence to the cross section \cite{deltamb}, but they can be absorbed
in an effective parameter tan$\beta_{\rm eff}$, since the dominant terms
which are enhanced by tan$\beta$ correspond to emission and reabsorption
of virtual heavy supersymmetric particles at the bottom Yukawa vertex,
which are confined to small space-time regions compared with QCD
subprocesses involving massless gluons. The subleading terms are small.
The dominant terms are universal contributions to the bottom Yukawa
coupling \cite{deltamb}.  This implies that the method described below
determines this effective parameter tan$\beta_{\rm eff}$ in the MSSM.
The extraction of the fundamental tan$\beta$ parameter requires
additional knowledge of the sbottom and gluino masses as well as the
$\mu$ parameter. These corrections are in general absent in a 2HDM so
that in these models the extracted value belongs to the fundamental
tan$\beta$ parameter. 
%The following analysis is valid for the 2HDM and the general MSSM, if
%Higgs boson decays into supersymmetric and other non-standard particles
%are kinematically forbidden or suppressed.
The H/A~$\ra\mu\mu$ \cite{bellucci}
and H/A~$\ra\tau\tau$ decay channels have been identified as the most
promising for the searches of the heavy neutral MSSM Higgs boson H and A
at large tan$\beta$. The final states $\rm e\mu$, $\ell\ell$ ($\rm
\ell\ell=e\mu, ee,\mu\mu$) \cite{2lepton}, lepton+jet \cite{hljet} and
two-jets \cite{2jets} have been investigated for the H/A~$\ra\tau\tau$
decay mode.  For heavy charged Higgs bosons the H$^{\pm} \ra \tau\nu$
decay channel in fully hadronic events (both $\tau$ and top decaying
into hadrons) has been found to yield the largest parameter reach and a
clean signature \cite{hplus}. For the MSSM SUSY parameters, the
following values are taken: M$_2$ = 200~GeV/$c^2$, $\mu$ =
300~GeV/$c^2$, M$_{\tilde{\rm g}}$ = 800~GeV/$c^2$, M$_{\tilde{\rm
q},{\tilde{\ell}}}$ = 1 TeV/$c^2$ and A$_{\rm t}$ is set to
2450~GeV/$c^2$. The top mass is set to 175~GeV/$c^2$.
The Higgs boson decays to SUSY particles are allowed.

In this work the theoretical uncertainty of
the cross section and the branching ratio, 
the uncertainty of the luminosity measurement and
statistical errors are taken into account. 
%The uncertainty of the tan$\beta$ measurement originates from the strong
%dependence of the event rates on the MSSM parameter space point, the
%precision of the luminosity measurement and the theoretical accuracy of
%the cross section. 
The uncertainty of the next-to-leading order (NLO)
cross sections for the gg~$\ra$~b$\overline{\rm b}$H/A/h and
gb~$\ra$ tH$^{\pm}$ processes has been shown to be 20--30\% for the
total rate \cite{hep-ph/0309204,plehn}. However, it depends on the
transverse momentum range of the spectator b quarks and reduces to
10--15\% with the requirement of p$_{\rm T}^{\rm b,\bar{\rm b}}\gsim$ 20
GeV/$c$ \cite{hep-ph/0309204,hep-ph/0204093,dawson}.
The uncertainty of the branching ratio BR(H/A$\rightarrow\tau\tau$) related to
the uncertainties of the SM input parameters is about 3\%.
We have not taken into account the uncertainties related to the MSSM parameters, but kept them fixed at our chosen values. We only vary the pseudoscalar mass m$_{\rm A}$ and tan$\beta$.
A 5\% uncertainty of the luminosity
measurement was taken.
The precision of the mass measurement in H/A$\rightarrow\tau\tau$ is estimated 
and taken into account
in the precision of the tan$\beta$ determination. The uncertainty of the background
estimation as well as the uncertainty of the signal selection efficiency
have not yet been taken into account in this study. 
%However, it is expected to be smaller than the present theoretical uncertainty. 
We expect, however, 
that the background uncertainty and uncertainty of the signal selection
will be of the order of 5 \%.
%, which is still smaller than the theoretical uncertainty.

In this work the accuracy of the tan$\beta$ measurement is estimated in
the H/A~$\ra\tau\tau$ and H$^{\pm} \ra \tau\nu$ decay channels by
exploiting the studies of Refs.~\cite{2lepton,2jets,hplus}. The
discovery reach for the lepton+jet final state from H/A~$\ra\tau\tau$,
described in Ref.~\cite{hljet}, is re-evaluated and used in the
tan$\beta$ measurement. The event rates for the gb~$\ra$ tH$^{\pm}$,
H$^{\pm} \ra \tau\nu$ channel are also updated according to the recent
theoretical calculations of the cross section \cite{plehn}.

\section{Transverse momentum of b quarks in \bf{$\rm g \rm g \rightarrow
\rm b \bar{\rm b}H/A$}}

Higgs boson production in the $\rm gg \rightarrow \rm b \bar{\rm b}H/A$
process was obtained with the PYTHIA \cite{Sjostrand:2000wi,Sjostrand:2000wi0}
two-to-three
processes 181 and 186 and with the PYTHIA6.158 default values for the
parton distribution functions and the renormalization and factorization
scales. No cut on the transverse momentum of the b quarks has been
applied at the generation level but $\rm E_{\rm T}^{\rm jet}>$~20 GeV is
used for the b-jet identification in the event analysis.  Therefore it
is important to know how well PYTHIA describes the $\rm p_{\rm T}$
spectrum of the b quarks compared to the NLO calculations \cite{mspira}
in order to estimate how well the efficiency of the event selections can
be trusted.  
Comparizon was made for the SM process $\rm gg\ra b\bar{b}h$ (PYTHIA process 121)
with Higgs mass of 120 GeV/$c^2$.
The PYTHIA and the NLO cross sections are compared in
Table~\ref{table:cross_section} as a function of a cut on the transverse
momentum of the b quark with highest $\rm p_{\rm T}$. In PYTHIA as well
as in the NLO calculations the b quark momentum was taken after gluon
radiation. The total PYTHIA cross sections (p$_{\rm T}>0$) were
normalized to the total NLO cross sections. The agreement between the
PYTHIA and the NLO values turns out to be at the level of 5--10\%. The
statistical uncertainties of the PYTHIA cross sections are shown, too.
For completeness the PYTHIA LO cross sections are also compared to the
corresponding theoretical LO calculation (the lower two rows in
Table.~\ref{table:cross_section}). In this case the PYTHIA b quark was
taken before gluon radiation. Good agreement within 1--2 \% has been
obtained.

\begin{table}[h]
  \vskip 0.1 in
  \centering
  \begin{tabular}{|l|c|c|c|c|c|c|}
    \hline
     p$_{\rm T}$ cut           & 0 GeV/c & 10 GeV/c & 20 GeV/c & 30 GeV/c & 40 GeV/c & 50 GeV/c \\
    \hline
    $\sigma_{\rm NLO}$ (pb)    & 734 & 507 &  294 & 173 & 106 &  68 \\
    $\sigma_{\rm PYTHIA}$ (pb) & 734 %$\pm$ 0 
                               & 523 $\pm$ 3
                               & 275 $\pm$ 3
                               & 156 $\pm$ 3
                               &  92 $\pm$ 2
                               &  60 $\pm$ 2 \\
    \hline
    $\sigma_{\rm LO}$ (pb)     & 528 & 393 &  241 & 152 & 102 &  71 \\
    $\sigma_{\rm PYTHIA}$ (pb) & 528 %$\pm$ 0
                               & 407 $\pm$ 2
                               & 245 $\pm$ 3
                               & 154 $\pm$ 2
                               & 101 $\pm$ 2
                               &  70 $\pm$ 2 \\
    \hline
  \end{tabular}
  \caption{Comparison of the NLO and LO cross sections to the PYTHIA
  cross sections as a function of the cut on the transverse momentum of
  the b quark with highest $\rm p_{\rm T}$. The total PYTHIA cross
  sections ($\rm p_{\rm T}>$ 0) are normalized to the corresponding
  NLO(LO) cross sections.}
  \label{table:cross_section}
\end{table}

\begin{figure}[p]
\begin{center}
\mbox{\epsfig{file=\FIG_PATH/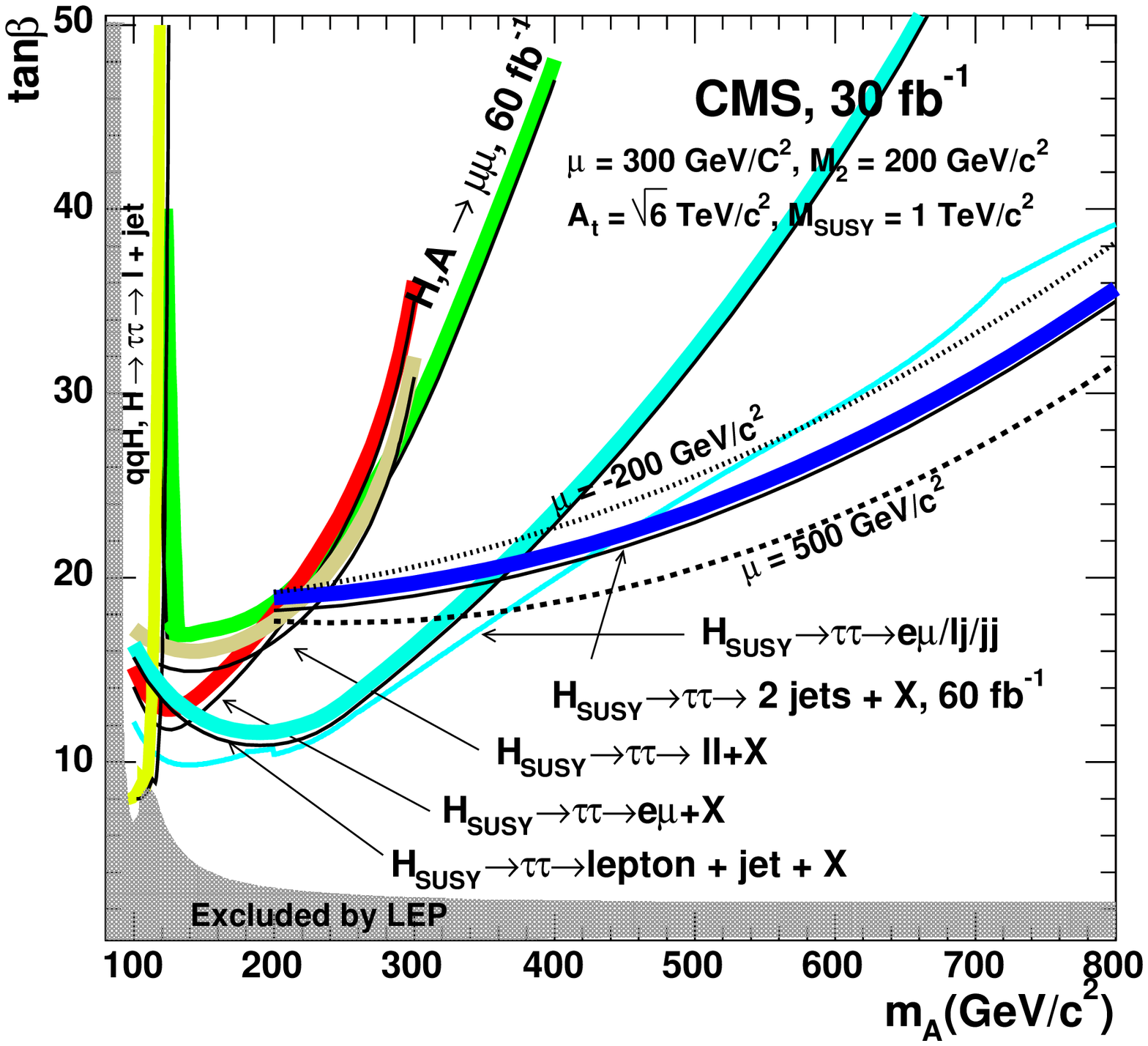,height=9cm,width=12cm}}
\end{center}
\caption{The 5$\sigma$-discovery potential for the heavy neutral MSSM
Higgs bosons as a function of $\rm m_{\rm A}$ and tan$\beta$ with maximal
stop mixing for 30~fb$^{-1}$.  The H/A~$\ra\mu\mu$ and H/A~$\ra\tau\tau
\ra$~two-jet channels are shown for 60~fb$^{-1}$.}
\protect\label{fig:discovery_HA}
%\end{figure}
%\begin{figure}[p]
\begin{center}
\mbox{\epsfig{file=\FIG_PATH/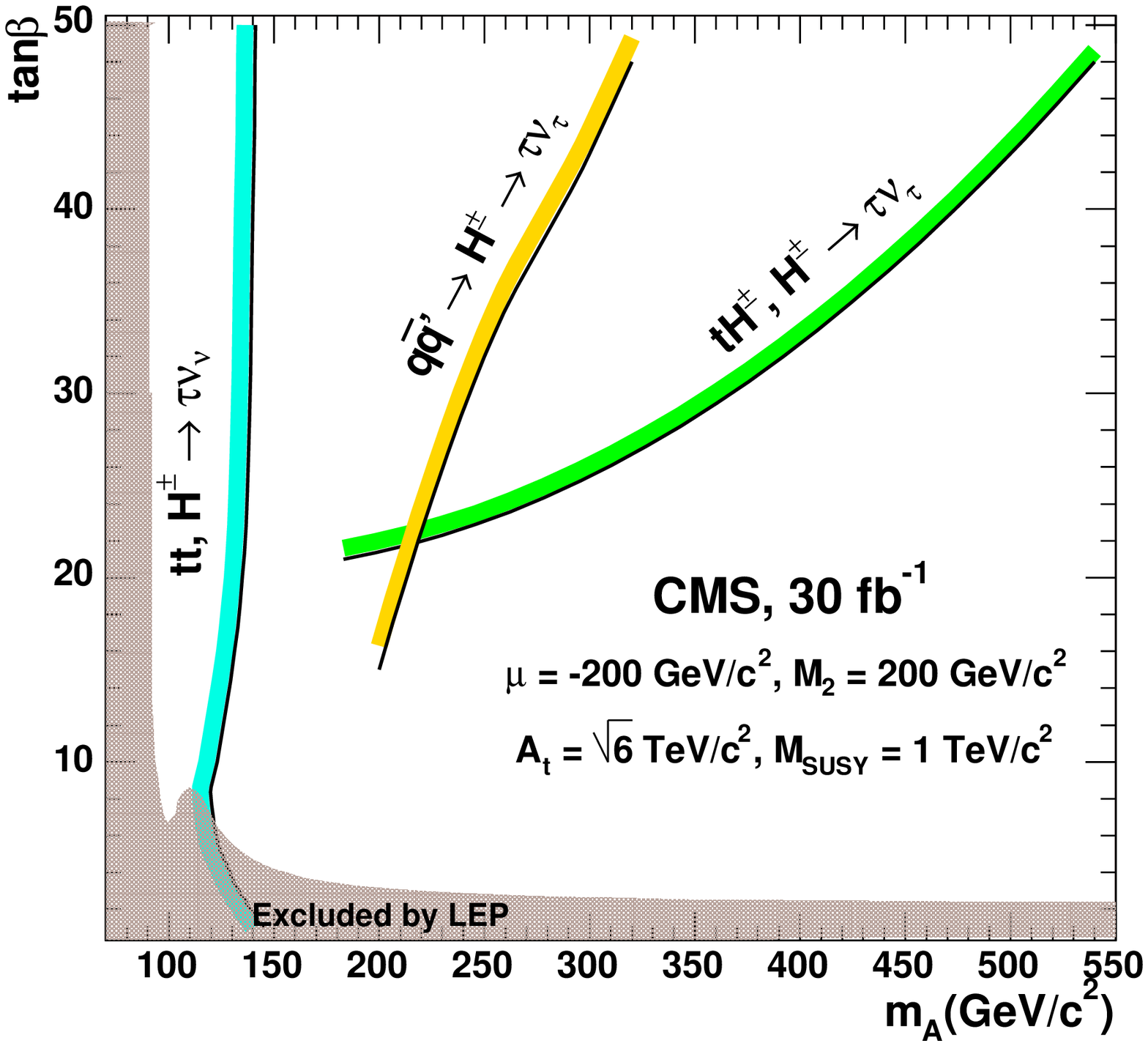,height=9cm,width=12cm}}
\end{center}
\caption{The 5$\sigma$-discovery potential for the charged Higgs bosons
in the H$^{\pm} \ra \tau\nu$ decay channel as
a function of $\rm m_{\rm A}$ and tan$\beta$ with maximal stop mixing for
30~fb$^{-1}$.} \protect\label{fig:discovery_Hplus}
\end{figure}

\section{Event selections and expected discovery reaches}

If the Higgs boson is detected with high enough signal significance, it is
possible to count events in order to measure the value of tan$\beta$.
%Figure ~\ref{fig:discovery_HA} shows the region of the (m$_{\rm
%A},\tan\beta$) parameter space accessible with
%H,A,h$\rightarrow\tau\tau$ and H,A$\rightarrow\mu\mu$ decay channels at
%low luminosity. 
The 5$\sigma$-discovery potential for the H/A/h~$\ra\tau\tau$ decay
channels with the e$\mu$, $\ell\ell$ and lepton+jet final states for
30~fb$^{-1}$ and with the two-jet final state for 60~fb$^{-1}$ is shown
in Figure~\ref{fig:discovery_HA}. The 5$\sigma$-discovery reaches of the
H/A~$\ra\mu\mu$ decay channel with 60~fb$^{-1}$ and of the
H~$\ra\tau\tau \ra$~lepton+jet channel in the weak gauge boson fusion
with 30~fb$^{-1}$ are also depicted in the figure.
Figure~\ref{fig:discovery_Hplus} presents the 5$\sigma$-discovery
potential for the charged Higgs bosons in the H$^{\pm} \ra \tau\nu$
decay channel for 30~fb$^{-1}$.  In these regions
of the parameter space the value of tan$\beta$ can be determined by
counting the neutral and charged Higgs bosons.

%The precision of the mass measurement and the uncertainty of the background
%estimation have not yet been taken into account in this study.

\subsection{Neutral Higgs bosons}

The event selections for the two-lepton (e$\mu$ and $\ell\ell$),
lepton+jet ($\ell$j) and two-jet (jj) final states from H/A/h$\ra\tau\tau$ 
are described in detail in
Refs.~\cite{2lepton,hljet,2jets}. The branching ratios into these final
states are shown in Table~\ref{table:BR}.  The discovery potential in
the H/A/h$\ra\tau\tau \ra$~lepton+jet channel is re-evaluated
using the cross sections of Ref.~\cite{spira_web} and with updated
$\tau$ selection and b-tagging efficiencies. Unlike
Ref.~\cite{hljet}, the recent analysis is extended to large Higgs boson
masses,
%. The 5$\sigma$-discovery potential shown in
%Fig.~\ref{fig:discovery_HA} is from this study. A
and a 5$\sigma$ reach up to m$_{\rm A}\sim$ 650~GeV/$c^2$ at tan$\beta
\sim$~50 is obtained.  The details will be described in an upcoming
note. 

\begin{table}[h]
  \vskip 0.1 in
  \centering
  \begin{tabular}{|l|c|}
    \hline
     Final state & branching ratio \\
    \hline
     H/A/h$\ra\tau\tau \ra$ e$\mu$+X   & $\sim$6.3\%  \\
     H/A/h$\ra\tau\tau \ra \ell\ell$+X & $\sim$12.5\% \\
     H/A/h$\ra\tau\tau \ra\ell$j+X  & $\sim$45.6\% \\
     H/A/h$\ra\tau\tau \ra$~jj+X & $\sim$41.5\% \\
    \hline
  \end{tabular}
  \caption{The branching ratios into final states $\tau\tau\ra$X.}
  \label{table:BR}
\end{table}

\begin{figure}[t]
  \centering
  \vskip 0.1 in
  \begin{tabular}{cc}
  \begin{minipage}{7.5cm}
    \centering
    \resizebox{\linewidth}{60 mm}{\includegraphics{\FIG_PATH/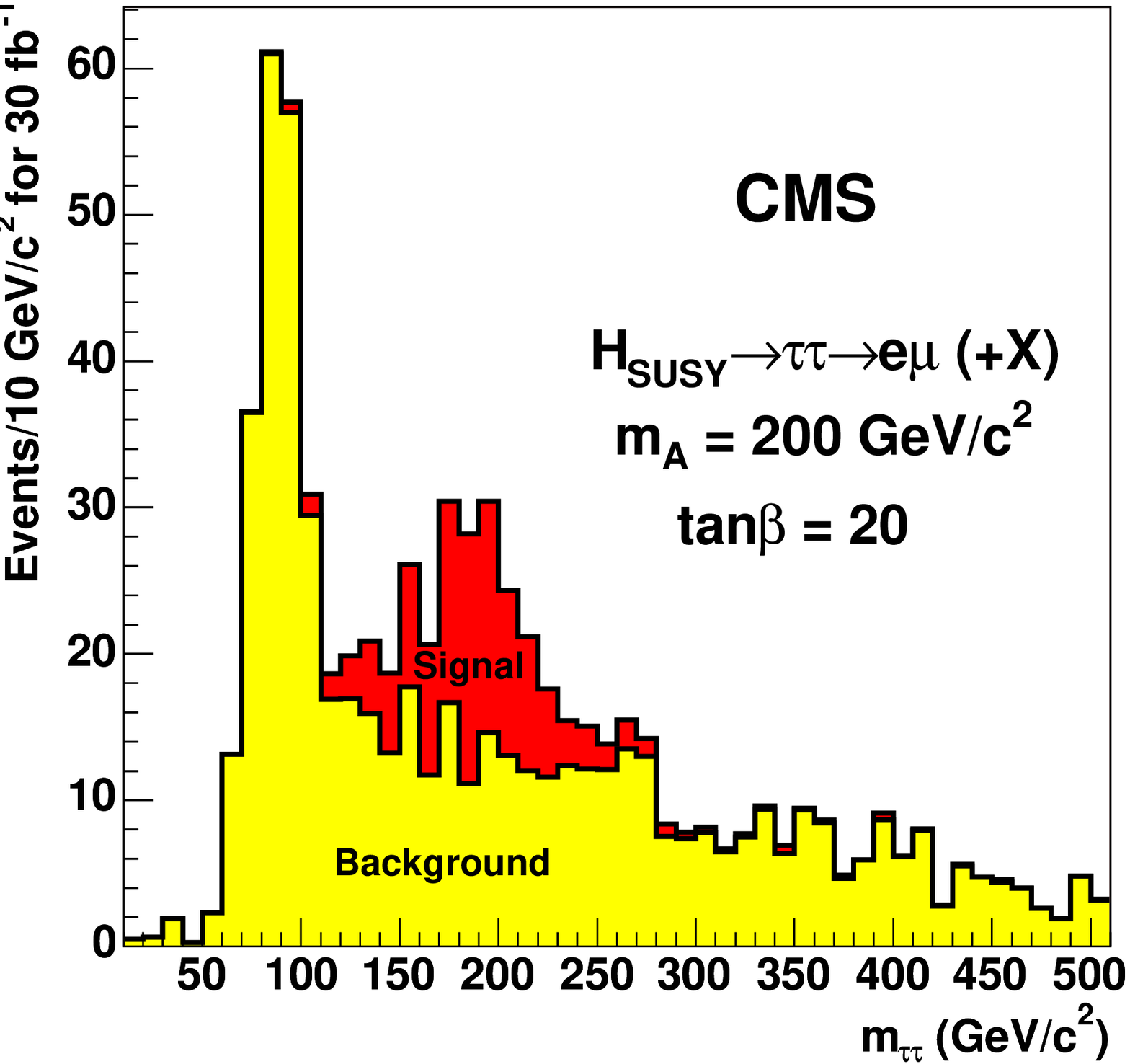}}
    \caption{Reconstructed $\tau\tau$ invariant mass in the e$\mu$ final state
             in the H/A/h~$\rightarrow\tau\tau$ signal (dark) and in the total 
             background (light) with m$_{\rm A}$ = 200 GeV/$c^2$ and tan$\beta$ 
             = 20 for 30~fb$^{-1}$.}
    \label{fig:efmass_emu}
  \end{minipage}
  &
  \begin{minipage}{7.5cm}
    \centering
    \resizebox{\linewidth}{60 mm}{\includegraphics{\FIG_PATH/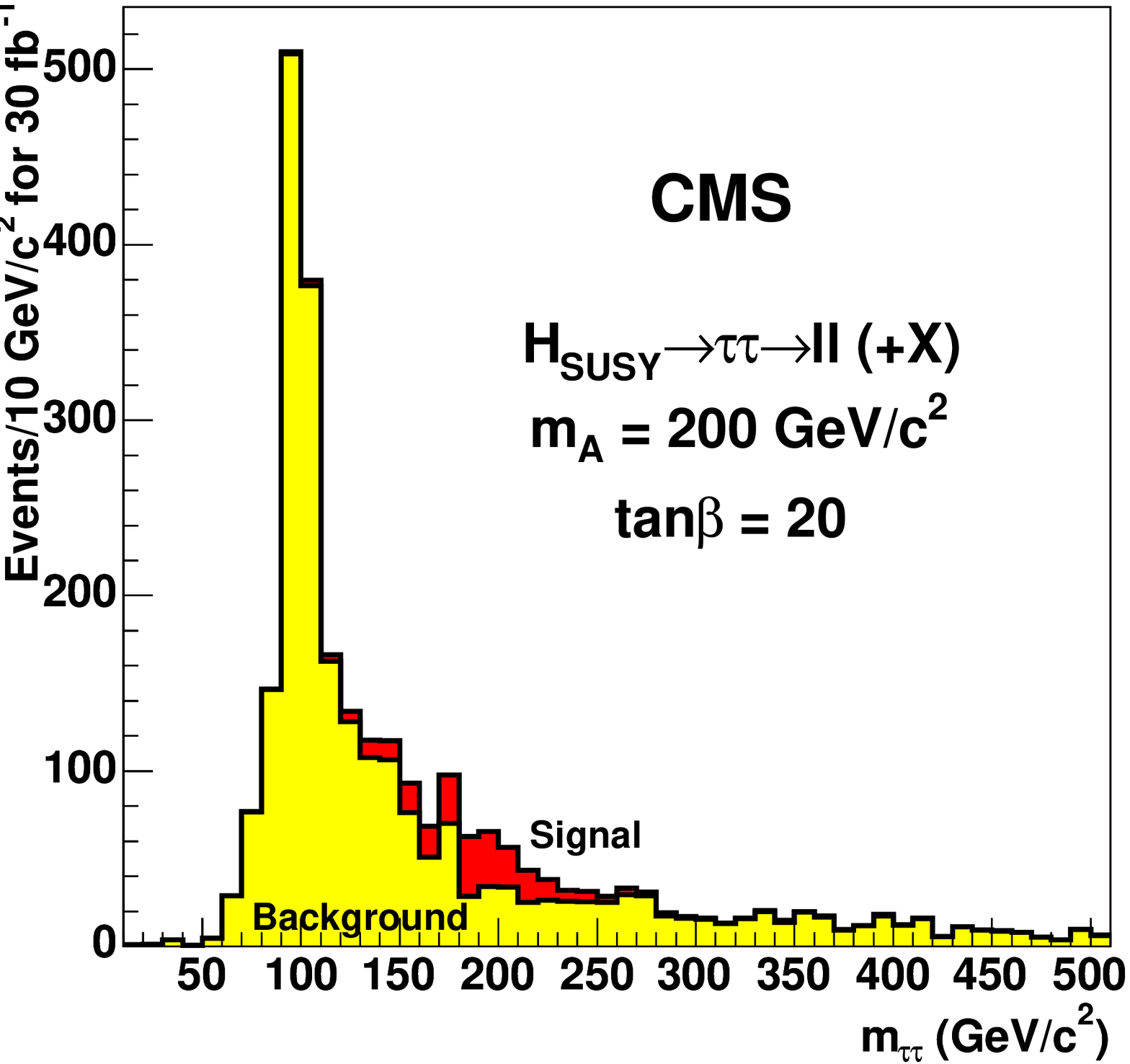}}
    \caption{Reconstructed $\tau\tau$ invariant mass in the $\ell\ell$ final state
             in the H/A/h~$\rightarrow\tau\tau$ signal (dark) and in the total 
             background (light) with m$_{\rm A}$ = 200 GeV/$c^2$ and tan$\beta$ = 
             20 for 30~fb$^{-1}$.}
    \label{fig:efmass_ll} 
  \end{minipage}
\\
  \begin{minipage}{7.5cm}
    \centering  
    \resizebox{\linewidth}{60 mm}{\includegraphics{\FIG_PATH/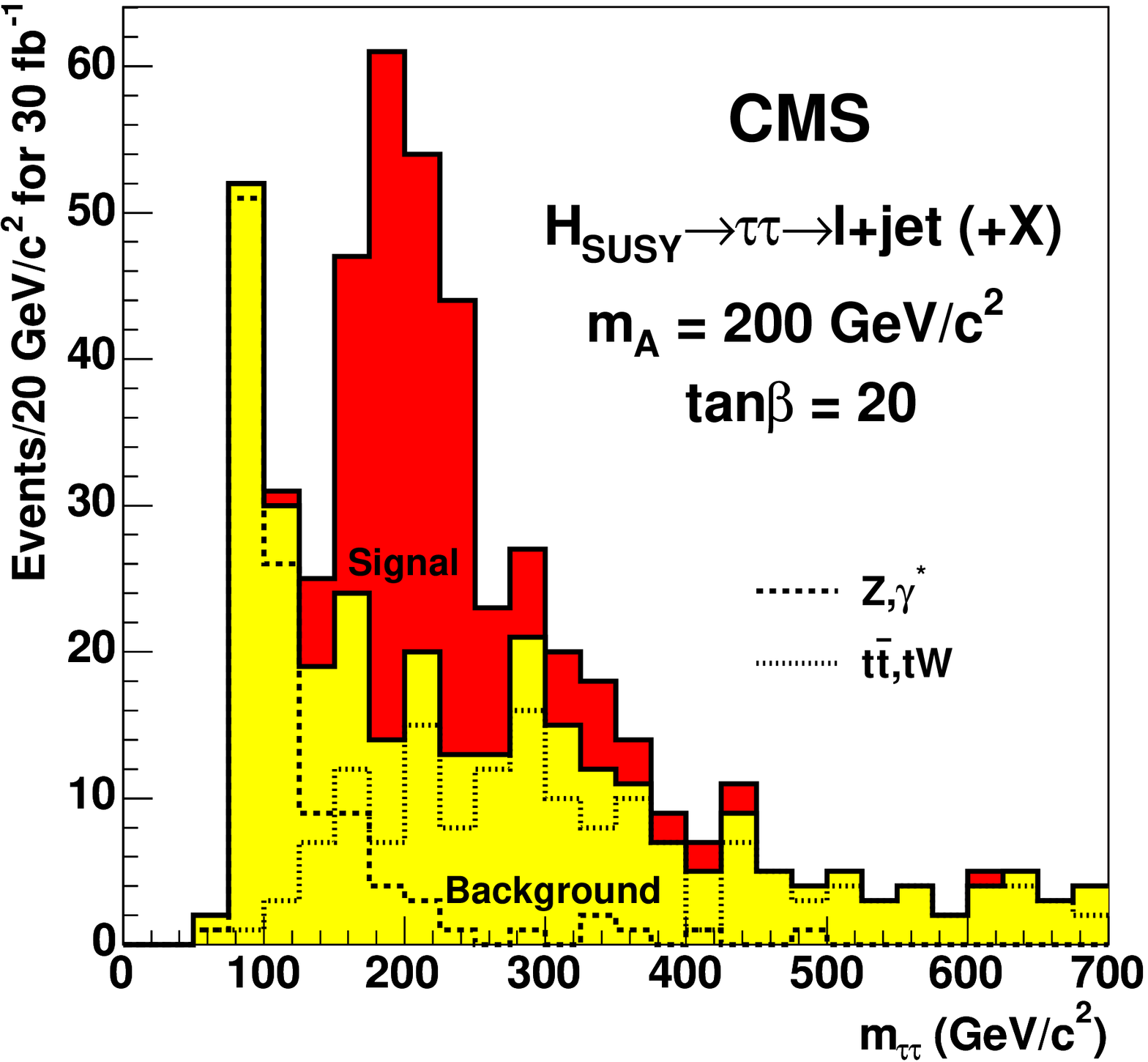}}
    \caption{Reconstructed $\tau\tau$ invariant mass in the lepton+jet final state
             in the H/A/h~$\rightarrow\tau\tau$ signal (dark) and in the total 
             background (light) with m$_{\rm A}$ = 200 GeV/$c^2$ and tan$\beta$ = 
             20 for 30~fb$^{-1}$.}
    \label{fig:efmass_lj}
  \end{minipage}
  &
  \begin{minipage}{7.5cm}
    \centering
    \resizebox{\linewidth}{60 mm}{\includegraphics{\FIG_PATH/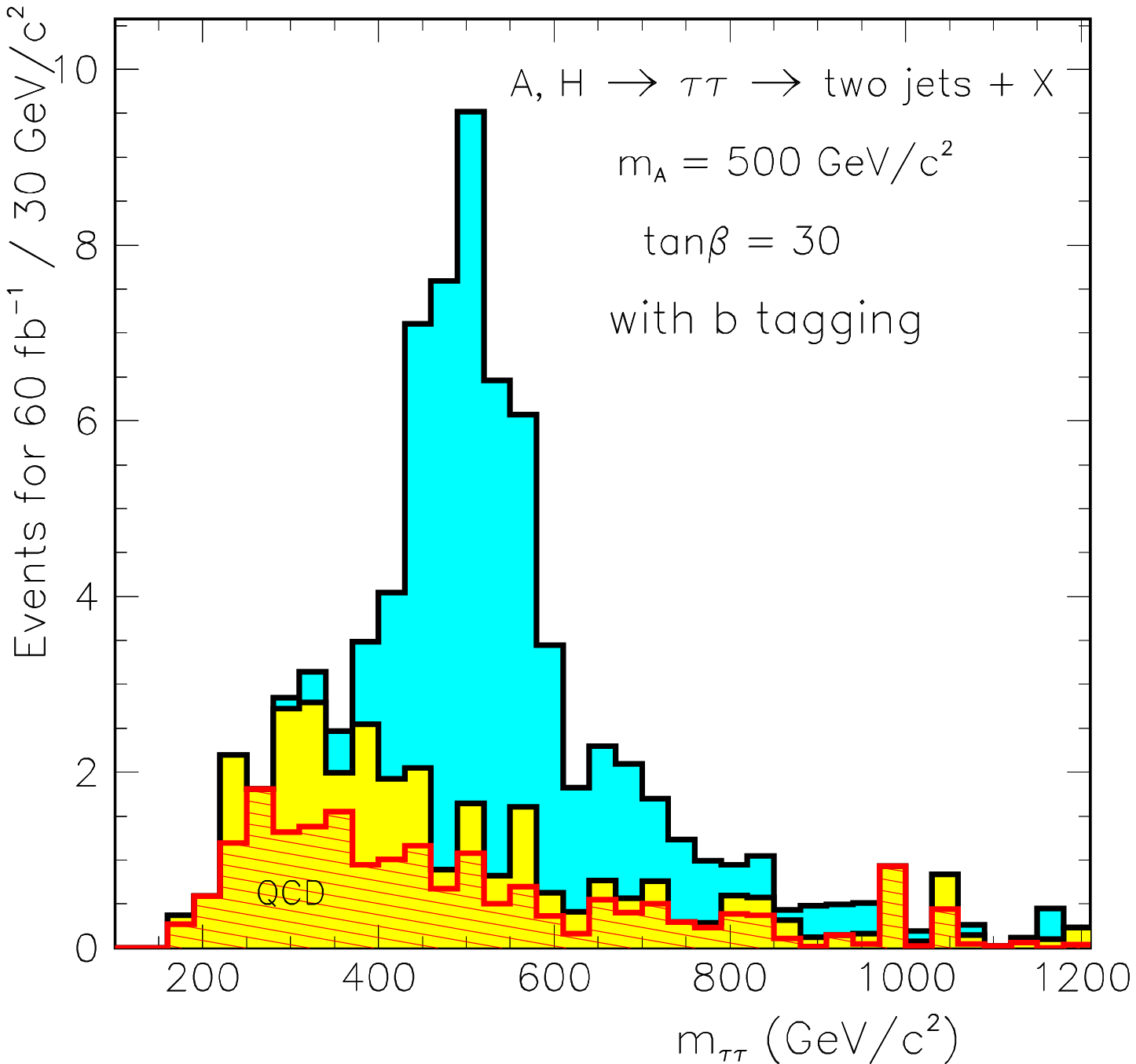}}
    \caption{Reconstructed $\tau\tau$ invariant mass in H/A~$\rightarrow\tau\tau\rightarrow$ 
             2 jets (dark), in the total background
             (light) and in the multi-jet background (dashed) with m$_{\rm A}$ = 
             500 GeV/$c^2$ and tan$\beta$ = 30 for 60~fb$^{-1}$.}
    \label{fig:efmass_jj}
  \end{minipage}
  \end{tabular}
\end{figure}

%\begin{2figures}{t}
%  \resizebox{\linewidth}{60 mm}{\includegraphics{\FIG_PATH/effmass_h2tauemu_a200b20_btagging.eps}}&
%  \resizebox{\linewidth}{60 mm}{\includegraphics{\FIG_PATH/effmass_h2tau2l_a200b20_btagging.eps}}\\
%  \caption{Reconstructed $\tau\tau$ invariant mass in the e$\mu$ final state
%   in the H/A/h~$\rightarrow\tau\tau$ signal (dark) and in the total background (light)
%   with m$_{\rm A}$ = 200 GeV/$c^2$ and $\tan\beta$ = 20 for 30~fb$^{-1}$.}
%  \label{fig:efmass_emu} &
%  \caption{Reconstructed $\tau\tau$ invariant mass in the $\ell\ell$ final state
%   in the H/A/h~$\rightarrow\tau\tau$ signal (dark) and in the total background (light)
%   with m$_{\rm A}$ = 200 GeV/$c^2$ and $\tan\beta$ = 20 for 30~fb$^{-1}$.}
%  \label{fig:efmass_ll} \\
%  \resizebox{\linewidth}{60 mm}{\includegraphics{\FIG_PATH/effmass_a200b20_lj_truestat.eps}} &
%  \resizebox{\linewidth}{60 mm}{\includegraphics{\FIG_PATH/ip5_btag_h500_enu_orig.eps}} \\
%  \caption{Reconstructed $\tau\tau$ invariant mass in the lepton+jet final state
%   in the H/A/h~$\rightarrow\tau\tau$ signal (dark) and in the total background (light)
%   with m$_{\rm A}$ = 200 GeV/$c^2$ and $\tan\beta$ = 20 for 30~fb$^{-1}$.}
%  \label{fig:efmass_lj} &
%  \caption{Reconstructed $\tau\tau$ invariant mass in the two-jet final state
%   in the H/A~$\rightarrow\tau\tau$ signal (dark), in the total background (light)
%  and in the QCD multi-jet background (dashed)
%   with m$_{\rm A}$ = 500 GeV/$c^2$ and $\tan\beta$ = 30 for 60~fb$^{-1}$.}
%  \label{fig:efmass_jj} \\
%\end{2figures}

 The common backgrounds for all the H/A$\ra\tau\tau$ channels
are the Z,$\gamma^*\rightarrow\tau\tau$ Drell-Yan process, $\rm
t\bar{\rm t}$ production with real and fake $\tau$'s and single top
production (Wt). The channels with leptons in the final state suffer
from the $\rm b\bar{\rm b}$ background, and the final states with
hadronic $\tau$ decays are plagued by the W+jet background. For fully
hadronic final states with both $\tau$'s decaying hadronically there is
in addition the QCD multi-jet background with jets faking $\tau$'s, and
for the H/A$\ra\tau\tau \ra \ell\ell$+X channel there is the
additional background from Z,$\gamma^*$ decaying to electron and muon
pairs.

%The signal and backround events were first filtered by triggers
%\cite{DA-HLT-TDR}. 
The hadronic Tau Trigger for the two-jet final state 
was studied with full simulation in Ref. \cite{2jets,DA-HLT-TDR}. For the
e$\mu$, $\ell\ell$ and the $\ell$j final states the trigger was simulated
by selecting the kinematic cuts above the trigger thresholds, and taking
the trigger efficiencies from Ref. \cite{DA-HLT-TDR}. The used triggers
were the Inclusive muon trigger with efficiency
0.9*0.97*0.97 (trigger threshold effect*$\mu$ reconstruction efficiency* 
calorimetric isolation), the Di-electron trigger with efficiency 0.95*0.872*0.946 
per electron (trigger threshold effect*Level-1 e efficiency*Level-2.5 e efficiency)
and e-$\tau$jet trigger with efficiency 0.95*0.872*0.77*0.95 (e trigger threshold
effect*Level-1 e efficiency*HLT e efficiency*$\tau$ trigger threshold effect).
The backgrounds were
suppressed with lepton tracker isolation, $\tau$ jet identification, $\tau$
tagging with impact parameter, b tagging and jet veto.  The $\tau$ jet
identification \cite{2jets} selects collimated low multiplicity jets
with high p$_{\rm T}$ charged particles.  The hadronic jets are
suppressed by a factor of $\sim$~1000. Tau tagging
\cite{2lepton} exploits the short but measurable lifetime of the $\tau$:
the decay vertex is displaced from the primary vertex. For
b tagging the B hadron lifetime is used to distinguish the associated b
jets from c jets and light quark/gluon jets. B tagging suppresses 
efficiently the Drell-Yan and QCD multi-jet backgrounds by a factor of
$\sim$~100, but it also suppresses the Higgs boson production with no 
associated b jets.
% for which the tan$\beta$ dependence is different.
The jet veto is
directed against the $\rm t\bar{\rm t}$ and Wt backgrounds, in which the
jets are more energetic and easier to reconstruct and to b-tag compared
to jets associated with the signal. Rejecting events with more than one
jet (including the b jet and not counting $\tau$'s) suppresses the $\rm
t\bar{\rm t}$ background by a factor of $\sim$~5 \cite{2jets}.

Despite several neutrinos in all the H/A$\rightarrow\tau\tau$
final states the Higgs boson mass can be reconstructed. The effective
$\tau\tau$ mass is evaluated assuming that the neutrinos are emitted
along the measured $\tau$ decay products. The neutrino energies are
estimated by projecting the missing energy vector onto the neutrinos.
Uncertainties in the missing energy measurement can lead to negative
neutrino energies.
% Requiring positive neutrino energies a significant fraction of the
% signal 
%events can be lost 
A significant fraction of the signal events is lost, when positive
neutrino energies are required, but the backgrounds from $\rm t\bar{\rm
t}$, Wt and QCD multi-jet events are suppressed, since for these
backgrounds the neutrinos are generally not emitted along the true or
fake $\tau$'s.  The mass resolution can be improved efficiently with a
cut in the $\Delta\phi$ or space angle between the two $\tau$'s.  For
the two-jet final state in  Ref.~\cite{2jets} a mass reconstruction
efficiency of 53\% has been obtained with the cut $\Delta\phi <$~175$^o$
and requiring one b jet with $\rm E_{\rm T}>$~30~GeV.  The
reconstructed Higgs boson mass is shown in Figs.~\ref{fig:efmass_emu} to
\ref{fig:efmass_jj} for the four $\tau\tau$ final states. The
reconstructed mass peak is a superposition of the H and A signals. In
the region m$_{\rm A}\lsim$ 130 GeV/$c^2$ the contribution from the
lightest Higgs boson h cannot be separated in these channels and is also
included in the signal event rates.

A 5$\sigma$-discovery reach combining the $\rm e \mu$, lepton+jet and
two-jet final states from the H/A$\ra \tau\tau$ decay channel  is also
shown in Fig.~\ref{fig:discovery_HA}.  The combined reach is evaluated
by adding the number of signal and background events from the three
final states in a given  (m$_{\rm A}$, tan$\beta$) point.  However, this
method can lead to an unsatisfactory result as the analysis of these
final states has been optimized to reach the best possible signal
significance which has led to different background levels. For example
at low values of m$_{\rm A}$ and tan$\beta$ the signal for the
H/A$\ra\tau\tau \ra \ell\ell$+X channel \cite{2lepton} suffers from a
significantly larger Drell-Yan backgound than that for the
H/A$\ra\tau\tau \ra$~lepton+jet channel. If the $\ell\ell$ final state
is included, the combined reach is smaller than that from the lepton+jet
final state alone. 

\subsection{Charged Higgs bosons}

The production of heavy charged Higgs bosons has been studied in the
gb~$\ra$ tH$^{\pm}$, H$^{\pm} \ra \tau\nu$ channel with hadronic $\tau$
decays in Ref.~\cite{hplus}.  The W+jet and QCD multi-jet backgrounds
have been suppressed by b tagging and reconstruction of the associated
top quark. To suppress the t$\overline{\rm t}$ and Wt backgrounds with
genuine $\tau$'s, the helicity correlations have been exploited by
requiring at least 80\% of the $\tau$-jet energy to be carried by a
single charged pion.  In purely hadronic final states the transverse
Higgs boson mass can be reconstructed from the $\tau$ jet and the
missing transverse energy, with an endpoint at m$_{\rm H^{\pm}}$ for the
signal and at m$_{\rm W}$ for the t$\overline{\rm t}$, Wt and W+jet
backgrounds.

\begin{figure}[h]
  \vskip 0.1 in
  \centering
  \mbox{\epsfig{file=\FIG_PATH/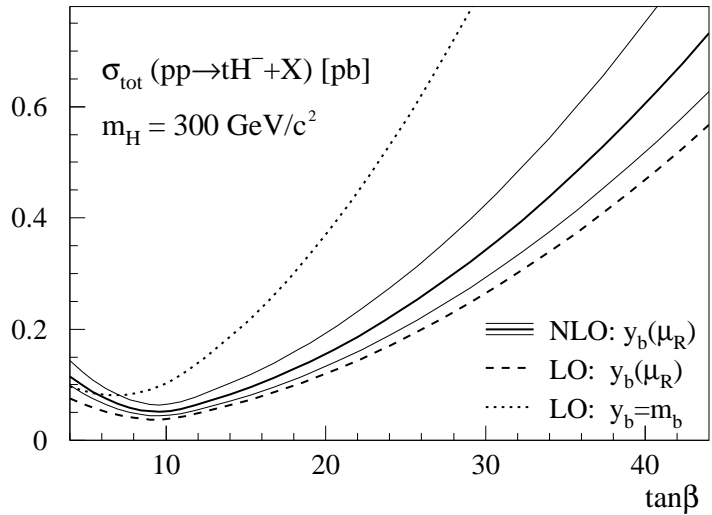,height=7cm,width=10cm}}
  \caption{The inclusive production cross section $gb\to\rm tH^\pm$ at the
   LHC for m$_{\rm H^\pm}=300$ GeV/$c^2$ as a function of tan$\beta$ from
   Ref.~\cite{plehn}. The dashed and solid curves present the consistent
   leading order and next-to-leading order results. The dotted line depicts
   the cross section with the bottom Yukawa coupling defined
   (inappropriately) in terms of the bottom pole mass and thus illustrate
   the enhancement through large logarithms. The range for the
   next-to-leading order result is given for $\mu_F=\mu_R=m_{\rm av}/4
   \cdots 4 m_{\rm av}$ with m$_{\rm av} = (\rm m_{\rm H^\pm}+m_{\rm t})/2$ 
   denoting the average mass of the produced particles.}
\protect\label{fig:plehn}
\end{figure}

%\begin{figure}[h]
%  \centering
%  \vskip 0.1 in
%  \includegraphics[height=70mm,width=100mm]{\FIG_PATH/tmass_h200b30_phi.eps}
%      \caption{Transverse mass reconstructed from the $\tau$ jet and
%               $\rm E_{\rm T}^{\rm miss}$ in the $\rm gg \ra \rm tH^{\pm}$,
%               $\rm H^{\pm} \ra \tau\nu_{\tau}$ signal (dark) and in the
%               background (light) with $\rm m_{\rm H^{\pm}}$ = 200~GeV/$c^2$ 
%               and tan$\beta$ = 30 for 30~fb$^{-1}$.}
%  \label{fig:hplus_mt}
%\end{figure}

In Ref.~\cite{hplus} the PYTHIA estimates were used for the cross
sections and branching ratios.  Recently, detailed theoretical LO and NLO
calculations have been published on the charged Higgs boson production
at the LHC~\cite{plehn}. These calculations show significantly lower LO
production cross sections especially for light charged Higgs bosons than
those used in Refs.~\cite{hplus,nikita}. Figure \ref{fig:plehn} shows
the cross section for gb~$\ra$ tH$^{\pm}$ at m$_{\rm
H^{\pm}}$~=~300~GeV/$c^2$ as a function of tan$\beta$ from
Ref.~\cite{plehn}.  The scale dependence of the LO cross section
originating from the Yukawa coupling in the production mechanism 
is large at large tan$\beta$. The
LO cross section of Ref.~\cite{plehn} at m$_{\rm H^{\pm}} \sim$
200~GeV/$c^2$ calculated with running b quark mass is a factor of
$\sim$ 1.7 lower than that used in Ref.~\cite{hplus}.  At m$_{\rm
H^{\pm}} \sim$ 600~GeV/$c^2$ the difference reduces to 10\%.  The
H$^{\pm} \ra \tau\nu$ branching ratio was also overestimated in
Ref.~\cite{hplus} by about 20\% at m$_{\rm H^{\pm}}$~=~200~GeV/$c^2$.
The 5$\sigma$-discovery reach in Fig.~\ref{fig:discovery_Hplus} is shown
with updated cross sections \cite{plehn} and branching ratios
\cite{Djouadi:1998yw}.  The statistical significance is calculated with Poisson
statistics. The reconstructed transverse mass is shown in 
Fig. \ref{fig:hplus_mt}.
%The 5$\sigma$-discovery reach for the for gb~$\ra$
%tH$^{\pm}$, H$^{\pm} \ra$~tb channel from Ref.~\cite{nikita} is also
%updated to take into account the correct production cross section
%\cite{plehn}.

\begin{figure}[h]
  \centering
  \vskip 0.1 in
  \begin{tabular}{cc}
  \begin{minipage}{7.5cm}
    \centering
  \includegraphics[height=70mm,width=80mm]{\FIG_PATH/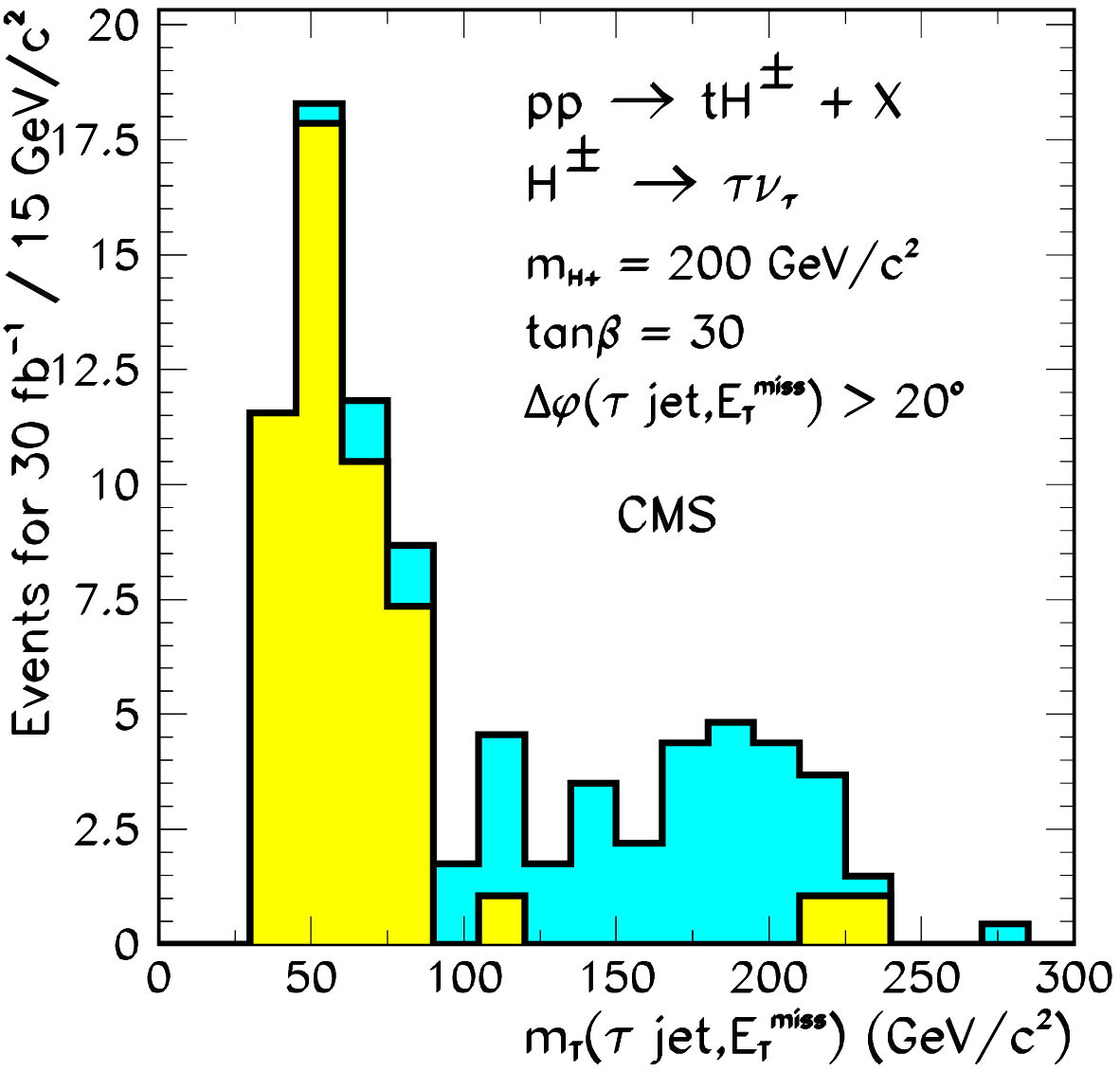}
      \caption{Transverse mass reconstructed from the $\tau$ jet and
               $\rm E_{\rm T}^{\rm miss}$ in the $\rm gg \ra \rm tH^{\pm}$,
               $\rm H^{\pm} \ra \tau\nu_{\tau}$ with $\rm m_{\rm H^{\pm}}$ = 200~GeV/$c^2$
               and tan$\beta$ = 30 for 30~fb$^{-1}$.}
% signal (dark) and in the
%               background (light) with $\rm m_{\rm H^{\pm}}$ = 200~GeV/$c^2$
%               and tan$\beta$ = 30 for 30~fb$^{-1}$.}
  \label{fig:hplus_mt}
  \end{minipage}
  &
  \begin{minipage}{7.5cm}
    \centering
  \includegraphics[height=70mm,width=80mm]{\FIG_PATH/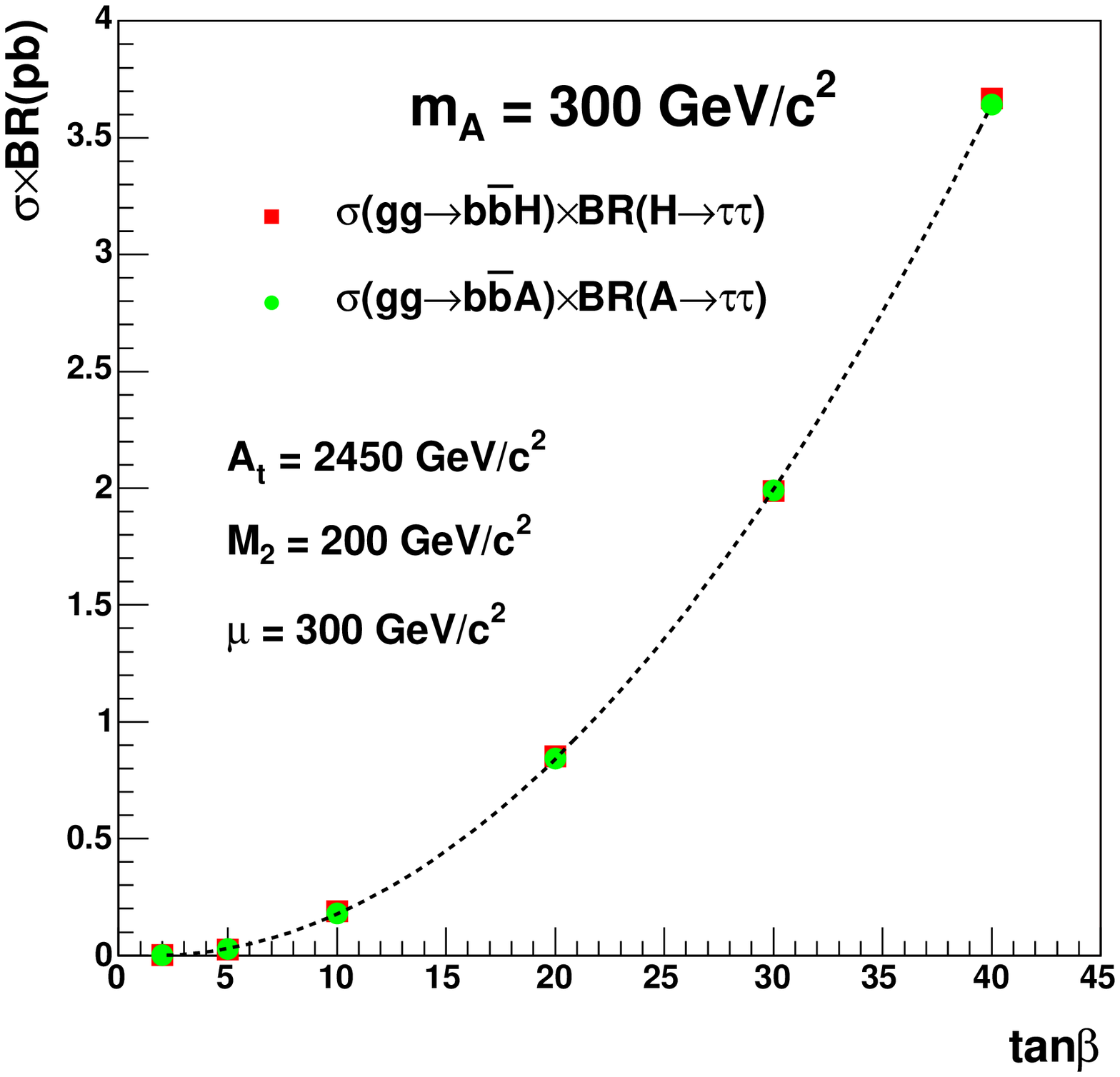}
      \caption{Cross section times branching ratio for
               $\rm gg\rightarrow b\bar{b}H/A, H/A\rightarrow\tau\tau$
               calculated with the programs of Ref.~\cite{spira_web}. }
  \label{fig:bbh_crosssection}
  \end{minipage}
  \end{tabular}
\end{figure}

\section{Calculation of the tan$\beta$ measurement uncertainty}

The accuracy of the tan$\beta$ measurement is due to the statistical
uncertainty of event rates, the systematic uncertainty from the
luminosity measurement and the theoretical uncertainty of the cross
section calculation.  The associated Higgs boson production cross
sections for $\rm gg\rightarrow b\bar{\rm b}H/A$ and  gb~$\ra$
tH$^{\pm}$ are approximately proportional to tan$^2\beta$ at large
tan$\beta$ due to the dominance of the bottom Yukawa coupling. The loop
corrections introduce some additional tan$\beta$ dependence to the cross
section, but they can be absorbed in an effective parameter
tan$\beta_{eff}$ \cite{deltamb}. The results obtained in this analysis
correspond to this effective parameter. The extraction of the
fundamental MSSM tan$\beta$ value is beyond the scope of this work. 

The branching ratio BR(H/A~$\rightarrow\tau\tau$) is approximately constant
at large tan$\beta$. At large tan$\beta$ the total decay width is
dominated by Higgs boson decays to heavy down type fermions, $\tau^+
\tau^-$ and $b\bar b$ pairs, for which the decay widths have similar
tan$\beta$ dependence. If the SUSY corrections, which are different for
the bottom and $\tau$ Yukawa couplings, are not large, the tan$\beta$
dependence cancels out in the ratio
$\Gamma$(H/A~$\rightarrow\tau\tau$)/$\Gamma_{\rm tot}$, which becomes
approximately constant. The branching ratio
BR(H$^{\pm}\rightarrow\tau\nu$) is also approximately constant at large
tan$\beta$.  The counting of signal events measures the total rate
$\sigma\times$BR into the chosen final state, which is therefore
approximately proportional to tan$^2\beta$. The total rate for
the neutral Higgs boson production as a function of tan$\beta$
is shown in Fig. \ref{fig:bbh_crosssection} for 
m$_{\rm A}$ = 300 GeV/$c^{2}$.

At large tan$\beta$ the production rate can be written as 
\begin{equation}
\sigma = \rm tan^2\beta \times\rm X,
\end{equation}
where X is the tan$\beta$ independent part of the production rate.  The
number of signal events after experimental selections is therefore
\begin{equation} 
\rm N_{\rm S} = \sigma\times\rm L\times \varepsilon_{\rm sel} = \tan^2\beta \times\rm X \times\rm L \times \varepsilon_{\rm sel}, \label{eq:NS}
\end{equation}
where L is the luminosity and $\varepsilon_{\rm sel}$ is the selection efficiency.
The value of tan$\beta$ is given by
\begin{eqnarray}
  \tan\beta & = & \tan\beta_0 \pm \Delta\rm stat \pm \Delta syst  %\nonumber \\
%            & = & \tan\beta_0 \pm \Delta\rm stat \pm (\Delta\rm lum + \Delta\rm theor)
\end{eqnarray}
where tan$\beta_0$ is the measured value of tan$\beta$.
In this work we consider systematic uncertainties $\Delta\rm syst$ due to 
the luminosity uncertainty and the uncertainty of the cross section.
%, and $\Delta\rm
%stat$, $\Delta\rm lum$, $\Delta\rm theor$ are the uncertainties which
%can be derived from equation (\ref{eq:NS}). 
The maximum error is the sum of the statistical and systematic uncertainties

\begin{eqnarray}
  \Delta \rm tan\beta / tan\beta & = & \frac{1}{2}\Delta N_{\rm S}/ \rm N_{\rm S} 
+ \frac{1}{2}\Delta\rm L/\rm L +
\frac{1}{2}\Delta\rm X/\rm X \nonumber \\
  & = & \frac{1}{2}\sqrt{\rm N_{\rm S} + \rm
N_{\rm B}}/ \rm N_{\rm S} + \frac{1}{2}\Delta\rm L/\rm L +
\frac{1}{2}\Delta\rm X/\rm X,
\end{eqnarray}

where $\rm N_{\rm S}$ and $\rm N_{\rm B}$ are the number of the signal
and background events, $\Delta \rm L / \rm L$ is the luminosity
error and $\Delta \rm X / \rm X$ consist of the theoretical uncertainties of
the cross section and the branching ratio, and the uncertainty of the cross section
due to uncertainty of the measured Higgs boson mass.
%is the theoretical uncertainty. 

The statistical errors from different H/A/h$\rightarrow\tau\tau$ final 
states are combined
using the standard weighted least-squares procedure \cite{Hagiwara:2002fs}.
The measurements are assumed to be uncorrelated and the weighted 
%average and 
error is calculated as

\begin{equation}
\overline{\rm tan\beta} \pm \overline{\Delta\rm stat}
= \frac{\Sigma_i w_i\rm tan\beta_i}{\Sigma_i w_i} \pm (\Sigma_i w_i)^{-1/2}, \label{combined}
\end{equation}

where

\begin{equation}
w_i = 1/(\Delta{\rm stat}_i)^2. \label{weight}
\end{equation}

Since the theoretical uncertainty of the associated production cross
section decreases for p$_{\rm T}^{\rm b,\bar{\rm b}}\gsim$ 20 GeV/$c$,
the question about requiring two b jets per event with jet E$_{\rm T}>$
20~GeV arises naturally.  Table \ref{table:2btagging} shows the number
of signal and background events for the H/A~$\ra\tau\tau
\ra$~lepton+jet+X channel with m$_{\rm A}$ = 200 GeV/$c^2$ and
tan$\beta$ = 20 for one b tagged jet in the event (plus a veto on
additional jets) and for events with two b tagged jets.  Although the
theoretical error is smaller for the events with two b tagged jets with
jet E$_{\rm T}>$ 20 GeV, the decrease of the signal statistics increases
the error of the measurement.  This is due to low b tagging efficiency
for soft b jets \cite{CMSNote2001/019}. The jet reconstruction is also
more difficult, if the jets are very soft.  Therefore, only one b jet
per event is assumed to be tagged in this study.  The theoretical
uncertainty of about 20\% is adopted according to
Refs.~\citer{hep-ph/0309204,dawson} for both the neutral and charged
Higgs boson production cross sections and 3\% for the branching ratios.
The error of the luminosity measurement is assumed to be 5\%.

\begin{table}[h]
  \vskip 0.1 in
  \begin{center}
  \begin{tabular}{|l|c|c|c|c|c|c|}
   \hline
    m$_{\rm A}$ = 200 GeV/$c^2$, tan$\beta$ = 20 & N$_{\rm S}$ & N$_{\rm B}$ & signif.
    & $\sqrt{\rm N_{\rm S} + N_{\rm B}}/\rm N_{\rm S}$ & $\Delta\sigma/\sigma$ & $\Delta\rm tan\beta/\rm tan\beta$$^*$ \\
   \hline
%    1b-tagging+jet veto & 205 & 91  & 21.5$\sigma$ & 8.4\%  & 20\% & 16.7\%      \\
%    2b-tagging          & 12  & 57  & 1.6$\sigma$  & 69.2\% & 10-15\%    & 42.1-44.6\% \\
    1b-tagging+jet veto & 157 & 70  & 18.8$\sigma$ & 9.6\%  & 20\%    & 17.3\%      \\
    2b-tagging          & 9   & 44  & 1.3$\sigma$  & 80.9\% & 10-15\% & 48.0-50.5\% \\
   \hline
\multicolumn{7}{l}{ \small $^{*)}$ Statistical + theoretical cross section errors only }
  \end{tabular}
  \end{center}
  \caption{The uncertainty of the tan$\beta$ measurement for the
           H/A~$\ra\tau\tau \ra$~lepton+jet+X channel for 30 fb$^{-1}$
           with one or two b tagged jets with jet E$_{\rm T}>$ 20 GeV.}
  \label{table:2btagging}
\end{table}

%\begin{figure}[h]
%  \centering
%  \vskip 0.1 in
%  \begin{tabular}{ccc}
%  \begin{minipage}{5cm}
%    \centering
%  \includegraphics[height=70mm,width=50mm]{\FIG_PATH/massFitGauss_H_a200b20.eps}
%      \caption{}
%  \label{fig:fit_H}
%  \end{minipage}
%  &
%  \begin{minipage}{5cm}
%    \centering
%  \includegraphics[height=70mm,width=50mm]{\FIG_PATH/mreso_h2tau.eps}
%      \caption{}
%  \label{fig:mreso}
%  \end{minipage}
%  &
%  \begin{minipage}{5cm}
%    \centering
%  \includegraphics[height=70mm,width=50mm]{\FIG_PATH/massFit_a200b20_2Landau1Gauss.eps}
%      \caption{}
%  \label{fig:fit_SB}
%  \end{minipage}
%  \end{tabular}
%\end{figure}

Since the value of the cross section depends on the Higgs boson mass, the 
uncertainty of the mass measurement leads to uncertainty in the signal
rate. The Higgs mass is measured using the different final states,
and the cross section uncertainty due to mass measurement errors
are combined using equations \ref{combined} and \ref{weight}
which give smallest weight to the channels with largest error. The mass 
resolution is almost constant as a function of m$_{\rm A}$, $\sim$ 24\%
for the leptonic final states, $\sim$ 17 \% for the lepton+jet final state
and $\sim$ 12 \% for the hadronic final state \cite{2lepton}. The 
uncertainty of the mass measurement is calculated from the gaussian fit of the
mass peak as $\sigma_{Gauss}/\sqrt{\rm N_{\rm S}}$, and the
error induced to the cross section ($\Delta\sigma(\Delta\rm m)$) 
is estimated by varying the cross section 
for Higgs masses
m$_0$ and m$_0\pm\sigma_{Gauss}/\sqrt{\rm N_{\rm S}}$. At 5$\sigma$ limit 
where the signal statistics is lowest, the uncertainty of the mass measurement
brings 5 - 6\% uncertainty to the tan$\beta$ measurement.

\section{Measurement of tan$\beta$} \label{sec:Results}
%in H/A $\ra\tau\tau$ and H$^{\pm}\ra\tau\nu$ } \label{sec:Results}

\subsection{H/A $\ra\tau\tau$}

Table \ref{table:errors} shows the statistical uncertainty of the
tan$\beta$ measurement and the uncertainty of the mass measurement 
for each individual final state and for the
combined final states from H/A~$\ra \tau\tau$ for 30~fb$^{-1}$.  
The total estimated uncertainty including theoretical and luminosity errors 
are shown for the combined final states.
The results are shown for
the region of the (m$_{\rm A}$, tan$\beta$) parameter space where the
statistical significance exceeds 5$\sigma$. Close to the 5$\sigma$ limit
 the statistical uncertainty is of the order of 11 - 12\%, but it
decreases rapidly for increasing tan$\beta$.

\begin{table}[h]
\centering
\vskip 0.1 in
\begin{tabular}{|c||c|c|c|c|c|c|c|c|}
\hline
\multirow{2}{2cm}{\centering \Large 30 fb$^{-1}$}
%& \multicolumn{4}{|c|}{Stat.+lumi.+theor. error\% (stat. error\%)} \\
%%%%& \multicolumn{4}{|c|}{$\Delta \rm tan\beta / tan\beta$ \% ($\Delta$stat \%)} \\
%%%% \cline{2-5}
& \multicolumn{2}{|c|}{\begin{minipage}{2.5cm}
m$_{\rm A}$~=~200~GeV/$c^2$ \\
tan$\beta$ = 20
\end{minipage}}
& \multicolumn{2}{|c|}{\begin{minipage}{2.5cm}
m$_{\rm A}$~=~200~GeV/$c^2$ \\
tan$\beta$ = 30
\end{minipage}}
& \multicolumn{2}{|c|}{\begin{minipage}{2.5cm}
m$_{\rm A}$~=~500~GeV/$c^2$ \\
tan$\beta$ = 30
\end{minipage}}
& \multicolumn{2}{|c|}{\begin{minipage}{2.5cm}
m$_{\rm A}$~=~500~GeV/$c^2$ \\
tan$\beta$ = 40
\end{minipage}}\\
\cline{2-9}
 & $\Delta$stat & $\Delta\sigma(\Delta\rm m)$ & $\Delta$stat & $\Delta\sigma(\Delta\rm m)$
 & $\Delta$stat & $\Delta\sigma(\Delta\rm m)$ & $\Delta$stat & $\Delta\sigma(\Delta\rm m)$ \\
\hline
H/A$\rightarrow\tau\tau\rightarrow$e$\mu$   & 8.95\% & 4.82\% & 4.85\% & 3.27\% & - & - & - & -  \\
H/A$\rightarrow\tau\tau\rightarrow\ell\ell$ & 7.96\% & 3.50\% & 4.08\% & 2.37\% & - & - & - & -  \\
H/A$\rightarrow\tau\tau\rightarrow\ell$j    & 4.81\% & 2.46\% & 2.84\% & 1.65\% & - & - & 8.40\% & 4.82\% \\
H/A$\rightarrow\tau\tau\rightarrow$jj       & 13.7\% & 4.73\% & 8.25\% & 3.21\% & 12.4\% & 5.82\% & 8.45\% & 4.44\% \\
\hline
\hline
\multirow{2}{3cm}{\begin{minipage}{2.cm}\bf Combined \\ e$\mu$+$\ell$j+jj\end{minipage}}
& 4.05\% & 1.99\% & 2.35\% & 1.34\% & 9.09\% & 4.28\% & 5.96\% & 3.26\% \\
\cline{2-9}
\cline{2-9}
& \multicolumn{2}{|c|}{$\Delta \rm tan\beta / tan\beta$  }
& \multicolumn{2}{|c|}{$\Delta \rm tan\beta / tan\beta$  }
& \multicolumn{2}{|c|}{$\Delta \rm tan\beta / tan\beta$  }
& \multicolumn{2}{|c|}{$\Delta \rm tan\beta / tan\beta$  } \\
\cline{2-9}
& \multicolumn{2}{|c|}{ 20.1\% }
& \multicolumn{2}{|c|}{ 17.7\% }
& \multicolumn{2}{|c|}{ 27.4\% }
& \multicolumn{2}{|c|}{ 23.3\% } \\
\hline
\multirow{2}{3cm}{\begin{minipage}{2.cm}\bf Combined \\ $\ell\ell$+$\ell$j+jj\end{minipage}}
& 3.94\% & 1.85\% & 2.24\% & 1.25\% & 9.09\% & 4.28\% & 5.96\% & 3.26\% \\
\cline{2-9}
\cline{2-9}
& \multicolumn{2}{|c|}{$\Delta \rm tan\beta / tan\beta$  }
& \multicolumn{2}{|c|}{$\Delta \rm tan\beta / tan\beta$  }
& \multicolumn{2}{|c|}{$\Delta \rm tan\beta / tan\beta$  }
& \multicolumn{2}{|c|}{$\Delta \rm tan\beta / tan\beta$  } \\
\cline{2-9}
& \multicolumn{2}{|c|}{ 19.9\% }
& \multicolumn{2}{|c|}{ 17.5\% }
& \multicolumn{2}{|c|}{ 27.4\% }
& \multicolumn{2}{|c|}{ 23.3\% } \\
\hline
%\multicolumn{5}{l}{ \small $^{*)}$ Significance $< 5\sigma$ }
\end{tabular}
\caption{Statistical uncertainties of the tan$\beta$ measurement and the
         uncertainties due to mass measurement for individual
         H/A$\rightarrow\tau\tau$ and combined final states in
         four (m$_{\rm A}$,tan$\beta$) parameter
         space point for 30 fb$^{-1}$. The total error includes statistical
         error, mass measurement error, theoretical uncertainty of the cross 
         section (20\%) and the branching ratio (3\%), and
         the luminosity uncertainty (5\%).}
\label{table:errors}
\end{table}

As shown in the table, the highest statistical accuracy, about 5\% for
m$_{\rm A}$ = 200 GeV/$c^2$ and tan$\beta$ = 20, is obtained with the
lepton+jet final state. Combining other channels with the lepton+jet
channel in this mass range improves the precision only slightly. 
%For m$_{\rm A}$ = 500 GeV/$c^2$ and tan$\beta$ = 30, the lepton+jet channel
%still yields higher statistical accuracy, about 10\%, than the two-jet
%channel ($\sim$~12\%). 
The difference between the fully leptonic
channels (e$\mu$ and $\ell\ell$) is small: the statistical uncertainty
is slightly smaller for the e$\mu$ channel, if m$_{\rm A}$ is close to
the Z peak, but already at m$_{\rm A}$ = 200 GeV/$c^2$ the final state
with any two leptons yields better statistics and lower uncertainties.
The combined $\ell\ell$+$\ell$j+jj channel yields a slightly smaller
statistical error at tan$\beta$ = 20 than the combined
e$\mu$+$\ell$j+jj channel despite the larger backgrounds in the
$\ell\ell$ final state.

\begin{figure}[h]
  \centering
  \vskip 0.1 in
    \includegraphics[width=100mm,height=80mm]{\FIG_PATH/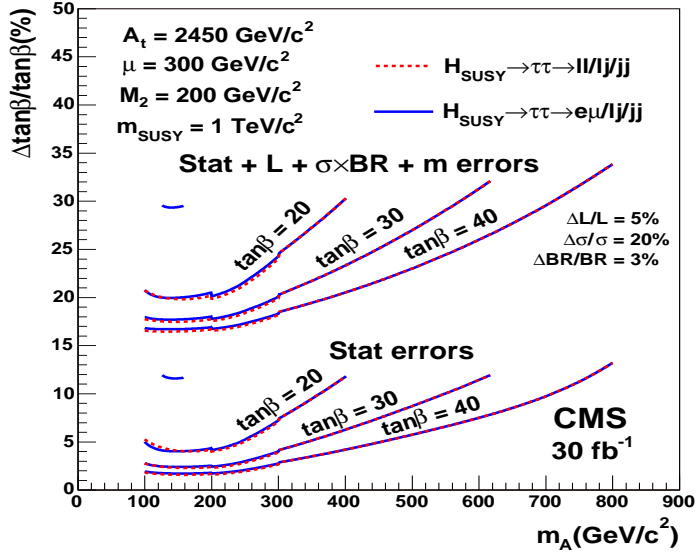}
    \caption{Three lower curves show the uncertainty of the
             tan$\beta$ measurement when statistical errors only are taken
             into account. The three upper curves show the uncertainty
             when statistical errors, the mass measurement uncertainties, 
             the luminosity uncertainty (5\%) 
             and the theoretical uncertainty of the production cross section (20\%) 
             and the branching ratio (3\%) are
             taken into account.}
  \label{fig:dtanb_mA_h2tau_emu_lj_jj}
\end{figure}

\begin{figure}[p]
  \centering
  \includegraphics[width=110mm]{\FIG_PATH/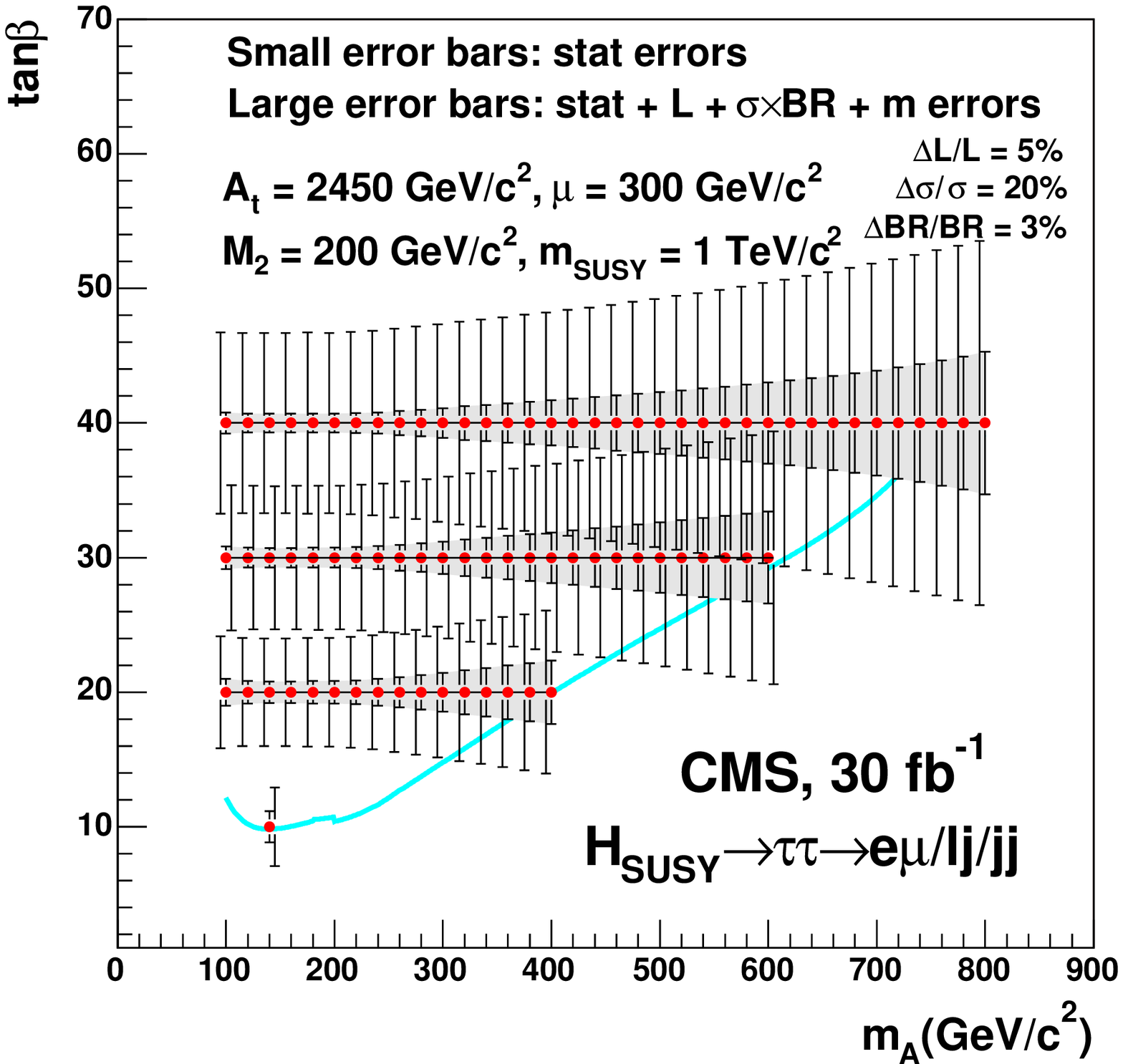}
  \caption{The uncertainty of the tan$\beta$ measurement shown as error bars.
           The small error bars and gray area show the statistical errors only.
           The large error bars show the uncertainty
           when statistical errors, the mass measurement uncertainties, 
           the luminosity uncertainty (5\%)
           and the theoretical uncertainty of the production cross section (20\%) 
           and the branching ratio (3\%) are
           taken into account.
%bars for the combined e$\mu$ + $\ell$j + jj final state from 
%H/A/h$\rightarrow\tau\tau$ as a function of m$_{\rm A}$ and tan$\beta$
%for 30 fb$^{-1}$. The gray area and smaller error bars present the
%statistical uncertainties, the larger error bars show the statistical plus
%systematic uncertainties. The error bars for the total uncertainties are
%slightly shifted due to visual reasons. 
            The solid curve corresponds to the
5$\sigma$-discovery contour of Fig.~\ref{fig:discovery_HA}.}
  \protect\label{fig:errorbars30fb}
  \centering
  \includegraphics[width=110mm]{\FIG_PATH/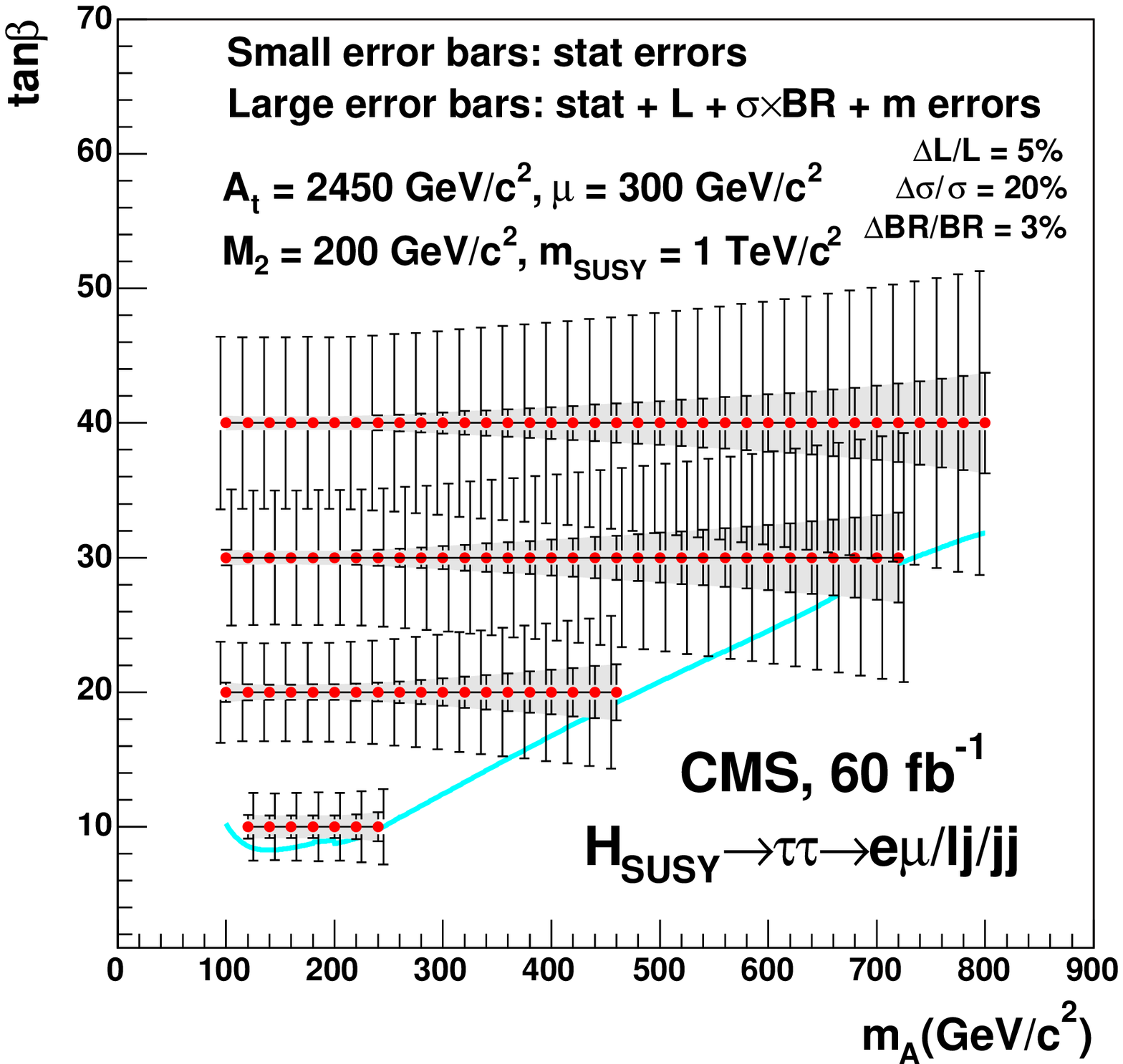}
  \caption{The same as in Fig.~\ref{fig:errorbars30fb} but for 60 fb$^{-1}$ taken at low luminosity.}
  \protect\label{fig:errorbars60fb}
\end{figure}

Figure \ref{fig:dtanb_mA_h2tau_emu_lj_jj} shows the statistical
error and the
statistical plus systematic uncertainty of tan$\beta$ for the combined
e$\mu$ + $\ell$j + jj and $\ell\ell$ + $\ell$j + jj
final states as a function of m$_{\rm A}$ for
tan$\beta$ = 20, 30 and 40, and for 30 fb$^{-1}$. 
%Figures
%\ref{fig:dtanb_mA_h2tau_ll_lj_jj} and
%\ref{fig:dtanbsyst_mA_h2tau_ll_lj_jj} show the same for the combined
%$\ell\ell$ + $\ell$j + jj final state. 
The drop in the curves at m$_{\rm
A}$ = 300 GeV/$c^2$ is due to fully leptonic final states (e$\mu$ and
$\ell\ell$) being accessible and included in the tan$\beta$ measurement
in the region from m$_{\rm A}$ = 100 GeV/$c^2$ to m$_{\rm A}$ = 300
GeV/$c^2$.  Similarly, a small decrease is visible at m$_{\rm A}$ = 200
GeV/$c^2$ due to the fully hadronic final state being analyzed only in
the region from m$_{\rm A}$ = 200 GeV/$c^2$ to m$_{\rm A}$ = 800
GeV/$c^2$.  
%Figures \ref{fig:dtanb_mA_h2tau_emu_lj_jj} to
%\ref{fig:dtanbsyst_mA_h2tau_ll_lj_jj} show the statistical and total
%uncertainties also without the contribution from the fully leptonic
%final states. At small tan$\beta$ the contributions from the e$\mu$ or
%$\ell\ell$ final states may increase the error due to the large
%Z,$\gamma^{\ast}$ background for these channels.

Figures \ref{fig:errorbars30fb} and \ref{fig:errorbars60fb} show the
error on the tan$\beta$ measurement with error bars for the combined
e$\mu$ + $\ell$j + jj channel for 30 and 60 fb$^{-1}$ at low luminosity.
The statistical uncertainties are depicted by the smaller error bars and
gray area, the uncertainties including the systematic errors are
presented with longer error bars. The errors are shown in the region
with signal significance larger than 5$\sigma$.  The combined 5$\sigma$
reach is plotted with the contribution from the e$\mu$ final state
included up to m$_{\rm A}$ = 180 GeV/$c^2$ in order to extend the
reach to lower tan$\beta$ values.  For the same reason at very high 
values of m$_{\rm A}$ only the two-jet final
state contributes to the reach. The errors are calculated within the
shown 5$\sigma$ reach using all available information, including
leptonic final states for m$_{\rm A}<$ 300 GeV/$c^2$, and
$\ell$j final state for m$_{\rm A}<$ 800 GeV/$c^2$.  
The statistical uncertainty is largest close to the
5$\sigma$ limit, where combining the different final states improves the
accuracy most. 
%The improvement is shown in Figures \ref{fig:errorbars30fb} and
%\ref{fig:errorbars60fb} as a drop in the error at m$_{\rm A}\lsim$  750
%GeV/$c^2$ (m$_{\rm A}\lsim$ 720 GeV/$c^2$) and at m$_{\rm A}\lsim$ 170
%GeV/$c^2$ . 

\subsection{H$^{\pm}\ra\tau\nu$}

\begin{table}[b]
  \vskip 0.1 in
  \begin{center}
  \begin{tabular}{|c|c|c|c|c|}
   \hline
    &  $\rm N_{\rm S}$ & $\rm N_{\rm B}$ & signif. & $\Delta \rm tan\beta / tan\beta$ \\
  \hline
   m$_{\rm H^{\pm}}$ = 200~GeV/$c^2$  &   27.5 &  3.1  & 9.2$\sigma$ & 24.1\% (10\%)  \\
   m$_{\rm H^{\pm}}$ = 400~GeV/$c^2$ &   10.4  &  2.1  & 4.9$\sigma$ & 31.0\% (17\%)  \\
  \hline
  \end{tabular}
  \end{center}
  \caption{Number of signal and background events for 30~fb$^{-1}$, statistical
           significance
           and the error of the tan$\beta$ measurement for
           gb~$\ra$tH$^{\pm}$, H$^{\pm} \ra \tau\nu$ for
           tan$\beta$ = 30 at m$_{\rm H^{\pm}}$ = 200 and 400~GeV/$c^2$.
           The statistical tan$\beta$ measurement uncertainty is shown in
           parenthesis.
           The statistical significance is calculated with poisson statistics.}
  \label{tab:hplus}
\end{table}

Table \ref{tab:hplus} contains the number of signal and background
events for 30~fb$^{-1}$, the statistical significance and the
uncertainty of the tan$\beta$ measurement for charged Higgs bosons in
the gb~$\ra$tH$^{\pm}$, H$^{\pm} \ra \tau\nu$ channel with tan$\beta$ =
30 at m$_{\rm H^{\pm}}$ = 200 and 400~GeV/$c^2$. The cross sections are
normalized to the results of Ref.~\cite{plehn} and the branching ratios
to those from HDECAY \cite{Djouadi:1998yw}. A statistical uncertainty of
$\sim$~10\% is reached at m$_{\rm H^{\pm}}$ = 200~GeV/$c^2$ and
tan$\beta$ = 30. 
For the charged Higgs bosons, the uncertainty of the the
H$^{\pm} \ra \tau\nu$ branching ratio (3\%) is taken into account
while the uncertainty of the mass measurement
has not yet been determined and is not included in the
tan$\beta$ measurement.
It should be noted that the comparison of the
tan$\beta$ measurements in the neutral and charged Higgs sectors could
yield information on the CP properties of the Higgs sector as only the
neutral sector is sensitive to CP-mixing \cite{filip}.

\subsection{H/A $\ra\mu\mu$}

In the region m$_{\rm A} \lsim$~300~GeV/$c^2$, the value of tan$\beta$
could also be measured in the H/A$\rightarrow \mu\mu$ channel
using event rates. In this channel, the Higgs mass resolution is about
2\% \cite{bellucci}. Therefore the total width of the Higgs boson could
be measured with good precision from the Higgs boson mass peak.  The
variation of the natural width as a function of tan$\beta$, from less
than 1 GeV/$c^2$ to about 20 GeV/$c^2$ in the expected tan$\beta$ range, 
could be
used to determine the value of tan$\beta$. This method, based on the
direct width measurement, would therefore be complementary to the method
explained in this note.

\section{Conclusions}

The precision of the tan$\beta$ measurement has been estimated in the
H/A~$\ra\tau\tau$ decay channel with two-lepton, lepton+jet and two-jet
final states and in the H$^{\pm} \ra \tau\nu$ decay channel of the
charged Higgs bosons. In the region of large tan$\beta$, the tan$^2
\beta$ dependence of the associated production processes gg~$\ra \rm b
\overline{\rm b}$H/A and gb~$\ra$tH$^{\pm}$ has been exploited to obtain
a statistical uncertainty being a factor of two smaller than that of the
event rates. Due to the presence of potentially large radiative
corrections to the bottom Yukawa coupling \cite{deltamb}, the results
obtained in this analysis correspond to an effective parameter
tan$\beta_{eff}$ which absorbs the leading universal part of these
corrections. A theoretical error of 20\% + 3\% (cross section and 
branching ratio) and a luminosity uncertainty of
5\% has been assumed.  If two b jets with $\rm E_{\rm T}>$~20~GeV are
tagged, the theoretical uncertainty of the cross section reduces to 
10-15\%, but the event
rates have been found to decrease significantly leading to a worse
accuracy of the tan$\beta$ measurement.
%The associated production processes were selected by tagging one b jet
%with jet $\rm E_{\rm T}>$~20~GeV/$c^2$. 
With one tagged b jet in the event the value of tan$\beta$ can be
determined in the H/A~$\ra\tau\tau$ decay channels after collecting
30~fb$^{-1}$ with an accuracy of better than $\sim$~35\%
% in the $\rm e\mu$ and $\ell\ell$ final states, 
%$\sim$~25\% in the $\ell$+jet final state  and $\sim$~27\% in the
%two-jet final state
within the 5$\sigma$-discovery reach.
% and with an integrated luminosity of 30~fb$^{-1}$.
A combination of the $\rm e\mu$, $\ell$+jet and two-jet or $\ell\ell$,
$\ell$+jet and two-jet final states develops an up to 4\% better
accuracy than the best individual final state.
%an accuracy better than $\sim$~21\% can be reached.
The uncertainty of the tan$\beta$ measurement with the charged Higgs
bosons in the H$^{\pm} \ra \tau\nu$ decay channel for tan$\beta$~=~30
has been found to be 24\% at m$_{\rm H^{\pm}}$ = 200~GeV/$c^2$ and 31\%
at m$_{\rm H^{\pm}}$ = 400~GeV/$c^2$ with an integrated luminosity of
30~fb$^{-1}$. 
The tan$\beta$ uncertainties for the charged Higgs boson do not include
the uncertainty of the mass measurement.
%A theoretical uncertainty of 20\% and a luminosity uncertainty of 5\%
%was assumed. A theoretical uncertainty of 10-15\% could be assumed if
%two b jets with $\rm E_{\rm T}>$~20~GeV/$c^2$ are tagged but the event
%rates were found to decrease significantly leading to larger
%uncertainty of the tan$\beta$ measurement.
Close to the 5$\sigma$-discovery limit the statistical uncertainty
ranges in the same order as the theoretical one, but for tan$\beta$
regions where the signal significance exceeds $5\sigma$ significantly
the theoretical error dominates.

}

%% file: mellado.tex
{
\topmargin 0cm
\textheight 24.0cm
\textwidth 16.0cm
\oddsidemargin 0cm
\evensidemargin 0cm

\renewcommand{\ssx}{\hspace*{0.6cm}}
\renewcommand{\textfraction}{-0.5}
%\renewcommand{\floatpagefraction}{-0.5}

% COMMAND DEFINITIONS
\newcommand{\tev}{{\rm Te}\kern-1.pt{\rm V}}
\newcommand{\gev}{{\rm Ge}\kern-1.pt{\rm V}}
\newcommand{\mev}{{\rm Me}\kern-1.pt{\rm V}}
\newcommand{\kev}{{\rm Ke}\kern-1.pt{\rm V}}
\newcommand{\gevsq}{\mbox{$\mathrm{{\rm Ge}\kern-1.pt{\rm V}}^2$}}
\newcommand{\gevmsq}{\mbox{$\mathrm{{\rm Ge}\kern-1.pt{\rm V}}^{-2}$}}
\newcommand{\mee}{M_{e^+e^-}}

\newcommand{\qsq}       {\mbox{$Q^{2}$}}
\newcommand{\Acce}      {\mbox{$\mathcal{A}$}}
\newcommand{\Pur}      {\mbox{$\mathcal{P}$}}
\newcommand{\Lumi}      {\mbox{$\mathcal{L}$}}
\newcommand{\BR}        {\mbox{$\mathcal{B}$}}
\newcommand{\Fracc}     {\mbox{$\mathcal{F}$}}
\newcommand{\muu}{M_{\mu^+\mu^-}}

\newcommand {\pom} {I\hspace{-0.2em}P}
\newcommand {\reg} {I\hspace{-0.2em}R}
\newcommand {\pomb} {I\hspace{-0.35em}P}
\newcommand {\regb} {I\hspace{-0.35em}R}
\newcommand {\alphapom} {\mbox{$\alpha_{_{\pom}}$}}
\newcommand {\alphappom} {\mbox{$\alpha^\prime_{_{\pom}}$}}

%pomeronin text
%\newcommand {\pom} {I\!\!P}
%pomeron as subscrript
\newcommand {\pomsub} {{\scriptscriptstyle \pom}}
\newcommand {\regsub} {{\scriptscriptstyle \reg}}
\newcommand {\ppp} {\pom\pom\pom}
\newcommand {\ppr} {\pom\pom\reg}
\newcommand {\xpom} {x_{\pomsub}}
\newcommand {\apom} {\alpha_{\pomsub}}
\newcommand {\areg} {\alpha_{\regsub}}
\newcommand {\rzfzz}     {\mbox{${r^{04}_{00}}$}}
\newcommand {\rzfpm}     {\mbox{${r^{04}_{1-1}}$}}

\newcommand\units{\,\mathrm}

\newcommand{\mayor} {\mbox{\raisebox{-0.4ex}
{$\;\stackrel{>}{\scriptstyle \sim}\;$}}}
\newcommand{\menor} {\mbox{\raisebox{-0.4ex}
{$\;\stackrel{<}{\scriptstyle \sim}\;$}}}
\newcommand{\igual} {\mbox{\raisebox{-0.4ex}
{$\;\stackrel{?}{\scriptstyle =}\;$}}}

%\newcommand{\sla}[1]{/\!\!\!#1}

%     special abrev. for ww-fusion
%
\newcommand{\qcdtau}{QCD \mbox{$\tau \tau + jets$}}
\newcommand{\ewtau}{EW \mbox{$\tau \tau + jets$}}
\newcommand{\tautau}{\mbox{$\tau \tau + jets$}}
\newcommand{\qcdww}{QCD \mbox{$W W + jets$}}
\newcommand{\ewww}{EW \mbox{$W W + jets$}}
\newcommand{\wwjets}{\mbox{$W W + jets$}}
\newcommand{\ifb}{\mbox{$\rm fb^{-1}$}}
\newcommand{\ipb}{\mbox{$\rm pb^{-1}$}}
\newcommand{\sig}{\mbox{$\sigma$}}
\newcommand{\ttb}{\mbox{$t \overline{t}$}}
\newcommand{\bbb}{\mbox{$b \overline{b}$}}
\newcommand{\qqb}{\mbox{$q \overline{q}$}}
\newcommand{\hwwll}{\mbox{$H \rightarrow W W^{(*)} \rightarrow \ell \nu \ell \nu$}}

%
%    some useful tex-constructs, to make surviving at Tex level
%    easier....
%    K.J.                                          Aug. 1994
%
%
%    definition of textcolors and colorboxes...
\newcommand{\tc}{\textcolor}
\newcommand{\tcr}{\textcolor{red}}
\newcommand{\tcg}{\textcolor{green}}
\newcommand{\tcb}{\textcolor{blue}}
\newcommand{\tcm}{\textcolor{magenta}}
%
%    important symbols for real physics (pp collider)
%
\newcommand{\ppbar}{\mbox{$p\overline{p}$}}
\newcommand{\rts}{\mbox{$\sqrt{s}$}}
\newcommand{\alphas}{\mbox{$\alpha_s$}}
\newcommand{\PT}{\mbox{$P_T$}}
\newcommand{\pt}{\mbox{$P_T$}}
\newcommand{\etmiss}{\mbox{$E_T^{miss}$}}
\newcommand{\ptmiss}{\mbox{$P_T^{miss}$}}
\newcommand{\ptw}{\mbox{$P_T(W)$}}
\newcommand{\ptz}{\mbox{$P_T(Z)$}}
\newcommand{\deleta}{\mbox{$\Delta \eta \times \Delta \phi$}}
\newcommand{\linteg}{\mbox{$\int{{\cal L} dt}$}}
\newcommand{\abseta}{\mbox{$\mid \eta \mid$}}
\newcommand{\lqcd}{\mbox{$\Lambda_{QCD}$}}
\newcommand{\QSQ}{\mbox{$Q^{2}$}}
\newcommand{\sla}[1]{/\!\!\!#1}
\newcommand{\mevp}{MeV/$c$}
\newcommand{\meve}{MeV/$c^2$}
\newcommand{\gevp}{GeV/$c$}
\newcommand{\geve}{GeV/$c^2$}
\newcommand{\Gcs}{${\rm{GeV/}c^2}$}
\newcommand{\Tcs}{${\rm{TeV/}c^2}$}
\newcommand{\Gc}{${\rm{GeV/}c}$}
\newcommand{\sw}{\sin\theta_W}
\newcommand{\cw}{\cos\theta_W}
\newcommand{\stbr}{\sigma \cdot BR}
% 

%
%   e+ e- physics
%
\newcommand{\epem}{\mbox{${\rm e}^+{\rm e}^-$}}
\newcommand{\mpmm}{\mbox{$\mu^+\mu^-$}}
\newcommand{\zzero}{\mbox{Z$^0$}}
%
%    B physics 
%
\newcommand{\bs}{\mbox{$B_s^0$}}
\newcommand{\bd}{\mbox{$B_d^0$}}
\newcommand{\bsb}{\mbox{$\bar{B}_s^0$}}
\newcommand{\bdb}{\mbox{$\bar{B}_d^0$}}
\newcommand{\dms}{\mbox{$\Delta m_s$}}
%
%    K physics
% 
\newcommand{\ks}{\mbox{$K_S$}}
\newcommand{\kl}{\mbox{$K_L$}}
%
%    general symbols
%
\newcommand{\cl}{\mbox{$95 \% \ CL$}}
\newcommand{\pizero}{\mbox{$\pi^0$}}
\newcommand{\xrad}{\mbox{$X_0$}}
\newcommand{\degr}{\mbox{$^{\circ}$}}
\newcommand{\dsdpt}{\mbox{$frac{d\sigma}{dP_T}$}}
\newcommand{\jpsi}{\mbox{$J/\psi K_s^0$}}
\newcommand{\btojpsi}{\mbox{$B^0_d \rightarrow  J/\psi K_s^0$}}
\newcommand{\btopipi}{\mbox{$B^0_d \rightarrow  \pi \pi$}}
\newcommand{\sbeta}{\mbox{$\sin 2 \beta$}}
\newcommand{\salpha}{\mbox{$\sin 2 \alpha$}}
\newcommand{\dsbeta}{\mbox{$\Delta \sin 2 \beta$}}
\newcommand{\dsalpha}{\mbox{$\Delta \sin 2 \alpha$}}
%
%     Higgs decays 
%
\newcommand{\hgg}{\mbox{$H \rightarrow  \gamma \gamma$}}
\newcommand{\hbb}{\mbox{$H \rightarrow  b \bar{b}$}}
\newcommand{\hzz}{\mbox{$H \rightarrow  Z Z$}}
\newcommand{\hfourl}{\mbox{$H \rightarrow Z Z^{*} \rightarrow 4\ell$}}
\newcommand{\hzzfourl}{\mbox{$H \rightarrow Z Z \rightarrow 4\ell$}}
\newcommand{\hww}{\mbox{$H \rightarrow  W W $}}
\newcommand{\hwws}{\mbox{$H \rightarrow  W W^{(*)}$}}
\newcommand{\hwwsll}{\mbox{$H \rightarrow  W W^{(*)} \rightarrow \ell \nu \ell \nu $}}
\newcommand{\hzzs}{\mbox{$H \rightarrow  Z Z^*$}}
\newcommand{\gamgam}{\mbox{$\gamma \gamma$}}
\newcommand{\htautau}{\mbox{$H \rightarrow \tau \tau$}}

%
%   Higgs production
%
\newcommand{\vbfprocess}{\mbox{$qq \rightarrow qq H $}}

%
%     W decays 
%
\newcommand{\wlnu}{\mbox{$W \rightarrow \ell \nu $}}
\newcommand{\wenu}{\mbox{$W \rightarrow e \nu $}}
\newcommand{\zll}{\mbox{$Z \rightarrow \ell \ell $}}
\newcommand{\zmumu}{\mbox{$Z \rightarrow \mu \mu $}}
%
%     other LHC processes
%
%
\newcommand{\ttbar}{\mbox{$t \overline{t} $}}
\newcommand{\bbbar}{\mbox{$b \overline{b} $}}
\newcommand{\qqbar}{\mbox{$q \overline{q} $}}
\newcommand{\Zbbbar}{\mbox{$Z b \overline{b} $}}
\newcommand{\ttjets}{\mbox{$t \overline{t} + jets$}}

%
%
%     LHC luminosities
%
\newcommand{\fbs}{\mbox{$\rm{fb}^{-1}$}}
\newcommand{\pbs}{\mbox{$\rm{pb}^{-1}$}}
\newcommand{\lhigh}{\mbox{${\cal L} = 10^{34} \ \rm{cm}^{-2} \rm{sec}^{-1}$}}
\newcommand{\llow}{\mbox{${\cal L} =  10^{33} \ \rm{cm}^{-2} \rm{sec}^{-1}$}}
\newcommand{\lintyear}{\mbox{$\int {\cal L} dt \ = \ 10 \ fb^{-1}$}}
\newcommand{\lintlow}{\mbox{$\int {\cal L} dt \ = \ 30 \ fb^{-1}$}}
\newcommand{\linthigh}{\mbox{$\int {\cal L} dt \ = \  100 \ fb^{-1}$}}
%
%     SUSY parameters
%
%  (i) symbols for particles 
%
\newcommand{\chiplus}{\mbox{$\chi^+$}}
\newcommand{\chiminus}{\mbox{$\chi^-$}}
\newcommand{\chizero}{\mbox{$\chi^0$}}
\newcommand{\chipm}{\mbox{$\chi^{\pm}$}}
\newcommand{\chione}{\mbox{$\chi^0_1$}}
\newcommand{\chitwo}{\mbox{$\chi^0_2$}}
\newcommand{\chithree}{\mbox{$\chi^0_3$}}
\newcommand{\sel}{\mbox{$\tilde{e}$}}
\newcommand{\smu}{\mbox{$\tilde{\mu}$}}
\newcommand{\stau}{\mbox{$\tilde{\tau}$}}
\newcommand{\slep}{\mbox{$\tilde{l}$}}
\newcommand{\slepton}{\mbox{$\tilde{l}$}}
\newcommand{\sfer}{\mbox{$\tilde{f}$}}
\newcommand{\sneu}{\mbox{$\tilde{\nu}$}}
\newcommand{\stauone}{\ensuremath{\tilde{\tau}_1 }}
\newcommand{\stautwo}{\ensuremath{\tilde{\tau}_2 }}
\newcommand{\sq}{\mbox{$\tilde{q}$}}
\newcommand{\sgl}{\mbox{$\tilde{g}$}}
\newcommand{\st}{\mbox{$\tilde{t}$}}
\newcommand{\sbot}{\mbox{$\tilde{b}$}}
%
% (ii) SUSY masses 
%
\newcommand{\msquark}{\mbox{$m_{\tilde{q}}$}}
\newcommand{\mgluino}{\mbox{$m_{\tilde{g}}$}}
\newcommand{\mzero}{\mbox{$m_{0}$}}
\newcommand{\msneu}{\mbox{$m_{\tilde{\nu}}$}}
\newcommand{\mchiplus}{\mbox{$m_{\tilde{\chi}^+}$}}
\newcommand{\mchi}{\mbox{$m_{\tilde{\chi^0}}$}}
%
%   SUSY parameters 
%
\newcommand{\tanb}{\mbox{$\tan \beta$}}
\newcommand{\matb}{\mbox{$(m_A,\tan \beta )$}}
\newcommand{\atau}{\ensuremath{A_{\tau} }}
\newcommand{\phitau}{\ensuremath{\varphi_{\tau} }}
\newcommand{\delm}{\ensuremath{\Delta M}}
\newcommand{\mumtwo}{\ensuremath{(\mu - M_2)}}

{

\noindent
{\Large \bf E. Prospects for Higgs Searches via VBF at the LHC with the ATLAS Detector} \\[0.5cm]
{\it K.~Cranmer, Y.Q.~Fang, B.~Mellado, S.~Paganis, W.~Quayle and
Sau~Lan~Wu}

\begin{abstract}
We report on the potential for the discovery of a Standard Model
Higgs boson with the vector boson fusion mechanism in the mass
range $115<M_H<500\,\gev$ with the ATLAS experiment at the LHC.
Feasibility studies at hadron level followed by a fast detector
simulation have been performed for $H\rightarrow
W^{(*)}W^{(*)}\rightarrow l^+l^-\sla{p_T}$,
$H\rightarrow\gamma\gamma$ and $H\rightarrow ZZ\rightarrow
l^+l^-q\overline{q}$. The results obtained show a large discovery
potential in the range $115<M_H<300\,\gev$. Results obtained with
multivariate techniques are reported for a number of channels.
\end{abstract}

\section{Introduction}
\label{sec:introduction}

In the Standard Model (SM) of electro-weak and strong
interactions, there are four types of  gauge vector bosons (gluon,
photon, W and Z) and twelve types of fermions (six quarks and six
leptons)~\cite{np_22_579,prl_19_1264,sal_1968_bis,pr_2_1285}.
These particles have been observed experimentally. The SM also
predicts the existence of one scalar boson, the Higgs
boson~\cite{pl_12_132,prl_13_508,pr_145_1156,prl_13_321,prl_13_585,pr_155_1554}.
The observation of the Higgs boson remains one of the major
cornerstones  of the SM. This is a primary focus of the ATLAS
Collaboration~\cite{LHCC99-14}.

The  Higgs at the LHC is produced predominantly via gluon-gluon
fusion~\cite{prl_40_11_692}. For Higgs masses, $M_H$,  such that
$M_H>100\,\gev$, the second dominant process is vector boson
fusion (VBF)~\cite{pl_136_196,pl_148_367}.

Early  analyses performed at the parton level with the decays
$H\rightarrow W^{(*)}W^{(*)}$, $\tau^+\tau^-$  and $\gamma\gamma$
via VBF indicated that this mechanism could produce strong
discovery modes in the range of the Higgs mass
$115<M_H<200\,\gev$~\cite{pr_160_113004,pl_503_113,pr_61_093005,JHEP_9712_005}.
The ATLAS collaboration has performed feasibility studies for
these decay modes including more detailed detector description and
the implementation of initial-state and final-state parton showers
(IFSR), hadronization and multiple
interactions~\cite{SN-ATLAS-2003-024}.

Here, we present an update of the potential of observing the Higgs
boson via VBF with $H\rightarrow W^{(*)}W^{(*)}\rightarrow
l^+l^-\sla{p_T}$, where $\sla{p_T}$ stands for missing transverse
momentum carried by neutrinos, reported
in~\cite{SN-ATLAS-2003-024}.  This analysis has been extended to
larger Higgs masses. Also, we investigated the prospects of
observing a SM Higgs boson with $H\rightarrow\gamma\gamma$ and
$H\rightarrow ZZ\rightarrow l^+l^-q\overline{q}$. Results obtained
with multivariate techniques are reported for a number of
channels. Finally, the status of the overall SM Higgs discovery
potential of the ATLAS detector is presented.

\section{Experimental Signatures}
\label{sec:expsig}

The VBF mechanism displays a number of distinct features, which
may be exploited experimentally to suppress SM backgrounds: Higgs
decay products are accompanied by two energetic forward jets
originating from incoming quarks and suppressed jet production in
the central region is expected due to the lack of color flow
between the initial state quarks. In this paper, tagging jets are
defined as the highest and next highest transverse momentum,
$P_T$, jets in the event. A number of variables were used in the
VBF analyses reported in this paper: $P_T$ of the leading and the
sub-leading jets, $P_{Tj_1}$ and $P_{Tj_2}$, pseudorapidity,
$\eta$, of the leading and sub-leading jets, $\eta_{j_1}$ and
$\eta_{j_1}$, with
$\Delta\eta_{j_1j_2}=\left|\eta_{j_1}-\eta_{j_2}\right|$, the
difference of tagging jets' azimuthal angles,
$\Delta\phi_{j_1j_2}$ and tagging jets' invariant mass,
$M_{j_1j_2}$. The tagging jets are required to be in opposite
hemispheres ($\eta_{j_1}\eta_{j_2}<0$).

In Sections~\ref{sec:hww} and~\ref{sec:hzz} a number of variables
are used: pseudorapidity and azimuthal angle difference between
leptons, $\eta_{ll}$ and $\phi_{ll}$, and di-lepton invariant
mass, $M_{ll}$. In Section~\ref{sec:hgg} the following variables
are used: $P_T$ of the leading and sub-leading $\gamma$'s,
$P_{T\gamma_1}$ and $P_{T\gamma_2}$, pseudorapidity and azimuthal
angle difference between $\gamma$'s, $\eta_{\gamma\gamma}$ and
$\phi_{\gamma\gamma}$, and di-$\gamma$ invariant mass,
$M_{\gamma\gamma}$.

The following decay chains have been considered in the analysis:
$H\rightarrow W^{(*)}W^{(*)}\rightarrow l^+l^-\sla{p_T}$, $H
\rightarrow \gamma \gamma$ and $H\rightarrow ZZ\rightarrow
l^+l^-q\overline{q}$. A number of relevant experimental aspects
have been addressed in detail
in~\cite{LHCC99-14,SN-ATLAS-2003-024} and will not be touched upon
in this work: triggering, lepton and photon identification, fake
lepton and photon rejection, jet tagging, central jet veto and
b-jet veto efficiencies.\footnote{The central jet and b-jet vetoes
are applied if a jet (b-jet) with $P_T>20\,\gev$ is found in the
range $\left|\eta\right|<3.2$ and $\left|\eta\right|<2.5$,
respectively.}

\section{Signal}
\label{sec:signal}

 The Born level cross-sections for the VBF process have been
calculated using the PYTHIA
package~\cite{cpc_82_74,cpc_135_238}.\footnote{The results from
PYTHIA agree with the calculation provided by VV2H~\cite{VV2H} by
better than $3\,\%$.} The results are given in
Tables~\ref{t:sig_br}-\ref{t:sig_br2} for different Higgs masses.
The signal (and background) Born level cross-sections have been
computed using the CTEQ5L structure function
parametrization~\cite{epj_12_375}. The products of the signal
cross-sections and the branching ratios of the Higgs boson into
$W^{(*)}W^{(*)}$, $\gamma\gamma$, and $ZZ$, which have been
calculated using the programme HDECAY~\cite{Djouadi:1998yw}, are also
included in Table~\ref{t:sig_br}-\ref{t:sig_br2}.

\begin{table}[t]
\begin{center}
%\begin{minipage}{.75\linewidth}
\footnotesize
\begin{tabular}{l r || c  c  c  c  c  c  c  }
\hline \hline
$m_H$ &  $(\gev)$  & 120 & 130 & 140 & 150 & 160 & 170 & 180\\
\hline \hline
$\sigma (q\overline{q} H)$ & (pb) & 4.29 & 3.97 & 3.69 & 3.45 & 3.19 & 2.95 & 2.80 \\
\hline $\sigma \cdot BR (H \rightarrow W^{(*)}W^{(*)})$ & (fb) &
522 & 1107 & 1771 & 2363 & 2887 & 2850 & 2618 \\
$\sigma \cdot BR (H \rightarrow \gamma \gamma)$ & (fb) &
 9.38 & 8.89 & 7.14 & 4.76 & 1.71 & -  & -   \\
\hline \hline
\end{tabular}
\footnotesize \caption{\small \it Total vector boson fusion
production cross-sections, $\sigma (q\overline{q}H)$,  $\sigma
\cdot BR (H \rightarrow W^{(*)}W^{(*)})$ and $\sigma \cdot BR (H
\rightarrow \gamma \gamma )$ for a low mass Higgs. The
cross-sections have been computed using the CTEQ5L structure
function parametrization. }\label{t:sig_br}
%\end{minipage}
\end{center}
\end{table}

\begin{table}[t]
\begin{center}
%\begin{minipage}{.75\linewidth}
\footnotesize
\begin{tabular}{l r || c  c  c  c  c  c  c  c }
\hline \hline
$m_H$ &  $(\gev)$ & 190 & 200 & 250 & 300 & 350 & 400 & 450 & 500 \\
\hline \hline
$\sigma (q\overline{q} H)$ & (pb) & 2.58 & 2.44 & 1.82 & 1.38 & 1.06 & 0.83 & 0.66 & 0.53 \\
\hline $\sigma \cdot BR (H \rightarrow WW)$ & (fb) &
2005 & 1793 & 1276 & 954 & 721 & 488 & 363 & 289 \\
$\sigma \cdot BR (H \rightarrow ZZ)$ & (fb) &
 562 & 637 & 542 & 424 & 332 & 227 & 172 & 138 \\
\hline \hline
\end{tabular}
\footnotesize \caption{\small \it Total vector boson fusion
production cross-sections, $\sigma \cdot BR (H \rightarrow WW)$
and $\sigma \cdot BR (H \rightarrow ZZ )$ for a heavier Higgs. The
cross-sections have been computed using the CTEQ5L structure
function parametrization. }\label{t:sig_br2}
%\end{minipage}
\end{center}
\end{table}

The impact of initial and final state QCD radiation,
hadronization, multiple interactions and underlying event were
simulated with PYTHIA6.1~\cite{cpc_82_74,cpc_135_238}. The signal
and background simulation used the package ATLFAST~\cite{ATLFAST}
in order to simulate the response of the ATLAS detector.

\section{The $H\rightarrow W^{(*)}W^{(*)}\rightarrow l^+l^-\sla{p_T}$ Mode}
\label{sec:hww}

A study of this mode at hadron level followed by a fast simulation
of the ATLAS detector was first performed
in~\cite{ButtarHarperJakobs}. In this Section we report on a
re-analysis over a broader mass range $115<M_H<500\,\gev$.
Additionally, the treatment of the main background process is
improved in the present analysis.

\subsection{Background Generation}
\label{sec:wwbackround}

\subsubsection{$t\overline{t}$ Production Associated with Jets}\
\label{sec:tt} The production of $t\overline{t}$ associated with
one jet, $t\overline{t}j$, was identified as the main background
process for this mode~\cite{pr_160_113004,pl_503_113}. Early
parton level analyses were based on $t\overline{t}j$ Leading Order
(LO) Matrix Element (ME) calculation. In order to assess
hadronization and detector effects, it is necessary to interface
the fixed order ME calculations with a parton shower in a
consistent way. Here we use a Next-to-Leading-Order description of
the $t\overline{t}$ ME matched with parton shower provided within
the MC@NLO package, which avoids double-counting and allows for a
smooth matching between hard and soft/collinear emission
regions~\cite{JHEP_0206_029,JHEP_0308_007}.
\begin{figure}[t]
    \begin{center}
        \vspace{-0.5cm}
      \mbox{\psfig{figure=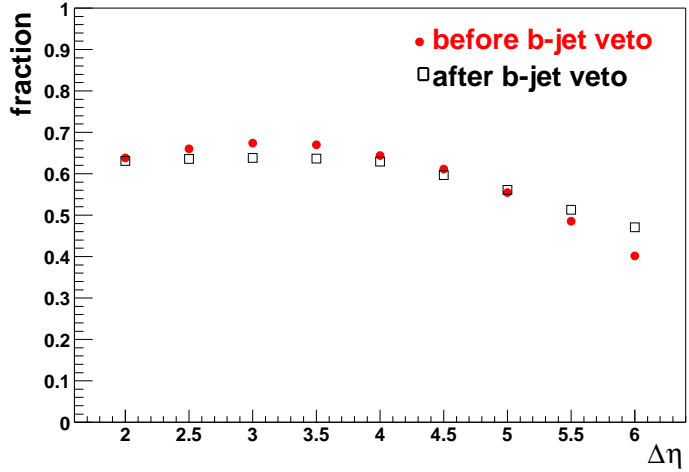,width=4.0in}}
\vspace{-0.2cm}
      \caption{
Fraction of events for which either the leading or the sub-leading
jet is a b-jet as a function of the cut on $\Delta\eta_{j_1j_2}$
before and after the application of a b-jet veto.}
        \label{1B}
    \end{center}
\end{figure}
In MC@NLO hard emissions are treated as in NLO calculations while
soft/collinear emissions are handled by the MC simulation
(HERWIG6.5 in this case) with the MC logarithmic accuracy: the
$t\overline{t}$ rates are known to NLO while the parton shower
part preserves unitarity. Comparisons between MC@NLO and LO event
generators PYTHIA6.2~\cite{cpc_82_74,cpc_135_238} and
HERWIG6.5~\cite{JHEP_0101_010,HERWIG6.5} show that, within the
MC@NLO approach, the low $P_T$ region is dominated by the parton
shower prescription, while at higher $P_T$ the NLO calculation
dominates predicting a significantly higher $P_T$ for the
$t\overline{t}$ system.

PYTHIA6.2 predicts a softer $P_T$ distribution with strong
differences in the high $P_T$ region ($P_T>100\,\gev$) with
respect to the NLO prediction. It was also found that all three
models give similar b-jet $P_T$ distribution.

The MC@NLO  description of the second jet from the
$t\overline{t}jj$ process was tested against a LO
$t\overline{t}jj$ ME calculation using
MadGraphII~\cite{pc_81_357,hep-ph_0208156} interfaced to
HERWIG6.5~\cite{Wisc_soft}. To reduce the double-counting in the
HERWIG6.5 interface with MadGraphII, the parton shower cutoff was
set to the $P_T$ of the lowest $P_T$ QCD parton in the ME
calculation. The resulting $P_T$ distribution comparison showed
that MC@NLO predicts a sub-leading non-b jet which is in good
agreement for $P_T>50\,\gev$ with the MadGraphII $t\overline{t}jj$
ME calculation. In conclusion, MC@NLO also provides a
reasonable description of the sub-leading radiation.

MC@NLO was used to define a  $t\overline{t}j$ control sample via
an event selection similar to the one used
in~\cite{pr_160_113004,pl_503_113,pr_61_093005,JHEP_9712_005}.
The dependence of various kinematic distributions on
$\Delta\eta_{j_1j_2}$ was evaluated. In a large fraction
($\simeq20\%$) of events with small values of
$\Delta\eta_{j_1j_2}$, both leading jets are b-jets. For
$\Delta\eta_{j_1j_2}>3.5$ about $65\%$ of the events have just one
of the two leading jets being a b-jet (see Figure~\ref{1B}). This fraction is clearly
dominated by $t\overline{t}j$ where the extra jet is hard. The
rest of the events were examined and about $30\%$ were found to have two
leading jets that are non-b-jets. These events are dominated by
$t\overline{t}jj$ where the two radiated partons are hard.

The results presented here show a small dependence of the jet
topology on the b-jet veto.  Only the third most energetic jet is
affected but the reduction of the fraction of events for which the
third jet is a b-jet is nearly constant as a function of the cut
on $\Delta\eta_{j_1j_2}$. According to these results, it is
possible to define a control sample in the early stages of data
taking with ATLAS to study properties of the $t\overline{t}$
process (for instance, normalization, central jet veto, b-jet
veto). One would like to use the part of the phase space which is
dominated by $t\overline{t}j$ and this is clearly the region for
which the separation of the tagging jets is
$\Delta\eta_{j_1j_2}\mayor3.5$. For a $<10\%$ systematic error in
the normalization of the $t\overline{t}j$ background about
$300-500$ pb$^{-1}$ of integrated luminosity will be
needed.\footnote{More details on this work are available
in~\cite{ATL-COM-PHYS-2003-043}.}

\subsubsection{Other Background Processes}
Other background processes were
considered~\cite{SN-ATLAS-2003-024}:
\begin{itemize}
\item Electro-weak $WWjj$ production; a quark scattering process
mediated by a vector boson, where the W bosons are produced and
decay leptonically.  This process is the second-dominant
background for most masses. To model this process, we use a
ME~\cite{zeppenfeld_1} that has been interfaced to
PYTHIA6.1~\cite{Mazini}. \item QCD $WWjj$ production. For this
process, we use the generator provided in PYTHIA6.1. \item
Electro-weak $Zjj$ production.  A $Z$ boson is produced in a
weak-boson-mediated quark-scattering process and decays into
$\tau$'s, which in turn decay leptonically.  This process was
modelled using a LO ME from the MadCUP project~\cite{MadCUP}.
\item QCD $Zjj$ production.  For this process, we use a LO ME from
the MadCUP project. As before, we consider events where
$Z\rightarrow\tau^+\tau^-$, $\tau\rightarrow l\nu\nu$. \item QCD
$Zjj$ production with $Z\rightarrow l^+l^-$ and $l=e,\mu$. This
background can be reduced substantially by requiring a minimum
missing $P_{T}$. However, it cannot be ignored because of its
large cross-section. We model this process with the generator
provided within PYTHIA6.1.\footnote{In the final version of this
work this process will be treated with a LO ME provided within
MadCUP.}
\end{itemize}

\subsection{Event Selection}
\label{sec:hwwevsel}

\begin{table}[t]
\begin{center}
\begin{tabular}{l || c c c c c c}
\hline \hline
Cut               & VBF       & $t\overline{t}$       & EW $WW$ & QCD $WW$        & EW $Zjj$        & QCD $Zjj$ \\
\hline \hline
{\bf a}     & 33.2  & 3.34$\times 10^3$      & 18.2  & 670   & 11.6  & 2.15$\times 10^3$     \\
{\bf b}     & 13.1  & 128   & 11.1  & 3.58  & 3.19  & 66.9  \\
{\bf c}     & 12.4  & 117   & 10.5  & 3.31  & 1.13  & 19.6  \\
{\bf d}     & 10.1  & 85.1  & 7.74  & 0.95 & 0.96  & 8.55  \\
{\bf e}     & 7.59  & 13    & 5.78  & 0.69 & 0.90 & 6.01  \\
{\bf f}     & 5.67  & 2.26  & 1.03  & 0.16 & 0.27 & 0.92 \\
{\bf g}     & 4.62  & 1.12  & 0.44 & 0.1        & 0.01        & 0.02        \\
\hline \hline
\end{tabular}
\caption{Cut flow for $M_{H}=160\,\gev$ in the $e-\mu$ channel.
Effective cross-sections are given in fb.  The event selection
presented in Section~\ref{sec:hwwevsel} is used. MC@NLO was used
to estimate the contribution from $t\overline{t}$ production (see
Section~\ref{sec:tt}) }\label{tab:mcanloCuts}
\end{center}
\end{table}

Our procedure for optimizing the cuts is as follows:  Begin with a
set of loose (pre-selection) cuts and choose cuts on
$\Delta\eta_{j_1j_2}$, $\Delta\eta_{ll}$, $\Delta\phi_{ll}$,
$M_{j_1j_2}$, $M_{ll}$, and the transverse mass,
$M_{T}$,\footnote{The transverse mass is defined as
in~\cite{pr_160_113004,pl_503_113,pr_61_093005,JHEP_9712_005}.}
that optimize $S/\sqrt{B}$, where S and B are the expected number
of signal and background events for $30\,$fb$^{-1}$ of luminosity,
respectively. We perform this optimization with a genetic
algorithm~\cite{GAlib}. We perform this procedure for several
masses and find a parametrization for the optimal cut as a
function of the Higgs mass.

The following event selection was chosen:
\begin{itemize}
\item[{\bf a.}] Topology cuts. Require two charged leptons
($e,\mu$) that pass the single or double charged lepton trigger in
ATLAS. Here, a veto on b-jets is applied (see
Section~\ref{sec:expsig} and~\cite{SN-ATLAS-2003-024}). \item[{\bf
b.}] Forward jet tagging with $P_{Tj_1},P_{Tj_2}>20\,\gev$ and
$\Delta\eta_{j_1j_2}^{min}<\Delta\eta_{j_1j_2}$ according to
\begin{equation}
\Delta\eta_{j_1j_2}^{min}={a\over(M_{H}-b)}+cM_{H}+d,
\end{equation}
where $a=2861$, $b=-327$, $c=9.6\times 10^{-3}$, and $d=-3.44$.
Leptons are required to be in between jets in pseudorapidity.
\item[{\bf c.}]  Tau
rejection~\cite{pr_160_113004,pl_503_113,pr_61_093005,JHEP_9712_005,SN-ATLAS-2003-024}.
\item[{\bf d.}]  Tagging jets invariant mass:
$520\,\gev<M_{j_1j_2}<3325\,\gev$ \item[{\bf e.}] Central jet veto
(see Section~\ref{sec:expsig} and~\cite{SN-ATLAS-2003-024}).
\item[{\bf f.}] Lepton angular cuts: We require
$\Delta\eta_{ll}<\Delta\eta_{ll}^{max}$ with
\begin{equation}
\Delta\eta_{ll}^{max}=a+bM_H+cM_H^2,
\end{equation}
where $a=6.25$, $b=-6.24\times 10^{-2}$, $c=1.99\times 10^{-4}$
for $M_H<200\,\gev$, and $a=3.88$, $b=-4.17\times 10^{-3}$, $c=0$
for $M_H>200\,\gev$. It is required that
$\Delta\phi_{ll}^{min}<\Delta\phi_{ll}<\Delta\phi_{ll}^{max}$ with
\begin{equation}
\Delta\phi_{ll}^{min}=a+bM_H,
\end{equation}
where $a=-2.20$, $b=7.54\times 10^{-3}$, and
\begin{equation}
\Delta\phi_{ll}^{max}=a+bM_H+cM_H^2+dM_H^3,
\end{equation}
where $a=-4.07$, $b=0.156559$, $c=-1.310956\times 10^{-3}$, and
$d=3.42011\times 10^{-6}$. As one would expect, the minimum cut is
only important at high Higgs masses, and the upper bound is only
relevant at low Higgs masses. It is
 required that $M_{ll}^{min}<M_{ll}<M_{ll}^{max}$ with
 \begin{equation}
M_{ll}^{min}=a (M_{H}-b)^{2}+c,
 \end{equation}
 where $a=-2.82\times 10^{-3}$, $b=464$, $c=129$, and
\begin{equation}
M_{ll}^{max}={{a (M_{H}-b)^{2}} \over {d + (M_{H}-b)^{2}}} +c,
\end{equation}
where $a=310$, $b=114$, $c=47.6$, and $d=13290$. In order to
further reduce the
 contribution from Drell-Yan, we
 require $85<M_{ll}<95\,\gev$ and $\sla{p_T}>30\,\gev$, if leptons are of same flavor.
\item[{\bf g.}] Transverse mass cuts. We require that
 $M_{T}^{min}<M_{T}<M_{T}^{max}$,  with
\begin{equation}
M_{T}^{min}=a+bM_{H},
\end{equation}
where $a=-17$ and $b=0.73$ and
\begin{equation}
M_{T}^{max}=a+bM_{H}+cM_{H}^{2}+dM_{H}^3,
\end{equation}
where $a=106$, $b=-0.83$, $c=9.46\times 10^{-3}$, and
$d=-9.49\times 10^{-6}$. We also require $m_T(ll\nu\nu)>30\,\gev$,
with
$m_T(ll\nu\nu)=\sqrt{2P^{ll}_T\sla{p_T}(1-\cos{\Delta\phi})}$,
where $P^{ll}_T$ is the $P_T$ of the di-lepton system and
$\Delta\phi$ corresponds to the angle between the di-lepton vector
and the $\sla{p_T}$ vector in the transverse plane.

\end{itemize}

\begin{table}[t]
\begin{center}
\begin{tabular}{c || c c c c}
\hline \hline
$M_{H}(\gev)$ & $e-\mu$ & $e-e$ & $\mu-\mu$ & Combined\\
\hline \hline
115     & 0.9   & 0.4   & 0.5   & 1.4\\
130     & 3.0   & 1.5   & 2.2   & 4.3\\
160     & 8.2   & 5.1   & 6.3   & 11.6\\
200     & 4.4   & 2.6   & 3.0   & 6.0\\
300     & 2.3   & 1.4   & 1.5   & 3.1\\
500     & 1.0   & 0.6   & 0.6   & 1.5\\
\hline \hline
\end{tabular}
\caption{Expected Poisson significance for the parameterized cuts
listed in Section~\ref{sec:hwwevsel} with 10\,fb$^{-1}$ of
integrated luminosity. A 10\% systematic uncertainty is applied to
all backgrounds when calculating the significance.
}\label{table:mcatnloSig}
\end{center}
\end{table}

\subsection{Results and Discovery Potential}
\label{sec:hwwres}

Table~\ref{tab:mcanloCuts} displays effective cross-sections for
signal and background after application of successive cuts
presented in Section~\ref{sec:hwwevsel}. Cross-sections are
presented for $M_H=160\,\gev$ in the $e-\mu$ channel. It is worth
noting that the central jet veto survival probability for
$t\overline{t}$ production is significantly lower than that
reported in~\cite{SN-ATLAS-2003-024}. However, this is compensated by a lower rejection due to requiring two tagging jets (see cut {\bf b} in the previous Section). As a result, the relative contribution to the background from $t\overline{t}$ production obtained here is similar to the one reported in~\cite{SN-ATLAS-2003-024}.
Table~\ref{table:mcatnloSig}
reports the expected Poisson significance for 10\,fb$^{-1}$ of
integrated luminosity. Simple event counting is used and a
$10\,\%$ systematic error on the background determination was
assumed. In order to incorporate the systematic errors we
incorporated~\cite{ATL-PHYS-2003-008,physics_03_12050} the
formalism developed in~\cite{nim_A320_331}. The implementation of
MC@NLO to simulate the $t\overline{t}$ background has not changed
the conclusions drawn in~\cite{SN-ATLAS-2003-024} for the $M_H$
considered there. The $H\rightarrow W^{(*)}W^{(*)}\rightarrow
l^+l^-\sla{p_T}$ mode has a strong potential in a wide rage of
Higgs masses. A significance of or greater than $5\,\sigma$ may be
achieved with 30\,fb$^{-1}$ of integrated luminosity for
$125<M_H<300\,\gev$.

\section{The $H\rightarrow\gamma\gamma$ Mode}
\label{sec:hgg}

\subsection{Generation of Background Processes}
\label{sec:ggbackground} The relevant background processes for
this mode are subdivided into two major groups. Firstly, the
production of two $\gamma$'s associated with two jets (real photon
production). Secondly, a sizeable contribution is expected from
events in which at least one jet is misidentified as a photon
(fake photon production). Despite the impressive jet rejection
rate after the application of $\gamma$ selection criteria expected
to be achieved by the ATLAS detector~\cite{LHCC99-14} ($\mayor
10^3$ for each jet), the contribution from fake photons will not
be negligible due to the large cross-sections of QCD processes at
the LHC.

LO ME based MC's were used to simulate $\gamma\gamma jj$ (both QCD
and EW diagrams), $\gamma jjj$ and $jjjj$ production. For this
purpose MadGraphII~\cite{pc_81_357,hep-ph_0208156} interfaced
with PYTHIA6.2 was used~\cite{Wisc_soft}. The factorization and
re-normalization scales were set to the $P_T$ of the lowest $P_T$
parton.

After the application of a number of basic cuts at the generator
level (see~\cite{ATL-PHYS-2003-036}) the QCD and EW $\gamma jjj$
diagrams correspond to 6.32 nb and 1.21 pb, respectively. Assuming
an effective jet rejection of the order of $10^{3}$, the starting
cross-section for the EW $\gamma jjj$ process would be about
$1\,$fb. This small cross-section will be severely reduced after
the application of further selection cuts (see
Section~\ref{sec:ggevsel}). In the analysis EW $\gamma jjj$ and
$jjjj$ diagrams were neglected.\footnote{More details of MC
generation for background processes are available
in~\cite{ATL-PHYS-2003-036}.}

\subsection{Event Selection}
\label{sec:ggevsel}

A number of pre-selection cuts are applied which are similar to
those used in the multivariate analysis of VBF $H\rightarrow
W^{(*)}W^{(*)} \rightarrow
l^{+}l^{-}\sla{p_{T}}$~\cite{ATL-PHYS-2003-007}:
\begin{itemize}
\item [{\bf a.}] $P_{T\gamma 1}, P_{T\gamma 2}>25\,\gev$. The
$\gamma$'s are required to fall in the central region of the
detector excluding the interface between the barrel and end-cap
calorimeters ($1.37<\left|\eta\right|<1.52$). The latter
requirement reduces the acceptance by about 10$\%$. \item [{\bf
b.}] Tagging jets with $P_{Tj_1}, P_{Tj_2}>20\,\gev$ and
$\Delta\eta_{j_1j_2}>3.5$. \item [{\bf c.}] The $\gamma$'s should
be in between the tagging jets in pseudorapidity. \item [{\bf d.}]
Invariant mass of the tagging jets, $M_{j_1j_2}>100\,\gev$. \item
[{\bf e.}] Central jet veto~\cite{SN-ATLAS-2003-024}. \item [{\bf
f.}] Invariant mass window:
$M_H-2\,\gev<M_{\gamma\gamma}<M_H+2\,\gev$.
\end{itemize}

The final event selection is obtained by means of maximizing the
Poisson significance for 30 fb$^{-1}$ of integrated luminosity for
$M_H=120\,\gev$. The maximization procedure is performed with the
help of the MINUIT program~\cite{cpc_10_343}. The following
variables are chosen: $P_{Tj_1}$, $P_{Tj_2}$,
$\Delta\eta_{j_1j_2}$, $\Delta\phi_{j_1j_2}$, $M_{j_1j_2}$,
$P_{T\gamma_1}$, $P_{T\gamma_2}$, and $\Delta\eta_{\gamma\gamma}$.

Due to the implementation of parton shower and hadronization
effects, the kinematics of the final state  will be somewhat
different from that  of the parton level analysis performed
in~\cite{Rainwaterthesis}. After the application of cut {\bf f} in
the pre-selection, the dominant background corresponds to QCD
$\gamma\gamma jj$ and the fake photon production, therefore, the
optimization process will be mainly determined by the kinematics
of these process together with that of the VBF signal.

\begin{table}[t]
\begin{center}
\begin{tabular}{l || c c c c}
%\begin{tabular}{||c|c|c|c|}
\hline \hline
  Cut   & Pre-selection  & Parton Level & Optimization  \\
  \hline\hline
 {\bf a} &  $P_{T\gamma 1}, P_{T\gamma 2} >25\,\gev$  &     $P_{T\gamma 1}>50\,\gev$    &     $P_{T\gamma 1}>57\,\gev$   \\

& &  $P_{T\gamma 2}>25\,\gev$ &  $P_{T\gamma 2}>34\,\gev$ \\

& & & $\Delta\eta_{\gamma\gamma}<1.58$,
$\Delta\phi_{\gamma\gamma}<3$~rad \\ \hline

 {\bf b} &   $P_{Tj_1}, P_{Tj_2} >20\,\gev$  &  $P_{Tj_1}>40\,\gev$
 & $P_{Tj_1}>40\,\gev$ \\

 & & $P_{Tj_2}>20\,\gev$ & $P_{Tj_2}>29.5\,\gev$ \\

& $\Delta\eta_{j_1j_2}>3.5$ &  $\Delta\eta_{j_1j_2}>4.4$ &
$\Delta\eta_{j_1j_2}>3.9$  \\ \hline

{\bf d} &   $M_{j_1j_2}>100\,\gev$    &     -  &     $M_{j_1j_2}>610\,\gev$ \\
\hline\hline

  \end{tabular}
 \caption{Values of the cuts applied for different event selections (see Section~\ref{sec:ggevsel}).}
 \label{tab:cuts}
\end{center}
\end{table}

Initially, it has been verified that the inclusion of variables
additional to those considered in~\cite{Rainwaterthesis} improves
the signal significance. The addition of the photon related
variables $\Delta\eta_{\gamma\gamma}$ and
$\Delta\phi_{\gamma\gamma}$ improves the signal significance by
some $10-20\%$ depending on the Higgs mass. The implementation of
those two variables separately proves more efficient than the
combined $\Delta R_{\gamma\gamma}$. The inclusion of the hadronic
variable $\Delta\phi_{jj}$ does not noticeably increase the signal
significance.

Table~\ref{tab:cuts} shows the results of the optimization
together with the values of the cuts placed at the pre-selection
level and for the parton level analysis performed
in~\cite{Rainwaterthesis}. Due to the significant increase in the
background contribution compared to the parton level analysis, the
optimized event selection is significantly tighter, resulting into
reduced signal and background rates (see Section~\ref{sec:ggres}).
The increase of the background comes from the different choice of
the width of the mass window, the implementation of parton showers
for the estimation of the central jet veto probability and the
inclusion of fake photon events.

\subsection{Results and Discovery Potential}
\label{sec:ggres}

Here, we use  the event selection obtained in the optimization
procedure performed in Section~\ref{sec:ggevsel} (see
Table~\ref{tab:cuts}). The expected  signal and background
cross-sections corrected for acceptance and efficiency corrections
are shown in Table~\ref{tab:cross120} for a mass window around
$M_H=120\,\gev$ after the application of successive cuts.

The contribution from the fake photon background has been severely
reduced thanks to the inclusion of the photon angular variables.
The contribution from this background, however, is important. The
normalization of the fake photon background is subject to sizeable
systematic uncertainties. This is partly due to the uncertainty on
the determination of the fake photon rejection
rate~\cite{LHCC99-14}.

Figure~\ref{fig:mgg} shows the expected signal and background
effective cross-section (in fb) as a function of
$M_{\gamma\gamma}$ for $M_H=130\,\gev$. The dashed line shows the
total background contribution whereas the dotted line corresponds
to the real $\gamma\gamma$ background. The solid line displays the
expected contribution of signal plus background.
\begin{figure}[t]
{\centerline{\epsfig{figure=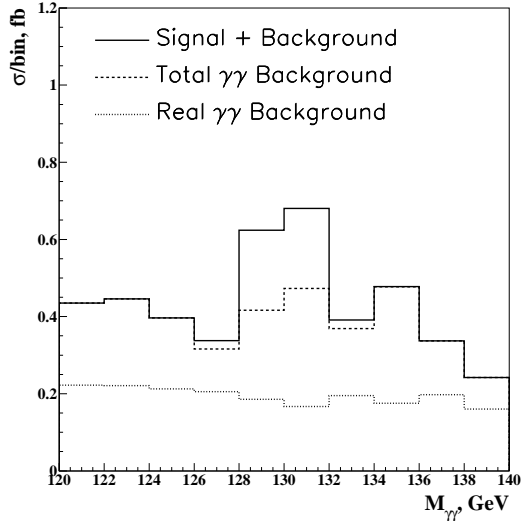,width=7.cm}}}
\vspace{-0.2cm} \caption[]{Expected signal and background
effective cross-section (in fb) as a function of
$M_{\gamma\gamma}$ for $M_H=130\,\gev$. The dashed line shows the
total background contribution whereas the dotted line corresponds
to the real $\gamma\gamma$ background. The solid line displays the
expected contribution of signal plus background.} \label{fig:mgg}
\end{figure}
In Table~\ref{tab:ggresults}, results are given in terms of $S$
and $B$, for 30 fb$^{-1}$ of integrated luminosity. The signal
significance was calculated with a Poissonian calculation. The
signal significance expected with this VBF mode alone reaches
2.2$\,\sigma$ for 30 fb$^{-1}$ of integrated luminosity.

The QCD $\gamma\gamma jj$ has been estimated with QCD
$\gamma\gamma jj$ ME based MC alone. The rate of additional (non
tagging) jets has been estimated with the help of the parton
shower. This approach yields a central jet veto survival
probability significantly smaller than that calculated
in~\cite{Rainwaterthesis}. Both effects go in the direction of
overestimating of the $\gamma\gamma jj$ background. Similar
discussion applies to the estimation of the fake photon background
performed here. This background estimation may be improved with
the implementation of a more realistic MC for the simulation of
the real photon background. This mode is considerably more sensitive to the understanding of fake photon rejection than the inclusive analysis~\cite{LHCC99-14}.

 \begin{table}[t]
 \begin{center}
% \begin{tabular}{||c||c|c||c|c||c|c||}
 \begin{tabular}{l ||  c c c c c c}
 \hline \hline
 Cut & VBF H & g-g Fusion H & QCD $\gamma\gamma jj$ &      EW $\gamma\gamma jj$ & $\gamma jjj$ & $jjjj$ \\
\hline\hline
 {\bf a} &       2.25 &       5.45 &     246.90 &       7.97 &     172.60 &     691.06  \\
 {\bf b} &       0.73 &       0.08 &      31.83 &       4.39 &      28.30 &      35.22  \\
 {\bf c} &       0.70 &       0.07 &      16.81 &       4.20 &      21.76 &      30.06  \\
 {\bf d} &       0.57 &       0.04 &       7.43 &       3.69 &      12.77 &      16.99  \\
 {\bf e} &       0.42 &       0.02 &       5.41 &       2.50 &       8.52 &       8.49  \\
 {\bf f} &       0.38 &       0.02 &       0.28 &       0.14 &       0.22 &       0.25  \\
 \hline\hline
 \end{tabular}
 \caption{Expected signal and background cross-sections (in fb) corrected for acceptance and efficiency corrections after the application of successive cuts (see Section~\ref{sec:ggevsel}). Here $M_H=120\,\gev$.}
 \label{tab:cross120}
 \end{center}
 \end{table}

 \begin{table}[t]
 \begin{center}
 %\begin{tabular}{||c|c|c|c|c|c||}
\begin{tabular}{c || c c c c}
 \hline\hline
 $M_H$ & $S$ & $B$ & $S/B$ & $\sigma_P$     \\
\hline\hline
  110 &      10.05 &      30.69 &       0.33 &     1.56  \\
  120 &      12.06 &      26.54 &       0.45 &     2.02  \\
  130 &      12.52 &      23.97 &       0.52 &     2.19  \\
  140 &      10.91 &      22.90 &       0.48 &     1.94  \\
  150 &       7.69 &      20.15 &       0.38 &     1.42  \\
  160 &       2.89 &      17.21 &       0.17 &     0.44  \\
 \hline \hline
 \end{tabular}
 \caption{Expected number of signal and background events, $S/B$ and the corresponding signal significance for 30~fb$^{-1}$ of integrated luminosity (see Section~\ref{sec:ggres}).}
 \label{tab:ggresults}
 \end{center}
 \end{table}

\section{The $H\rightarrow ZZ\rightarrow l^+l^-q\overline{q}$ Mode}
\label{sec:hzz}

\subsection{Generation of Background Processes}
\label{sec:zzback}

Cross-section for the QCD $Z+4j, Z\rightarrow l^+l^-, l=e,\mu$
process were calculated with two independent packages:
ALPGEN~\cite{hep-ph_02_06293} and
MadGraphII~\cite{pc_81_357,hep-ph_0208156}. Both calculations
include the $Z/\gamma^{\star}$ interference effects. The following
cuts at the generator level were used for the cross-section
calculation for the nominal event generation:
\begin{itemize}
\item QCD parton's transverse momentum, $P_T>20\,\gev$,
pseudorapidity, $\left|\eta\right|<5$. Separation between QCD
partons, $\Delta R>0.5$. \item Minimal transverse momentum cuts on
leptons, $P_T>3\,\gev$ with $\left|\eta\right|<3$. The angle
separation between leptons and leptons and jets were set to
$\Delta R>0.2$
\end{itemize}

The Born level cross-section of QCD $Z+4j$ production is subject
to large uncertainties. Some properties of jets in association
with  $W$ and $Z$ bosons have been studied and have been compared
with QCD predictions at the
Tevatron~\cite{prl_77_448,prl_79_4760}. The measured
cross-sections of $W/Z+n~jets$ where $n=1,2,3,4$ lie in between
the LO predictions calculated using the re-normalization and
factorization scales equal to the average transverse momentum of
the partons, $\langle P_T\rangle$,  and the transverse energy of
the weak boson, $E_T^{WB}$, respectively. The LO prediction
calculated with the first choice of scale systematically
undershoots the measured cross-section. At the LHC $\langle
P_T\rangle>100\,\gev$, due to the large phase space. Thus, the
scale was set to the mass of the weak boson.

After the application of the cuts at the generator level and the
choice of scales mentioned above both ALPGEN and MadGraphII yield
$47.5\,$pb. 8.5 million un-weighted events were generated with
MadGraphII. The output from MadGraphII was interfaced to the
HERWIG6.5 package~\cite{Wisc_soft}. In order to avoid severe
double counting in the generation of hadronic jets, the scale of
the parton shower evolution was set to the $P_T$ of the lowest
transverse momentum parton in the event.

The cross-section for  $Z+4j, Z\rightarrow l^+l^-, l=e,\mu$
production with one EW boson in the internal lines was evaluated
with MadGraphII. These diagrams include QCD $ZZjj$ and $ZW^\pm
jj$. A cross-section of $1.6\,$pb was obtained after cuts at
generator level and by applying the same choice of scales as for
the QCD $Z+4j$ case. The impact of these diagrams is small, hence,
they were not included in the final results reported in
Section~\ref{sec:hzzresults}. Diagrams with two EW bosons in the
internal lines were not considered, as they are expected to be
negligible.

A sample of events for $t\overline{t}$ production was used. These
events were generated with the MC@NLO package (see
Section~\ref{sec:wwbackround}).

\subsection{Event Selection} \label{sec:hzzevsel}

The event selection presented in this Section is obtained by
maximizing the signal significance for a Higgs for $M_H=300\,\gev$
with $30\,$fb$^{-1}$ of integrated luminosity.

\begin{figure}[t]
{\centerline{\epsfig{figure=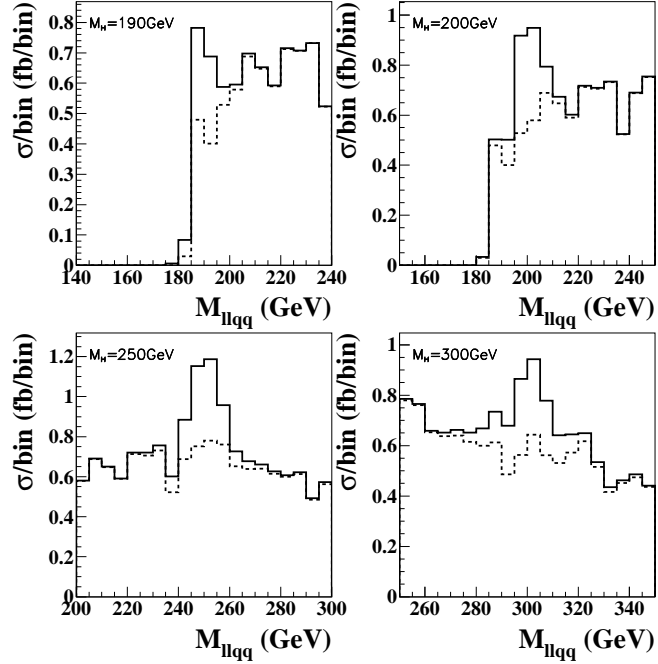,width=9.cm}}}
\vspace{-0.2cm} \caption[]{Invariant mass of the Higgs candidates
after the application of kinematic fits. The solid lines
correspond to the sum of the signal (VBF $H\rightarrow
ZZ\rightarrow l^+l^-q\overline{q}$) and the main background (QCD
$Z+4j, Z\rightarrow l^+l^-, l=e,\mu$). The dashed lines show the
contribution of the main background alone. Here $M_H=190, 200,
250, 300\,\gev$.} \label{fig:massplots_af_1}
\end{figure}

A number of basic features common to VBF modes remain.  A feature
specific to the mode under study is the additional ambiguity in
the definition of tagging jets introduced by the presence of
relatively hard jets produced from the decay of the $Z$'s.  A
search for two jets with an invariant mass close to $Z$ mass,
$M_Z$, is performed. After reconstructing the $Z$ decaying
hadronicaly, the event looks like a ``typical'' VBF candidate.

The following event selection was chosen:
\begin{itemize}
\item[{\bf a}.] Two isolated, oppositely charged, of equal flavor
leptons in the central detector region, $\left|\eta\right|<2.5$.
\item[{\bf b}.] The event is required to pass the single or double
lepton trigger in ATLAS. \item[{\bf c}.] Two hadronic jets ($j_3,
j_4$) with transverse momentum, $P_T>30\,\gev$  with $M_{j_3j_4}$
close to $M_Z$ were required in the fiducial region of the
calorimeter, $\left|\eta\right|<4.9$. The relative invariant mass
resolution of two jets is expected to be approximately $10\,\%$.
The following mass window was chosen: $75<M_{j_3j_4}<101\,\gev$.
These jest were ``masked out'' from the list of jets. \item[{\bf
d}.] Tagging jets with $P_{Tj_1}>40\,\gev$, $P_{Tj_2}>30\,\gev$
and $\Delta\eta_{j_1j_2}>4.4$. \item[{\bf e}.] Both leptons were
required to lie in between the tagging jets in pseudorapidity.
\item[{\bf f}.] Leptonic cuts. It was required that
$M_Z-10<M_{ll}<M_Z+10\,\gev$. This cut is expected to suppress
di-lepton final states with $W^+W^-\rightarrow ll\nu\nu$. It is
particularly important to suppress the contribution from
$t\overline{t}$ production associated with jets. No b-tagging
rejection algorithms were applied in this analysis due to the
large branching ratio of $Z$ decaying into heavy quarks.
\item[{\bf g}.] The invariant mass of the tagging jets was
required to be greater than $900\,\gev$. \item[{\bf h}.] Central
jet veto. Extra jets with $P_T>20\,\gev$ are looked for in the
central region of the detector ($\left|\eta\right|<3.2$). However,
high $P_T$ quarks from the decay of one of the $Z$'s are expected
to radiate hard gluons with a high probability, thus, faking
hadronic jets produced prior to the decay. If $\Delta R$ between
the extra jet and the jets of the Higgs candidate is larger than
one unit, the event is vetoed. \item[{\bf i}.] In order to further
reduce the contribution from events with $W^+W^-\rightarrow
ll\nu\nu$, it is required that $\sla{p_T}<30\,\gev$.
\end{itemize}

The $M_{llj_3j_4}$ spectrum could be distorted due to the
ambiguity in defining tagging jets. The distortion of the
$M_{llj_3j_4}$ spectrum, however, is not sizeable.
Figure~\ref{fig:massplots_af_1} displays the $M_{llj_3j_4}$
spectra for signal and background after the application of the
event selection presented in this Section. A Higgs mass resolution
of approximately $2.5\%$ is obtained for
$2M_Z<M_H<300\,\gev$~\cite{ATL-COM-PHYS-2003-035}.

\subsection{Results and Discovery Potential}
\label{sec:hzzresults}

Table~\ref{tab:hzzcrosec} shows the expected signal effective
cross-sections (in fb) for a Higgs mass of $M_H=300\,\gev$.
Table~\ref{tab:hzzcrosec} also displays the effective
cross-sections for the major background processes. Cross-sections
are given after successive cuts (see Section~\ref{sec:hzzevsel}).
The background is largely dominated by the QCD $Z+4j, Z\rightarrow
l^+l^-, l=e,\mu$ production. Diagrams with one or two EW boson in
the internal lines were neglected. The contribution from
$t\overline{t}$ is small and it is also neglected in the final
results.

\begin{table}[t]
 \begin{center}
 \begin{tabular}{l || c c c c c c c c c}
 \hline\hline
 Process & {\bf a} & {\bf b} & {\bf c} & {\bf d} & {\bf e} & {\bf f} & {\bf g} &{\bf h}& {\bf i}\\
\hline\hline
 VBF ($ M_H=300\,\gev$) &      31.69 &      31.50 &      12.63 &       3.39 &       3.26 &       2.93 &       2.24 &       1.72 &       1.66\\
QCD $Z+4j$ &   25930 &      25902 &  10345   &   277  &  205   &    205 &   116 &  36.6 &  34.6  \\
  $t\overline{t}$ &  14793 &   14268 &   4233 &   135 &   106 &    10.5  &   6.4   &   2.3 &  0.3  \\
 \hline\hline
 \end{tabular}
\caption{Expected effective cross-sections (in fb) for
$H\rightarrow ZZ\rightarrow llq\overline{q}$ produced via VBF
$(M_H=300\,\gev$) and the main background processes.
Cross-sections are given after successive cuts presented in
Section~\ref{sec:hzzevsel}.}
 \label{tab:hzzcrosec}
 \end{center}
 \end{table}

Table~\ref{tab:results_af} reports results in terms of $S$, $B$,
$S/B$ and signal significance, $\sigma_L$, with 30~fb$^{-1}$ of
integrated luminosity for different values of $M_H$. The effective
signal and background cross-sections are evaluated in a
$4\,\sigma_M$ (where $\sigma_M$ is the mass resolution) wide mass
window. The signal significance was calculated with a likelihood
ratio technique using the invariant mass of the Higgs candidate as
a discriminant
variable~\cite{ATL-PHYS-2003-008,physics_03_12050}. A signal
significance of $3.75\,\sigma$ may be achieved for $M_H=300\,\gev$
with 30\,fb$^{-1}$ of integrated luminosity. It should be noted
that the cross-sections for the main background reported here are
subject to large theoretical uncertainty. Fortunately, the
background may be determined from side bands for Higgs searches
with $M_H>200\,\gev$.

 \begin{table}[t]
 \begin{center}
 \begin{tabular}{ c || c c c c}
 \hline \hline
$M_H (\gev)$ & $S$ & $B$ & $S/B$ & $\sigma_L$ \\
\hline\hline
 190 & 18.9 & 31.2 & 0.61 & 3.47 \\
 200 & 27.3 & 52.8 & 0.52 & 3.76 \\
 300 & 39.3 & 116.1 & 0.34 & 3.75 \\
 500 & 20.1 & 124.2 & 0.16 & 1.98 \\
\hline\hline
 \end{tabular}
 \caption{Expected number of signal and background events,  ratio of signal to
background and signal significance (in $\sigma$) for a SM Higgs
produced via VBF using the decay mode $H\rightarrow ZZ\rightarrow
l^+l^-q\overline{q}$ with 30~fb$^{-1}$ of integrated luminosity
for different values of $M_H$. The effective signal and background
cross-sections are evaluated in a $4\,\sigma_M$ (where $\sigma_M$
is the mass resolution) wide mass window. The signal significance,
$\sigma_L$, was calculated with a likelihood ratio technique using
the invariant mass of the Higgs candidate as a discriminant
variable.}
 \label{tab:results_af}
 \end{center}
 \end{table}

\section{Multivariate Analysis}
\label{sec:NN}

Results reported in~\cite{SN-ATLAS-2003-024} and the present paper
were based on classical cut analyses. Multivariate techniques have
been extensively used in physics analyses, for instance, in LEP
experiments. Neural Networks (NN) are the most commonly used tools
in multivariate analyses. NN training has been performed on the
$H\rightarrow W^{(*)}W^{(*)}\rightarrow
l^+l^-\sla{p_T}$~\cite{ATL-PHYS-2003-007} and $H\rightarrow
\tau^+\tau^-\rightarrow l^+l^-\sla{p_T}$~\cite{NNVBFtautaull}
modes. NN training was performed with a relatively small number of
variables. It was required that these variables are infra-red safe
and their correlations do not depend strongly on detector effects:
$\Delta\eta_{j_1j_2}$, $\Delta\phi_{j_1j_2}$, $M_{j_1j_2}$,
$\Delta\eta_{ll}$, $\Delta\phi_{ll}$, $M_{ll}$, and $M_{T}$ (or
the invariant mass of the $\tau^+\tau^-$ system in the case of the
$H\rightarrow \tau^+\tau^-\rightarrow l^+l^-\sla{p_T}$ mode). The
signal significance was calculated with a likelihood ratio
technique using the NN output as a discriminant variable. An
enhancement of approximately $30-50\,\%$ of the signal
significance with respect to the classical cut analysis was
obtained for both modes under consideration.

\section{Conclusions}
\label{sec:conclusions}

The discovery potential for the SM Higgs boson produced with VBF
in the range $115<M_H<500\,\gev$ has been reported. An updated
study at hadron level followed by a fast detector simulation of
the $H\rightarrow W^{(*)}W^{(*)}\rightarrow l^+l^-\sla{p_T}$ mode
has been presented: the main background, $t\overline{t}$
associated with jets, has been modelled with the MC@NLO program
and the Higgs mass range has been extended to $500\,\gev$. This
mode has a strong potential: a signal significance of more than
$5\,\sigma$ may be achieved with 30\,fb$^{-1}$ of integrated
luminosity for $125<M_H<300\,\gev$. The discovery potential of the
$H\rightarrow\gamma\gamma$ and $H\rightarrow ZZ\rightarrow
l^+l^-q\overline{q}$ modes have also been reported with analyses
at hadron level followed by a fast detector simulation.

The discovery potential of the modes presented in this work was
combined with results reported in past studies performed for the
ATLAS
detector. Results from recent studies~\cite{ATL-PHYS-2003-001,ATL-PHYS-2003-024,ATL-PHYS-2003-025}, which were not used in~\cite{SN-ATLAS-2003-024}, were added here.
Likelihood ratio techniques have been used to perform the
combination~\cite{ATL-PHYS-2003-008,physics_03_12050}. In order
to incorporate systematic errors, the formalism developed
in~\cite{nim_A320_331} was implemented. A 10\,\%
systematic error on the background estimation has been assumed for
modes related to VBF~\cite{SN-ATLAS-2003-024}. Figure~\ref{fig:discoveryPlot} displays the
overall discovery potential of the ATLAS detector with
10\,fb$^{-1}$ of integrated luminosity. Results from NN
based analyses and discriminating variables have not been included
in the combination. The present study confirms the results
reported
in~\cite{pr_160_113004,pl_503_113,pr_61_093005,JHEP_9712_005,SN-ATLAS-2003-024},
that the VBF mechanism yields a strong discovery potential at the
LHC in a wide range of the Higgs boson mass.

\begin{figure}[t]
\begin{center}
\epsfig{file=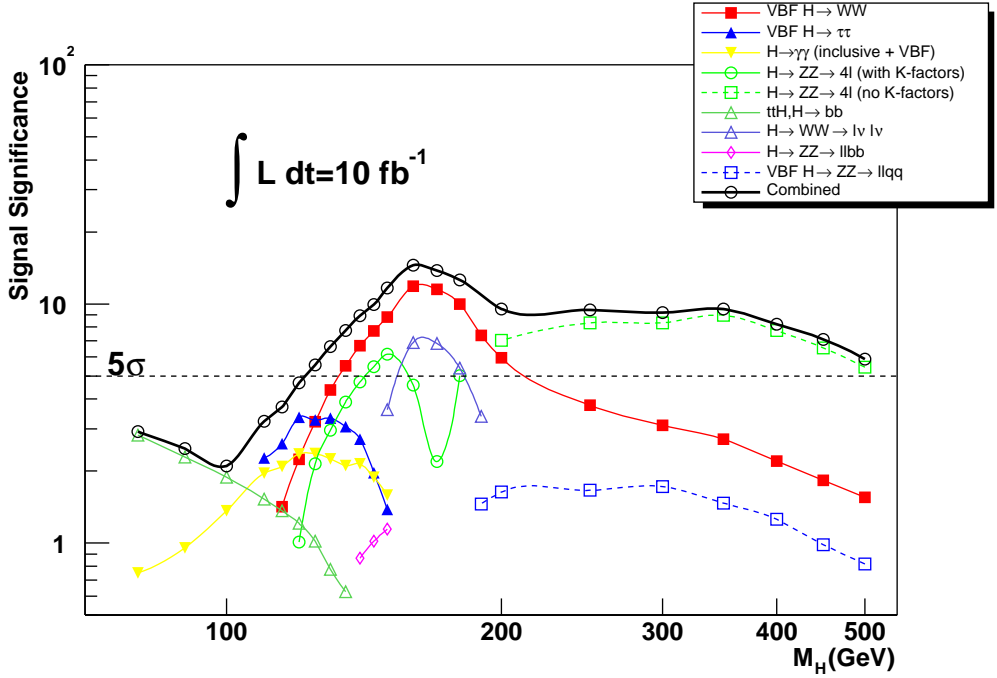, width=13.5cm}
\vspace{-0.2cm} \caption{Expected significance for ATLAS as a
function of Higgs mass for 10\,fb$^{-1}$ of integrated
luminosity.} \label{fig:discoveryPlot}
\end{center}
\end{figure}

}

%% file: bisset.tex
{
\def\s#1{{\small#1}}
\def\lsim{\:\raisebox{-0.5ex}{$\stackrel{\textstyle<}{\sim}$}\:}
\def\gsim{\:\raisebox{-0.5ex}{$\stackrel{\textstyle>}{\sim}$}\:}
\def\PD{\s{PDG}}
\def\TA{{\small TAUOLA}}
\def\HW{\s{HERWIG}}
\def\JS{\s{JETSET}}
\def\PY{\s{PYTHIA}}
\def\IS{\s{ISAJET}}
\def\IW{\s{ISAWIG}}
\def\SM{\s{SM}}
\def\MSSM{{MSSM}}
\def\SY{\s{SUSY}}
\def\QCD{\s{QCD}}
\def\QED{\s{QED}}
\def\DIS{\s{DIS}}
\def\LEP{\s{LEP}}
\def\LHC{\s{LHC}}
\def\OPAL{\s{OPAL}}
\def\PDF{\s{PDFLIB}}
\def\CERN{\s{CERN}}
\def\RPV{\rlap{/}{R}$_{\mbox{\scriptsize p}}$}
\def\BNV{\rlap{/}{B}}
\def\Ord{\buildrel{\scriptscriptstyle <}\over{\scriptscriptstyle\sim}}
\def\OOrd{\buildrel{\scriptscriptstyle >}\over{\scriptscriptstyle\sim}}
\def\gh{\Gamma_{\scriptscriptstyle \rm H}}
\def\gtap{\raisebox{-.4ex}{\rlap{$\sim$}} \raisebox{.4ex}{$>$}}
\def\ltap{\raisebox{-.4ex}{\rlap{$\sim$}} \raisebox{.4ex}{$<$}}
\def\ycut{$y_{\mbox{\tiny cut}}$}
\def\mw{m_{\scriptscriptstyle \rm W}}
\def\mh{m_{\scriptscriptstyle \rm H}}
%                  Define Lambda MS bar
\def\lms{\Lambda_{\overline{\rm MS}}}
\def\half{\mbox{\small $\frac{1}{2}$}}
\def\thlf{\mbox{\small $\frac{3}{2}$}}
\def\as{\alpha_{\mbox{\tiny S}}}
\def\ee{e^+e^-}
\def\MC{Monte Carlo}
\def\VEV#1{\langle{#1}\rangle}
%                  Define antiparticles
\def\qbar{\bar{q}}
\def\Qbar{\bar{Q}}
\def\dbar{\bar{d}}
\def\ubar{\bar{u}}
\def\sbar{\bar{s}}
\def\cbar{\bar{c}}
\def\bbar{\bar{b}}
\def\tbar{\bar{t}}
\def\pbar{\bar{p}}
\def\B0bar{\overline{B^0}}
\def\lbar{\bar{\l}}
\def\l{\ell}
\def\lsim{\raisebox{-.1em}{$
\buildrel{\scriptscriptstyle <}\over{\scriptscriptstyle\sim}$}}
\def\gsim{\raisebox{-.1em}{$
\buildrel{\scriptscriptstyle >}\over{\scriptscriptstyle\sim}$}}
\def\preprint{{preprint}}

%%%%%%%%%%%%%%%%%%%%%%%%%%%%%%%%%%%%%%%%%%%%%
%\begin{flushright}
%{TUHEP-TH-03???}\\
%{SHEP-04-02}\\
%{\today}\\
%\end{flushright}
%%%%%%%%%%%%%%%%%%%%%%%%%%%%%%%%%%%%%%%%%%%%%

\noindent
{\Large \bf F. Four-Lepton Signatures at the LHC of heavy neutral MSSM
Higgs Bosons via Decays into Neutralino/Chargino Pairs} \\[0.5cm]
{\it M.\,Bisset, N.\,Kersting, J.\,Li, S.\,Moretti and F.\,Moortgat}

\begin{abstract}
{We investigate the scope of heavy neutral Higgs boson decays
into chargino/neutralino pairs yielding four-lepton signatures
in the context of the Minimal Supersymmetric Standard Model (MSSM),
by exploiting all available modes. A preliminary analysis points to
the possibility 
of detection at intermediate values of $\tan\beta$ and $H^0/A^0$ masses in the
region of 400 GeV and above, provided MSSM parameters associated to
the Supersymmetric (SUSY) sector of the model are favourable.}
\end{abstract}

\section{Introduction}

There have been four previous studies \cite{Baer:1992kd,Baer:1994fx,Moortgat:2001pp,Bisset:1995dc}
focusing on the discovery potential of the
decays of the heavier neutral MSSM Higgs bosons into neutralinos and
charginos (henceforth, --inos):
\begin{eqnarray}
H^0,A^0 \rightarrow \tilde{\chi}_a^+\tilde{\chi}_b^-,
\tilde{\chi}_i^0\tilde{\chi}_j^0 \;\;\;\;\;\;\;
(a,b = 1,2, \;\; i,j = 1,2,3,4).
\label{allproc}
\end{eqnarray}
However, all such works dealt only with the
channels\footnote{The decays $H^0,A^0 \rightarrow
\tilde{\chi}_1^+\tilde{\chi}_1^-,
\tilde{\chi}_1^0\tilde{\chi}_2^0$ were also studied in 
\cite{Baer:1992kd} but found to be unproductive due to large backgrounds to 
the resulting `di-lepton signals'.}
$H^0,A^0 \rightarrow \tilde{\chi}_2^0\tilde{\chi}_2^0$.
Furthermore, the subsequent neutralino decays
$\tilde{\chi}_2^0 \rightarrow \tilde{\chi}_1^0 \ell^+ \ell^-$
($\ell=e$ or $\mu$)
were always assumed to proceed via three-body decays with an off-shell
intermediate $Z^{0*}$ or slepton, neglecting the possibility of the
intermediate slepton being on-mass-shell. The novelty of the present
analysis is that we incorporate {\sl all} the decays in (\ref{allproc})
and allow for intermediate sleptons to be on mass-shell.

The importance of investigating the potential of SUSY decays of Higgs
bosons in covering the so-called LHC wedge region of the MSSM parameter
space
($4~\lsim~\tan\beta~\lsim~10$ and $M_{A^0}\gsim~200$
GeV) -- where only the lightest MSSM Higgs boson $h^0$ can be detected and this
is indistinguishable from the Standard Model (SM) state 
(decoupling regime) -- has already been
stated in several papers, for the case of both neutral \cite{Moortgat:2001pp} 
and charged  \cite{Bisset:2000ud,Bisset:2003ix,Moretti:2002ht,Assamagan:2001yz,Cavalli:2002vs} 
Higgs states. The reason is
twofold. Firstly, for consistency: it is rather 
unnatural in the MSSM to assume a heavy SUSY particle (or sparticle) 
spectrum as implied by
only considering decays of MSSM Higgs bosons into (visible) ordinary matter,
when the mechanism of Electro-Weak Symmetry Breaking (EWSB) does require
sparticle masses to be at or below the TeV scale.
Secondly, for the benefits: several Higgs $\to$  SUSY signals 
are  indeed detectable at the LHC and provide the means of
distinguishing the SM from the MSSM Higgs sector in the wedge region
even independently of the discovery of the SUSY particles themselves
in other channels \cite{Moortgat:2001pp}, 
\cite{Bisset:2000ud,Bisset:2003ix,Moretti:2002ht,Assamagan:2001yz,Cavalli:2002vs}.
The rationale for looking at leptonic signatures ($\ell=e,\mu$)
is clearly the difficult LHC environment 
(due to the large QCD activity). Finally, the reason for  
including Higgs decays to the heavier --inos is to extend the reach 
to larger Higgs masses.

It is also worthwhile to investigate the role played by sleptons,
by allowing the latter to participate
in the decays as both on- and off-mass-shell objects with 
`optimal' masses, hence
maximising the leptonic --ino Branching Ratios (BRs). That is, sleptons are here made as 
light as possible, compatibly with LEP2 \cite{W1LEP2} limits
(i.e., $m_{
 {\tilde{   e}_1}
({\tilde{ \mu}_1})
[{\tilde{\tau}_1}]} \ge 99(91)[85]\, \hbox{GeV}$
and $m_{\tilde{\nu}} \ge 43.7\, \hbox{GeV}$, 
assuming that no slepton is nearly-degenerate with the LSP,
i.e., the lightest neutralino, $\tilde{\chi}_1^0$).
Flavour-diagonal inputs are adopted from the slepton sector for
both the left/right soft slepton masses 
(selectrons, smuons and staus) and the trilinear `$A$-terms'. (Also
notice that we adopt $m_{\tilde{e}_R} \simeq m_{\tilde{\mu}_R}$ and 
$m_{\tilde{e}_L} \simeq m_{\tilde{\mu}_L}$). 
If all three generations have the same soft inputs
(with $A_\ell = 0$, including $A_\tau$), then the  
slepton sector is effectively reduced to one optimal input value (which 
we identify with $m_{\tilde{\ell}_{\scriptscriptstyle R}}\equiv
m_{\tilde{\ell}}$).  
However, since --ino decays to tau-leptons are generally not anywhere near 
as beneficial as are --ino decays to electrons or muons,  it would be 
even better if the stau inputs were significantly above those of the 
first two generations, which reflects our approach here (we set the soft stau mass 
inputs $100\, \hbox{GeV}$ above those of the other sleptons). Unless
stated otherwise, we take
$m_{\tilde{\ell}}=$ 150 GeV ($\ell=e,\mu$), 
hence $m_{\tilde{\tau}}=$ 250 GeV, throughout. 

\section{MSSM Parameter Space}

\begin{figure}[!t]
\begin{center}
\epsfig{file=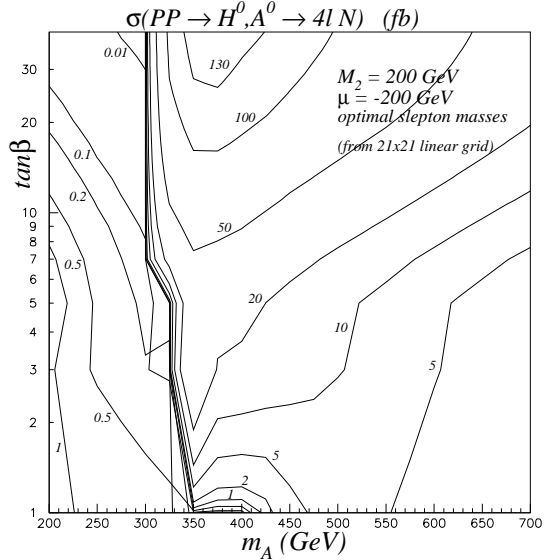,height=80mm,
width=80mm}
\end{center}
\vskip -1.0cm
\caption{
$\sigma(pp \rightarrow A^0,H^0)$ $\times$
BR$(A^0/H^0 \rightarrow 4\ell N)$ (in fb), where $\ell = e^{\pm}$ or
${\mu}^{\pm}$ and $N$ represents invisible final state particles:
$M_2=200$ GeV, $\mu = -200\, \hbox{GeV}$ with
optimised slepton masses, plus 
$m_t = 175\, \hbox{GeV}$, $m_b = 4.25\, \hbox{GeV}$, 
$m_{\tilde q} = 1\, \hbox{TeV}$, $m_{\tilde g} = 800\, \hbox{GeV}$ and
$A_{\ell} = 0$. % The black shaded areas are excluded by LEP.
}
\label{MSSMplane}
\end{figure}

The total event rate for all
possible four-lepton channels is found in Fig.~\ref{MSSMplane}, 
over the customary $(M_{A^0},\tan\beta)$ plane, for the representative
choice of SUSY parameters $M_2=200$ GeV ($M_{1}
=\frac{5}{3}\tan^2\theta_W M_{2}$)
and $\mu=-200$ GeV (and
optimal slepton masses as defined above). The normalisation of the
$gg\to H^0/A^0$ and $gg,q\bar q\to b\bar b H^0/A^0$ production
channels is from the SUSY implementation \cite{Moretti:2002eu}
of HERWIG \cite{Corcella:2000bw} in v6.4 \cite{Corcella:2001wc} 
default configuration
-- except for the choice of CTEQ6L Parton Distribution Functions
(PDFs) -- whereas decay rates are extracted from ISASUSY/ISASUGRA
routines \cite{Baer:1999sp} interfaced to HERWIG via
the ISAWIG module \cite{ISAWIG}. While the maximum of the cross section is
found at $M_{A^0}\approx 300$ to 400 GeV for large $\tan\beta$, in the
critical wedge region the production and decay rates are still favourable,
ranging from 500 to 5000 four-lepton events per year at high luminosity
(before any cuts).

The plots in Fig.~\ref{tb8color} show 
the event rates in the plane $(M_2,\mu)$ for
a choice in $\tan\beta$, equal to 8, and a selection of
$M_{A^0}$ values (400, 500 and 600 GeV) in the core of the wedge region.
The distribution of decay rates as highlighted by the colour scheme 
proves that the main source of the four-lepton signals emerging from
the decays in (\ref{allproc}) at such optimal points in the SUSY parameter
space shifts from $\tilde{\chi}_2^0\tilde{\chi}_2^0$ 
to heavier --ino pairs as $M_{A^0}$ rises from ${\lsim}~400~\hbox{GeV}$ to
${\lsim}~600~\hbox{GeV}$.  Also, a new region of the --ino parameter space
is opened up: the $\tilde{\chi}_2^0\tilde{\chi}_2^0$ decays
favour moderate to high $| \mu |$ and low $M_2$ whereas the decays to
heavier --inos prefer lower $| \mu |$ values and extend up to fairly high
values of $M_2$.

\begin{figure}[!t] 
\vspace*{-2.25cm} 
\begin{center}
\epsfig{file=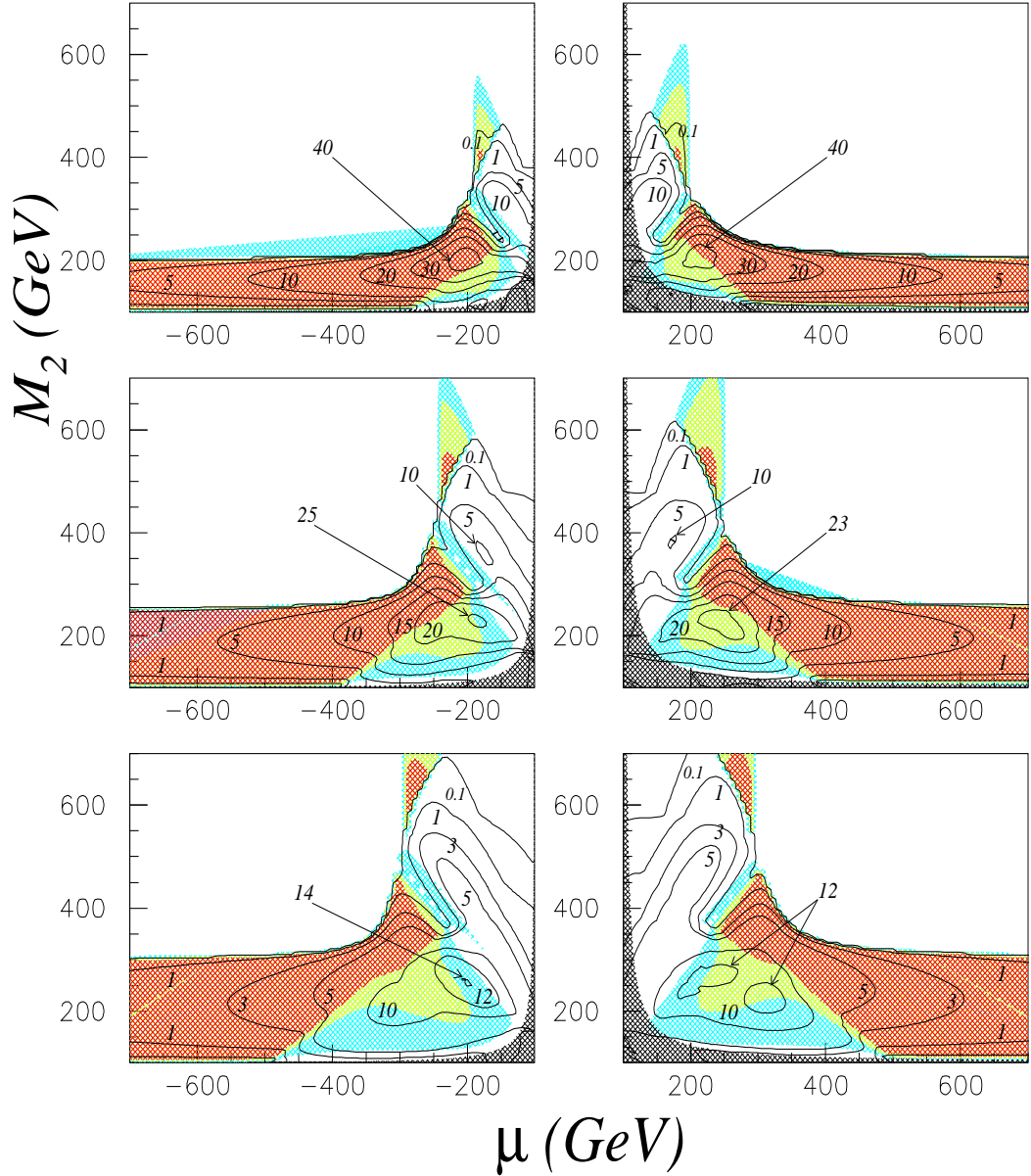,height=180mm,
width=160mm}
\end{center}
\vskip -1.0cm
\caption{
$\sigma(pp \rightarrow A^0,H^0)$ $\times$
BR$(A^0/H^0 \rightarrow 4\ell N)$ (in fb), where $\ell = e^{\pm}$ or
${\mu}^{\pm}$ and $N$ represents invisible final state particles:
percentage from $A^0,H^0 \rightarrow 
\tilde{\chi}^0_2 \tilde{\chi}^0_2$
$>$ 90\% (red), 50\% -- 90\% (yellow), 10\% -- 50\% (light blue),
$<$ 10\% (white). Parameters are:
$\tan\beta = 8$, $M_{A^0} = 400\, \hbox{GeV}$ (top),
$500\, \hbox{GeV}$ (middle), $600\, \hbox{GeV}$ (bottom).
Optimised slepton masses are assumed, plus
$m_t = 175\, \hbox{GeV}$, $m_b = 4.25\, \hbox{GeV}$,
$m_{\tilde q} = 1\, \hbox{TeV}$, $m_{\tilde g} = 800\, \hbox{GeV}$ and
$A_{\ell} = 0$. The black shaded areas are excluded by LEP.
}
\label{tb8color}
\end{figure}

\section{Detector Simulation Analysis}

The MC used for the LHC analysis is again HERWIG v6.4 in the configuration
described in the previous section. The detector simulation assumes a typical
LHC experiment, as provided by private programs checked against
results in literature.
The event selection criteria used in the analysis are as follows.
\begin{itemize}
\item $4\ell$ events: we select exactly four leptons ($\ell=e$ or $\mu$)
detected within $|\eta^\ell|<2.4$ and with initial thresholds at
 $E_T^\ell >7,4$~GeV for $e,\mu$, 
respectively.
Leptons are isolated by
requiring no tracks (of charged particles) with $p_T >
1.5$ GeV in a cone of 0.3 radians around the lepton direction.
\item $Z$-veto: no opposite-charge same-flavour lepton pairs may reconstruct 
$M_Z \pm 10$~GeV.
\item  $E_T^{\ell}$: all leptons must finally have $20~{\rm{GeV}}<E_T^\ell<80$~GeV.
\item $E_T^{\rm{miss}}$: events must have $20~{\rm{GeV}}<E_T^{\rm{miss}}<130$~GeV (in missing transverse momentum).
\item  $E_{T}^{\rm{jet}}$: all jets must have $E_T^{\rm{jet}}<50$~GeV.
\item  $4\ell$ inv. m.: the four-lepton invariant mass must be
 $\leq M_{H^0}-2M_{{\tilde\chi}^0_1}=360$~GeV.
\end{itemize}

\noindent
We have chosen two representative points for the MSSM parameter 
space\footnote{It is worth noticing
that the location in the SUSY parameter space where the 
raw signal rate is largest may differ from the location where the 
signal-to-background ratio  
is largest.} (recall that $A_\ell=0$):

\begin{enumerate}

\item $M_{A^0}$ = 600 GeV, $\tan\beta$ = 10, $M_1$ = 125 GeV, $M_2$ =
250 GeV, $\mu$ = + 450 GeV, $m_{\tilde{\ell}/\tilde{\tau}}$ = 150/250 GeV, 
$m_{\tilde{q}/\tilde{g}}$ = 1000/800 GeV. This is a configuration in
which ${\tilde\chi}_2^0 {\tilde\chi}_2^0$ decays are dominant (basically
100\%).         

\item 
$M_{A^0}$ = 400 GeV, $\tan\beta$ = 5, $M_1$ = 100 GeV, $M_2$ =
200 GeV, $\mu = -150$ GeV, $m_{\tilde{\ell}/\tilde{\tau}}$ = 150/250 GeV,
$m_{\tilde{q}/ \tilde{g}}$ = 1000/800 GeV. This is a configuration in
which ${\tilde\chi}_2^0 {\tilde\chi}_2^0$ (50\%)
and ${\tilde\chi}_2^0 {\tilde\chi}_4^0$ (38\%)
decays have comparable rates.

\end{enumerate}

\noindent
The relevant -ino masses can be found 
in Tab.~21 (recall that $M_{H^0}\approx M_{A^0}$).

\begin{table}[h]
\label{tab:masses}
    \caption{-ino masses (in GeV) for points 1.--2.}
    \begin{center}
    \begin{tabular}{|l||l|l|} \hline 
   &  1. & 2. \\ \hline 
%   $\mu$ &$450$ &$-150$ \\ \hline 
%   $M_2$  &$250$ &$200$ \\ \hline 
%$M_A$ &$600$ &$400$ \\ \hline 
% $\tan\beta$ &$10$ &$5$ \\ \hline 
 ${\tilde\chi}^0_1$ &$123$ &$94$ \\ \hline 
 ${\tilde\chi}^0_2$ &$236$ &$137$ \\ \hline 
 ${\tilde\chi}^0_3$  &$455$ &$167$ \\ \hline 
 ${\tilde\chi}^0_4$ &$471$ &$237$ \\ \hline
 ${\tilde\chi}^\pm_1$ &$236$ &$136$ \\ \hline 
 ${\tilde\chi}^\pm_2$ &$471$ &$238$ \\ 
\hline
      \end{tabular}
    \end{center}
\label{point-table}
\end{table}

\noindent
The results of our simulations are presented in Tabs.~22--23.
Despite our studies 
being preliminary, in the sense that a full simulation of the $t\bar t$
and  $\tilde{t} \tilde{t}^*$ backgrounds is still lacking, prospects of detecting
$H^0,A^0$ signatures at the LHC via the decays in (\ref{allproc}) in the
interesting wedge (or decoupling) region of the MSSM seem very good.
This applies not only to the already investigated
${\tilde\chi}_2^0 {\tilde\chi}_2^0$ modes (point 1.), but especially to the
additional ones (exemplified by the ${\tilde\chi}_2^0 {\tilde\chi}_4^0$
contribution in 2.). As for the $t\bar t$
and  $\tilde{t} \tilde{t}^*$ noises, we would expect these not to
undermine such conclusions, based on preliminary estimates for
the efficiency of extracting $4\ell$ events from top-antitop 
(also appearing from stop-antistop) decays. Another aspect to be mentioned
is the somewhat poor efficiency for the signal following the $Z$-veto and 
$E_T^{\ell}$ cuts, particularly for point 1. However, notice that both such
(or similar) requirements are needed: the former to reject the $ZZ$-noise
and the second against the squark/gluino background. Some
optimisation is in order in this case, though.
 
\begin{table}[h]\label{tab:1}
    \caption{Number of events after the successive cuts 
defined in the text for point 1. (at 100 fb$^{-1}$).}
    \begin{center}
    \begin{tabular}{|l||c|c|c|c|c|c|} \hline 
        Process           & $4\ell$ events & $Z$-veto & $E_T^{\ell}$ &
 $E_T^{\rm{miss}}$& $E_{T}^{\rm{jet}}$ & $4\ell$ inv. m. \\ \hline
    $\tilde{q}, \tilde{g}$  
      & 1218 & 417 & 76 & 14 &  0   & 0       \\ \hline 
%    $\tilde{t} \tilde{t}$   &       &     &           &        &     &    \\ \hline 
    $\tilde{\ell}$,$\tilde{\nu}$ & 1 & 0.2   & 0&   0 & 0  &  0      \\ \hline 
    $\tilde{\chi} \tilde{\chi}$& 240  &67 & 12   &4  &3 & 1 \\ \hline
    $tH^-$ + c.c.  & 1     &  0   &    0  &  0   & 0  &  0  \\ \hline 
%    total SUSY bkg.       &  &   &        &       &     &          \\ \hline
    $ZZ$                  & 807   & 16 & 10 &  2    &1  & 1    \\ \hline  
%   $Z~$+jet                & 0   &0  & 0        &  0    & 0  & 0     \\ \hline
    $ttZ$                & 46   & 1 &    0     &  0    & 0 &0   \\ \hline
   $tth^0$                & 3   & 1 &    1     &  0    & 0 & 0   \\ \hline
%    total bkg.            &     &   &           &   &       &        \\ \hline 
    $A^0,H^0$ signal          &  218(141+77)   & 81 & 29  & 24 & 20 &14 \\ \hline
    \end{tabular}
    \end{center}
\end{table}

\begin{table}[h]\label{tab:2}
    \caption{Number of events after the successive cuts 
defined in the text for point 2. (at 100 fb$^{-1}$).}
    \begin{center}
    \begin{tabular}{|l||c|c|c|c|c|c|} \hline 
        Process           & $4\ell$ events & $Z$-veto & $E_T^{\ell}$ &
 $E_T^{\rm{miss}}$& $E_{T}^{\rm{jet}}$ & $4\ell$ inv. m. \\ \hline
    $\tilde{q}$, $\tilde{g}$  
      & 547 & 203   &  55 &  4 & 0 &0 \\ \hline 
%    $\tilde{t} \tilde{t}$   &       &     &           &        &        &    \\ \hline 
    $\tilde{\ell}$,$\tilde{\nu}$ & 246 & 183 & 125 & 107 & 83 & 83      \\ \hline 
    $\tilde{\chi} \tilde{\chi}$& 304    & 123 & 79  & 66 & 50 & 48  \\ \hline
    $tH^-$ + c.c.   & 31      &  10   &    7  &  6   & 3  &  2  \\ \hline 
    $ZZ$                  & 1490 &32  & 25 & 5 &  4    &  4    \\ \hline 
%    $Z+$~jet                &    &  &      &   & &    \\ \hline
    $ttZ$                &49  &2  &  0 &0  &0  &0    \\ \hline
   $tth^0$                &  6 &1  & 1     &1    &0 & 0  \\ \hline   
    $H^0,A^0$ signal          &  147(68+79) & 98  & 85 & 68 & 54 & 54  \\ \hline
    \end{tabular}
    \end{center}
\end{table}

Finally, we should mention that further studies are in progress,
attempting to distinguish between `hadronically quiet' vs. `hadronically
active' events, in the hope of further increasing the signal-to-background
rates and the statistical significances of the signals. 
In fact, recall that the
signal is produced via $gg\to H^0/A^0$ and $gg,q\bar q\to b\bar b H^0/A^0$.
The former tend to produce additional jets which are rather forward and
soft, so that they tend to not enter the detector. In the latter, the
$b$-jets are somewhat harder, yet most of them will be missed by the
apparata. Hence, it may be worthwhile to veto jet activity above
$E_T$ values even lower than 50 GeV, 
as it is likely that jets from the background will  
populate the high $E_T$ region of the detector more often than 
those of the signal.

\section{Outlook}

Future work will develop along the two following directions.

\begin{enumerate}

\item {\underbar{Discovery regions}} Just like it was done in 
Refs.~\cite{Moortgat:2001pp} and \cite{Bisset:2003ix} (see also
\cite{Denegri:2001pn}), the next step of the analysis will be to 
express the discovery potential of the four-lepton mode over
the customary $(M_{A^0},\tan\beta)$ plane, corresponding to some
sample choices of SUSY parameters. Indeed, we foresee that
a significant part of the LHC wedge region will be covered, as
our results may be considered as un upgrade of those in 
\cite{Moortgat:2001pp,Denegri:2001pn} (e.g., see  Fig. 19(27) 
in \cite{Moortgat:2001pp}(\cite{Denegri:2001pn})\footnote{Some 
optimisation of cuts is also being currently investigated,
by exploiting the fact that the heavier the Higgs masses the higher 
the typical lepton momenta. In fact, so far
we have not tiered the selection cuts to the actual value
of $M_{H^0}$ and $M_{A^0}$. This approach should further strengthen
our considerations on the impact of $t\bar t$ and ${\tilde t}{\tilde t}$
background events.}).

\item {\underbar{-ino mass spectrum and -ino-ino-Higgs coupling determination}}
This is a new direction that was not exploited in Refs.~\cite{Moortgat:2001pp} 
and \cite{Bisset:2003ix}, so that we document it below
in some detail, also showing several preliminary results. In short, 
the idea is to ascertain information about the gaugino mass spectrum
and the Higgs-gaugino coupling strengths from a subset of our four-lepton
events, limited to those with two $e^{\pm}$'s and two ${\mu}^{\pm}$'s, out
of a signal enriched sample (or possibly after the SM 
and SUSY backgrounds have been subtracted) obtained via
our selection procedure.  The original idea is due to 
Refs.~\cite{Baer:1992kd,Baer:1994fx,Moortgat:2001pp}, where it was however applied
to $H^0,A^0 \rightarrow \tilde{\chi}_2^0\tilde{\chi}_2^0$ only.  Chief
differences in our context are: (i) additional
pairs of unlike neutralinos leading to the four-lepton events;
(ii) a mixture of different gaugino pairs (charged or neutral) yielding 
the same multi-lepton signature;
(iii) a cascade of decays leading from the original --inos generated
in the 
Higgs boson decays to the final four-lepton (plus LSP's) final states;
(iv) the potential of many more signal events (especially at higher
values of $\tan\beta$) than were anticipated in those analyses;
(v) generally stiffer lepton momenta since heavier Higgs decaying
into heavier --inos occur with non-negligible probability.

\end{enumerate}

\noindent
By making `Dalitz-type' plots of various combinations of leptons' 
invariant masses, we noticed some interesting structure. 
Consider point 2. and plot the di-lepton invariant masses $M_{e^+e^-}$
and $M_{\mu^+\mu^-}$ from the
same event against each other, for the signal and the dominant
SUSY backgrounds (see Tab.~22). 
Clearly the two-dimensional topology of the plots
contains kinematical information that should point to
the underlying decay. For example, one should see the
${\tilde\chi}^0_{2,4}-{\tilde\chi}^0_1$ mass differences from 
the edges of the mass distributions. (The forthcoming plots
will correspond to $H^0+A^0$ signal events and assume 
up to  $1200$~fb$^{-1}$, i.e., both ATLAS and
CMS statistics after six years of LHC running at high luminosity).
Fig.~\ref{fig:masses} (left plot), 
even if one only enforces the first 
two cuts (`$4\ell$ events' and
`$Z$-veto'), confirms these expectations and
also proves that  backgrounds do not seem to alter tragically the 
pattern expected in the di-lepton invariant mass plots for the signal.
In fact, to implement the next cut (`$E_T^{\ell}$') leads to the
backgrounds {\sl fortifying} the pattern expected from the signal alone,
see Fig.~\ref{fig:masses} (right plot). The study of this kind of 
plots should enable us not only to say which decays
are happening but also to extract
the relevant masses and couplings entering the channels in (\ref{allproc}). Work is
in progress in this direction.

\begin{figure}[t]
\epsfig{file=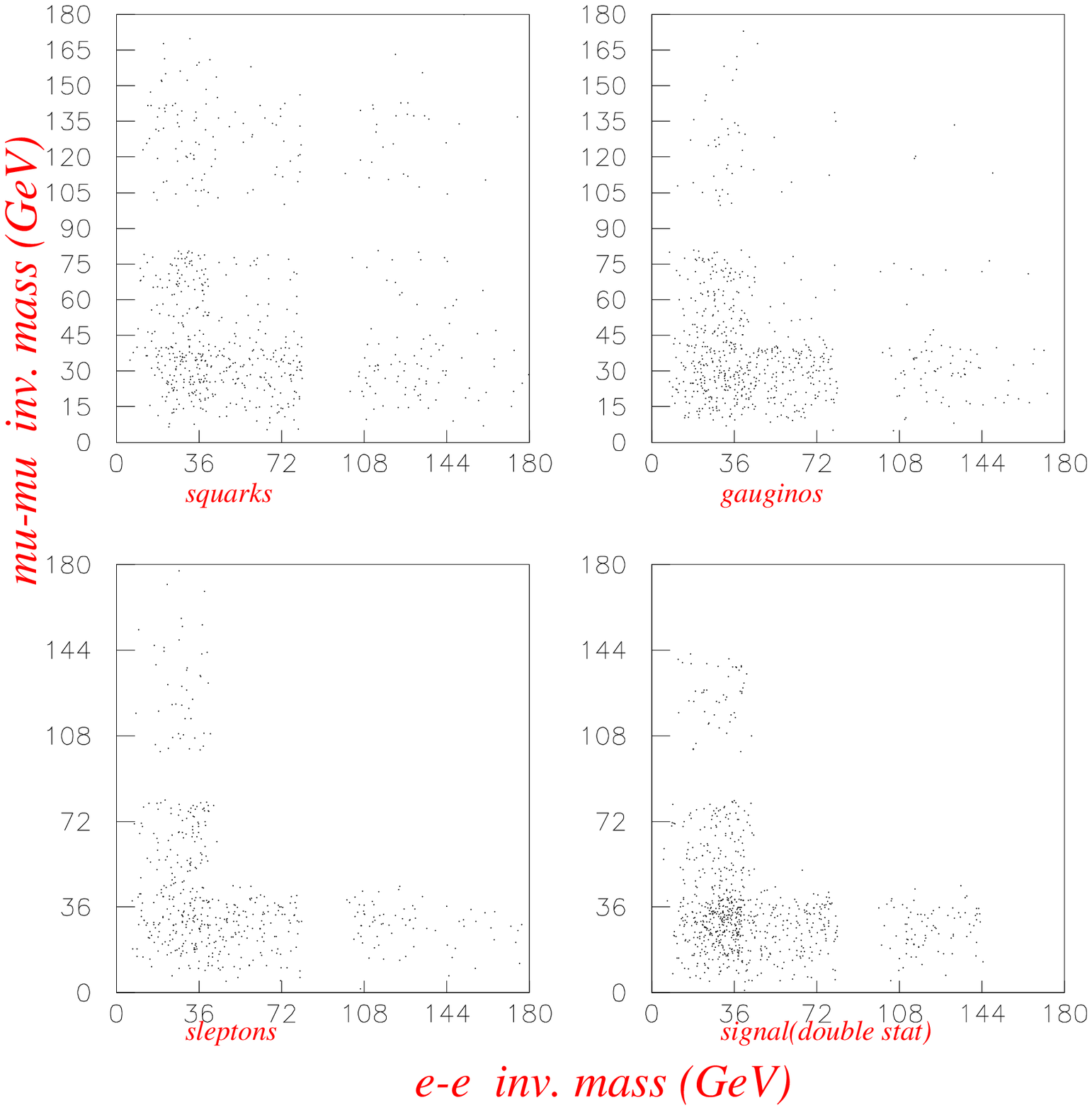,height=80mm,width=80mm}
\epsfig{file=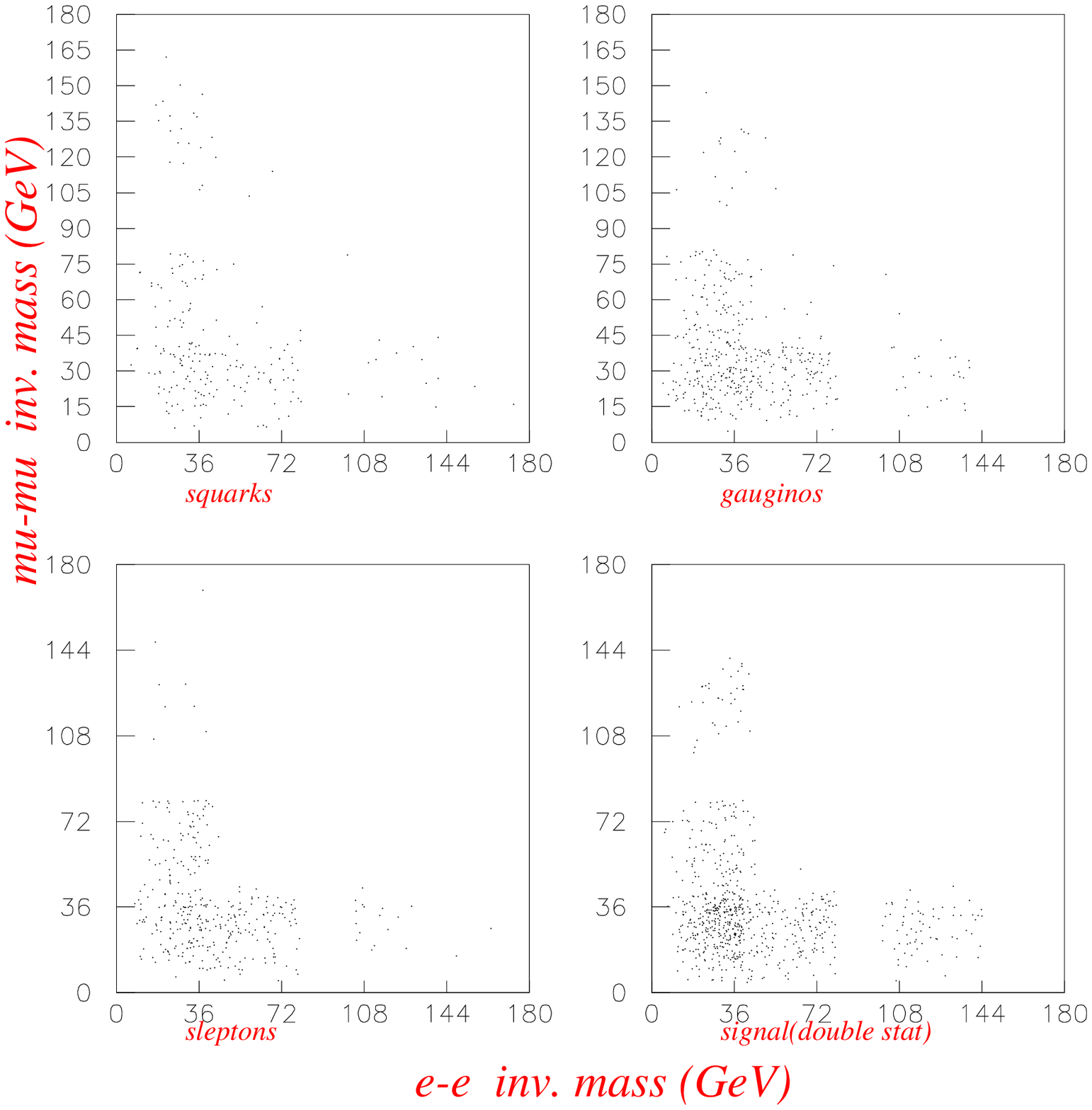,height=80mm,width=80mm}
\caption{Left: Di-lepton invariant mass plots for the most
significant backgrounds and signal corresponding to point 2., 
at $600$~fb$^{-1}$ (except the signal, which is twice this). Here the
only cuts made are the `$4\ell$ events' and
`$Z$-veto' ones. Right: Same as previous plot, but in presence of
the additional `$E_T^{\ell}$' cut.
\label{fig:masses} }
\end{figure}

\section{Conclusions}

We have reported on work currently being done into furthering the 
scope of Higgs $\to$ SUSY particle decays in 
the LHC quest for MSSM Higgs bosons. In particular, we have
highlighted the potential of $H^0/A^0$ decays into all possible
-ino pairs (both neutral and charged), finally yielding four-lepton
signatures of electron and muons (plus missing energy and some
hadronic activity), in covering part of the 
so-called wedge region of the $(M_{A^0},\tan\beta)$ plane expected 
at the LHC, where -- so long that only Higgs decays into ordinary SM
objects are considered -- the lightest
MSSM Higgs boson is the only accessible Higgs state 
and indistinguishable from the SM one
(the so-called decoupling scenario). Our analysis expands
upon previous ones  \cite{Baer:1992kd,Baer:1994fx,Moortgat:2001pp,Bisset:1995dc} which were
limited to 
$H^0/A^0 \rightarrow \tilde{\chi}_2^0\tilde{\chi}_2^0$
decays. Our conclusions of course assume
that the relevant SUSY parameters are in favourable
configurations, yet the latter are found to be those
that will be probed first by the LHC (as they lay close
to current experimental limits, see Fig.~\ref{tb8color}), 
so that our analysis
is of immediate phenomenological impact at the CERN machine.
Refinements of this work are in progress and possible
outlooks in this respect were described in some detail.

}

%% file: ravat.tex
{
\noindent
{\Large \bf G. The $H\to\gamma\gamma$ in associated Production
Channel} \\[0.5cm]
{\it O.\,Ravat and M.\,Lethuillier}

\begin{abstract}
  This paper describes a study of the Higgs associated production with a gauge
  boson, W or Z, in the Standard Model framework. The W and Z decay
  leptonically. Higgs Boson masses  from 115 to 150 GeV
  and backgrounds have been generated with the CompHEP generator, and the fast
  detector simulation CMSJET is used. Results are presented for an integrated luminosity corresponding
  to 1 year of LHC running at high luminosity.
 \end{abstract}

\section{Introduction}

The observation of a light Higgs boson decaying to two photons in the inclusive channel is not an easy task.
QCD backgrounds are important and are very demanding on ECAL performances. In this paper will be considered the
$pp \to \gamma\gamma + lepton(s)$ channel. In this channel the cross section is much smaller but the strong 
background suppression makes a discovery less demanding on the ECAL. Another interesting feature
of this channel is the presence of at least one charged lepton giving the location of the Higgs production
vertex easily, with a good precision, and hence could improve the resolution on $M_{Higgs}$.
Finally, it has been 
shown in \cite{denegri} that the $\gamma\gamma + leptons$ channel could be a rescue channel in case of maximal 
stop mixing. In this paper will be presented the simulation tools for the generation of the events, the K-factors obtained, and the discovery potential for the $\gamma\gamma + l + E_{T_{Mis}}$ and $\gamma\gamma l^++l^-$ final states.

\section{Simulation tools}

The calculation of cross sections is done using the V2HV program \cite{v2hv}, which performs NLO calculations. The K-factors are extracted and applied to the cross-section given by the Leading
Order generator : CompHEP \cite{comphep}.
The 4-vectors generation is done with the CompHEP package, wich allows to calculate squared matrix elements corresponding to the complete set of SM tree level diagrams of the considered processes, and performs the convolution with the parton distributions. The branching ratios concerning the Higgs decay are taken from HDECAY \cite{Djouadi:1998yw}.The final state is then processed by
Pythia \cite{Sjostrand:2000wi,Sjostrand:2000wi0}, 
which performs the hadronization and the fragmentation of the jets. 
Pythia is also used for the generation of underlying events and minimum-bias events.
The fast detector simulation is done with CMSJET \cite{cmsjet}

\section{WH production}

The Next to Leading Order calculations program V2HV gives the cross sections. CompHEP, a LO generator at the parton level, gives 1.33 times less and a K factor of 1.33 is applied for the signal. WH samples have been generated for Higgs Bosons masses from 115 GeV to 150 GeV. As no NLO generator for the background is available yet, the same factor is used for irreducible background.

\section{ZH production}

Again V2HV is used to determine the cross sections of the process, from $M_H$=115 GeV to $M_H$
=150 GeV. CompHEP is also used, and the K-factor is 1.27.

\section{Backgrounds}

For the generation of the backgrounds, only CompHEP is used. The $\gamma\gamma+l+E_{T_{Miss}}$ and $\gamma\gamma+
l^+l^-$ final states are considered. The irreducible backgrounds are not taken into account yet, but according to \cite{dubinin} they shouldn't be important. In order to have an efficient production, kinematical cuts were applied :
both photons are required to have $p_T>$ 20 GeV, as well as leptons. An isolation criterium is imposed between
generated particles: they have to be separated by a cone of $\Delta R = \sqrt{\Delta\phi^2+\Delta y^2} >0.3$.
Finally there is a cut on the rapidity of the particles : $|y|<2.7$.
The same cuts are applied to the signal and the following cross sections are found : 

The cuts applied at generation don't represent a serious handicap, as the thresholds for the Level-2 trigger are 31 GeV and 16.9 GeV GeV for di-photons, or di-electrons (12GeV for di-muons) : see \cite{tdr}.  

\section{CMSJET parametrisation}

The processing of 4-vectors is done with CMSJET, a fast simulation tool. The pile-up is included 
at this level of the simulation, the average number of envents per bunch crossing is 17.3. Only particles satisfying the following cuts are processed by CMSJET :
The minimum transverse momentum values required are the same as the ones from CompHEP. An isolation criterium is applied on photons and leptons. No charged track with $p_T>2$ GeV in a cone of $\Delta R = \sqrt{\Delta\phi^2+\Delta\eta^2}=0.3$ ( {\it tracker  isolation} ), no cluster with more than 5 GeV in the same cone ( {\it calorimeter isolation} ). The $\eta$ cut is harder, the particles are asked to have $|\eta|<2.4$. Electronic noise is set to 50 MeV in the ECAL Barrel, 150 MeV in the endcaps. It is assumed that 0.5$\%$ of the calorimeter cells are dead. Trigger efficiencies are taken from the Data Acquisition $\&$ High Level Trigger TDR : 
59$\%$ for electrons from Ws, 83.7$\%$ for the Higgs photons, and 42$\%$ for muons from Ws;
We hence get a trigger efficiency of 93.32$\%$ for the $e^\pm \nu\gamma\gamma$ channel, and 90.55$\%$ for the $\mu^\pm \nu\gamma\gamma$ one.

\section{Results after 1 year at high luminosity}

As only irreducible backgrounds are considered in this study, a simple first analysis is used.
Simple cuts on transverse momentum are applied.  

The figure \ref{fig:opti_hard} shows $s/\sqrt{b}$ as a function of the cut on the photons momentum. We first seek 
the best value in terms of $s/\sqrt{b}$ for the hardest photon. Then the best cut found is applied, and 
the cut on the soft photon is looked for. The best set of cut is the following : 55 GeV for the hardest photon, 30 for the softest.
\begin{figure}[hbtp]
  \begin{center}
    \resizebox{5cm}{4cm}{\includegraphics{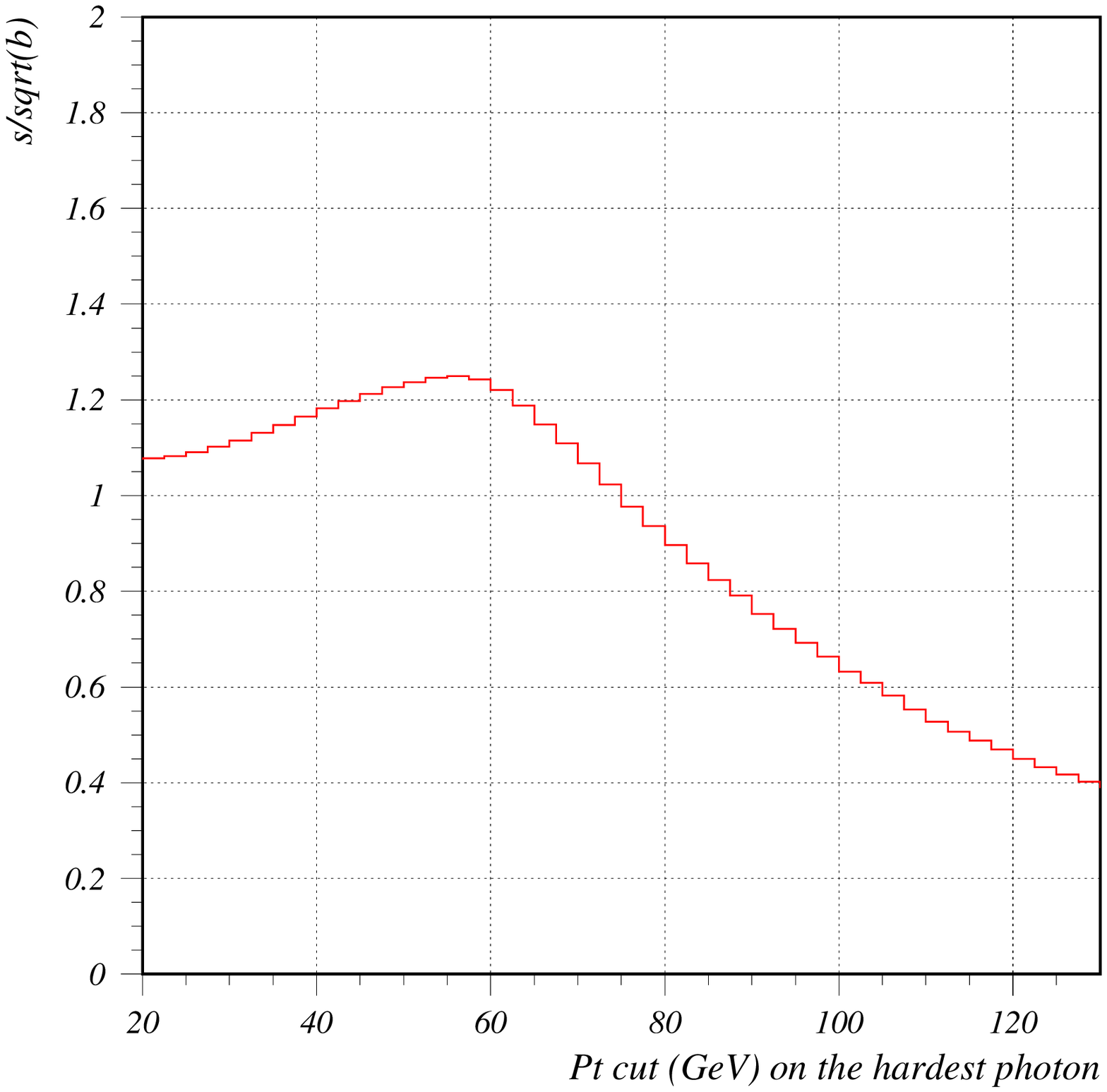}}\resizebox{5cm}{4cm}{\includegraphics{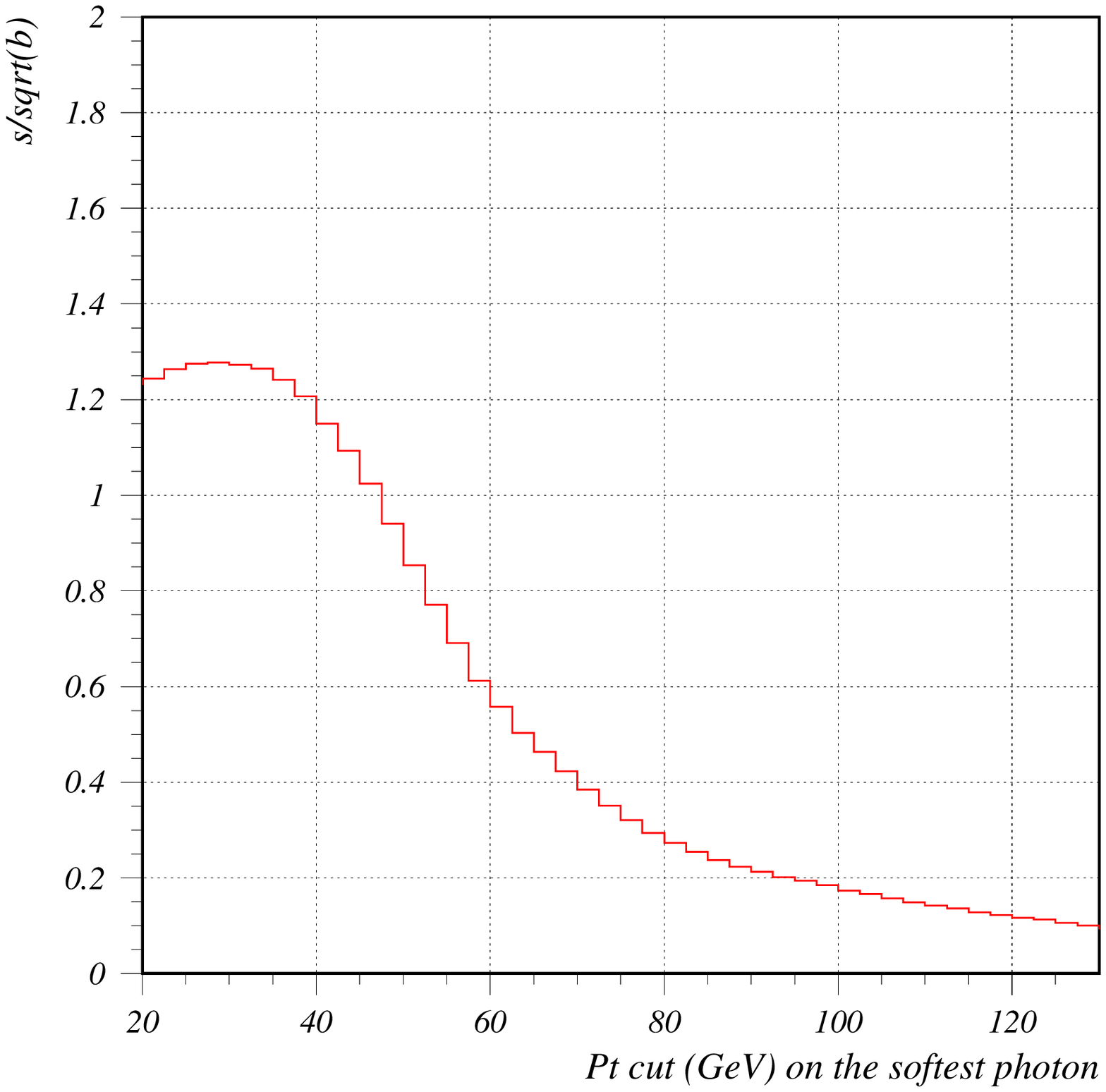}}
    \caption{Choice of the cut on the hardest photon.}
    \label{fig:opti_hard}
  \end{center}
\end{figure}

After simple cuts on photons of 55 GeV and 30 GeV we get the following number of events in a mass
window of $M_H \pm 1.6 GeV$ ({\it ie} : $M_H \pm 2\sigma$) : 13 for signal and 1.4 for background in the WH case, 1.13 and 1.73 
in the ZH case.

\begin{figure}[hbtp]
  \begin{center}
    \resizebox{4cm}{3cm}{\includegraphics{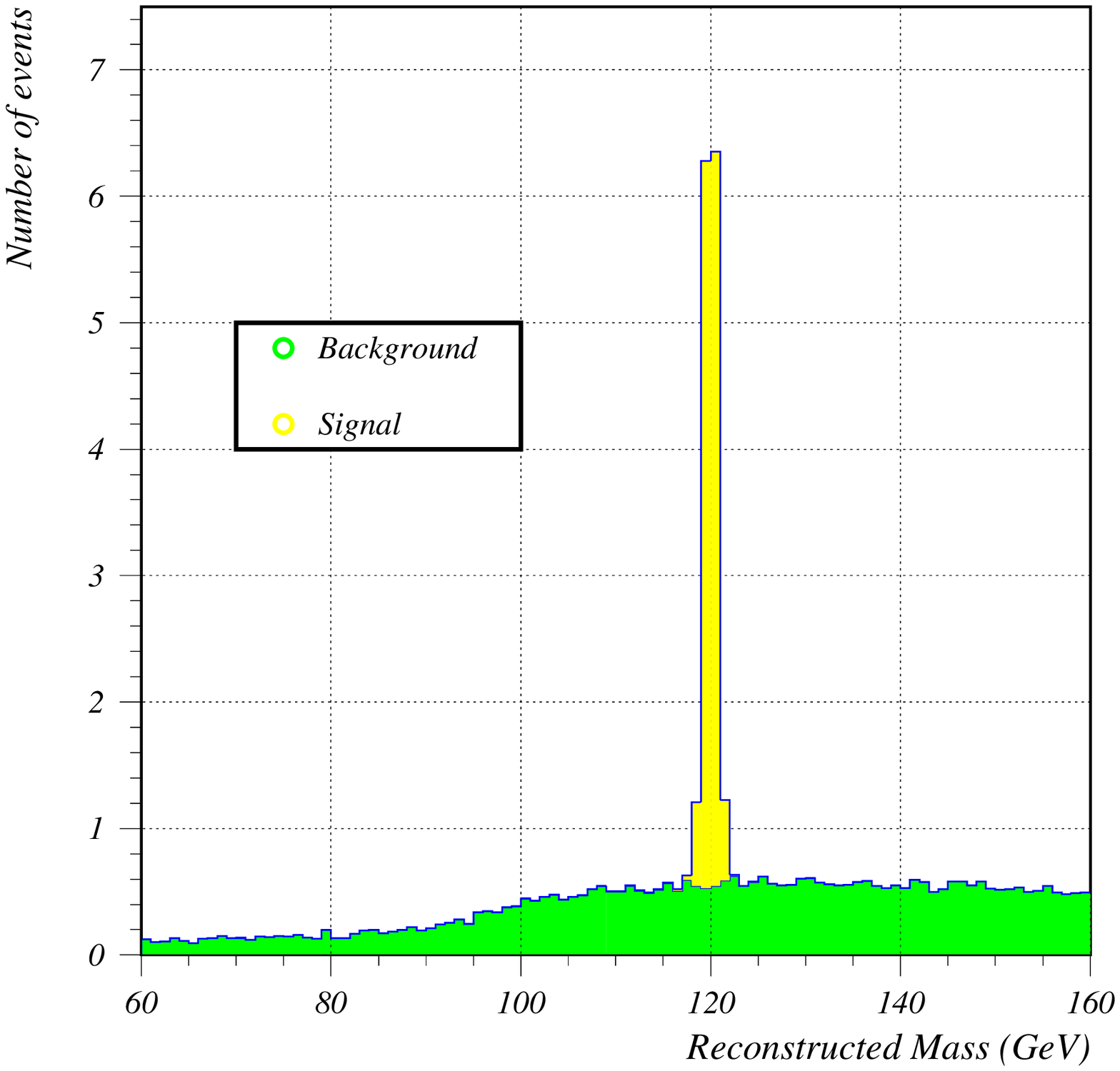}}\resizebox{4cm}{3cm}{\includegraphics{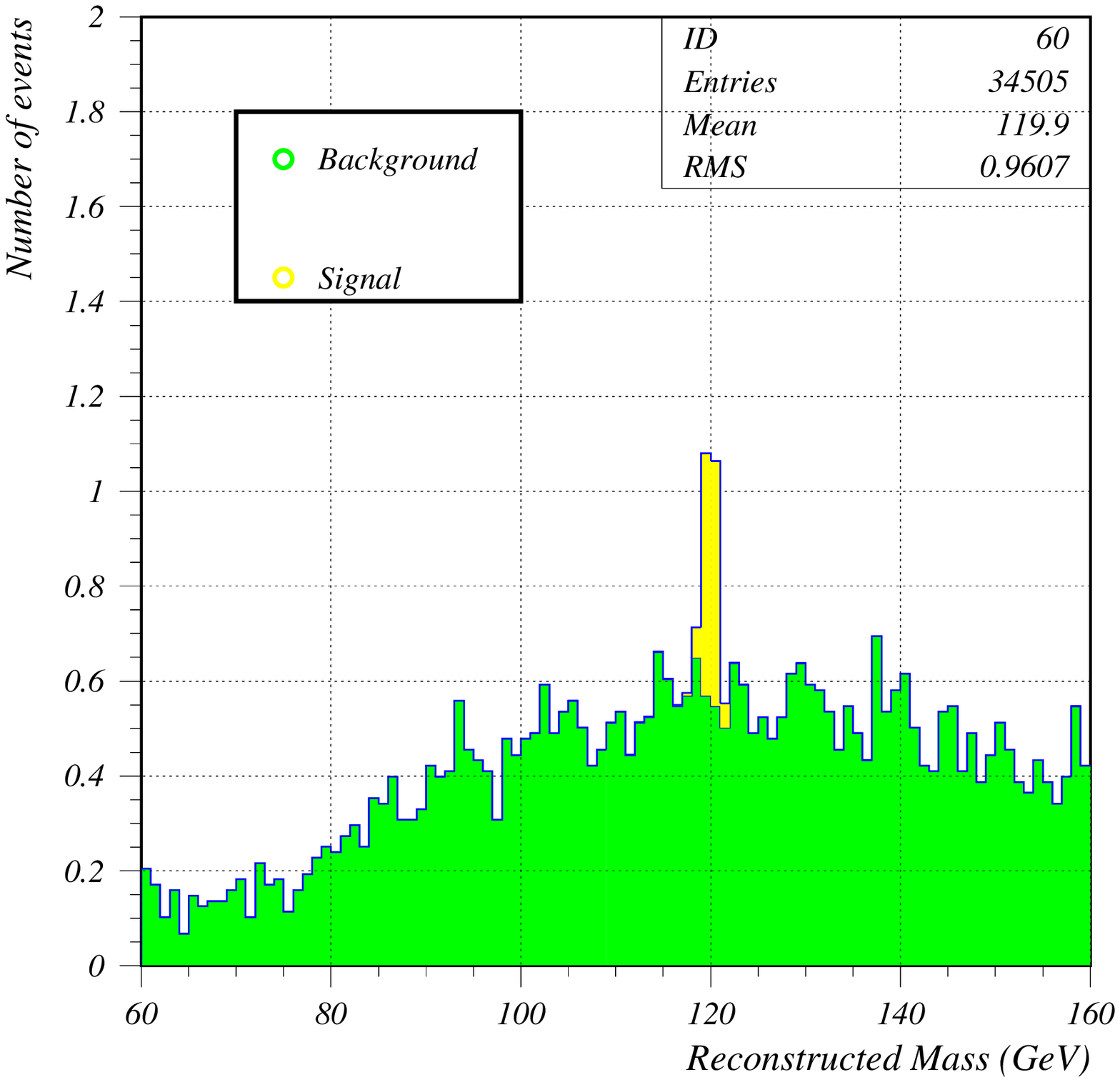}}\resizebox{4cm}{3cm}{\includegraphics{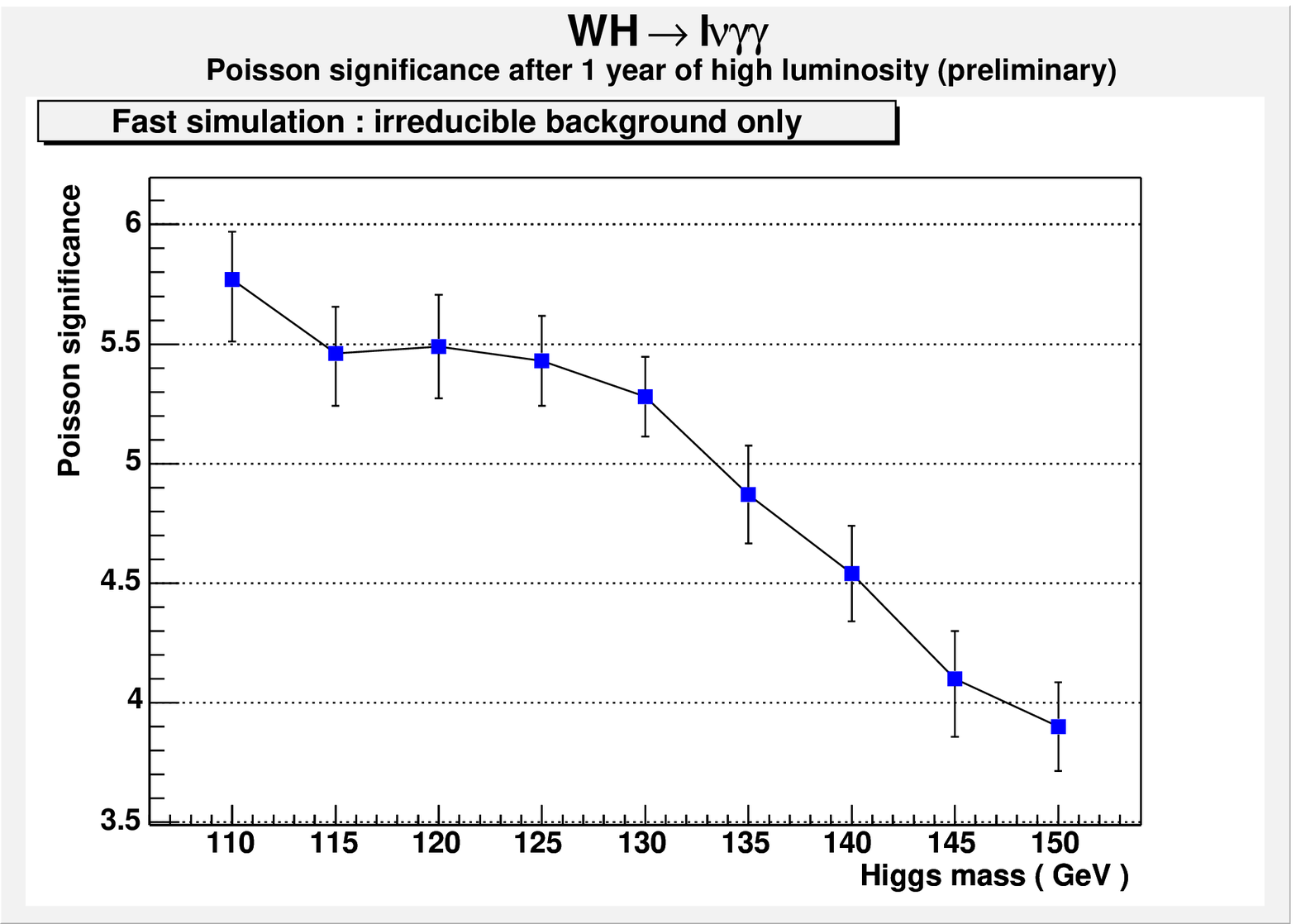}}

    \caption{$M_{\gamma\gamma}$ for both channels.}
    \label{fig:ex2}
  \end{center}
\end{figure}

 Those numbers being definitely not compatible with gaussian distributions, Poisson 
significances have been calculated. Results are given in figure \ref{fig:ex2}.

\section{Conclusions and Outlook}

The associated production of Higgs Bosons together with gauge bosons, despite its small cross section,
looks as an interesting way of completing a discovery scheme. It is less demanding on calorimeter 
performances and provides easily the Higgs vertex, improving the mass resolution. This first analysis considered the feasibility of a study of this channel. Only irreducible background has been considered. The promising results from fast simulation lead us to start a complete study, using full simulation of the detector. Besides these considerations, many models beyond the Standard Model show an enhancement of the discovery potential of this channel, and will be soon studied . 

}

%% file: boos.tex
{

\newcommand{\lsim}{\raisebox{-0.13cm}{~\shortstack{$<$ \\[-0.07cm] $\sim$}}~}
\newcommand{\gsim}{\raisebox{-0.13cm}{~\shortstack{$>$ \\[-0.07cm] $\sim$}}~}
\newcommand{\dx}{\mbox{\rm d}}
\newcommand{\ra}{\rightarrow}
\newcommand{\ee}{e^+e^-}
\newcommand{\tb}{\tan \beta}
\newcommand{\s}{\smallskip}
\newcommand{\nn}{\noindent}
\newcommand{\non}{\nonumber}
\newcommand{\beq}{\begin{eqnarray}}
\newcommand{\eeq}{\end{eqnarray}}
\newcommand{\miss}{\not\hspace*{-1.8mm}E}
\newcommand{\ct}[1]{c_{\theta_#1}}
\renewcommand{\st}[1]{s_{\theta_#1}}
\newcommand{\pn}[1]{\not{p}_#1}
\newcommand{\charpmi}[0]{\chi^\pm_i}
\newcommand{\charpmj}[0]{\chi^\pm_j}
\newcommand{\ctw}{c_W}
\newcommand{\stw}{s_W}
\newcommand{\ctwn}[1]{c^#1_W}
\newcommand{\stwn}[1]{s^#1_W}

%\begin{center}

\noindent
{\Large \bf H. MSSM Higgs Bosons in the Intense-Coupling Regime at the
LHC} \\[0.5cm]
{\it E.\,Boos, A.\,Djouadi and A.\,Nikitenko}

%\vspace*{2mm}

%\vspace*{1mm}
 
\begin{abstract}
Prospects for searching for the MSSM Higgs bosons in the intencse
coupling regime at the LHC are investigated.
\end{abstract}

 \section{The Intense-Coupling Regime}

In the MSSM Higgs sector, the intense-coupling 
regime \cite{Boos:2002ze,Boos:2003jt} is
characterized by a rather large value of $\tb$, and a pseudoscalar Higgs boson
mass that is close to the maximal (minimal) value  of the CP-even $h$ ($H$)
boson mass, $M_A \sim M_h^{\rm max}$, almost leading to a mass degeneracy of
the neutral Higgs particles, $M_h \sim M_A \sim M_H$. In the following, we will
summarize the main features of this scenario. For the numerical illustration,
we will use {\tt HDECAY}  \cite{Djouadi:1998yw}, fix the
parameter $\tb$ to the value
$\tb=30$ and choose the maximal mixing scenario, where the trilinear
Higgs--stop coupling is given by  $A_t \simeq \sqrt{6} M_S$ with the common
stop masses fixed to $M_S=1$ TeV; the other SUSY parameter will play only a
minor role. \s

Figure~\ref{fig:boos_a} (left) 
displays the masses of the MSSM Higgs bosons  as a function of
$M_A$. As  can be seen, for $M_A$ close to the maximal $h$ boson mass, which in
this case is $M_h^{\rm max}  \simeq 130$ GeV, the mass differences $M_A-M_h$
and $M_H - M_A$ are less than about 5 GeV. The $H^\pm$ boson mass, given by
$M_{H^\pm}^2 \sim M_A^2 +M_W^2$, is larger\,: in the range $M_A \lsim 140$ GeV,
one has $M_{H^\pm} \lsim 160$ GeV,  implying that charged Higgs bosons can
always be produced in top-quark decays, $t \to H^+ b$.  The couplings of the
CP-even Higgs bosons to fermions and gauge bosons  normalized to the SM Higgs
boson couplings are also shown in  Fig.~\ref{fig:boos_a} (right). 
For small $M_A$ values, the
$H$ boson has almost SM couplings, while the couplings of the $h$ boson to
$W,Z,t$ $(b)$  are suppressed (enhanced); for large $M_A$ values the roles of
$h$ and $H$ are interchanged. For medium values, $M_A \sim M_h^{\rm max}$, the
couplings of both $h$ and $H$ to gauge bosons $V=W,Z$  and top quarks are
suppressed, while the couplings to $b$ quarks  are strongly enhanced. The
normalized couplings of the CP-even Higgs particle are simply $g_{AVV}=0$ and
$g_{Abb} = 1/ g_{Att} =\tan\beta =30$.\s

These couplings determine the branching ratios of the Higgs particle, which are
shown in Fig.~\ref{fig:boos_b}. 
Because the enhanced couplings, the three Higgs particle
branching ratios to $b\bar{b}$ and $\tau^+\tau^-$  are the dominant ones,
with values $\sim 90$\% and $\sim 10$\% respectively. The decays $H \to WW^*$
do not exceed the level of 10\%, even for small $M_A$ values [where $H$ is
almost SM-like] and in most of the $M_A$ range the decays $H,h \to WW^*$ are
suppressed to the level where they are not useful.  The decays into $ZZ^*$ are
one order of magnitude smaller and the decays into $\gamma \gamma$ are very
strongly suppressed for the three Higgsses and cannot be used anymore. Finally,
note that the  branching ratios for the decays into muons, $\Phi \to \mu^+
\mu^-$, are constant in the entire $M_A$ range exhibited, at the level of $3
\times 10^{-4}$. \s

Summing up the partial widths for all decays, the total decay widths of the
three Higgs particles are shown in the left-hand side of Fig.~\ref{fig:boos3}. 
As can be
seen, for $M_A \sim 130$ GeV, they are at the level of 1--2 GeV, i.e. two
orders of magnitude larger than the width of the SM Higgs boson for this value
of $\tb$ [the total width increases as $\tan^2\beta$]. The right-hand side of
the figure shows the mass bands $M_\Phi \pm \Gamma_\Phi$ and, as can be seen, 
for the above value of $M_A$, the three Higgs boson masses are overlapping.
%\newpage

\begin{figure}[htbp]
\vspace*{-1.8cm}
\includegraphics[width=16cm]{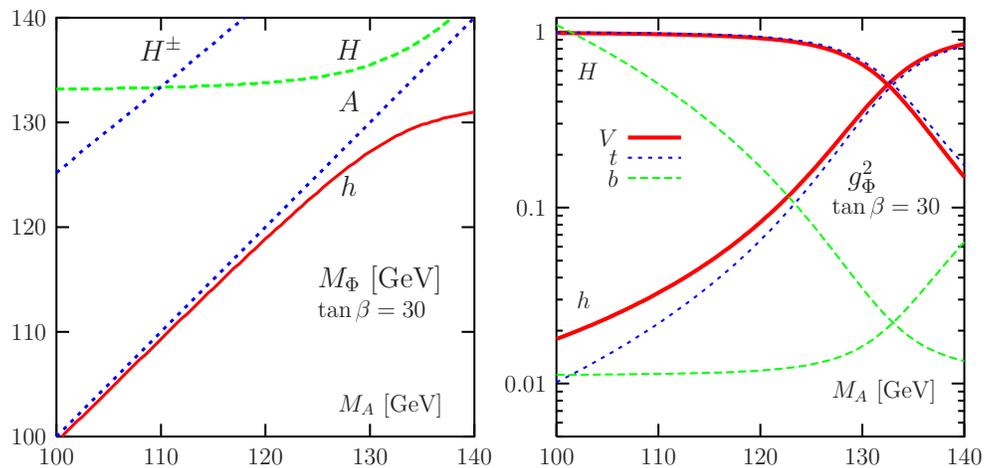}
%\begin{center}
%\centerline{\psfig{file=icr1.ps,width=16cm}}
\vspace*{-16.5cm}
\caption{ The masses of the MSSM Higgs bosons (left) and the normalized 
couplings of the CP-even Higgs bosons to vector bosons and third-generation 
quarks (right) as a function of $M_A$ and $tan \beta=30$. For the $b$-quark
couplings, the values $10 \times g_{\Phi bb}^{-2}$ are plotted.}
%\end{center}
\vspace*{-1.cm}
\label{fig:boos_a}
\end{figure}

\begin{figure}[htbp]
\vspace*{-1.7cm}
%\hspace*{-1.4cm}
\begin{center}
\vspace*{-2.2cm}
\hspace*{-1.4cm}
\includegraphics[width=18cm,height=16cm]{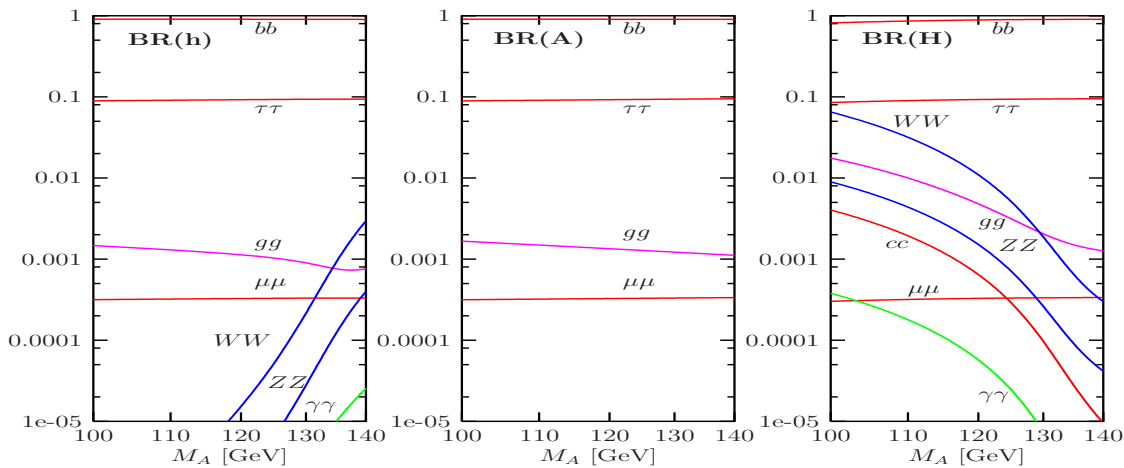}
%\begin{center}
%\vspace*{-2.cm}
%\centerline{\psfig{file=icr3.ps,width=18cm,height=16cm}}
\vspace*{-9.2cm}
\caption{ The branching ratios of the neutral MSSM Higgs bosons $h,A,H$ for 
the various decay modes as a function of $M_A$ and for $tan \beta=30$. }
\label{fig:boos_b}
\end{center}
\vspace*{-0.8cm}
\end{figure}

\begin{figure}[htbp]
\vspace*{-4.0cm}
\includegraphics[width=16cm]{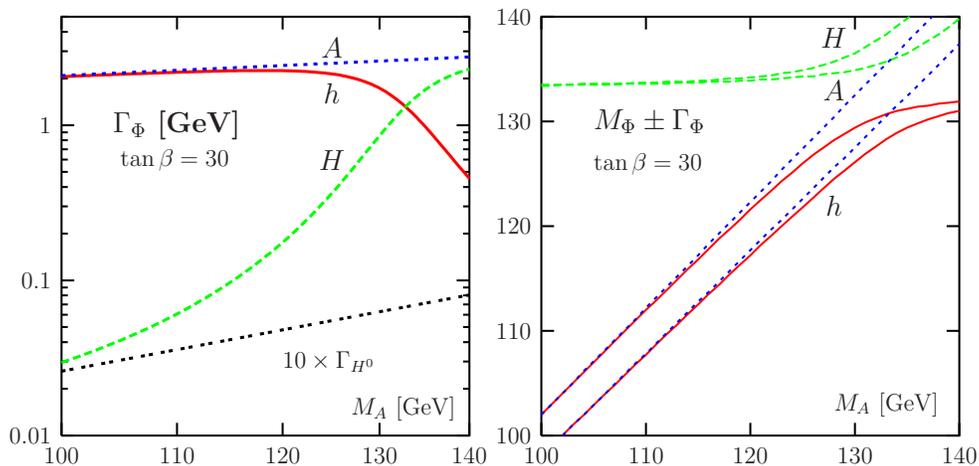}
\begin{center}
\vspace*{-1.0cm}
%\centerline{\psfig{file=icr2.ps,width=16cm}}
\vspace*{-15.4cm}
\caption{  Total decay widths $\Gamma_\Phi$ (left) and the mass bands $M_\Phi
\pm \Gamma_\Phi$ (right) for the neutral MSSM Higgs bosons as a function of 
$M_A$ and for $tan \beta=30$.}
\label{fig:boos3}
\end{center}
\vspace*{-3.0cm}
\end{figure}

%\newpage

\section{Discrimination of the three Higgs Bosons at the LHC}

The most difficult problem  we must face in the intense-coupling regime, is to
resolve between the three peaks of the neutral  Higgs bosons when their masses
are close to one another. The only decays with large branching ratios on which
we can rely are the $b\bar{b}$ and  $\tau^+ \tau^-$ modes. At the LHC, the
former has too large QCD background to be useful, while for the latter 
channel [which has been shown to be viable for  discovery] the expected
experimental resolution on the invariant mass of the  $\tau^+ \tau^-$ system is
of the order of 10 to 20 GeV, and thus clearly too large. One would then simply
observe a relatively wide resonance corresponding to $A$ and $h$ and/or $H$
production.  Since the branching ratios of the decays into $\gamma \gamma$ and
$ZZ^* \to  4\ell$ are too small, a way out is  to use the Higgs decays into
muon pairs: although the branchings ratio is rather small, BR($\Phi \to
\mu^+\mu^-) \sim 3.3 \times 10^{-4}$,  the resolution is expected to be as good
as 1 GeV, i.e. comparable to the total width,  for $M_\Phi \sim 130$ GeV. \s

Because of the strong enhancement of the Higgs couplings to bottom quarks, the
three Higgs bosons  will be produced at the LHC mainly\footnote{The  
Higgs-strahlung and vector-boson fusion  processes, as well as associated
production with top quarks, will have smaller cross sections since the Higgs
couplings to the involved particles are suppressed.} in the gluon--gluon
process, $gg \to \Phi=h,H,A \to \mu^+ \mu^-$,  which is dominantly mediated by
$b$-quark loops, and the associated production  with $b\bar{b}$ pairs,
$gg/q\bar{q} \to b\bar{b}+\Phi \to b\bar{b}+\mu^+ \mu^-$. We have generated
both the signals and backgrounds  with the program {\tt CompHEP}
\cite{Pukhov:1999gg}.  For the backgrounds to $\mu^+\mu^-$ production,
we have
included only the Drell--Yan process $pp \to  \gamma^*, Z^* \to \mu^+\mu^-$,
which is expected  to be the largest source. But for the $pp  \to \mu^+ \mu^-
b\bar{b}$ final state, however, we have included the full 4-fermion background,
which is mainly due to the process $pp \to b\bar{b} Z$ with $Z \to \mu^+
\mu^-$.  \s

\begin{figure}[htbp]
\vspace*{-1.1cm}
\includegraphics[width=16cm]{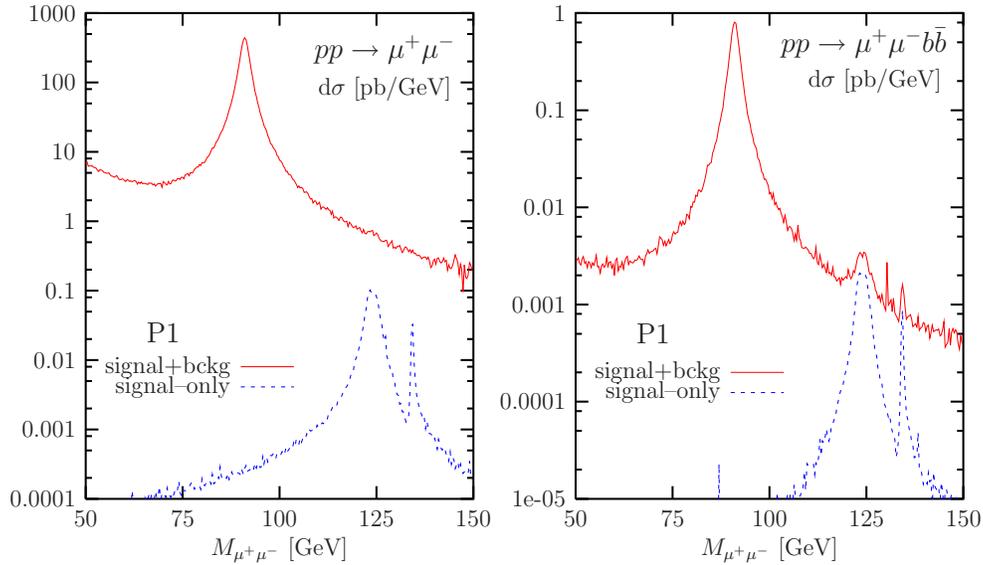}
%\begin{center}
%\vspace*{-1.5cm}
%\centerline{\psfig{file=icr4.ps,width=16cm}}
\vspace*{-15.0cm}
\caption{  The differential cross section in pb/GeV as a function of the
dimuon mass for the point P1, for both the signal and signal plus background 
in the processes $pp (\to \Phi) \to \mu^+ \mu^-$ (left figure) and $pp (\to 
\Phi b\bar{b}) \to \mu^+ \mu^- b\bar{b}$ (right figure).}
%\end{center}
%\vspace*{-0.5cm}
\label{fig:boos4}
\end{figure}

The differential cross sections are shown for the scenario $M_A=125$ GeV and
$\tb=30$, which leads to $M_h=123.3$ GeV and $M_H=134.3$ GeV, as a function of 
the invariant dimuon mass in Fig.~\ref{fig:boos4} (left), 
for  $pp (\to h,H,A) \to \mu^+
\mu^-$. As can be seen, the signal rate is fairly large but when put on top of
the huge Drell--Yan background, the signal becomes completely invisible. We
thus conclude, that already at the level of a ``theoretical simulation", the 
Higgs signal will probably be hopeless to  extract in this process for $M_H
\lsim 140$ GeV. In the right-hand side of Fig.~\ref{fig:boos4}, 
we display, again for the 
same scenario, the signal from $pp \to \mu^+\mu^- b\bar{b}$ and the complete
4-fermion SM  background  as a function of the dimuon system. The number of
signal events is an order of magnitude smaller than in the previous case, but
one can still see the two peaks, corresponding to $h/A$ and $H$ production, on
top of the background. \s

In order to perform a more realistic analysis, we have generated unweighted
events for the full 4-fermion background $pp \to \mu^+ \mu^- +b\bar{b}$ and for
the signal. With the help of the new {\tt CompHEP-PYTHIA} interface
\cite{Belyaev:2000wn},  the unweighted events have been processed
through {\tt PYTHIA 6.2} \cite{Sjostrand:2001yu} for fragmentation and
hadronization. To simulate
detector effects, such as acceptance, muon momentum smearing, and $b$--jet
tagging,  we take the example of the CMS detector. The details have been given
in Ref.~\cite{Boos:2003jt} and the main points are that: 1) the mass
resolution on
the dimuons  is about 1\%, and 2) the efficiency for $b$--jet tagging is  of
the order of 40\%. The results of the simulation for a luminosity of 100
fb$^{-1}$ are shown in Fig.~\ref{fig:boos5}, 
where the number of $\mu^+\mu^- b\bar{b}$ events
in bins of 0.25 GeV is shown as a function of the mass of the dimuon system.
The left-hand side shows the signals with and without the resolution smearing
as obtained  in the Monte-Carlo analysis, while the  figures in the right-hand
side show also the backgrounds, including the  detector effects. \s

For the point under consideration, the signal cross section for the
heavier  CP-even $H$ boson is significantly smaller than the signals from the
lighter CP-even $h$ and pseudoscalar $A$ bosons; the latter particles are too 
too close in mass to be resolved, and only one single broad peak for $h/A$ is 
clearly visible. To resolve also the peak for the $H$ boson, the   integrated
luminosity should be increased  by a factor of 3 to 4. We have also performed
the analysis for $M_A=130$ and 135 GeV. In the former case, it would be
possible to see also  the  second peak, corresponding to the $H$ boson signal
with a luminosity  of 100 fb$^{-1}$, but again the $h$ and $A$ peaks cannot be
resolved.  In the latter case, all three $h,A$ and $H$ bosons have comparable
signal rates, and the mass differences are large enough for us to hope to be
able to isolate the three different peaks, although with some difficulty. \s
  
\begin{figure}[htb]
%\begin{center}
\vspace*{-0.3cm}
\includegraphics[width=7.5cm]{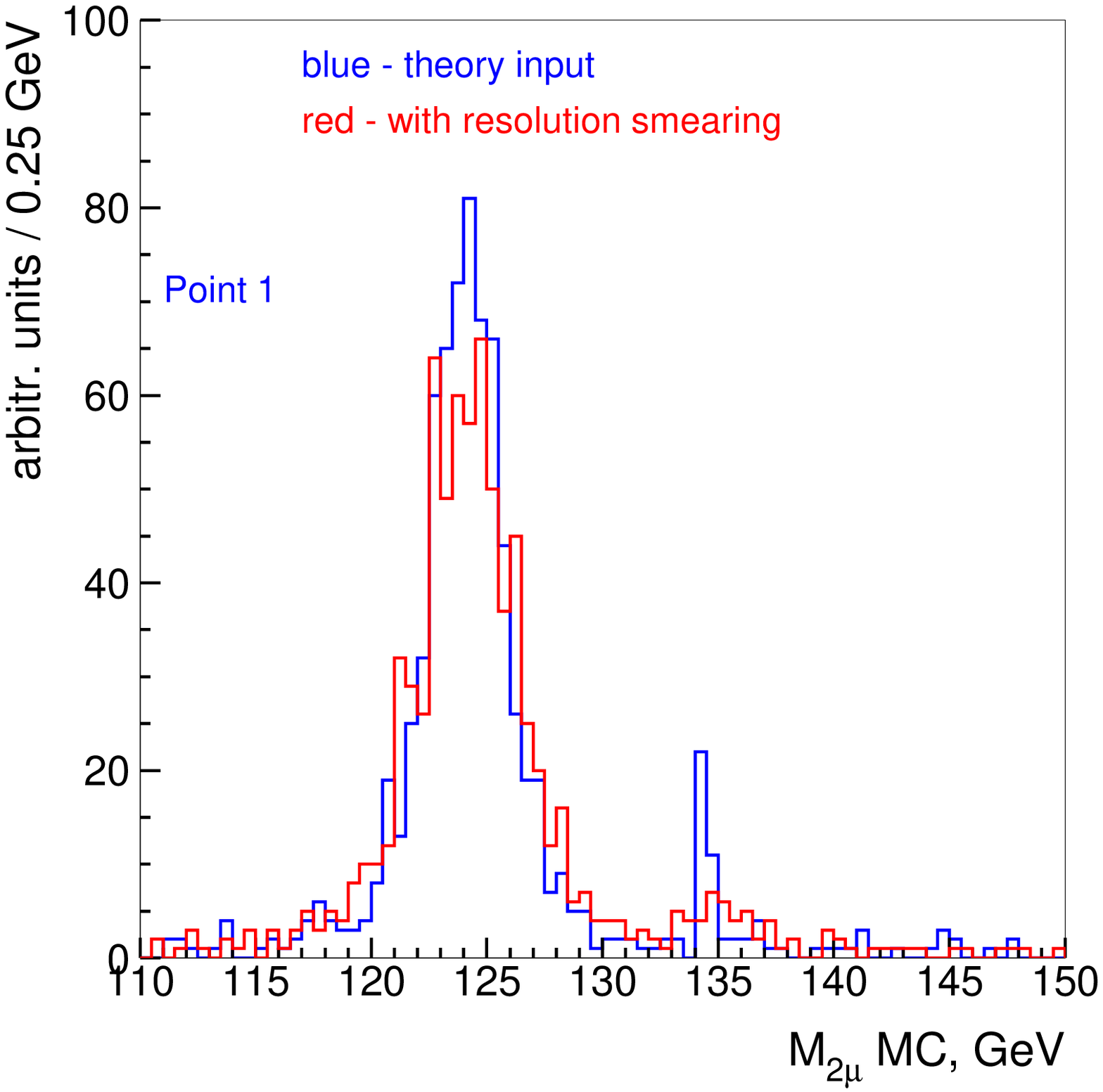}
\includegraphics[width=7.5cm]{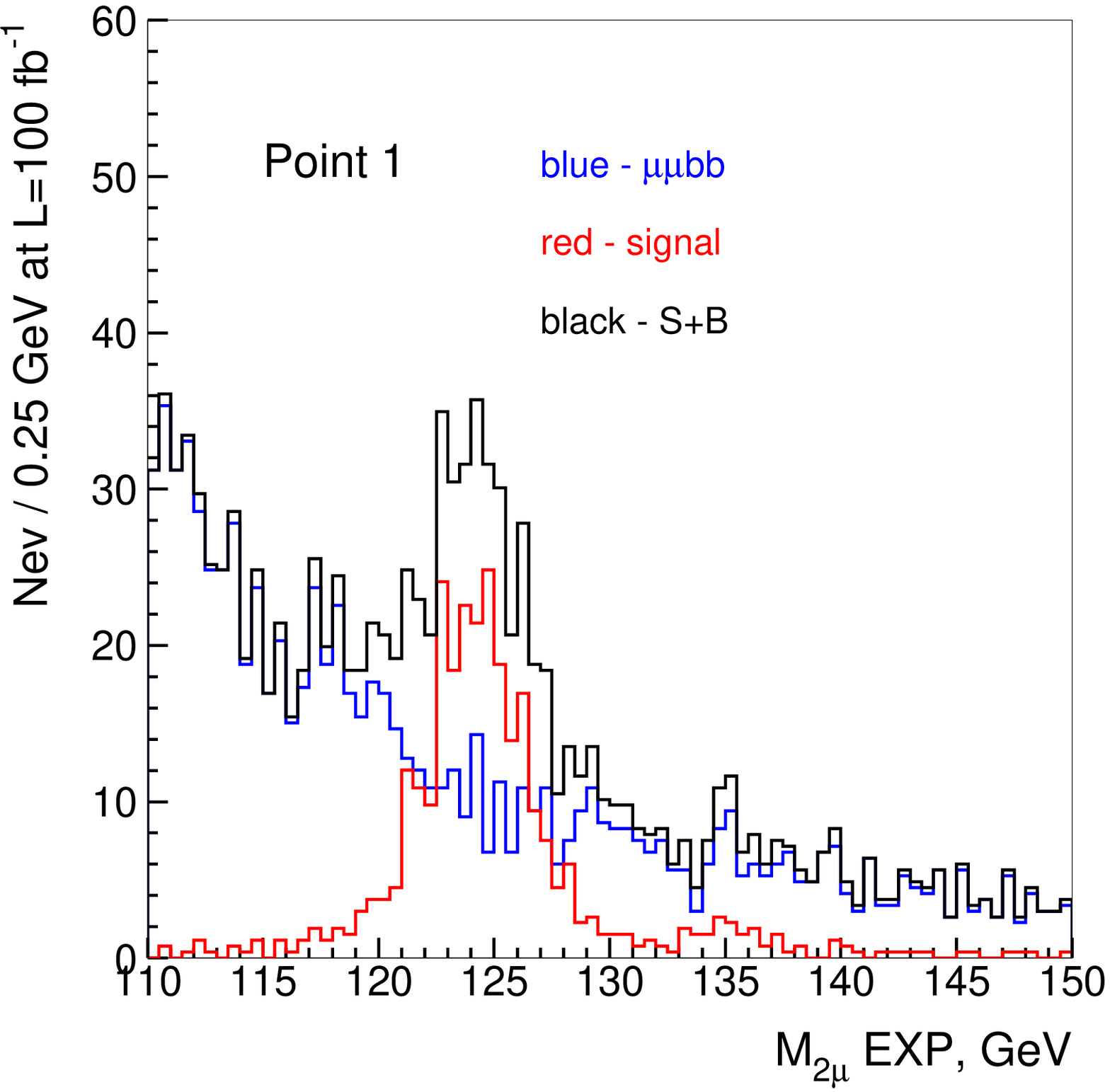}
%\begin{center}
%\vspace*{-1.1cm}
%\centerline{\epsfig{file=point1.eps,width=9cm}
%\epsfig{file=point1_sb.eps,width=9cm}}
\vspace*{-0.5cm}
\caption{ $\mu^+ \mu^-$ pair invariant mass distributions
for the signal before and after detector resolution smearing (left)
and for the signal and the background (right) for $M_A=125$ GeV.}
%\end{center}
\vspace*{-0.5cm}
\label{fig:boos5}
\end{figure}
%\newpage

\section{Summary} 

We have shown that in the intense-coupling regime, i.e. when the $h,H$ and $A$
MSSM bosons have masses too close to the critical point $M_h^{\rm max}$ and
when the value of $\tb$ is large, the detection of the individual Higgs boson
peaks is very challenging at the LHC. It is only in the associated Higgs
production  mechanism with $b\bar{b}$ pairs, with at least one tagged $b$-jet,
and with Higgs particles decaying  into the clean muon-pair final states, that
there is a  chance of observing the three signals and resolve between  
them\footnote{Recently, it has been argued that central diffractive 
Higgs production could allow to discriminate between $h$ and 
$H$ production  since a very small mass
resolution can be obtained \cite{Kaidalov:2003fw}.}.
This would be possible only if the Higgs  mass differences are larger than about 5
GeV. In this note, we mostly concentrated on the fully exclusive $b\bar{b}+
\mu^+\mu^-$ signature. In a more complete study,  one should consider the 
case where only one single $b$-jet is tagged \cite{Campbell:2002zm}, and
take into  
account also the large reducible backgrounds from
$pp \to Z^*/\gamma^* \to \mu^+\mu^-$ with mistagged jets. Furthermore, once the
signal peaks have been isolated, the $pp \to \mu^+ \mu^-$ process can possibly
be used to improve further the discrimination. Such a study  
is under way \cite{elsewhere}.

}

%% file: guasch.tex
{
%

%%%%%%%%%%%%%%%%%%%%%%%%%%%%%%%%%%%%%%%%%%%%%%%%%%

%%%%%%%%%%%%%%%%%%%%%%%%%%%%%%%%%%%%%%%%%%%%%%%%%%%%%%%%%%%%%%%%%%%%%%%%%
% Macro definitions
%%%%%%%%%%%%%%%%%%%%%%%%%%%%%%%%%%%%%%%%%%%%%%%%%%%%%%%%%%%%%%%%%%%%%%%%%

\newcommand{\gsim}{\stackrel{\scriptstyle >}{{ }_{\sim}}}
\newcommand{\lsim}{\stackrel{\scriptstyle <}{{ }_{\sim}}}
\newcommand{\mb}{\ensuremath{m_b}}
\newcommand{\mt}{\ensuremath{m_t}}
\newcommand{\Dmb}[1][]{\ensuremath{\Delta\mb^{#1}}}
\newcommand{\Dmq}[1][]{\ensuremath{\Delta m_q^{#1}}}
\newcommand{\Dmf}[1][]{\ensuremath{\Delta m_f^{#1}}}
\newcommand{\tb}[1][]{\ensuremath{\tan^{#1}\!\beta}}
\newcommand{\MSUSY}{\ensuremath{M_\mathrm{SUSY}}}
\newcommand{\mg}{\ensuremath{M_{\tilde{g}}}}
\newcommand{\msb}[1]{\ensuremath{M_{\tilde{b}_{#1}}}}
\newcommand{\mst}[1]{\ensuremath{M_{\tilde{t}_{#1}}}}
\newcommand{\mstau}[1]{\ensuremath{M_{\tilde{\tau}_{#1}}}}
\newcommand{\msf}[1]{\ensuremath{M_{\tilde{f}_{#1}}}}
\newcommand{\TeV}{\ensuremath{\,{\rm TeV}}}
\newcommand{\GeV}{\ensuremath{\,{\rm GeV}}}
\newcommand{\rsm}{\ensuremath{R^{\rm SM}}}
\newcommand{\rmssm}{\ensuremath{R^{\rm MSSM}}}
\newcommand{\brHtb}{\ensuremath{BR(H^{+} \to t\bar{b})}}
\newcommand{\brHtaunu}{\ensuremath{BR(H^{+}\to \tau^+\nu^-)}}
\newcommand{\brHbb}{\ensuremath{BR(H\to b\bar{b})}}
\newcommand{\brHtt}{\ensuremath{BR(H\to \tau^+\tau^-)}}
\newcommand{\mtau}{\ensuremath{m_\tau}}
\newcommand{\mbsH}{\ensuremath{\mb^2(M_H)}}
\newcommand{\hb}{\ensuremath{h_b}}
\newcommand{\htau}{\ensuremath{h_\tau}}
\newcommand{\sa}{\ensuremath{\sin\alpha}}
\newcommand{\ca}{\ensuremath{\cos\alpha}}
\newcommand{\ta}{\ensuremath{\tan\alpha}}
\newcommand{\sbt}{\ensuremath{\sin\beta}}
\newcommand{\cbt}{\ensuremath{\cos\beta}}
\newcommand{\ma}{\ensuremath{M_{A^0}}}
\newcommand{\mz}{\ensuremath{M_{Z}}}
\newcommand{\Dmtau}{\ensuremath{\Delta m_\tau}}
\newcommand{\mHc}{\ensuremath{m_{H^\pm}}}
%Units
\newcommand{\fb}{\ensuremath{\rm \,fb}}
\newcommand{\pb}{\ensuremath{\rm \,pb}}
%\newcommand{\GeV}{\ensuremath{\rm \,GeV}}
%%%%%%%%%%%%%%%%%%%%%%%%%%%%%%%%%%%%%%%%%%%%%%%%%%
% Joint bibliographies:
%\bibliographystyle{unsrthep}

\newcommand{\CGGJS}{Guasch:1995rn}
\newcommand{\jaume}{Belyaev:2001qm,Belyaev:2002eq,Belyaev:2002sa}
\newcommand{\logan}{Carena:2001bg}
\newcommand{\tauola}{Jadach:1990mz,Jezabek:1992qp,Jadach:1993hs}
\newcommand{\ketevi}{Assamagan:2002ne,Assamagan:2002hz}
\newcommand{\Roy}{Roy:1999xw,Moretti:1999bw,Miller:1999bm}
%%%%%%%%%%%%%%%%%%%%%%%%%%%%%%%%%%%%%%%%%%%%%%%%%%

\hyphenation{pa-ra-me-ter}
%\begin{titlepage} 

\section[ ]{Determining the ratio of the $H^{+} \rightarrow \tau \nu$ to
$H^{+} \rightarrow t \bar b$ decay rates for large $\tan \beta$ at the
LHC\footnote{K.A.\,Assamagan, J.\,Guasch, S.\,Moretti and S.\,Pe{\~n}aranda}
}

%\section{Introduction}
\vspace*{0.2cm}
\noindent
In this note we investigate the production of charged Higgs bosons in
association with top quarks at the LHC, from the experimental and
theoretical point of view, by studying hadronic ($H^+\to t\bar{b}$) and
leptonic ($H^+\to \tau^+\nu$) decay signatures. The
interest of this investigation is many-fold.
\begin{itemize}
\item The discovery of a charged Higgs boson will point immediately to
  the existence of some extension of the Standard Model (SM).
\item The associated production of a charged Higgs boson with a top quark
  ($pp\to H^+ \bar{t}+X$)~{\cite{Gunion:1994sv,Barger:1994th}} is 
  only relevant at large values of
  $\tb$\footnote{It is in principle
   also relevant at very low values of $\tb$ (say, $\lsim 1$).
   In practise, this $\tb$ regime is excluded in the Minimal Supersymmetric
Standard Model (MSSM) from the
    negative neutral Higgs search at LEP~\cite{lep2}.
    Hence, hereafter, we will refrain from investigating the low $\tan\beta$
    case.}, a regime where
  Higgs boson observables receive large Supersymmetric (SUSY) radiative
  corrections. 
\item While SUSY radiative effects might be difficult to discern in the
  production cross-sections separately, they will appear neatly in the
  following relation between the two mentioned channels:
  \begin{equation}
    \label{eq:relation}
    R\equiv\frac{\sigma(pp\to H^+\bar{t} + X \to \tau^+ \nu t + X)}{\sigma(pp\to H^+\bar{t} + X \to t\bar{b} \bar{t} +
    X)}\,\,.
  \end{equation}
\item In fact, in the ratio of~(\ref{eq:relation}), the dependence on the
production mode (and on its large sources of
  uncertainty deriving from parton luminosity, unknown QCD radiative
  corrections, scale choices, etc.) cancels out:
  \begin{equation}
    \label{eq:relation2}
    R=\frac{BR(H^+\to \tau^+\nu)}{BR(H^+\to t\bar{b})}=\frac{\Gamma(H^+\to \tau^+\nu)}{\Gamma(H^+\to t\bar{b})} \,\,.
  \end{equation}
\end{itemize}
From these remarks it is clear that the quantity
$R$ is extremely interesting both experimentally and
theoretically in investigating the nature of Electro-Weak Symmetry Breaking
(EWSB).

In the MSSM, Higgs boson couplings to down-type fermions receive
large quantum corrections, enhanced by \tb. These corrections
have been resummed to all orders in perturbation theory with the help of
the effective Lagrangian formalism  for the $t\bar{b}H^+$
vertex~\cite{Carena:1999py,Guasch:2003cv}. 
The $b$-quark Yukawa coupling, \hb\,, is related to the 
corresponding running mass at tree level by $\hb=\mb/v_1$.
Once radiative corrections are taken into account, due to the breaking of
SUSY, this relation is modified to
$\mb\equiv\hb v_1
\left(1+\Delta\mb\right)$~\cite{Carena:1999py,Guasch:2003cv}, 
% \begin{equation}
%   \label{eq:deffmb}
%   \mb\equiv\hb v_1 \left(1+\Delta\mb\right)\,,
% \end{equation}
where $v_i$ is the Vacuum Expectation Value (VEV)
 of the Higgs doublet  $H_i$ and
$\Dmb$ is a non-decoupling quantity that encodes the leading higher
order effects. Similarly to the $b$-quark case, 
the relation between $m_\tau$ and
the $\tau$-lepton Yukawa coupling, $h_\tau$, is also modified
by quantum corrections, $\Dmtau$. {We} adopt in our
analysis the effective Lagrangian approach by 
 relating the fermion mass to the Yukawa coupling via a generic
$\Dmf$ ($f=b,\tau$),
\begin{equation}
  \label{eq:deffhb}
  h_f=\frac{m_f(Q)}{v_1} \frac{1}{1+\Dmf}=
      \frac{m_f(Q)}{v \cos\beta}\frac{1}{1+\Dmf} \quad
    (v= (v_1^2+v_2^2)^{1/2},  \quad \tan\beta=\frac{v_1}{v_2})\,,
\end{equation}
in which the resummation of all possible $\tb$
enhanced corrections of the type $(\alpha_{s} \tb)^n$ is
included~\cite{Carena:1999py,Guasch:2003cv}. 
The leading part of the (potentially) non-decoupling
contributions proportional to soft-SUSY-breaking trilinear scalar
couplings ($A_f$) can be absorbed in the definition of
the effective Yukawa coupling at low energies and only subleading
effects survive~\cite{Guasch:2003cv}. Therefore,
the expression~(\ref{eq:deffhb}) contains all (potentially) large leading 
radiative effects. The SUSY-QCD contributions to $\Dmb$ are proportional 
to the Higgsino mass parameter $\Dmb\sim\mu$, while the leading SUSY-EW
contributions behave like $\Dmb\sim \mu A_t$~\cite{Guasch:2001wv}. Thus, 
they
can either enhance or screen each other, depending on the sign of $A_t$. 
It is precisely these effects that will allow us to  
distinguish between different Higgs mechanisms of EWSB.
For example, the analysis of 
these corrections in the ratio of neutral Higgs boson decay rates,
$R'={\brHbb}/{\brHtt}$, revealed large deviations from the SM values 
for several MSSM parameter combinations~\cite{Guasch:2001wv}. 
Extensive theoretical analyses of
one-loop corrections to both neutral and charged Higgs boson decays have
been performed in~\cite{\CGGJS,Guasch:1998jc,\jaume,Guasch:2001wv,\logan}. 
We now explore the one-loop MSSM contributions to the ratio of the 
branching ratios ($BR\,$s) of a charged Higgs boson
$H^{\pm}$ in~(\ref{eq:relation2}), which at leading order (and neglecting
kinematical factors) is given by $R=\htau^2/3 \hb^2$ in the large $\tb$
limit. The SUSY
corrections to the $H^+t\bar{b}$ vertex entering the
decay processes $t\to H^+b$ and $H^+\to
t\bar{b}$ have been analysed in~\cite{\CGGJS,Guasch:1998jc},
where it was shown that they change significantly the
Tevatron limits on $\mHc$~\cite{Guasch:1998jc}. They were further explored
in the production process $pp(p\bar{p})\to H^- t\bar{b}$
at LHC and Tevatron in~\cite{\jaume,Guchait:2001pi}, where they were shown to shift
significantly the prospects for discovery of a charged Higgs boson at both 
colliders. 

\medskip
Here, we have performed a detailed phenomenological analysis for the LHC
of charged Higgs boson signatures, by using the subprocess 
$g\bar{b}\to H^+\bar{t}$. The QCD corrections to this
channel are known to next-to-leading
(NLO)~\cite{Plehn:2002vy}. However, we have normalised our
production cross-section to the LO {result}, for consistency with the
tree-level treatment of the backgrounds\footnote{In all the analysis we
  disregard the subleading QCD and SUSY corrections which affect the
  signal and the background, and will take only into account the leading
  SUSY corrections to the signal cross-section, which are absent in the
  background processes.}. In our simulation,
we have let the 
top quarks decay through the SM-like channel
$t\to W^+b$. In the hadronic decay channel of the charged Higgs boson
($H^+\to t\bar{b}$) we require one of the two $W$'s
emerging from the decay chain $H^+\bar t\to (t\bar{b})\bar t
\to (W^+b)\bar b(\bar bW^-)$ to
decay leptonically, to provide an efficient trigger, while the other $W$
is forced to decay hadronically, since this mode provides
the largest rate and in order 
to avoid excessive missing energy. The $\tau$-lepton in the $H^+\to\tau^+\nu$
decay mode is searched for through hadronic one- and multi-prong channels.
In summary, the
experimental signatures of the two production channels under investigation are
($l=e,\mu$):
\begin{eqnarray}
\label{eq:leptonic} 
  pp(g\bar{b})&\to & H^+ \bar{t} \to (\tau^+\nu) \bar{t} \to \tau^+\nu\,
  (jj \bar{b})\,\,,\\
  \label{eq:hadronic}
  pp(g\bar{b})&\to & H^+ \bar{t} \to (t\bar{b}) \bar{t} \to (jj [l\nu]
  b)\, \bar{b} \, (l\nu [jj]\bar{b})\,\,. 
\end{eqnarray}
(In the numerical analysis we always combine the signals 
in~(\ref{eq:leptonic}) and (\ref{eq:hadronic}) with their 
charged-conjugated modes.)

The Monte Carlo (MC) simulation has been performed using  {\tt PYTHIA}
({v6.217})~\cite{Sjostrand:2000wi} for the signal and most of the background
processes. (We have cross-checked the signal cross-section
with~\cite{Plehn:2002vy}.) We have used {\tt HDECAY}\cite{Djouadi:1998yw} 
for the Higgs boson decay rates. One of the background processes (the 
single-top one: see below) has been 
generated with {\tt TopRex} \cite{Slabospitsky:2002ag} with a custom 
interface to {\tt PYTHIA}. We have used {\tt ATLFAST}~\cite{atlfast} for the 
detector simulation. (Further details of the detector can be found
in~\cite{\ketevi}.)  We have adopted the
CTEQ5L~\cite{Lai:1999wy} parton distribution functions in their default
{\tt PYTHIA} {v6.217} setup and we have used
running quark masses derived from the pole values $m_t^{\rm pole}=175\GeV$ and
$m_b^{\rm pole}=4.62\GeV$. The {\tt TAUOLA}~\cite{\tauola} package was 
interfaced 
to the {\tt PYTHIA} event generator for treatment of the  
$\tau$-lepton polarisation. 

\medskip
The leptonic decay channel of the charged Higgs boson provides the
best probe for the detection of such a state at the
LHC. In fact, it turns out that despite the small branching
ratio $BR(H^+\to\tau^+\nu)$, the $\tau$-lepton affords an efficient
trigger to observe this channel. The production rates $\sigma\times
BR(H^+\to \tau^+\nu)\times BR(W\to jj)$ are shown in
Tab.~\ref{tabletau}. The main background processes in this channel
are: top-pair
production with one of the $W$'s decaying into $\tau\nu$ ($gg\to
t\bar{t}\to jj\,b\, \tau\nu\bar{b}$) and $W^\pm t$ associated
production ($g\bar{b}\to W^+\bar{t}\to \tau^+\nu \bar{t}$). 

We have used 
the following trigger conditions: hadronic $\tau$-jet ($p_T^{\tau} >
30\GeV$); a $b$-tagged jet ($p_T^{b} > 30 \GeV$) and at least two
light jets  ($p_T^{j} > 30 \GeV$). We apply afterwards a $b$-jet veto
to reject the $t\bar{t}$ QCD background. 
As there is no isolated lepton (electron or muon) in the final state, 
the observation of this channel 
requires a multi-jet trigger with a $\tau$-trigger.
After reconstructing the jet-jet invariant mass $m_{jj}$ and 
retaining the candidates consistent with the $W$-boson mass,
 $|m_W-m_{jj}| < 25\GeV$, the jet four-momenta are rescaled and 
the associated top quark is reconstructed by minimising 
$\chi^2 \equiv (m_{jjb}-m_t)^2$. 
Subsequently, a sufficiently high threshold on the $p_T$ of the
$\tau$-jet is required, $p_T^{\tau} > 100 \GeV$. 
The background events satisfying this cut need 
a large boost from the $W$-boson. This results in a 
small azimuthal opening angle $\Delta\phi$ between the $\tau$-jet 
and the missing transverse momentum, ${p\!\!\!/}_T$. 
For background suppression we then have the cut 
$\Delta\phi({p\!\!\!/}_T,p_T^{\tau}) > 1$. Besides, the missing transverse
momentum is harder for the signal than for the background
while the differences between their
 distributions in azimuthal angle and missing transverse
momentum increase with increasing $m_{H^\pm}$. These 
effects are well cumulated in the transverse mass, 
$m_T = \sqrt{2p_T^\tau{p\!\!\!/}_T\left[1-\cos(\Delta\phi)\right]}$,
which provides good discrimination between the signal and the 
backgrounds, as shown in Fig.~\ref{fig:transversemass}. 
(Further details of this kind of studies are
available in~\cite{\ketevi}.) The discussed set of cuts reduces the
total
 background by six orders of magnitude while the signal is only suppressed
by  two orders.
The production rates and total detection efficiency (including detector
acceptance, $b$- and $\tau$-identification, {pileup} and 
the effect of cuts) are also shown in Tab.~\ref{tabletau} for an integrated
luminosity of $300 \fb^{-1}$. We can see that the signal rates are large
enough to indeed consider  
$H^{+} \rightarrow \tau \nu$ a {\it golden channel} for the $H^{+}$
discovery at large $\tan \beta$.
\begin{figure}[t]
\begin{center}\vspace*{-0.8cm}
\begin{tabular}{cc}
\resizebox{10cm}{!}{\includegraphics{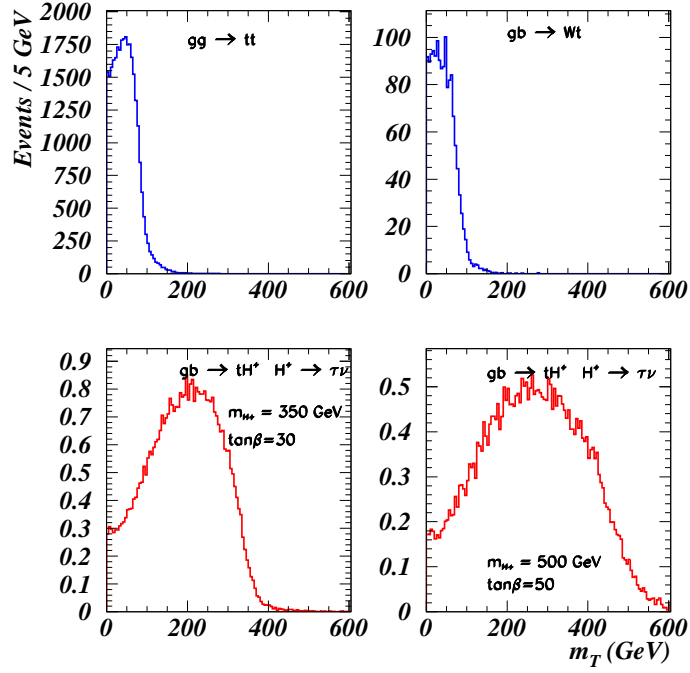}}
\end{tabular}
\end{center}\vspace*{-0.5cm}
\caption{Transverse mass $m_T$ distribution for signal and
  total background taking into account the polarisation of
  the $\tau$-lepton, for an integrated luminosity of $30 \fb^{-1}$. A final 
cut $m_T > 200$~GeV was used for the calculations of the signal-to-background
ratios and for the signal significances.}
\label{fig:transversemass}
\end{figure}
\begin{table}
\begin{center}
\begin{tabular}{||l|c|c||c|c||}
\hline
& $\mHc=350$ & $\mHc=500$ & $t\bar{t}$ & $W^\pm t$
\\\hline
$\sigma\times BR$ & $99.9\fb$ & $30.7\fb$ & $79.1\pb$ & $16.3\pb$
\\\hline
Events  & 29958 & 9219  & $2.3 \times 10^{6}$ & $4.89 \times
10^{5}$ \\\hline
Events after cuts & 174 & 96 & 17 & 3 \\\hline
Efficiency & 0.6\%& 1\% & $8\times 10^{-6}$ & $6\times10^{-6}$ \\ \hline
$S/B$ & 7.9 & 4.4  \\\cline{1-3}
$S/\sqrt{B}$ & 37.1 & 20.5  \\ \cline{1-3}
Poisson & 23.1 & 14.6 \\\cline{1-3}
% Error & 8 \% & 11 \% \\\cline{1-3}
\end{tabular}
\end{center}
\caption{The signal and background cross-sections, the number of events before cuts,
  the number of events after all cuts, the total efficiency, 
  the signal-to-background ratios ($S/B$), and the signal significances (Gaussian
  and Poisson) for the detection of the charged Higgs in the $\tau\nu$ channel at
  the LHC, for $300\fb^{-1}$ integrated luminosity and $\tb=50$.}
\label{tabletau}
\end{table}

\medskip
The production rates $\sigma\times
BR(H^+\to t \bar b)\times BR(W^+W^-\to jj l\nu)$ are shown in
Tab.~\ref{tabletb}. The decay mode
$H^\pm \to t b$ has large QCD backgrounds at hadron colliders that come
from $t\bar{t}q$ production with 
$t\bar{t}\rightarrow  Wb\,Wb\rightarrow l\nu b \,jj b$. 
However, the possibility of 
efficient $b$-tagging has considerably improved the situation
\cite{\Roy}. 
We search for an isolated lepton ($p_T^e > 20 \GeV$,  $p_T^{\mu} > 8 \GeV$),
three $b$-tagged jets ($p_T^b > 30 \GeV$) and 
at least two non $b$-jets ($p_T^j > 30 \GeV$). 
We retain the jet-jet combinations whose 
invariant masses are consistent with the $W$-boson mass, 
$|m_W-m_{jj}| < 25\GeV$, then 
we use the $W$-boson mass constraint to find the 
longitudinal component of the neutrino momentum in $W^\pm\to l\nu$, 
by assuming that the missing transverse 
momentum belongs only to the neutrino. Subsequently, 
the two top quarks entering the  $H^+\bar t\to (t\bar{b})\bar t\to (jj [l\nu]
  b)\, \bar{b} \, (l\nu [jj]\bar{b})$ 
decay chain are reconstructed, retaining the 
pairing whose invariant masses $m_{l\nu b}$ and 
$m_{jjb}$ minimise 
$\chi^2 \equiv (m_t-m_{l\nu b})^2 + (m_t - m_{jjb})^2$. 
The remaining $b$-jet can be paired with either top quark to give 
two charged Higgs candidates, one of which leads to a combinatorial 
background. The expected rates for
signal and
background (after the mentioned decays) are shown in Tab.~\ref{tabletb}. 
 (This analysis is presented extensively in~\cite{\ketevi}.)

\begin{figure}[t]
\begin{center}\vspace*{-0.8cm}
\begin{tabular}{cc}
\resizebox{10cm}{!}{\includegraphics{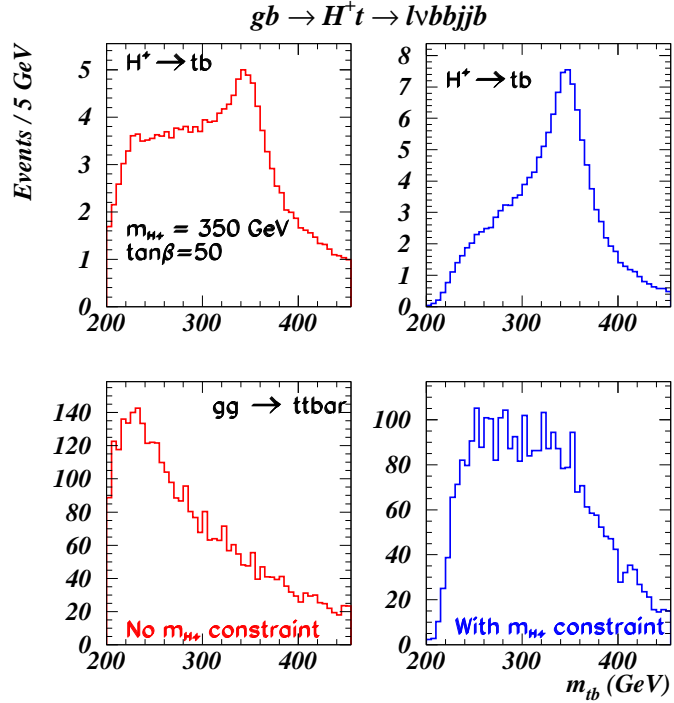}}\\
\end{tabular}
\end{center}\vspace*{-0.6cm}
\caption{The signal and the background distributions for the
  reconstructed invariant mass $m_{tb}$ of $\mHc=350 \GeV$, 
$\tan \beta =50$ and an integrated luminosity of $30 \fb ^{-1}$. Assuming that
the charged Higgs is discovered in the $H^\pm \to \tau\nu$ channel, one can use
$\mHc$ as a constraint to reduce the combinatorial background: this is 
shown on the right plots.}
\label{fig:Hmassrecons}
\end{figure}

At this point, we have two charged Higgs candidates: $t_1 b_3$ or $t_2 b_3$. 
Assuming that the charged Higgs is discovered through the $H^\pm \to \tau\nu$
channel and its mass determined from the $\tau\nu$ transverse mass 
distribution~\cite{\ketevi}, the correct charged Higgs candidate in the $tb$ 
channel can be selected by using the measured $\mHc$ as a constraint. This is 
done by selecting the candidate whose invariant mass is closest to the 
measured charged Higgs mass: $\chi^2 =(m_{tb}-\mHc)^2$.
The signal distribution 
for the reconstructed invariant mass $m_{tb}$ for a charged Higgs boson
weighing $350 \GeV$, with $\tan \beta=50$ and after integrated luminosity 
of $30 \fb ^{-1}$, is shown in 
Fig.~\ref{fig:Hmassrecons}. We can see that for the  
$H^{+}\rightarrow t\bar{b}$ decay some irreducible combinatorial 
noise still appears even 
when the $\mHc$ constraint is included. In addition,
for the background, we have found
that the $\mHc$ constraint reshapes the distributions in
$g g \rightarrow t \bar t X$ in such a way that no improvement in the 
signal-to-background ratio and signal significance is 
further observed. Finally, recall that
the knowledge of the shape and the normalisation of the reshaped background 
would be necessary for the signal extraction. For these reasons, we did not use 
the $\mHc$ constraint for the results shown in this {work}. 
The subtraction of the 
background can then
be done by fitting the side bands and extrapolating in the 
signal region which will be known from the $\mHc$ determination in the 
$H^\pm\to\tau\nu$ channel: however, 
this would be possible only for Higgs masses above 
$300\GeV$ -- see 
Fig.~\ref{fig:Hmassrecons}.  The signal and background results are 
summarised in Tab.~\ref{tabletb} at an
integrated luminosity of $300 \fb ^{-1}$ for different values of
$\mHc$ and $\tan \beta=50$. It is shown that it is
difficult to observe $H^{\pm}$ signals in this channel above $\sim 400 \GeV$, 
even 
with the $\mHc$ constraint. For masses above $\mHc\sim 400\GeV$ the signal 
significance
can be enhanced by using the kinematics of the three-body production 
  process $gg\to H^+\bar{t}b$ \cite{\jaume,\Roy}. 

We assume a theoretical uncertainty of $5\%$ on the branching ratios, $BR\,$s. Previous 
ATLAS studies have shown the residual $gg \to t\bar{t}$ shape and normalisation can be 
determined to $5\%$~\cite{\ketevi}. The scale uncertainties on jet and lepton energies are 
expected to be of the order $1 \%$ and $0.1 \%$ respectively~\cite{\ketevi}. As explained 
above, for $m_{H^\pm} > 300$~GeV, the side band procedure can be used the subtract the 
residual background under the $H^{+}\rightarrow t b$ signal: we
assume also a $5\%$ uncertainty in the background subtraction method. Thus, the
statistical uncertainties can be estimated as $\sqrt{1/S}$. 
The uncertainty in the ratio $R$ are dominated by the
reduced 
knowledge of the background shape and rate in the 
$H^{+} \rightarrow t b$ channel.  The cumulative
results for the two channels
are summarised in Tab.~\ref{tablefinal} at an integrated
luminosity of $300 \fb^{-1}$.  Here,
the final result for the ratio $R$ is
  obtained by correcting the visible production rates after cuts for the
  total detection efficiency in Tabs.~\ref{tabletau} and~\ref{tabletb}
  and by the decay $BR\,$s of the $W$-bosons.
The simulation shows that the above mentioned 
ratio can be measured with
an accuracy of $\sim 12-14 \%$ for $\tan \beta=50$, for
$\mHc=300-500\GeV$ and at an integrated luminosity of $300 \fb^{-1}$. 

\medskip
We turn now to the impact of the SUSY radiative corrections. Their role
is twofold. Firstly, 
by changing the
value of the Yukawa coupling they change the value of the observable
$R$. Secondly, they can change
the value of the production cross-section $\sigma(pp\to H^+\bar{t}+X)$,
hence shifting significantly (by as much as $100 \GeV$) the range of charged
Higgs masses accessible at the LHC~\cite{\jaume}. 

\begin{table}\begin{center}\vspace*{-0.5cm}
\begin{tabular}{||l|c|c||c||}
\hline
& $\mHc=350$ & $\mHc=500$ & $t\bar{t}q$ \\\hline
$\sigma\times BR$ & $248.4\fb$ &  $88\fb$ & $85\pb$
\\\hline
Events  & 74510 &  26389 & $2.55\times 10^{7}$ \\\hline
Events after cuts & 2100 & 784 &  59688
 \\\hline
Efficiency & 2.8\% & 3\% & 0.2\%
 \\ \hline
$S/B$ & 0.035 & 0.013   \\\cline{1-3}
 $S/\sqrt{B}$ & 8.6 & 3.2  \\ \cline{1-3}
% Error & 12\% &  31\% \\\cline{1-3}
\end{tabular}
\end{center}\vspace*{-0.3cm}
\caption{The signal and background cross-sections, the number of events before cuts,
  the number of events after all cuts, total efficiency, $S/B$, and the signal 
  significances for the detection of the charged Higgs in the $tb$ channel at
  the LHC, for $300\fb^{-1}$ integrated luminosity and $\tb=50$.
\label{tabletb}}
\end{table}

\begin{table}\begin{center}
\begin{tabular}{|l|c|c|}
\hline
& $\mHc=350$ & $\mHc=500$ \\\hline
Signals $\tau\nu/tb$ & 174 / 2100 = 0.08 &
96 / 784 = 0.12 \\ \hline
Signals (corrected) $\tau\nu/tb$ &  0.18 & 0.16 \\\hline
Systematics unc. & $\sim 9\%$ & $\sim 9\%$ \\ \hline
Total unc. & 12\% & 14\% \\\hline
Theory & 0.18 & 0.16 \\\hline
\end{tabular}
\end{center}\vspace*{-0.3cm}
\caption{Experimental determination of the ratio~(\ref{eq:relation})
   for $300\fb^{-1}$ and $\tb=50$. Shown are: the signal after cuts, the
   signal after correcting for efficiencies and branching ratios, the
   systematic uncertainty, the total combined uncertainty, and the
   theoretical prediction (without SUSY corrections).\label{tablefinal}}
\end{table}

To explore the second consequence, we rely on the fact that the
bulk of the SUSY corrections to the production cross-section is given by
the Yukawa coupling
redefinition in 
(\ref{eq:deffhb})~\cite{\jaume,Plehn:2002vy}. By neglecting the
kinematic effects, taking the large $\tb$ limit and assuming that the
dominant decay channel of the charged Higgs boson is $H^+\to t\bar{b}$
(large mass limit), we can estimate the corrected production rates. For
simplicity, we show only the contributions to $\Dmb$, since they are the
dominant ones:
\begin{eqnarray}
\sigma^{\rm{corr}}(g\bar{b}\to H^+\bar{t}\to t\bar{b}\bar{t}) &=&
\sigma^{\rm{corr}}(g\bar{b}\to H^+\bar{t}) \times BR^{\rm{corr}}(H^+\to
t\bar{b})\simeq
\frac{\sigma^{0}(g\bar{b}\to H^+\bar{t})}{(1+\Dmb)^2}\,\,,\nonumber\\
\sigma^{\rm{corr}}(g\bar{b}\to H^+\bar{t}\to \tau^+\nu \bar{t}) &=&
\sigma^{\rm{corr}}(g\bar{b}\to H^+\bar{t}) \times
\frac{\Gamma(H^+\to\tau^+\nu)}{\Gamma^{\rm{corr}}(H^+\to t\bar{b})}\nonumber\\
&\simeq&
\frac{\sigma^{0}(g\bar{b}\to H^+\bar{t})}{(1+\Dmb)^2}\times\frac{\Gamma(H^+\to\tau^+\nu)}{\Gamma^{0}(H^+\to t\bar{b})\times\frac{1}{(1+\Dmb)^2}} \,\,.
  \label{eq:correctedrates}
\end{eqnarray}
This very simple exercise shows that the production rate in the $\tau$-channel
 is fairly independent of the SUSY radiative corrections and
therefore the tree-level analysis performed above can (to a very good
approximation) be used for our original purposes. Actually, once we take 
into account kinematical
effects, the $\tau$-channel will receive small (negative)
corrections in the low
charged Higgs mass range. However, in this range, $BR(H^+\to \tau^+ \nu)$
is quite large and one should not fear to loose the signal. Quite the
opposite, the hadronic $t\bar{b}$ production channel receives large radiative
corrections. These corrections can be either positive (enhancing the
signal, and therefore the significance in Tab.~\ref{tabletb}) or negative
(reducing it, possibly below observable
levels)\footnote{Alternative analyses may permit the signal to
  be seen even in this unfavourable case \cite{\jaume}.}. In
Fig.~\ref{fig:theory}a, we show the discussed 
enhancement/suppression factors as a
function of $\tb$ for $\mHc=350\GeV$ and a SUSY mass spectrum defined as
SPS4 of the  {\sl Snowmass Points and Slopes}
in \cite{Allanach:2002nj}, but choosing different scenarios for the
sign of $\mu$ and $A_t$\footnote{The SPS4 spectrum is affected by
moderate SUSY radiative corrections.}.
(It is worth noting that the production rate for the
$tb$-channel can be enhanced by a factor larger than 3 in some SUSY
scenarios, which would enhance significantly the corresponding 
signal thus overcoming the low signal-to-background ratio 
of this channel.)

\begin{figure}[t]
\begin{tabular}{cc}
\resizebox{7cm}{!}{\includegraphics{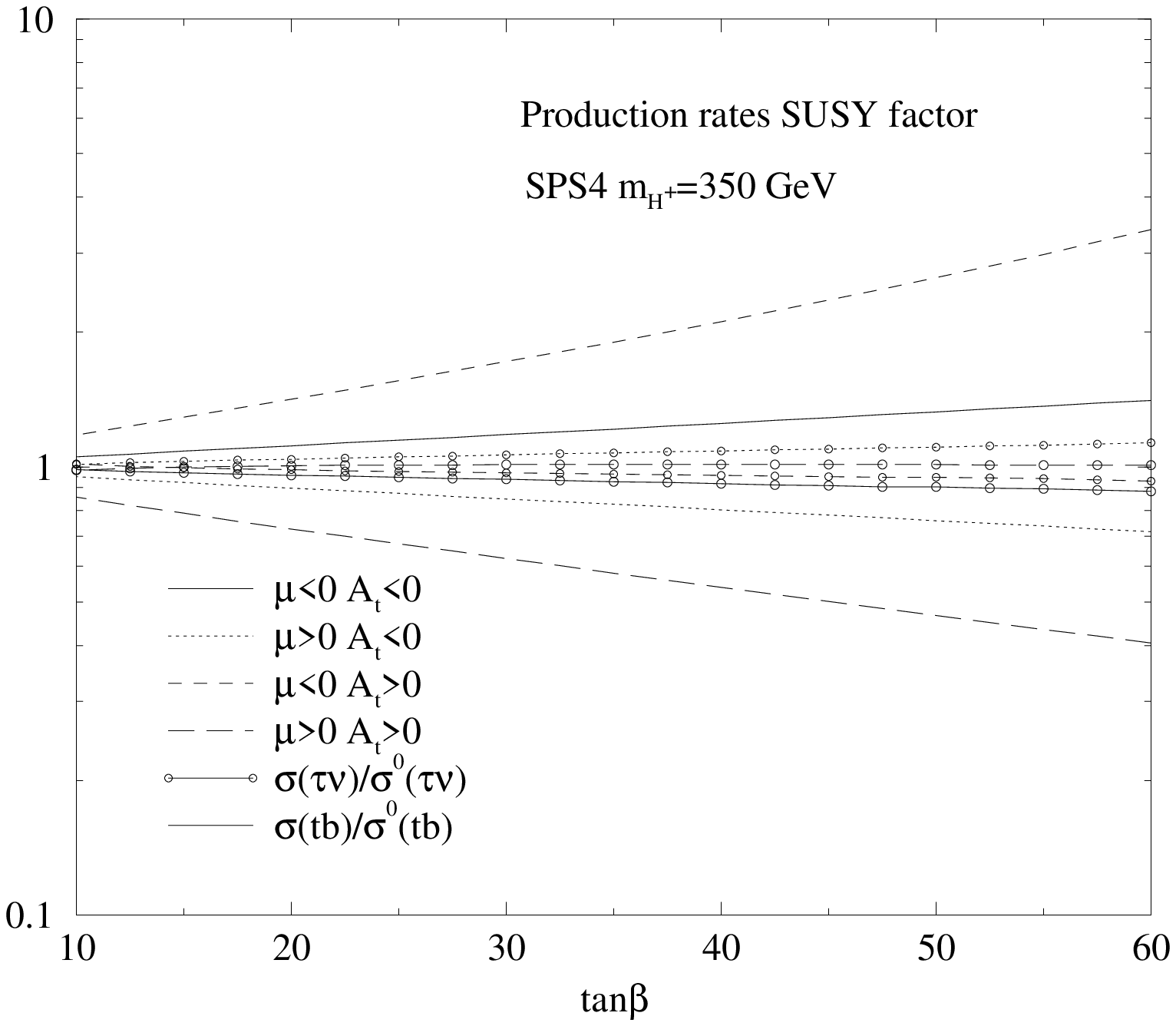}} &
\resizebox{7cm}{!}{\includegraphics{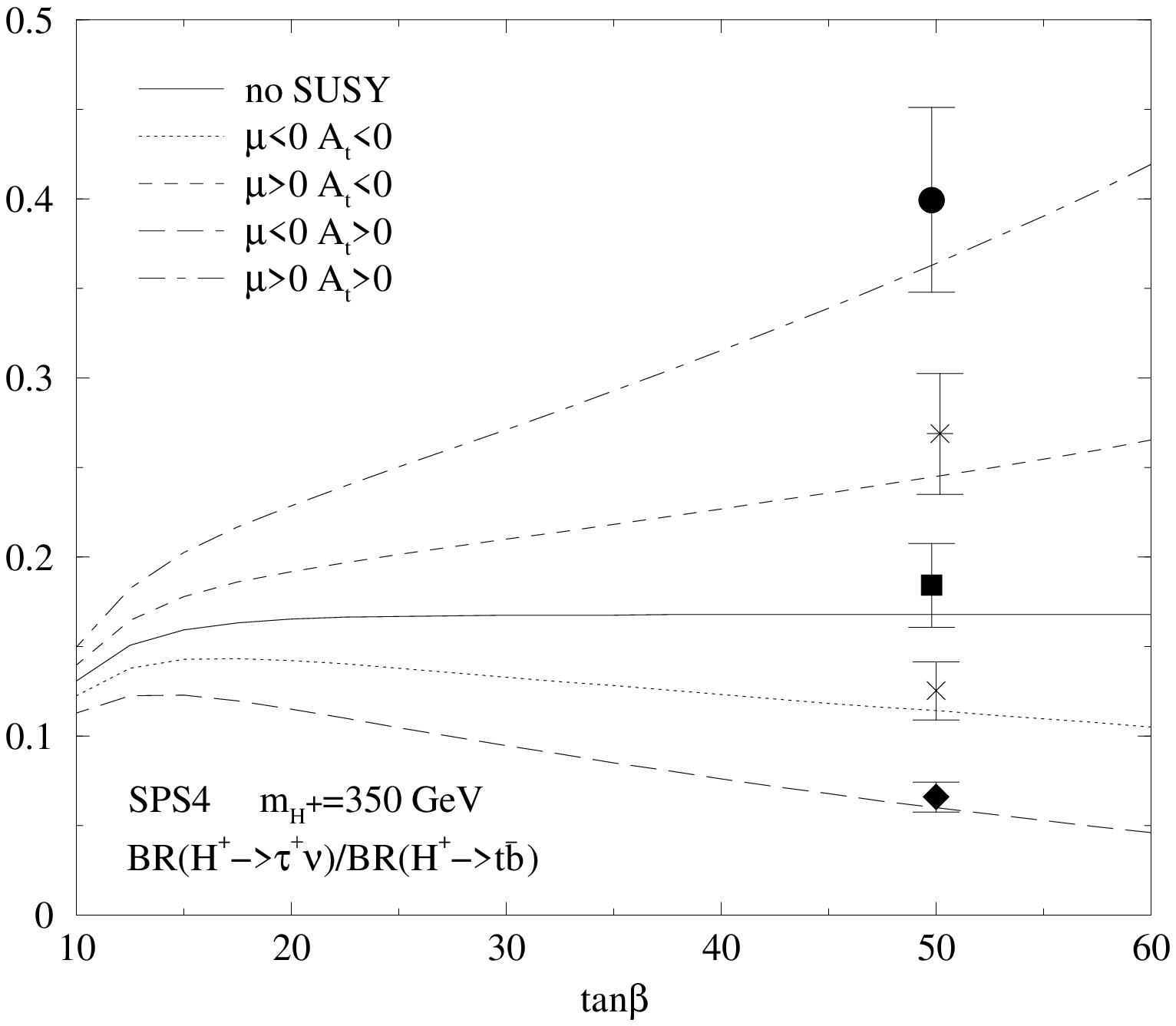}}\\
(a)&(b) 
\end{tabular}
\caption{a) Production rates enhancement/suppression factors
  for the $\tau$ and the $tb$ channels; b) the SUSY correction
  to the rate~(\ref{eq:relation2}). Plots as  functions of $\tb$ for
  $\mHc=350\GeV$ and a SUSY spectrum as in SPS4, but for different
  choices of the signs of $\mu$ and $A_t$. Shown is also the
  experimental determination for each scenario.\label{fig:theory}}
\end{figure}

We now turn our attention to the observable under analysis.
Fig.~\ref{fig:theory}b shows the prediction for the ratio
$R$ as a function of $\tb$ for the
SPS4 scenario with a charged
Higgs mass of $\mHc=350\GeV$. The value of $R$ only depends  on 
$\mHc$ through kinematical factors and the dependence is
weak for $\mHc\gsim 300\GeV$. In this figure, we also show the experimental
determination carried out as before and repeated for each SUSY setup.
From Fig.~\ref{fig:theory}b it is clear that radiative SUSY
effects are visible at the LHC at a large significance. 
In particular, {the $\mu<0$ scenarios can easily be discriminated, while the
  $\mu>0$ ones will be more difficult to establish, due to the lower
  signal rate of the hadronic channel.}
This feature then also allows for a measurement of the sign of the $\mu$
parameter. In contrast, since radiative corrections are independent of the
overall SUSY scale, the observable $R$ cannot provide us with an estimation
of the typical mass of SUSY particles. Nonetheless, the information obtained 
in other production channels (e.g., neutral Higgs bosons or SUSY particles 
direct production) can be used to perform precision tests of the MSSM.

\medskip
To summarise, we have used the observable $R\equiv
\frac{\sigma(pp\to H^+\bar{t} + X \to \tau^+ \nu t + X)}
     {\sigma(pp\to H^+\bar{t} + X \to t\bar{b} t +
    X)}=\frac{\brHtaunu}{\brHtb}\,$ 
 to discriminate between SUSY and non-SUSY Higgs models. This quantity
is a theoretically \textit{clean} observable. The experimental
uncertainties that appear in this ratio have been analysed in details
through detailed phenomenological simulations. In the MSSM, $R$ is affected 
by quantum contributions that do not decouple 
even in the heavy SUSY mass limit. We have quantitatively
shown that an LHC measurement of $R$ can give 
clear evidence for or against the SUSY nature of charged Higgs bosons. 

}

%% file: guchait.tex
{
\def\s#1{{\small#1}}
\def\lsim{\:\raisebox{-0.5ex}{$\stackrel{\textstyle<}{\sim}$}\:}
\def\gsim{\:\raisebox{-0.5ex}{$\stackrel{\textstyle>}{\sim}$}\:}
\def\PD{\s{PDG}}
\def\TA{{\small TAUOLA}}
\def\HW{\s{HERWIG}}
\def\JS{\s{JETSET}}
\def\PY{\s{PYTHIA}}
\def\IS{\s{ISAJET}}
\def\IW{\s{ISAWIG}}
\def\SM{\s{SM}}
\def\MSSM{{MSSM}}
\def\SY{\s{SUSY}}
\def\QCD{\s{QCD}}
\def\QED{\s{QED}}
\def\DIS{\s{DIS}}
\def\LEP{\s{LEP}}
\def\LHC{\s{LHC}}
\def\OPAL{\s{OPAL}}
\def\PDF{\s{PDFLIB}}
\def\CERN{\s{CERN}}
\def\RPV{\rlap{/}{R}$_{\mbox{\scriptsize p}}$}
\def\BNV{\rlap{/}{B}}
\def\Ord{\buildrel{\scriptscriptstyle <}\over{\scriptscriptstyle\sim}}
\def\OOrd{\buildrel{\scriptscriptstyle >}\over{\scriptscriptstyle\sim}}
\def\gh{\Gamma_{\scriptscriptstyle \rm H}}
\def\gtap{\raisebox{-.4ex}{\rlap{$\sim$}} \raisebox{.4ex}{$>$}}
\def\ltap{\raisebox{-.4ex}{\rlap{$\sim$}} \raisebox{.4ex}{$<$}}
\def\ycut{$y_{\mbox{\tiny cut}}$}
\def\mw{m_{\scriptscriptstyle \rm W}}
\def\mh{m_{\scriptscriptstyle \rm H}}
%                  Define Lambda MS bar
\def\lms{\Lambda_{\overline{\rm MS}}}
\def\half{\mbox{\small $\frac{1}{2}$}}
\def\thlf{\mbox{\small $\frac{3}{2}$}}
\def\as{\alpha_{\mbox{\tiny S}}}
\def\ee{e^+e^-}
\def\MC{Monte Carlo}
\def\VEV#1{\langle{#1}\rangle}
%                  Define antiparticles
\def\qbar{\bar{q}}
\def\Qbar{\bar{Q}}
\def\dbar{\bar{d}}
\def\ubar{\bar{u}}
\def\sbar{\bar{s}}
\def\cbar{\bar{c}}
\def\bbar{\bar{b}}
\def\tbar{\bar{t}}
\def\pbar{\bar{p}}
\def\B0bar{\overline{B^0}}
\def\lbar{\bar{\l}}
\def\l{\ell}

\section[ ]{Charged Higgs Bosons in the Transition Region 
\boldmath${M_{H^\pm} \sim m_t}$
 at the LHC\footnote{K.A.\,Assamagan, M.\,Guchait and S.\,Moretti}}

\subsection{The Threshold Region}

The detection of charged Higgs bosons ($H^\pm$) would
unequivocally imply the existence of physics beyond the Standard Model (SM),
since spin-less charged scalar states do not belong to its particle spectrum.
Singly charged Higgs bosons appear in any Two-Higgs Doublet Model (2HDM),
including a Type-II  in presence of minimal Supersymmetry (SUSY), namely, 
the Minimal Supersymmetric Standard Model (MSSM). Depending on its mass,
the machines that
are likely to first discover such a state are Tevatron 
$p\bar p$ ($\sqrt s=2$ TeV) and the LHC ($\sqrt s=14$ TeV). Current limits 
on the charged Higgs boson mass are set by LEP at about 
80~GeV. At the Tevatron a charged Higgs boson could be 
discovered for masses up to $m_t-m_b$, whereas the LHC
has a reach up to the TeV scale, if $\tan\beta$ is favourable
(i.e., either large or small).

For the LHC, 
the ATLAS discovery potential of $H^\pm$ bosons in a general Type-II 2HDM  
or MSSM 
(prior to the results of this study) is visualised in the left-hand side
of Fig.~\ref{fig:LHC}. (A similar CMS plot, also including neutral Higgs states,
is given for comparison.)
The existence of a gap in coverage for $M_{H^\pm}\approx m_t$ was already
denounced in Refs.~\cite{Cavalli:2002vs,Moretti:2002ht} as being due to
the fact that Monte Carlo (MC) simulations of $H^\pm$ production for
$M_{H^\pm}\sim m_t$ were flawed by a wrong choice of the hard scattering
process. In fact, for $M_{H^\pm}< m_t$, the  estimates in both plots in
Fig.~\ref{fig:LHC}
were made by assuming 
as  main production mode of $H^\pm$ scalars the decay 
of top
(anti)quarks produced via QCD in the annihilation
of gluon-gluon and quark-antiquark pairs (hence -- by definition -- 
the attainable Higgs mass is strictly confined to 
the region $M_{H^\pm}\le m_t-m_b$).
This should not be surprising (the problem was also
encountered by CMS, see right-hand side of
Fig.~\ref{fig:LHC}), since standard MC programs, such as
{\small PYTHIA} and \HW\ \cite{Sjostrand:2003wg,Corcella:2000bw}, have 
historically accounted for this process through the usual procedure of 
factorising
the production mode, $gg,q\bar q\to t\bar t$, times the
decay one, $\bar t\to \bar b H^-$, in the so-called Narrow Width
Approximation (NWA) \cite{Guchait:2001pi}. This description
fails to correctly account for the production 
phenomenology of charged Higgs bosons when their mass approaches or indeed
exceeds that of the top-quark (i.e., falls in the so called `threshold 
region'). This is evident from the left plot in 
Fig.~\ref{fig:threshold}. (The problem also occurs at
Tevatron, see right plot therein
and Refs.~\cite{Guchait:2001pi,Alwall:2003tc}.) As remarked in
Ref.~\cite{Guchait:2001pi}, the use of the $2\to 3$ hard scattering
process $gg,q\bar q\to t\bar b H^-$  
\cite{Gunion:1994sv}--\cite{Belyaev:2002eq}, in place of the
`factorisation' procedure in NWA, is mandatory in the threshold
region, as the former correctly keeps into account 
both effects of the finite width of the top quark
and the presence of other $H^\pm$ production mechanisms,
such as Higgs-strahlung and $b\bar t \to H^-$ fusion (and relative 
interferences). 
The differences seen between the two descriptions in 
Fig.~\ref{fig:threshold} are independent of $\tan\beta$ and 
also survive in, e.g., $p_T$ and $\eta$ spectra \cite{Guchait:2001pi}.

One more 
remark is in order, concerning the LHC plot in Fig.~\ref{fig:threshold}.
In fact, at the CERN hadron collider, the above $2\to3$ reaction is
dominated by the $gg$-initiated subprocesses, rather than by 
$q\bar q$-annihilation, 
as is the case at the Tevatron. This means that a potential 
problem of double counting arises in the simulation of $t H^- X$ + c.c. 
events at the LHC, if one considers that Higgs-strahlung can also be 
emulated through the $2\to2$ process $bg\to t H^-$ + c.c., 
as was done in assessing the ATLAS (and CMS) discovery reaches 
in the $H^+\to t\bar b$ and $H^+\to \tau^+\nu_\tau$ 
channels for $M_{H^\pm}> m_t$ (see 
Refs.~\cite{Assamagan:2002ne,Denegri:2001pn}
for reviews). The difference between the
two approaches is well understood, and prescriptions exist for
combining the two, either through the subtraction of a common logarithmic
term \cite{Borzumati:1999th,Moretti:1999bw}
or by means of a cut in phase space \cite{Belyaev:2002eq}. 

\begin{figure}[!h]
\vspace{-0.25cm}
\begin{center}
\epsfig{file=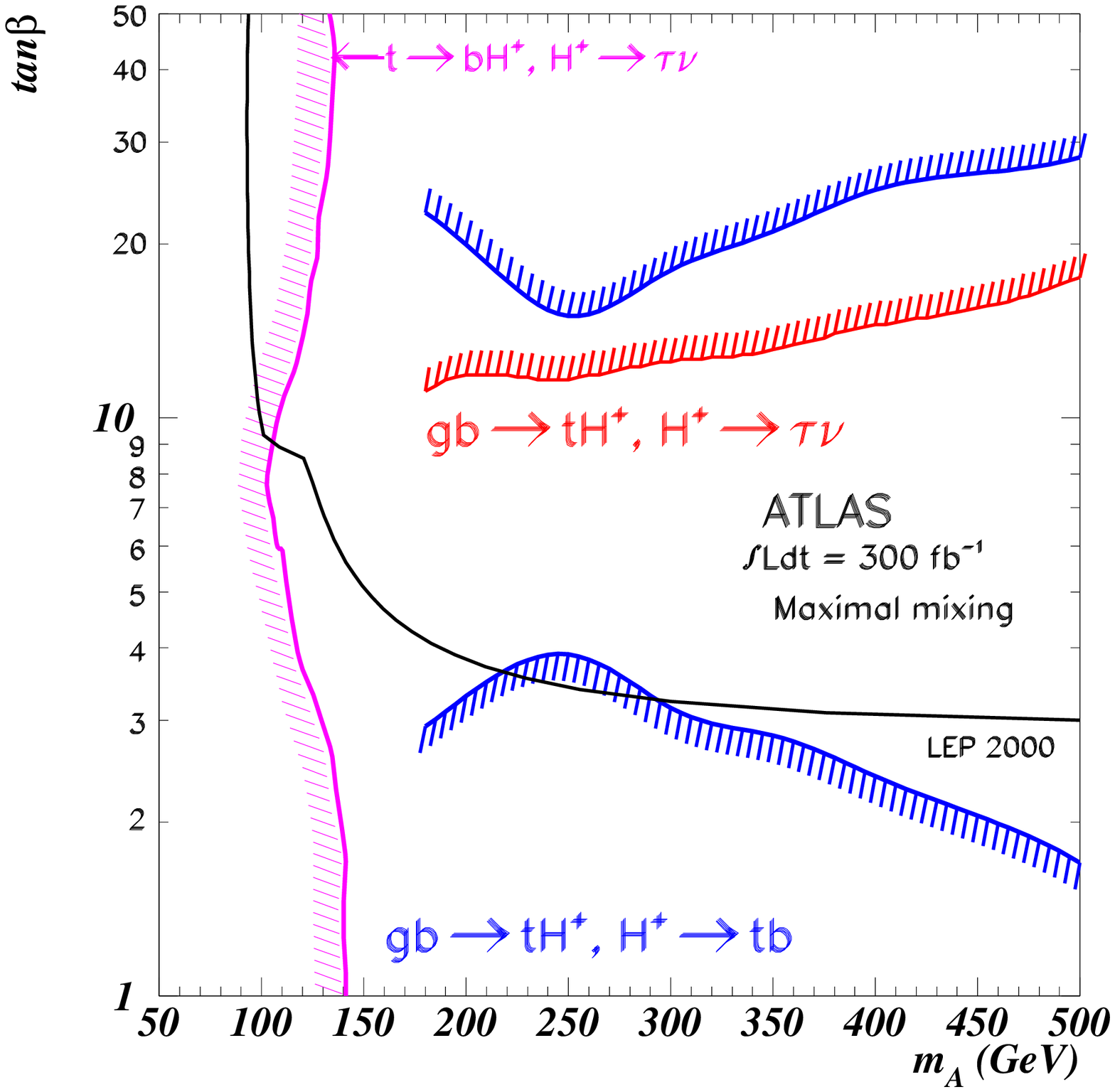,width=6.5cm,height=5.0cm}
\epsfig{file=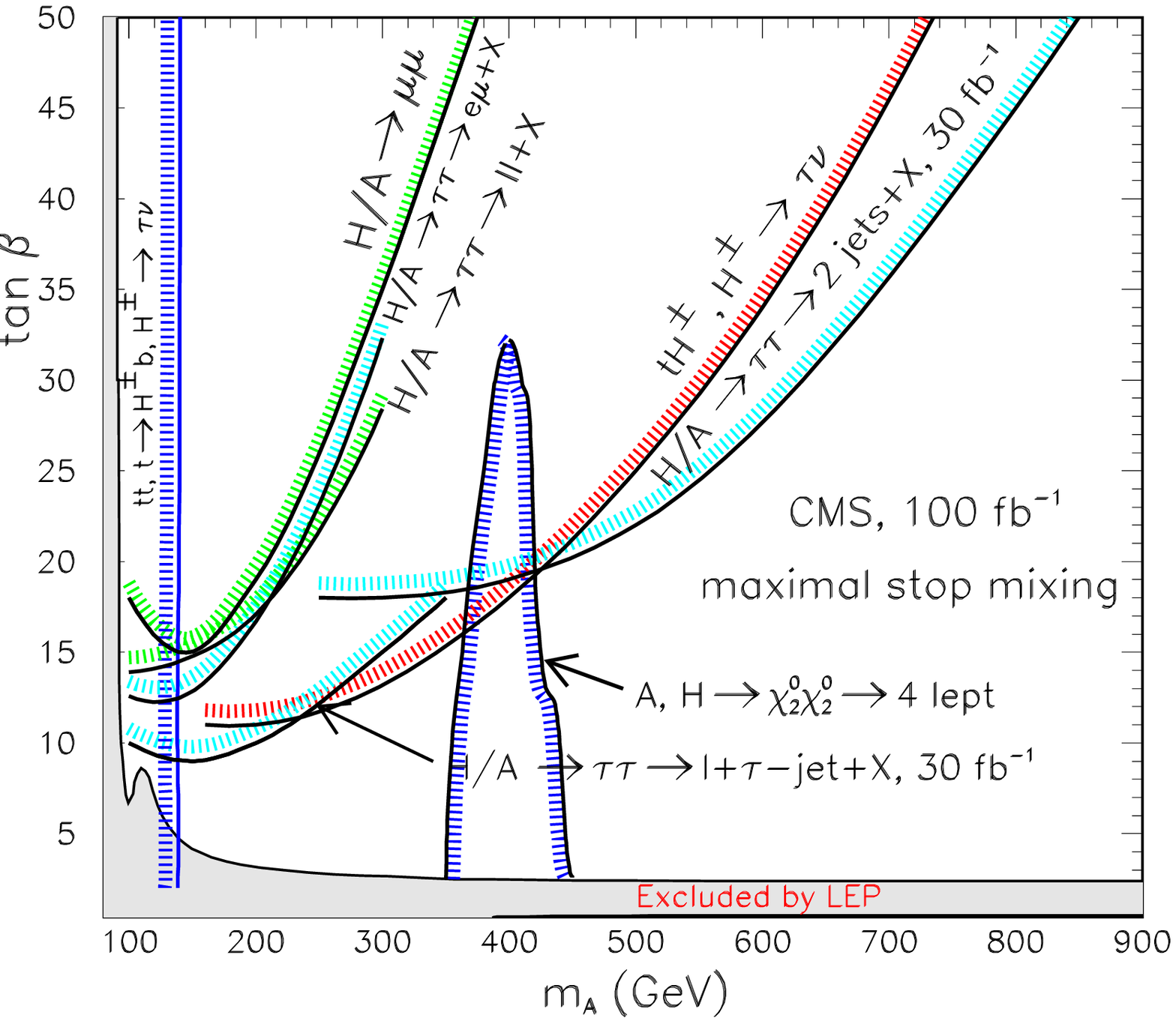,width=6.0cm,height=5.0cm}
\caption{\small The ATLAS 5-$\sigma$ discovery contours of 
2HDM charged Higgs 
bosons for 
300~fb$^{-1}$ of luminosity, only including the reach of SM decay modes  (left plot).
The CMS 5-$\sigma$ discovery contours of MSSM Higgs bosons for 100~fb$^{-1}$ of 
luminosity, also including the reach of $H,A \to \chi^0_2 \chi^0_2 \to 4l^{\pm}$ decays, 
assuming $M_1$ = 90~GeV, $M_2$ = 180~GeV, $\mu$ = 500~GeV, $M_{\tilde{\ell}}$ = 250~GeV, 
$M_{\tilde{q},\tilde{g}}$ = 1000~GeV (right plot). }
\label{fig:LHC}
%{\vskip-6.6cm\hskip-6.0cm{\tiny{\Blue{$gb\to tH^+$, $H^+\to t\bar b$}}}}
%\vskip+5.6cm
\end{center}
\end{figure}

If one then
looks at the most promising (and cleanest) charged Higgs boson decay 
channel, i.e., $H^\pm\to\tau^\pm\nu_\tau$~\cite{Moretti:1995ds}, while 
using the $gg,q\bar q\to t\bar bH^-$ + c.c. 
description and reconstructing the 
accompanying top quark hadronically, the prospects of $H^\pm$ detection 
should improve significantly for $M_{H^\pm}$ values close to $m_t$, eventually
leading to the closure of the mentioned gap. The $2\to3$ 
description of the $H^\pm$ 
production dynamics (as well as the spin correlations in $\tau$-decays
usually exploited in the ATLAS  $H^\pm\to\tau^\pm\nu_\tau$ analysis) have
been made available in version 6.4 \cite{Corcella:2001wc} of 
the \HW\  event generator (the latter also 
through an interface to {\small TAUOLA} \cite{Jadach:1990mz}), 
so that detailed
simulations of $H^\pm$ signatures
at both the Tevatron and the LHC are now possible
for the threshold region, including fragmentation/hadronisation and detector
effects. In the next section we will discuss
the details of an ATLAS analysis based on such tools that has lead 
to the closure of the mentioned gap through the discussed charged Higgs
decay channel. This analysis was initiated 
in the context of the 2003 Les Houches workshop.

\begin{figure}[!h]
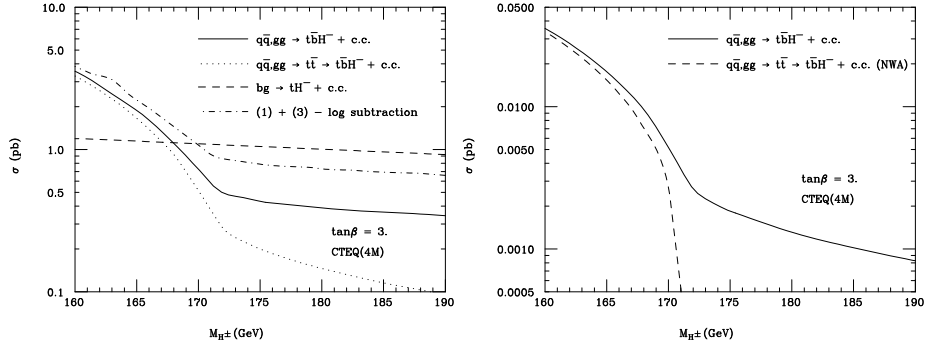

%\vspace{-0.25cm}
\begin{center}
\epsfig{file=LHCthreshold3_LH.eps,angle=90,height=4.5cm}
\epsfig{file=TEVthreshold3_LH.ps,angle=90,height=4.5cm}
\caption{\small  
Cross section for $gg,q\bar q\to t\bar b H^-$,
$gg,q\bar q \to t\bar t \to t\bar b H^-$ with finite top quark width, 
$bg\to tH^-$ and
the combination of the first and the last, at the LHC with $\sqrt s=14$ TeV
(left plot). Cross section for $gg,q\bar q\to t\bar b H^-$ and
$gg,q\bar q \to t\bar t \to t\bar b H^-$ in NWA, at 
the Tevatron with $\sqrt s=2$ TeV (right plot). Rates are
function of $M_{H^\pm}$ for a representative value of $\tan\beta$.}
\label{fig:threshold}
\vspace{-0.25cm}
\end{center}
\end{figure}

\subsection{Analysis}
\label{sec:ana}

The signal $gg \to tbH^\pm \to jjbb\tau\nu$ and the major backgrounds, 
$gg \to t\bar{t} \to jjb\tau\nu b$ and $q\bar q,qg,\bar qg \to W+\mbox{jets}$, are generated 
with {\small HERWIG} v6.4 in the default implementation 
except for
{\small CTEQ5L}~\cite{PDFs} Parton Distribution Functions
(PDFs). The detector is
simulated with {\small ATLFAST}~\cite{ATLFAST}. The 
{\small TAUOLA} package~\cite{Jadach:1990mz} is used for the polarisation of the $\tau$-lepton. 
The selection of the final state requires a multi-jet trigger with a $ \tau$-trigger:
\begin{description}
  \item[(1)~~] We search for one hadronic $\tau$-jet, two $b$-tagged jets and at least two 
light-jets, all with $p_T > 30$~GeV. Furthermore, the $\tau$-jet and the $b$-tagged jets are required
to be within the tracking range of the ATLAS Inner Detector, $|\eta| < 2.5$. We assume a 
$\tau$-tagging efficiency of 30\% and a $b$-tagging efficiency of 60\%(50\%) at low(high) 
luminosity. The efficiency of this selection is at the level of 1.31\% for the signal (e.g., 
at $M_{H^\pm}=170$~GeV), 1.25\% for $gg \to t\bar{t} \to jjb\tau\nu b$ events and 
$(0.36\,\times\,10^{-3})$\% for $W^\pm$+jets events.
   \item[(2)~~] We reconstruct the invariant masses of pairs of light-jets, $m_{jj}$, and keep 
those consistent with the $W^\pm$ mass: $|m_{jj}-M_W| < 25$~GeV. The associated 
top-quark is then reconstructed requiring $|m_{jjb}-m_t| < 25$~GeV. For the signal with a 
charged Higgs mass of 170~GeV, 0.68\% of signal events pass this selection criteria compared to 
0.73\% and $(0.45\,\times\,10^{-6})$\% for the $t\bar{t}$ and $W^\pm$+jets backgrounds, respectively. 
   \item[(3)~~] We require that the transverse momentum of the $\tau$-jet be greater than 100~GeV, the 
transverse missing momentum be greater than 100~GeV and the azimuthal opening angle 
between the $\tau$-jet and the missing momentum vector be greater than one radian. 
Indeed, in the signal, the $\tau$-lepton originates from a scalar particle ($H^\pm$) 
whereas in the background the $\tau$-lepton comes from the decay of a vector particle ($W^\pm$). 
This difference reflects in the polarisation state of the $\tau$ and leads to harder $\tau$-jets 
in the signal compared to the backgrounds \cite{Assamagan:2002ne}--\cite{Roy:1999xw}. 
Furthermore, to satisfy 
the large cut on the transverse missing momentum and because the charged Higgs is heavier 
than the $W^\pm$-boson, a much larger boost is required from the $W^\pm$- in the background than 
from the $H^\pm$-boson in the signal. As a result, the spectra of the 
azimuthal opening angle between the 
$\tau$-jet and the missing transverse momentum are different for signals
and backgrounds, as shown in 
Fig.~\ref{fig:angle_mT} (left plot). 

\begin{figure}[!htbp]
\begin{center}
\epsfig{file=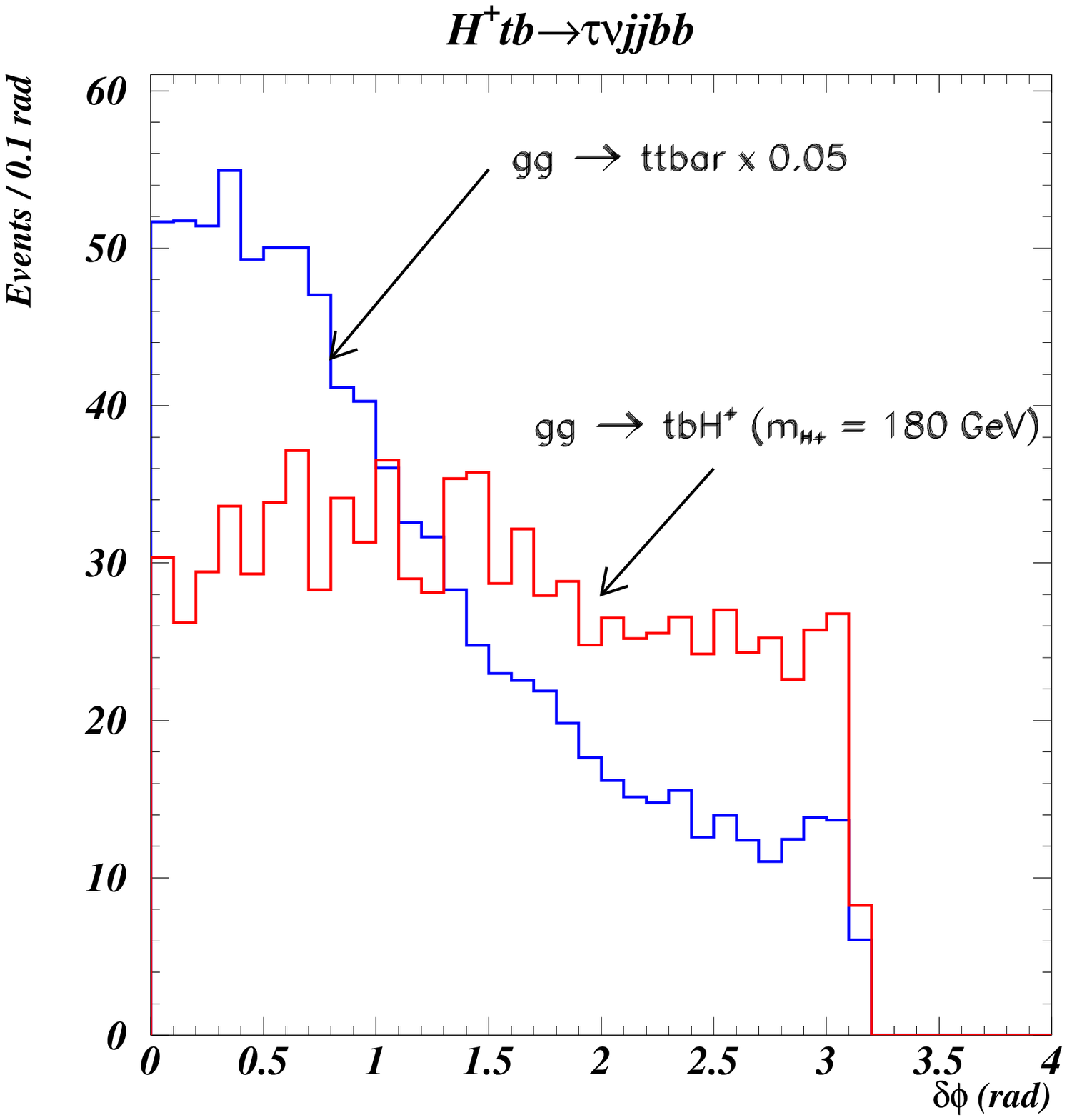,height=4.5cm}
\epsfig{file=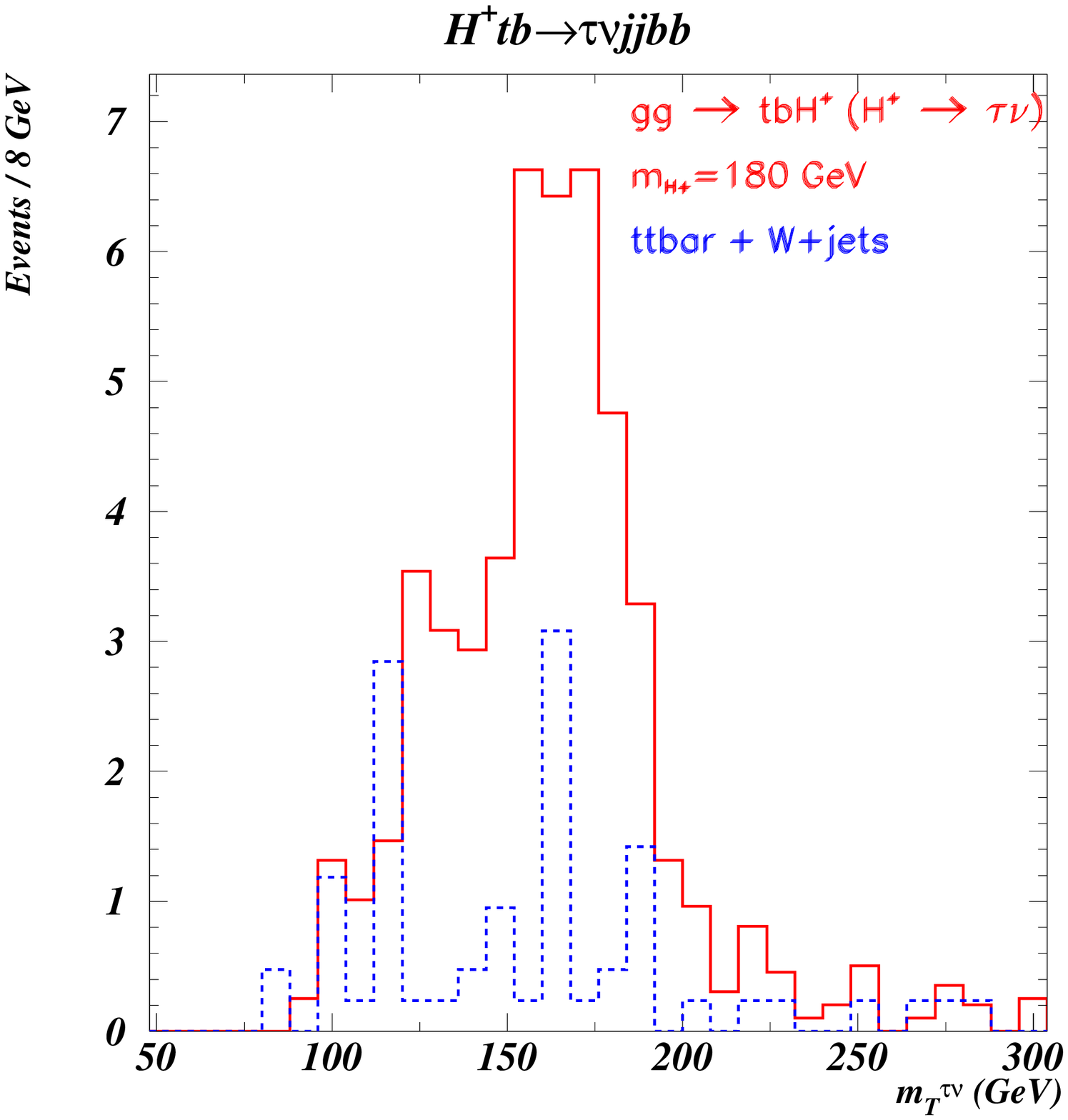,height=4.5cm}
\caption{\small The plot on the left shows the azimuthal opening angle between the $\tau$-jet 
and the transverse missing momentum. It peaks forward in the background and 
more and more backward in the 
signal, as the charged Higgs mass increases. The 
right plot shows the reconstructed transverse mass for a 180~GeV Higgs. (Both plots are shown 
for an integrated luminosity of 30~fb$^{-1}$.)}
\label{fig:angle_mT}
\vspace{-0.25cm}
\end{center}
\end{figure}

Although the full invariant mass of the $H^\pm \to \tau\nu$ system cannot be 
reconstructed because of the neutrino in the final state, the transverse mass (which is 
kinematically constrained to be below the $W^\pm$-mass in the backgrounds and below the 
$H^\pm$-mass in the signal)
\begin{equation}
\label{eq:trans}
m_T = \sqrt{2p_T^{\tau-{\rm jet}}{p\!\!\!/}_T\left[1-\cos(\Delta\phi)\right]}
\end{equation}
combines the benefits of both the polarisation effects and the kinematic boost, thus providing a 
good discriminating observable, as shown in Fig.~\ref{fig:angle_mT} (right plot). (The residual 
background under the signal is due to the experimental $E_T^{\rm{miss}}$ resolution.) 
   \item[(4)~~] We also apply a combination of other cuts on: the invariant mass and the azimuthal opening angle of the $\tau b$-jet 
system, where $b$-jet is here the remaining one after 
the reconstruction of the top quark ($m _{\tau b-{\rm{jet}}} >$ 100 GeV and $\Delta\Phi(\tau-{\rm {jet}},b-{\rm{jet}})>1.25$ radians); 
the invariant mass 
of the 
$b\bar{b}$ pair ($m_{bb-{\rm{jet}}}>225$ GeV) and the 
transverse mass of the $\tau b$-jet system ($p_T^{\tau b-{\rm{jet}}}>190$ GeV). 
The cumulative effect of these cuts is the reduction of the $W^\pm$+jets background by 
more than one order of magnitude, while the signal 
($M_{H^\pm} =$ 170~GeV) and the $t\bar{t}$ background are 
suppressed by only a factor of two.
   \item[(5)~~] Finally, we require $m_T > 100$ GeV for the calculation of the 
signal-to-background ratios and the signal significances in Tab.~\ref{tab:table1}. 
This cut is very 
efficient against the $t\bar{t}$ noise (the efficiency is 0.06\% for a 
$M_{H^\pm} =170$ GeV Higgs signal, $1.9\,\times
\,10^{-3}$ and $0.42\,\times\,10^{-6}$ for the $t\bar{t}$ and the $W^\pm$+jets backgrounds, respectively).
\begin{table*}
\begin{center}
\begin{minipage}{.75\linewidth} 
\caption{\label{tab:table1} Sensitivity of the ATLAS detector to the observation of 
charged Higgs bosons through $H^\pm\rightarrow \tau\nu$ decays in the transition region, 
for an integrated luminosity of 30~fb$^{-1}$ and $\tan\beta=50$.}
\end{minipage}
\vspace*{0.25cm}
\vbox{\offinterlineskip 
\halign{&#& \strut\quad#\hfil\quad\cr  
%\colrule
& $M_{H^\pm}$ (GeV)                && 160  && 170  && 180   && 190  & \cr
%\colrule
& Signal ($S$)                     && 35   && 46   &&  50   &&  35  & \cr
& Backgrounds ($B$)                && 13   && 13   &&  13   &&  13  & \cr
& $S/B$                            && 2.7  && 3.5  &&  3.8  &&  2.7 & \cr
& $S/\sqrt{B}$                     && 9.7  && 12.8 &&  13.9 && 9.7  & \cr
& Poisson Significance             && 7.3  && 9.1  &&  9.8  && 7.3  & \cr
& Poisson Significance$+$5\% syst. && 7.1  && 8.9  &&  9.5  && 7.1  & \cr
%\colrule
}}
\end{center}
\end{table*}
\end{description}

\begin{figure}[!htbp]
\begin{center}
\epsfig{file=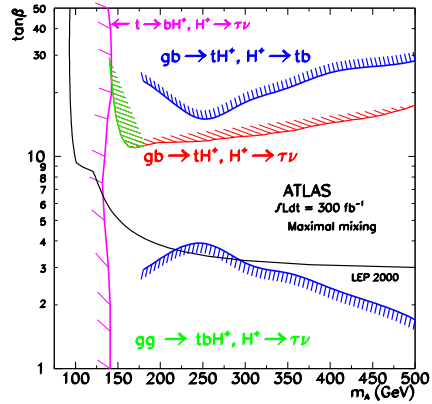,height=6cm}
\caption{\small The new ATLAS discovery potential for charged Higgs bosons. The results
of the current analysis are shown in green.}
\label{fig:newcon}
\vspace{-0.25cm}
\end{center}
\end{figure}

\subsection{Results}
The discovery contour in the transition region resulting from this new analysis
is shown in Fig.~\ref{fig:newcon}. Notice that, at lower 
masses, the signal reconstruction efficiency decreases (although the rate is higher),
thus explaining the upward turn of the discovery reach.  

Before closing, some additional information is in order regarding the
interplay between the new curve and the two old ones. In fact, recall that
above the top-quark mass, the $2\to 2$ process, $bg \to tH^-$, with $H^\pm \to\tau\nu$, 
was used while below it the charged Higgs was searched for in top-quark decays, 
$t \to bH^\pm$, counting the excess of $\tau$-leptons over the SM expectations. 
Furthermore, in the analysis above the top-quark mass, {\small CTEQ2L} 
PDFs~\cite{PDFs} were used and the charged Higgs production cross sections were obtained from
another generator, 
{\small PYTHIA} v5.7. These differences complicate the matching of the various contours at 
their boundaries, especially between the transition region and the high mass region 
($M_{H^\pm} > m_t$). In the result shown, the normalisation cross sections for the transition 
region were matched to the {\small PYTHIA} v5.7 numbers above $m_t$, for consistency with the previous analysis 
of the high mass region~\cite{Assamagan:2002ne}. A second stage of this analysis is currently underway to 
update all the discovery contours by adopting the same $2\to3$ production process throughout.

\subsection{Conclusions}
Meanwhile, as {\sl ad interim} conclusion, we would like to claim that the LHC
discovery potential of charged Higgs bosons has been extended further by our preliminary analysis. 

}

%% file: lowette.tex
{
\section[ ]{Heavy Charged MSSM Higgs Bosons in the $H^{\pm} \to tb$ Decay in
CMS\footnote{S.\,Lowette, P.\,Vanlaer and J.\,Heyninck}}

\subsection{Introduction}
One of the most straightforward ways to extend the Standard Model, is to add an extra complex Higgs doublet to the theory, thus giving rise to five physical Higgs bosons after electroweak symmetry breaking~\cite{Gunion:HiggsHunter}, two of which are the charged scalars $H^{\pm}$. A particular example of a model containing two Higgs doublets, is the much investigated Minimal Supersymmetric Standard Model (MSSM).

The production cross section and decay modes of the charged Higgs $H^{\pm}$, can be described in the MSSM by two parameters at tree level. These are usually taken as the ratio of the vacuum expectation values of the two Higgs doublets $\tan \beta = v_2/v_1$, and the mass of the pseudoscalar Higgs $m_A$. This mass $m_A$ is, again at tree level, related to the charged Higgs boson mass as $m_A^2 = m_{H^{\pm}}^2 - m_{W^{\pm}}^2$. The branching ratios for the decay channels of the charged Higgs, depend mainly on its mass. As shown in Fig.~\ref{fig:chHbr}, for masses above $m_t+m_b$, the channel $H^{\pm} \to tb$ dominates. In the main production channel $gb \to tH^{\pm}$, it will result in complex final states, the most interesting being the semileptonic one,
\begin{equation}\label{eqn:finalstate}
gb \to tH^{\pm} \to ttb \to W^+W^-bbb \to qq^{\prime}\ell \nu bbb,
\end{equation}
because the Higgs boson mass can still be reconstructed, while an isolated lepton is present to trigger on.

\begin{figure}[ht]
  \begin{center}
    \begin{minipage}{7.5cm}
      \begin{center}
        \includegraphics[scale=0.25]{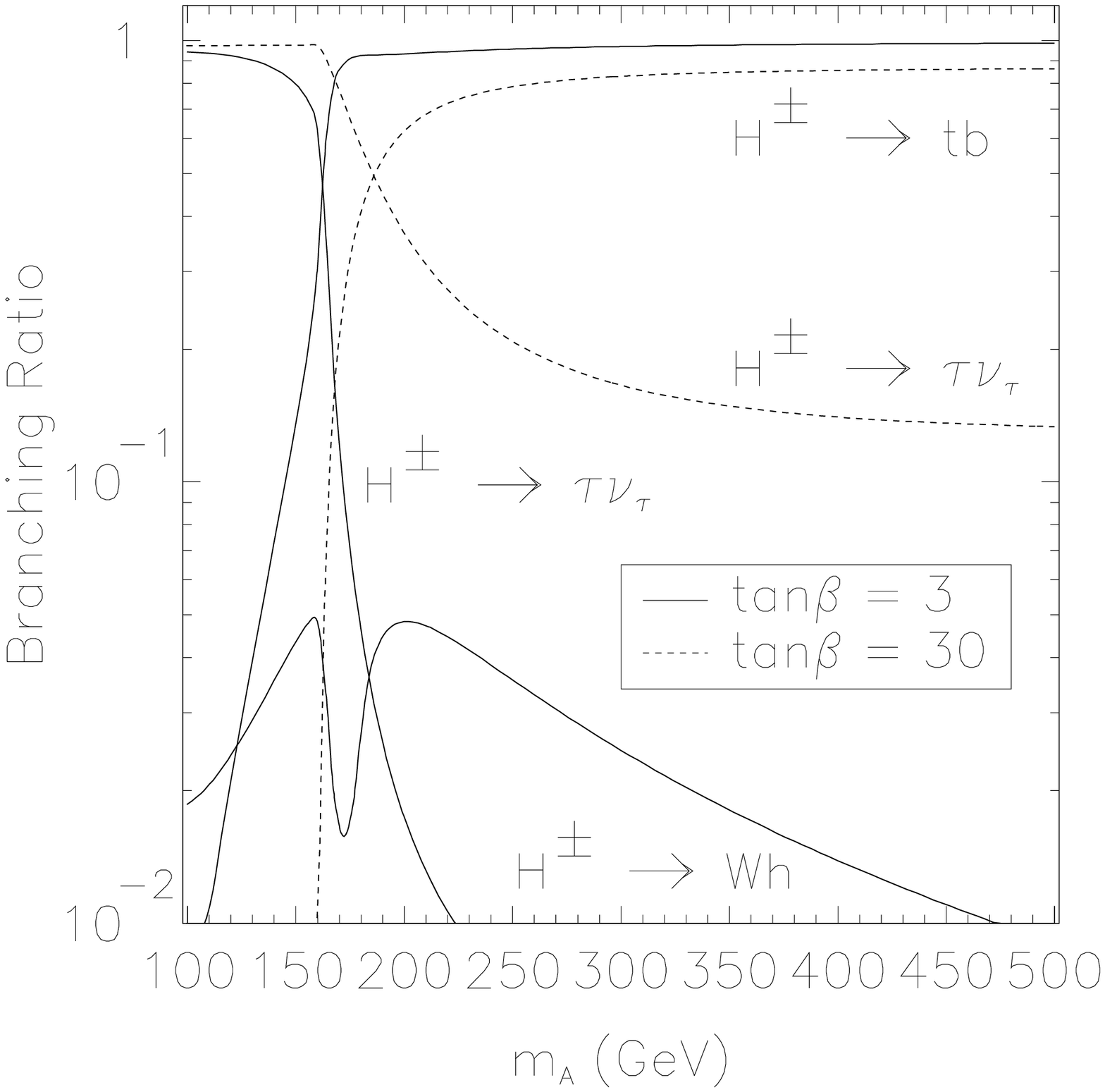}
        \vspace{-4mm}
        \caption{Charged Higgs boson branching ratios in function of $m_A$, generated with HDECAY.}
        \label{fig:chHbr}
      \end{center}
    \end{minipage}
    \hspace{\stretch{1}}
    \begin{minipage}{7.5cm}
      \begin{center}
        \includegraphics[scale=.5]{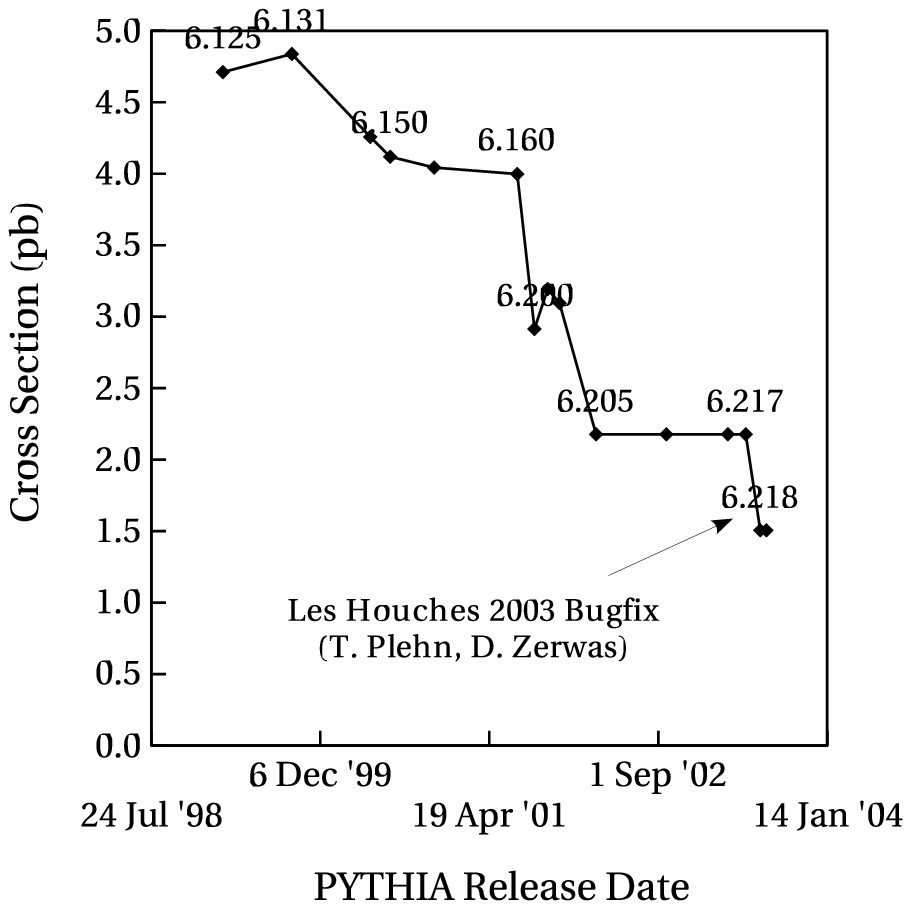}
        \vspace{-4mm}
        \caption{Evolution through time of the PYTHIA cross section value for $gb \to tH^{\pm}$. The PYTHIA version is shown in the labels on the curve. ($m_A = 300 \, \mathrm{GeV}$, $\tan \beta = 50$)}
        \label{fig:pythiacrosssecevol}
      \end{center}
    \end{minipage}
  \end{center}
  \vspace{-6mm}
\end{figure}

The potential of the decay channel $H^{\pm} \to tb$ for high Higgs boson masses at LHC, has been considered before at parton level in several phenomenological studies~\cite{Barger:1994th,Gunion:1994sv,Moretti:1999bw,Miller:1999bm,Cavalli:2002vs}. These studies showed the possibility of detecting the charged Higgs in certain regions of $(m_A,\tan\beta)$ parameter space during the low luminosity run of LHC, with both three or four $b$--jets tagged, provided good $b$--tagging capabilities of the detectors to suppress the large $t\bar{t} + \mathrm{jets}$ background. Fast simulation studies, taking into account parametrized detector performances, have also been carried out for CMS~\cite{Salmi:CMS-NOTE-2002-024} and ATLAS~\cite{Assamagan:ATLAS-PHYS-99-013,Assamagan:ATLAS-PHYS-2001-017}.

In this analysis, charged Higgs detection has been studied for the final state~(\ref{eqn:finalstate}) using triple $b$--tagging, during the low luminosity period where LHC will acquire an integrated luminosity $L = \int \mathcal{L} dt = 60 \, \mathrm{fb}^{-1}$ of data. Supersymmetric particles are supposed heavy enough, so that decays into them can be neglected. The main improvement of this analysis, is the inclusion of the most recent theoretical calculation\cite{Plehn:2002vy} for the signal cross section at leading order (LO) and next--to--leading order (NLO), leading to sizable effects compared to the previous discovery prospects. Indeed, the leading order cross section values, predicted by the Monte--Carlo program PYTHIA~\cite{Sjostrand:2000wi}, have decreased by a factor $\sim 3$ over the last 5 years\footnote{During this Les Houches workshop, a bug was fixed in version 6.218, such that the PYTHIA output now corresponds to the theoretical calculation.}, as shown in Fig.~\ref{fig:pythiacrosssecevol}. Other improvements are the use of a new dedicated background simulation, the inclusion of CMS trigger acceptances, the introduction of a likelihood based method to suppress the combinatorial background and the estimation of the influence of systematic uncertainties on the background cross section.

\subsection{Signal and Background Simulation} \label{sec:signbacksim}

%\subsection{Signal simulation} \label{sec:signsim}
The production of the charged Higgs boson is considered in the dominant inclusive channel $pp\to tH^{\pm}\! X$. The cross section for this process should be evaluated at leading order in the channel $gb \to tH^{\pm}$~\cite{Plehn:2002vy}. Its dependency on $\tan\beta$ and $m_A$ has been visualized in Figs.~\ref{fig:crosssecma} and~\ref{fig:crosssectanb}. The cross section decreases exponentially with rising $m_A$, and is enhanced at low and high values of $\tan\beta$, with a minimum at $\tan\beta = \sqrt{m_t/m_b}$. The calculation of the signal cross section has also been performed at NLO~\cite{Plehn:2002vy}. The resulting increase in cross section depends on the value of the MSSM parameters. In the mass region considered and for $\tan\beta > 30$, however, the signal $k$--factor keeps the constant value $k = 1.3$.

\begin{figure}[ht]
  \begin{center}
    \begin{minipage}{7.5cm}
      \begin{center}
        \includegraphics[scale=0.25]{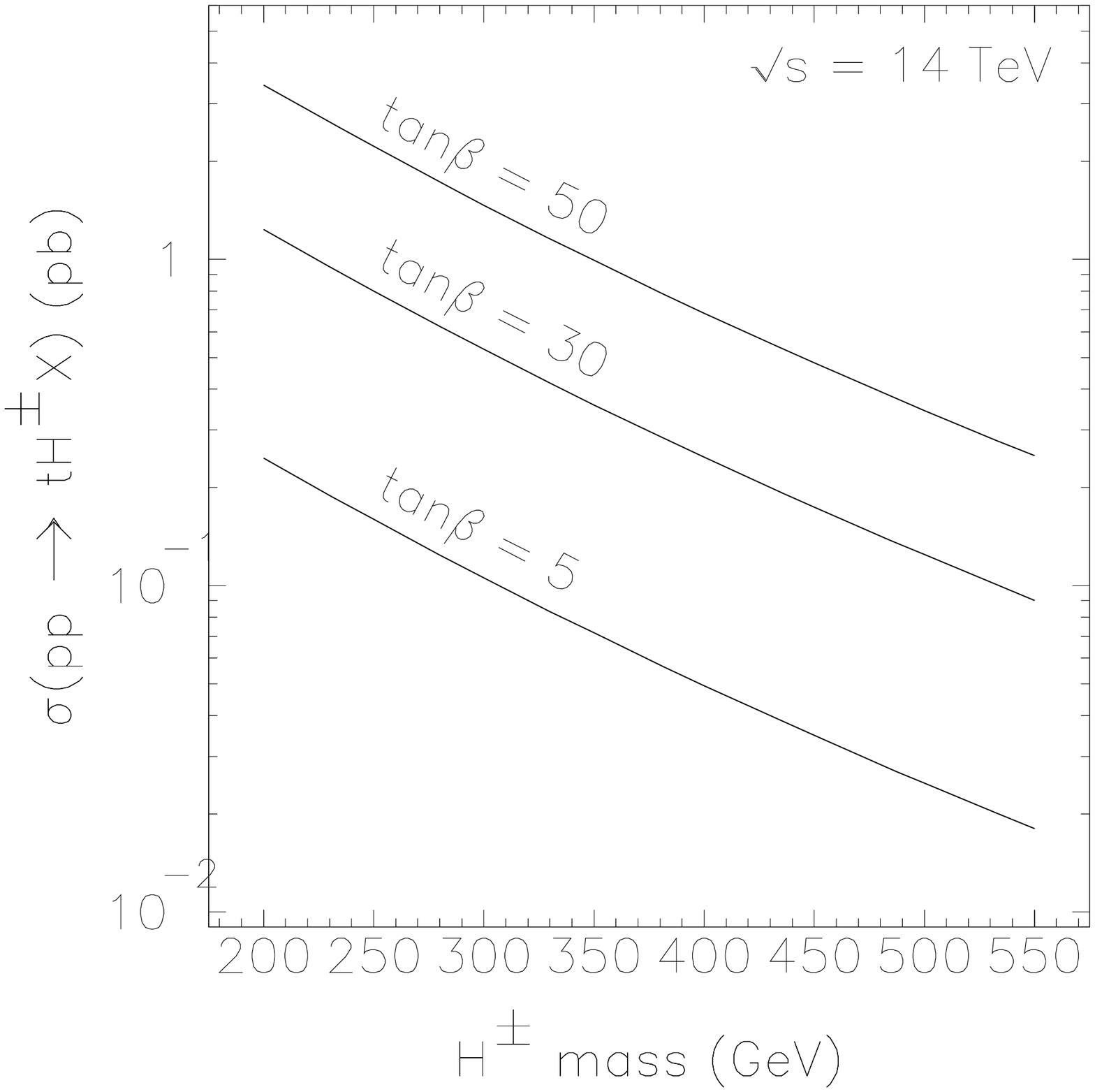}
        \vspace{-4mm}    
        \caption{$pp \to tH^{\pm}X$ cross section dependence on $m_A$.}
        \label{fig:crosssecma}
      \end{center}
    \end{minipage}
    \hspace{\stretch{1}}
    \begin{minipage}{7.5cm}
      \begin{center}
        \includegraphics[scale=0.25]{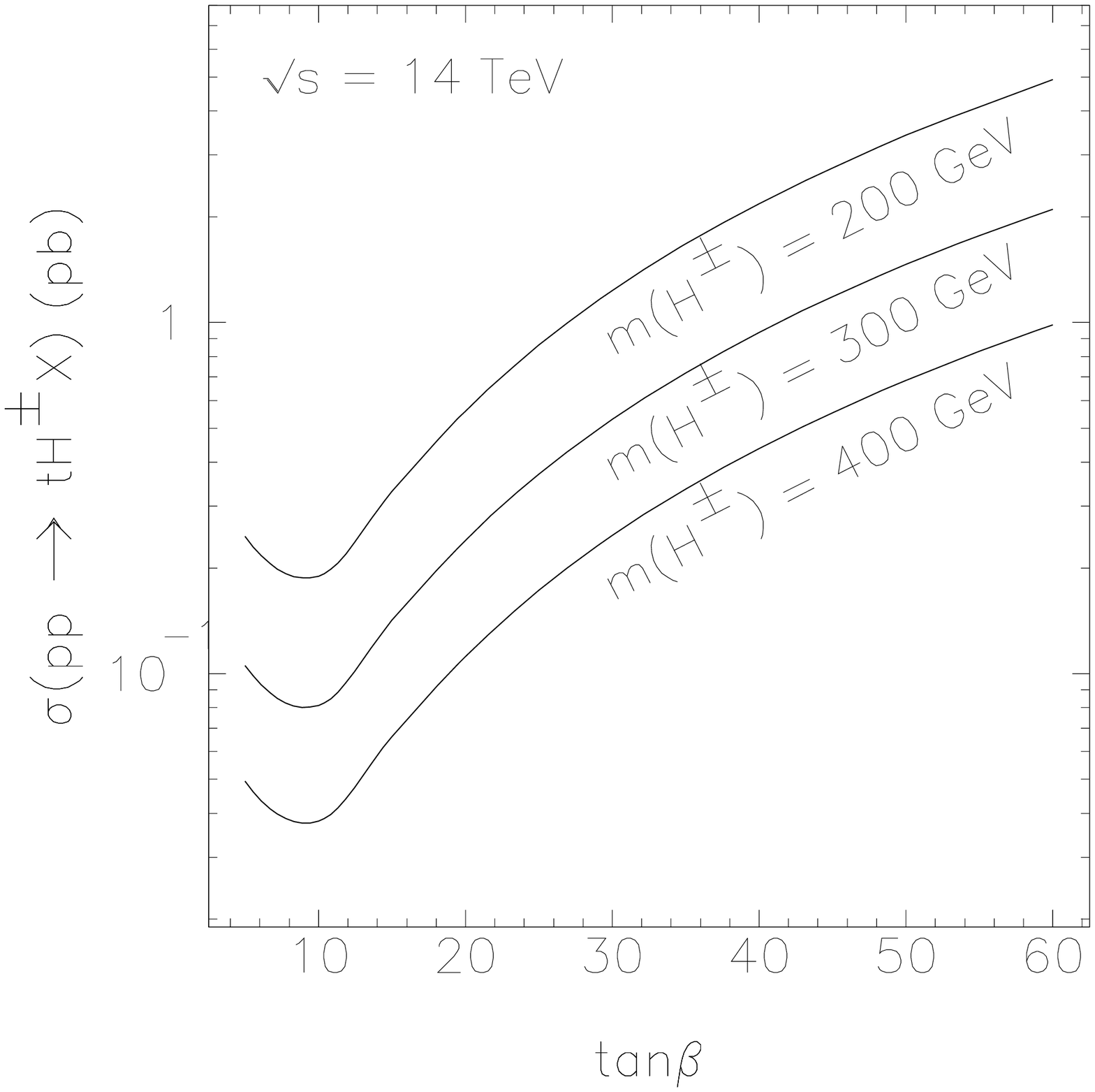}
        \vspace{-4mm}    
        \caption{$pp \to tH^{\pm}X$ cross section dependence on $\tan\beta$.}
        \label{fig:crosssectanb}
      \end{center}
    \end{minipage}
  \end{center}
  \vspace{-6mm}
\end{figure}

The generation of the signal has been performed with PYTHIA, using the cross section values from~\cite{Plehn:2002vy,Plehn:privcomm}, and forcing the $H^{\pm} \to tb$ decay. The branching ratios for this decay process were calculated with HDECAY~\cite{Djouadi:1998yw}, ranging from $\sim 80 \%$ for low $m_A$ and high $\tan\beta$ up to $\sim 100 \%$ for high $m_A$ and low $\tan\beta$, as also shown in Fig.~\ref{fig:chHbr}. Six samples have been generated at $\tan\beta = 50$ and masses $m_A$ ranging from $250 \, \mathrm{GeV}$ to $500 \, \mathrm{GeV}$. This yields a number of signal events before the event selection of almost $55\,000$ for $m_A = 250 \, \mathrm{GeV}$ down to about $7\,500$ for $m_A = 500 \, \mathrm{GeV}$.

%\subsection{Background simulation} \label{sec:backsim}
At leading order, the dominant background comes from Standard Model $gb \to t\bar{t}b$ and $t\bar{t} + \mathrm{jet}$ production, where in the latter case the accompanying quark or gluon jet is misidentified as being a $b$--jet. Other potential multi--jet backgrounds are much smaller~\cite{Barger:1994th,Sonnenschein:CMS-NOTE-2001-001} and neglected. The aforementioned background processes cannot be generated with PYTHIA. Therefore the simulation of the background has in the first place been performed by generating $t\bar{t}$ events with PYTHIA, where the parton shower generates additional jets. An overall LO cross section of $560\, \mathrm{pb}$ was used, resulting in about $17\times 10^6$ events before the event selection. This background will further be referred to, as the $t\bar{t}$ background.

The background simulation has also been performed using the matrix element generator MadGraph/MadEvent \cite{Maltoni:2002qb}, in order to simulate directly the hard interactions $pp \to t\bar{t}b$ and $pp \to t\bar{t}j$. A cut on the transverse momentum $p_T > 10 \, \mathrm{GeV}$ and the pseudorapidity $\left| \eta \right| < 2.5$ of the extra jet accompanying the tops was applied, resulting in a total cross-section of $678 \, \mathrm{pb}$, or over $20\times 10^{6}$ events before selection. After the simulation of the hard interaction, the events were interfaced to PYTHIA for parton showering, decay and hadronisation. This background will further be called the $t\bar{t}b / t\bar{t}j$ background.

When looking at next--to--leading order, the cross section for the $t\bar{t}$ background scales up to about $800 \, \mathrm{pb}$~\cite{Altarelli:2000ye}. This rise has been taken into account by using a $k$--factor of $k = 1.43$ for the $t\bar{t}$ background when quoting results at NLO. The calculation for the processes $pp\to t\bar{t}b$ and $pp \to t\bar{t}j$ at NLO has not been performed yet, however, and therefore no NLO comparison has been made for this background.

\subsection{Event Selection and Triggering} \label{sec:evseltrig}

%\subsection{Minimal selection criteria}
To simulate CMS detector performance, the programs CMSJET~\cite{Abdullin:CMS-TN-94-180} and FATSIM~\cite{Karimaki:CMS-IN-2000-034} have been used for detector response parametrization. In this study, $b$--tagging is performed with a method based on impact parameter significance. For both the background samples, the $b$--tagging efficiency was found to be about 45\%, while for the signal this efficiency grows from 44\% to 48\%, due to the harder event kinematics, with $m_A$ going from $250 \, \mathrm{GeV}$ to $500 \, \mathrm{GeV}$. This behaviour is also observed in the light quark mistag rate, ranging from 1.10\% to 1.18\% with rising $m_A$. For the background, this mistag rate is significantly different for both background samples. For the $t\bar{t}$ and $t\bar{t}b/t\bar{t}j$ background, the mistag probability was found to be 0.92\% and 1.02\% respectively. This is a result of the harder event kinematics in the $t\bar{t}b/t\bar{t}j$ background compared to the $t\bar{t}$ one.

In order to be able to reconstruct an event, a selection is performed, accepting only those events for which the reconstruction yields at least
\begin{itemize}
\item 1 isolated lepton (electron or muon) with $|\eta|<2.4$ and $p_T > 19 \, \mathrm{GeV}$ for muons and $p_T > 29 \, \mathrm{GeV}$ for electrons.
\item 5 jets ($b$ or non $b$) with $p_T > 20 \, \mathrm{GeV}$ and $|\eta|<2.4$. Jets are reconstructed using a cone algorithm with $\Delta R = \sqrt{\Delta \phi^2 + \Delta \eta^2} = 0.5$.
\item 3 $b$--tagged jets.
\end{itemize}
\vspace{1mm}

The total efficiency of these criteria on the different background and signal samples, is shown in the third column of Table~\ref{tab:evsel} for $\tan\beta = 50$ and $30 \, \mathrm{fb}^{-1}$ of integrated luminosity. The low efficiency is mainly due to the demand for three $b$--tagged jets, allowing for the suppression of the background one order of magnitude more than the signal. When comparing the efficiency for the $t\bar{t}b/t\bar{t}j$ background with the $t\bar{t}$ background, the difference is found to be mainly due to the higher mistag rate.

%\subsection{High Level Trigger (HLT) selection} \label{sec:trigacc}
In order to estimate the influence of the CMS trigger acceptances on the event rate of reconstructable events, the High Level Trigger (HLT) cuts are applied only after these minimal selection criteria. As an isolated lepton is present in the final state, high triggering efficiencies are expected with only the inclusive electron and muon triggers. The HLT cuts at low luminosity are taken at $29 \, \mathrm{GeV}$ for single electrons and $19 \, \mathrm{GeV}$ for single muons~\cite{CMS:TriDAS-TDR2}. Additionally, a correction factor of 68.9\% was applied on the events with an electron, to account for inefficiencies in the online electron reconstruction~\cite{CMS:TriDAS-TDR2}.

For as well the backgrounds as the different signal samples, about 86\% of the events fulfilling the minimal selection criteria, are accepted by the CMS HLT. Of these events passing the HLT cuts, about 65\% come from the muon trigger chain, while the remaining 35\% passed the electron trigger.

\begin{table}[ht]
  \vspace{-3mm}
  \begin{center}
  \caption{Selection and solution finding efficiencies.}
  \vspace{1mm}
  \begin{tabular}{|r|r|r|r|}
    \hline
    & \makebox[2.7cm][c]{\# events}
    & \makebox[3.2cm][c]{\# events after minimal}
    & \makebox[3.2cm][c]{\# events after HLT}\\
    \makebox[3.3cm][l]{$\tan\beta = 50$, $30 \, \mathrm{fb}^{-1}$}
    & \makebox[2.7cm][c]{before cuts}
    & \makebox[3.2cm][c]{selection criteria}
    & \makebox[3.2cm][c]{and with $\ge 1$ solution}\\
    \hline
    \rule{0mm}{4mm} $t\bar{t}$ background
    & 16\,800\,000 & 15\,736\,\, (0.09\%) & 4\,932\,\, (31\%) \\
    \rule{0mm}{4mm} $t\bar{t}b/t\bar{t}j$ background
    & 20\,340\,000 & 23\,593\,\, (0.12\%) & 7\,872\,\, (33\%) \\
    \rule{0mm}{4mm} $tH^{\pm}$ ($m_A=250\, \mathrm{GeV}$)
    &      54\,644 &     769\,\, (1.41\%) &    314\,\, (41\%) \\
    \rule{0mm}{4mm} $tH^{\pm}$ ($m_A=300\, \mathrm{GeV}$)
    &      36\,681 &     659\,\, (1.80\%) &    235\,\, (36\%) \\
    \rule{0mm}{4mm} $tH^{\pm}$ ($m_A=350\, \mathrm{GeV}$)
    &      23\,988 &     492\,\, (2.05\%) &    173\,\, (35\%) \\
    \rule{0mm}{4mm} $tH^{\pm}$ ($m_A=400\, \mathrm{GeV}$)
    &      16\,176 &     381\,\, (2.36\%) &    116\,\, (30\%) \\
    \rule{0mm}{4mm} $tH^{\pm}$ ($m_A=450\, \mathrm{GeV}$)
    &      10\,888 &     270\,\, (2.48\%) &     86\,\, (32\%) \\
    \rule{0mm}{4mm} $tH^{\pm}$ ($m_A=500\, \mathrm{GeV}$)
    &       7\,472 &     198\,\, (2.65\%) &     72\,\, (36\%) \\
    \hline
  \end{tabular}
  \label{tab:evsel}
  \end{center}
  \vspace{-7mm}
\end{table}

\subsection{Analysis Strategy} \label{sec:anastrat}

\subsubsection{Event Reconstruction}
Starting from the complex final state~(\ref{eqn:finalstate}), it is possible to reconstruct the charged Higgs boson mass. First, as the $z$--component of the missing energy is not measured, the longitudinal momentum of the neutrino is calculated using the $W^{\pm}$ mass constraint, giving rise to none or two real solutions. Then the hadronically decayed $W^{\pm}$ candidate is reconstructed, followed by the top candidates, using the constraints
\begin{equation}\label{eqn:mWconstr}
| m_{qq^{\prime}} - m_{W^{\pm}} | < 30 \, \mathrm{GeV} \quad \mathrm{,} \quad | m_{qq^{\prime}b} - m_{t} | < 50 \, \mathrm{GeV} \quad \mathrm{and} \quad | m_{\ell \nu b} - m_{t} | < 50 \, \mathrm{GeV}.
\end{equation}
Each reconstructed top quark is now combined with a remaining $b$--tagged jet, giving rise to a charged Higgs candidate. As the Higgs boson mass is a priori not known, both combinations are possible correct solutions, leading to an extra irreducible background from wrong $t$--$b$ combinations. The only further kinematical difference observed between the signal and background, is the $p_T$ spectrum of this $b$--tagged jet, used to reconstruct the charged Higgs candidates. This is visualized in Fig.~\ref{fig:ptbhiggs}. An additional cut $p_T (b_{H^{\pm}}) > 50 \, \mathrm{GeV}$ has therefore been introduced.

In general, there will exist several reconstruction solutions fulfilling the cuts, due to the combinatorics of the $b$'s and the extra jets. If no solution is found, the event is discarded. The final number of events, passing the HLT cuts, as well as having at least one solution, is shown in the last column of Table~\ref{tab:evsel}.

\subsubsection{Determination of the best solution} \label{sec:bestsol}
As there is no constraint on the Higgs boson mass, the only way to distinguish good from bad solutions, is by using the information on the $H^{\pm}$'s decay products. For this analysis, the following likelihood function is defined, starting from the reconstructed masses $m_{qq^{\prime} b}$ and $m_{\ell \nu b}$ of both the top quark candidates, and from the reconstructed mass $m_{qq^{\prime}}$ of the hadronic $W^{\pm}$:
\begin{equation} \label{eqn:likelihood}
\mathcal{L} = \exp \left[-\frac{1}{2} \left( \frac{m_{qq^{\prime}} - m_{qq^{\prime}}^*}{\sigma_{m_{qq^{\prime}}}^*} \right)^2 - \frac{1}{2} \left( \frac{m_{qq^{\prime} b} - m_{qq^{\prime} b}^*}{ \sigma_{m_{qq^{\prime} b}}^*} \right)^2 - \frac{1}{2} \left( \frac{m_{\ell \nu b} - m_{\ell \nu b}^*}{\sigma_{m_{\ell \nu b}}^*} \right)^2 \right].
\end{equation}
The values of the different masses $m^*$ and widths $\sigma^*_m$ are obtained from the distributions of the reconstructed masses, for those events where the jets and/or lepton are matched to the particles at generator level they come from, within a cone $\Delta R = 0.2$. This will take into account, for example, the fact that the resolution of the leptonically decaying top is larger than the resolution of the hadronic one, due to inefficiencies in the neutrino reconstruction.

For each event, the best solution is now determined as the one maximizing the likelihood function~(\ref{eqn:likelihood}). The distribution of this value $\mathcal{L}$ for the best solution, however, has a similar shape for background and signal. Therefore, no cut is applied on $\mathcal{L}$, in order not to further reduce the signal statistics.

\subsection{Results and Discussion} \label{sec:results}

\subsubsection{Mass distributions} \label{sec:massdistr}
Because of the ambiguity between the Higgs boson mass solutions, built from either the leptonically or the hadronically decaying top, they have to be added up. The resulting distribution of the reconstructed Higgs boson mass is shown in Fig.~\ref{fig:hmasstot}, for the signal, the $t\bar{t}$ background, and the sum of both, for $m_A = 300 \, \mathrm{GeV}$ and $\tan\beta = 50$. For the $t\bar{t}b/t\bar{t}j$ background, this distribution is very similar to the one for $t\bar{t}$, apart from an overall increase of the number of events.

\begin{figure}[ht]
  \begin{center}
    \begin{minipage}{7.5cm}
      \begin{center}
        \includegraphics[scale=0.25]{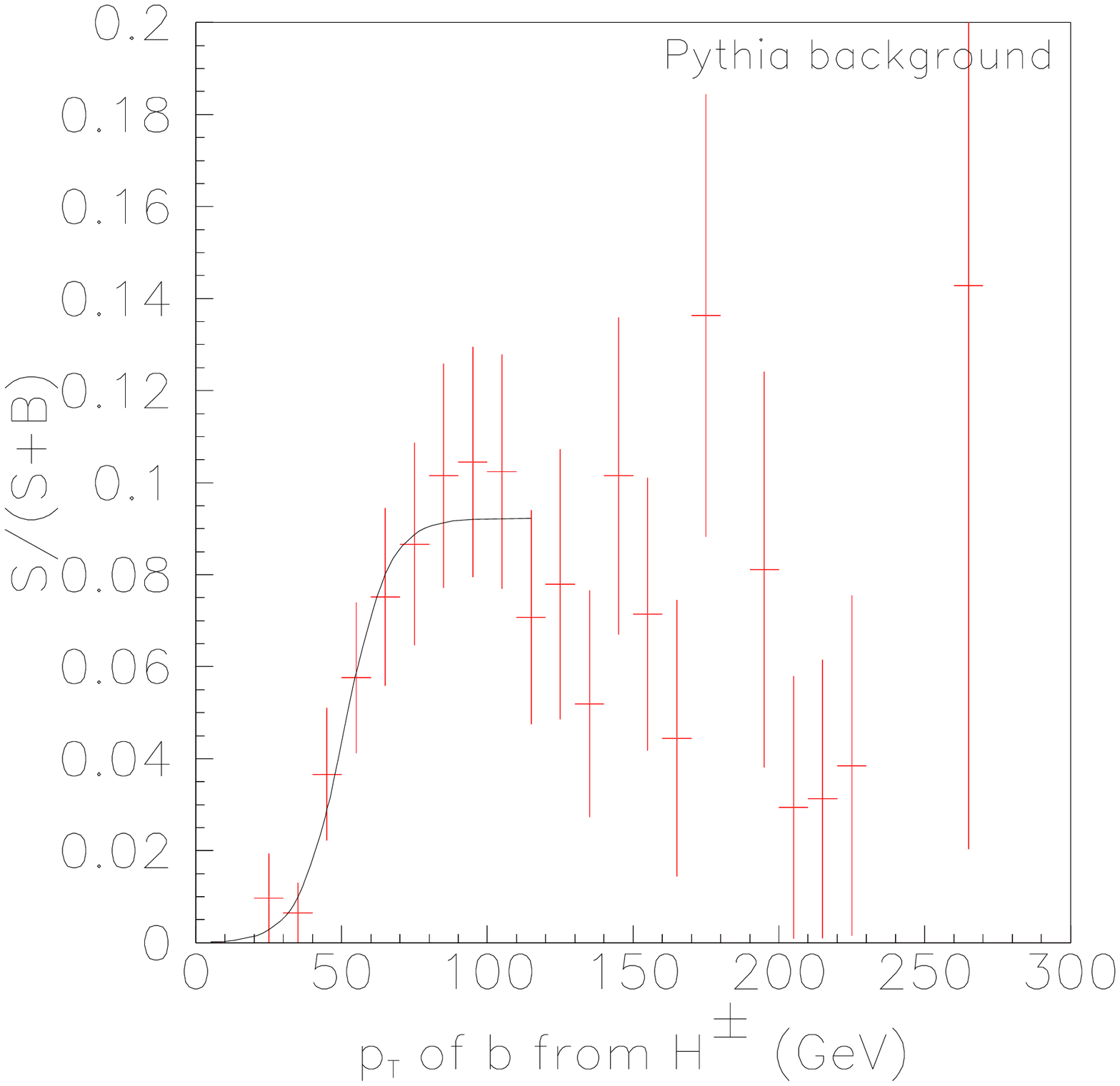}
        \vspace{-4mm}
        \caption{Bin by bin values of $S/(S+B)$ for the $p_T$ of the $b$--tagged jet, considered to come from the $H^{\pm}\to t b$ decay. $S$ and $B$ are the numbers of events for signal and background respectively. ($30 \, \mathrm{fb}^{-1}$, $m_A = 300 \, \mathrm{GeV}$, $\tan\beta=50$)}
        \label{fig:ptbhiggs}
      \end{center}
    \end{minipage}
    \hspace{\stretch{1}}
    \begin{minipage}{7.5cm}
      \begin{center}
        \includegraphics[scale=0.25]{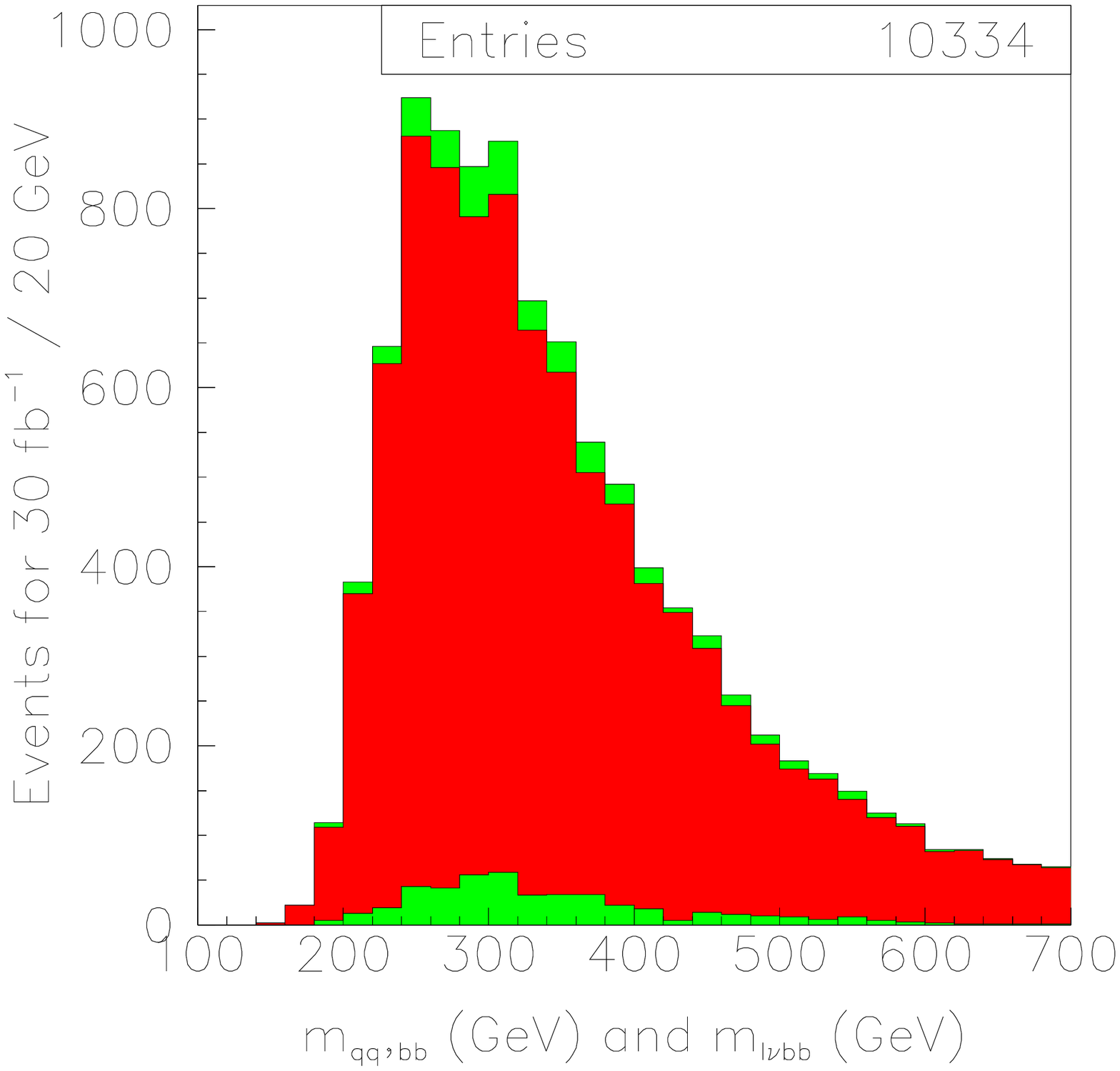}
        \vspace{-4mm}    
        \caption{Sum of the leptonic and hadronic solutions of the charged Higgs boson mass for the signal, the background and the sum of the background and the signal. ($m_A = 300 \, \mathrm{GeV}$, $\tan\beta = 50$, $30 \, \mathrm{fb}^{-1}$)}
        \label{fig:hmasstot}
      \end{center}
    \end{minipage}
  \end{center}
  \vspace{-7mm}
\end{figure}

\subsubsection{Signal significance and discovery contours} \label{sec:disccontours}
The discovery potential for this analysis at low luminosity in the CMS experiment has been estimated, using the statistical significance of the signal defined as \mbox{$\sigma = S / \sqrt{B}$}, with $S$ and $B$ the number of signal and background events respectively. Discovery contours have been constructed in the MSSM parameter space for $\sigma = 5$. For an integrated luminosity of $30 \, \mathrm{fb}^{-1}$ and $60 \, \mathrm{fb}^{-1}$, the result is shown in Fig.~\ref{fig:signif}, using the $t\bar{t}$ background. In the same plot, also the current discovery contour is shown for the subdominant decay channel, $H^{\pm} \to \tau \nu$, after $30 \mathrm{fb}^{-1}$ of integrated luminosity~\cite{Kinnunen:CMS-NOTE-2000-045}. The large difference with the previous CMS result~\cite{Salmi:CMS-NOTE-2002-024} was found to be due to the large drop in the prediction of the signal cross section, described in Section~\ref{sec:signbacksim} In Fig.~\ref{fig:signifpythiavsnlovsmgme} the comparison is shown of the LO $t\bar{t}$ background with the NLO result, and with the LO $t\bar{t}b/t\bar{t}j$ background, for $30 \, \mathrm{fb}^{-1}$.

\begin{figure}[ht]
  \begin{center}
    \begin{minipage}{7.8cm}
      \begin{center}
        \includegraphics[scale=0.25]{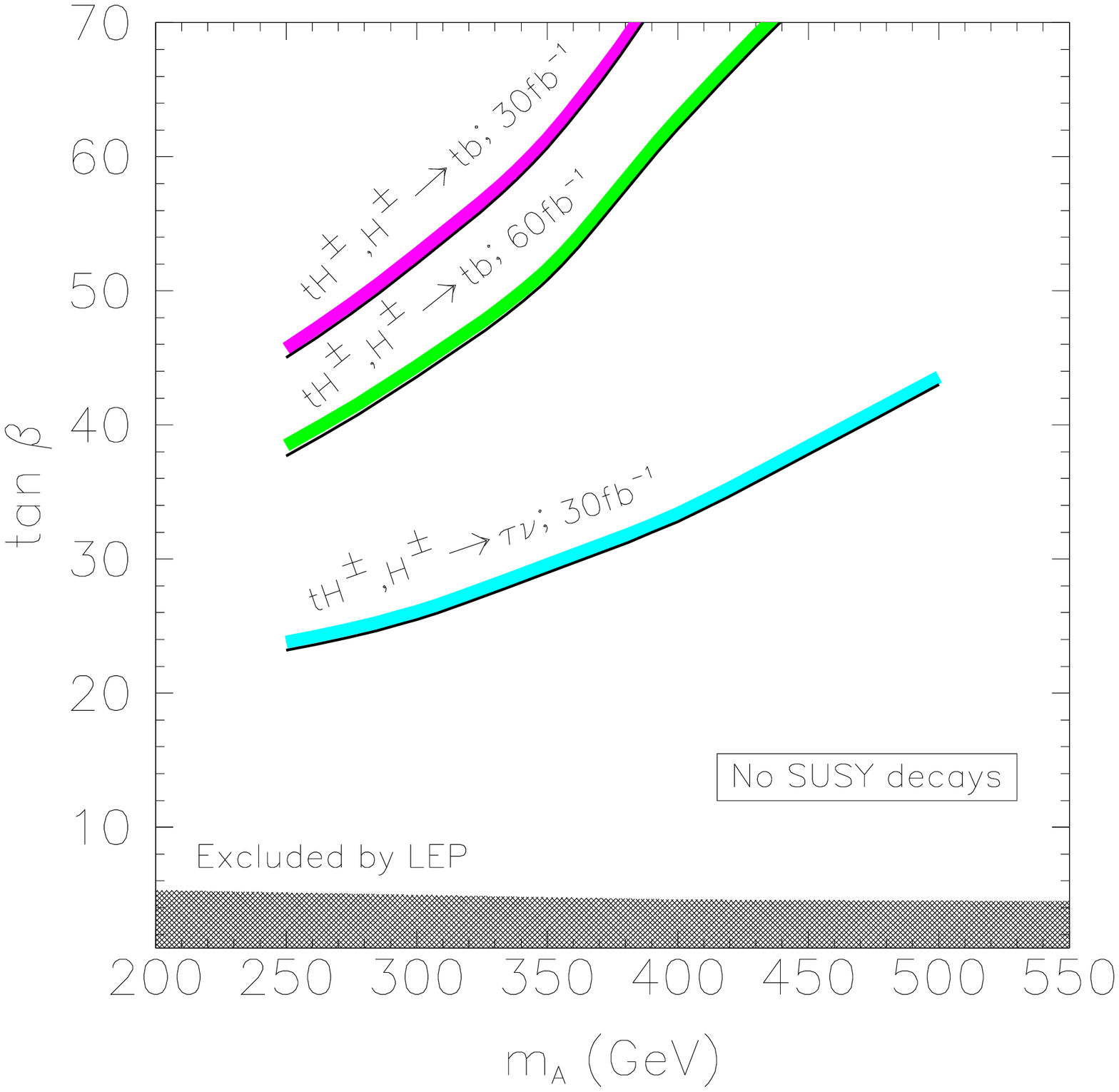}
        \vspace{-4mm}    
        \caption{Discovery contours for the $tH^{\pm},H^{\pm}\to tb$ channel with $30 \, \mathrm{fb}^{-1}$ and $60 \, \mathrm{fb}^{-1}$, and for the $tH^{\pm},H^{\pm}\to \tau\nu$ channel with $30 \, \mathrm{fb}^{-1}$.}
        \label{fig:signif}
      \end{center}
    \end{minipage}
    \hspace{\stretch{1}}
    \begin{minipage}{7.8cm}
      \begin{center}
        \includegraphics[scale=0.25]{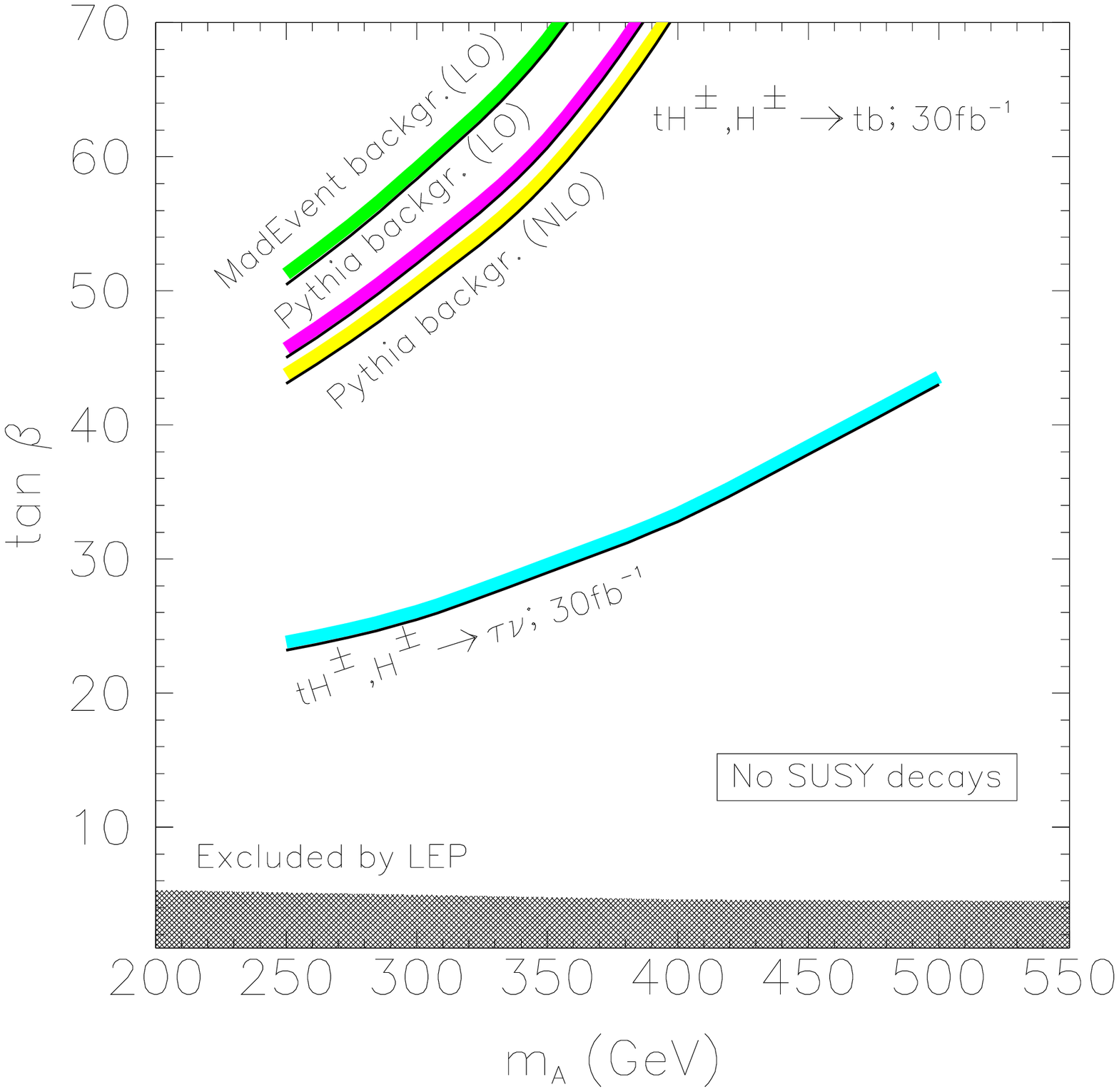}
        \vspace{-4mm}    
        \caption{Discovery contours for the $tH^{\pm},H^{\pm}\to tb$ channel with $30 \, \mathrm{fb}^{-1}$, for the backgrounds $t\bar{t}$ from PYTHIA at LO and NLO, and $t\bar{t}b/t\bar{t}j$ from MadGraph/MadEvent at LO.}
        \label{fig:signifpythiavsnlovsmgme}
      \end{center}
    \end{minipage}
  \end{center}
  \vspace{-7mm}
\end{figure}

\subsubsection{Influence of systematic uncertainties on the background cross section} \label{sec:systuncert}
So far in this study, the significance has been calculated in the ideal case of perfect knowledge of the background cross section. The background is large, however, and the combinatorial background limits the analysis to a counting experiment. Therefore, the effect of systematical uncertainties on the knowledge of the background has been estimated.

Two methods are proposed to extract the background from data. First, the difference between the signal and the background for the $p_T$ spectrum of the $b$-jet from the Higgs decay was used, looking at the low $p_T$ region. The background can in this way be measured with an uncertainty of about 5\% from statistics and remaining signal, plus an additional, possibly sizeable, contribution from the uncertainty on the shape of this $p_T$ distribution. Another possibility is the measurement of the background, tagging 1 $b$--jet less. Using a measured $b$--tagging efficiency and purity, one can then calculate the background when tagging three $b$--jets. A 7\% systematic uncertainty was estimated this way, due to the uncertainty on the $b$--mistag probability, which was taken as 10\%, as found as systematical uncertainty in CDF for the secondary vertex tag technique~\cite{Affolder:2001wd}. Additionally, a possibly sizeable uncertainty is introduced from the ratio of $t\bar{t}b$ to $t\bar{t}j$ events.

In the absence of an experimental measurement of the background, it can be estimated from the theoretically calculated cross section, the luminosity and the reconstruction and analysis efficiencies. One should then add a typical 10 to 15\% uncertainty from the not yet available NLO calculation, an expected 5\% from the luminosity measurement, and additional contributions from the event selection.

With these systematical uncertainty estimations, the effects on the visibility of the signal can be evaluated. Considering a systematical uncertainty of $\epsilon B$ background events after full analysis, a total uncertainty $\Delta B = \sqrt{B+\epsilon^2 B^2}$ on the number of background events $B$ is obtained. The signal significance for $S$ signal events now becomes $\sigma = S / \sqrt{B + \epsilon^2 B^2}$. For this channel with large background, the $S/B$ value is small, and cannot be improved without losing too much signal statistics. In Fig.~\ref{fig:signifsyserrb} the discovery contours are plotted, when supposing perfect knowledge of the $t\bar{t}$ cross section ($\epsilon = 0$), a $1\%$ ($\epsilon = 0.01$) and a $3\%$ uncertainty ($\epsilon = 0.03$). A value of $5 \%$ doesn't show up anymore on the plot.

\begin{figure}[ht]
  \begin{center}
    \includegraphics[scale=0.25]{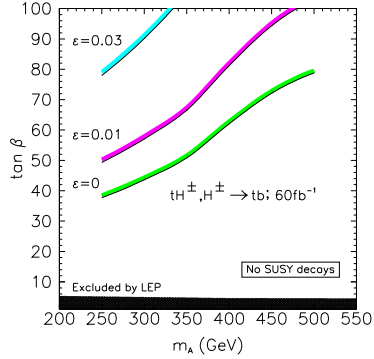}
    \vspace{-4mm}
    \caption{Influence on the discovery contour of systematical uncertainties $\epsilon$ ($0\%$, $1\%$ and $3\%$) on the $t\bar{t}$ background, for $60 \, \mathrm{fb}^{-1}$.}
    \label{fig:signifsyserrb}
  \end{center}
  \vspace{-7mm}
\end{figure}

\subsection{Fully Hadronic Channel}
The fully hadronic $tH^{\pm} \to ttb$ decay, where both $W^{\pm}$'s from the $t$ decay hadronically, has also been studied. In this case there is no lepton to trigger on, however, and it was found that the CMS jet trigger acceptances at HLT alone, already reduce the signal to 1 to 7\% for $250 \, \mathrm{GeV} < m_A < 500 \, \mathrm{GeV}$. Without an HLT $b$--trigger, no hope is left for this decay channel.

\subsection{Conclusion}
In this paper the prospects have been presented to discover, in CMS at low luminosity, a heavy charged MSSM Higgs boson in the $H^{\pm} \to tb$ decay channel, asking for three $b$--tagged jets. The latest signal cross section values were used, along with a matrix element simulation of the $t\bar{t}b/t\bar{t}j$ background. This analysis includes HLT acceptances. The background was rejected with a factor 2\,600, while the signal efficiency ranged from 0.6\% to 1.0\% for $250 \, \mathrm{GeV} < m_A < 500 \, \mathrm{GeV}$. Discovery contours were constructed, and the effects of systematic uncertainties on the background were investigated. An uncertainty $\epsilon$ of at least 10\% on the background level was estimated, starting from data or theoretical calculations. For $\epsilon = 0.03$, however, the reach is limited to $\tan \beta > 80$. Therefore, no visibility for this channel is left in the MSSM parameter space.

}

%% file: gunion.tex
{
\noindent
{\Large \bf J. NMSSM Higgs Discovery at the LHC} \\[0.5cm]
{\it U.\,Ellwanger, J.F.\,Gunion, C.\,Hugonie and S.\,Moretti}

%\vspace{1em}
\def\cnone{\widetilde \chi_1^0}
\def\cpone{\widetilde \chi_1^+}
\def\cmone{\widetilde \chi_1^-}
\def\mcnone{m_{\cnone}}
\def\mcpmone{m_{\widetilde\chi_1^{\pm}}}
\def\tev{~{\rm TeV}}
\def\gev{~{\rm GeV}}
\def\mgut{M_{U}}
\def\what{\widehat}
\def\hpm{H^{\pm}}
\def\ie{{\it i.e.}}
\def\anti{\overline}
\def\br{BR}
\def\gam{\gamma}

\begin{abstract}
We demonstrate that Higgs discovery at the LHC is possible
in the context of the NMSSM even for those scenarios such
that the only strongly produced Higgs boson is a very SM-like
CP-even scalar
which decays almost entirely
to a pair of relatvely light CP-odd states.
In combination with other search channels, we are on
the verge of demonstrating that detection of at least
one of the NMSSM Higgs bosons is guaranteed at the LHC
for accumulated luminosity of $300~{\rm fb}^{-1}$.
\end{abstract}

\section{Introduction}

One of the most attractive supersymmetric models
is the Next to Minimal Supersymmetric Standard
Model (NMSSM) (see
\cite{Ellis:1989er,Ellwanger:2001iw} and
references therein) which extends the MSSM by the
introduction of just one singlet superfield, $\what
S$. When the scalar component of $\what S$
acquires a TeV scale vacuum expectation value (a
very natural result in the context of the model),
the superpotential term $\what S \what H_u \what
H_d$ generates an effective $\mu\what H_u \what
H_d$ interaction for the Higgs doublet
superfields.  Such a term is essential for
acceptable phenomenology. No other SUSY model
generates this crucial component of the
superpotential in as natural a fashion. Thus, the
phenomenological implications of the NMSSM at
future accelerators should be considered very
seriously.  One aspect of this is the fact that
the $h,H,A,\hpm$ Higgs sector of the MSSM is
extended so that there are three CP-even Higgs
bosons ($h_{1,2,3}$, $m_{h_1}<m_{h_2}<m_{h_3}$),
two CP-odd Higgs bosons ($a_{1,2}$,
$m_{a_1}<m_{a_2}$) (we assume that CP is not
violated in the Higgs sector) and a charged Higgs
pair ($h^\pm$). An important question is then the
extent to which the no-lose theorem for MSSM Higgs
boson discovery at the LHC (after LEP constraints)
is retained when going to the NMSSM; \ie\ is the
LHC guaranteed to find at least one of the
$h_{1,2,3}$, $a_{1,2}$, $h^\pm$? The first exploration
of this issue appeared in \cite{Gunion:1996fb},
with the conclusion that for substantial portions
of parameter space the LHC would be unable to
detect any of the NMSSM Higgs bosons.
Since then, there have been improvements in many
of the detection modes and the addition of
new ones. These will be summarized below
and the implications reviewed.  However, these
improvements and additions do not address the 
possibly important $h\to aa$ type decays that could
suppress all other types of signals \cite{Gunion:1996fb,Dobrescu:2000jt}.

One of the key ingredients in the no-lose theorem
for MSSM Higgs boson discovery is the fact that
relations among the Higgs boson masses are such
that decays of the SM-like Higgs boson to $AA$ are
only possible if $m_A$ is quite small, a region
that is ruled out by LEP by virtue of the fact
that $Z\to hA$ pair production was not detected
despite the fact that the relevant coupling is
large for small $m_A$.  In the NMSSM, the lighter
Higgs bosons, $h_1$ or $h_2$, can be SM-like (in
particular being the only Higgs with substantial
$WW/ZZ$ coupling) without the $a_1$ necessarily
being heavy.  In addition, this situation is not
excluded by LEP searches for $e^+e^-\to Z^*\to
h_{1,2}a_1$ since, in the NMSSM, the $a_1$ can
have small $Zh_2 a_1$ ($Zh_1 a_1$) coupling when
$h_1$ ($h_2$) is SM-like. [In addition, sum rules
require that the $Zh_1 a_1$ ($Zh_2 a_1$) coupling
is small when the $h_1WW$ ($h_2WW$) couplings are
near SM strength.]  As a result, NMSSM parameters that are
not excluded by current data can be chosen so that
the $h_{1,2}$ masses are moderate in size ($\sim
100-130$~GeV) and the $h_1\to a_1a_1$ or $h_2\to
a_1a_1$ decays are dominant.  Dominance of such
decays falls outside the scope of the usual
detection modes for the SM-like MSSM $h$ on which
the MSSM no-lose LHC theorem largely relies.

In Ref.~\cite{Ellwanger:2001iw}, a partial no-lose
theorem for NMSSM Higgs boson discovery at the LHC
was established.  In particular, it was shown that
the LHC would be able to detect at least one of
the Higgs bosons (typically, one of the lighter
CP-even Higgs states) throughout the full
parameter space of the model, excluding only those
parameter choices for which there is sensitivity
to the model-dependent decays of Higgs bosons to
other Higgs bosons and/or superparticles.  Here,
we will address the question of whether or not
this no-lose theorem can be extended to those
regions of NMSSM parameter space for which Higgs
bosons can decay to other Higgs bosons.  We find
that the parameter choices such that the
``standard'' discovery modes fail {\it would}
allow Higgs boson discovery if detection of $h\to
aa$ decays is possible. (When used generically,
the symbol $h$ will now refer to $h=h_1$, $h_2$ or
$h_3$ and the symbol $a$ will refer to $a=a_1$ or
$a_2$).  Detection of $h\to aa$ will be difficult
since each $a$ will decay primarily to 
$b\anti b$ (or 2 jets if $m_a<2m_b$),
$\tau^+\tau^-$, and, possibly, $\cnone\cnone$,
yielding final states that will typically have
large backgrounds at the LHC.

In \cite{Ellwanger:2001iw} we scanned the
parameter space, removing parameter choices ruled
out by constraints from LEP on Higgs boson
production, $e^+ e^- \to Z h$ or $e^+ e^- \to h a$
\cite{LEPLEPHA}, and eliminating parameter choices
for which one Higgs boson can decay to two other
Higgs bosons or a vector boson plus a Higgs boson.
For the surviving regions of parameter space, we
estimated the statistical significances
($N_{SD}=S/\sqrt B$) for all Higgs boson detection
modes so far studied at the 
LHC~\cite{CMS,unknown:1999fr,Zeppenfeld:2000td,Zeppenfeld:2002ng}.
%,ATLAS,6.2r,Zeppenfeld:2002ng}. 
These are (with $\ell=e,\mu$)

1) $g g \to h/a \to \gamma \gamma$;\par
2) associated $W h/a$ or $t \bar{t} h/a$ production with 
$\gamma \gamma\ell^{\pm}$ in the final state;\par
3) associated $t \bar{t} h/a$ production with $h/a \to b \bar{b}$;\par
4) associated $b \bar{b} h/a$ production with $h/a \to \tau^+\tau^-$;\par
5) $g g \to h \to Z Z^{(*)} \to$ 4 leptons;\par
6) $g g \to h \to W W^{(*)} \to \ell^+ \ell^- \nu \bar{\nu}$;\par
7) $W W \to h \to \tau^+ \tau^-$;\par
8) $W W \to h\to W W^{(*)}$.\par

\noindent
For an integrated luminosity of $300~{\rm
  fb}^{-1}$ at the LHC, all the surviving points
yielded $N_{SD}>10$ after combining all modes,
including the $W$-fusion modes. Thus, NMSSM Higgs
boson discovery by just one detector with
$L=300~{\rm fb}^{-1}$ is essentially guaranteed
for those portions of parameter space for which
Higgs boson decays to other Higgs bosons or
supersymmetric particles are kinematically
forbidden.

In this work, we investigate the complementary
part of the parameter space, where {\it at least
  one} Higgs boson decays to other Higgs bosons.
To be more precise, we require at least one of the
following decay modes to be kinematically allowed:
\begin{eqnarray}
& i) \ h \to h' h' \; , \quad ii) \ h \to a a \; , \quad iii) \ h \to h^\pm
h^\mp \; , \quad iv) \ h \to a Z \; , \nonumber \\
& v) \ h \to h^\pm W^\mp \; , \quad vi) \ a' \to h a \; , \quad vii) \ a \to h
Z \; , \quad viii) \ a \to h^\pm W^\mp \; .
\end{eqnarray}
After searching those regions of parameter space
for which one or more of the decays $ i) - viii)$
is allowed, we found that the only subregions for
which discovery of a Higgs boson in modes 1) -- 8)
was not possible correspond to NMSSM parameter
choices for which (a) there is a light CP-even
Higgs boson with substantial doublet content that
decays mainly to two still lighter CP-odd Higgs
states, $h\to aa$, and (b) all the other Higgs
states are either dominantly singlet-like,
implying highly suppressed production rates, or
relatively heavy, decaying to $t\anti t$, to one
of the ``difficult'' modes $i) - viii)$ or to a
pair of sparticles. In such cases, the best
opportunity for detecting at least one of the
NMSSM Higgs bosons is to employ $WW\to h$
production and develop techniques for extracting a
signal for the $h\to aa\to jj\tau^+\tau^-$ 
(including $jj=b\anti b$) process.  We
have performed a detailed simulation of 
the $aa\to jj \tau^+\tau^-$ 
final state and find that its detection may be possible
after accumulating $300~{\rm fb}^{-1}$ in both the
ATLAS and CMS detectors.  

\section{The model and scanning procedures}

We consider the simplest version of the NMSSM
\cite{Ellis:1989er}, where the term $\mu \widehat
H_1 \widehat H_2$ in the superpotential of the
MSSM is replaced by (we use the notation $\widehat
A$ for the superfield and $A$ for its scalar
component field)
\begin{equation}\label{2.1r}
\lambda \widehat H_1 \widehat H_2 \widehat S\ + \ \frac{\kappa}{3} \widehat S^3
\ \ ,
\end{equation}
\noindent so that the superpotential is scale invariant. 
We make no assumption on ``universal'' soft terms.
Hence, the five soft supersymmetry breaking terms
\begin{equation}\label{2.2r}
m_{H_1}^2 H_1^2\ +\ m_{H_2}^2 H_2^2\ +\ m_S^2 S^2\ +\ \lambda
A_{\lambda}H_1 H_2 S\ +\ \frac{\kappa}{3} A_{\kappa}S^3
\end{equation}
\noindent are considered as independent. 
The masses and/or couplings of sparticles will
be such that their contributions to the
loop diagrams inducing Higgs boson production by
gluon fusion and Higgs boson decay into $\gamma
\gamma$ are negligible. 
In the gaugino sector, we chose $M_2=1\tev$ (at low scales).
Assuming universal gaugino masses at the coupling
constant unification scale,
this yields $M_1\sim 500\gev$ and $M_3\sim 3\tev$.
In the squark sector, as particularly relevant
for the top squarks which
appear in the radiative corrections to the Higgs
potential, we chose the soft masses $m_Q = m_T
\equiv M_{susy}= 1$ TeV, and varied the stop
mixing parameter
\begin{equation}\label{2.4r}
X_t \equiv 2 \ \frac{A_t^2}{M_{susy}^2+m_t^2} \left ( 1 -
\frac{A_t^2}{12(M_{susy}^2+m_t^2)} \right ) \ .
\end{equation} 
\noindent As in the MSSM, 
the value $X_t = \sqrt{6}$ -- so called maximal
mixing -- maximizes the radiative corrections to
the Higgs boson masses, and we found that it leads
to the most challenging points in the parameter
space of the NMSSM.  We adopt the convention
$\lambda,\kappa > 0$, in which $\tan\beta$ can
have either sign. We require $|\mu_{\rm eff}|\ >\ 
100$~GeV; otherwise a light chargino would have
been detected at LEP. The only possibly light SUSY particle
will be the $\cnone$.  A light $\cnone$ is a frequent
characteristic of parameter choices that yield a 
light $a_1$.

We have performed a numerical scan over the free
parameters.  For each point, we computed the
masses and mixings of the CP-even and CP-odd Higgs
bosons, $h_i$ ($i=1,2,3$) and $a_j$ ($j=1,2$),
taking into account radiative corrections up to
the dominant two loop terms, as described in
\cite{Ellwanger:1999ji}.  We eliminated parameter
choices excluded by LEP
constraints~\cite{LEPLEPHA} on $e^+ e^- \to Z h_i$
and $e^+ e^- \to h_i a_j$. The latter provides an
upper bound on the $Zh_ia_j$ reduced coupling,
$R'_{ij}$, as a function of $m_{h_i}+m_{a_j}$ for
$m_{h_i} \simeq m_{a_j}$.  Finally, we calculated
$m_{h^\pm}$ and required $m_{h^\pm} > 155$~GeV, so
that $t \to h^\pm b$ would not be seen.

In order to probe the complementary part of the
parameter space as compared to the scanning of
Ref. \cite{Ellwanger:2001iw}, we required that at
least one of the decay modes $i) - viii)$ is
allowed.  For each Higgs state, we calculated all
branching ratios including those for modes $i) -
viii)$, using an adapted version of the FORTRAN
code HDECAY \cite{Djouadi:1998yw}. We then
estimated the expected statistical significances
at the LHC in all Higgs boson detection modes 1)
-- 8) by rescaling results for the SM Higgs boson
and/or the MSSM $h, H$ and/or $A$. The
rescaling factors are determined by $R_i$, $t_i$
and $b_i=\tau_i$, the ratios of the $VVh_i$,
$t\anti t h_i$ and $b\anti b h_i,\tau^+\tau^- h_i$
couplings, respectively, to those of a SM Higgs
boson.  Of course $|R_i| < 1$, but $t_i$ and $b_i$
can be larger, smaller or even differ in sign with
respect to the SM. For the CP-odd Higgs bosons,
$R_i'=0$ at tree-level; $t'_j$ and $b'_j$ are the
ratios of the $i\gamma_5$ couplings for $t\bar{t}$
and $b\bar{b}$, respectively, relative to SM-like
strength.  A detailed discussion of the procedures
for rescaling SM and MSSM simulation results for
the statistical significances in channels 1) -- 8)
will appear elsewhere.

\begin{table}[p]
\begin{center}
\footnotesize
\vspace*{-.2in}
\hspace*{-.5in}
\begin{tabular} {|l|l|l|l|l|l|l|} 
\hline
Point Number & 1 & 2 & 3 & 4 & 5 & 6  \\
\hline \hline
Bare Parameters &\multicolumn{6}{c|}{} \\
\hline
$\lambda$            & 0.2872 & 0.2124 & 0.3373 & 0.3340 & 0.4744 & 0.5212 \\
\hline
$\kappa$             & 0.5332 & 0.5647 & 0.5204 & 0.0574 & 0.0844 & 0.0010 \\
\hline
$\tan\beta$          &   2.5  &   3.5  &   5.5  &    2.5 &    2.5 & 2.5 \\
\hline
$\mu_{\rm eff}~({\rm GeV})$&    200 &    200 &    200 &    200 &    200 & 200 \\
\hline
$A_{\lambda}~({\rm GeV})$  &    100 &      0 &     50 &    500 &    500 & 500 \\
\hline
$A_{\kappa}~({\rm GeV})$   &      0 &      0 &      0 &      0 &      0 & 0 \\
\hline \hline
CP-even Higgs Boson Masses and Couplings &\multicolumn{6}{c|}{} \\
\hline \hline
$m_{h_1}$~(GeV)      &    115 &    119 &    123 &     76 &     85 &  51\\
\hline
$R_1 $               &   1.00 &   1.00 &  -1.00 &   0.08 &   0.10 &  -0.25\\
\hline
$t_1 $               &   0.99 &   1.00 &  -1.00 &   0.05 &   0.06 &  -0.29\\
\hline
$b_1 $               &   1.06 &   1.05 &  -1.03 &   0.27 &   0.37 &  0.01\\
\hline
Relative 
gg Production Rate   &   0.97 &   0.99 &   0.99 &   0.00 &   0.01 &  0.08\\
\hline
$\br(h_1\to 
b\anti b)$           &   0.02 &   0.01 &   0.01 &   0.91 &   0.91 &  0.00\\
\hline
$\br(h_1\to 
\tau^+\tau^-)$      &   0.00 &   0.00 &   0.00 &   0.08 &   0.08 &  0.00\\
\hline
$\br(h_1\to a_1 a_1)$&   0.98 &   0.99 &   0.98 &   0.00 &   0.00 &  1.00\\
\hline \hline

$m_{h_2}$~(GeV)      &    516 &    626 &    594 &    118 &    124 &  130\\
\hline
$R_2 $               &  -0.03 &  -0.01 &   0.01 &  -1.00 &  -0.99 &  -0.97\\
\hline
$t_2 $               &  -0.43 &  -0.30 &  -0.10 &  -0.99 &  -0.99 &  -0.95\\
\hline
$b_2 $               &   2.46 &  -3.48 &   3.44 &  -1.03 &  -1.00 &  -1.07\\
\hline
Relative
gg Production Rate   &   0.18 &   0.09 &   0.01 &   0.98 &   0.99 &  0.90\\
\hline
$\br(h_2\to 
b\anti b)$           &   0.01 &   0.04 &   0.04 &   0.02 &   0.01 &  0.00\\
\hline
$\br(h_2\to 
\tau^+\tau^-)$      &   0.00 &   0.01 &   0.00 &   0.00 &   0.00 &  0.00\\
\hline
$\br(h_2\to a_1 a_1)$&   0.04 &   0.02 &   0.83 &   0.97 &   0.98 &  0.96\\
%\hline
%$\br(h_2\to h_1 h_1)$&   0.01 &   0.01 &   0.00 &   0.00 &   0.00 &  0.04\\
\hline \hline

$m_{h_3}$~(GeV)      &    745 &   1064 &    653 &    553 &    554 &  535\\
\hline \hline

CP-odd Higgs Boson Masses and Couplings &\multicolumn{6}{c|}{} \\
%and Couplings &\multicolumn{6}{c|}{} \\
\hline \hline
$m_{a_1}$~(GeV)      &     56 &      7 &     35 &     41 &     59 &  7\\
\hline
$t_1' $               &   0.05 &   0.03 &   0.01 &  -0.03 &  -0.05 &  -0.06\\
\hline
$b_1' $               &   0.29 &   0.34 &   0.44 &  -0.20 &  -0.29 &  -0.39\\
\hline
Relative
gg Production Rate   &   0.01 &   0.03 &   0.05 &   0.01 &   0.01 &  0.05\\
\hline
$\br(a_1\to 
b\anti b)$           &   0.92 &   0.00 &   0.93 &   0.92 &   0.92 &  0.00\\
\hline
$\br(a_1\to 
\tau^+\tau^-)$      &   0.08 &   0.94 &   0.07 &   0.07 &   0.08 &  0.90\\
\hline \hline

$m_{a_2}$~(GeV)      &    528 &    639 &    643 &    560 &    563 &  547\\
\hline 
Charged Higgs  
Mass (GeV)           &    528 &    640 &    643 &    561 &    559 &  539\\
\hline\hline
Most Visible of the LHC Processes 1)-8) &  2 ($h_1$) &  2 ($h_1$) &  8
                  ($h_1$) &  2 ($h_2$) &  8 ($h_2$)  &  8 ($h_2$)\\
\hline            
$N_{SD}=S/\sqrt B$ 
Significance of this process at $L=$300~${\rm fb}^{-1}$
                     &   0.48 &   0.26 &   0.55 &   0.62 &  0.53  & 0.16\\
\hline
\hline
$N_{SD}(L=300~{\rm fb}^{-1})$  for
$WW\to h\to aa\to jj \tau^+\tau^-$ at LHC & 50 &  22 &  69 &  63&  62 &  21 \\
\hline
\end{tabular}
\end{center}
\vspace*{-.2in}\caption{\label{tpoints}\footnotesize
Properties of selected scenarios that could escape detection
at the LHC. In the table, $R_i=g_{h_i VV}/g_{h_{SM} VV}$, 
$t_i=g_{h_i t\anti t}/g_{h_{SM} t\anti t}$ and $b_i=g_{h_ib\anti b}/g_{h_{SM} b\anti b}$ 
for $m_{h_{SM}}=m_{h_i}$; $t_1'$ and $b_1'$
are the $i\gam_5$ couplings of $a_1$ 
to $t\anti t$ and $b\anti b$ normalized
relative to the scalar 
$t\anti t$ and $b\anti b$ SM Higgs couplings.
We also give the $gg$ fusion production rate ratio,
$gg\to h_i/gg\to h_{SM}$, for $m_{h_{SM}}=m_{h_i}$. 
Important absolute branching
ratios are displayed. For points 2 and 6, the decays
$a_1\to jj$ ($j\neq b$) have 
$\br(a_1\to jj)\simeq 1-\br(a_1\to \tau^+\tau^-)$.
For the heavy $h_3$ and $a_2$, we give only their masses.
For all points 1 -- 6, the statistical
significances for the detection of any 
Higgs boson in any of the channels 1) --
8) are tiny; the next-to-last row gives their maximum 
together with the process number and 
the corresponding Higgs state.
The last row gives the statistical significance
of the new $WW\to h \to aa\to jj \tau^+\tau^-$
[$h=h_1$ ($h=h_2$) for points 1--3 (4--6)] LHC
signal explored here. 
}
\end{table}

In our set of randomly scanned points, we selected
those for which all the statistical significances
in modes 1) -- 8) are below $5\sigma$. We obtained
a lot of points, all with similar characteristics.
Namely, in the Higgs spectrum, we always have a
very SM-like CP-even Higgs boson with a mass
between 115 and 135~GeV ({\it i.e.} above the LEP
limit), which can be either $h_1$ or $h_2$, with a
reduced coupling to the gauge bosons $R_1 \simeq
1$ or $R_2\simeq 1$, respectively. This state
decays dominantly to a pair of (very) light CP-odd
states, $a_1a_1$, with $m_{a_1}$ between 5 and
65~GeV.  The singlet component of $a_1$ cannot be
dominant if we are to have a large $h_1 \to a_1
a_1$ or $h_2\to a_1a_1$ branching ratio when the
$h_1$ or $h_2$, respectively, is the SM-like Higgs
boson.  Further, when the $h_1$ or $h_2$ is very
SM-like, one has small $Zh_1a_1$ or $Zh_2a_1$ coupling,
respectively, so that $e^+ e^- \to
h_1 a_1$ or $e^+e^-\to h_2 a_1$ associated
production places no constraint on the light
CP-odd state at LEP. We have selected six
difficult benchmark points, displayed in
Table~\ref{tpoints}.  These are such that
$a_1\to\cnone\cnone$ decays are negligible or
forbidden.  (Techniques for cases such that 
$\cnone\cnone$ decay modes are important
are under development.)
For points 1 -- 3, $h_1$ is the
SM-like CP-even state, while for points 4 -- 6 it
is $h_2$. We have selected the points so that there is
some variation in the 
$h_{1,2}$ and $a_1$ masses. The
main characteristics of the benchmark points are
displayed in Table~\ref{tpoints}. Note the large
$\br(h\to a_1 a_1)$ of the SM-like $h$ ($h=h_1$
for points 1 -- 3 and $h=h_2$ for points 4 --6).
For points 4 -- 6, with $m_{h_1}<100\gev$, the
$h_1$ is mainly singlet.  As a result, the $Zh_1a_1$
coupling is very small, implying no LEP constraints on the
$h_1$ and $a_1$ from $e^+e^-\to h_1 a_1$
production.

We note that in the case of the points 1 -- 3, the
$h_2$ would not be detectable either at the LHC or
at a Linear Collider (LC). 
For points 4 -- 6, the $h_1$, though
light, is singlet in nature and would not be
detectable.  Further, the $h_3$ or $a_2$ will only
be detectable for points 1 -- 6 if a super high
energy LC is eventually built so that $e^+e^-\to
Z\to h_3 a_2$ is possible.  Thus, we will focus on
searching for the SM-like $h_1$ ($h_2$) for points
1 -- 3 (4 -- 6) using the dominant $h_1(h_2)\to
a_1a_1$ decay mode.

In the case of points 2 and 6, the $a_1\to
\tau^+\tau^-$ decays are dominant. The final state
of interest will be $jj\tau^+\tau^-$, where the
$jj$ actually comes primarily from
$a_1a_1\to\tau^+\tau^-\tau^+\tau^-$ followed by jet
decays of two of the $\tau$'s: $\tau^+\tau^-\to
jj+\nu's$.  (The contribution from direct $a_1\to jj$ decays
to the $jj\tau^+\tau^-$ final state is relatively
small for points 2 and 6.)
In what follows, when we speak of
$\tau^+\tau^-$, we refer to those $\tau$'s that
are seen in the $\tau^+\tau^-\to
\ell^+\ell^-+\nu's$ final state ($\ell=e,\mu$).
  For points 1 and 3
-- 5 $\br(a_1\to b\anti b)$ is substantial.  The
relevant final state is $b\anti b \tau^+\tau^-$.
Nonetheless, we begin with a study of the backgrounds and
signals without requiring $b$-tagging.  
With our latest cuts, 
we will see that $b$-tagging is not necessary 
to overcome the apriori large Drell-Yan
$\tau^+\tau^-$+jets background.  It is eliminated
by stringent cuts for finding the highly energetic
forward / backward
jets characteristic of the $WW$ the fusion  process.
As a result, we will find good signals for all
6 of our points.

\def\cO#1{{\cal{O}}\left(#1\right)}
\def\nn {\nonumber}
\newcommand{\bn}{\begin{enumerate}}
\newcommand{\en}{\end{enumerate}}
\newcommand{\bc}{\begin{center}}
\newcommand{\ec}{\end{center}}
\newcommand{\ul}{\underline}
\newcommand{\ol}{\overline}
\newcommand{\ar}{\rightarrow}
\newcommand{\sm}{${\cal {SM}}$}
\newcommand{\as}{\alpha_s}
\newcommand{\aem}{\alpha_{em}}
\newcommand{\ycut}{y_{\mathrm{cut}}}
\newcommand{\susy}{{{SUSY}}}
\newcommand{\Dir}{\kern -6.4pt\Big{/}}
\newcommand{\Dirin}{\kern -10.4pt\Big{/}\kern 4.4pt}
\newcommand{\DDir}{\kern -10.6pt\Big{/}}
\newcommand{\DGir}{\kern -6.0pt\Big{/}}
\def\Ecm{\ifmmode{E_{\mathrm{cm}}}\else{$E_{\mathrm{cm}}$}\fi}
\def\gluino{\ifmmode{\mathaccent"7E g}\else{$\mathaccent"7E g$}\fi}
\def\photino{\ifmmode{\mathaccent"7E \gamma}\else{$\mathaccent"7E \gamma$}\fi}
\def\mgluino{\ifmmode{m_{\mathaccent"7E g}}
             \else{$m_{\mathaccent"7E g}$}\fi}
\def\taugluino{\ifmmode{\tau_{\mathaccent"7E g}}
             \else{$\tau_{\mathaccent"7E g}$}\fi}
\def\mphotino{\ifmmode{m_{\mathaccent"7E \gamma}}
             \else{$m_{\mathaccent"7E \gamma}$}\fi}
\def\ML{\ifmmode{{\mathaccent"7E M}_L}
             \else{${\mathaccent"7E M}_L$}\fi}
\def\MR{\ifmmode{{\mathaccent"7E M}_R}
             \else{${\mathaccent"7E M}_R$}\fi}

\def\lsim{\buildrel{\scriptscriptstyle <}\over{\scriptscriptstyle\sim}}
\def\gsim{\buildrel{\scriptscriptstyle >}\over{\scriptscriptstyle\sim}}
\def\Jnl #1#2#3#4 {#1 {\bf #2} (#3) #4}
\def\NPB {{\rm Nucl. Phys.} {\bf B}}
\def\PLB {{\rm Phys. Lett.}  {\bf B}}
\def\PRL {\rm Phys. Rev. Lett.}
\def\PRD {{\rm Phys. Rev.} {\bf D}}
\def\ZPC {{\rm Z. Phys.} {\bf C}}
\def\EPJC {{\rm Eur. Phys. J.} {\bf C}}
\def\Ord{\lower .7ex\hbox{$\;\stackrel{\textstyle <}{\sim}\;$}}
\def\OOrd{\lower .7ex\hbox{$\;\stackrel{\textstyle >}{\sim}\;$}}
\def\eps{\epsilon}

In principle, one could explore final states other than 
$b\anti b \tau^+\tau^-$ (or $jj\tau^+\tau^-$ for points 2
and 6). However, all other channels will be much more
problematical at the LHC. A $4b$-signal would
be burdened by a large QCD background even after
implementing $b$-tagging.  A $4j$-signal would be
completely swamped by QCD background.  Meanwhile,
the $4\tau$-channel (by which we mean that all
taus decay leptonically) would not allow one to
reconstruct the $h_1,h_2$ resonances.  

In the case of the $2b2\tau$ (or
$2j2\tau$) signature, we identify the $\tau$'s
through their leptonic decays to electrons
and muons. Thus, they will yield some amount of
missing (transverse) momentum, $p_{\rm miss}^T$.
This missing transverse momentum can be projected
onto the visible $e,\mu$-momenta in an attempt to
reconstruct the parent $\tau$-direction.

\section{Monte Carlo Results for the LHC}

Let us now focus on the $WW\to h\to aa$
channel that we believe provides 
the best hope for Higgs detection in these
difficult NMSSM cases.  (We reemphasize that the
$h_1$ [cases 1 -- 3] or $h_2$ [cases 4 -- 6] has
nearly full SM strength coupling to $WW$.)
The $b\anti b\tau^+\tau^-$ (or
$2j\tau^+\tau^-$, for points 2 and 6) final state
of relevance is complex and subject to large
backgrounds, and the $a_1$ masses of interest are
very modest in size.  In order to extract the $WW$ fusion
$2j2\tau$ NMSSM Higgs boson signature, it is crucial
to strongly exploit forward
and backward jet tagging on the light quarks
emerging after the double $W$-strahlung preceding
$WW$-fusion.  We also require two additional central
jets (from one of the $a$'s) and two opposite sign
central leptons ($\ell=e,\mu$) coming from the 
the $\tau^+\tau^-$ emerging from the decay of the other
$a$. By imposing stringent
 forward / backward jet tagging cuts, we remove the
otherwise very large background from Drell-Yan
$\tau^+\tau^-+jets$ production. 
In the end, the most important background is due
to $t\anti t$ production and decay via the purely
SM process, $gg\to t\bar t\to b\bar b W^+W^-\to
b\bar b \tau^+\tau^- + p_{\rm miss}^T,$ in
association with forward and backward jet radiation.

We have employed numerical simulations based on a version of
{\tt HERWIG v6.4}~\cite{Moretti:2002eu,Corcella:2001wc,Corcella:2000bw}
modified to allow for appropriate NMSSM couplings
and decay rates. Calorimeter emulation
was performed using the {\tt GETJET} code
\cite{GETJET}. 
Since the $a_1$ will not have been detected
  previously, we must assume a value for
  $m_{a_1}$.  In dealing with 
actual experimental data, it will be necessary to
  repeat the analysis for densely spaced $m_{a_1}$
  values and look for the $m_{a_1}$ choice that
  produces the best signal.
 We look among the central jets for the
  combination with invariant mass $M_{jj}$ closest
  to $m_{a_1}$. In Fig.~\ref{MH}, we show
the $M_{jj\tau^+\tau^-}$ invariant mass distribution
obtained after cuts, but before $b$-tagging
or inclusion of $K$ factors 
--- the plot presented assumes
that we have hit on the correct $m_{a_1}$ choice.

\begin{figure}[h]
\begin{center}
\centerline{$~~~~~~~~~~~~~~~$LHC, $\sqrt{s_{pp}}=14$ TeV}
\centering\epsfig{file=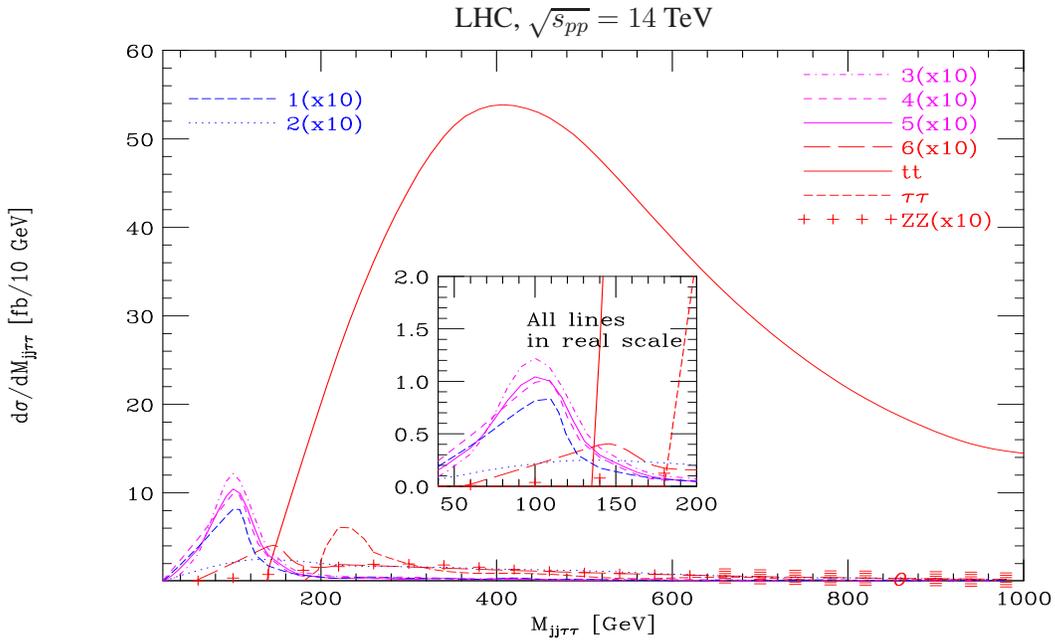,angle=90,height=8cm,width=14cm}

\vspace*{1.0truecm}

\noindent
\vspace{-1.0cm}
\caption{\footnotesize We plot $d\sigma/dM_{jj\tau^+\tau^-}$ [fb/10~GeV] vs $M_{jj\tau^+\tau^-}$~[GeV]
for signals and backgrounds after basic
event selections, but before $b$ tagging. 
The lines corresponding to points 4 and 5
are visually indistinguishable. No $K$ factors are included.
}
\label{MH}
\end{center}
\end{figure}

The selection strategy adopted is 
a more refined (as regards
forward / backward jet tagging)
version of that summarized in \cite{Ellwanger:2003jt}.
It is clearly
efficient in reconstructing the $h_1$ (for points
1--3) and $h_2$ (for points 4--6) masses from the
$jj \tau^+\tau^-$ system, as one can appreciate by
noting the peaks appearing  at
$M_{jj\tau^+\tau^-}\approx100$~GeV. In contrast,
the heavy Higgs resonances at $m_{h_2}$ for points
1--3 and the rather light resonances at $m_{h_1}$
for points 4--6 (recall Table~\ref{tpoints}) do
not appear, the former mainly because of the very
poor production rates and the latter due to the
fact that either the $h_1\to a_1 a_1$ decay mode
is not open (points 4, 5) or -- if it is -- the
jets and $e/\mu$-leptons eventually emerging
from the $a_1$ decays are too soft to pass the
acceptance cuts (point 6, for which
$m_{a_1}=7$~GeV and $m_{h_1}=51$~GeV).  For all
six NMSSM setups, the Higgs resonance produces a
bump below the end of the low mass tail of the
$t\bar t$ background (see the insert in 
Fig.~\ref{MH}).  Note how small the DY
$\tau^+\tau^-$ background is after strong
forward / backward jet tagging.  Since the main
surviving background is from $t\anti t$ production,
$b$ tagging is not helpful.  For points 2 and 6,
for which the signal has no $b$'s in the final state,
anti-$b$-tagging might be useful, but has not been
considered here.

To estimate $S/\sqrt B$, we assume $L=300~{\rm
  fb}^{-1}$, a $K$ factor of 1.1 for the $WW$
fusion signal and $K$ factors of 1, 1 and 1.6 for
the DY $\tau^+\tau^-$, $ZZ$ production and $t\anti
t$ backgrounds, respectively.  (These $K$ factors
are not included in the plot of Fig.~\ref{MH}.)
We sum events over the region $40\leq
M_{jj\tau^+\tau^-}\leq 150$~GeV.  
(Had we only included masses below $130$~GeV,
we would have had no $t\anti t$ background,
and the $S/\sqrt B$ values would be enormous.  However,
we are concerned that this absence of $t\anti t$
background below $130\gev$ might be a reflection
of limited Monte Carlo statistics.  As a result
we have taken the more conservative approach of
at least including the first few bins for which
our Monte Carlo does predict some $t\anti t$ background.)

For points 1, 2, 3,
4, 5 and 6, we obtain signal rates of about $S=1636$, 702,
  2235,  2041, 2013, and  683, respectively.
The $t\anti t$+jets background rate is $B_{tt}\sim
795$. The $ZZ$ background rate is $B_{ZZ}\sim 6$.
The DY $\tau^+\tau^-$ background rate is
negligible. (We are continuing to increase our statistics
to get a fully reliable estimate.)
The resulting $N_{SD}=S/\sqrt B$ values for points 1-6
are 50,  22, 69,  63,  62, and  21, respectively.
The smaller values for points 2 and 6 are simply
a reflection of the difficulty of
isolating and reconstructing
the two jets coming from the decay of a very light $a_1$.
Overall, these preliminary results are very encouraging
and suggest that a no-lose theorem for NMSSM Higgs
detection at the LHC is close at hand.

\section{Conclusions}

In summary, we have obtained a statistically very significant
LHC signal in the $jj\tau^+\tau^-$ final
state of $WW$ fusion for cases in which the NMSSM parameters 
are such that
the most SM-like of the CP-even Higgs bosons, $h$,
is relatively light and decays primarily to a pair
of CP-odd Higgs states, $h\to aa$ with $a\to b\anti b,\tau^+\tau^-$ if $m_a>2m_b$ or $a\to jj,\tau^+\tau^-$
if $m_a<2m_b$. 
 The statistical significances are (at least)
of order 50 to 70 for points with $m_a>2m_b$
and of order 20 for points with $m_a<2m_b$.
These high significances were obtained by
imposing stringent cuts requiring
highly energetic forward / backward jets
in order to isolate the $WW$ fusion signal process
from backgrounds such as DY $\tau^+\tau^-$ pair 
production. Still, this signal will
be the only evidence for Higgs bosons at the LHC. 
A future LC will probably be 
essential in order to confirm that
the enhancement seen at the LHC really does
correspond to a Higgs boson. At the LC, discovery
of a light SM-like $h$ is guaranteed to be
possible in the $Zh$ final state using the recoil
mass technique~\cite{Gunion:2003fd}.  

In the present study, we have not
explored the cases in which the
$a_1\to\cnone\cnone$ decay has a large branching
ratio.  Detecting a Higgs signal in such cases
will require a rather different procedure.
Work on the $WW\to h\to {\rm invisible}$ signal
is in progress~\cite{preparation}.

As we have stressed, for parameter space points of
the type we have discussed here, detection of any
of the other MSSM Higgs bosons is likely to be
impossible at the LHC and is likely to require an
LC with $\sqrt{s_{e^+e^-}}$ above the relevant
thresholds for $h'a'$ production, where $h'$ and
$a'$ are heavy CP-even and CP-odd Higgs bosons,
respectively.  

Although results for the LHC
indicate that Higgs boson discovery will be possible
for the type of situations we have
considered, it is clearly important
to refine and improve the techniques for extracting a
signal. This will almost certainly be possible 
once data is in
hand and the $t\anti t$ background can be more
completely modeled.  

Clearly, if SUSY is
discovered and $WW\to WW$ scattering is found to
be perturbative at $WW$ energies of 1 TeV (and
higher), and yet no Higgs bosons are detected in
the standard MSSM modes, a careful search for the
signal we have considered should have a high
priority.  

Finally, we should remark that the
$h\to aa$ search channel considered here in the
NMSSM framework is also highly relevant for a
general two-Higgs-doublet model, 2HDM.  It is
really quite possible that the most SM-like
CP-even Higgs boson of a 2HDM will decay primarily
to two CP-odd states.  This is possible even if
the CP-even state is quite heavy, unlike the NMSSM
cases considered here. If CP violation is introduced
in the Higgs sector, either at tree-level
or as a result of one-loop corrections (as, for example,
is possible in the MSSM),
$h\to h' h''$ decays will generally be present. The 
critical signal will be the same as
that considered here.

}

%% file: barklow.tex
{
\newcommand{\ee}       {\mbox{$e^+ e^-$}}
\def\ra{\rightarrow}

\noindent
{\Large \bf K. Higgs Coupling Measurements at a 1 TeV Linear
Collider} \\[0.5cm]
{\it T.\,Barklow}

\begin{abstract}
Methods for extracting Higgs boson signals at  a 1~TeV center-of-mass energy
$\ee$ linear collider are described.  In addition, estimates are given for the
accuracy with which branching fractions can be measured for Higgs boson decays
to $b\bar b$, $WW$, $gg$,  and $\gamma \gamma $.
\end{abstract}

\section{Introduction}

The precision measurement of the Higgs boson couplings to fermions and gauge bosons is one of the most important goals of 
an $\ee$ linear collider.  These measurements will distinguish between different models of electroweak symmetry breaking,
and can be used to extract parameters within a specific model, such as supersymmetry.  
Most linear collider Higgs studies have
been made assuming a center-of-mass energy of 0.35 TeV, where the Higgsstrahlung cross-section is not too far from its
peak value for Higgs boson masses less than 250~GeV .
%, and where top quark threshold studies can be performed simultaneously.   
Higgs branching fraction measurements with errors of $2-10$\% can be achieved at $\sqrt{s}=0.35$~TeV for many Higgs decay modes, and 
the total Higgs width can be measured with an accuracy of $5-13$\% if the $\sqrt{s}=0.35$~TeV data is combined with $WW$ fusion 
production at $\sqrt{s}=0.50$~TeV\cite{Aguilar-Saavedra:2001rg}.
These measurement errors are very good, but is it possible to do better?  
%For example, to convert most of the Higgs branching fractions to partial widths one needs the
%total Higgs width, and so it is important to ask if the Higgs total width measurement can be improved. 

In the CLIC study of physics at a 3~TeV $\ee$ linear collider it was recognized  that rare Higgs decay modes such as $h\ra\mu^+\mu^-$ could be observed using
Higgs bosons produced through $WW$ fusion\cite{Battaglia:2001vf, Battaglia:2002gq}.  This is possible because
the cross-section for Higgs production through $WW$ fusion  rises with center-of-mass energy, while the design luminosity
of a linear collider also rises with energy.
%to compensate for the $1/s$ cross-section dependence of $s$-channel proceeses.  
One doesn't have to wait for 
a center-of-mass energy of 3~TeV, however, to take advantage of this situation.   Already at $\sqrt{s}=1$~TeV the cross-section for Higgs boson production through $WW$ fusion
is two to four times larger than the Higgsstrahlung cross-section at $\sqrt{s}=0.35$~TeV, and the linear collider design luminosity is two times larger at $\sqrt{s}=1$~TeV than at $\sqrt{s}=0.35$~TeV\cite{unknown:2003na}.
Table~\ref{tab:higgs-sample_comparison} summarizes the Higgs event rates at $\sqrt{s}=0.35$ and 1~TeV for several Higgs boson masses. 
%and for several values for the initial state positron polarization,
%assuming the integrated luminosity is 1~ab$^{-1}$ at $\sqrt{s}=1$~TeV and 0.5~ab$^{-1}$ at $\sqrt{s}=0.35$~TeV.

\begin{table}[b]
\begin{center}
\caption{Number of inclusive Higgs events assuming an initial state electron polarization of -80\% and integrated luminosities 
of 500 (1000) fb$^{-1}$ for $\sqrt{s}=350$ (1000) GeV.
Effects from beamstrahlung and initial state radiation are included assuming the NLC machine design.  
}
\begin{tabular}{rr|rrrr}
     &  &                  \multicolumn{4}{c}{Higgs Mass (GeV)} \\
 $\sqrt{s}$ (GeV) & $e^+_{\rm pol}$ (\%)  & 120 & 140 & 160 & 200 \\
\hline
& & & & & \\
 350  & 0 &  110280  & 89150    & 69975   &  37385   \\
 350  & +50 &  159115  &  128520   & 100800   &  53775   \\
 1000 & 0 &  386550  &  350690  &  317530 &  259190  \\
 1000 & +50 &  569750  &  516830   & 467900   & 382070   \\

\hline
\end{tabular}
\label{tab:higgs-sample_comparison}
\end{center}
\end{table}

In this report methods for extracting Higgs boson signals at 
a 1~TeV center-of-mass energy $\ee$ linear collider are presented, along with estimates 
of the accuracy with which the Higgs boson cross-section times branching fractions, $\sigma\cdot B_{xx}$, can be measured.  
All results and figures at $\sqrt{s}=1$~TeV assume 1000~fb$^{-1}$ luminosity, -80\% electron polarization, and +50\% positron polarization.

%The
%error estimates for $\sigma\cdot B_{xx}$ at $\sqrt{s}=1000$~GeV are then combined with the expected measurement errors for $B_{bb}$ and $B_{WW}$ at
%$\sqrt{s}=350$~GeV to obtain new errors estimates for Higgs branching fractions and the total Higgs boson decay width.

\section{Event Simulation}

The Standard Model backgrounds from all 0,2,4,6-fermion processes and the top quark-dominated 8-fermion processes are generated 
at the parton level using the WHIZARD Monte Carlo\cite{Kilian:2002cg}. 
%(The 0-fermion processes include reactions such as $\ee\ra \gamma \gamma$.)
  In the case of processes such as 
$\ee\ra \ee f \bar{f}$ the photon flux from real beamstrahlung photons is included along with the photon flux from Weisz\"{a}cker-Williams low-$q^2$ virtual photons.
The production of the Higgs boson and its subsequent decay to $b\bar{b}$ and $\tau^+\tau^-$ is automatically included in WHIZARD in the generation of the 4-fermion
processes
$\ee\ra f\bar{f} b\bar{b}$ and $\ee\ra f\bar{f} \tau^+\tau^-$.  For other Higgs decay modes the WHIZARD Monte Carlo is used to simulate $\ee\ra f\bar{f} h$ and
the decay of the Higgs boson is then simulated using PYTHIA\cite{Sjostrand:2000wi}.
%, with branching fractions calculated by HDECAY.  
The PYTHIA program is also used for final state QED and QCD radiation and for hadronization.
The CIRCE parameterization\cite{Ohl:1997fi} of the
NLC design\cite{unknown:2003na} at $\sqrt{s}=1$~TeV is used to simulate the effects of beamstrahlung.   For
the detector Monte Carlo the SIMDET V4.0 simulation\cite{Pohl:2002vk} of the TESLA detector\cite{Behnke:2001qq} is utilized.

\section{Measurement of $\bf \sigma\cdot B_{xx}$ at $\bf \sqrt{s}=1$~TeV}

Results will be presented for the Higgs decay modes $h\ra b\bar{b},\  WW,\ gg,\ \gamma\gamma$.  The $h\ra c\bar{c}$ decay is not studied 
since a detailed charm-tagging analysis is beyond the scope of this paper;  however it might be interesting for charm-tagging experts 
to pursue this decay mode at $\sqrt{s}=1$~TeV.  The  $h\ra \tau^+\tau^-$ decay is not considered since the neutrinos from the decays of the taus
severely degrade the Higgs mass reconstruction.

%Approximately 90\% of the inclusive Higgs boson production at $ \sqrt{s}=1$~TeV is due to the $WW$ fusion process $\ee\ra \nu_e\bar{\nu_e} h$, assuming 
%an initial state electron
%polarization of -80\%.
%; for 0 electron polarization the $WW$ fusion fraction drops to 83\%.  
%Such events have large missing energy,
%large missing transverse momentum, and a large boost in the forward direction.   

Higgs events are preselected by requiring that there be no isolated electron or muon, and that the angle of the thrust axis $\theta_{\rm thrust}$,
visible energy $E({\rm visible})$, and total visible transverse momentum $p_T({\rm visible})$ satisfy
\begin{eqnarray}
	|\cos{\theta_{\rm thrust}}|<0.95,\quad && \nonumber \\
	 100 <E({\rm visible})<400\ {\rm GeV}, && 20<p_T({\rm visible})<500{\rm GeV}.
\end{eqnarray}
Other event variables which will be used in the Higgs event selection include the total visible mass $M({\rm visible})$, the number of charged tracks  $N({\rm chg})$,
the number of large impact parameter charged tracks $N({\rm imp})$, and
the number of jets $N({\rm jet})$ as determined by the PYCLUS algorithm of PYTHIA with parameters MSTU(46)=1 and PARU(44)=5.

\subsection{$\bf h\ra b\bar{b}$ }
      Decays of  Higgs bosons to $b$ quarks are selected by requiring:
% that there be between 7 and 19 large impact parameter charged tracks, inclusive, and
%that the total visible mass $M({\rm visible})$ satisfy 
\begin{eqnarray}
      6\le N({\rm chg}) \le 19,\quad &&  7\le N({\rm imp}) \le 19, \nonumber \\
      2\le N({\rm jet}) \le 3,\quad  &&M_h-10\ {\rm GeV}<M({\rm visible})<M_h+6\ {\rm GeV},
\end{eqnarray}
where $M_h$ is the Higgs boson mass measured at $\sqrt{s}=350$~GeV.
Histograms of $M({\rm visible})$ are shown in Fig.~\ref{fig-mvis_bb} assuming Higgs boson masses of 120 and 200~GeV.   
%The peaks in the non-Higgs SM background distribution at $M({\rm visible})=80$~GeV and 90~GeV in the left hand plot are due to $\ee\ra e\nu W,\ eeZ,\ \nu \nu Z$.
Most of the non-Higgs SM background in the left-hand plot is due to $\ee\ra e\nu W,\ eeZ,\ \nu \nu Z$, while the non-Higgs background in the right-hand plot
is mostly $\gamma\gamma \ra WW$.
%The spikes in the non-Higgs SM background at $M({\rm visible})=162,174,\ {\rm and}\ 182$~GeV in the 
%right-hand plot of Fig.~\ref{fig-mvis_bb}  are not meaningful and are just an artifact of low Monte Carlo statistics 
%for processes such as $\gamma\gamma \ra WW$.   
%Note that the cut on $M({\rm visible})$ is a viable method for isolating a Higgs boson signal since 
%the Higgs boson mass $M_h$ will have been measured to an accuracy of
%better than 100~MeV in the Higgsstrahlung process at $\sqrt{s}=0.35$~TeV.    
%In total there are 94910 (348) signal events and
%6455  (730) background events passing the $b\bar{b}$ selection criteria for $M_h=120\ (200)$~GeV.   
%The corresponding statistical accuracy for cross-section times branching
The statistical accuracy for cross-section times branching
ratio, $\sigma\cdot B_{bb}$,  is shown in the first row of Table~\ref{tab:higgs-sigma_br}, along with results for $M_h=115,\ 140,\ {\rm and}\ 160$~GeV.    

The Higgs background makes up 1.2\% of the events in the left-hand plot that pass all cuts, and of these
70\% are $c\bar{c}$, 20\%  are $gg$, 5\% are $WW^*$, and 5\% are $ZZ^*$.   The Higgs background is small enough that Higgs branching fraction measurements from
$\sqrt{s}=350$~GeV can be used to account for this background without introducing a significant systematic error.  The non-Higgs background should be calculated with an accuracy
of 1 to 2\% to keep the non-Higgs background systematic error below the statistical error.

\begin{figure}
\begin{center}
\includegraphics[width=7.9cm]{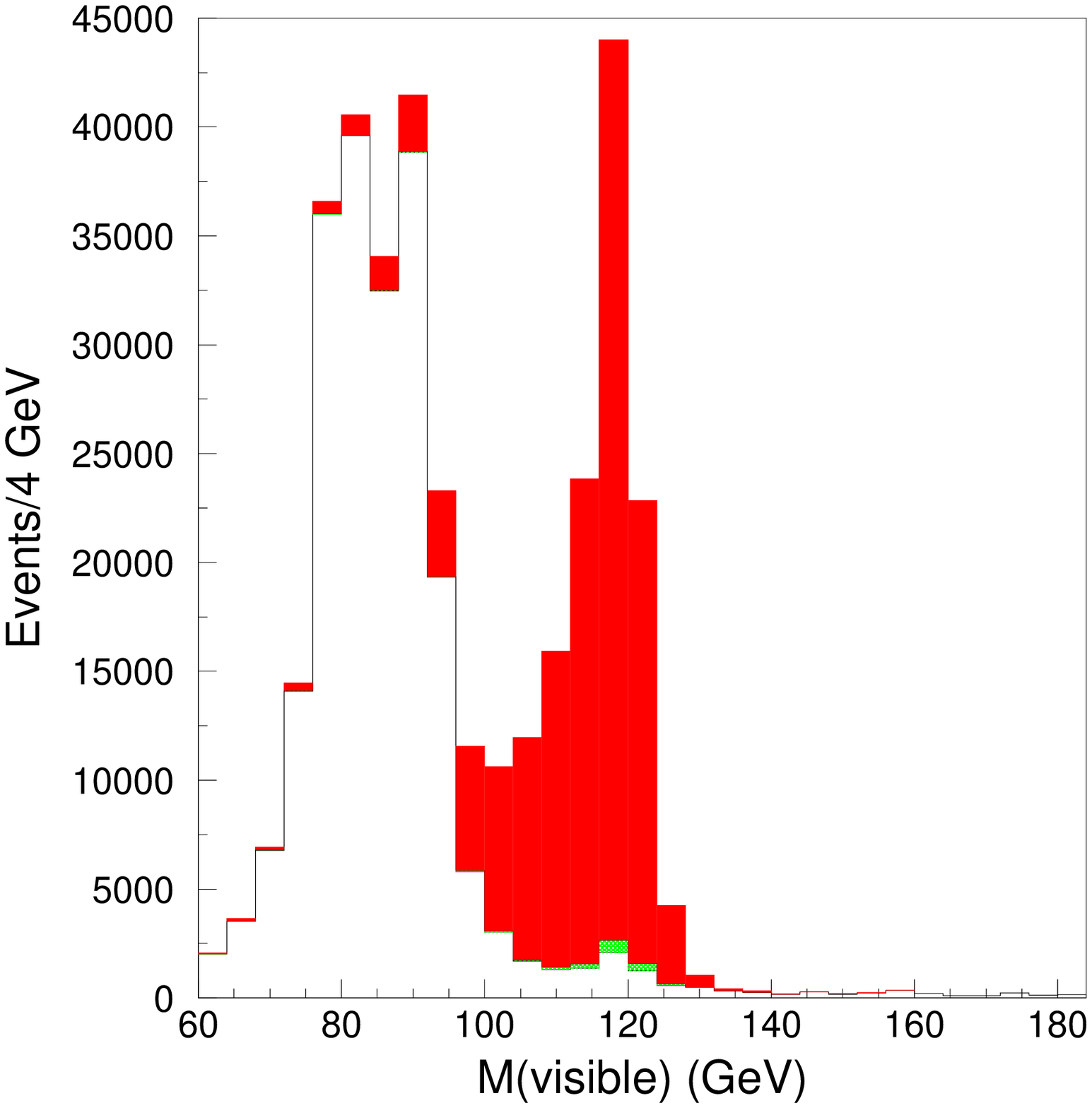} 
\includegraphics[width=7.9cm]{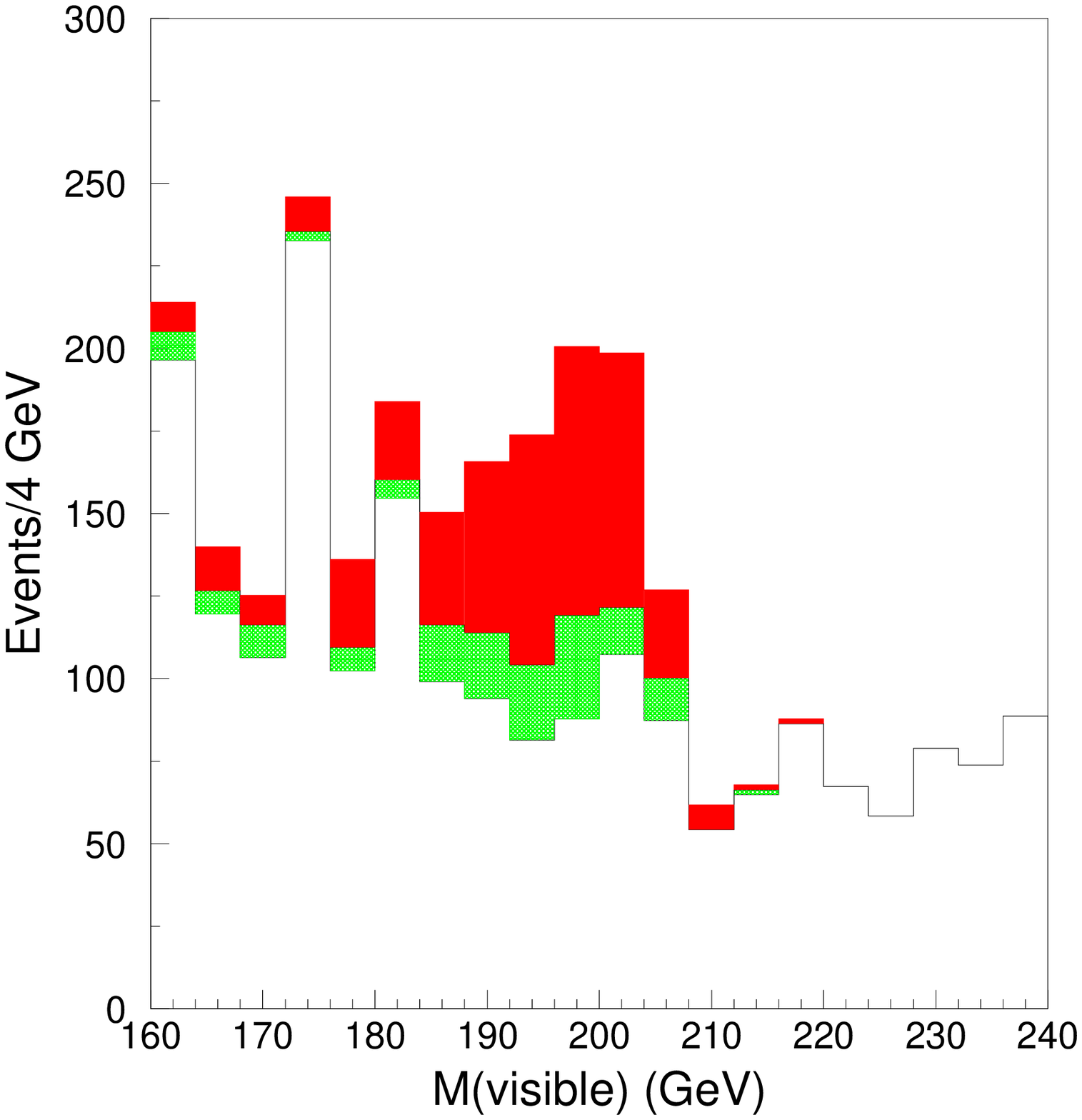} 
\caption{Histograms of $M({\rm visible})$ following $b\bar{b}$ selection cuts for background and signal assuming 
$M_h=120$~GeV (left) and $M_h=200$~GeV (right).  The histograms contain
non-Higgs SM background (white), $h\ra b\bar{b}$ (red) and other Higgs decays (green). 
%The luminosity and center-of-mass energy are 
%1000 fb$^{-1}$ at $\sqrt{s}=1000$~GeV.
}
\label{fig-mvis_bb}
\end{center}
\end{figure}

\subsection{$\bf h\ra \gamma\gamma$ }

      Decays of  Higgs bosons to photon pairs are selected by requiring:
\begin{eqnarray}
      N({\rm chg})=0,\quad &&  N({\rm imp})=0, \nonumber \\
      N({\rm jet})=2,\quad  &&M_h-2\ {\rm GeV}<M({\rm visible})<M_h+1\ {\rm GeV}.
\end{eqnarray}
Histograms of $M({\rm visible})$ are shown in Fig.~\ref{fig-mvis_aa} assuming Higgs boson masses of 120 and 160~GeV.   
The SM background is almost entirely $\ee\ra \nu\nu\gamma\gamma$.
%The statistical accuracy for $\sigma\cdot B_{\gamma\gamma}$  is shown in the fourth row of Table~\ref{tab:higgs-sigma_br}.

\begin{figure}
\begin{center}
\includegraphics[width=7.9cm]{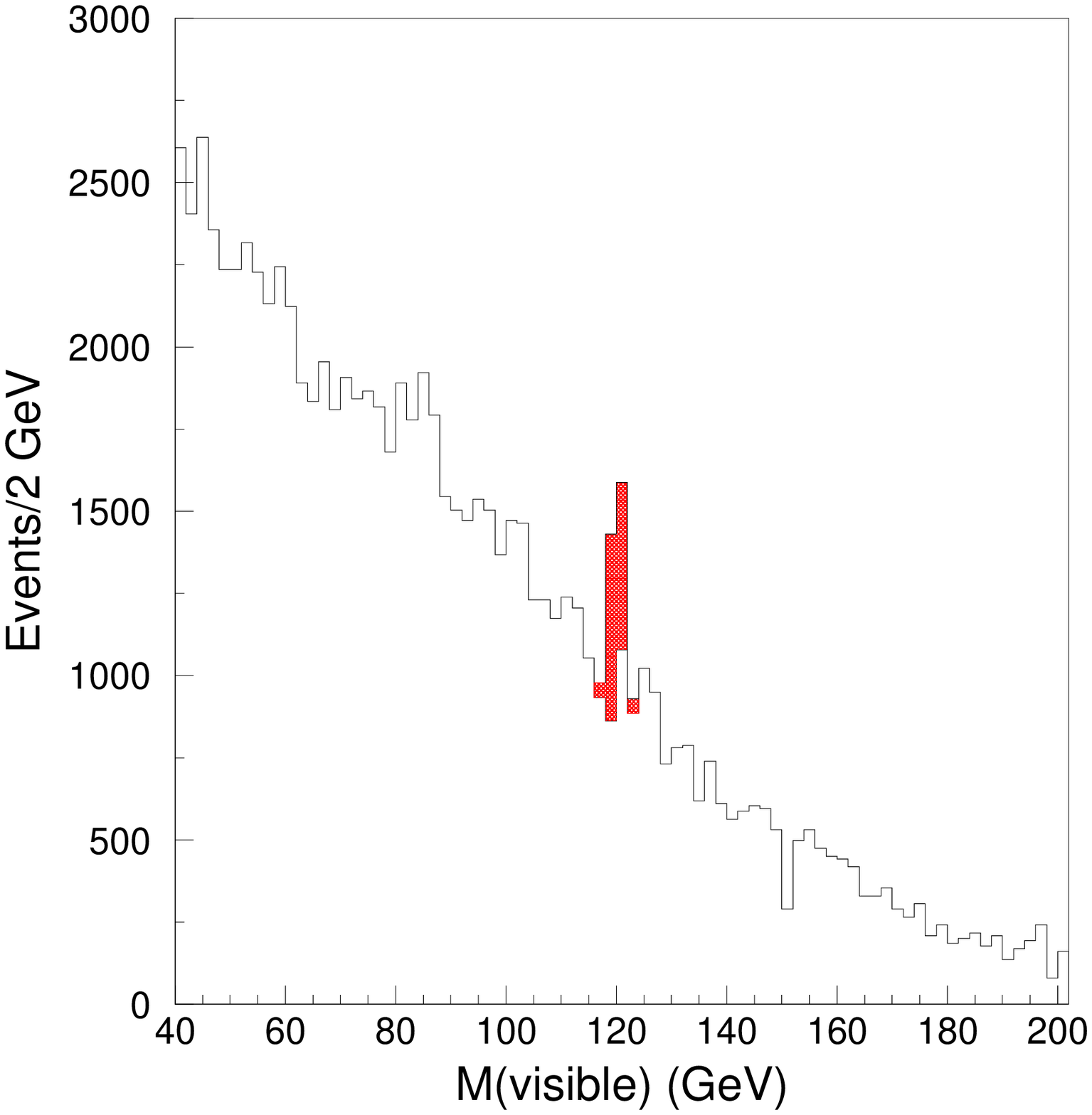} 
\includegraphics[width=7.9cm]{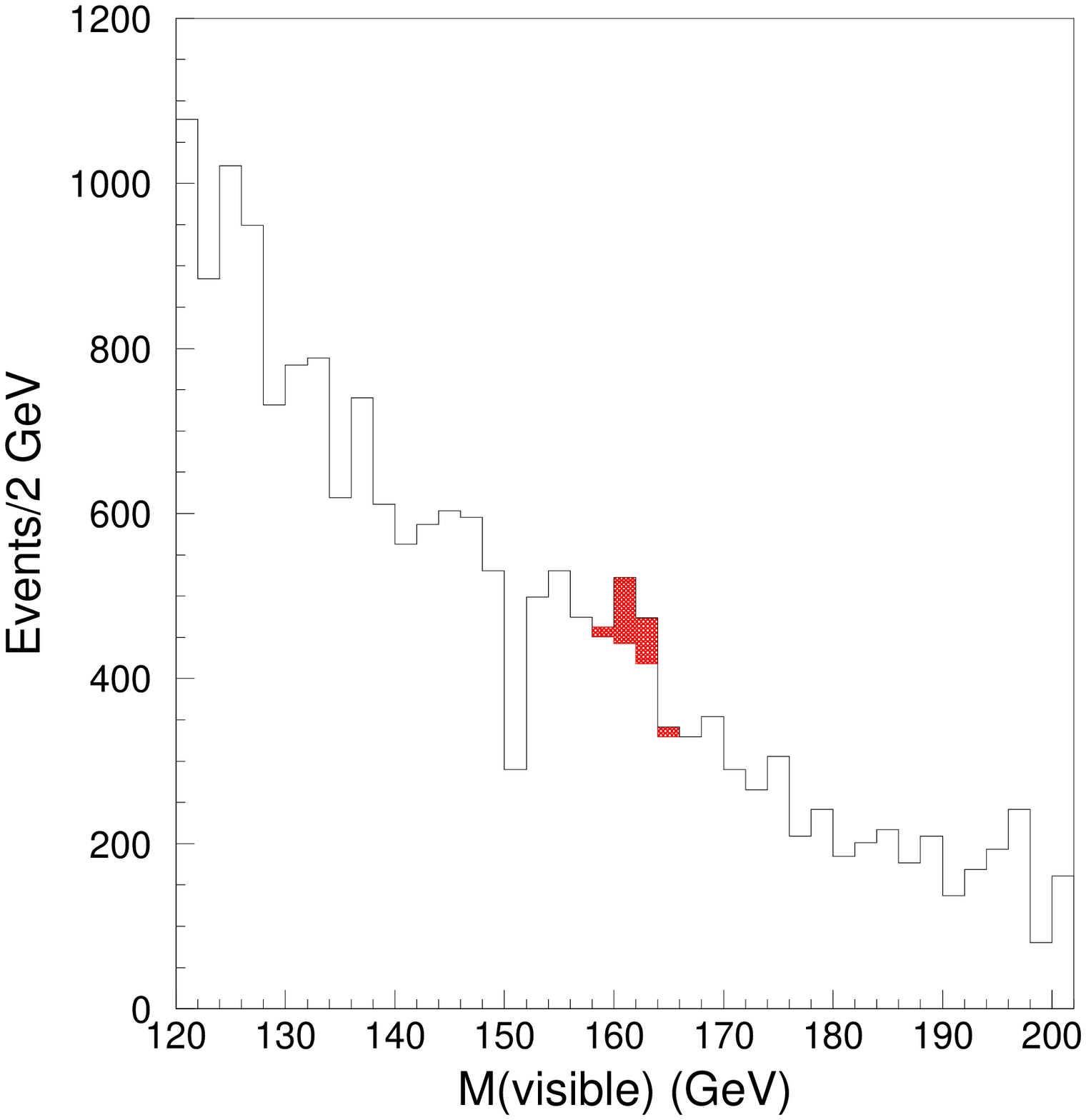} 
\caption{Histograms of $M({\rm visible})$ following $\gamma\gamma$ selection cuts for background and signal assuming 
$M_h=120$~GeV (left) and $M_h=160$~GeV (right).  The histograms contain
non-Higgs SM background (white) and $h\ra \gamma\gamma$ (red). 
%The luminosity and center-of-mass energy are 
%1000 fb$^{-1}$ at $\sqrt{s}=1000$~GeV.
}
\label{fig-mvis_aa}
\end{center}
\end{figure}

\subsection{$\bf h\ra WW,\ \ gg$}

      Decays of  Higgs bosons to $WW$ or $WW^*$ are selected by requiring:
\begin{eqnarray}
      16\le N({\rm chg}) \le 44,\quad &&  N({\rm imp}) \le 6, \nonumber \\
      4\le N({\rm jet}) \le 5,\quad  &&M_h-10\ {\rm GeV}<M({\rm visible})<M_h+6\ {\rm GeV}.
\end{eqnarray}
The histogram of $M({\rm visible})$ following the $WW$ cuts is shown in the left-hand side of Fig.~\ref{fig-mvis_ww} for a Higgs boson mass of 120~GeV.   
The non-Higgs SM background is mostly $\ee\ra e\nu W$.  There is also a substantial Higgs boson background  consisting of $h\ra gg$ (63\%), $h\ra bb$ (14\%),
$h\ra cc$ (12\%) and $h\ra ZZ^*$(12\%).
In order to isolate the $h\ra WW$ signal from the other Higgs decay modes, events are forced into 4 jets and a neural net analysis is performed using 
the 4-momentum dot products between pairs of jets and the event variables $E({\rm visible})$, $p_T({\rm visible})$,
$N({\rm chg})$,  $N({\rm imp})$, and $N({\rm jet})$.  The results of this neural net analysis are
shown in the right-hand side of Fig.~\ref{fig-mvis_ww}.
%The background from $h\ra bb$
%can be estimated using the measured value of $\sigma\cdot B_{bb}$, but the background from $h\ra gg$ can only be dealt with by fitting simultaneously for $\sigma\cdot B_{WW}$
%and $\sigma\cdot B_{gg}$.

%The  $h\ra WW$ neural net analysis reduces the background from $h\ra gg$ and $h\ra bb$, but does not eliminate it.  
The background from $h\ra b\bar{b},\ c\bar{c},\ ZZ^*$ is small enough that Higgs branching fraction results from $\sqrt{s}=350$~GeV can be used to account for these decays
without introducing significant systematic errors.
However, the contribution from $h\ra gg$ can only be dealt with by measuring
 $\sigma\cdot B_{WW}$ and $\sigma\cdot B_{gg}$ simultaneously.   To that end the decay  $h\ra gg$ is selected by requiring:
\begin{eqnarray}
      11\le N({\rm chg}) \le 49,\quad &&  N({\rm imp}) \le 6, \nonumber \\
      2\le N({\rm jet}) \le 4,\quad  &&M_h-10\ {\rm GeV}<M({\rm visible})<M_h+6\ {\rm GeV}.
\end{eqnarray}
An  $h\ra gg$ neural net analysis is performed with a set of variables identical to that used in the $h\ra WW$ neural net analysis.  The results of the simultaneous fit
of $\sigma\cdot B_{WW}$ and $\sigma\cdot B_{gg}$ for $M_h=115,120,140,160$~GeV are shown in rows 2 and 3 of Table~\ref{tab:higgs-sigma_br}.  For $M_h=200$~GeV
the $h\ra gg$ decay mode is negligible and so
a simultaneous fit of $\sigma\cdot B_{WW}$ and $\sigma\cdot B_{ZZ}$ is made where the $ZZ$ selection cuts are the same as the $WW$ selection cuts and
an $h\ra ZZ$ neural net analysis is performed to separate $h\ra ZZ$ from $h\ra WW$.

\begin{figure}
\begin{center}
\includegraphics[width=7.9cm]{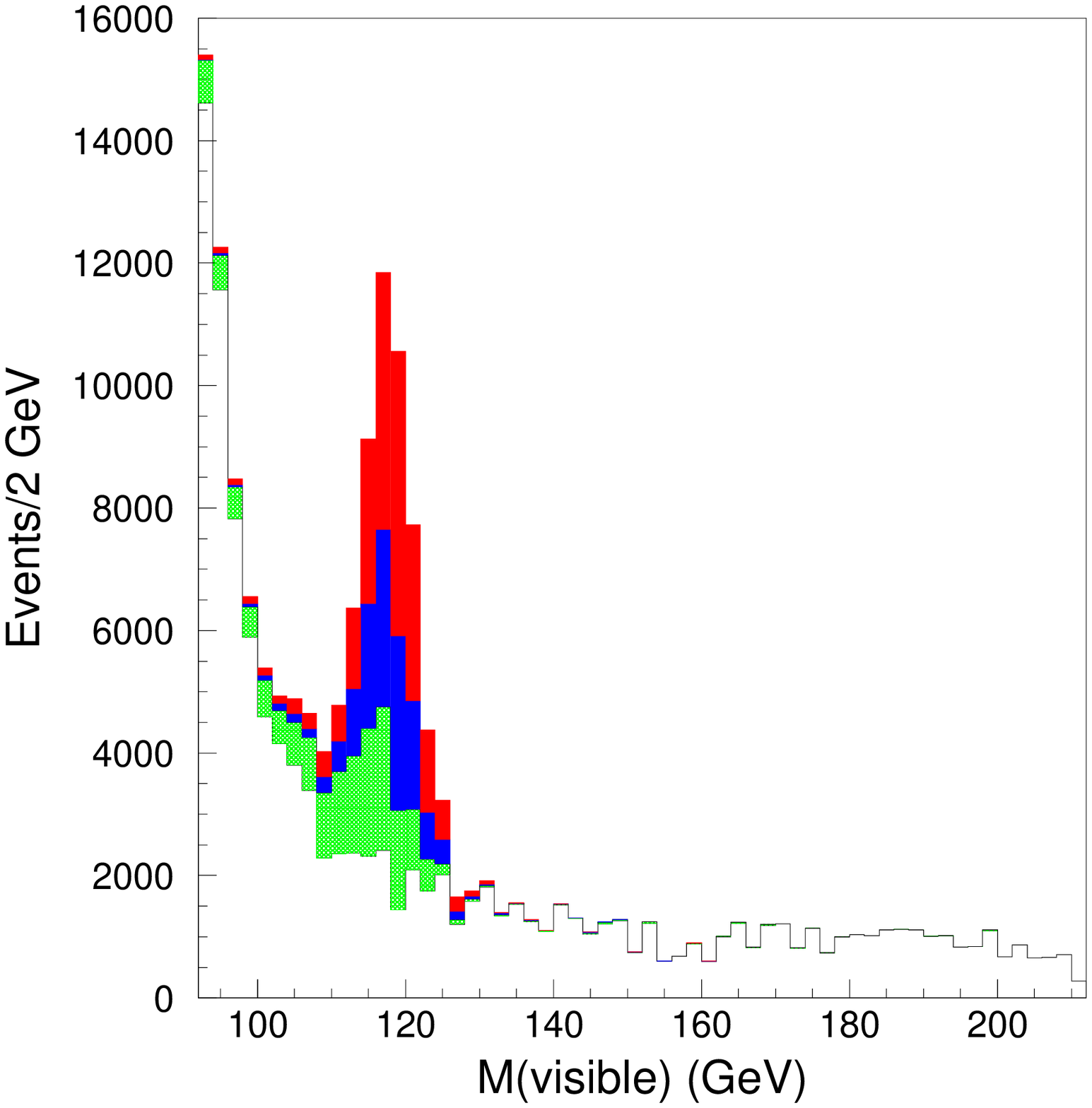} 
\includegraphics[width=7.9cm]{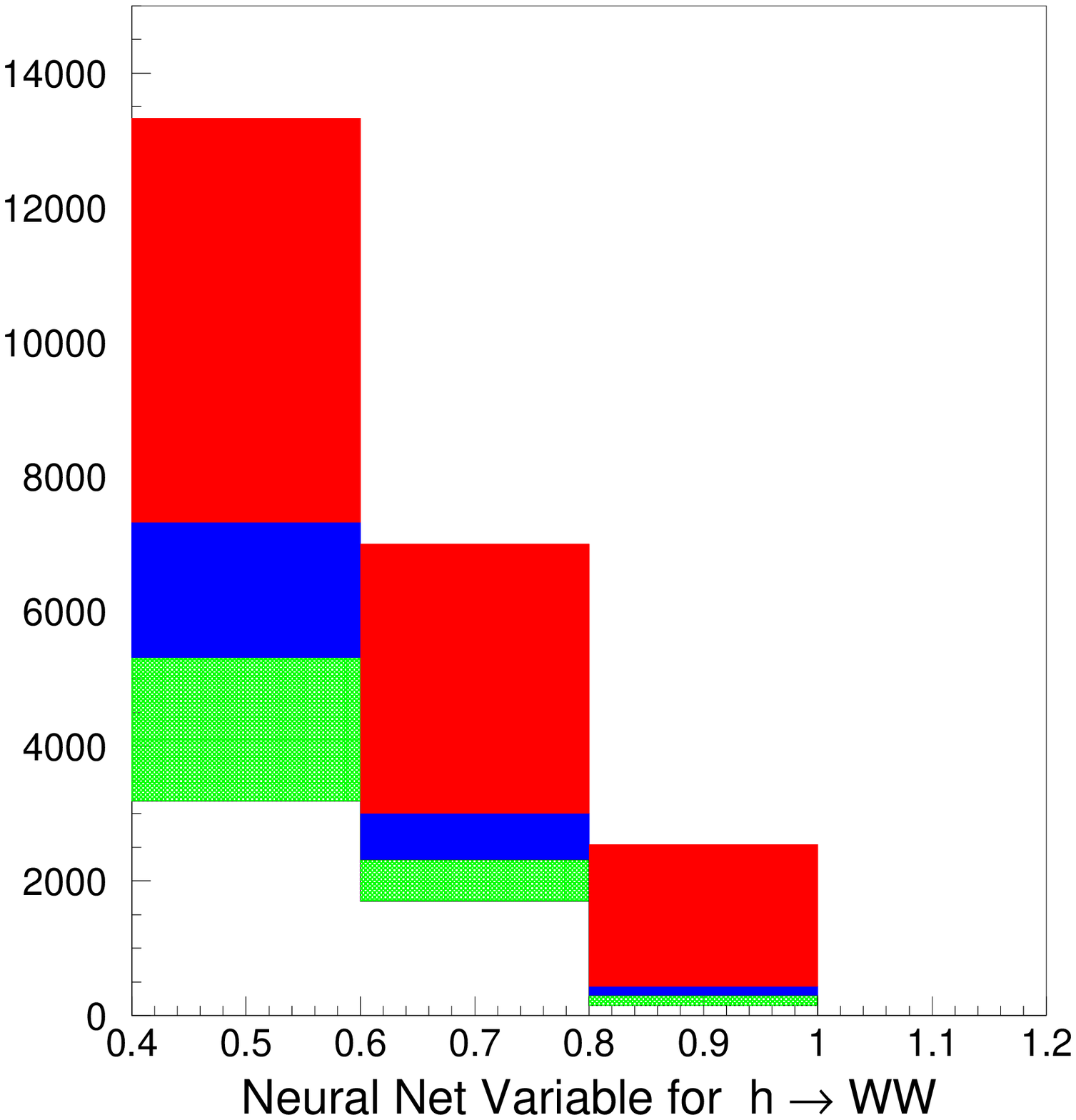} 
\caption{Histograms of $M({\rm visible})$ (left) and the $h\ra WW$ neural net variable (right) following $WW$ selection cuts assuming 
$M_h=120$~GeV.  The histograms contain
non-Higgs SM background (white), $h\ra W\bar{W}$ (red), $h\ra gg$ (blue), and $h\ra b\bar{b},\ c\bar{c},\ ZZ^*$ (green). 
%The luminosity and center-of-mass energy are 
%1000 fb$^{-1}$ at $\sqrt{s}=1000$~GeV.
}
\label{fig-mvis_ww}
\end{center}
\end{figure}

\begin{table}[]
\begin{center}
\caption{Statistical accuracies for the measurement of $\sigma\cdot B_{xx}$ for different Higgs decay 
modes $h\ra xx$ 
%assuming 1000 fb$^{-1}$ 
%is collected 
at $\sqrt{s}=1000$~GeV.
}
\begin{tabular}{l|rrrrr}
     &                    \multicolumn{5}{c}{Higgs Mass (GeV)} \\
   & 115 & 120 & 140 & 160 & 200 \\
\hline
& & & & & \\
 $\Delta (\sigma\cdot B_{bb})/ (\sigma\cdot B_{bb})$ & $\pm 0.003$ &  $\pm 0.004$  & $\pm 0.005$    & $\pm 0.018$  & $\pm 0.090$   \\
 $\Delta (\sigma\cdot B_{WW})/ (\sigma\cdot B_{WW})$ & $\pm 0.021$ &  $\pm 0.013$  & $\pm 0.005$    & $\pm 0.004$  & $\pm 0.005$   \\
 $\Delta (\sigma\cdot B_{gg})/ (\sigma\cdot B_{gg})$ & $\pm 0.014$ &  $\pm 0.015$  & $\pm 0.025$    & $\pm 0.145$  &               \\
 $\Delta (\sigma\cdot B_{\gamma\gamma})/ (\sigma\cdot B_{\gamma\gamma})$ & $\pm 0.053$ &  $\pm 0.051$  & $\pm 0.059$    & $\pm 0.237$  &                \\
$\Delta (\sigma\cdot B_{ZZ})/ (\sigma\cdot B_{ZZ})$ &             &               &                &              & $\pm 0.013$    \\

\hline
\end{tabular}
\label{tab:higgs-sigma_br}
\end{center}
\end{table}

\section{Measurement of Higgs Branching Fractions and the total Higgs Decay
Width}

The measurements of $\sigma\cdot B_{xx}$ in Table~\ref{tab:higgs-sigma_br} can be converted into model independent measurements of Higgs branching fractions and the total Higgs decay width if they
are combined with measurements of the branching fractions $B_{bb}^*$ and $B_{WW}^*$ from $\sqrt{s}=350$~GeV:
\begin{eqnarray}
	B_{xx} &=& (\sigma\cdot B_{xx})(\sigma\cdot B_{WW})^{-1}B_{WW}^*=(\sigma\cdot B_{xx})(\sigma\cdot B_{bb})^{-1}B_{bb}^*             \nonumber \\
	  \Gamma_{tot} & \propto & (\sigma\cdot B_{bb})(B_{bb}^*)^{-1}(B_{WW}^*)^{-1}=(\sigma\cdot B_{bb})^2(\sigma\cdot B_{WW})^{-1}(B_{bb}^*)^{-2}.
\label{eqn-sig_br_to_br}
\end{eqnarray}
The assumed values for the errors on $B_{bb}^*$ and $B_{WW}^*$  are shown in Table~\ref{tab:higgs-assumed_low_energy}.  The errors are taken from the TESLA TDR\cite{Aguilar-Saavedra:2001rg}
when the branching fractions are small.  For large branching fractions, however, it is better to use the direct method\cite{Brient:2002qz} for measuring branching fractions because binomial statistics 
reduce the error by a factor of $\sqrt{1-B_{xx}}$.  
%Direct method errors are therefore used for  $B_{bb}^*$  when $M_h=115,120$~GeV, and for  $B_{WW}^*$ when
%$M_h=160,200$~GeV. 
%(Ref.xxx only has results for $M_h=120$~GeV, so that the direct method errors are rough estimates for $M_h=115,160,200$~GeV.)

Utilizing the relations in Eq.(6) a least squares fit is performed to obtain measurement errors for $B_{bb}$, $B_{WW}$, $B_{gg}$, $B_{\gamma\gamma}$,
and $\Gamma_{tot}$ at a fixed value of $M_h$.   The results are summarized in Table~\ref{tab:higgs-final_errors}.  Compared to branching fraction measurements at 
 $\sqrt{s}=350$~GeV\cite{Aguilar-Saavedra:2001rg} the results of Table~\ref{tab:higgs-final_errors} provide a significant improvement for Higgs decay modes with small branching fractions,
such as $B_{bb}$ for $160<M_h<200$~GeV,  $B_{WW}$ for $115<M_h<140$~GeV and $B_{gg}$ and  $B_{\gamma\gamma}$ for all Higgs masses.  
%The error on the total width is 
%improved by a factor of two for $115<M_h<140$~GeV.

\begin{table}[]
\begin{center}
\caption{Assumed branching fraction errors  for Higgs boson decays to $\ bb\ $ and $WW$  from measuements made at
$\sqrt{s}=350$~GeV  with 500 fb$^{-1}$ luminosity.  
%These numbers are either taken directly from the literature or are extrapolations of numbers in the literature; see text for details.
}
\begin{tabular}{l|rrrrr}
     &                    \multicolumn{5}{c}{Higgs Mass (GeV)} \\
   & 115 & 120 & 140 & 160 & 200 \\
\hline
& & & & & \\
 $\Delta B_{bb}^*/B_{bb}^*$ & $\pm 0.015$ &  $\pm 0.017$  & $\pm 0.026$    & $\pm 0.065$  & $\pm 0.340$   \\
 $\Delta B_{WW}^*/B_{WW}^*$ & $\pm 0.061$ &  $\pm 0.051$  & $\pm 0.025$    & $\pm 0.010$  & $\pm 0.025$   \\

\hline
\end{tabular}
\label{tab:higgs-assumed_low_energy}
\end{center}
\end{table}

\begin{table}[]
\begin{center}
\caption{Relative accuracies for the measurement of Higgs branching fractions and the Higgs boson total decay width 
obtained by combining results from Tables \ref{tab:higgs-sigma_br} and \ref{tab:higgs-assumed_low_energy}.
%assuming 
%500 fb$^{-1}$ is collected at $\sqrt{s}=350$~GeV and  and 1000 fb$^{-1}$ is collected at $\sqrt{s}=1000$~GeV
}
\begin{tabular}{l|rrrrr}
     &                    \multicolumn{5}{c}{Higgs Mass (GeV)} \\
   & 115 & 120 & 140 & 160 & 200 \\
\hline
& & & & & \\
 $\Delta B_{bb}/B_{bb}$ & $\pm 0.015$ &  $\pm 0.016$  & $\pm 0.018$    & $\pm 0.020$  & $\pm 0.090$   \\
 $\Delta B_{WW}/B_{WW}$ & $\pm 0.024$ &  $\pm 0.020$  & $\pm 0.018$    & $\pm 0.010$  & $\pm 0.025$   \\
 $\Delta B_{gg}/B_{gg}$ & $\pm 0.021$ &  $\pm 0.023$  & $\pm 0.035$    & $\pm 0.146$  &               \\
 $\Delta B_{\gamma\gamma}/B_{\gamma\gamma}$ & $\pm 0.055$ &  $\pm 0.054$  & $\pm 0.062$    & $\pm 0.237$  &               \\
 $\Delta \Gamma_{tot}/\Gamma_{tot}$ & $\pm 0.035$ &  $\pm 0.034$  & $\pm 0.036$    & $\pm 0.020$  & $\pm 0.050$   \\

\hline
\end{tabular}
\label{tab:higgs-final_errors}
\end{center}
\end{table}

\section{Conclusion}

The couplings of Higgs bosons in the mass range $115<M_h<200$~GeV can continue to be measured as the
energy of an $\ee$ linear collider is upgraded to $\sqrt{s}=1000$~GeV.  The Higgs event rate is so large that
some of the rarer decay modes that were inaccessible at $\sqrt{s}=350$~GeV can be probed at $\sqrt{s}=1000$~GeV, 
such as $h\ra b\bar{b}$ for  $M_h=200$~GeV, and  $h\ra gg,\ \gamma\gamma$  for  $M_h=140$~GeV.  The Higgs physics results
from $\sqrt{s}=1000$~GeV will help provide a more complete picture of the Higgs boson profile.
%than the one obtained 
%from $\sqrt{s}=350-500$~GeV alone.

%The measurement of the Higgs cross-section times branching fractions at $\sqrt{s}=1000$~GeV will provide  information about Higgs decay modes with small branching fractions
%that will not be accessible at $\sqrt{s}=350$~GeV.  

%The smallest relative error for a branching fraction was 1\%. Detector and theory systematic errors have not been discussed.  The results presented here
%can be used to help set targets for detector design and radiative correction calculations.

}